%% file: LH_NewPhysics_Report.tex
\documentclass[hyperref,12pt]{cernyrep}
\usepackage{ifpdf}
\usepackage{url}
\usepackage{subfigure}
\usepackage{rotating}
\usepackage{units}
\usepackage[english]{babel}
\usepackage{feynmp,xcolor}
\usepackage{graphicx}
\usepackage{cite}
\usepackage{color}
\usepackage[utf8]{inputenc}
\usepackage{enumitem}
\usepackage[export]{adjustbox}
\usepackage{xspace}
\usepackage{booktabs}
\usepackage[compat=1.1.0]{tikz-feynman}
\usepackage{cleveref}
\usepackage{listings}

\hypersetup{
  colorlinks,
  citecolor=blue,
  linkcolor=red,
  urlcolor=blue}

\usepackage{tikz}
\usetikzlibrary{trees}
\usetikzlibrary{decorations.pathmorphing}
\usetikzlibrary{decorations.markings}

  \definecolor{jblue}  {RGB}{20,50,100}
  \definecolor{npurple}  {RGB} {153, 51, 204}
  \definecolor{wred}   {RGB}{217,0,56}
  \definecolor{white}   {RGB}{255,255,255}
  
  \definecolor{korange}   {RGB}{235, 80,  43}
  \definecolor{korange2}   {RGB}{245, 100,  63}
  \definecolor{kyelloworange}   {RGB}{255, 210,  110}
  \definecolor{kyelloworange2}   {RGB}{240, 170,  90}
  \definecolor{kred}   {RGB}{204,  102, 153}
  \definecolor{kpurple}   {RGB}{153,  61, 190}
  \definecolor{kpurplelight}   {RGB}{213,  161, 230}

	\tikzset{
	  photon/.style={decorate, decoration={snake}, draw=npurple,very thick},
	  boson/.style={decorate, decoration={snake}, draw=npurple,very thick},
	  electron/.style={draw=jblue,very thick, postaction={decorate},
	           decoration={markings,mark=at position .55 with {\arrow[draw=jblue]{>}}}
	  },
	  electron2/.style={draw=jblue,very thick, postaction={decorate},
	           decoration={markings,mark=at position .55 with {\arrow[draw=jblue]{<}}}
	  },
	  fermion/.style={draw=jblue,very thick, postaction={decorate},
	            decoration={markings,mark=at position .55 with {\arrow[draw=jblue]{}}}
	  },
	  gluon/.style={decorate, draw=korange,very thick, 
	    decoration={coil,amplitude=4pt, segment length=6pt}},
	  higgs/.style={draw=wred,very thick, postaction={decorate},
	           decoration={markings,mark=at position .55 with {\arrow[draw=wred]{>}}}
	  },
	  graviton/.style={draw=wred,very thick, postaction={decorate},
	           decoration={snake}
	  },
	  nothing/.style={draw=white,very thick}
	}

\input Commands.tex

\begin{document}

\setcounter{tocdepth}{0}
\thispagestyle{empty}

\vspace{1cm}

\begin{center}
{\Large {\bf LES HOUCHES 2019: PHYSICS AT TEV COLLIDERS \\[4mm]}}
{\Large {\bf NEW PHYSICS WORKING GROUP REPORT}}
\end{center}

\vspace{0.1cm}
\input all_authors.tex

 \vspace{1cm}
\begin{center}
{\large {\bf Abstract}}\\[.2cm]
\end{center}
This report presents the activities of the `New Physics' working group for the `Physics at
TeV Colliders' workshop (Les Houches, France, 10--28 June, 2019).  These activities include studies
of direct searches for new physics, approaches to exploit published data to constrain new physics,
as well as the development of tools to further facilitate these investigations.  Benefits of machine
learning for both the search for new physics and the interpretation of these searches are also presented.
\vspace{1cm}
\begin{center}
{\bf Acknowledgements}\\[.2cm]
\end{center}
We would like to heartily thank all funding bodies, the organisers
(N.~Berger, F.~Boudjema, C.~Delaunay, M.~Delmastro, B.~Fuks, S.~Gascon, M.-H.~Genest, P.~Gras,
J.P.~Guillet, B.~Herrmann, S.~Kraml, N.~Makovec, G.~Moreau, E.~Re), the staff and all participants of the Les
Houches workshop for providing a stimulating and lively atmosphere in which to
work.

\newpage

\vspace{1cm}

\thispagestyle{empty}
\setcounter{page}{2}


\newpage

\tableofcontentscern

\newpage

\input intro.tex

\addtocontents{toc}{\protect\contentsline{part}{\protect\numberline{} \hspace{-2cm}Introduction}{6}{}}
\AddToContent{G.~Brooijmans, A.~Buckley, S.~Caron, A.~Falkowski, B.~Fuks, A.~Gilbert, W.~J.~Murray, M.~Nardecchia, J.~M.~No, R.~Torre, T.~You, G.~Zevi Della Porta}

\setcounter{figure}{0}
\setcounter{table}{0}
\setcounter{section}{0}
\setcounter{equation}{0}
\setcounter{footnote}{0}
\clearpage



\superpart{ New physics }

\input lfu/lfu.main.tex

\AddToContent{T.~Berger-Hryn'ova,    D.~Buttazzo, J.~Dutta,    D.~Marzocca, M.~Son}

\input dilepton/dilepton.main.tex
\AddToContent{J.~Bernigaud, A.~Bharucha,  D.~Buttazzo,  J.~Dutta,  D.~Marzocca, P.~Pani,  G.~Polesello,  M.~Son,  E.~Venturini,  N.~Vignaroli}
\renewcommand{\thesection}{\arabic{section}}

\input singlelq/singlelq.main.tex

\AddToContent{D.~Buttazzo, D.~Marzocca,  M.~Nardecchia, P.~Pani, G.~Polesello}

\input rdm/RDM.main.tex

\AddToContent{G.~B\'elanger, J.~Bernigaud, A.~Bharucha, B.~Fuks, A.~Goudelis, J.~Heisig, A.~Jueid, A.~Lessa, D.~Marzocca, M.~Nardecchia, G.~Polesello, P.~Pani, A.~Pukhov, D.~Sengupta and J.~Zurita}

\input vqlqvl/vq_lq_vl.main.tex

\AddToContent{J.~Butterworth, G.~Cacciapaglia, L.~Corpe, T.~Flacke, B.~Fuks, L.~Panizzi, W.~Porod, D.~Yallup}

\input filc/FILC.main.tex

\AddToContent{ G.~B\'elanger, A.~Bharucha,  N.~Desai,   B.~Fuks,  A.~Goudelis,  J.~Heisig,  A.~Lessa,  A.~Mjallal,  S.~Sekmen, D.~Sengupta, J. Zurita}

\input llp-dm/LLP_DM.main.tex

\AddToContent{B.~Fuks, J.~Heisig, A.~Lessa, J.\,M.~No, S.~Sekmen, D.~Sengupta, J.~Zurita}

\superpart{ The Higgs Boson }

\input gwc/GWC.main.tex
\AddToContent{J.~M.~Butterworth, K.~Lane, D.~Sperka}
\renewcommand{\thesection}{\arabic{section}}

\input 2hdm/2hdma.main.tex

\AddToContent{J.~M.~Butterworth, M.~Habedank,  P.~Pani}

\input HiggsZa/HiggsZa.main.tex

\AddToContent{A. Bharucha, J. M. Butterworth, N. Desai, S. Gascon-Shotkin, S. Jain, A. Lesauvage, 
G. Moreau, S. Mutzel, J. M. No, J. Quevillon, C. Smith, K. Tobioka, N. Vignaroli, S. L. Williamson, J. Zurita}
\renewcommand{\thesection}{\arabic{section}}

\input vbf/VBF.main.tex

\AddToContent{S.~Fichet, S.~Gascon-Shotkin, A.~Lesauvage, G.~Moreau}

\input wwtj/wwtj.main.tex

\AddToContent{A. Falkowski, S. Ganguly, P. Gras, J. M. No, K. Tobioka, E. Venturini, N. Vignaroli, T. You}

\input ttHZ/tthz.main.tex

\AddToContent{S.~Banerjee, R.~S.~Gupta, S.~Jain,  E.~Venturini}

\superpart{ Tools and Methods}

\input higgseftobs/higgseftobs.main.tex

\AddToContent{A.~Gilbert, J.~Langford, N.~Wardle}
\renewcommand{\thesection}{\arabic{section}}

\input likelihood/likelihoodinfo.main.tex

\AddToContent{G.~Alguero, J.~Y.~Araz, B.~Fuks, S.~Kraml, W.~Waltenberger}

\input anacor/BSM_AnalysisCorrelations.main.tex
\AddToContent{A.~Buckley, B.~Fuks, H.~Reyes-Gonz\'alez, W.~Waltenberger, S.~L.~Williamson}

\input adl/LH19ADLoverlap.main.tex

\AddToContent{H.~B.~Prosper, S.~Sekmen, W.~Waltenberger}

\input{contur/contur-update.main.tex}

\AddToContent{A.~Buckley, J.~M.~Butterworth, L.~Corpe, G.~Watt, D.~Yallup}

\input recastcomp/recastcomp.main.tex

\AddToContent{J.~Y.~Araz, A.~Buckley, N.~Desai, B.~Fuks,
  T.~Gonzalo, P.~Gras, A.~Kvellestad, A.~Raklev, R.~Ruiz de Austri, S.~Sekmen}

\input searchmeasurement/SearchMeasurementComplementarity.main.tex
\AddToContent{T.~Berger-Hryn'ova, A.~Buckley, J.~M. Butterworth, L.~Corpe,
  D.~Yallup, M.~van Beekveld, B.~Fuks, T.~Gonzalo, A.~Kvellestad}

\input 4top/4top.main.tex

\AddToContent{D.~Kar, L.~Corpe}

\superpart{Machine Learning}

\input ml-1/PhenoAI.main.tex
\AddToContent{S.~Caron, A.~Coccaro, S.~Ganguly, S.~Kraml, A.~Lessa, S.~Otten, R.~Ruiz, H.~Reyes-Gonz\'{a}lez, R.~Ruiz de Austri, B.~Stienen, R.~Torre}

\input darkml/darkmachines.main.tex
\AddToContent{M.~van Beekveld, S.~Caron, A.~De Simone, A.~Farbin,  L.~Hendriks, A.~Jueid,  A.~Leinweber,  J.~Mamuzic,  E.~Mer\'enyi,  A.~Morandini,  C.~Nellist,  S.~Otten,  M.~Pierini,  R.~Ruiz de Austri, S.~Sekmen,  J.~Schouwenberg, R.~Vilalta, M.~White}

\clearpage


\bibliography{LH_NewPhysics_Biblio}

\end{document}

%% file: Commands.tex
\addto\captionsenglish{}

\makeatletter
\def\myaddcontentsline@toc#1#2#3{%
   \addtocontents{#1}{\protect\thispagestyle{empty}}%
   \addtocontents{#1}{\protect \vspace{-.5cm}}
   \addtocontents{#1}{\protect\contentsline{#2}{#3}{}{}}
   \addtocontents{#1}{\protect \vspace{-.3cm}}}
\def\myaddcontentsline#1#2#3{%
  \expandafter\@ifundefined{addcontentsline@#1}%
  {\addtocontents{#1}{\protect \vspace{-.5cm}}
  \addtocontents{#1}{\protect\contentsline{#2}{#3}{}{}}
  \addtocontents{#1}{\protect \vspace{-.3cm}}}
  {\csname addcontentsline@#1\endcsname{#1}{#2}{#3}}{}}
\makeatother

\makeatletter
\def\mysuperaddcontentsline@toc#1#2#3{%
   \addtocontents{#1}{\protect\thispagestyle{empty}}%
   \addtocontents{#1}{\protect \vspace{.1cm}}
   \addtocontents{#1}{\protect\contentsline{#2}{#3}{\thepage}{}}
   \addtocontents{#1}{\protect \vspace{-.1cm}}}
\def\mysuperaddcontentsline#1#2#3{%
  \expandafter\@ifundefined{addcontentsline@#1}%
  {\addtocontents{#1}{\protect \vspace{.1cm}}
  \addtocontents{#1}{\protect\contentsline{#2}{#3}{\thepage}{}}
  \addtocontents{#1}{\protect \vspace{-.1cm}}}
  {\csname addcontentsline@#1\endcsname{#1}{#2}{#3}}{}}
\makeatother

\newcommand{\superpart}[1]{\clearpage \thispagestyle{empty}  \vphantom{blabla}\vspace{5cm} \centerline{\huge #1} 
\mysuperaddcontentsline{toc}{part}{\protect\numberline{} \hspace{-2cm}#1}%
\clearpage}
   
\newcommand{\AddToContent}[1]{\myaddcontentsline{toc}{chapter}{\protect  \textit{\hspace{1.5cm} \small#1}}}

\def\etmiss{E_T^{miss}}

\catcode`\@=11
\font\manfnt=manfnt
\def\Watchout{\@ifnextchar [{\W@tchout}{\W@tchout[1]}}
\def\W@tchout[#1]{{\manfnt\@tempcnta#1\relax%
  \@whilenum\@tempcnta>\z@\do{%
    \char"7F\hskip 0.3em\advance\@tempcnta\m@ne}}}
\let\foo\W@tchout
\def\dubious{\@ifnextchar[{\@dubious}{\@dubious[1]}}

\def\@dubious[#1]{%
  \setbox\@tempboxa\hbox{\@W@tchout#1}
  \@tempdima\wd\@tempboxa
  \list{}{\leftmargin\@tempdima}\item[\hbox to 0pt{\hss\@W@tchout#1}]}
\def\@W@tchout#1{\W@tchout[#1]}
\catcode`\@=12

\renewcommand{\ie}[0]{\textit{i.e.}}

\newcommand{\leer}[1]{}

\newcommand{\mt}[1]{\ensuremath{m_{t'}}}
\newcommand{\mb}[1]{\ensuremath{m_{b'}}}
\newcommand{\me}[1]{\ensuremath{m_{\tau'}}}
\newcommand{\mv}[1]{\ensuremath{m_{\nu'}}}

%% file: all_authors.tex
\begin{center}
  \textbf{G.~Brooijmans}$^{1}$,
  \textbf{ A.~Buckley}$^{2}$,
  \textbf{ S.~Caron}$^{3,4}$,
  \textbf{ A.~Falkowski}$^{5}$,\\
  \textbf{ B.~Fuks}$^{6,7}$,
  \textbf{ A.~Gilbert}$^{8}$,
 \textbf{W.~J.~Murray}$^{9,10}$,
  \textbf{ M.~Nardecchia}$^{11}$,\\
  \textbf{ J.~M.~No}$^{12}$,
  \textbf{ R.~Torre}$^{13,14}$,
  \textbf{ T.~You}$^{14,15}$,
  \textbf{ G.~Zevi Della Porta}$^{16}$,
 \textbf{(convenors)}\\
G.~Alguero$^{17}$, 
J.~Y.~Araz$^{18}$, 
S.~Banerjee$^{19}$, 
G.~B\'elanger$^{20}$, 
T.~Berger-Hryn'ova$^{21}$, 
J.~Bernigaud$^{22,23}$,
A.~Bharucha$^{24}$, 
D.~Buttazzo$^{25}$, 
J.~M.~Butterworth$^{26}$, 
G.~Cacciapaglia$^{27}$,
A.~Coccaro$^{13}$, 
L.~Corpe$^{26}$, 
N.~Desai$^{28}$,
A.~De Simone$^{29,30}$,
J.~Dutta$^{31,32}$,
A.~Farbin$^{33}$,
S.~Fichet$^{34}$, 
T.~Flacke$^{35}$,
S.~Ganguly$^{36}$, 
S.~Gascon-Shotkin$^{27}$, 
T.~Gonzalo$^{37}$, 
A.~Goudelis$^{38}$, 
P.~Gras$^{39}$, 
R.~S.~Gupta$^{19}$, 
M.~Habedank$^{40}$, 
J.~Heisig$^{41}$,
L.~Hendriks$^{3}$,
S.~Jain$^{42}$,
A.~Jueid$^{43}$, 
D.~Kar$^{44}$, 
S.~Kraml$^{17}$, 
A.~Kvellestad$^{45,46}$, 
K.~D.~Lane$^{47}$, 
J.~Langford$^{45}$,
A.~Leinweber$^{48}$,
A.~Lesauvage$^{27}$,
A.~Lessa$^{49}$,
J.~Mamuzic$^{50}$,
D.~Marzocca$^{30}$, 
E.~Mer\'enyi$^{51}$,
A.~Mjallal$^{20}$, 
K.~A.~Mohan$^{52}$,
A.~Morandini$^{29,30}$, 
G.~Moreau$^{5}$, 
S.~Mutzel$^{24}$, 
C.~Nellist$^{3,4}$,
S.~Otten$^{3,53}$, 
P.~Pani$^{54}$, 
L.~Panizzi$^{55}$,
M.~Pierini$^{56}$,
G.~Polesello$^{57}$, 
W.~Porod$^{58}$,
H.~B.~Prosper$^{59}$, 
A.~Pukhov$^{60}$, 
J.~Quevillon$^{17}$,  
A.~Raklev$^{46}$, 
H.~Reyes-Gonz\'alez$^{17}$,
R.~Ruiz$^{41}$,
R.~Ruiz de Austri$^{50}$, 
J.~Schouwenberg$^{3,4}$,
S.~Sekmen$^{61}$, 
D.~Sengupta$^{62}$, 
C.~Smith$^{17}$, 
M.~Son$^{63}$, 
D.~Sperka$^{47}$,
B.~Stienen$^{3,4}$,
K.~Tobioka$^{59}$,
M.~van Beekveld$^{3,4}$, 
E.~Venturini$^{64}$, 
N.~Vignaroli$^{25,65}$,
R.~Vilalta$^{66}$,
W.~Waltenberger$^{67}$, 
N.~Wardle$^{45}$,
G.~Watt$^{19}$,
M.~White$^{48}$,
S.~L.~Williamson$^{7,68}$, 
D.~Yallup$^{26}$,
J.~Zurita$^{22,23}$
\end{center}

\begin{itemize}
\vspace*{-.3cm}\item[$^1$] Physics Department, Columbia University, New York, NY 10027, USA
\vspace*{-.3cm}\item[$^{2}$] School of Physics and Astronomy, University of Glasgow, Glasgow, UK
\vspace*{-.3cm}\item[$^{3}$] IMAPP, Radboud University, Nijmegen, the Netherlands
\vspace*{-.3cm}\item[$^{4}$] Nikhef, Amsterdam, the Netherlands
\vspace*{-.3cm}\item[$^{5}$] Laboratoire de Physique des 2 Infinis Ir\`ene Joliot-Curie d'Orsay (IJCLab), B\^at. 100, CNRS / PARIS-SACLAY, F-91898 ORSAY Cedex, France
\vspace*{-.3cm}\item[$^{6}$] Institut Universitaire de France, 103 Boulevard Saint-Michel,  75005 Paris, France
\vspace*{-.3cm}\item[$^{7}$] Laboratoire de Physique Th\'eorique et Hautes Energies (LPTHE), UMR 7589, Sorbonne Universit\'e et CNRS, 4 place Jussieu, 75252 Paris Cedex 05, France
\vspace*{-.3cm}\item[$^{8}$] Northwestern University, Evanston, USA
\vspace*{-.3cm}\item[$^{9}$] Department of Physics, University of Warwick, Coventry, United Kingdom
\vspace*{-.3cm}\item[$^{10}$] Particle Physics Department, Rutherford Appleton Laboratory, Didcot, United Kingdom
\vspace*{-.3cm}\item[$^{11}$] Sapienza - Universita di Roma, Piazzale Aldo Moro 2, 00185, Roma, Italy
\vspace*{-.3cm}\item[$^{12}$] Departamento de Fisica Teorica and Instituto de Fisica Teorica, IFT-UAM/CSIC, Cantoblanco, 28049, Madrid, Spain,
\vspace*{-.3cm}\item[$^{13}$] INFN, Sezione di Genova, Via Dodecaneso 33, I-16146 Genova, Italy
\vspace*{-.3cm}\item[$^{14}$] CERN, Theoretical Physics Department, Geneva, Switzerland
\vspace*{-.3cm}\item[$^{15}$] DAMTP, University of Cambridge, Wilberforce Road, Cambridge, CB3 0WA, UK; Cavendish Laboratory, University of Cambridge, J.J. Thomson Avenue, Cambridge, CB3 0HE, UK
\vspace*{-.3cm}\item[$^{16}$] CERN, Beams Department, Geneva, Switzerland
\vspace*{-.3cm}\item[$^{17}$] Laboratoire de Physique Subatomique et de Cosmologie, Universit\'{e} Grenoble-Alpes, \\ CNRS/IN2P3, 53 Avenue des Martyrs, F-38026 Grenoble, France
\vspace*{-.3cm}\item[$^{18}$] Concordia University, 7141 Sherbrooke St. West, Montr\'eal, QC, Canada H4B 1R6
\vspace*{-.3cm}\item[$^{19}$] Institute for Particle Physics Phenomenology, Department of Physics, Durham University, Durham DH1 3LE, United Kingdom
\vspace*{-.3cm}\item[$^{20}$] LAPTh, Univ. Grenoble Alpes, USMB, CNRS, 9 Chemin de Bellevue, F-74940 Annecy, France
\vspace*{-.3cm}\item[$^{21}$] Univ.  Grenoble Alpes, Univ.  Savoie Mont Blanc, CNRS, IN2P3-LAPP, Annecy, France
\vspace*{-.3cm}\item[$^{22}$] Institute for Nuclear Physics (IKP), KIT Karlsruhe Institute of Technology,  Hermann-von-Helmholtz-Platz 1, D-76344 Eggenstein-Leopoldshafen, Germany
\vspace*{-.3cm}\item[$^{23}$] Institute for Theoretical Particle Physics (TTP), Karlsruhe Institute of Technology,  Engesserstra{\ss}e 7, D-76128 Karlsruhe, Germany
\vspace*{-.3cm}\item[$^{24}$] Aix Marseille Univ, Universit\'e de Toulon, CNRS, CPT, Marseille, France
\vspace*{-.3cm}\item[$^{25}$] INFN Sezione di Pisa, Largo Bruno Pontecorvo 3, 56127 Pisa, Italy
\vspace*{-.3cm}\item[$^{26}$] Department of Physics \& Astronomy, University College London, Gower St., London, WC1E 6BT, UK
\vspace*{-.3cm}\item[$^{27}$] Univ. Lyon, Universit\'e Claude Bernard Lyon 1, CNRS/IN2P3, UMR5822 IPNL F-69622 Villeurbanne, France
\vspace*{-.3cm}\item[$^{28}$] Department of Theoretical Physics, Tata Institute of Fundamental Research, Mumbai 400005, India
\vspace*{-.3cm}\item[$^{29}$] SISSA, Trieste, Italy
\vspace*{-.3cm}\item[$^{30}$] INFN Sezione di Trieste, SISSA, Via Bonomea 265, 34136, Trieste, Italy
\vspace*{-.3cm}\item[$^{31}$] Regional Centre for Accelerator-based Particle Physics, Harish-Chandra Research Institute, HBNI, Chhatnag Road, Jhunsi, Allahabad-211019, India
\vspace*{-.3cm}\item[$^{32}$] II. Institut f\"ur Theoretische Physik, Universit\"at Hamburg, Luruper Chaussee 149, 22761 Hamburg, Germany,
\vspace*{-.3cm}\item[$^{33}$] University of Texas Arlington, USA
\vspace*{-.3cm}\item[$^{34}$] International Institute of Physics, UFRN, Av. Odilon Gomes de Lima 1722 - Capim~Macio - 59078-400 - Natal-RN, Brazil
\vspace*{-.3cm}\item[$^{35}$] Center for Theoretical Physics of the Universe, Institute for Basic Science (IBS), Daejeon, 34126, Korea
\vspace*{-.3cm}\item[$^{36}$] Weizmann Institute of Science, Israel
\vspace*{-.3cm}\item[$^{37}$] School of Physics and Astronomy, Monash University, Melbourne, VIC 3800, Australia
\vspace*{-.3cm}\item[$^{38}$] Laboratoire de Physique de Clermont (UMR 6533), CNRS/IN2P3, Univ.\ Clermont Auvergne, 4 Av.\ Blaise Pascal, F-63178 Aubi\`ere Cedex, France
\vspace*{-.3cm}\item[$^{39}$] CEA Institut de Recherche sur les lois Fondamentales de l'Univers, Universit\'e Paris-Saclay, Gif-sur-Yvette, France
\vspace*{-.3cm}\item[$^{40}$] Department of Physics, HU Berlin, Germany
\vspace*{-.3cm}\item[$^{41}$] Centre for Cosmology, Particle Physics and Phenomenology (CP3), Universit\'e catholique de Louvain, Chemin du Cyclotron 2, B-1348 Louvain-la-Neuve, Belgium
\vspace*{-.3cm}\item[$^{42}$] School of Physics \& Astronomy, University of Minnesota, Minneapolis MN 55455, United States of America
\vspace*{-.3cm}\item[$^{43}$] Department of Physics, Konkuk University, Seoul 05029, Republic of Korea
\vspace*{-.3cm}\item[$^{44}$] School of Physics, University of Witwatersrand, Johannesburg, South Africa
\vspace*{-.3cm}\item[$^{45}$] Department of Physics, Imperial College London, UK
\vspace*{-.3cm}\item[$^{46}$] Department of Physics, University of Oslo, Norway
\vspace*{-.3cm}\item[$^{47}$] Department of Physics, Boston University, Boston, MA 02215, USA
\vspace*{-.3cm}\item[$^{48}$] University of Adelaide, Adelaide, Australia
\vspace*{-.3cm}\item[$^{49}$] Centro de Ci$\hat{e}$ncias Naturais e Humanas, Universidade Federal do ABC, Santo Andr\'e, 09210-580 SP, Brazil
\vspace*{-.3cm}\item[$^{50}$] Instituto de F\'isica Corpuscular, CSIC-Universitat de Val\`encia, Spain
\vspace*{-.3cm}\item[$^{51}$] Rice University, Houston, Texas, USA
\vspace*{-.3cm}\item[$^{52}$] Department of Physics and Astronomy, Michigan State University,  East Lansing, Michigan-48824, USA
\vspace*{-.3cm}\item[$^{53}$] GRAPPA, University of Amsterdam, the Netherlands
\vspace*{-.3cm}\item[$^{54}$] Deutsches Elektronen Synchrotron, DESY, 15738 Zeuthen, Germany
\vspace*{-.3cm}\item[$^{55}$] Department of Physics and Astronomy, Uppsala University, Box 516, SE-751 20 Uppsala, Sweden
\vspace*{-.3cm}\item[$^{56}$] CERN, Experimental Physics Department, Geneva, Switzerland
\vspace*{-.3cm}\item[$^{57}$] INFN, Sezione di Pavia, Via Bassi 6, 27100 Pavia, Italy
\vspace*{-.3cm}\item[$^{58}$] Uni.~W\"urzburg, Campus Hubland Nord, D-97074 W\"urzburg, Germany
\vspace*{-.3cm}\item[$^{59}$] Department of Physics, Florida State University, Tallahassee, Florida 32306, USA
\vspace*{-.3cm}\item[$^{60}$] Skobeltsyn Institute of Nuclear Physics, Moscow State University, Moscow 119992, Russia
\vspace*{-.3cm}\item[$^{61}$] Department of Physics, Kyungpook National University, Daegu, South Korea
\vspace*{-.3cm}\item[$^{62}$] Department of Physics and Astronomy, 9500 Gilman Drive, University of California, San Diego, USA
\vspace*{-.3cm}\item[$^{63}$] Department of Physics, Korea Advanced Institute of Science and Technology, 291 Daehak-ro, Yuseong-gu, Daejeon 34141, Republic of Korea
\vspace*{-.3cm}\item[$^{64}$] Physik-Department, Technische Universit\"{a}t M\"{u}nchen, 85748 Garching, Germany
\vspace*{-.3cm}\item[$^{65}$] Dipartimento di Fisica "E. Fermi", Università di Pisa, Largo Bruno Pontecorvo 3, 56127 Pisa, Italy
\vspace*{-.3cm}\item[$^{66}$] University of Houston, Houston, Texas, USA
\vspace*{-.3cm}\item[$^{67}$] Institut f\"ur Hochenergiephysik, \"Osterreichische Akademie der Wissenschaften, Nikolsdorfer Gasse 18, 1050 Wien, Austria
\vspace*{-.3cm}\item[$^{68}$] Institute for Theoretical Physics (ITP), Karlsruhe Institute of Technology, 76128 Karlsruhe, Germany
\end{itemize}

%% file: intro.tex
\noindent {\Large {\bf Introduction}}
\vspace{.5cm}

{\it G.~Brooijmans, A.~Buckley, S.~Caron, A.~Falkowski, B.~Fuks, A.~Gilbert, W.~J.~Murray, M.~Nardecchia, J.~M.~No, R.~Torre, T.~You, G.~Zevi Della Porta}\\[.4cm]

This document is the report of the New Physics session of the 2019 Les Houches
Workshop `Physics at TeV Colliders'. The workshop brought together theorists
and experimenters who discussed a significant number of novel ideas related to
beyond the Standard Model and Higgs physics.  New computational methods
and techniques, including machine learning, were considered, with the aim of improving the technology
available for theoretical, phenomenological and experimental new physics studies.

A first set of contributions uses the precise measurements of high mass Drell-Yan $e^+e^-$ and $\mu^+\mu^-$ production to constrain new
physics, either using their ratio to investigate signals related to the hints for lepton flavor universality violation observed by LHCb, or
using a SMEFT analysis to put model-independent constraints on all the dimension-six effective operators which can interfere with the
Standard Model contribution.  A second set of contributions examines leptoquarks and vector-like quarks: one evaluates whether 
a leptoquark solution to the $R_{D^{(*)}}$ deviations can be tested at the LHC through final states including the hadronic decay of a
$\tau$ lepton, a heavy quark ($b$ or $t$) and missing transverse momentum.  Another
uses {\sc Contur} to study bounds on cascade decays of leptoquarks and vector-like quarks from existing searches.
A last contribution in this area tests if in a simplified model a 
vanilla leptoquark solution to the $R_{D^{(*)}}$ deviations can meet relic abundance constraints if the leptoquark is the mediator to a dark sector.
Two other contributions are oriented towards dark sectors: there is a  study of freeze-in scenarios where the mediator to the dark sector is a heavy vector-like fermion
with a mass that is higher than the reheating temperature, making prompt LHC dark matter searches become relevant again, unlike in conventional scenarios of freeze-in.
This is followed by a study of long-lived particle signatures of freeze-out at the LHC, identifying blind spots in the current analyses and making
suggestions for improvement.

The next series of contributions are directed at the Higgs boson.  Two of these use {\sc Contur} to study in one case
bounds on Gildener-Weinberg models of electroweak symmetry breaking and in a second
the sensitivity of ATLAS, CMS and LHCb measurements  to a Dark Matter model involving two Higgs doublets and an additional pseudoscalar mediator. The Higgs Yukawa couplings for charm and top are studied in the context of off-shell Higgs measurements, with sensitivity projections for the HL-LHC in these processes. 
There is an estimate of the LHC sensitivity to exotic Higgs decays to $Za$, with $a \to \gamma\gamma, \mu\mu, \tau\tau$ and its interpretation
when $a$ is an axion-like particle, as well as one that examines the potential for discovery and measurement of CP properties for a light scalar decaying
to a pair of photons produced through vector boson fusion.  A final Higgs-focused contribution considers probing the $Zt_R\bar{t}_R$ and $hZt_R\bar{t}_R$  interactions
using $p p \to t \bar{t} h Z$ and $p p \to t \bar{t} h j$ processes at future hadron colliders.

This is followed by reports on development of a variety of tools:  {\sc EFT2Obs} is a tool to map an effective field theory parametrization to experimental constraints,
demonstrated using the Higgs Effective Lagrangian.  With ATLAS and CMS having made progress towards publishing more details of their results, the
usage of  simplified or full likelihood information in two public re-interpretation tools, {\sc MADANALYSIS 5}  and {\sc SMODELS} is tested.  A complementary
set of contributions evaluate on the one hand a probabilistic approach to identify pairs of analyses that can safely be considered uncorrelated in global BSM
reinterpretation fits, and on the other hand  the potential of analysis description languages to assist, and even automate, the comparison of multiple analyses.
There is an overview of  functional and technical updates to {\sc Contur}, and a description and comparison of public tools that can utilise experimental information
to reproduce and recast the LHC results with acceptable accuracy.  Two final topics in this section examine whether it is possible to preserve searches in the
same way as measurements are preserved, and thus automate the re-interpretation process for searches to be as fast and efficient as it is for measurements,
and for a given measurement expected to be performed at the LHC, what precision needs to be achieved in order to exclude a certain region of parameter space
of a specific new physics model.

Finally, in a section on machine learning, a first report evaluates whether Machine Learning (ML)  can be used to examine models of new physics in their full dimensionality,
and a second studies model-independent signal detection algorithms, with a particular focus on approaches that are based on unsupervised machine learning.  This
last contribution also provides a benchmark LHC dataset, for example for the comparison of signal detection algorithms.

The meeting in Les Houches has fostered a large number of discussions between
theorists and experimenters.  On the timescale required for these proceedings, only a
fraction of the ideas could be examined in depth.
Even those that could not converge to a written contribution have paid off
through the breadth of searches conducted by experimenters and the understanding 
of the challenges placed on an experiment by the ever-changing theoretical landscape.
We expect that
many more future results will  benefit from the discussions held at the
workshop.

%% file: lfu/lfu.main.tex
\graphicspath{{lfu/}}

\newcommand{\Herwig}{H\protect\scalebox{0.8}{ERWIG}\xspace}
\newcommand{\Pythia}{P\protect\scalebox{0.8}{YTHIA}\xspace}
\newcommand{\Sherpa}{S\protect\scalebox{0.8}{HERPA}\xspace}
\newcommand{\Rivet}{R\protect\scalebox{0.8}{IVET}\xspace}
\newcommand{\Professor}{P\protect\scalebox{0.8}{ROFESSOR}\xspace}
\newcommand{\eps}{\varepsilon}
\newcommand{\mc}[1]{\mathcal{#1}}
\newcommand{\mr}[1]{\mathrm{#1}}
\newcommand{\tm}[1]{\scalebox{0.95}{$#1$}}


\chapter{Lepton flavor universality ratio in high-mass dilepton production at LHC}
{\it T.~Berger-Hryn'ova, D.~Buttazzo, J.~Dutta,        D.~Marzocca, M.~Son}

\label{sec:BSM_lfvr}



While the Standard Model has proven tremendously successful, much 
experimental evidence points to it not being a complete description of our universe. 
The searches for new phenomena are therefore an important component of the LHC 
experimental program, where a number of analyses are looking for signs 
of new heavy particles decaying to different final states. 
Although up to date there is no direct evidence for BSM phenomena at the LHC, 
there are several 2--3$\sigma$ hints of deviations of from the SM predictions in decays of B mesons, 
in $b\rightarrow sll$ and $b\rightarrow c\tau\nu_{\tau}$ 
transitions~\cite{PhysRevD.88.072012,PhysRevLett.118.211801,Aaij:2014ora,Aaij2017}.

If these and other deviations of a similar kind are confirmed, 
they signal a potential departure from lepton flavour universality (LFU). 
LFU follows directly from the assumption that the SM gauge group, 
SU(3)$\times$SU(2)$\times$U(1), 
is one and the same for all three generations of fermions. 
This implies that all leptons couple universally to the gauge bosons, 
and that the only difference in their interactions is caused by the 
difference in mass, e.g. in their interactions with the Higgs field. 
Thus any departure from LFU is a sign of the existence new phenomena 
at the next energy scale, similar to nuclear beta decay and other 
weak transitions pointing to the existence of electroweak W and Z bosons. 

The $b\rightarrow sll$ and $b\rightarrow c\tau\nu_{\tau}$ transitions
are connected to $b\bar{s}\rightarrow \ell\ell$ and $b\bar{c}\rightarrow\ell\nu$ scattering 
processes by crossing symmetry~\cite{Greljo:2018tzh}. 
Thus any new phenomena which affect these B meson decays should 
lead to a deviation from the SM prediction in the dilepton mass spectra, in a LFU-violating manner.

The most precise differential high-mass Drell-Yan (HMDY) 
cross-section measurement to date is based on the 20.3$\,fb^{-1}$ of proton--proton ($pp$)
collision data at centre-of-mass energy of 8$\,$TeV collected by the ATLAS experiment in 2012~\cite{Aad:2016zzw}. 
It covers dielectron and dimuon final states in mass range between 
116 and 1500$\,$GeV and electron and muon pseudo-rapidities ($|\eta|$) up to 2.5. 
A total experimental precision of better than 2 (11)$\%$ in each channel 
below 200 (1000)$\,$GeV. Dominant systematic uncertainties are due to $t\bar{t}$ 
and multijet backgrounds. Above approximately 250$\,$GeV statistical 
uncertainties become dominant in each of the channels. 

\begin{figure}[h]
  \begin{center}
    \includegraphics[width=0.49\textwidth]{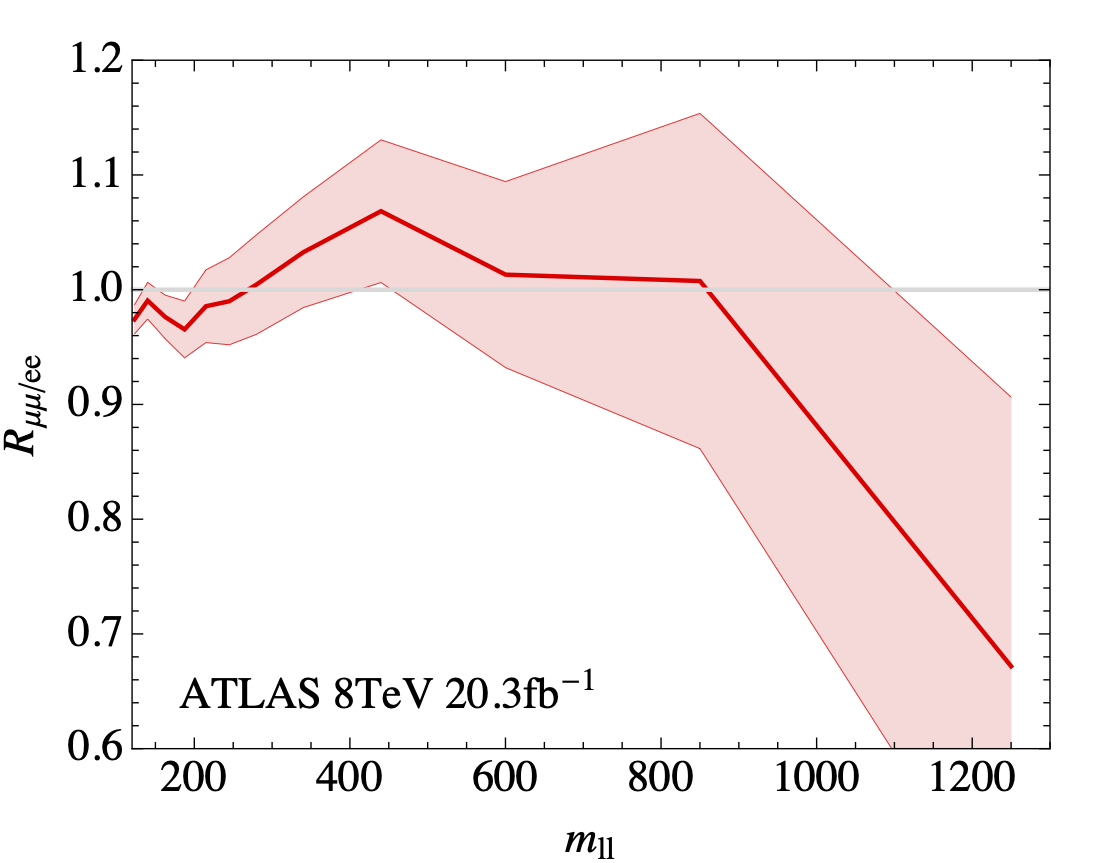}
    \includegraphics[width=0.49\textwidth]{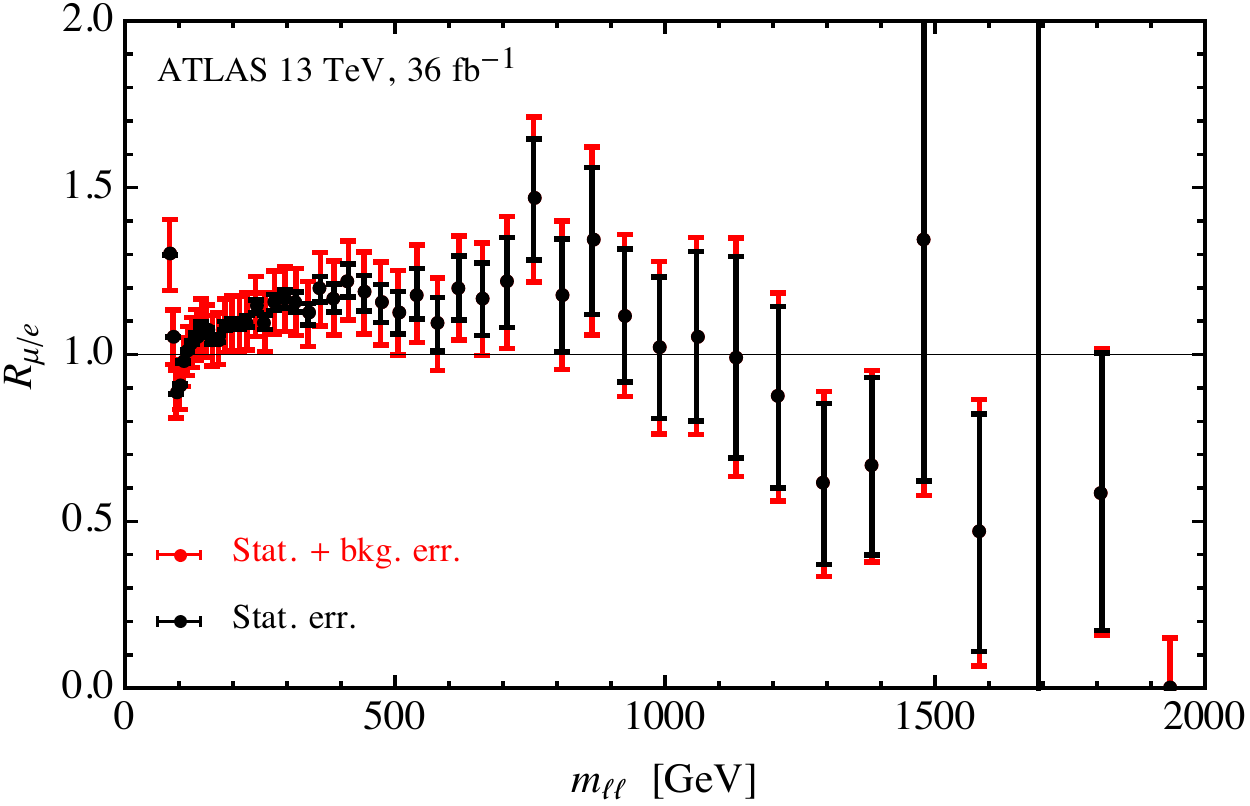}
  \end{center}
  \caption{Ratio of dimuon to dielectron high-mass Drell-Yan cross-sections
as a function of dilepton mass. Input data from (right) Ref.~\cite{Aad:2016zzw}
and (left) Ref.~\cite{Aaboud:2017buh}.
}
  \label{fig:BSM_lfvr}
\end{figure}

Inspired by the LFU tests done in B-physics at low energy, We propose to study the differential LFU ratio
$$
	R_{\mu\mu/ee}(m_{\ell\ell}) \equiv \left( d\sigma(p p \to \mu^+ \mu^-) / d m_{\ell\ell} \right) \left( d\sigma(p p \to e^+ e^-) / d m_{\ell\ell} \right)^{-1} ~,
$$
as a both theoretically and experimentally cleaner observable, 
although in the current implementation of the analysis only the luminosity and pdf uncertainties 
are expected to cancel in the ratio.  
It is important for future measurements to reduce uncertainties on this ratio as much as possible to 
investigate any potential discrepancy with respect to the SM expectation of one.  
The measured $R_{\mu\mu/ee}(m_{ll})$ from the ATLAS 8 TeV analysis, with the $1\sigma$ band, is shown in Fig.~\ref{fig:BSM_lfvr} (left). 
All uncertainties in each channel have been summed in quadrature.
Fig.~\ref{fig:BSM_lfvr} (right) shows that a similar precision is reached in narrow resonance search with 
36$\,fb^{-1}$ of $pp$ collision data at centre-of-mass 
energy of 13$\,$TeV collected by the ATLAS experiment in 2015--2016~\cite{Aaboud:2017buh}. 
Potential impact on this ratio from various BSM scenarios is discussed in Ref.~\cite{Greljo:2017vvb}.

The experimental results in the dielectron and dimuon channels 
will be improved by addition of the full Run 2 data set
which has about 14--21 times higher statistics (depending on dilepton mass) 
and should decrease the available statistical uncertainties approximately by a factor of 4 (per experiment),
making the measurement statistically limited only above 500$\,$GeV assuming the current binning. 
For the ATLAS measurement in the dielectron final state, there is also possibility 
to double the studied pseudo-rapidity range up to 4.9 (for one of the two electrons), 
which would provide better 
sensitivity to BSM due to additional coverage in the angular distributions of the final states. 

\let\Herwig\undefined
\let\Pythia\undefined
\let\Sherpa\undefined
\let\Rivet\undefined
\let\Professor\undefined
\let\eps\undefined
\let\mc\undefined
\let\mr\undefined
\let\mb\undefined
\let\tm\undefined

%% file: dilepton/dilepton.main.tex
\graphicspath{{dilepton/}}

\newcommand{\be}{\begin{equation}}
\newcommand{\ee}{\end{equation}}
\newcommand{\MeV}{\textrm{ MeV}}
\newcommand{\GeV}{\textrm{ GeV}}
\newcommand{\TeV}{\textrm{ TeV}}
\newcommand{\SM}{\textrm{SM}}
\newcommand{\OO}{\mathcal{O}}
\newcommand{\LL}{\mathcal{L}}

\newcommand{\fr}{{\sc \small FeynRules} }
\newcommand{\ma}{{\sc\small MadAnalysis}~5 }
\newcommand{\mg}{{\sc\small MG5\_aMC} }
\newcommand{\rot}{{\sc\small ROOT} }

\newcommand{\DM}[1]{{\color{red}\textbf{[#1]}}}

\chapter{Global SMEFT analysis of LHC dilepton tails}

{\it J.~Bernigaud,
  A.~Bharucha,
  D.~Buttazzo,
  J.~Dutta,
  D.~Marzocca,
  P.~Pani,
  G.~Polesello,
  M.~Son,
  E.~Venturini,
  N.~Vignaroli}


\label{sec:dilepton}


\begin{abstract}
	The high-energy tail of the Drell-Yan production of two leptons at the LHC is a powerful probe of new physics. In this proceedings the ATLAS 8~TeV data is used to put model-independent constraints on all the dimension-six effective operators which can interfere with the Standard Model contribution, without  assumptions on the flavor structure. 
The ATLAS analysis provides fiducial measurements of differential cross sections in invariant mass and in the two-dimensional plane of invariant mass and dilepton rapidity. Likelihoods are derived for both, as well as using only the linear or also the quadratic dependence of the cross section on the effective field theory coefficients.
Code containing these likelihoods is provided,\footnote{The likelihoods can be downloaded at this address: \href{https://people.sissa.it/~dmarzocc/dileptonATLAS8TeVchiSQ.zip}{\tt people.sissa.it/\textasciitilde{}dmarzocc/dileptonATLAS8TeVchiSQ.zip}.} allowing for a straightforward derivation of the limits in any specific model.
\end{abstract}

\section{Introduction}

It is well known that the high-energy tail of the dilepton invariant mass distribution in $p p \to \ell^+ \ell^-$ at a hadron collider is potentially sensitive to very high New Physics (NP) scales, possibly even above the on-shell production threshold.
In this case the effect of heavy NP can be parametrised in a model-independent way using an effective field theory (EFT) approach, adding to the Standard Model (SM) a set of higher-dimension effective operators, the leading contribution to this process being generated by those of dimension six.
Specifically, semi-leptonic four-fermion effective operators contribute to the Drell-Yan process with a leading contribution growing, with respect to the SM one, as $\delta \mathcal{A}/\mathcal{A}_{\SM} \sim s / (g_{\SM}^2 \Lambda^2)$, where $s$ is the invariant mass of the dilepton system, $g_{\SM}$ is a typical SM electroweak coupling, and $1/\Lambda^2$ is the coefficient of the effective operators.
It is then clear that the high-energy tail of the dilepton invariant mass distribution has an enhanced sensitivity to the effect of these operators. 
The constraints derived from this process have been shown to be competitive with some limits from flavour physics \cite{Cirigliano:2012ab,deBlas:2013qqa} and electroweak precision measurements at LEP, even though the LHC precision is worse than that of LEP \cite{Farina:2016rws,Alioli:2017nzr}. Ref.~\cite{Greljo:2017vvb} used the limits from this process to put constraints on some specific flavor structures in relation to possible solutions of the experimental anomalies in $B$-physics.

While several new physics models can potentially be constrained by this process, a general, model-independent, and easily applicable set of constraints is still not available. In Ref.~\cite{Greljo:2017vvb} a fit in the Gaussian approximation was performed including all semi-leptonic dimension-six operators with an energy-growing interference with the SM, however the analysis used a 13 TeV ATLAS analysis \cite{Aaboud:2017buh} (now updated in \cite{Aad:2019fac}) which was more focussed on resonance searches than precision SM measurements, therefore treating the uncertainties less carefully, and didn't provide the general global likelihood from the fit.

In this work we use the ATLAS 8 TeV analysis of $p p \to \ell^+ \ell^-$ (with $\ell = e, \mu$) with 20.3 fb$^{-1}$ of luminosity \cite{Aad:2016zzw}, which provides both the unfolded differential cross section in invariant mass $\frac{d \sigma}{d m_{\ell \ell}}$, as well as the double-differential cross section in invariant mass and (pseudo-)rapdity $\frac{d^2 \sigma}{d m_{\ell \ell} d|y_{\ell\ell}|}$ (or $\frac{d^2 \sigma}{d m_{\ell \ell} d|
\Delta \eta_{\ell\ell}|}$) at the Born-level in QED, separately for the electron and muon channels, in bins of invariant mass from 116 to 1500 GeV. 
Our goal is to provide general constraints on EFT coefficients from LHC dilepton tails, studying also the effect of adding information on the rapidity distribution to the fit.

Much stronger constraining power is expected from 13 TeV data, which however is still not analysed at the same level of sophistication as the 8 TeV data, see however \cite{CMS:2016pkm} for an early study by CMS and \cite{Sirunyan:2018ipj,Aad:2019fac} for NP resonance searches in dilepton tails by CMS and ATLAS.

\section{EFT dependence}
\label{sec:dilepton:EFT}

The SM effective field theory (SMEFT) is constructed by adding to the SM a set of gauge-invariant higher-dimension operators, suppressed by powers of the cutoff scale $\Lambda$, corresponding to the mass scale of the new physics states.
Such an effective theory can be used to study in a model-independent way new physics effects in processes with a typical energy $E \ll \Lambda$.
Assuming baryon and lepton number conservation, the leading effects are described by dimension-six operators
\be
	\LL_{\rm EFT} = \sum_j \frac{c_j}{\Lambda^2} \OO_j + O(\Lambda^{-4}).
\ee
In the Warsaw basis \cite{Grzadkowski:2010es}, those which contribute at the tree-level to the high-energy tails of $p p \to \ell^+ \ell^- (+j)$, with $\ell = e, \mu$, and interfere with the Standard Model Drell-Yan contribution growing like $E^2 / \Lambda^2$, are the following semileptonic four-fermion operators: \footnote{Another set of operators contributing to the $p p \to \ell^+ \ell^-$ process are those which affect the $Z$ couplings to leptons and quarks. These, however, are already well constrained by LEP data \cite{ALEPH:2005ab}, and their effect does not increase with the energy but scales as $v^2 / \Lambda^2$, therefore it remains always too small to have a sizable impact in the high-energy tails.}
\be
\begin{array}{|l | l |}
	\hline
	(\OO_{lq}^{(1)})_{\alpha i} = (\bar l_\alpha \gamma_\mu l_\alpha) (\bar q_i \gamma^\mu q_i) & 
	(\OO_{lq}^{(3)})_{\alpha i} = (\bar l_\alpha \gamma_\mu \sigma^a l_\alpha) (\bar q_i \gamma^\mu \sigma^a q_i) \\
	( \OO_{qe} )_{i \alpha} = (\bar q_i \gamma^\mu q_i) (\bar e_\alpha \gamma^\mu e_\alpha) &  \\
	(\OO_{lu})_{\alpha i} = (\bar l_\alpha \gamma_\mu l_\alpha) (\bar u_i \gamma^\mu u_i) & 
	(\OO_{ld})_{\alpha i} = (\bar l_\alpha \gamma_\mu l_\alpha) (\bar d_i \gamma^\mu d_i) \\ 
	(\OO_{eu})_{\alpha i} = (\bar e_\alpha \gamma_\mu e_\alpha) (\bar u_i \gamma^\mu u_i) & 
	(\OO_{ed})_{\alpha i} = (\bar e_\alpha \gamma_\mu e_\alpha) (\bar d_i \gamma^\mu d_i) \\ 
	\hline
\end{array}~,
\label{eq:dilepton:EFTops}
\ee
where $l$ and $q$ are the $SU(2)_L$ left-handed doublets, $e, u$ and $d$ are the right-handed singlets, and $\alpha$ ($i$) is a lepton (quark) flavor index. In order to work with dimensionless parameters, we can reabsorb the EFT scale $\Lambda$ redefining the coefficients as:
\be
	C_i \equiv c_i \frac{v^2}{ \Lambda^2 } = \frac{c_i}{\sqrt{2} G_f \Lambda^2}~,
	\label{eq:dilepton:defC}
\ee
with $v=246 \GeV$. Excluding operators with $t_R$ or $t_L$ (which cannot be probed in $p p \to \ell\ell$), there are 36 independent flavor-diagonal operators (18 for muons and 18 for electrons) \cite{Greljo:2017vvb}. The complete list of coefficients is given in Tab.~\ref{tab:dilepton:EFTcoeff}.

\begin{table}[t]
\centering
\begin{tabular}{| c | c | c |}\hline
	$(C_{lq}^{(1)})_{\alpha 1}$ & $(C_{lq}^{(1)})_{\alpha 2}$ & $(C_{lq}^{(+)})_{\alpha 3}$ \\
	$(C_{lq}^{(3)})_{\alpha 1}$ & $(C_{lq}^{(3)})_{\alpha 2}$ &  \\
	$(C_{qe})_{1 \alpha}$ & $(C_{qe})_{2 \alpha}$ & $(C_{qe})_{3 \alpha}$ \\
	$(C_{lu})_{\alpha 1}$ & $(C_{lu})_{\alpha 2}$ &  \\
	$(C_{ld})_{\alpha 1}$ & $(C_{ld})_{\alpha 2}$ & $(C_{ld})_{\alpha 3}$ \\
	$(C_{eu})_{\alpha 1}$ & $(C_{eu})_{\alpha 2}$ &  \\
	$(C_{ed})_{\alpha 1}$ & $(C_{ed})_{\alpha 2}$ & $(C_{ed})_{\alpha 3}$ \\\hline
\end{tabular}
\caption{\label{tab:dilepton:EFTcoeff} List of the EFT coefficients included in our analysis, the index $\alpha = 1,2$ describes lepton flavor. We defined $(C_{lq}^{(+)})_{\alpha 3}  \equiv  (C_{lq}^{(1)})_{\alpha 3} + (C_{lq}^{(3)})_{\alpha 3} $ since the dilepton process is not sensitive to the top quark component.}
\end{table}
%

%
The EFT coefficients $C_i$ enter linearly in the scattering amplitude of this process, therefore the cross section in any given bin is necessarily given by a quadratic function of the EFT coefficients:
\be
	\sigma_{\rm bin}(\vec C) = \sigma_{\rm SM, bin} + 2 \Re[ C_i \, \sigma^i_{\rm SM-EFT, bin}] + C_i C^*_j \, \sigma^{ij}_{\rm EFT^2, bin}~.
	\label{eq:dilepton:sigmaEFT}
\ee
Since operators with different flavor content (different quarks or leptons) or different chirality do not interfere with each other (in the limit of massless fermions), the matrix $\sigma^{ij}_{\rm EFT^2}$ is diagonal.\footnote{A non-vanishing interference is present between singlet and triplet operators $\OO_{lq}^{(1)}$ and $\OO_{lq}^{(3)}$ since they share the same field content (both have $\bar d_L^i \gamma_\mu d_L^i$ and $\bar u_L^i \gamma_\mu u_L^i$ terms). However also in this case there are only two ways in which the coefficients enter the quadratic terms: $(C_{lq}^{(1)} \pm C_{lq}^{(3)})^2$.} The formula above thus reduces to:
\be
	\sigma_{\rm bin}(\vec C) = \sigma_{\rm SM, bin} + 2 C_i \, \sigma^i_{\rm SM-EFT, bin} + (C_i)^2 \, \sigma^{i}_{\rm EFT^2, bin}~.
	\label{eq:dilepton:EFTdep1}
\ee
In order to obtain this dependence we implemented the operators in Eq.~\eqref{eq:dilepton:EFTops} in \fr \cite{Alloul:2013bka} and used the exported UFO model files to generate SM and new physics events in \mg \cite{Alwall:2014hca}. Using the \mg ability to generate events using only the linear or quadratic terms, ({\tt NP\^{}2==1} or {\tt NP\^{}2==2}, respectively) it is straightforward to obtain the complete EFT dependence.

\begin{table}[t]
\centering
\begin{tabular}{| c || c | c | c | c | c |}\hline
	$m_{\ell\ell}$ [GeV] & [300-380] & [380-500] & [500-700] & [700-1000] & [1000-1500] \\[2pt]\hline\hline
	$\frac{1}{\Delta m_{\ell\ell}}\sigma^{\rm SM~NNLO}_{\rm bin}$ [fb/GeV] & 2.00 & 0.651 & 0.161 & $2.9 \times 10^{-2}$ & $3.64 \times 10^{-3}$ \\
	$\delta^{\rm th}$ [\%] & 3.7 & 4.6 & 5.7 & 7.6 & 11.6 \\\hline
	$\frac{1}{\Delta m_{\ell\ell}}\sigma^{ee, {\rm exp}}_{\rm bin}$ [fb/GeV] & 1.84 & 0.599 & 0.152 & $2.64 \times 10^{-2}$ & $3.23 \times 10^{-3}$ \\
	$\delta^{\rm stat}$ [\%] & 2.6 & 3.6 & 5.3 & 10.2 & 22.5 \\
	$\delta^{\rm syst}$ [\%] & 2.5 & 2.7 & 2.6 & 3.3 & 5.8 \\\hline
	$\frac{1}{\Delta m_{\ell\ell}}\sigma^{\mu\mu, {\rm exp}}_{\rm bin}$ [fb/GeV] & 1.90 & 0.640 & 0.154 & $2.66 \times 10^{-2}$ & $2.17 \times 10^{-3}$ \\
	$\delta^{\rm stat}$ [\%] & 2.3 & 3.2 & 5.0 & 9.6 & 26 \\
	$\delta^{\rm syst}$ [\%] & 1.9 & 1.8 & 2.0 & 2.1 & 2.7 \\\hline
\end{tabular}
\caption{\label{tab:dilepton:mllBins} Invariant mass bins with the associated predicted and measured cross section and uncertainties (theoretical, statistical, and total systematic) from the ATLAS 8 TeV analysis \cite{Aad:2016zzw} .}
\end{table}

The results in the ATLAS 8 TeV analysis \cite{Aad:2016zzw} are unfolded in the fiducial region
\be
(p_T^\ell)_{\rm leading} > 40 \GeV, \quad
(p_T^\ell)_{\rm sub-leading} > 30 \GeV, \quad
|\eta^\ell| < 2.5, \quad
m_{\ell\ell} < 1500 \GeV~.
\ee
Final state QED radiation effects are also unfolded, and the Born-level cross section in each fiducial bin is presented.
We generate all events at the leading order (LO) with the same cuts the same fiducial region. Since we are interested in the high-energy tail, we also require $m_{\ell\ell} > 300 \GeV$.
The five bins of invariant mass distribution we consider are: $[300 - 380 - 500 - 700 - 1000 - 1500]~\GeV$, see Table~\ref{tab:dilepton:mllBins} for the values of the predicted and measured cross sections and uncertainties.
For the two-dimensional distribution we choose $d^2 \sigma/ (d m_{\ell \ell} d|y_{\ell\ell}| )$, since the dilepton rapidity shows a stronger sensitivity to the different EFT operators when compared to the distribution in $|\Delta \eta_{\ell\ell}|$, as is discussed in Sec.~\ref{sec:dilepton:rapd}. In this case the invariant mass bins are merged into two large bins $[300 - 500 - 1500]~\GeV$ for each $|y_{\ell\ell}|$ bin in $[0 - 0.4 - 0.8 - 1.2 - 1.6 - 2.0 - 2.4]$.
Unfortunately, the additional information gained from the rapidity comes with a reduced sensitivity from the invariant mass, since the potential new physics effects, largest in the highest invariant mass bin, are partially washed out by the merging with the lower invariant mass bins, which have a larger absolute cross section but smaller sensitivity to the EFT coefficients.
This aspect is analysed quantitatively in Sec.~\ref{sec:dilepton:fits}

\begin{figure}[t]
  \centering
  \includegraphics[width=0.49\textwidth]{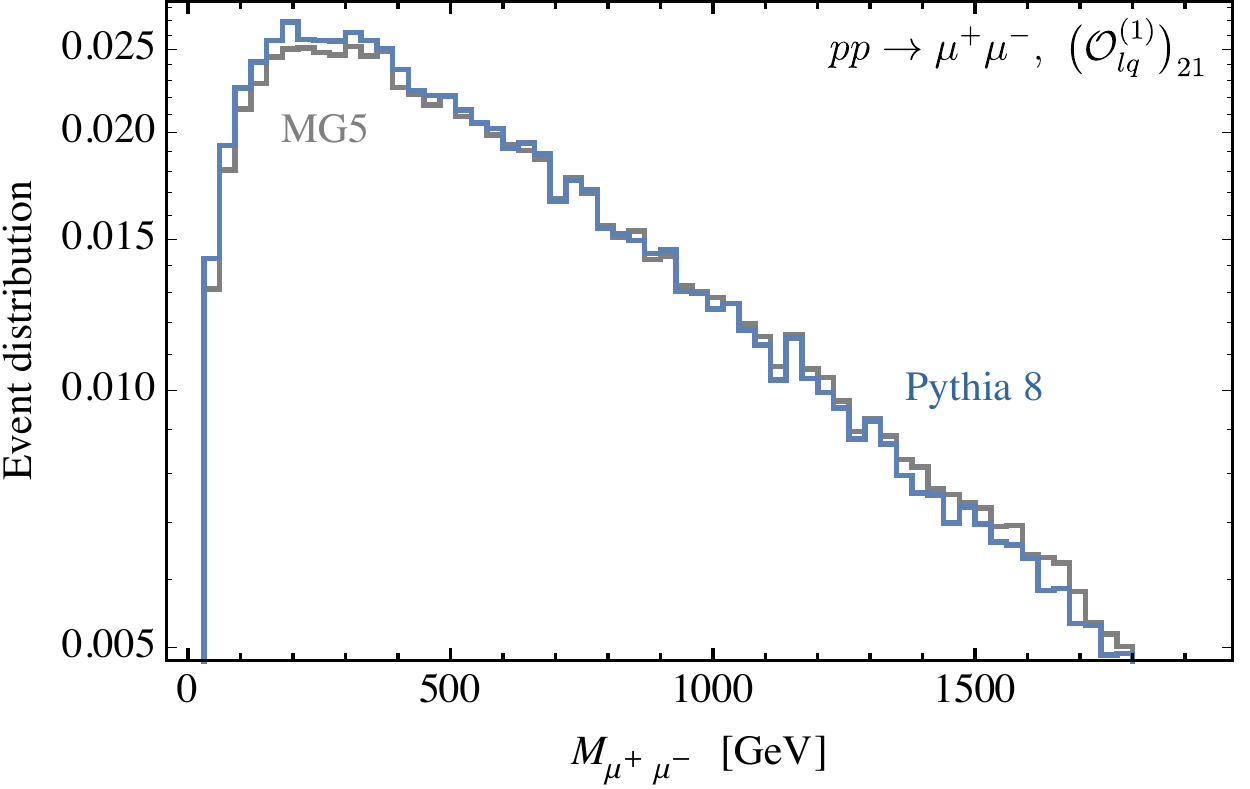}\hfill%
  \includegraphics[width=0.48\textwidth]{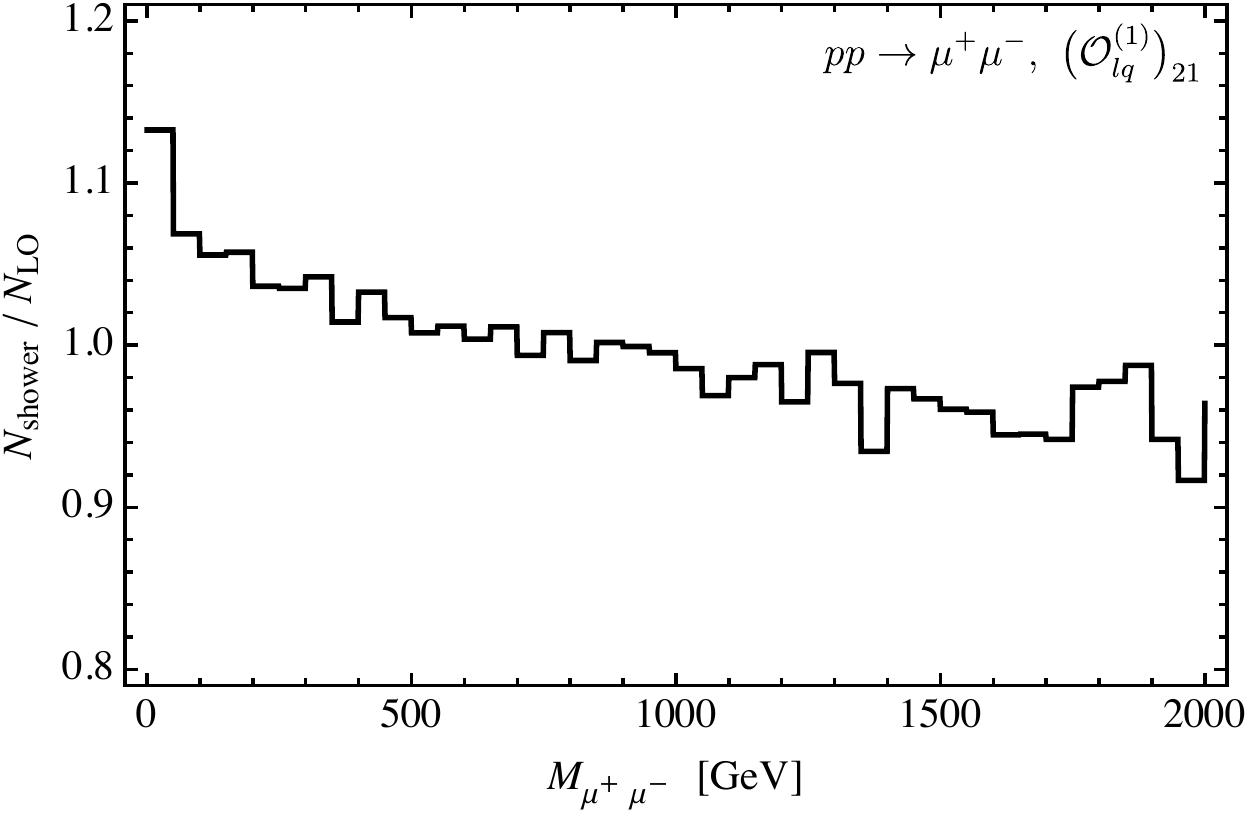}
  \caption{Left: Invariant mass distribution of dimuon signal events; the gray line is generated with {\sc\small mg5\_aMC}, the blue line includes {\sc\small pythia8} parton showering. Right: relative correction due to parton showering as function of dimuon invariant mass.}
\label{fig:dilepton:invmass}
\end{figure}

The events for each run are analysed in \ma \cite{Conte:2012fm} to obtain the cross section in each bin. The \ma analysis has been cross-checked with a custom \rot analysis code. We generated 2M events for each new physics run, and 5M events for the SM one, reducing the statistical uncertainty in the determination of the coefficients in the EFT dependence in each bin to be $\lesssim 1\%$.
These LO cross sections are then rescaled by the ratio of the SM cross section computed at NNLO  QCD and NLO EW, taken directly from \cite{Aad:2016zzw}, over the LO SM one:
\be
	\sigma^{\rm NNLO}_{\rm bin}(\vec C) \approx \frac{\sigma^{\rm NNLO}_{\rm SM, bin}}{\sigma^{\rm LO}_{\rm SM, bin}} \; \sigma_{\rm bin}(\vec C) ~.
	\label{eq:dilepton:EFTdep2}
\ee
In this way the very precise SM prediction for the cross section is reproduced in the limit of zero new physics, and only  possible small differences in how QCD corrections apply to new physics vs. the SM are neglected. We checked explicitly that the effects on QCD showering amounts to at most a (few)\% correction on the new physics effect. This can be seen in the right panel of Fig.~\ref{fig:dilepton:invmass}, which shows the ratio of the invariant mass distributions of signal events before and after {\sc\small pythia8} showering.

\subsection{New physics in the rapidity and pseudo-rapidity distributions}
\label{sec:dilepton:rapd}

\begin{figure}[t]
\centering
 \includegraphics[width=0.48\textwidth]{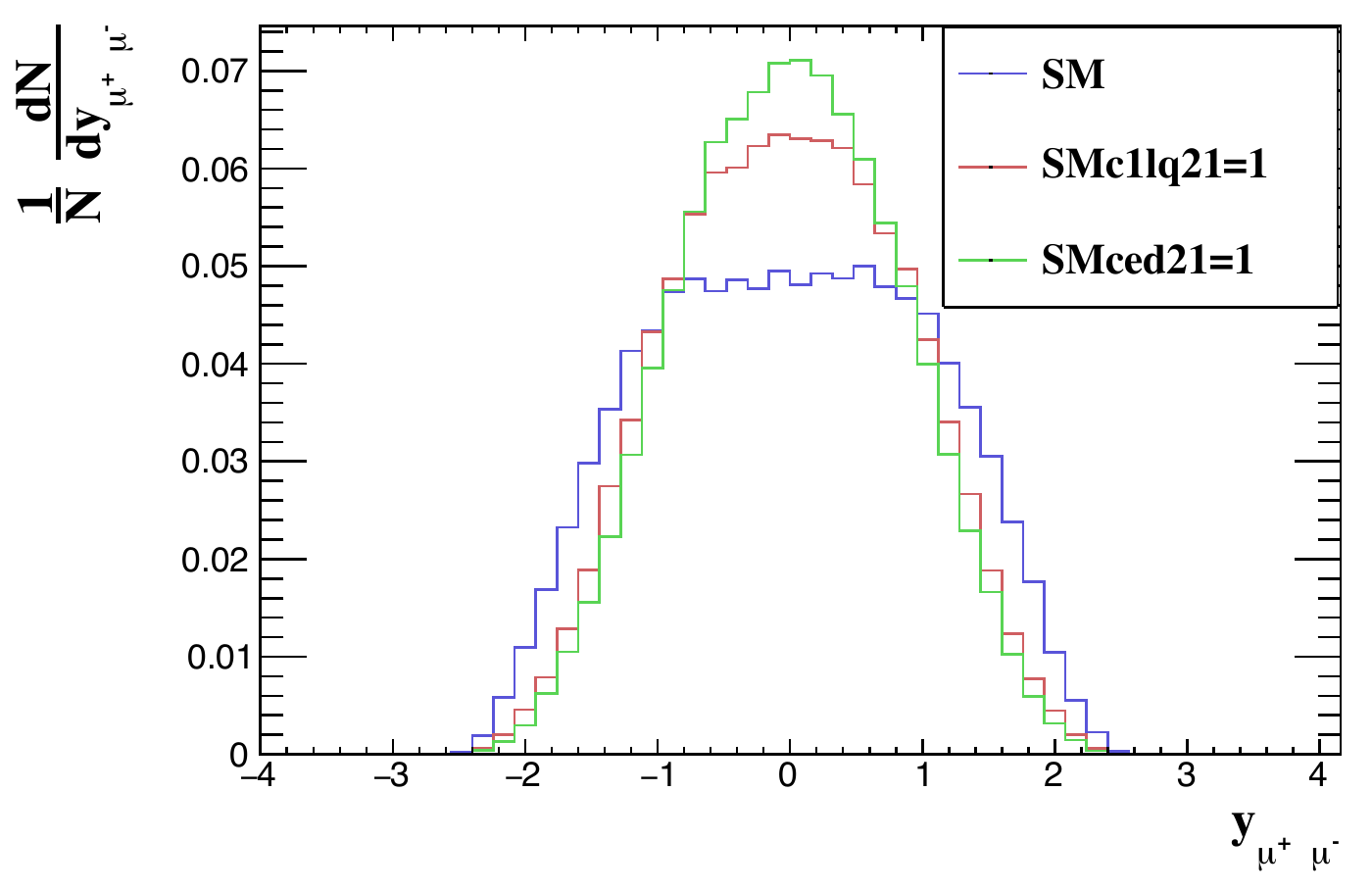}\hfill%
 \includegraphics[width=0.49\textwidth]{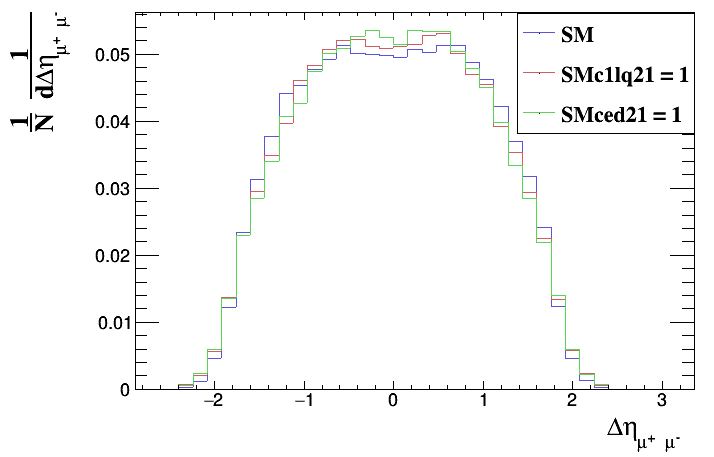}
 \caption{Distributions for the dimuon rapidity $y_{\mu\mu}$ and differences of the pseudo-rapidity $\Delta \eta_{\mu\mu}$ for some benchmark points: SM (blue), $(C^{(1)}_{lq})_{21} = 1$ (red), and $(C_{ed})_{21} = 1$ (green).}
 \label{fig:dilepton:rapd}
 \end{figure}

Fig.~\ref{fig:dilepton:rapd} shows the distribution in the rapidity of the dimuon pair (left plot) and in the difference of pseudo-rapidity between the two muons (right plot) for SM and for some NP benchmarks, after using the kinematic cuts outlined in Sec~\ref{sec:dilepton:EFT}. One observes some differences in the rapidity $y_{\ell\ell}$ and pseudorapidity $\Delta \eta_{\ell\ell}$ for the effective operators as compared to the SM, with the former being more sensitive to the presence of the effective operators. 

\section{Building the likelihood}

From the full EFT dependence in Eq.~\eqref{eq:dilepton:EFTdep2} and the fiducial measurement in each bin we build the likelihood $\mathcal{L}$ in the Gaussian approximation:
\be
	\chi^2(\{C\}) = -2 \log \mathcal{L}(\{C\}) = \sum_{\alpha,\beta \in {\rm bins}}  (X_\alpha(\vec C) - X_{\alpha}^{\rm exp}) (\Sigma^{-1})_{\alpha\beta} (X_\beta(\vec C) - X_{\beta}^{\rm exp})~,
	\label{eq:dilepton:ChiSq}
\ee
where $\alpha,\beta$ are indices which run over all bins (either the 1D invariant mass distribution or the 2D one), $X_\alpha(\vec C) \equiv \sigma^{\rm NNLO}_{\rm bin_\alpha}(\vec C)$ is the EFT dependence for that bin as computed in Eq.~\eqref{eq:dilepton:EFTdep2},  $X_{\alpha}^{\rm exp}$ the fiducial measurement, and $\Sigma$ the full variance matrix.
In the case of uncorrelated uncertainties $\Sigma_{\alpha\beta} = \delta_{\alpha\beta} \sigma_{\alpha}^2$, and $\sigma_{\alpha}^2$ contains the sum in quadrature of all uncertainties (theory, statistical, and systematic) in that bin.
As can be seen in Tab.~\ref{tab:dilepton:mllBins} the dominant uncertainties are statistical, thus  correlations between the small systematic uncertainties can be neglected in first approximation.
We are further obliged to neglect these correlations due to the fact that the correlations between systematic uncertainties (or, alternatively, the response functions in each bin) are not provided by the experiment.\footnote{The only systematic uncertainty for which we do know the response function is due to the luminosity, and amounts to 1.9\%. This can be treated as a nuisance parameter $\delta_{\rm lumi}$ with a response function corresponding to a simple overall rescaling of all bins, $X_{\alpha}^{\rm exp} \to X_{\alpha}^{\rm exp}(1+\delta_{\rm lumi})$, and a given uncertainty $\sigma_\delta=1.9\%$. We thus add to the $\chi^2$ in Eq.~\eqref{eq:dilepton:ChiSq} the term $\delta_{\rm lumi}^2 / \sigma_\delta^2$. Thanks to the simplicity of the response function for this uncertainty we can profile over $\delta_{\rm lumi}$ analytically, obtaining a modified expression for $\Sigma_{\alpha\beta}^{-1}$ which takes the form
\be
(\Sigma^{-1})_{\alpha\beta} =\,\delta_{\alpha\beta}\, \frac{1}{\sigma_{\alpha}^2}-\frac{X_\alpha^{\rm exp} X_\beta^{\rm exp}}{\sigma_{\alpha}\sigma_{\beta}} \,\bar{\sigma}^2~,
\qquad \text{where} \qquad
\frac{1}{\bar{\sigma}^2}= \frac{1}{\sigma_\delta^2}+\left(\frac{X_\alpha^{\rm exp}}{\sigma_\alpha}\right)^2 ~.
\ee
Note that in this case, in the uncrorrelated uncertainty of each bin, $\sigma_\alpha$, the luminosity uncertainty is not included.
}
Substituting in Eq.~\eqref{eq:dilepton:ChiSq} $X_\alpha^{\rm exp}$ with the SM prediction $X_\alpha^{\rm SM} = X_\alpha (\vec 0)$, we obtain the expected likelihoods.

We obtain the $\chi^2$ function for the datasets corresponding to the differential cross section measurement in the dilepton invariant mass $\frac{d \sigma}{d m_{\ell \ell}}$ or in the dilepton invariant mass and rapidity $\frac{d^2 \sigma}{d m_{\ell \ell} d|y_{\ell\ell}|}$, for both electrons and muons. We combine the measurements for electrons and muons but keep the separate dependence on the EFT coefficients for the two lepton flavors, so that lepton flavour universality tests can be performed.
Furthermore, one can choose to limit the dependence on the EFT coefficients in Eq.~\eqref{eq:dilepton:EFTdep1} to the linear terms, or to include also the quadratic terms.
We thus have four $\chi^2$ functions, which we make available alongside this work in the ancillary files listed in Tab.~\ref{tab:dilepton:chi2func} (downloadable \href{https://people.sissa.it/~dmarzocc/dileptonATLAS8TeVchiSQ.zip}{here}). These are functions of the coefficients in Tab.~\ref{tab:dilepton:EFTcoeff}. The notation used in the ancillary files is: $(C_{lq}^{(1)})_{2 1} \equiv$ {\tt C1lq[2, 1]}, $(C_{lq}^{(+)})_{1 3} \equiv$ {\tt Cplq[1, 3]}, $(C_{lq}^{(3)})_{2 2} \equiv$ {\tt C3lq[2, 2]}, $(C_{qe})_{3 2} \equiv$ {\tt Cqe[3, 2]}, and so on.

\begin{table}[t]
\centering
\begin{tabular}{|c | c |}\hline
	$\chi^2_{m_{\ell\ell}, {\rm lin}}$ & 			{\tt Chi2MllLin.txt} \\
	$\chi^2_{m_{\ell\ell}, {\rm quad}}$ & 			{\tt Chi2MllQuadr.txt} \\
	$\chi^2_{m_{\ell\ell} |y_{\ell\ell}|, {\rm lin}}$ &	{\tt Chi2MYllLin.txt} \\
	$\chi^2_{m_{\ell\ell} |y_{\ell\ell}|, {\rm quad}}$ &	 {\tt Chi2MYllQuadr.txt} \\\hline
\end{tabular}
\caption{\label{tab:dilepton:chi2func}$\chi^2$ functions obtained with our analysis, and the corresponding ancillary file name.}
\end{table}

\section{Some example fits and discussion}
\label{sec:dilepton:fits}

In this Section we show some examples of how the four likelihoods we obtain from dilepton data can be used to constrain the SMEFT  coefficients.
We start by considering each of the coefficients in Tab.~\ref{tab:dilepton:EFTcoeff} one at a time and derive the $95\%$ CL\footnote{In practice we impose $\chi^2 - \chi^2_{\rm min} = 3.84$ or 5.99 for 1D and 2D distributions, respectively.} constraint for that coefficient with each likelihood. The expected limits are collected in Tab.~\ref{tab:dilepton:1DlimitsExp} while the observed ones are in Tab.~\ref{tab:dilepton:1Dlimits}.

\begin{table}[p]
\centering
\begin{tabular}{| c | cc | cc | cc | cc |}
\hline
	95\%CL int. & 
	\multicolumn{2}{ c |}{$\chi^2_{m_{\ell\ell}, {\rm quad}}$} &  
	\multicolumn{2}{ c |}{$\chi^2_{m_{\ell\ell}, {\rm lin}}$} &  
	\multicolumn{2}{ c |}{$\chi^2_{m_{\ell\ell} |y_{\ell\ell}|, {\rm quad}}$} &
	\multicolumn{2}{ c |}{$\chi^2_{m_{\ell\ell} |y_{\ell\ell}|, {\rm lin}}$} 	\\ 
	& min & max & min & max & min & max & min & max \\ \hline
 $(C^{(1)}_{lq})_{11}$ & -0.0019 & 0.0035 & -0.0037 & 0.0037 & -0.0024 & 0.0055 & -0.0040 & 0.0040 \\
 $(C^{(1)}_{lq})_{12}$ & -0.016 & 0.015 & -0.22 & 0.22 & -0.020 & 0.018 & -0.20 & 0.20 \\
 $(C^{(+)}_{lq})_{13}$ & -0.066 & 0.051 & -0.17 & 0.17 & -0.083 & 0.056 & -0.15 & 0.15 \\
 $(C^{(3)}_{lq})_{11} $& -0.0034 & 0.0020 & -0.0043 & 0.0043 & -0.0053 & 0.0026 & -0.0047 & 0.0047 \\
 $(C^{(3)}_{lq})_{12}$ & -0.017 & 0.015 & -0.11 & 0.11 & -0.021 & 0.018 & -0.097 & 0.097 \\
 $(C_{ed})_{11}$ & -0.0085 & 0.0043 & -0.0072 & 0.0072 & -0.014 & 0.0052 & -0.0077 & 0.0077 \\
 $(C_{ed})_{12}$ & -0.032 & 0.026 & -0.11 & 0.11 & -0.041 & 0.030 & -0.097 & 0.097 \\
 $(C_{ed})_{13}$ & -0.062 & 0.055 & -0.38 & 0.38 & -0.075 & 0.063 & -0.33 & 0.33 \\
 $(C_{ld})_{11}$ & -0.0076 & 0.0052 & -0.015 & 0.015 & -0.011 & 0.0068 & -0.016 & 0.016 \\
 $(C_{ld})_{12}$ & -0.030 & 0.027 & -0.22 & 0.22 & -0.038 & 0.033 & -0.20 & 0.20 \\
 $(C_{ld})_{13}$ & -0.061 & 0.057 & -0.78 & 0.78 & -0.072 & 0.066 & -0.69 & 0.69 \\
 $(C_{lu})_{11}$ & -0.0030 & 0.0069 & -0.0045 & 0.0045 & -0.0037 & 0.012 & -0.0050 & 0.0050 \\
 $(C_{lu})_{12}$ & -0.035 & 0.041 & -0.17 & 0.17 & -0.040 & 0.051 & -0.16 & 0.16 \\
 $(C_{eu})_{11}$ & -0.0019 & 0.0028 & -0.0022 & 0.0022 & -0.0022 & 0.0028 & -0.0024 & 0.0024 \\
 $(C_{eu})_{12}$ & -0.032 & 0.044 & -0.085 & 0.085 & -0.035 & 0.057 & -0.076 & 0.076 \\
 $(C_{qe})_{11}$ & -0.0028 & 0.0050 & -0.0055 & 0.0055 & -0.0036 & 0.0080 & -0.0060 & 0.0060 \\
 $(C_{qe})_{21}$ & -0.021 & 0.024 & -0.13 & 0.13 & -0.025 & 0.031 & -0.12 & 0.12 \\
 $(C_{qe})_{31}$ & -0.057 & 0.061 & -0.66 & 0.66 & -0.064 & 0.071 & -0.57 & 0.57 \\
 $(C^{(1)}_{lq})_{21}$ & -0.0017 & 0.0031 & -0.0032 & 0.0032 & -0.0023 & 0.0054 & -0.0037 & 0.0037 \\
 $(C^{(1)}_{lq})_{22}$ & -0.014 & 0.014 & -0.20 & 0.20 & -0.019 & 0.018 & -0.19 & 0.19 \\
 $(C^{(+)}_{lq})_{23}$ & -0.059 & 0.046 & -0.16 & 0.16 & -0.081 & 0.054 & -0.14 & 0.14 \\
 $(C^{(3)}_{lq})_{21}$ & -0.0030 & 0.0018 & -0.0037 & 0.0037 & -0.0051 & 0.0025 & -0.0044 & 0.0044 \\
 $(C^{(3)}_{lq})_{22}$ & -0.015 & 0.013 & -0.098 & 0.098 & -0.020 & 0.017 & -0.092 & 0.092 \\
 $(C_{ed})_{21}$ & -0.0077 & 0.0037 & -0.0062 & 0.0062 & -0.013 & 0.0049 & -0.0072 & 0.0072 \\
 $(C_{ed})_{22}$ & -0.028 & 0.023 & -0.096 & 0.096 & -0.040 & 0.029 & -0.091 & 0.091 \\
 $(C_{ed})_{23}$ & -0.056 & 0.050 & -0.35 & 0.35 & -0.073 & 0.061 & -0.32 & 0.32 \\
 $(C_{ld})_{21}$ & -0.0067 & 0.0046 & -0.013 & 0.013 & -0.011 & 0.0065 & -0.015 & 0.015 \\
 $(C_{ld})_{22}$ & -0.027 & 0.024 & -0.20 & 0.20 & -0.037 & 0.032 & -0.19 & 0.19 \\
 $(C_{ld})_{23}$ & -0.055 & 0.052 & -0.72 & 0.72 & -0.070 & 0.064 & -0.65 & 0.65 \\
 $(C_{lu})_{21}$ & -0.0026 & 0.0062 & -0.0039 & 0.0039 & -0.0035 & 0.011 & -0.0046 & 0.0046 \\
 $(C_{lu})_{22}$ & -0.031 & 0.037 & -0.16 & 0.16 & -0.039 & 0.050 & -0.15 & 0.15 \\
 $(C_{eu})_{21}$ & -0.0017 & 0.0024 & -0.0019 & 0.0019 & -0.0021 & 0.0025 & -0.0022 & 0.0022 \\
 $(C_{eu})_{22}$ & -0.028 & 0.039 & -0.078 & 0.078 & -0.033 & 0.056 & -0.072 & 0.072 \\
 $(C_{qe})_{12}$ & -0.0025 & 0.0045 & -0.0048 & 0.0048 & -0.0034 & 0.0078 & -0.0056 & 0.0056 \\
 $(C_{qe})_{22}$ & -0.019 & 0.022 & -0.12 & 0.12 & -0.024 & 0.030 & -0.11 & 0.11 \\
 $(C_{qe})_{32}$ & -0.052 & 0.055 & -0.62 & 0.62 & -0.062 & 0.069 & -0.54 & 0.54  \\\hline
\end{tabular}
\caption{\label{tab:dilepton:1DlimitsExp} Expected 95\%CL intervals for all the EFT coefficients when taken one at a time, for the four different $\chi^2$ functions. We recall that $(C_{lq}^{(+)})_{\alpha 3}  \equiv (C_{lq}^{(1)})_{\alpha 3} + (C_{lq}^{(3)})_{\alpha 3}$.}
\end{table}

\begin{table}[p]
\centering
\begin{tabular}{| c | cc | cc | cc | cc |}
\hline
	95\%CL int. & 
	\multicolumn{2}{ c |}{$\chi^2_{m_{\ell\ell}, {\rm quad}}$} &  
	\multicolumn{2}{ c |}{$\chi^2_{m_{\ell\ell}, {\rm lin}}$} &  
	\multicolumn{2}{ c |}{$\chi^2_{m_{\ell\ell} |y_{\ell\ell}|, {\rm quad}}$} &
	\multicolumn{2}{ c |}{$\chi^2_{m_{\ell\ell} |y_{\ell\ell}|, {\rm lin}}$} 	\\ 
	& min & max & min & max & min & max & min & max \\ \hline
 $(C^{(1)}_{lq})_{11}$ & -0.00083 & 0.0033 & -0.00027 & 0.0071 & -0.00060 & 0.0048 &   0.00039 & 0.0084 \\
 $(C^{(1)}_{lq})_{12}$ & -0.012 & 0.011 & -0.47 & -0.029 & -0.014 & 0.011 & -0.47 &   -0.070 \\
 $(C^{(+)}_{lq})_{13}$ & -0.055 & 0.029 & -0.36 & -0.027 & -0.064 & 0.025 & -0.35 &   -0.057 \\
 $(C^{(3)}_{lq})_{11}$ & -0.0031 & 0.0010 & -0.0084 & 0.00033 & -0.0045 & 0.00088 &   -0.0099 & -0.00044 \\
 $(C^{(3)}_{lq})_{12}$ & -0.013 & 0.010 & -0.23 & -0.014 & -0.015 & 0.010 & -0.23 &   -0.034 \\
 $(C_{ed})_{11}$ & -0.0083 & 0.0013 & -0.014 & 0.00042 & -0.012 & 0.00025 & -0.017   & -0.0012 \\
 $(C_{ed})_{12}$ & -0.026 & 0.016 & -0.22 & -0.012 & -0.031 & 0.015 & -0.23 &   -0.033 \\
 $(C_{ed})_{13}$ & -0.048 & 0.037 & -0.81 & -0.058 & -0.053 & 0.036 & -0.79 &   -0.13 \\
 $(C_{ld})_{11}$ & -0.0067 & 0.0031 & -0.028 & 0.0010 & -0.0092 & 0.0031 & -0.034   & -0.0023 \\
 $(C_{ld})_{12}$ & -0.024 & 0.019 & -0.45 & -0.023 & -0.027 & 0.019 & -0.47 &   -0.066 \\
 $(C_{ld})_{13}$ & -0.045 & 0.040 & -1.7 & -0.11 & -0.049 & 0.040 & -1.6 & -0.25   \\
 $(C_{lu})_{11}$ & -0.00060 & 0.0069 & -0.00046 & 0.0086 & 0.00010 & 0.011 &   0.00016 & 0.010 \\
 $(C_{lu})_{12}$ & -0.022 & 0.033 & 0.023 & 0.37 & -0.021 & 0.038 & 0.055 & 0.37   \\
 $(C_{eu})_{11}$ & $1.7 \times 10^{-6}$ & 0.0087 & -0.00020 & 0.0043 & 0.00020 & 0.0087 & 0.00011   & 0.0050 \\
 $(C_{eu})_{12}$ & -0.016 & 0.038 & 0.012 & 0.18 & -0.013 & 0.046 & 0.027 & 0.18   \\
 $(C_{qe})_{11}$ & -0.0013 & 0.0047 & -0.00040 & 0.011 & -0.00092 & 0.0069 &   0.00063 & 0.013 \\
 $(C_{qe})_{21}$ & -0.014 & 0.020 & 0.017 & 0.27 & -0.014 & 0.023 & 0.041 & 0.27   \\
 $(C_{qe})_{31}$ & -0.039 & 0.046 & 0.12 & 1.4 & -0.038 & 0.048 & 0.23 & 1.4 \\
 $(C^{(1)}_{lq})_{21}$ & -0.00057 & 0.0022 & 0.00094 & 0.0073 & -0.0010 & 0.0046 &   -0.00077 & 0.0067 \\
 $(C^{(1)}_{lq})_{22}$ & -0.0086 & 0.0077 & -0.43 & -0.032 & -0.014 & 0.012 & -0.37 &   0.0029 \\
 $(C^{(+)}_{lq})_{23}$ & -0.038 & 0.023 & -0.33 & -0.016 & -0.061 & 0.028 & -0.28 &   -0.0012 \\
 $(C^{(3)}_{lq})_{21}$ & -0.0021 & 0.00068 & -0.0086 & -0.0011 & -0.0043 & 0.0012 &   -0.0079 & 0.00092 \\
 $(C^{(3)}_{lq})_{22}$ & -0.0090 & 0.0073 & -0.21 & -0.015 & -0.015 & 0.011 & -0.18 &   0.0015 \\
 $(C_{ed})_{21}$ & -0.0056 & 0.0010 & -0.014 & -0.0018 & -0.012 & 0.0015 & -0.014   & 0.00086 \\
 $(C_{ed})_{22}$ & -0.018 & 0.012 & -0.21 & -0.016 & -0.030 & 0.017 & -0.18 &   0.0020 \\
 $(C_{ed})_{23}$ & -0.034 & 0.027 & -0.74 & -0.037 & -0.052 & 0.037 & -0.64 &   -0.0027 \\
 $(C_{ld})_{21}$ & -0.0045 & 0.0021 & -0.029 & -0.0036 & -0.0087 & 0.0035 & -0.028   & 0.0018 \\
 $(C_{ld})_{22}$ & -0.016 & 0.013 & -0.43 & -0.034 & -0.027 & 0.020 & -0.37 &   0.0047 \\
 $(C_{ld})_{23}$ & -0.032 & 0.029 & -1.5 & -0.079 & -0.048 & 0.041 & -1.3 &   -0.0063 \\
 $(C_{lu})_{21}$ & -0.00044 & 0.0048 & 0.0012 & 0.0089 & -0.0012 & 0.010 & -0.0012   & 0.0080 \\
 $(C_{lu})_{22}$ & -0.017 & 0.023 & 0.022 & 0.34 & -0.023 & 0.036 & -0.0012 & 0.30   \\
 $(C_{eu})_{21}$ & 0.00040 & 0.0057 & 0.00057 & 0.0044 & -0.00057 & 0.0050 &   -0.00059 & 0.0039 \\
 $(C_{eu})_{22}$ & -0.014 & 0.026 & 0.011 & 0.17 & -0.016 & 0.043 & -0.00057 &   0.14 \\
 $(C_{qe})_{12}$ & -0.00087 & 0.0032 & 0.0014 & 0.011 & -0.0015 & 0.0066 & -0.0011   & 0.010 \\
 $(C_{qe})_{22}$ & -0.010 & 0.013 & 0.019 & 0.25 & -0.015 & 0.022 & -0.0016 & 0.22   \\
 $(C_{qe})_{32}$ & -0.029 & 0.033 & 0.042 & 1.3 & -0.039 & 0.048 & 0.0054 & 1.1 \\\hline
\end{tabular}
\caption{\label{tab:dilepton:1Dlimits} 95\%CL intervals for all the EFT coefficients when taken one at a time, for the four different $\chi^2$ functions built with the published 8TeV dilepton data. We recall that $(C_{lq}^{(+)})_{\alpha 3}  \equiv (C_{lq}^{(1)})_{\alpha 3} + (C_{lq}^{(3)})_{\alpha 3}$.}
\end{table}

When reading these results one should keep in mind the definition of the $C_i$ coefficients in Eq.~\eqref{eq:dilepton:defC}, $C_i \equiv c_i \frac{v^2}{ \Lambda^2 }$, therefore requiring a NP scale much above the EW scale implies $C_i \ll 1$. Furthermore, since the dilepton invariant mass goes up to 1.5 \TeV, the EFT expansion is valid if $\Lambda \gg 1.5 \TeV$. For reference, fixing $c_i = 1$ values of $C_i = \{ 10^{-1}, 10^{-2}, 10^{-3} \}$ correspond to $\Lambda = \{ 0.78, 2.5, 7.8 \} \TeV$, respectively. Larger values of $c_i$, i.e. a strongly coupled UV completion, imply also higher values of the NP scale for a given $C_i$, thus improving the EFT expansion validity.

Quadratic terms in the EFT dependence of the cross section, proportional to $(c^{(6)}_i)^2 / \Lambda^4$, are formally of the same order in $\Lambda$ as the interference of dimension-eight operators with the SM, which would be proportional to $c_i^{(8)} / \Lambda^2$. For weakly coupled UV completions, $c_i \lesssim 1$, it is thus inconsistent to include the former without the latter. Expecting the leading contribution to come from the linear dimension-6 terms, quadratic terms should not be included. This is the approach of the linear fits.
On the other hand, for strongly-coupled UV completions, $1 \ll c_i \lesssim 4\pi$, quadratic dimension-6 terms are enhanced compared to linear dimension-8 ones and it is thus possible to include them in the fit in a consistent way.
Unless some degeneracy is present in the fit, one would expect that if the EFT expansion is valid, linear and quadratic fits should give approximately the same results.

Comparing the expected (Tab.~\ref{tab:dilepton:1DlimitsExp}) and observed limits (Tab.~\ref{tab:dilepton:1Dlimits}) we observe that the latter are often quite asymmetric, especially for the linear fits. This is due to the deviations between the SM prediction and the data (see Tab.~\ref{tab:dilepton:mllBins}), which can push the coefficients in a specific direction depending on the sign of the interference term, while the quadratic terms are always necessarily positive.
In case of operators with first-generation quarks, the linear and quadratic fits differ only by some $\mathcal{O}(1)$ factor and the limits are at the per-mille level on the $C_i$'s, implying NP scales above few TeV, well above the partonic collision energy.
In the case of operators with heavier quarks, instead, the difference between linear and quadratic fits becomes very large, with the limits from the linear fits becoming very weak, hinting a possible breakdown of the EFT validity.
Some of these limits could possibly still be applied in strongly coupled UV completions, however a more detailed analysis would require a direct comparison with some explicit UV models, which is beyond the scope of this work.

We also observe that the constraints from the 2D distribution are of the same order (or somewhat weaker) than those obtained using only the invariant mass distribution. As already mentioned in Sec.~\ref{sec:dilepton:EFT}, this is due to the merging of the high-invariant mass bins with lower invariant masses, which dilutes the sensitivity on the EFT coefficients of the bins for which it is would be the strongest. In this case, the gain from the extra information on the rapidity distribution is not enough to offset this effect. On the other hand, we expect that the main benefit of the rapidity distribution is in disentangling the contributions from operators with different chiral structures.

 \begin{figure}[p]
  \centering
  \includegraphics[width=0.38\textwidth]{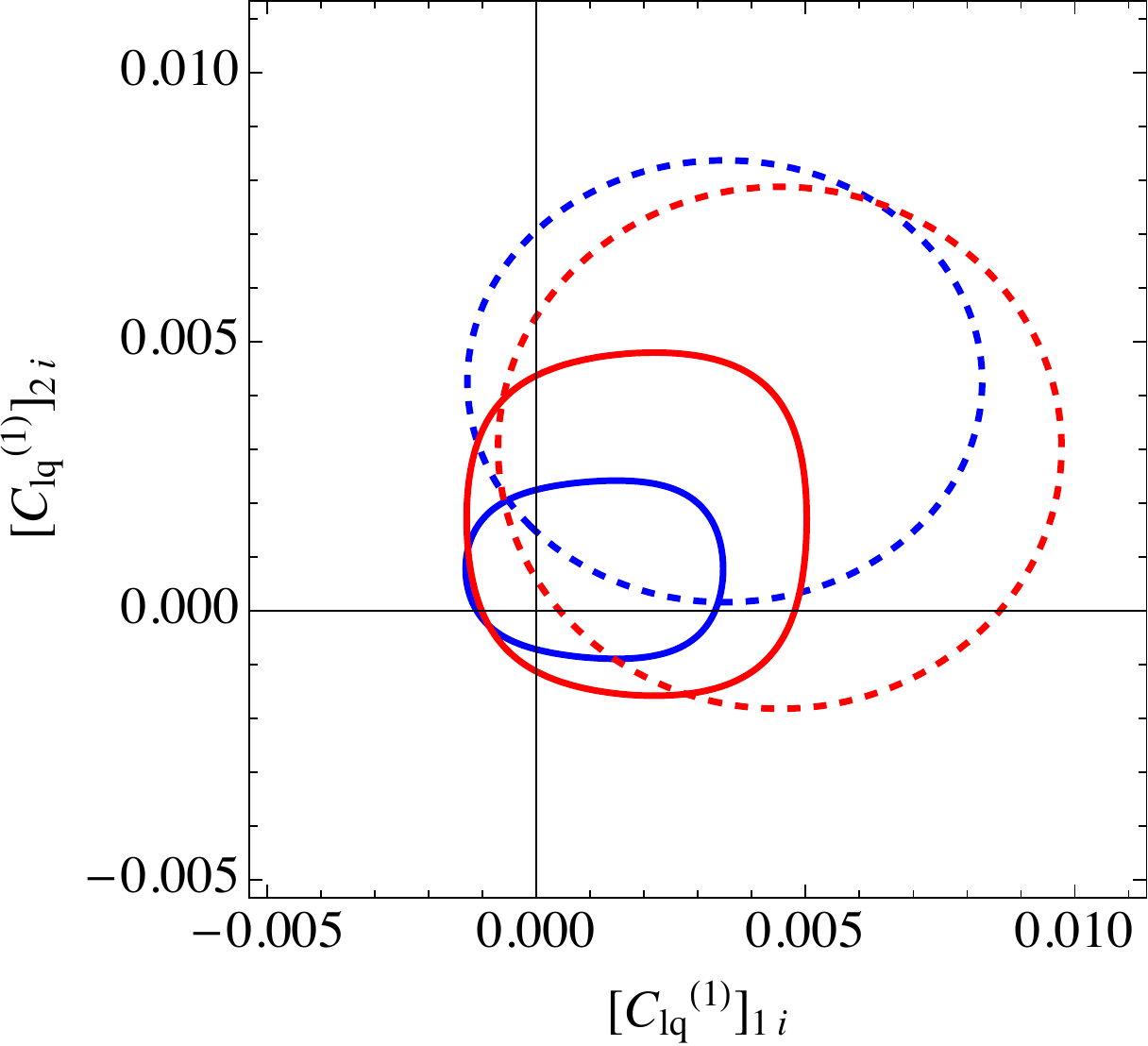} \hspace{0.11\textwidth}
  \includegraphics[width=0.18\textwidth]{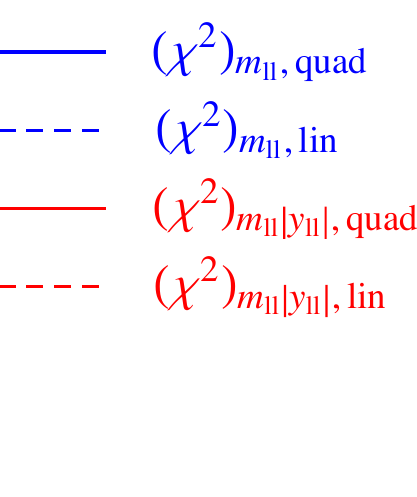} \hspace{0.1\textwidth} \text{ }\\[5pt]
  \includegraphics[width=0.4\textwidth]{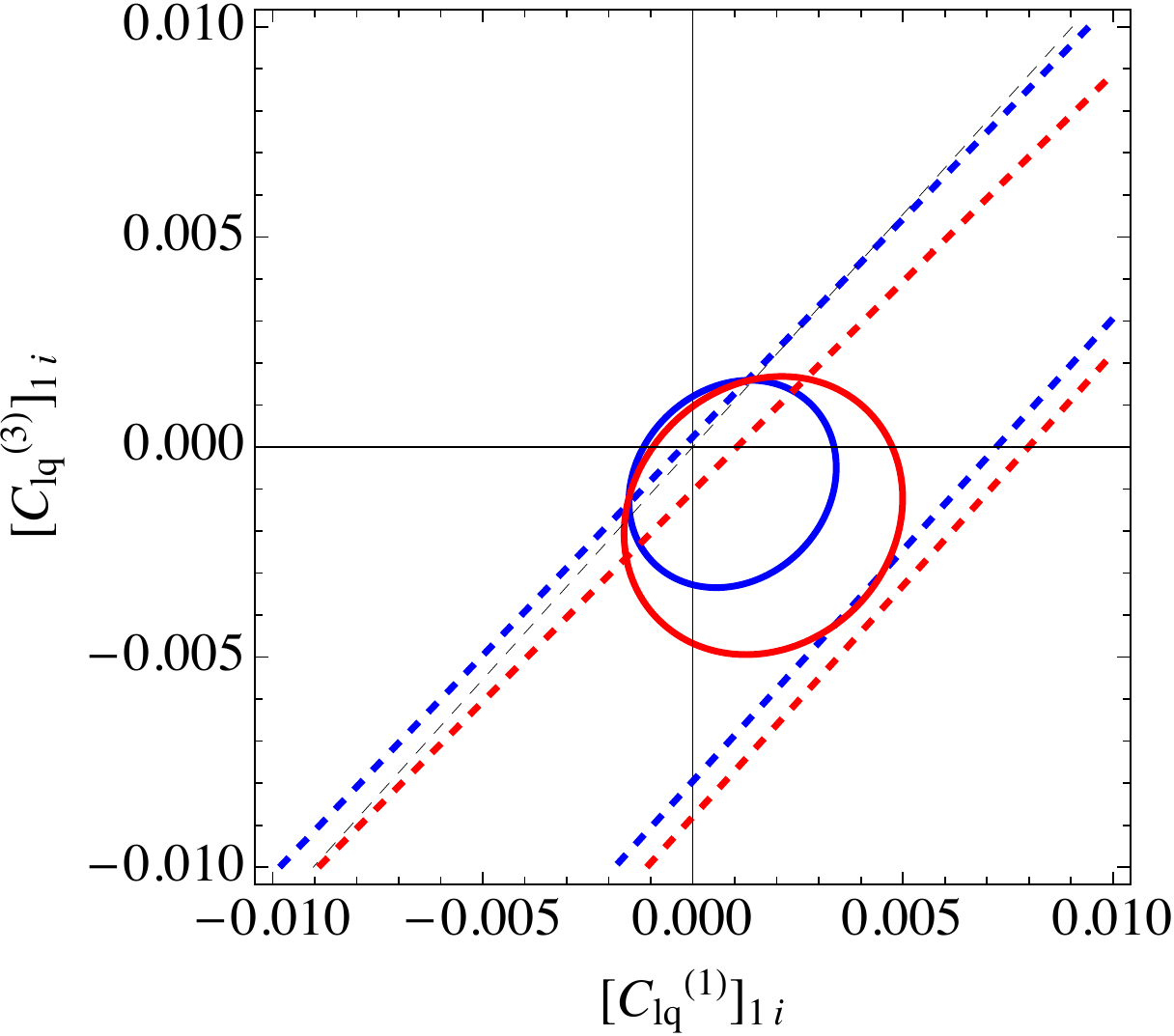} 
  \includegraphics[width=0.39\textwidth]{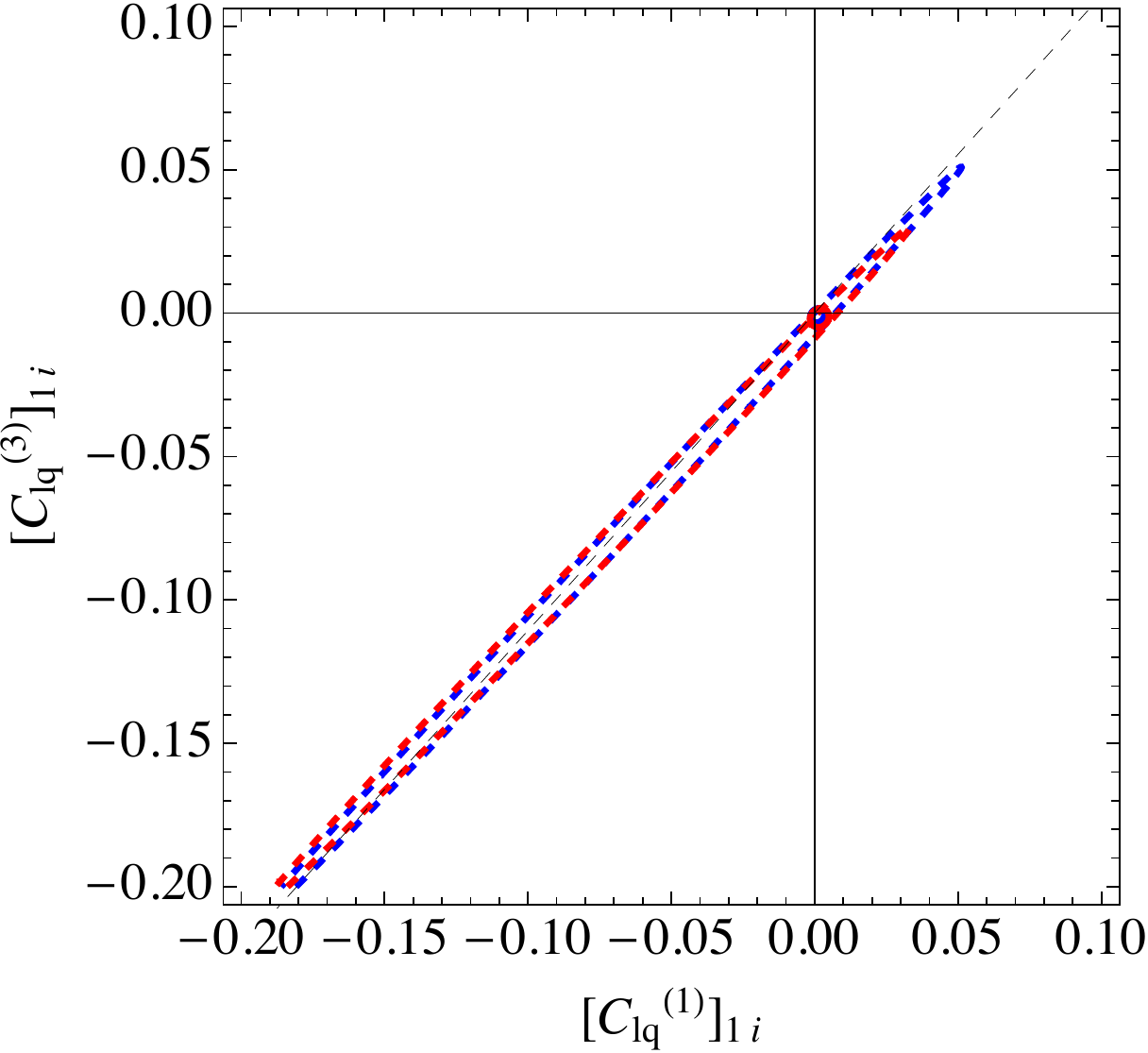} \\[5pt]
  \includegraphics[width=0.4\textwidth]{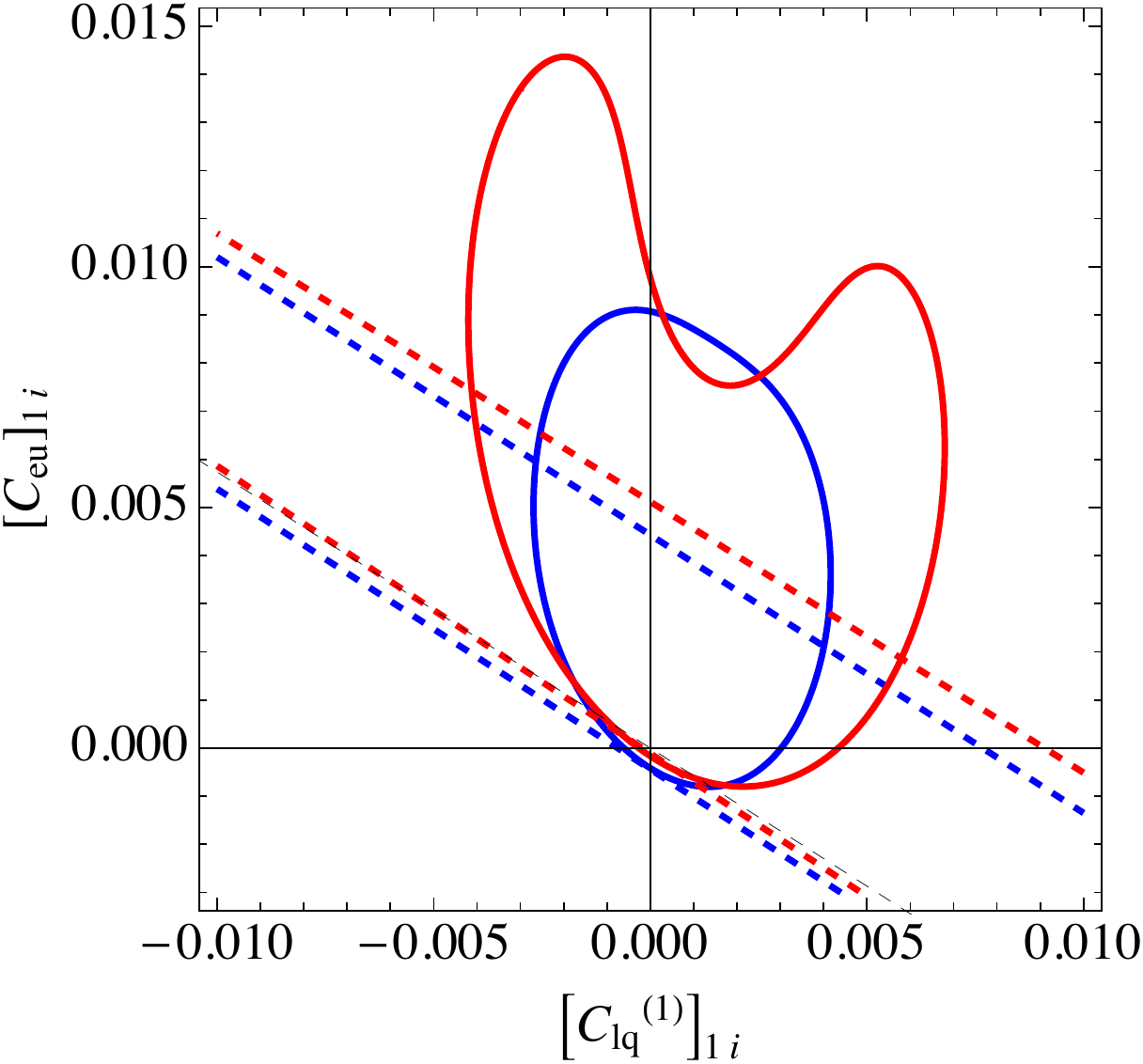} 
  \includegraphics[width=0.4\textwidth]{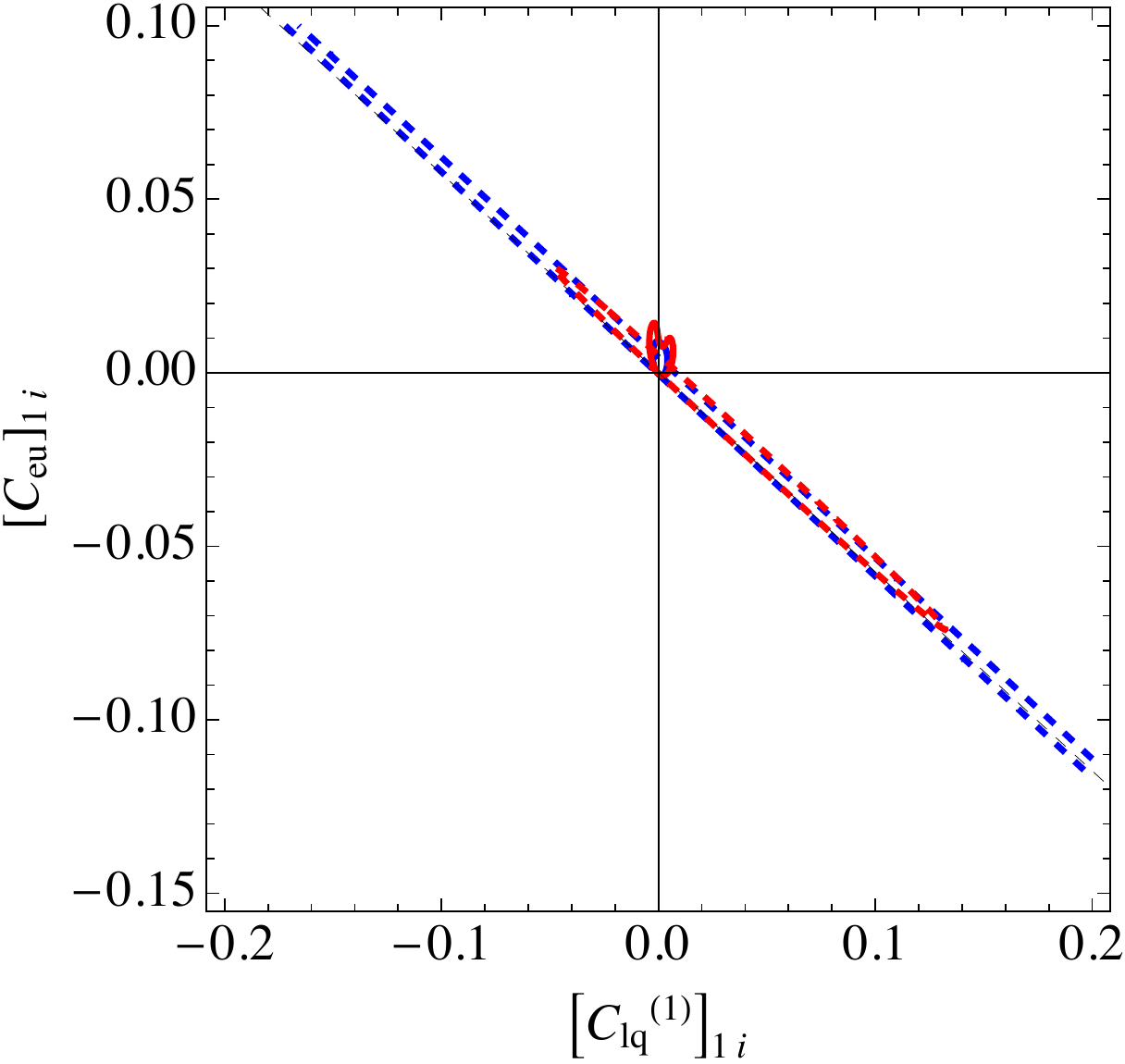} 
  \caption{95\%CL contours in some 2D spaces of EFT coefficients with quadratic (solid) or linear (dashed) EFT dependence using only the dilepton invariant mass distribution (blue) or also the information on the absolute rapidity of the dilepton system (red). The thin dashed black line corresponds to the direction in EFT space which leaves the linear dependence of the highest invariant mass bin unchanged, Eq.~\eqref{eq:dilepton:flatDir}. Right plots are zoomed-out version of the left ones.}
\label{fig:dilepton:2Dfits1}
\end{figure}

 \begin{figure}[t]
  \centering
   \includegraphics[width=0.18\textwidth]{./plots_dilepton/legend} \\[-25pt]
  \includegraphics[width=0.4\textwidth]{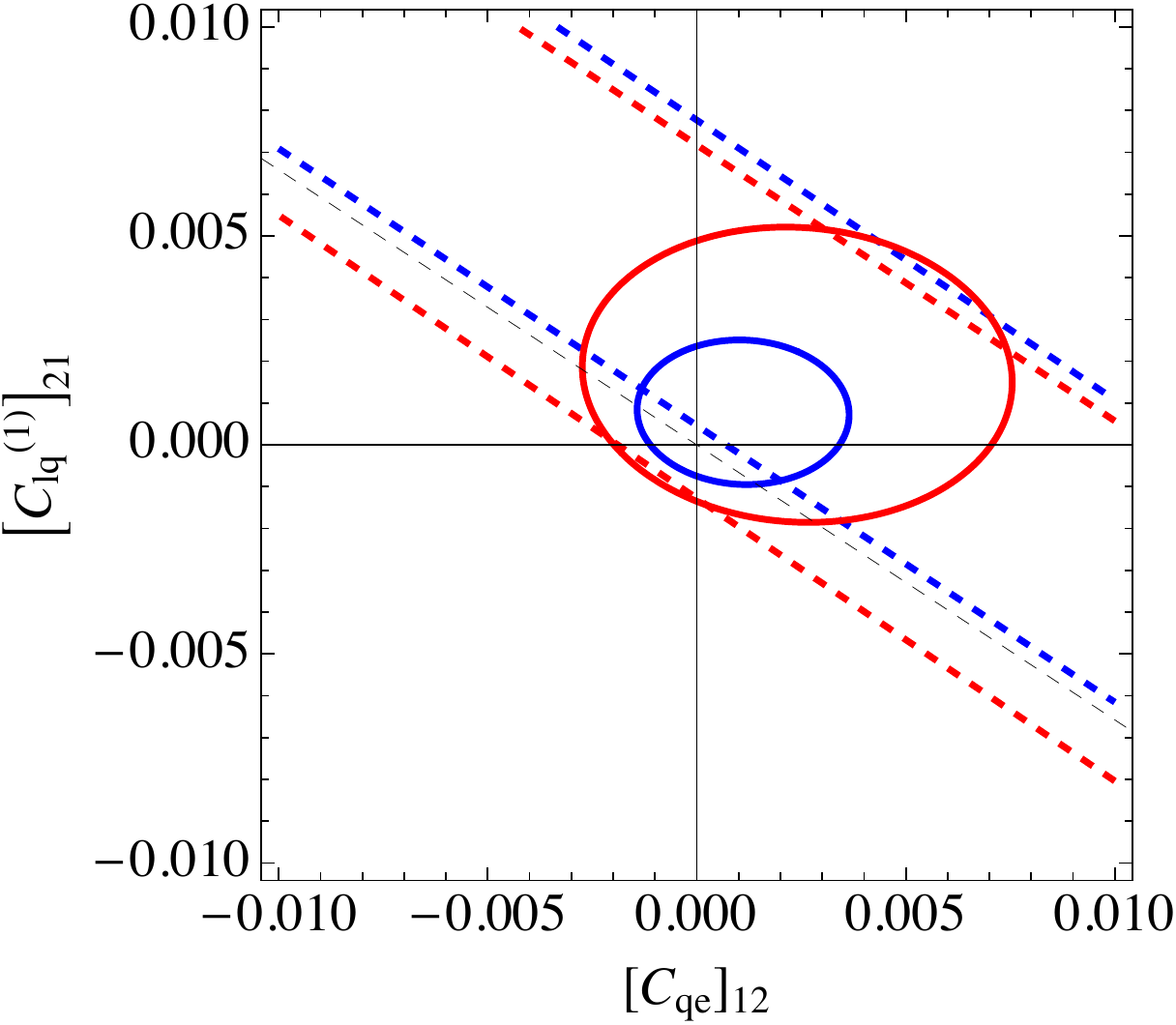} 
  \includegraphics[width=0.36\textwidth]{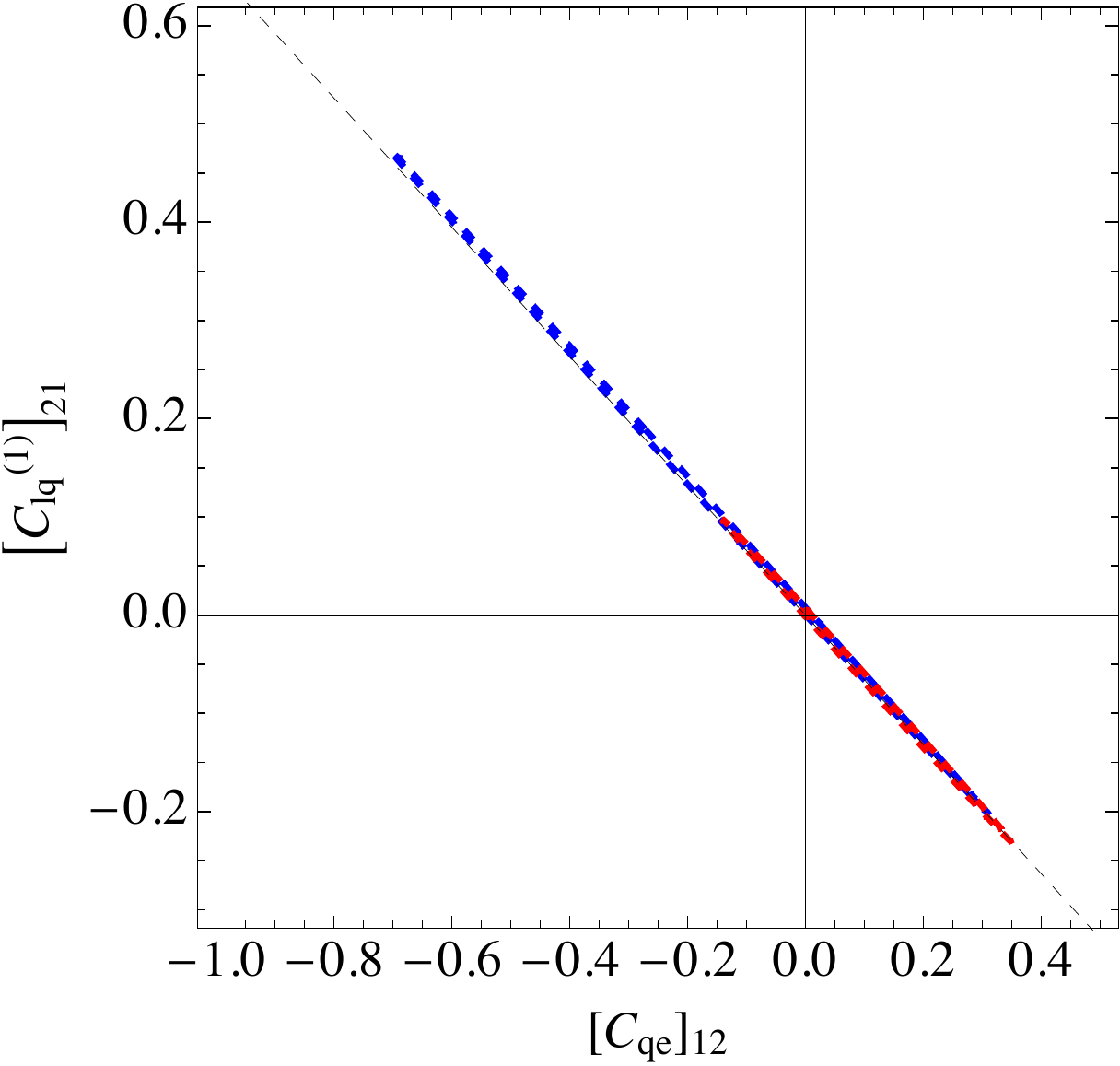} \\[5pt]
  \includegraphics[width=0.4\textwidth]{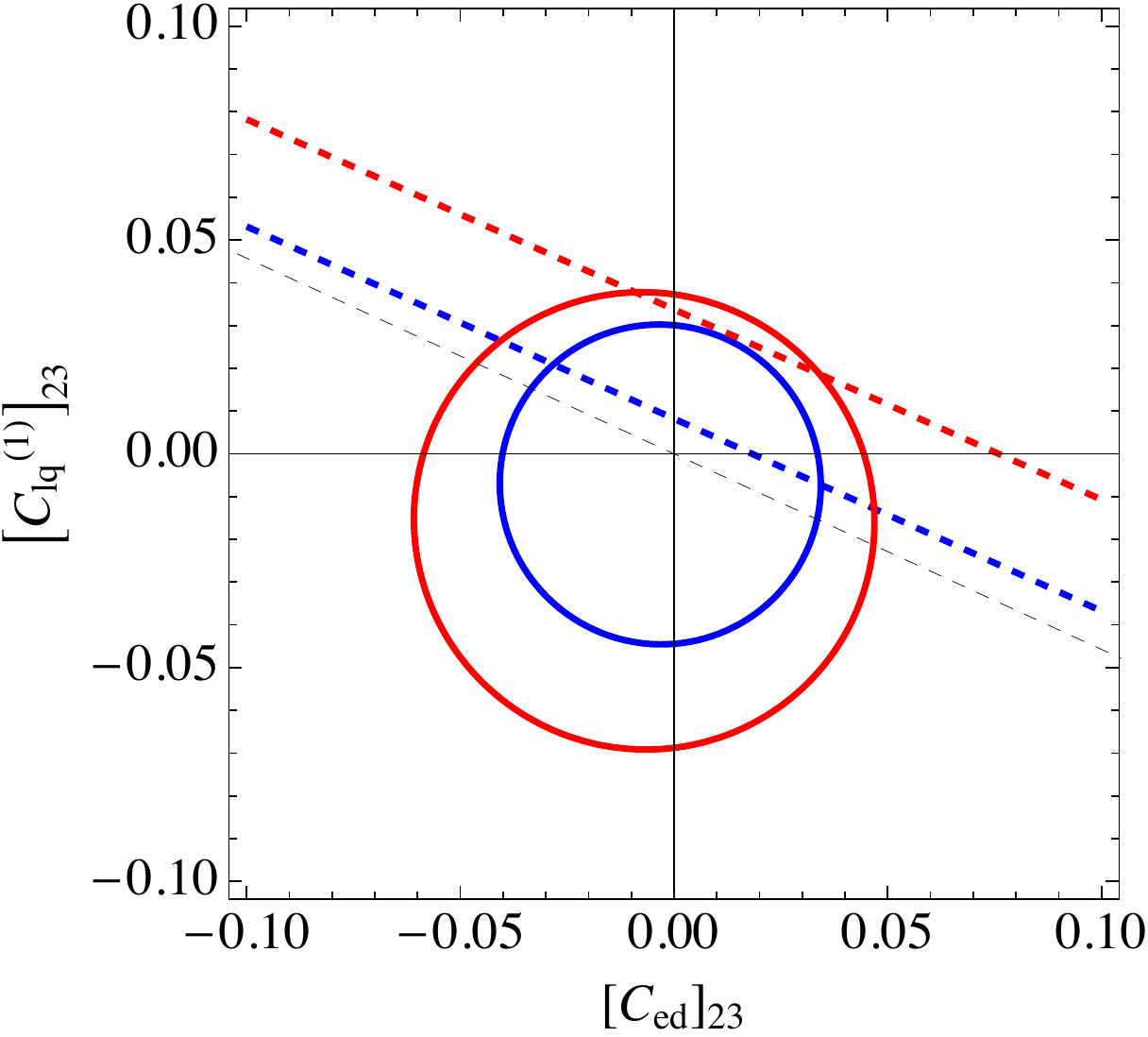} 
  \includegraphics[width=0.38\textwidth]{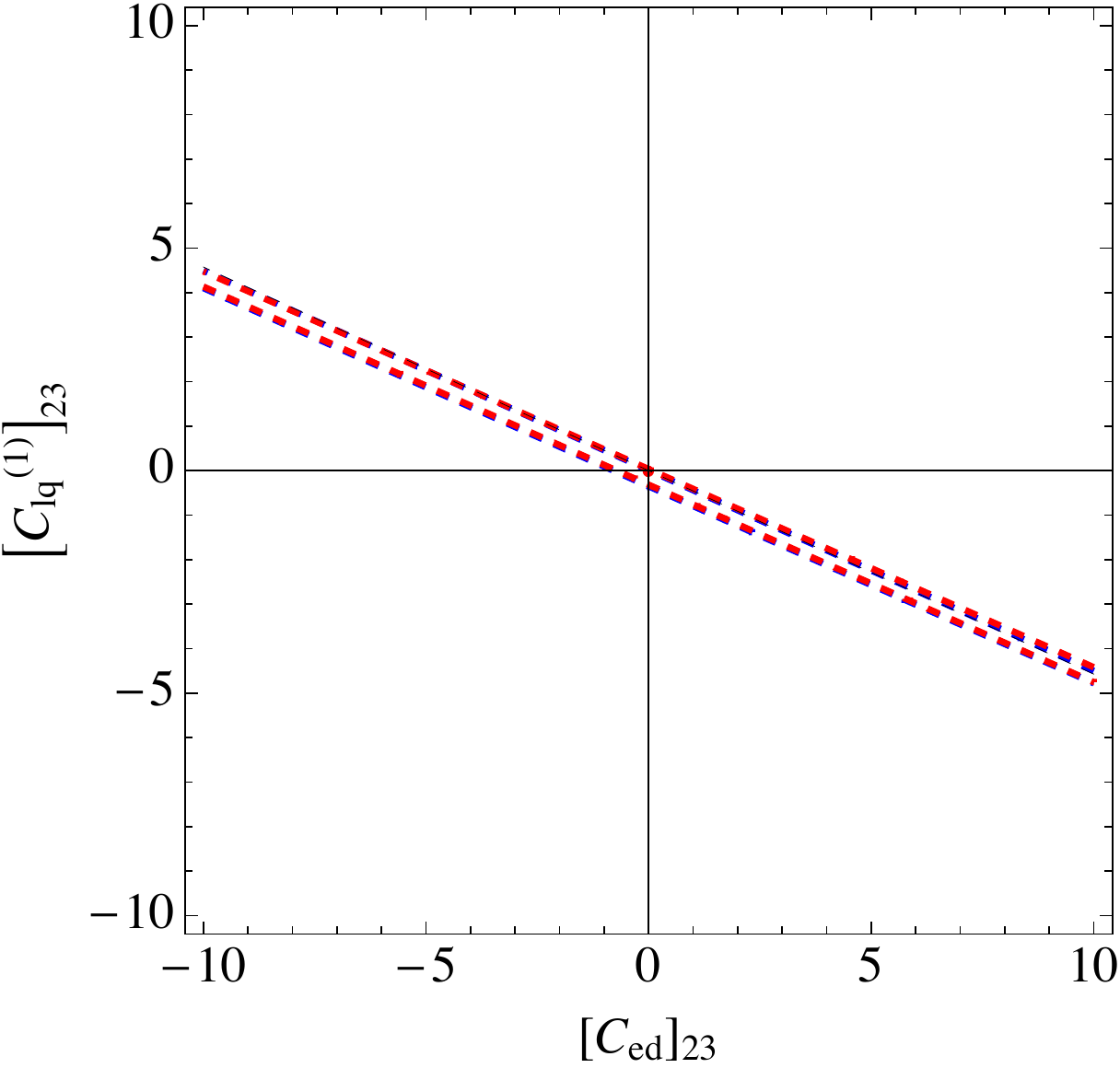}
  \caption{95\%CL contours in some 2D spaces of EFT coefficients with quadratic (solid) or linear (dashed) EFT dependence using only the dilepton invariant mass distribution (blue) or also the information on the absolute rapidity of the dilepton system (red). The thin dashed black line corresponds to the direction in EFT space which leaves the linear dependence of the highest invariant mass bin unchanged, Eq.~\eqref{eq:dilepton:flatDir}. Right plots are zoomed-out version of the left ones.}
\label{fig:dilepton:2Dfits2}
\end{figure}

In order to show this, and to show the presence of large degeneracies in the linear likelihoods, we perform a few fits keeping two EFT coefficients at the same time. The observed 95\% CL limits in the plane of the two coefficients are shown in Figs.~\ref{fig:dilepton:2Dfits1},\ref{fig:dilepton:2Dfits2}.

The top-left plot in Fig.~\ref{fig:dilepton:2Dfits1} shows a test of lepton-flavor-universality assuming flavor-independent coefficients in the quark sector. The two combinations of coefficients we test are $(C_{lq}^{(1)})_{1 i}$ vs. $(C_{lq}^{(1)})_{2 i}$. In this case we see that the linear and quadratic fits differ only by a $\mathcal{O}(1)$ factor (constraints from the linear fit being the weakest). The deviation from the SM in the linear fits is due to the deficit in the measured cross section in both channels compared to the SM prediction seen in Tab.~\ref{tab:dilepton:mllBins}. Since the quadratic contributions are strictly positive and are observed to dominate the quadratic fit, in this case the observed deviation cannot be completely saturated and thus the resulting fits are more compatible with the SM.

In all the other plots we see that the quadratic fits provide quite strong constraints and that those from the 2D distribution offer weaker limits, due to the wash-out effect discussed above.
The linear fits, on the other hand, show very shallow flat directions, along which the constraining power is mostly lost.
These flat directions can be understood by looking at the linear EFT dependence of the highest invariant mass bin, of $m_{\ell \ell} \in [1000 -1500] \GeV$, which is proportional to (normalised to the SM contribution):
\be\begin{split}
	K_{\text{flat dir}} &= - 73.0 (C^{(1)}_{lq})_{\alpha 1} 
		+ 0.765 (C^{(1)}_{lq})_{\alpha 2}
		+ 62.1 (C^{(3)}_{lq})_{\alpha 1}
		+ 1.55 (C^{(3)}_{lq})_{\alpha 2}
		+ 0.830 (C^{(+)}_{lq})_{\alpha 3} \\
	&	+ 36.6 (C_{ed})_{\alpha 1}
		+ 1.66 (C_{ed})_{\alpha 2}
		+ 0.380 (C_{ed})_{\alpha 3}
		- 124 (C_{eu})_{\alpha 1}
		- 1.85 (C_{eu})_{\alpha 2} \\
	&	+ 18.1 (C_{ld})_{\alpha 1}
		+ 0.822 (C_{ld})_{\alpha 2}
		+ 0.188 (C_{ld})_{\alpha 3}
		- 61.5 (C_{lu})_{\alpha 1}
		- 0.922 (C_{lu})_{\alpha 2} \\
	&	- 48.0 (C_{qe})_{1 \alpha}
		- 1.31 (C_{qe})_{2 \alpha}
		- 0.189 (C_{qe})_{3 \alpha}~.
	\label{eq:dilepton:flatDir}
\end{split}\ee
The $K_{\text{flat dir}} = 0$ direction is shown in the plots as dashed gray line and we see that it describes very well the flat direction of the fits done in the linear approximation, also when including the rapidity distribution information.
This reflects the expectation that most of the sensitivity on the EFT operators comes from the highest invariant mass bin.
Comparing the dashed blue (linear fit from the 1D distribution) with the dashed red (linear fit with including rapidity information) contours, we see that the extra information from the rapidity is able to partially alleviate the flat directions, thanks to the different dependence of different operators, as shown in Fig.~\ref{fig:dilepton:rapd} (left). We expect this positive effect to substantially improve with more precision.

\section{Conclusions}

The high-mass dilepton invariant mass distribution in Drell-Yan production $p p \to \ell^+ \ell^-$ is a particularly powerful probe of New Physics at the LHC. If the scale of NP is much higher than the energy scale of the process, a model-independent approach to describe the effect of heavy states is provided by Effective Field Theories.
We used ATLAS 8 TeV data with 20.3 fb$^{-1}$ of luminosity \cite{Aad:2016zzw} to constrain the complete set of dimension-six operators with non-vanishing interference with the SM amplitude, listed in Tab.~\ref{tab:dilepton:EFTcoeff}.
We obtain, and make public, four likelihoods: using the differential cross section in invariant mass or in invariant mass and dilepton rapidity, and for both including or excluding the quadratic dependence of the cross section on the EFT coefficients. The four files are listed in Tab.~\ref{tab:dilepton:chi2func} and can be downloaded at \href{https://people.sissa.it/~dmarzocc/dileptonATLAS8TeVchiSQ.zip}{this link}.

To illustrate the possible uses of these likelihoods we derive the 95\%CL limits for each EFT coefficient when taken one at a time, Tab.~\ref{tab:dilepton:1Dlimits}, and for some examples of pairs of coefficients, Figs.~\ref{fig:dilepton:2Dfits1}~and~\ref{fig:dilepton:2Dfits2}.
These examples show that the fits performed using the linear dependence provide constraints often much weaker than those obtained with the quadratic dependence, a behaviour that suggests possible issues with the EFT validity. More precise data, such as that collected at 13 TeV, is expected to alleviate this problem.
We also observe that in the single-operator fits and in all the quadratic fits, using the two-dimensional distribution provides worse constraints than using only the dilepton invariant mass. This is due to the merging of $m_{\ell\ell}$ bins which is done in the 2D distribution, which washes out some of the EFT sensitivity. Again, a larger dataset with more granular binning in invariant mass should address this issue. On the other hand, the extra information coming from the rapidity distribution is able to lift part of the degeneracy of the linear fits done with pair of EFT coefficients.
For these reasons we strongly advice the experimental collaborations against merging the highest invariant-mass bins in future analyses. A better solution, for the highest invariant-mass bins, could rather be to merge the angular distributions bins.

The likelihoods obtained in this work can be easily used to put constraints from the dilepton distribution in a wide class of NP models. We plan to update the results with 13 TeV data when the corresponding experimental analyses are available.

\section*{Acknowledgements}

The authors thank the organizers of the Les Houches workshop for the hospitality and for fostering a very friendly and productive atmosphere.
DM acknowledges partial support from the INFN grant SESAMO and MIUR grant PRIN\_2017L5W2PT. DB is supported by the MIUR grant PRIN\_2017L5W2PT and the INFN grant FLAVOR. The work of JD is partially supported by funding available from the Department of Atomic Energy, Government of India, for the Regional Centre for Accelerator-based Particle Physics (RECAPP), Harish-Chandra Research Institute (HRI) and by the INFOSYS scholarship for senior students at HRI. JD also acknowledges support by the Deutsche Forschungsgemeinschaft (DFG, German Research Foundation) under Germany's Excellence Strategy EXC 2121 "Quantum Universe"- 390833306.


\let\be\undefined
\let\ee\undefined
\let\MeV\undefined
\let\GeV\undefined
\let\TeV\undefined
\let\SM\undefined
\let\OO\undefined
\let\LL\undefined
\let\fr\undefined
\let\ma\undefined
\let\mg\undefined
\let\rot\undefined
\let\DM\undefined

%% file: singlelq/singlelq.main.tex
\graphicspath{{singlelq/}}


\def\be{\begin{equation}}
\def\ee{\end{equation}}
\def\bsp#1\esp{\begin{split}#1\end{split}}
\def\ttbar{\ensuremath{t\bar{t}}\;}
\def\ttbarZ{\ensuremath{t\bar{t}Z}\;}
\def\etmiss{\ensuremath{E_{\mathrm{T}}^{\mathrm{miss}}}\;}
\def\pt{\ensuremath{p_{\mathrm{T}}}\;}
\def\W{\ensuremath{W}\;}
\def\btagged{\ensuremath{b}-tagged\;}
\def\lag{{\cal L}}
\def\laR{{\bf \lambda_{\sss R}}}
\def\sss{\scriptscriptstyle}
\def\eR{\ell_{\sss R}}
\def\uRbar{{\bar u}_{\sss R}}
\def\laLuno{{\bf \lambda_{\sss 1L}}}
\def\Ll{L_{\sss L}}
\def\QLbar{\bar Q_{\sss L}}
\def\laLtre{{\bf \lambda_{\sss 3L}}}
\def\fr{{\sc \small FeynRules}}


\chapter{LHC sensitivity to leptoquark production in $\tau$+heavy quark final states}
{\it D.~Buttazzo,
  D.~Marzocca,
  M.~Nardecchia,
  P.~Pani,
  G.~Polesello}


\label{sec:singlelq}

\begin{abstract}
The anomalies observed by LHCb in flavor observables suggest the breakdown of lepton flavor universality. We study how a leptoquark solution can be tested at the LHC through final states including a the hadronic decay of a $\tau$ lepton, a heavy quark ($b$ or $t$) and \etmiss
\end{abstract}

\section{INTRODUCTION}

The deviations from the SM predictions observed by Babar, Belle, and LHCb in the Lepton-Flavour Universality (LFU) ratios $R_D$ and $R_{D^*}$ \footnote{See e.g. the HFLAV combination updated in the Spring of 2019 \cite{Amhis:2016xyh} \\ \href{https://hflav-eos.web.cern.ch/hflav-eos/semi/spring19/html/RDsDsstar/RDRDs.html}{{\tt https://hflav-eos.web.cern.ch/hflav-eos/semi/spring19/html/RDsDsstar/RDRDs.html}}.} are compelling hints for the presence of New Physics (NP) beyond the SM at a scale close to the TeV scale. The most successful explanations consist in scalar or vector leptoquark (LQ) states, with a larger coupling to taus than to muons.

The low scale of NP hinted by the observed deviations in flavour physics implies that direct searches at the LHC could be potentially sensible to these LQ. The processes most sensitive to such states are: i) pair-production followed by on-shell decays, ii) single production in association with a charged lepton or neutrino and iii) off-shell exchange in the t-channel of a LQ in the partonic process $p p \to \tau \nu (+j)$ or $\tau \tau (+j)$. See e.g. \cite{Dorsner:2018ynv} for a review.

At present, the limits from pair-production of third-generation scalar (vector) LQ -- i.e. LQ coupled mainly to third-generation fermions -- range between $\sim 0.7 - 1$~TeV ($\sim 1-1.5$~TeV) depending on the specific decay channel considered.
Single production is potentially more sensitive for larger masses and larger couplings. Experimental searches in this mode are much more limited, consisting only in a CMS search in the $b\tau$ decay channel with another associated $\tau$ \cite{Sirunyan:2018jdk}.
It should be noted that a LQ of electric charge $Q = 1/3$, such as the $S_1$ introduced below, does not contribute to this process.
For this reason in this work we consider the $b \nu$ (with associated $\tau$) or the $c \tau$ (with associated $\nu$) channels, both included in the process $p p \to \tau \nu + b{\rm -jet}$.\footnote{Note that the $b$-tag also final states with a $c$-jet will be selected.}
We devise such an analysis for LHC searches and obtain prospects for HL-LHC with 3~ab$^{-1}$ of luminosity.
Although in a different kinematical region, corresponding to a forward $b$-jet, the same process is also sensitive to the same partonic transition which takes place in the anomalous flavour process: $b \to c \tau \nu$. It was shown in Ref.~\cite{Greljo:2018tzh} that in several NP scenarios, LHC searches of $p p \to \tau \nu$ should probe the parameter region responsible for the anomalies. The $b$-tagging requirement should further increase the signal to background ratio and therefore the sensitivity to such NP compared to that work.

Due to $SU(2)_L$ gauge invariance, a coupling of a LQ to the left-handed bottom also implies a coupling to the left-handed top quark. It is thus also interesting to consider single production of scalar or vector LQ in final states with one top. In the second part of this work we devise such an analysis and obtain prospects for LHC searches with 3000~fb$^{-1}$  of luminosity.

\section{THE MODELS AND SOLUTIONS TO THE $R_{D^{(*)}}$ ANOMALIES}
\label{sec:singlelq:model}
In this work we consider a model with a pair of scalar LQs.

The scalar LQ model we implement comprise $S_1 \equiv (\overline{\mathbf{3}},\mathbf{1},1/3)$ and $S_3 \equiv (\overline{\mathbf{3}},\mathbf{3},1/3)$ with masses $M_{S_1}$ and $M_{S_3}$, where we specify transformation properties under the SM gauge group. First (second) integer in the brackets corresponds to the dimension of the irreducible representation of $SU(3)$ ($SU(2)$) that the LQ belongs to whereas the rational number is the LQ $U(1)$ hypercharge (normalised as $Q = T^{3}_L + Y$). We assume that lepton number and baryon number are conserved quantities.
The interaction of the two LQs with SM fermions is described by the Lagrangian
\be\bsp
 \lag_{\rm int}^{S_1 + S_3} = & 
    (\laR)_{i \alpha} (\uRbar^c)_i (\eR^{\phantom{c}})_\alpha \ S_1
  + (\laLuno)_{i \alpha} (\QLbar^c)_i \epsilon (\Ll^{\phantom{c}})_\alpha \ S_1
  + (\laLtre)_{i \alpha} (\QLbar^c)_i \epsilon \sigma^a (\Ll^{\phantom{c}})_\alpha \ S_3^{a}
   + {\rm h.c.} ~, 
\esp\label{eq:singlelq:lagS1S3}\ee
where $\epsilon$ is the antisymmetric matrix in $SU(2)_L$ indices and $i$ ($\alpha$) is a quark (lepton) flavor index.
All gauge indices are understood for simplicity.



For the event generation we use the \fr\ implementation of both models from \cite{Dorsner:2018ynv}.

Various combination of the couplings in the two LQ model can be used to solve the $R_{D^{(*)}}$ anomalies.
$S_1$ by itself can successfully address the anomaly with the couplings $(\laLuno)_{33} \lesssim 1$, $(\laR)_{23} \sim \mathcal{O}(1)$, and a mass $M_{S_1} \sim 1$~TeV. This solution is considered in another Les Houches proceeding so we will not consider it in the following.

Another possibility to address the flavour anomalies is to use only the $S_1$ couplings to LH fermions. In this case it is necessary to also include $S_3$ in order to evade a strong constraint from $B \to K^* \nu \nu$ \cite{Buttazzo:2017ixm}. A good parameter region able to fit the anomalies in this setup is given by $M_{S_1} \sim M_{S_3} \sim 1$~TeV, $(\laLuno)_{33} \sim (\laLtre)_{33} \sim 1$, and $(\laLuno)_{23} \sim - (\laLtre)_{23} \sim 1$. This model is particularly interesting since the $S_3$ LQ can easily address also the flavour anomalies in neutral-current $B$ decays via the $(\laLtre)_{22}$ and $(\laLtre)_{32}$ couplings \cite{Buttazzo:2017ixm}.\footnote{However, the couplings needed to fit those anomalies are too small to have an impact in the LHC searches considered here.}
\begin{figure}[t]
  \centering
  \includegraphics[width=0.47\textwidth]{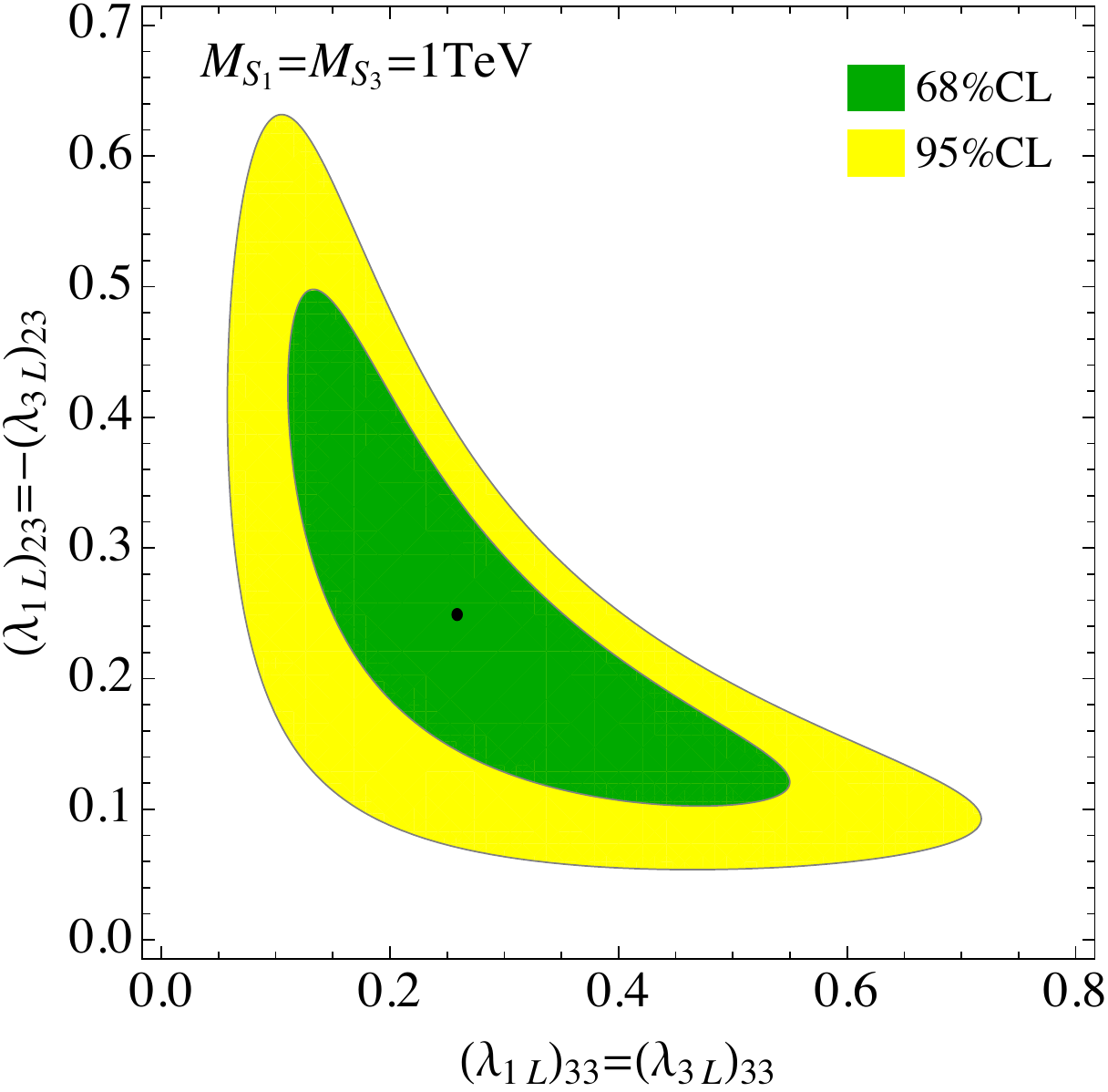} 
  \caption{\it Flavor fit to the $R_{D^{(*)}}$ anomalies and related observables in the $S_1 + S_3$ model.}
\label{fig:singlelq:fitS1S3}
\end{figure}
Other relevant observables putting strong constraints in this setup are: $Z \to \tau\tau, \nu\nu$, $\tau$ LFU tests, and $B_s$ mixing. The latter, in particular, is a very strong bound which puts a mild tension with $R_{D^{(*)}}$ \cite{Marzocca:2018wcf}.
The region in the  ($(\lambda_{\sss 1(3)L})_{33}$, $(\lambda_{\sss 1(3) L})_{23}$) plane compatible with the flavour anomalies and these other observables at 68 and 95\%CL from the best fit point is shown in Fig.~\ref{fig:singlelq:fitS1S3} for a simplified model comprising both $S_1$ and $S_3$, with approximately degenerate masses and equal (or opposite) couplings. In our collider analysis we focus on such a parameter region, but we include only the contribution from one LQ (e.g. $S_1$). This is motivated by the fact that in a more realistic scenario it is unlikely the two LQ will be exactly degenerate, but indeed a splitting of a few hundred GeV is to be expected (e.g. due to radiative EW corrections to the mass \cite{Marzocca:2018wcf}) and therefore the contribution to collider searches will be dominated by whichever of the two states is the lighter.\par


\section{SIMULATION OF LHC EVENTS.}
The simulation of the LQ signal and the SM background follows the
strategy introduced in Ref.~\cite{Pani:2017qyd}, to which we refer 
for additional details.\\ 
We simulate a full set of SM processes
leading to the presence of one or two final-state leptons ($e$, $\mu$, $\tau$)
originating from the decay of a $W$-boson, a $Z$-boson or a $\tau$ lepton.
Events are generated within either the {\tt POWHEG~BOX}
framework~\cite{Alioli:2010xd} or the {\tt MadGraph5\_aMC@NLO}
platform~\cite{Alwall:2014hca}, and the simulation of the QCD environment
(parton showering and hadronisation) has been performed with
{\tt PYTHIA 8}~\cite{Sjostrand:2014zea}.
Hard-scattering signal events have been produced with {\tt MadGraph5\_aMC@NLO}
on the basis of the UFO model~\cite{Degrande:2011ua} provided together with
Ref.~\cite{Dorsner:2018ynv}, and parton showering and hadronisation have been
again simulated with {\tt PYTHIA 8}.
We finally include detector effects by smearing the properties of
the final-state physics objects ({\it i.e.} electrons, muons, jets and
$\etmiss$) in a way reproducing the measured performance of the
ATLAS detector. We provide here some more details on the simulation
of $\tau$-tagging and flavour-tagging, as they are of key importance
for the following discussion. The parametrisation of the ATLAS $b$-tagging
algorithm is based on the performance figures provided in Ref.~\cite{ATLAS-CONF-2014-046}. For the analyses presented in the following  
a working point is chosen 
yielding a 77\% efficiency for $b$-jets, the efficiency for $c$-jets
will be 22\%, whereas the light jets are suppressed by a factor of 140.
The parametrisation of the $\tau$-tagging performance is taken 
from \cite{ATL-PHYS-PUB-2019-005}. The working point used in the analysis
has an average efficiency for the identification of hadronic tau decays
of approximately 80\% (70\%) for 1(3)-prong $\tau$ decays. The 
assumed rejection factor for light jets is taken as 80 (500) for 1(3)-prongs.\par
\section{ANATOMY OF SIGNAL: $\tau$+HEAVY QUARK+\etmiss}
Considering only the production of $S_1$ leptoquarks, 
the final state $\tau$ + heavy quark + \etmiss can be 
produced in the model described in Section~\ref{sec:singlelq:model}
in three different ways:
\begin{itemize}
\item
Through the pair production of two leptoquarks, with subsequent mixed decay:\\
$pp \rightarrow S_1 S_1^\dag, S_1\to b \nu_{\tau}, S_1^\dag\to \tau t$.
\item
Through the single production processes with a $b/c$ quark in final state:
\vskip 2mm
\begin{itemize}
\item
$gc \to \tau \nu_{\tau} b$
\item
$gs/gb \to \nu_{\tau} \tau c$\\ 
as discussed above, for the implementation 
of the ATLAS $b$-tagging used in this analysis, a significant fraction
of these events will pass the analysis requirement of the presence of 
a $b$-tagged jet in the event.

The dominant signal Feynman diagrams for respectively the $gc$ and $gs(b)$
partonic initial states  are shown in Figures~\ref{fig:singlelq:gc} and \ref{fig:singlelq:gs}.
\begin{figure}
\centering
\includegraphics[width=0.30\textwidth]{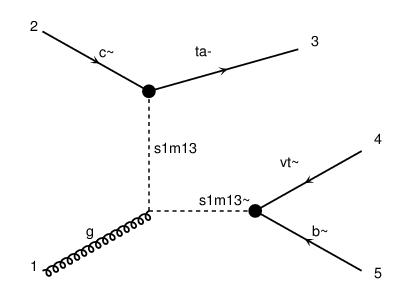}
\includegraphics[width=0.30\textwidth]{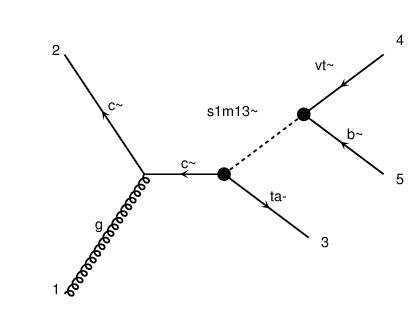}
\includegraphics[width=0.30\textwidth]{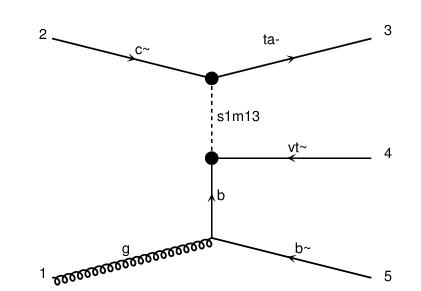}
\caption{\it Representative Feynman diagrams for the $gc$ initial state.}
\label{fig:singlelq:gc}
\end{figure}
\begin{figure}
\centering
\includegraphics[width=0.30\textwidth]{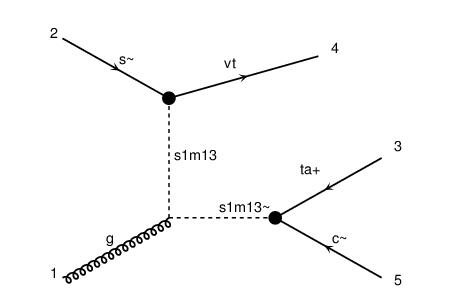}
\includegraphics[width=0.30\textwidth]{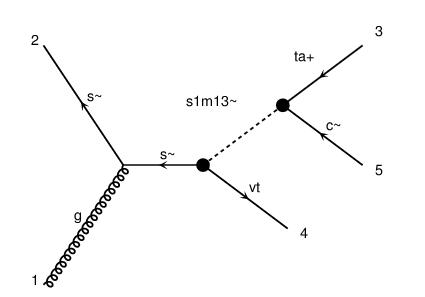}
\includegraphics[width=0.30\textwidth]{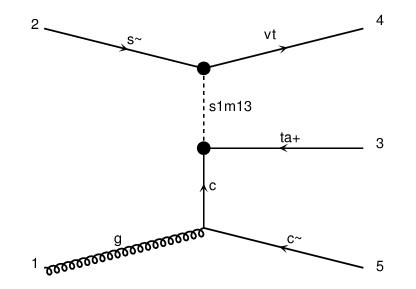}
\caption{\it Representative Feynman diagrams for the $gs$ initial state.}
\label{fig:singlelq:gs}
\end{figure}
In both cases the rightmost two diagrams correspond to associated
production of a resonant leptoquark and a lepton, and the rightmost 
one corresponds to the $t$-channel exchange of a leptoquark.\par 
\end{itemize}
\vskip 1mm
\item
Through the single production processes with a top quark in final state:\\
$gs/gb \to \nu_{\tau} \tau t$
\end{itemize}
The processes with a top quark in the final state contribute
to searches requiring a $b$-tagged jet in the final state, which
will be described in Section~\ref{sec:singlelq:btaunu}, but their discovery 
potential is optimised by searches which require an additional 
isolated lepton ($e$, $\mu$) from the semileptonic decay of the top,
and exploit the distinctive final-state kinematics of the top decay
described in Section~\ref{sec:singlelq:toptaunu}.
The relevant Feynman diagrams are the ones of Fig.~\ref{fig:singlelq:gs} with
the $c$ quark replaced by a $t$ quark.

\section{$\tau$+$b$-tagged jet+\etmiss final state.}
\label{sec:singlelq:btaunu} 
We produced a grid in the three-dimensional space defined by the leptoquark
mass $m_{S_1}$ (MS1 parameter in generator), and by the left-handed couplings 
$(\lambda_{\sss L})_{23}$ (yLL2x3)  and $(\lambda_{\sss L})_{33}$ (yLL3x3). 
All other couplings were put to zero.\par
For each model point only the initial states including a gluon 
or a  second or third generation quark were produced. 
Initial states including a first generation quark do not contribute
to the signal, and for them the generation was performed only once
with all couplings set to zero to be used as a background.\par
The corresponding MG5 process card is given below for reference.
\begin{verbatim}
import model LO_LQ_S1  
define tau = ta+ ta- 
define nutau = vt vt~ 
define top = t t~
define j = g u c d s b u~ c~ d~ s~ b~ 
define p1 = g c s b c~ s~ b~ 
generate p1 p1 > tau nutau j QED=4 NP=4 
\end{verbatim}
The signal thus generated also includes SM $Wc$-production, which
in the analysis will be subtracted using a sample produced
with all of the couplings set to zero to obtain a pure signal sample.
In addition the final state with a top in the final state produced 
as:
\begin{verbatim}
generate p1 p1 > tau nutau top QED=4 NP=4
\end{verbatim}
was incorporated in the signal sample after subtraction of the $Wt$
background.
The analysis strategy is based on requiring one and only
one $b$-tagged jet accompanying a $\tau$ and $\etmiss$. 
The production of a single top quark accompanied either by 
a $W$ boson or a light jet will present this final state,
the veto on a second $b$-tagged is used to reduce the very 
large \ttbar background. The background from SM $W$ production will 
affect this search through two main sources:
\begin{itemize}
\item
$W\rightarrow\tau\nu$ accompanied by a light jet, where 
in the jet showering a $b$-jet is produced 
\item
$W\rightarrow\tau\nu$ accompanied by a heavy-flavour jet ($b$ or $c$).
At leading order, the associated production of a $W$ with a single
$b$-jet does not exist, and the  associated production of
a $W$ and two $b$-jets will be suppressed by the veto on a second 
$b$-jet in the event. The dominant contribution will thus be the case in 
which the $W$ is produced with a $c$-jet, and the 
$c$-jet is misidentified as a $b$-jet.  
\end{itemize}
In order to restrict the event generation to the relevant kinematic
region, the following parton-level cuts were applied both on 
the signal and on the $W$+jets background at generation level 
in {\tt MadGraph5\_aMC@NLO}:
$m_T(\tau\nu_{\tau})>150$~GeV, where the $\tau$ is undecayed,
and $m_T$ is the transverse mass. This cut is meant to remove
the bulk of the resonant $W\rightarrow\tau\nu_{\tau}$ production,
which is the dominant background, and is peaked at 80~GeV with a
long tail corresponding to the Breit-Wigner tail of the $W$ boson.
The transverse mass instead of the mass of the
tau-neutrino system is chosen in order to have a selection which can
be reasonably approximated with experimental cuts.
Additionally we require $p_T(\tau\nu_{\tau})>150$~GeV, to define
a kinematic region where the tau-neutrino system recoils against
a hadronic system with high transverse momentum.

\begin{figure}
\centering
\includegraphics[width=0.45\textwidth]{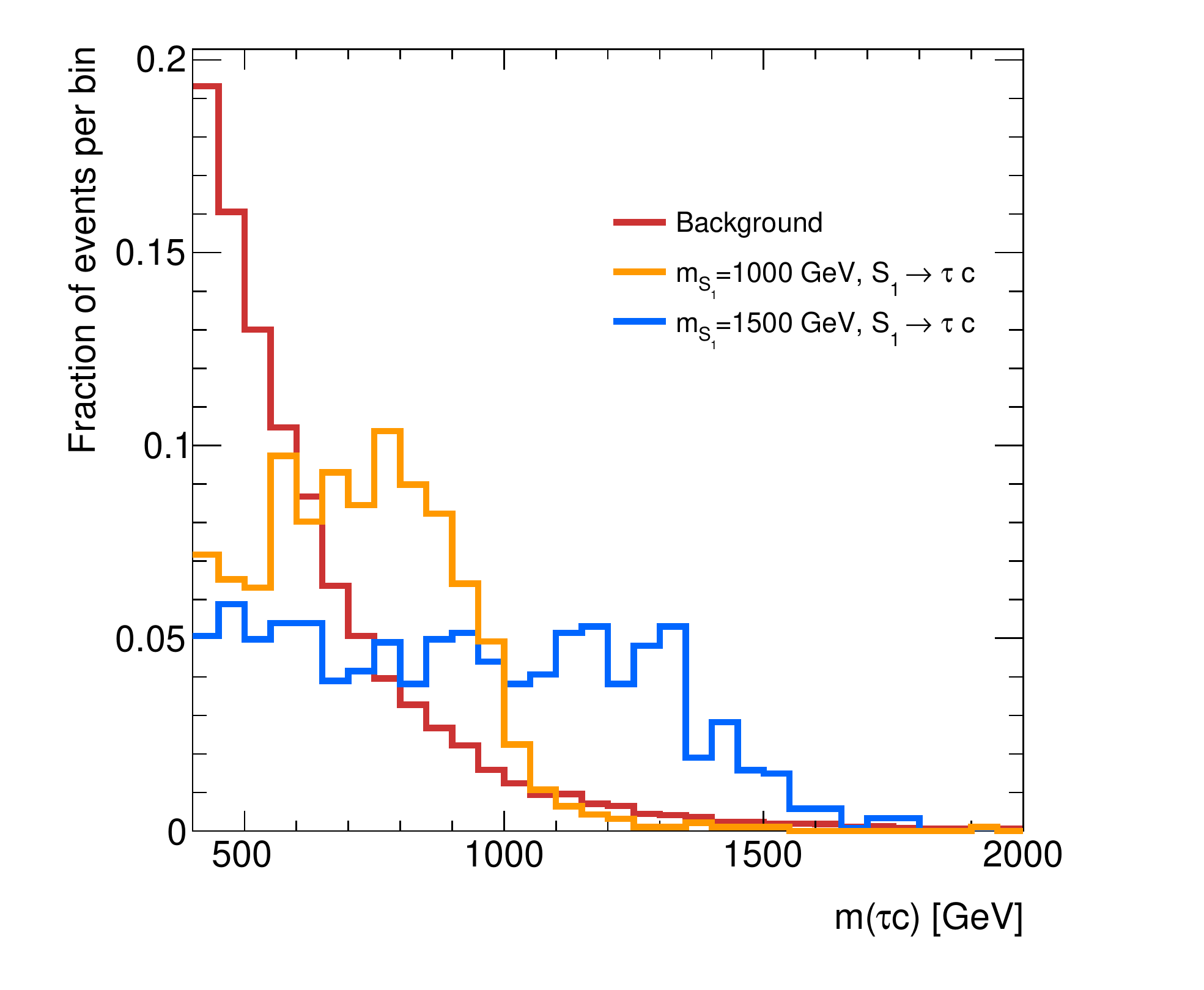} 
\includegraphics[width=0.45\textwidth]{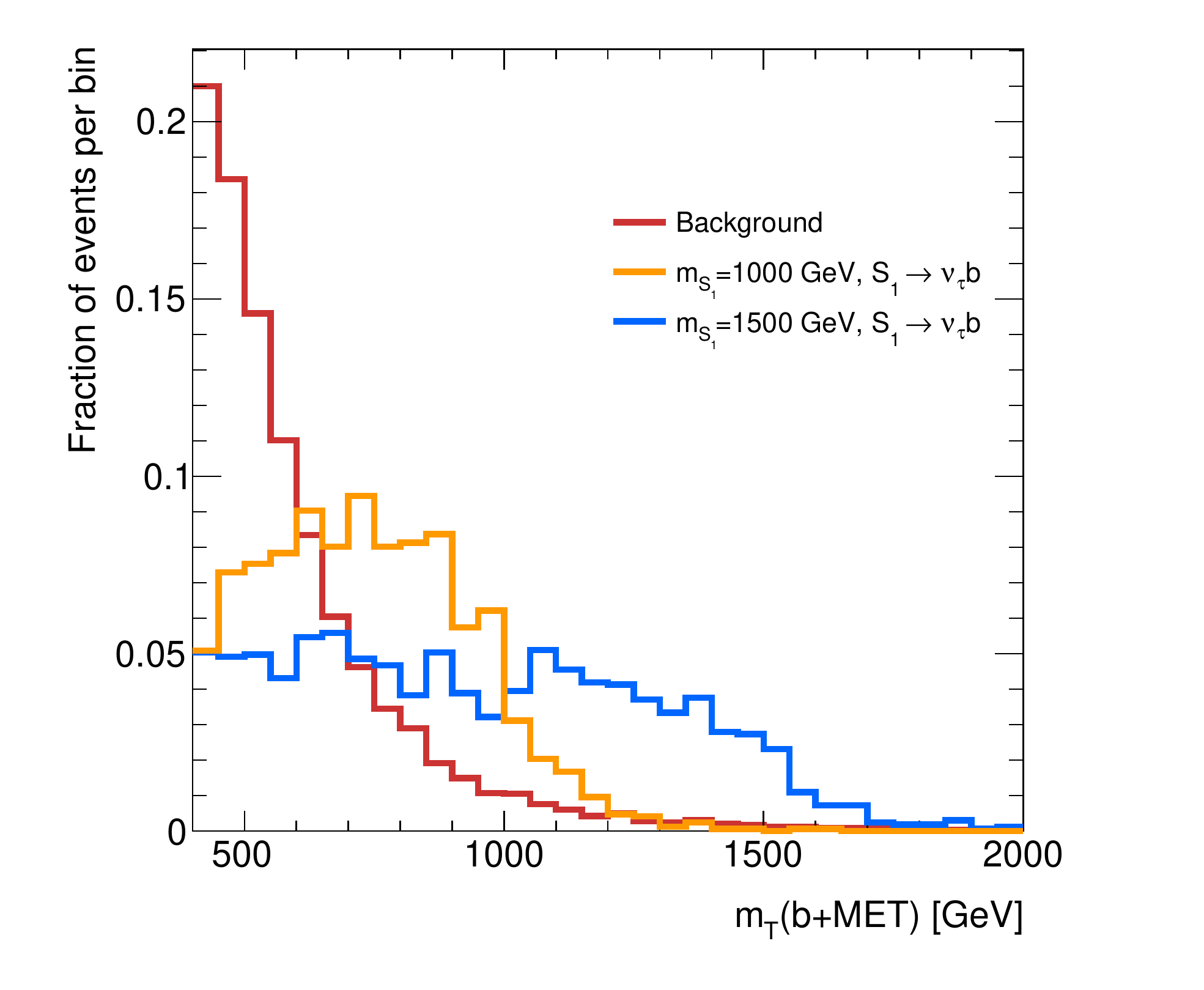} 
\caption{\it Normalised distributions of the variables $m_{\tau c}$ (left) and 
$m_T^b$ (right) for the Drell-Yan background and for two $S_1$ masses, 1000 and 1500 GeV.}
\label{fig:singlelq:kine}
\end{figure}

For the case where only the $(\lambda_{\sss L})_{23}$ coupling is different from zero, 
only the $\tau+c+\nu_{\tau}$ will be observable, and the invariant mass
of the visible part of the hadronic $\tau$ decay and of the $c$-jet ($m_{\tau c}$) 
will approximately reconstruct  the mass of the leptoquark in case of resonant production.
When the $(\lambda_{\sss L})_{33}$ coupling is different from zero as well, the final 
state $\tau+b+\nu_{\tau}$ will also be available, with the transverse
mass built out of the $b$-quark and \etmiss ($m_T^b$) presenting an edge 
at $m_{S_1}$. In both cases, however, the mass peak/edge
is smeared by the fact that the event includes additional undetected
neutrinos from the $\tau$ decay. This feature is illustrated in 
Fig.~\ref{fig:singlelq:kine}, where the normalised distributions for $m_{\tau c}$ (left)
and $m_T^b$ (right)  are shown for the Drell-Yan background and for 
two  $S_1$ masses, 1000 and 1500 GeV.
A clear enhancement for both distributions is visible for the signal
at values corresponding to the nominal $m_{S_1}$. We will therefore 
use these two variables as the main discriminants for the analysis.

As a first step, events with a jet tagged as the hadronic decay of
a $\tau$ lepton ($\tau$-jet) and at least one $b$-tagged  jet and some 
\etmiss are selected. Events are required to contain one and only 
one $\tau$-jet with $\pt>180$~GeV within $|\eta_{\tau}|<2.5$, 
and one and only one $b$-tagged jet with $\pt>250$~GeV within 
$|\eta|<1.5$, and to satisfy the requirement $\etmiss>150$~GeV. 
Events with additional \btagged jets with $\pt>30$~GeV
or containing isolated charged leptons ($e/\mu$) are rejected.\par
To match the generation-level cuts designed to
reject the bulk of the Drell-Yan background we require 
that the  transverse mass of the $\tau$-jet with \etmiss, 
$m_T^{\tau\nu}>300$~GeV, and the transverse momentum of the $\tau\nu$ system
is required to be $p_T^{\tau\nu}>200$~GeV. Finally, in order to 
reject events where \etmiss is generated by a mismeasured jet, 
a cut on the minimal azimuthal distance between a reconstructed
jet and \etmiss $\Delta\phi_{min}>0.6$ is applied.

Based on the final discriminant variables $m_{\tau c}$ and 
$m_{\tau c}$ defined above, 
two different orthogonal signal regions are defined, requiring
either $m_{\tau c}>m_T^b$ (SR1) or $m_{\tau c}>m_T^b$ (SR2), targeting the two 
different decays of the $S_1$. In each of the region a fit is performed 
respectively on the $m_{\tau c}$ ($m_T^b$) distribution for SR1 (SR2) 
over 4 bins starting from 600 GeV. The results of the two 
signal regions are then statistically combined to  get the final result.

\begin{figure}
\centering
\includegraphics[width=0.45\textwidth]{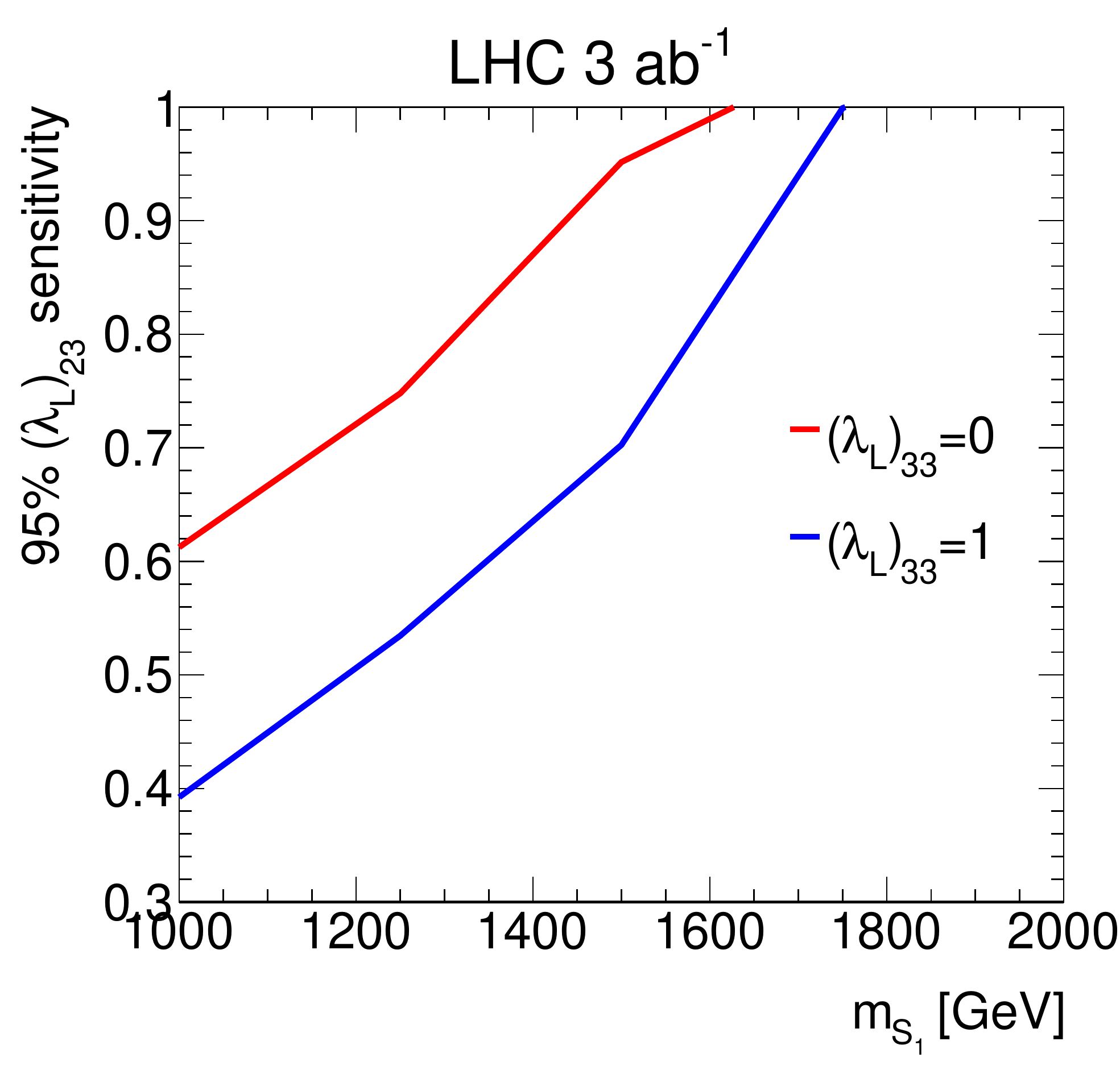}
\includegraphics[width=0.52\textwidth]{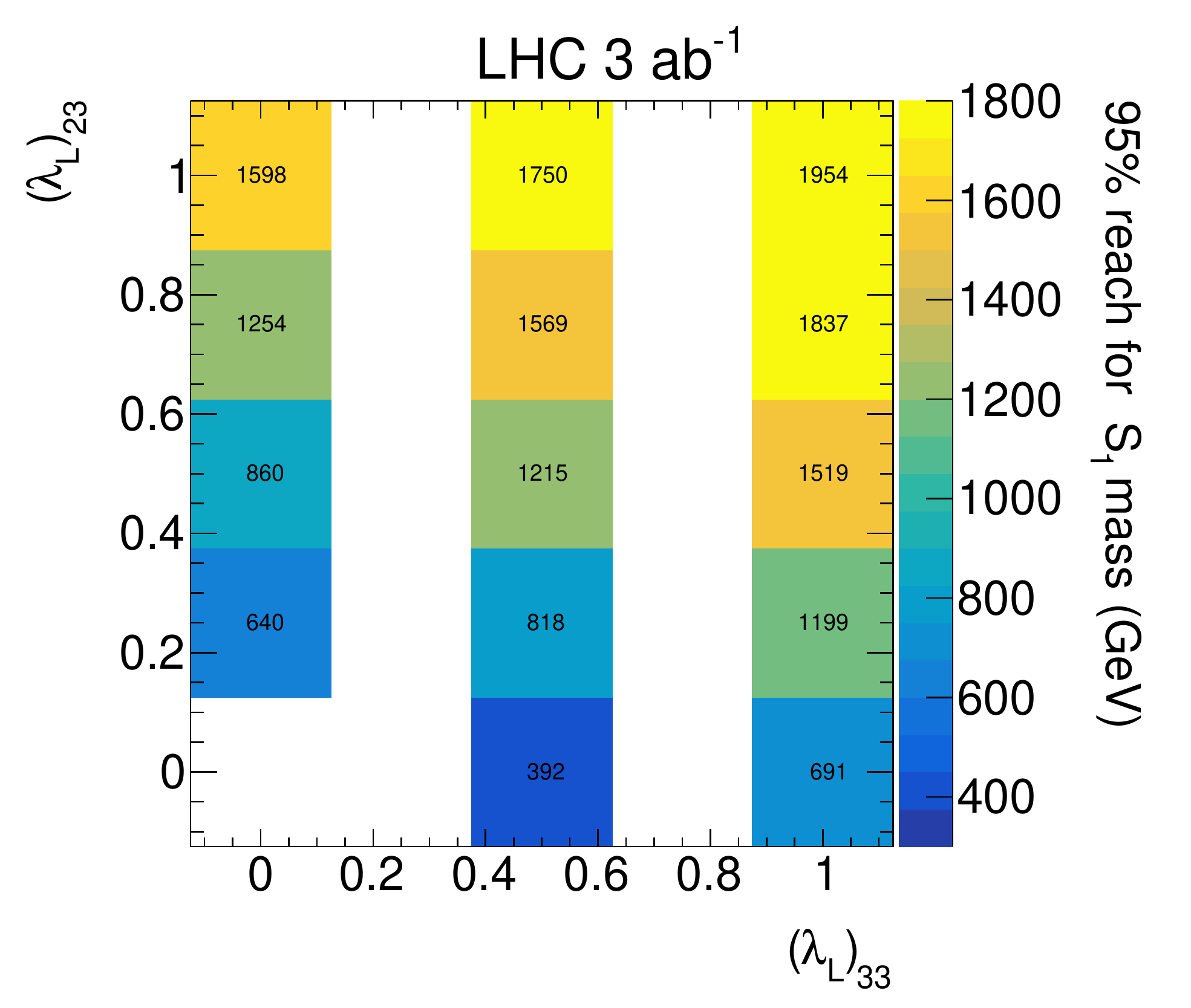}
\caption{\it Left panel: value of $(\lambda_{\sss L})_{23}$ which can be excluded
at 95\% CL at HL-LHC as a function of $m_{S_1}$. Right panel: $m_{S_1}$ which 
can be excluded at 95\% CL in the ($(\lambda_{\sss L})_{33}$,$(\lambda_{\sss L})_{23}$) plane}
\label{fig:singlelq:res}
\end{figure}

The 2-$\sigma$ sensitivity of the $\tau$+$b$-tagged jet analysis at the HL-LHC 
is shown in Fig.~\ref{fig:singlelq:res}. In the left panel the sensitivity 
on the $(\lambda_{\sss L})_{23}$ coupling is shown as a function of
$m_{S_1}$ for two values of $(\lambda_{\sss L})_{33}$. 
In the right panel the value of $m_{S_1}$ which can be 
excluded at the 95\% level is shown in the 
($(\lambda_{\sss L})_{33}$, $(\lambda_{\sss L})_{23}$) plane.

\section{Final state: $\tau$+top+\etmiss}
\label{sec:singlelq:toptaunu} 
\subsection{Search for leptoquark pair production}
As discussed above, the pair production of leptoquarks yields this
signature in the case where one of the two leptoquarks decays
$S_1 \to \tau t$ and the other decays in the channel $S_1 \to \nu_{\tau} b$.
Given that the two decays have approximately the same branching fraction,
the search for the mixed decay is in principle more powerful that
the searches for the $\tau t \tau t$ and $\nu b \nu b$ existing in literature,
see e.g \cite{Aaboud:2019bye}.
We generated for this process the complete decay pattern of the $S_1$,
through the card
\begin{verbatim}
define l3g = ta+ ta- vt vt~
define q23g =  c s b  c~ s~ b~ t t~
generate p p > S1m13 S1m13~, (S1m13 > l3g q23g), (S1m13~ > l3g q23g)
\end{verbatim}
to take into account the possible contribution of all of the channels
in the signal region. The generation was performed at the fixed values
of yLL2x3=1  and yLL3x3=1 for a grid of $S_1$ masses between 750 and 2500 GeV.

We concentrate on the semileptonic decay of the top.
Events are required to contain one and only one
isolated lepton $\ell$ ($e$, $\mu$) with $\pt>20$~GeV,
one $\tau$-jet with $\pt>30$~GeV
and at least one $b$-tagged jet with $\pt>30$~GeV all of them within
$|\eta|<2.5$, and to satisfy the requirement $\etmiss>200$~GeV.

The presence of at least an  additional jet with $\pt>20$~GeV is also required. 
The first step in the analysis is to select among the additional jets
the one most likely to be a $b$-jet, which is taken as the second $b$-jet
candidate. To reject the \ttbar and \ttbarZ backgrounds, the kinematics
of the two $b$-jet candidates, the lepton and the $\tau$-jet is made
to be incompatible with the presence of two semileptonic top decays
by requiring that $m_{b \ell}^t  >160 \, {\rm GeV}$, where
\begin{equation}
m_{b \ell}^t =\mathrm{min} \hspace{0.5mm} \Big  
(\mathrm{max} \hspace{0.5mm}  \big ( m_{\ell j_1}, m_{\tau j_2} \big  ) \Big ) \,,
\end{equation}
and $j_1$ and $j_2$ run over the two $b$-jet candidates. 

Each of the two $b$-jet candidates is then assigned either
to the decay either of the top ($b_t$) or of the leptoquark ($b_{lq}$).
The jet which has invariant mass with the lepton $m_{bl}<160$ GeV
is defined as $b_t$. If the condition is satisfied for both jets, the softer one is
defined as $b_t$. Based on this assignation, two variables sensitive to the mass of the
leptoquark in the two decay legs are respectively  $m_{b_t\ell\tau}$ and $m_{T}^{b_{lq}}$.
These are required to be $m_{b_t\ell\tau}>200$~GeV, $m_{T}^{b_{lq}}>200$~GeV. 

The final discrimination is obtained using $m^{\tau}_{\mathrm T2}$, 
which is the stranverse mass between the \etmiss and the $\tau$-lepton on one leg 
and the \etmiss and the electron or muon on the other. 
This distribution has an end-point for semi-leptonic \ttbar\ decays where one \W boson decays
into $\tau\nu$ and the other decays into $e\nu$ or $\mu\nu$, 
which can be then effectively suppressed with a minimal requirement of 100 GeV. 

\begin{figure}
\centering
\includegraphics[width=0.6\textwidth]{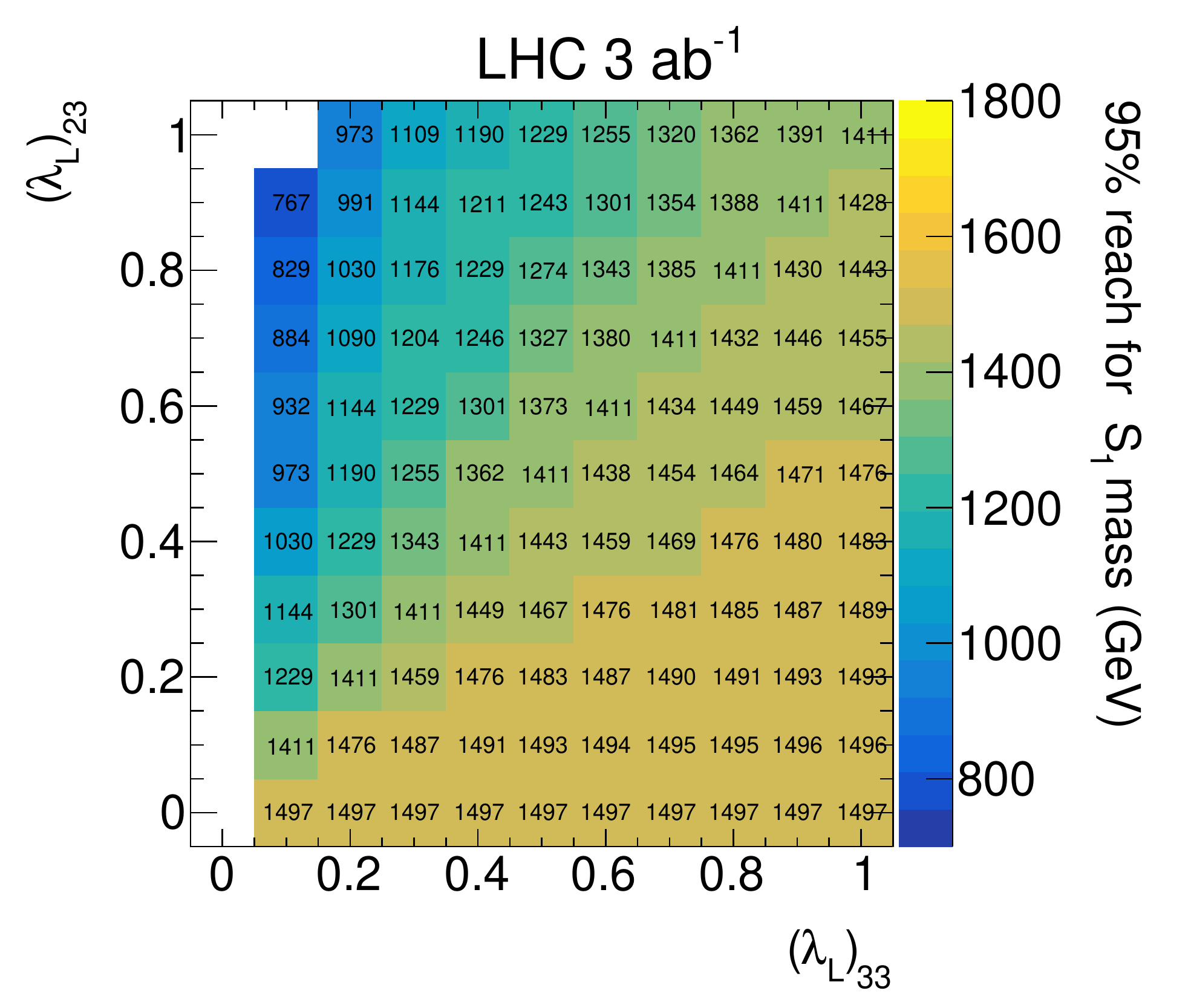}
\caption{\it $m_{S_1}$ which can be excluded at 95\% CL in the 
($(\lambda_{\sss L})_{33}$,$(\lambda_{\sss L})_{23}$) plane on the basis of the search 
for pair production of $S_1$ described in the text.}
\label{fig:singlelq:reachpair}
\end{figure}

The value of $m_{S_1}$ which can be excluded at the 95\% level 
on the basis of this analysis is shown in the 
($(\lambda_{\sss L})_{33}$, $(\lambda_{\sss L})_{23}$) plane in 
Fig.~\ref{fig:singlelq:reachpair}. The dependence on $(\lambda_{\sss L})_{23}$ is partially determined
by the interplay of branching ratios, and partially by the fact that, 
since only one $b$-tagged jet is required in the analysis, the 
case  where one leg is $S_1 \to s \nu$ is selected very efficiently.\par

\subsection{Search for single leptoquark production} 
The signature studied in this case corresponds to the first two diagrams
on the left of Fig.~\ref{fig:singlelq:gs}, where the quark in the initial
state can be either a $b$ of an $s$-quark, and the $c$ in the decay 
of the resonant $S_1$ is replaced by a top quark.\par
The signal is generated as
\begin{verbatim}
define nutau = vt vt~
define p1 = g c s b c~ s~ b~
define LQ = S1m13 S1m13~
generate p1 p1 > nutau LQ, (LQ > top tau)
\end{verbatim}
A grid is generated in the ($(\lambda_{\sss L})_{33}$, 
$(\lambda_{\sss L})_{23}$) plane for $m_{S_1}$ between 750~GeV and 2~TeV.\par
The selected final state is the same as for the analysis in the previous section.
The  $b$-tagged jet is required to have $\pt > 50$ GeV, and the lepton $\ell$ and the
$tau$-jet  are required to have $\pt > 25$ GeV. Additionally, a minimum azimuthal 
separation between the direction of the 
missing momentum and all jets is required ($\Delta\Phi_{\mathrm min}(\etmiss, jets) > 0.5$).

The \etmiss\ magnitude is required to be larger than 200 GeV. In addition, 
a minimal transverse invariant mass between the light lepton 
and the \etmiss, $m_{\mathrm T}^{\ell}$, of at least 100 GeV. 

\begin{figure}
\centering
\includegraphics[width=0.45\textwidth]{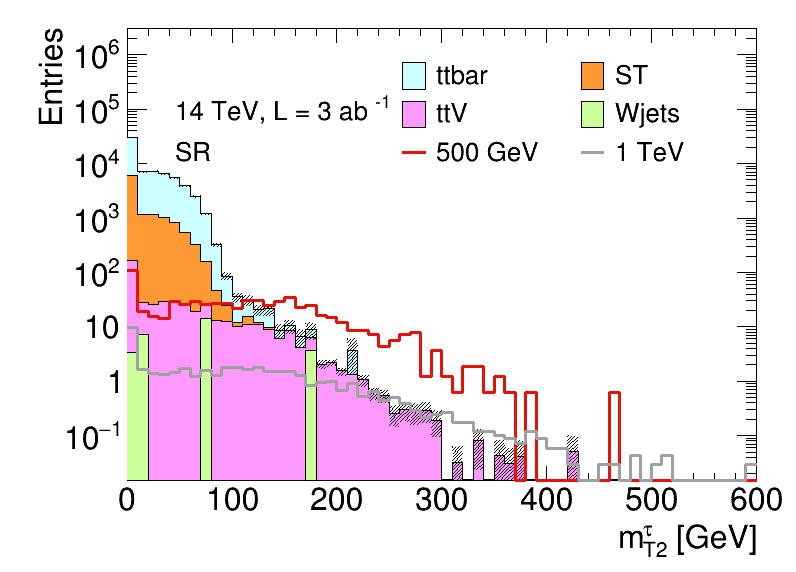}
\includegraphics[width=0.45\textwidth]{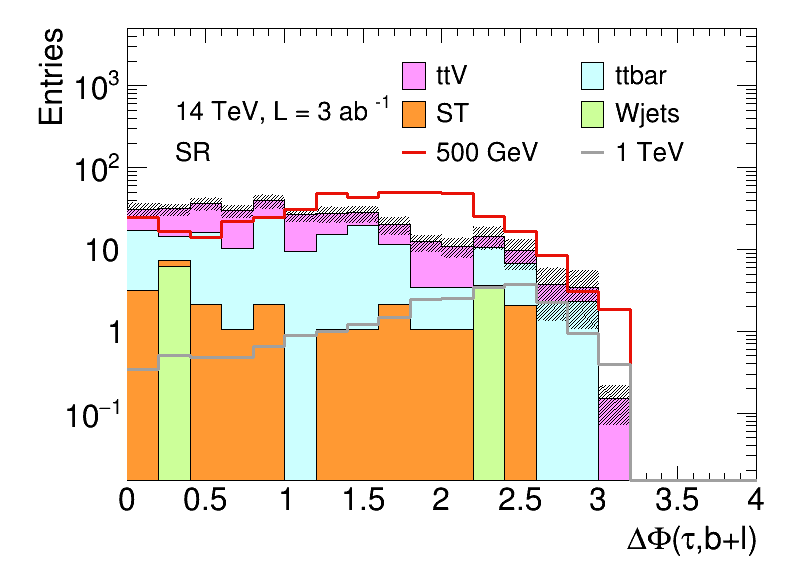}
\caption{\it Distributions after all requirements of $m^{\tau}_{\mathrm T2}$ 
and $\Delta\Phi(\tau,b+\ell)$ for representative signal models and the backgrounds.}
\label{fig:singlelq:Nm1pri}
\end{figure}

Two variables are finally used to discriminate the residual background and the signal.
One is $m^{\tau}_{\mathrm T2}$ described above, for which we require $m^{\tau}_{\mathrm T2}>100$~GeV.
The second variable is $\Delta\Phi(\tau,b+\ell)$, the azimuthal angle between the $\tau$-lepton 
and the vectorial sum of the momenta of the $b$-jet and the light lepton. If you assume that the sum of 
the $b$-jet and the light lepton is a good proxy for the direction of the top-quark, this angle is
a proxy for the opening angle between the decay product of the leptoquark, which tends to be larger than
in the case of the background processes. See an example of these two distributions in Fig.~\ref{fig:singlelq:Nm1pri}. 
All requirements except the ones on the variables shown in the plots are applied. 

\begin{figure}
\centering
\includegraphics[width=0.52\textwidth]{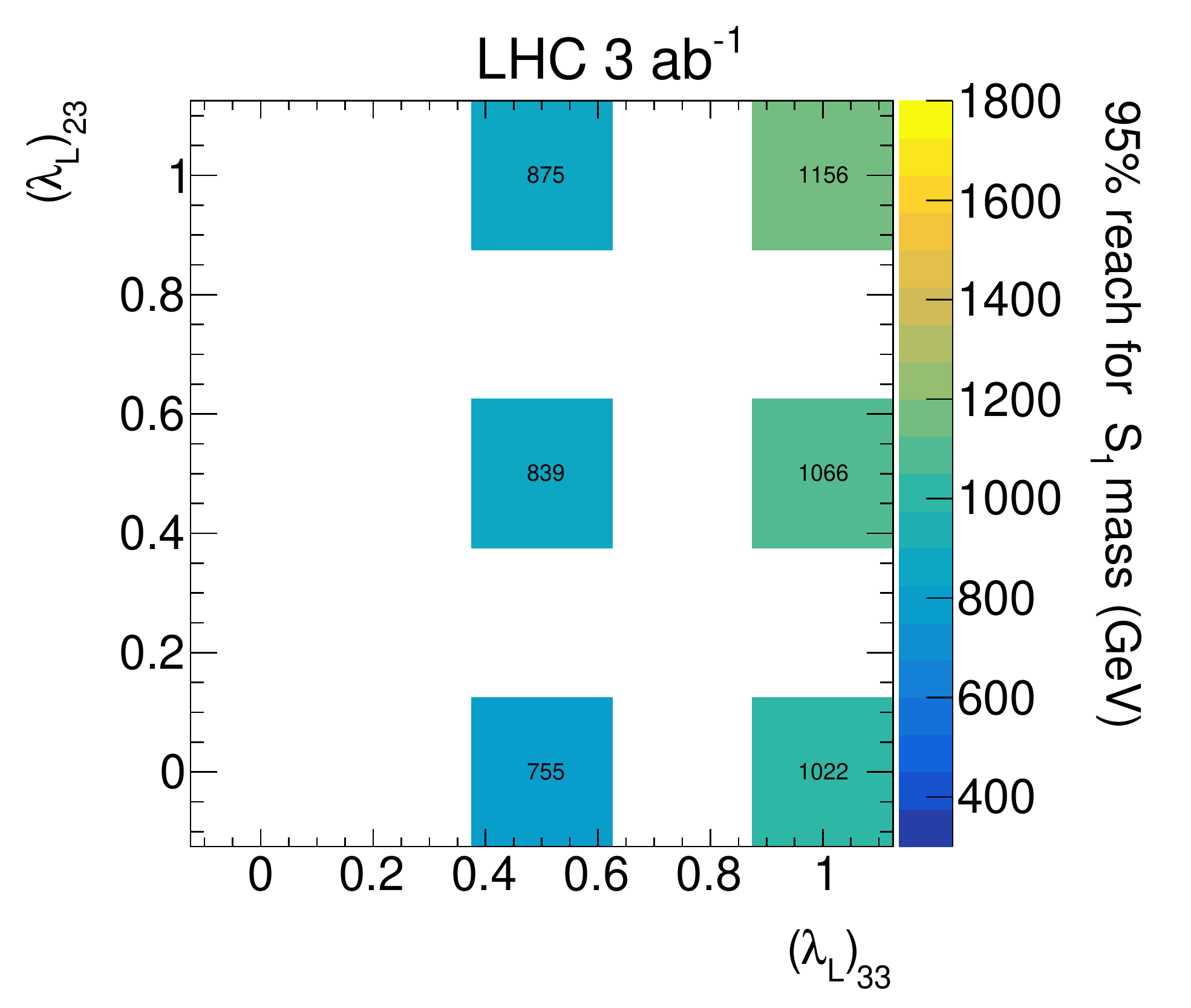}
\includegraphics[width=0.45\textwidth]{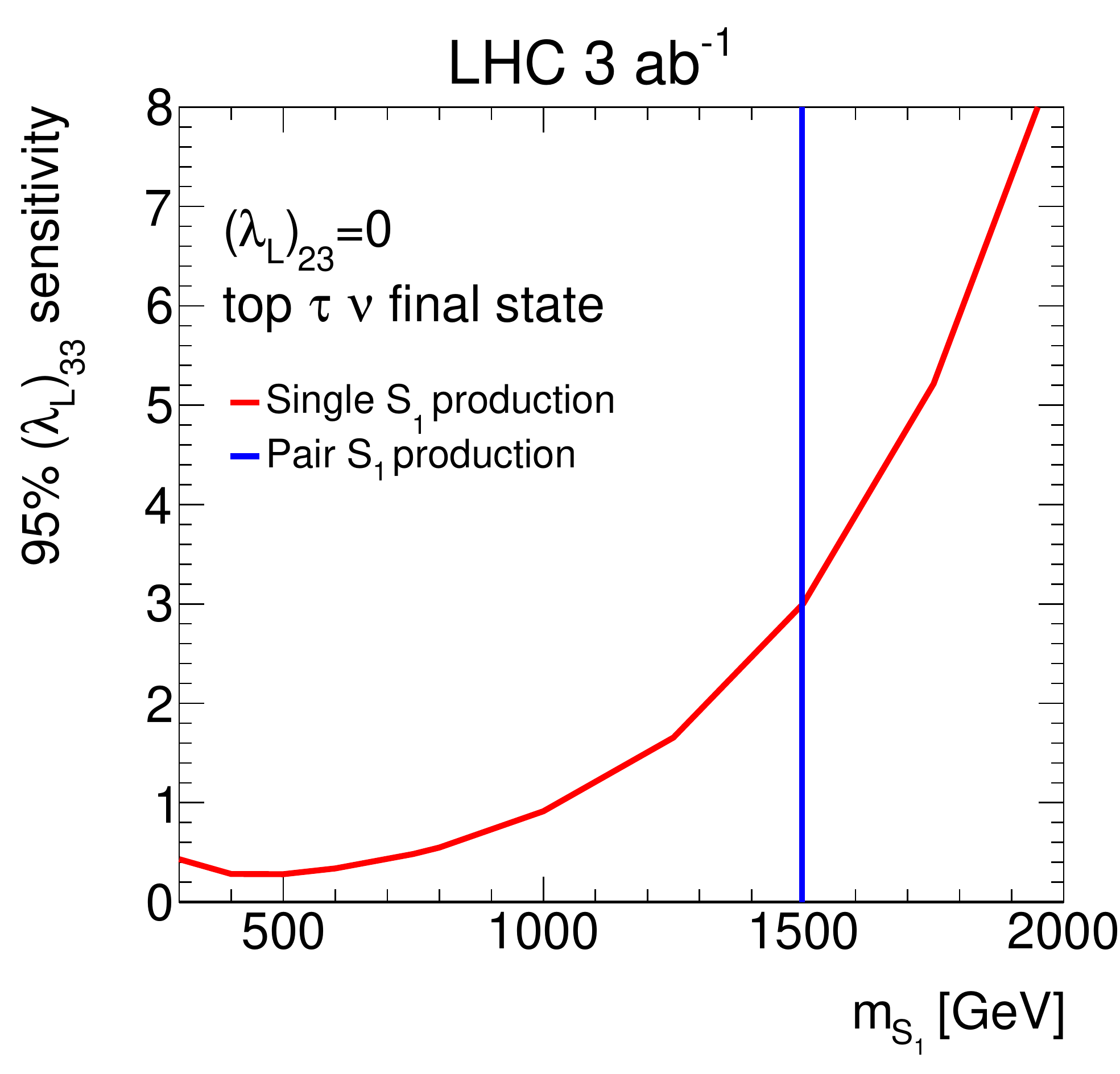}
\caption{\it Results of the search
for  production of $S_1$ in the final state top+$\tau$+$\nu$ 
at the LHC with 3~ab$^{-1}$ of integrated luminosity
Left panel: $m_{S_1}$ which can be excluded at 95\% CL in the 
($(\lambda_{\sss L})_{33}$,$(\lambda_{\sss L})_{23}$) plane 
through the study of single production. 
Right panel: Value of the $(\lambda_{\sss L})_{33}$
coupling which can be excluded at 95\% CL  as a function of the leptoquark mass
for $(\lambda_{\sss L})_{23}=0$. The results for single (red) and pair (blue) LQ production
are shown.}
\label{fig:singlelq:reachpri}
\end{figure}

The value of $m_{S_1}$ which can be excluded at the 95\% level
on the basis of this analysis is shown in the
($(\lambda_{\sss L})_{33}$, $(\lambda_{\sss L})_{23}$) plane in
the left panel of Fig.~\ref{fig:singlelq:reachpri}.  The reach rises
with $(\lambda_{\sss L})_{23}$ because of the contribution of 
the $gs$ initial state to leptoquark production. For $(\lambda_{\sss L})_{23}=0$
the top final state has higher sensitivity than the $b$-tagged jet final
state discussed in Sec.~\ref{sec:singlelq:btaunu}. 

In the right panel of Fig.~\ref{fig:singlelq:reachpri} the reach in the coupling 
$(\lambda_{\sss L})_{33}$ as a function of the leptoquark mass for 
$(\lambda_{\sss L})_{23}=0$ is shown. The reach for single production (red line) is 
compared to the one for pair production (blue line).
The single production overtakes the pair production for a coupling of approximately 3.

\section{CONCLUSIONS}
We studied the potential of the HL-LHC for the discovery of leptoquarks
in final states involving a heavy quark ($b/t$), a $\tau$ lepton 
decaying hadronically and \etmiss, produced either as the result 
of the pair production of two leptoquarks where the two legs 
decay in different final states, or of the associate production 
of a leptoquark with a $\tau$ or a $\nu_{\tau}$, with the leptoquark 
decaying into a heavy quark and a third generation lepton.

From the detailed simulation study involving a model with a 
scalar $S_1$ leptoquark, and only couplings of type 
$(\lambda_{\sss L})_{23}$ and  $(\lambda_{\sss L})_{33}$
different from zero we assess the parameter space covered 
by three dedicated searches for the final states of interest
at the HL-LHC for an integrated luminosity of 3~ab$^{-1}$.

Leptoquark masses above 1.5~TeV will be accessible over all
of the parameter space for both couplings $(\lambda_{\sss L})<1$.
The top+$\tau$+$\nu$ final state from leptoquark pair production
dominates the sensitivity for lower values of  $(\lambda_{\sss L})_{23}$,
whereas for $(\lambda_{\sss L})_{23}~1$ the 
reach is dominated by the $b/c$+$\tau$+$\nu$ final state,
yielding a reach up to 2~TeV. The studies were based on requiring
in the final state passing the $b$-tagging requirements of the ATLAS
experiment. It turns out that a significant part of the sensitivity
is provided by final states involving the fragmentation of a $c$-quark
mistagged as a $b$. A generic characteristic of the models
able to explain the $R_{D^{(*)}}$ anomalies is indeed a significant
production of LHC final states involving a $\tau$ lepton and a $c$-quark.
Analyses exploiting dedicated $c$-tagging algorithms will therefore
significantly improve the LHC reach for this kind of models,
and should be vigorously pursued by the experimental Collabirations.

\section*{ACKNOWLEDGEMENTS}
We would like to thank the organisers of the Les Houches 2019 workshop
for the friendly and very productive atmosphere of the workshop
and all of the colleagues taking part for the useful discussions.
DM acknowledges partial support from the INFN grant SESAMO and MIUR grant PRIN\_2017L5W2PT. DB is supported by the MIUR grant PRIN\_2017L5W2PT and the INFN grant FLAVOR.  

\let\be\undefined
\let\ee\undefined
\let\bsp\undefined
\let\ttbar\undefined
\let\ttbarZ\undefined
\let\etmiss\undefined
\let\pt\undefined
\let\W\undefined
\let\btagged\undefined
\let\lag\undefined
\let\laR\undefined
\let\sss\undefined
\let\eR\undefined
\let\uRbar\undefined
\let\laLuno\undefined
\let\Ll\undefined
\let\QLbar\undefined
\let\laLtre\undefined
\let\fr\undefined

%% file: rdm/RDM.main.tex
\graphicspath{{rdm/}}

\newcommand{\Herwig}{H\protect\scalebox{0.8}{ERWIG}\xspace}
\newcommand{\Pythia}{P\protect\scalebox{0.8}{YTHIA}\xspace}
\newcommand{\Sherpa}{S\protect\scalebox{0.8}{HERPA}\xspace}
\newcommand{\Rivet}{R\protect\scalebox{0.8}{IVET}\xspace}
\newcommand{\Professor}{P\protect\scalebox{0.8}{ROFESSOR}\xspace}
\newcommand{\eps}{\varepsilon}
\newcommand{\mc}[1]{\mathcal{#1}}
\newcommand{\mr}[1]{\mathrm{#1}}
\newcommand{\tm}[1]{\scalebox{0.95}{$#1$}}
\newcommand{\be}{\begin{equation}}
\newcommand{\ee}{\end{equation}}
\newcommand{\tev}{{\rm TeV}}

\def\RD{\ifmmode R_{D^{(*)}} \else $R_{D^{(*)}}$ \fi}
\def\RK{\ifmmode R_{K^{(*)}} \else $R_{K^{(*)}}$ \fi}
\def\mLQ{\ifmmode m_{S_1} \else $m_{S_1}$ \fi}

\newcommand{\fr}{{\sc \small FeynRules}}
\newcommand{\micromegas}{{\sc\small MicrOMEGAs}}
\newcommand{\mg}{{\sc\small MG5\_aMC}}
\newcommand{\ma}{{\sc\small MadAnalysis}~5}
\def\mxOne{\ifmmode m_{\chi_1} \else $m_{\chi_1}$ \fi}
\def\mxZero{\ifmmode m_{\chi_0} \else $m_{\chi_0}$ \fi}

\def\bsp#1\esp{\begin{split}#1\end{split}}

\def\lag{{\cal L}}
\def\sss{\scriptscriptstyle}
\def\Ll{L_{\sss L}}
\def\eR{\ell_{\sss R}}
\def\QL{Q_{\sss L}}
\def\QLbar{\bar Q_{\sss L}}
\def\uR{u_{\sss R}}
\def\uRbar{{\bar u}_{\sss R}}
\def\ydm{y_{\sss \chi}}


\chapter{A common solution to the $R_D$ anomalies and Dark Matter}

{\it G.~B\'elanger, J.~Bernigaud, A.~Bharucha, B.~Fuks, A.~Goudelis, J.~Heisig, A.~Jueid, A.~Lessa, D.~Marzocca, M.~Nardecchia, G.~Polesello, P.~Pani, A.~Pukhov, D.~Sengupta and J.~Zurita}


\label{sec:RDM}
  
\begin{abstract}
The anomalies observed by LHCb in flavor observables suggest the breakdown of lepton flavor universality. We study how a vanilla leptoquark solution to the \RD anomalies can also explain the measured relic abundance, if the leptoquark mediates between the visible sector and the dark sector. 
\end{abstract}

\section{Introduction}
\label{sec:RDM:intro}

The current LHC dataset has not shown any significant deviation from the Standard Model (SM) expectations, except for the \RK~\cite{Aaij:2017vbb,Aaij:2019wad} and \RD~\cite{Aaij:2015yra,Aaij:2017uff,Aaij:2017deq} anomalies. While many models have been considered to explain these observations, the consensus is that two classes of models can accommodate these results: leptoquarks and Z' models (for a recent review see e.g~\cite{Blanke:2019pek} and references therein). It is thus interesting to consider the interplay between these solutions and dark matter (DM), which requires to allow an additional decay of the leptoquark into the dark sector, studied previously in~\cite{Queiroz:2014pra,Baker:2015qna}. As this will reduce the branching ratios of the visible decays of the leptoquarks, it naively allows one to lower the current constraint on $m_{LQ} \gtrsim 1~\rm{TeV}$ with ${\cal O} (1) $ couplings, thus opening up novel parameter space and increasing the opportunities to probe this model at the LHC in the near future.

For our study we built a simplified model featuring the minimal ingredients to explain the charged current flavor anomalies \RD  and dark matter simultaneously, limiting ourselves to the couplings strictly necessary to explain the anomaly. We study how the different phenomenological constrains affect the parameter space: LHC searches for leptoquarks and dark matter, relic density, direct and indirect detection. We pay particular attention to the existing CMS search for a resonant leptoquark plus missing energy~\cite{Sirunyan:2018xtm}: while this search might not always be the golden channel for discovery, its importance lies in the fact that this is the only channel probing the $\RD$-DM connection (RDM).

\section{The model}
\label{sec:RDM:model}

We consider a simplified model in which we supplement the Standard Model by one
scalar leptoquark field $S_1$ and two extra fermionic fields, a Majorana
fermion  $\chi_0$ and a Dirac fermion $\chi_1$. The $S_1$ and $\chi_1$ fields
are electrically-charged coloured weak singlets lying in the
$({\bf 3}, {\bf 1})_{-1/3}$ representation of the Standard Model gauge group. In
contrast, $\chi_0$ is a dark matter candidate and therefore a non-coloured
electroweak singlet. In our \fr~\cite{Alloul:2013bka} implementation, we consider all potential
interactions of the new sector with the Standard Model sector, the corresponding
Lagrangian being written as
\be\bsp
 \lag = &\ \lag_{\rm SM} + \lag_{\rm kin}
   + \bigg[
    {\bf \lambda_{\sss R}}\ \uRbar^c\ \eR^{\phantom{c}} \ S_1^\dag
  + {\bf \lambda_{\sss L}}\ \QLbar^c \!\cdot\! \Ll^{\phantom{c}} \ S_1^\dag
  + \ydm \bar\chi_1\chi_0 S_1
   + {\rm h.c.} \bigg]\ .
\esp\label{eq:RDM:lag}\ee
In this expression, all flavour indices are understood for simplicity and the
dot stands for the $SU(2)$-invariant product of two fields lying in the
fundamental representation. In addition,
$\lag_{\rm SM}$ is the Standard Model Lagrangian and $\lag_{\rm kin}$ contains
gauge-invariant kinetic and mass terms for all new fields, the $\chi_1$ state
being assumed vector-like. The $\QL$ and $\Ll$ spinors stand for the $SU(2)_L$
doublets of left-handed quarks and leptons respectively, whilst $\uR$ and $\eR$
stand for the $SU(2)_L$ singlets of right-handed up-type quarks and charged
leptons, respectively. The
${\bf \lambda_{\sss L}}$ and ${\bf \lambda_{\sss R}}$ couplings are $3\times 3$
matrices in the flavour space, that are considered real in the following.
Moreover, in our
conventions, the first index $i$ of any $\lambda_{ij}$ element refers to the
quark generation whilst the second one $j$ refers to the lepton generation.

The field content of the new physics sector of our simplified model is
presented in table~\ref{tab:RDM:fields}, together with the corresponding
representation under the gauge and Poincar\'e groups, the potential Majorana
nature of the different particles, the adopted symbol in the \fr\
implementation and the PDG identifier that has been chosen for
each particle. The new physics coupling parameters are collected in
table~\ref{tab:RDM:params}, in which we additionally include the name used in the
\fr\ model and the Les Houches blocks~\cite{Skands:2003cj} in which the numerical values of
the different parameters can be changed by the user when running tools like
\mg~\cite{Alwall:2011uj} or \micromegas~\cite{Belanger:2014vza}.

\begin{table}[th!]
\renewcommand{\arraystretch}{1.4}
\setlength\tabcolsep{8pt}
\begin{tabular}{c c c c c c}
  Field & Spin & Repr. & Self-conj. & \fr\ name & PDG code\\
  \hline\hline
  $S_1$    & 0   & $({\bf 3}, {\bf 1})_{-1/3}$ & no  & {\tt LQ} & 42\\
  $\chi_0$ & 1/2 & $({\bf 1}, {\bf 1})_0$      & yes & {\tt chi0} & 5000521\\
  $\chi_1$ & 1/2 & $({\bf 3}, {\bf 1})_{-1/3}$ & no  & {\tt chi1} & 5000522\\
\end{tabular}
\caption{\it New particles supplementing the Standard Model, given
  together with the representations under $SU(3)_c\times SU(2)_L \times U(1)_Y$.
  We additionally indicate whether the particles are Majorana particles,
  their name in the \fr\ implementation and the associated Particle Data Group
  (PDG) identifier. Note that the PDG code 42 is the official PDG value for our leptoquark~\cite{Tanabashi:2018oca} and hence there are no issues when running showering and/or hadronizing the event sample.}
\label{tab:RDM:fields}
\end{table}

\begin{table}[th!]
\renewcommand{\arraystretch}{1.4}
\setlength\tabcolsep{15pt}
\begin{tabular}{c c c c}
  Coupling & \fr\ name & Les Houches block\\
  \hline\hline
  $(\lambda_{\sss L})_{ij}$ & {\tt lamL} & {\tt LQLAML}\\
  $(\lambda_{\sss R})_{ij}$ & {\tt lamR} & {\tt LQLAMR}\\
  $\ydm$ & {\tt yDM} & {\tt DMINPUTS}\\
\end{tabular}
\caption{\it Couplings of the new particles, given together with the associated
  \fr\ symbol and the Les Houches block of the parameter card.}
\label{tab:RDM:params}
\end{table}
In a nutshell, our model needs three new masses and several new couplings.The $R_D$ anomalies can be explained with only non-zero $(\lambda_{\sss L})_{33} \equiv \lambda_L$\footnote{The second index is not constrained, as the neutrino flavor is not tagged. For simplicity we restrict it here to the third generation.} and $(\lambda_{\sss R})_{23} \equiv \lambda_R$. Finally, the dark matter phenomenology requires the coupling $\ydm$. 
Hence these 6 parameters span our parameter space. In what follows we will study how the flavor anomalies, dark matter phenomenology and collider searches constrain the parameter space of this model, showing the currently allowed regions and commenting on the future prospects. 

\section{Leptoquark solutions to the $R_D$ anomalies}

\begin{figure}[!thp]
  \centering
  \includegraphics[width=0.49\textwidth]{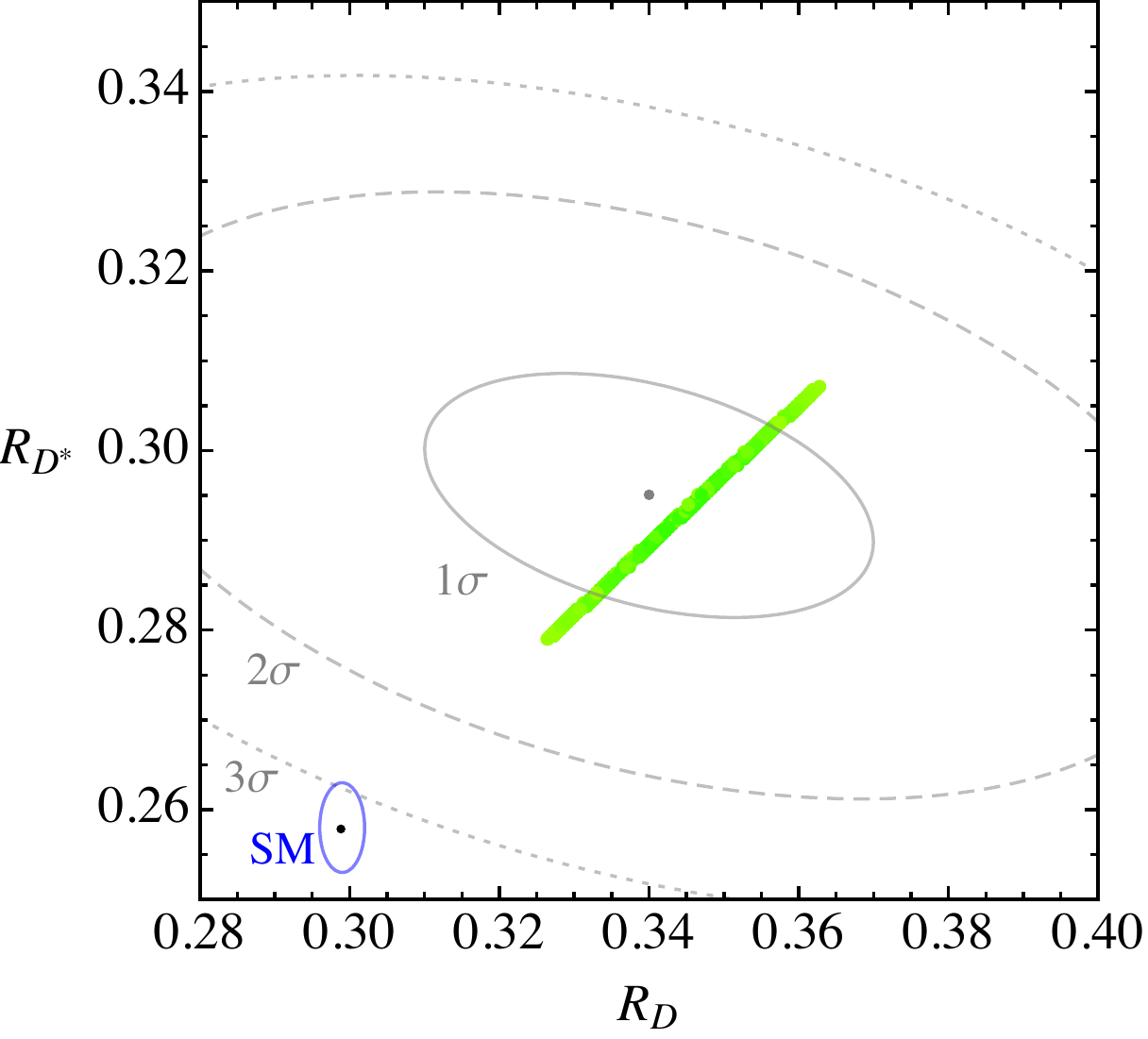} 
  \includegraphics[width=0.44\textwidth]{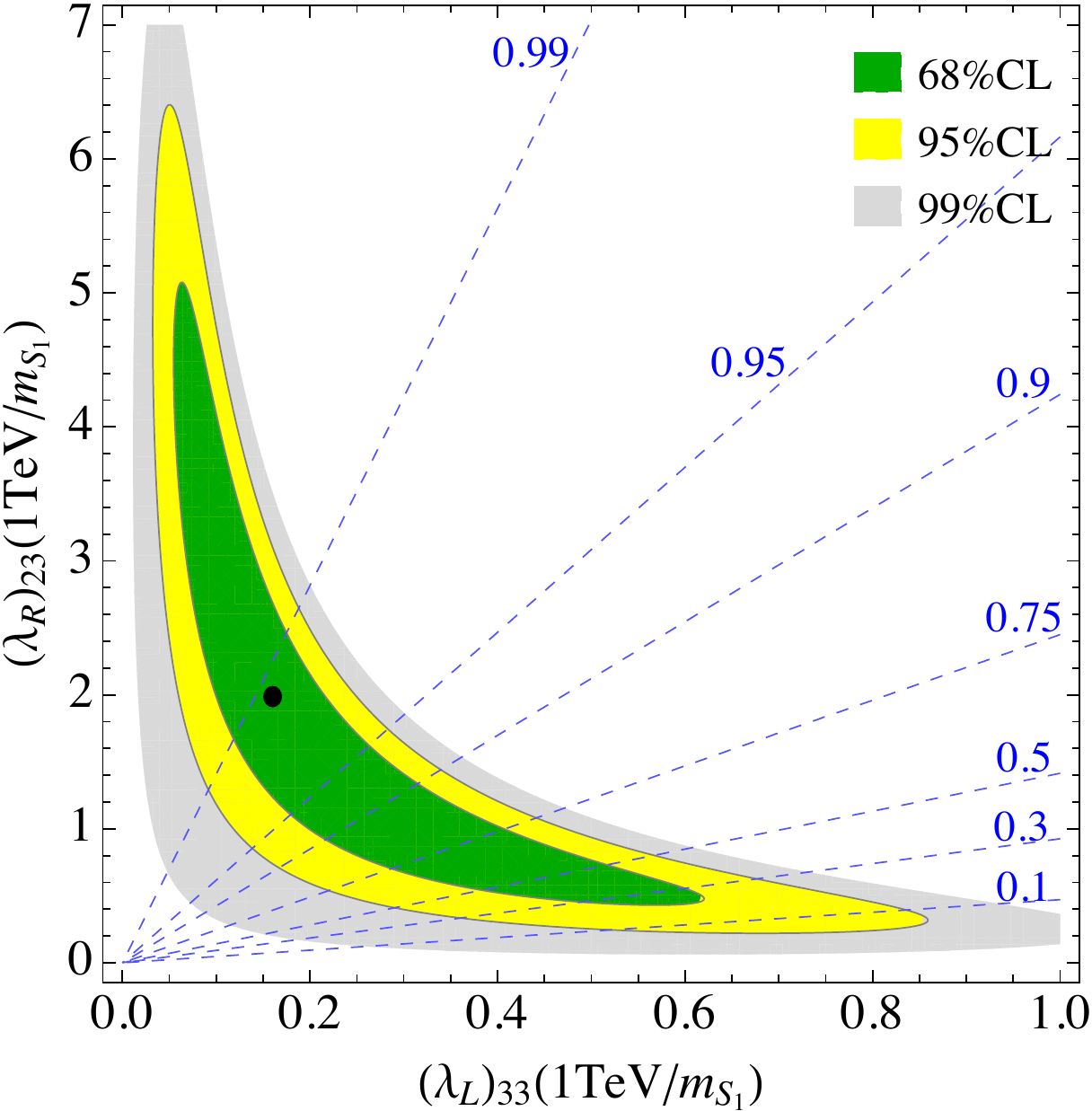} 
  \caption{\it Left: flavor fit to the charged current anomalies for $\mLQ=1.5$ TeV ($1\sigma$ region). The gray lines represent the present status in the measurements of the two observables, compared with the SM point (blue point and contour). Right: Viable parameter space from the fit of flavor and precision observables as function of $\lambda / \mLQ$. Blue dashed lines are iso-lines of the branching ratio $B(S_1 \to c_R \tau_R)$ assuming $y_\chi \ll (\lambda_L)_{33}, (\lambda_R)_{23}$ and $\mLQ \gg m_t$. The black dot represents the best-fit point.}
\label{fig:RDM:FlaFit}
\end{figure}

As mentioned in the previous chapter, a minimal solution of the $\RD$ anomalies can be obtained with just a left-handed ($(\lambda_L)_{33}$)  and a right-handed ($(\lambda_R)_{23}$) couplings to $S_1$. We note that the ($(\lambda_L)_{33}$) coupling is forced to be left-handed due to the presence of the neutrino\footnote{Alternatively one could explicitly introduce a right-handed neutrino that could serve as a dark matter candidate, see e.g~\cite{Azatov:2018kzb}.}. The choice of having a non-vanishing ($(\lambda_R)_{23}$) is due to the large contributions that $(\lambda_{\sss L})_{23}$ would generate to the $B \to X_s \nu \nu$ process, which could only be reduced by extending the model, e.g: considering a destructive interference with another leptoquark state~\cite{Crivellin:2017zlb,Buttazzo:2017ixm,Marzocca:2018wcf,Crivellin:2019dwb}.
We perform a global fit of the model with the two couplings $(\lambda_{\sss L})_{33}$ and $(\lambda_{\sss R})_{23}$ as free parameters. The relevant observables included in the fit are: $R_D$, $R_D^*$, $\text{Br}(B_c \to \tau \nu)$, deviations in the $Z$ couplings to $\tau_L$ and $\tau_R$, effective number of neutrinos from $Z$ decays, lepton flavour universality tests in $\tau$ decays. For more details on this fit see e.g.~\cite{Buttazzo:2017ixm,Marzocca:2018wcf,Azatov:2018kzb}. A contribution to $D^0 - \bar{D}^0$ mixing is generated at one loop, but is strongly CKM-suppressed and does not provide a relevant constraint.
The results of a scan in the two parameters of the model, for $\mLQ = 1.5~\text{TeV}$, where all points are within $1\sigma$ of the best-fit point, is shown in the plane of the $R_D$ and $R_D^*$ anomalies in the left panel of figure~\ref{fig:RDM:FlaFit}. The corresponding preferred region is presented in the right panel of figure~\ref{fig:RDM:FlaFit} in the $x,y$ plane, with $x= \lambda_L (\tev / \mLQ)$ and $y= \lambda_R (\tev / \mLQ)$. \footnote{The \RD anomalies scale with the product $xy$, while other constraints scale instead with $x^2$ or $y^2$.}

From the figure we can establish two working points. We pick the first one to obtain the lowest possible  $BR (S_1 \to c \tau)$, which corresponds to  $(x,y) = (0.6,0.5) = P_1$. For the second one we do not want to pick extreme values of $\lambda_R$, as we could spoil the validity of the narrow-width approximation for the leptoquark, as well as pertubativity. So we restrict ourselves to the best fit point  $(0.16,2.0) = P_2$ that already gives a negligible branching fraction into the third generation states. We thus use these working points to generate concrete benchmarks by picking two values for the leptoquark mass: $\mLQ= 500$ GeV and $\mLQ= 1000$ GeV. We show the corresponding values of the couplings and branching ratios in Table~\ref{tab:RDM:benchmark_points}, where we have assumed a negligible coupling of $S_1$ to the dark sector. We note that the solution of the $\RD$ anomalies, being an off-shell interference with the SM amplitude, will not be affected by diluting the leptoquark visible decays, but it would rather loose the constraints coming from direct searches of $S_1$, which we discuss in detail in Section~\ref{sec:RDM:lhc}.
\begin{table}[]
\begin{tabular}{lllllll}
$\mLQ$ {[}GeV{]} & $\lambda_L$ & $\lambda_R$ & $BR(S_1 \to b \nu)$ & $BR(S_1 \to t \tau)$ & $BR(S_1 \to c \tau)$ & $\Gamma(S_1)/\mLQ$ \\
\hline 
500                 & 0.30        & 0.25        & 0.4049              & 0.3139               & 0.2812               & 4.42E-3               \\
1000                & 0.6         & 0.5         & 0.3794              & 0.3571               & 0.2635               & 1.89E-2               \\
500                 & 0.08        & 1.0         & 0.0063              & 0.0049               & 0.9887               & 3.4E-2                \\
1000                & 0.16        & 2.0         & 0.0063              & 0.0059               & 0.9877               & 8.06E-2              
\end{tabular}
\caption{Benchmark points compatible with the flavour anomalies. Here we have assumed negligible decays into the dark sector. Note the wide range of possibilities for the $c-\tau$ channel (from 25 \% to roughly 100\%) while the b and top decays have an upper bound of about 40 and 30 \% respectively, and can actually become at the sub-percent level.}
\label{tab:RDM:benchmark_points}
\end{table}

\begin{figure}[!h]
	\centering
	\includegraphics[width=0.48\textwidth]{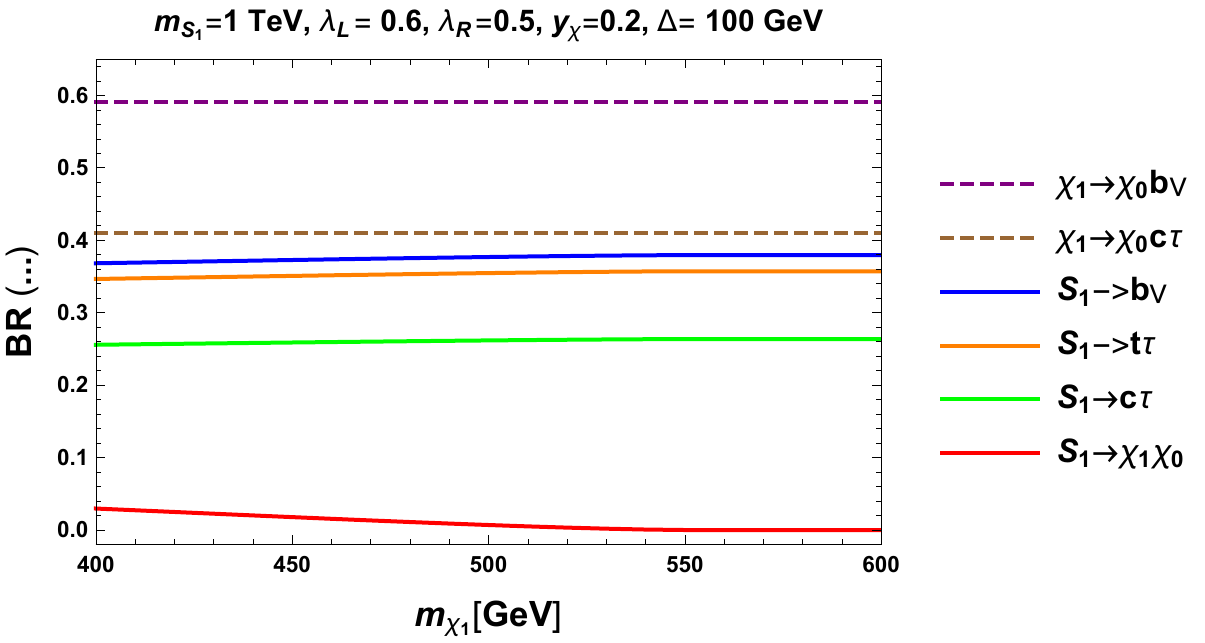}
	\includegraphics[width=0.48\textwidth]{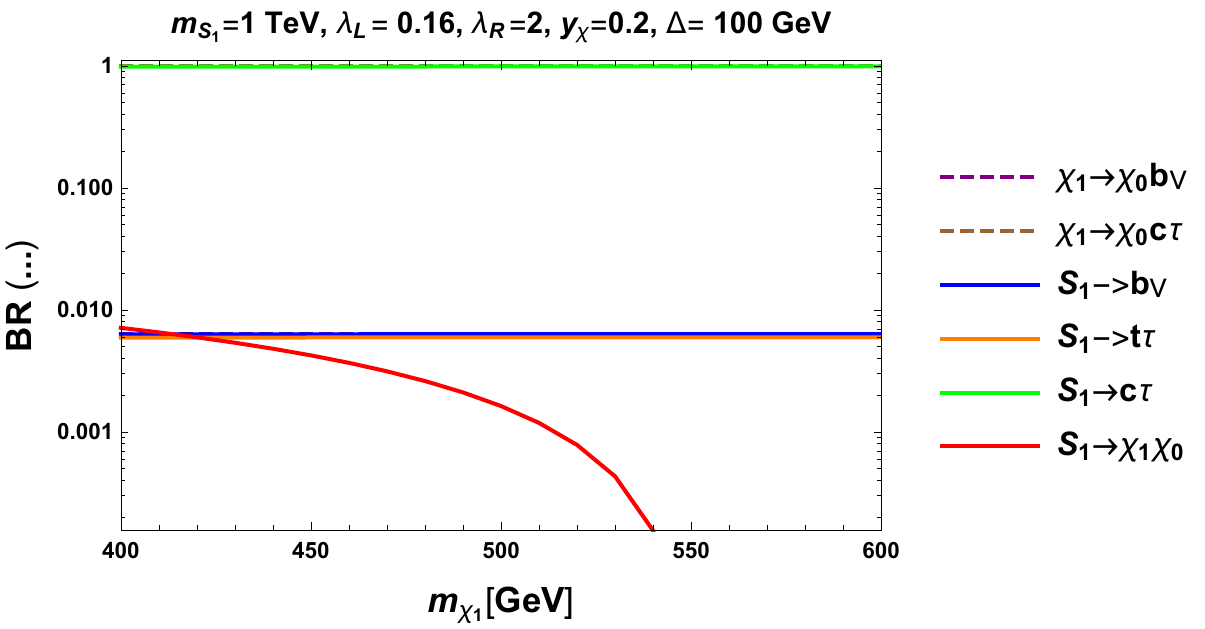}
	\caption{Branching ratios (top) for the $S_1$ and $\chi_1$ as a function of the $\chi_1$ mass. The mass difference between $\chi_1$ and $\chi_0$ is kept as 100~GeV, $y_{\chi} = 0.2$ and the other parameters correspond the second and fourth line of Table~\ref{tab:RDM:benchmark_points}, called $P_1$ and $P_2$ in the main text.}
		\label{fig:RDM:BRs}
\end{figure}

\begin{figure}[!h]
	\centering
	\includegraphics[width=0.48\textwidth]{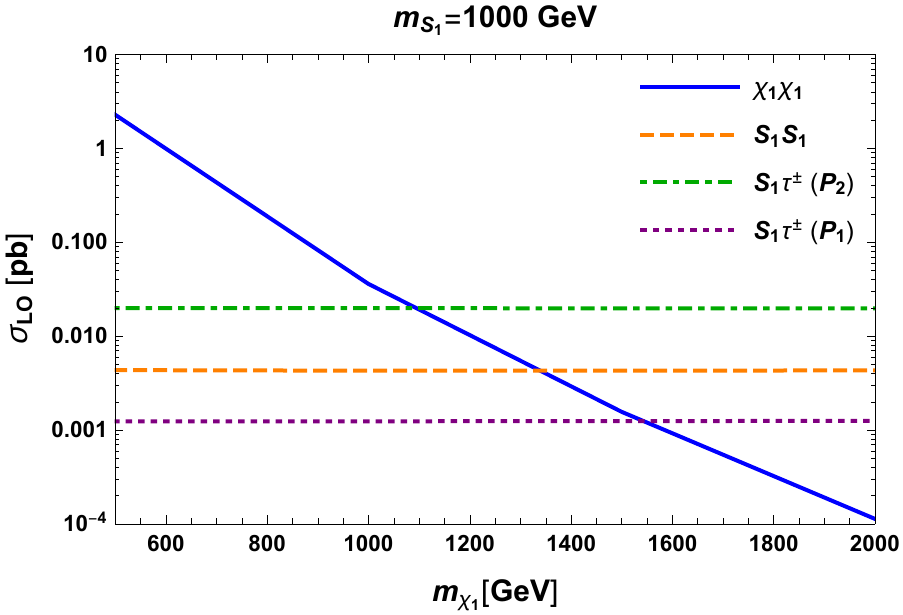}
	\includegraphics[width=0.48\textwidth]{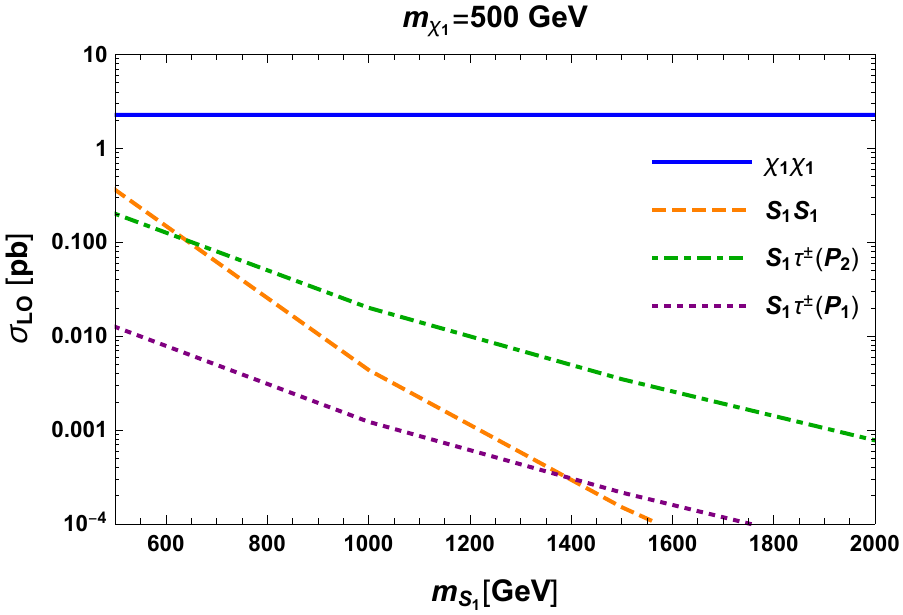}
	\caption{Cross-sections for $S_1$ and $\chi_1$ production at the LHC as a function of  $\mxOne$ (left panel) and $\mLQ$ (right panel), for pair production of $\chi_1$ (solid blue), pair production of $S_1$ (dashed orange) and associated production of $S_1$ and a tau lepton, which is shown for the benchmark points $P_2$ (dot-dashed green) and $P_1$ (dotted green). In the right panel we kept fixed $x=\lambda_L \tev / \mLQ$ and $y = \lambda_R \tev / \mLQ$ to the corresponding $P_1$ and $P_2$ values, see main text for details.}.
		\label{fig:RDM:LHCxsec}
\end{figure}

Before starting with the relevant collider and dark matter phenomenology, it is instructive to take a further insight into the phenomenology of our benchmarks. In figure~\ref{fig:RDM:BRs} we show the branching fractions of $S_1$ and $\chi_1$ for each of our working points, setting $\mLQ = 1$ TeV. For concreteness we fixed the dark coupling $y_{DM} = 0.2$ and $\Delta=\mxOne - \mxZero = 100 $ GeV, but the phenomenology does not strongly depend on the actual values. Note that in our model the $\chi_1$ always decays into a 3-body final state via an intermediate $S_1$.
The phenomenology for the second case (right panel) is fairly simple: the dominant decay of $S_1$ is into a charm-quark and a tau-lepton, hence $\chi_1$ decays via an off-shell $S_1$ almost exclusively into $c~\tau~\chi_0$. In the first scenario, where $\lambda_L$ and $\lambda_R$ are comparable, $S_1$ decays with appreciable ratios into both the three ``visible" channels and the invisible one. As a consequence, $\chi_1$ decays into the $c~\tau \chi_0$ and $b \nu \chi_0$ final states with similar ratios while the top final state is kinematically closed. This information is not enough to anticipate the collider phenomenology, as we will also need the production cross sections for $S_1$ and $\chi_1$ states. We present the leading-order (LO) QCD cross sections as a function of $\mxOne$ and $\mLQ$ in figure~\ref{fig:RDM:LHCxsec} for the three main production processes: pair production of $S_1$ and $\chi_1$ via gauge interactions, and associated production of a leptoquark and a tau-lepton. We have used as a benchmark the fourth line from Table~\ref{tab:RDM:benchmark_points}, hence when scanning over $m_{S_1}$ we also vary $\lambda_{L,R}$ as to keep $x$ and $y$ fixed to the benchmarked values.

It comes as no surprise that the pair production of $\chi_1$ is the process with the largest cross section, as both $S_1$ and $\chi_1$ have the same colour charges but $\chi_1$ is a fermion, and also because we are assuming that $S_1 \to \chi_1 \chi_0$ occurs on-shell, hence $\chi_1$ is lighter than $S_1$\footnote{If we abandon this hypothesis the take-home message is exactly the same: the largest event rate corresponds to the lightest BSM colored particle in the spectrum.}. The single LQ cross section becomes relevant for large values of $\lambda_R$: the $S_1 \tau$ cross section intercepts $S_1$ pair production at about 700 GeV (1.4 TeV) for $P_1 (P_2)$, while the $S_1$ pair production cross section gets suppressed very fast by the large-x parton distribution functions. In the case of $P_1$, where $\lambda_L$ and $\lambda_R$ are comparable, then we should also include associated production of $S_1$ with bottom and top quarks, but these are PDF suppressed and give a negligible contribution to the total cross section at the LHC. These could however be important when considering future hadron colliders with larger energies~\cite{Benedikt:2018csr}.

\section{LHC constraints}
\label{sec:RDM:lhc}
Our model contains three new fields that are probed by several LHC searches, which generically fall into two categories. On one hand, we have direct searches for leptoquarks ($S_1$), both for single and double production. On the other hand we have the full spectrum of missing transverse energy (MET) searches targeting invisible final states, in our case driven by pair production of $\chi_1$ via gauge interactions. As a hybrid of both categories, the CMS search for leptoquark plus dark matter~\cite{Sirunyan:2018xtm}, inspired by the Coannihilation Codex~\cite{Baker:2015qna}, gives not only interesting constraints on the parameter space, but moreover it would allow us to establish the connection between the flavor anomalies and dark matter, in a post-discovery scenario. 

In each of the categories described above, we will consider both a) existing searches and b) novel searches (for instance, the aforementioned LQ + MET search of ref~\cite{Sirunyan:2018xtm} only applies if the leptoquark decays to muons, but it is not applicable for e.g: electron decays). Given the large number of searches and the many processes they target, we summarize the ones we will use in this section in Table~\ref{tab:RDM:searches}.

\begin{table}[th!]
\begin{tabular}{|l|l|l|l|l|llll}
\cline{1-5}
Category & Reference                                                           & Target process                                  & Remarks                                                                                                                                  \\ \cline{1-5}
         & \cite{Aaboud:2019bye}                                    & $S_1 S_1 \to t \tau t \tau$             &                                                                                                                                      \\
         & \cite{Aaboud:2019bye}                                                                 & $S_1 S_1 \to b \nu b \nu$               &                                                                                                                                           \\
     LQ    & \cite{Aaboud:2018kya}                                                       & $S_1 S_1 \to c \tau c \tau$             & Recasting of $b \tau b \tau$ study.                                                                                                   \\
         &Buttazzo et al. this volume                                                        & $S_1 S_1 \to b \nu t \tau$              &                                                                                                                                             \\
         & \cite{Sirunyan:2018jdk} & $S_1 \tau \to c \tau \tau$           &             Rescaling of $b \tau \tau$ study                                                                 \\ \cline{1-5}
         & \cite{Sirunyan:2017kqq,Sirunyan:2017kiw}                            & $\chi_1 \chi_1 \to b b \chi_0 \chi_0$   &                                                                                                                                           \\
         & \cite{Aad:2019byo}                                                       & $\chi_1 \chi_1 \to \tau \tau \chi_0 \chi_0$   &                                                                                                                                      \\
MET      & \cite{ATLAS:2019vcq}                                           & $\chi_1 \chi_1 \to \chi_0 \chi_0 + n j$ & $n \geq 2$ only.                                                                                                                          \\
         & \cite{Aaboud:2017phn}                                                       & $\chi_1 \chi_1 \to \chi_0 \chi_0 + n j $  & $1 \leq n \leq 4$                                                                                                                               \\ \cline{1-5}
LQ + MET & \cite{Sirunyan:2018xtm}                                                       & $S_1 S_1 \to l q \chi_1 \chi_0$         & \begin{tabular}[c]{@{}l@{}}CMS requires $\mu$, \\ we use $ \tau \to \mu$ for interpretation\end{tabular}       \\ \cline{1-5}
\end{tabular}
\caption{\it List of searches considered in this work. We split the studies in three categories, depending if they correspond to standard leptoquark searches, MET searches, and also the CMS search for a resonant LQ + MET, belonging to the hybrid of the two first categories.}
\label{tab:RDM:searches}
\end{table}

As anticipated in the previous section, the collider bounds will be dominated by $\chi_1$ pair production, hence we expect the $b \bar{b} $ + MET and $\tau \tau $ + MET searches to set tight constraints on the parameter space. These would lose stem as the final state becomes inefficient (e.g: in our second working point the branching ratio of $S_1 \to b \nu$ is negligible) or if the dark sector becomes compressed, giving rise to soft decay products that would not pass the reconstruction requirements. In that case we can expect a bound from the mono-jet search, where $\chi_1$ gives only rise to MET and one invokes initial state radiation to boost the $\chi_1 \chi_1$ system while allowing for an efficient trigger. This bound should however, be looser than the one from the  $b \bar{b} / \tau \tau$ + MET and hence the LQ searches become competitive in probing the parameter space. The hybrid search is necessary to fingerprint the connection between the flavour anomalies and the dark sector, but naively we would not expect it to set the leading bounds. 

\subsection{Leptoquark searches}

\begin{figure}[!h]
	\centering
	\includegraphics[width=0.7\textwidth]{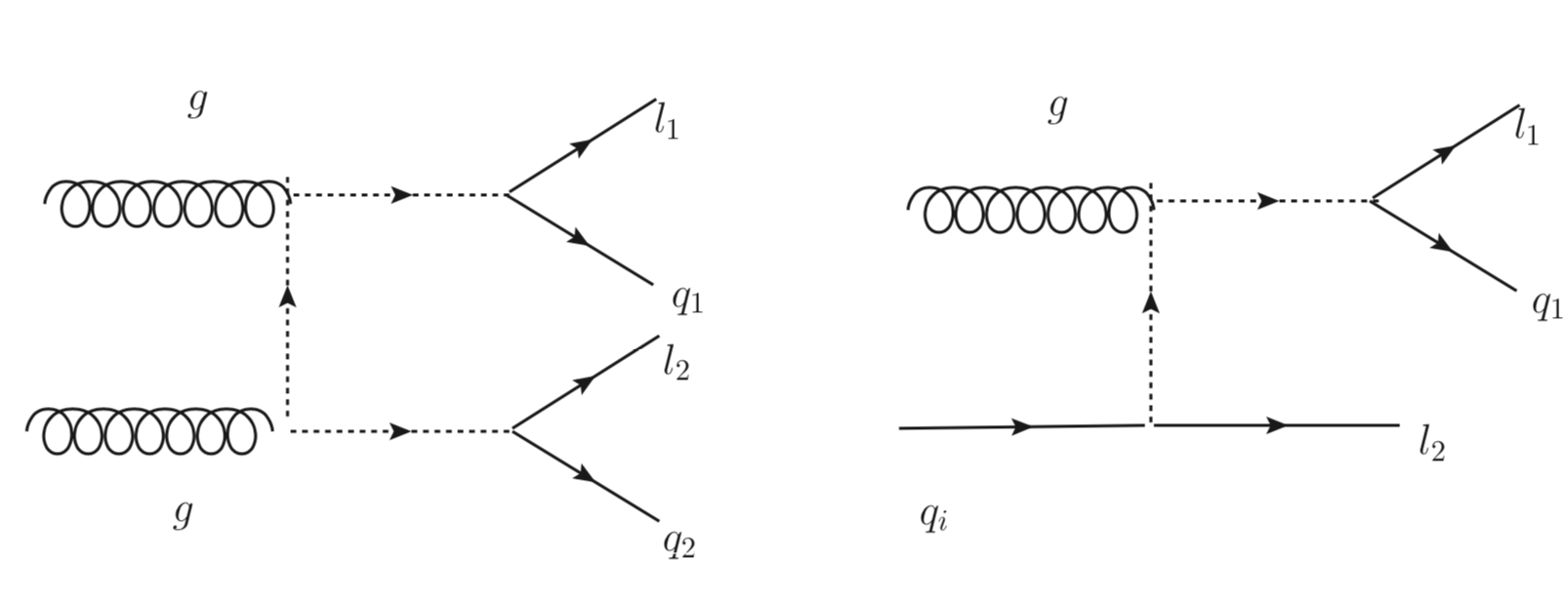}
	\caption{Representative Feynman diagrams for pair (left) and single (right) leptoquark production at hadron colliders.}
		\label{fig:RDM:LQvisible}
\end{figure}

There are two different production mechanisms for leptoquarks at the LHC, which are shown in figure~\ref{fig:RDM:LQvisible}: pair production (left panel) due to gauge interactions\footnote{In principle also t-channel leptonic exchange can be relevant given the size of the leptoquark couplings, and would be taken into account in the next-to-leading order calculation in a future publication.} and single production in association with a SM fermion (right panel). The pair production is a guaranteed mode, since its cross section is only a function of the leptoquark mass. The single production cross section depends both on the leptoquark couplings to the SM fermions and with the leptoquark masses. The experimental searches for leptoquarks are organized by these production mechanisms, and we will follow the same taxonomy here.

\subsubsection{LQ pair searches}

Searches for visible decays of LQ pairs at ATLAS~\cite{Aad:2015caa,Aaboud:2019jcc,Aaboud:2019bye} and CMS~\cite{Sirunyan:2018nkj,Sirunyan:2018kzh,Sirunyan:2018vhk}, focus on both leptoquarks decaying in the same final state. In Ref.~\cite{Aaboud:2019bye} five different searches are described. The first one is a search for LQ decaying into the $b \tau b \tau$ final state, reoptimising an existing di-Higgs search in the same final state~\cite{Aaboud:2018sfw}, to the different final state  kinematics. The other four searches are reinterpretations of existing studies of supersymmetric particles: searches for stop pair production in the zero-lepton~\cite{Aaboud:2017ayj} and one-lepton channel~\cite{Aaboud:2017aeu} ($t \bar{t} +$ MET), stop decays into staus~\cite{Aaboud:2018kya} and a sbottom search~\cite{Aaboud:2017wqg} ($b\bar{b} + $ MET ). 
Our model has four leptoquark decay channels, three into SM particles and one into the DM mediator, 
as shown in figure~\ref{fig:RDM:BRs}. Two of the decay channels are directly covered by the ATLAS searches,
namely the decays into $t \tau$ and into $b \nu$, through the searches respectively 
for the $b \tau b \tau$ final states, and for $b \bar{b} +$ MET. 
For any choice of the coupling, and barring effects from the top mass, we would have that $BR (S_1 \to t \tau) \approx BR (S_1 \to b \nu)$, in contrast to the ATLAS study where $BR(S_1 \to b \nu) = 1 - BR(S_1 \to t \tau)$ is assumed. 

We note that even if this search is restricted to the third generation final states, there is no search for the ``mixed" final state, namely  $b \nu t \tau$. In order to achieve an optimal coverage this channel should be included, and we have indeed carried out a preliminary simulation study. A detailed analysis of this final state for a 14 TeV LHC based on a parametrised simulation 
of the ATLAS detector, with full consideration of all relevant backgrounds is described 
elsewhere\footnote{See Buttazzo et al. in the same proceedings volume.}. In this study a dedicated selection is developed to separate the final state of interest from the overwhelming $t\bar{t}$ background through novel variables which optimally exploit the kinematical features of leptoquark pair production, and its difference from the $t\bar{t}$ kinematics. 

In addition we also try to address
the $c \tau c \tau$ final state through a rough rescaling of 
the existing ATLAS $b \tau b \tau$ limits.~\footnote{To be fully comprehensive one should also consider the mixed final states $c \tau b \nu$ and $c \tau t \tau$, but these are beyond the scope of this work. We expect them to be subdominant: given the lesser quality of $c$-tagging with respect to bottom- and top-tagging, $c$ final states will only matter when the 3rd generation branching fraction is very small, ${\cal O} (5 \%)$. In that case the largest branching fraction would occur in the $c \tau c \tau$ channel, which would be at least about an order of magnitude above the ``mixed" rate. }

In the model 
under consideration this decay is dominant over a large part of the parameter space 
allowed by the consideration of the flavour anomalies, as shown in figure~\ref{fig:RDM:FlaFit},
where dashed blue lines represent the isovalues of $BR (S_1 \to c \tau)$  in the 
$\lambda_{L}-\lambda_{R}$ plane. This final state is not targeted by the ATLAS searches which
are focused on the third generation. 

In absence of a dedicated analysis, 
a rough evaluation of the coverage for the $c \tau c \tau$ final state can 
be obtained by the consideration of the ATLAS results for the $b \tau b \tau$ final 
state. The experimental algorithm for tagging the $b$-jets used in 
Ref.~\cite{Aaboud:2019bye} has an average efficiency for tagging a jet from 
the fragmentation of a $b$-quark of 70\%. The same algorithm tags the jet 
from the fragmentation of a $c$-quark with an efficiency of $\sim8$\%
(\cite{ATL-PHYS-PUB-2016-012}). 
Therefore  the results for the $b \tau b \tau$ final can be 
recasted for the $c \tau c \tau$ final state by scaling down the excluded 
branching fraction by a factor $\sim$8.5. This scaling is conservative, as 
it assumes that the analysis requires two $b$-tagged jets in the event,
whereas both events with 1 and 2 $b$-tags are considered in \cite{Aaboud:2019bye}.
A dedicated analysis, based on e.g. the $c$-tagging algorithms recently
developed by ATLAS \cite{ATL-PHYS-PUB-2015-001} should provide a much better coverage
for the $c\tau$-dominated region and
the inclusion of such a study within the leptoquark search programme 
should be high in the list of priorities of the experimental collaborations.

\begin{figure}[!thp]
  \centering
  \includegraphics[width=0.46\textwidth]{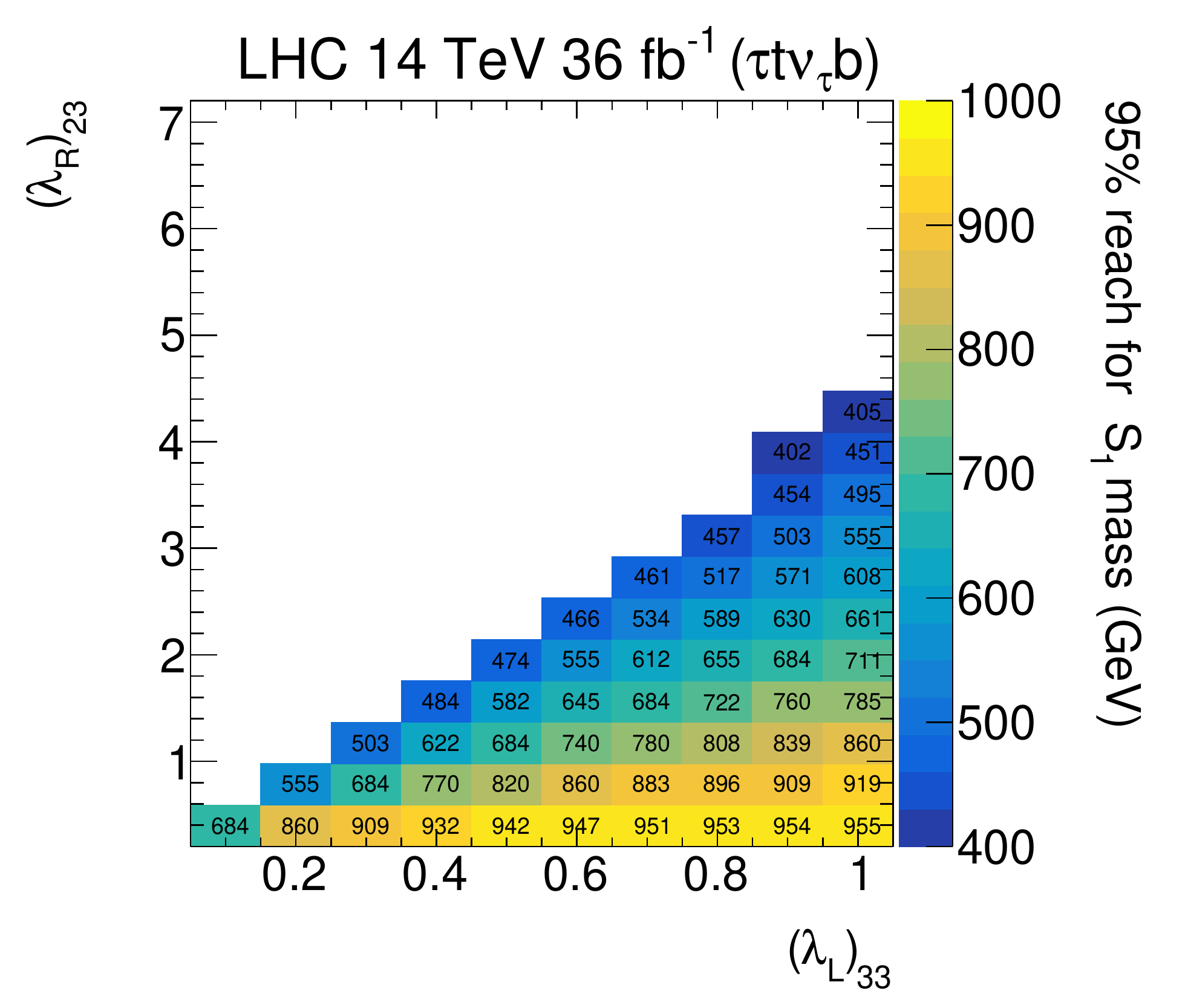} 
  \includegraphics[width=0.46\textwidth]{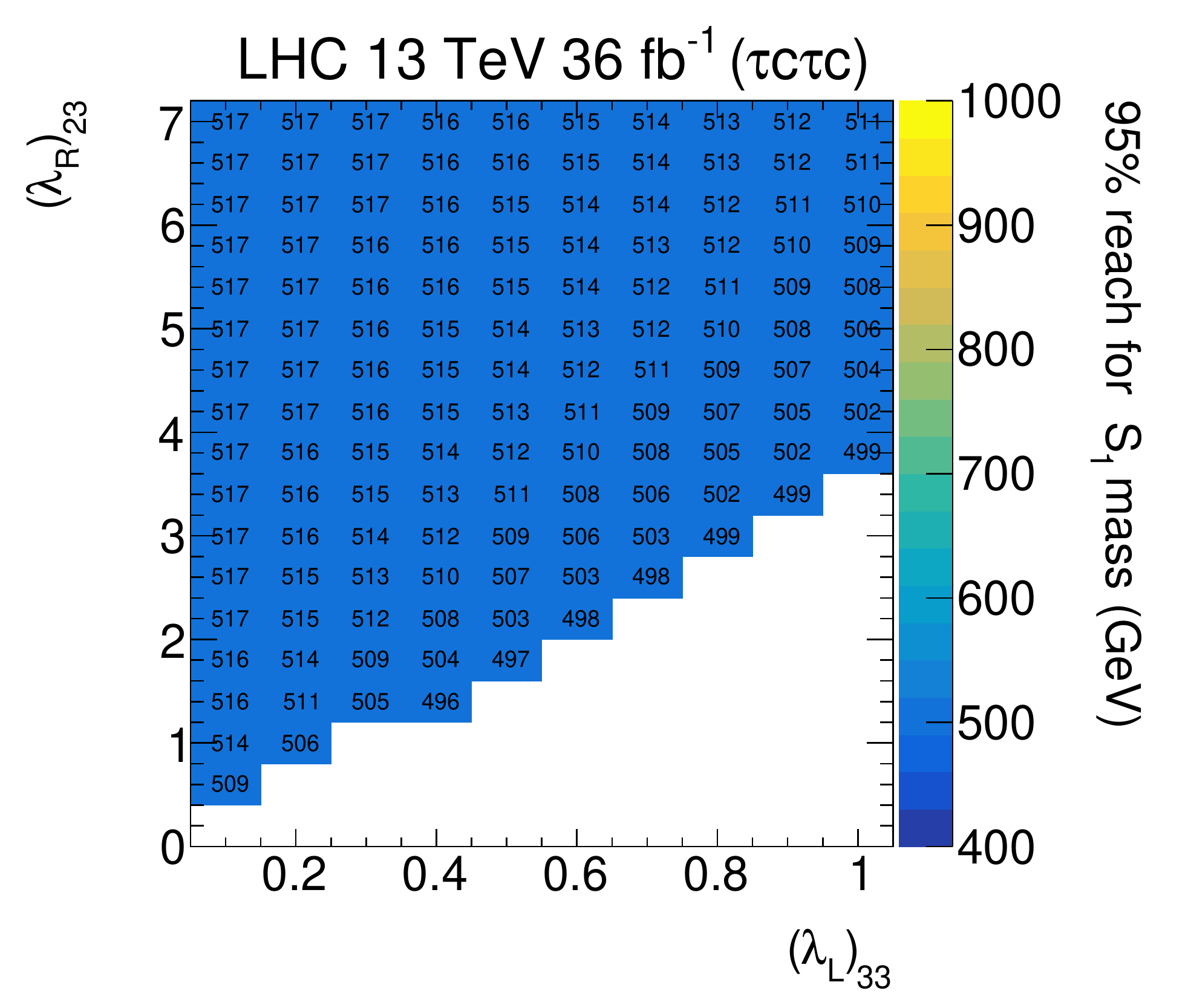} 
  \caption{\it Excluded values of $\mLQ$ at 95\% CL in the plane defined by the 
$(\lambda_{\sss L})_{33}-(\lambda_{\sss R})_{23}$ couplings for an integrated luminosity
of 36~fb$^{-1}$. In the right panel the 
14 TeV mixed-decay $b \nu_{\tau} t \tau$ analysis (described in Buttazzo et al within this proceeding volume) and on the right the rescaling of the ATLAS $b\tau b\tau$ analysis (described in the main text).}
\label{fig:RDM:recastlq}
\end{figure}

Both for the proposed 
mixed analysis and for the $c \tau c \tau$ rescaling of the ATLAS $b \tau b \tau$ analysis  
we evaluated, for each point in the $\lambda_{L}-\lambda_{R}$ plane the excluded
mass based on the considerations above. The result is shown in figure~\ref{fig:RDM:recastlq},
where the results for the mixed ( $c \tau c \tau$) analyses are shown in the left
(right) panel. For the uncovered points in the two panels the exclusion on the 
$S_1$ mass is below 400~GeV, which is the lowest mass for which the two considered
analyses provide an exclusion. 

By combining the results from the two recasts, 
with the currently published LHC statistics of 36~fb$^{-1}$ a value of $\mLQ$ of 
at least 500~GeV is (or can be) excluded at 95\% CL, but significantly 
larger discovery potential can be expected with a dedicated $c\tau c\tau$ analysis.

\begin{figure}[!thp]
  \centering
  \includegraphics[width=0.8\textwidth]{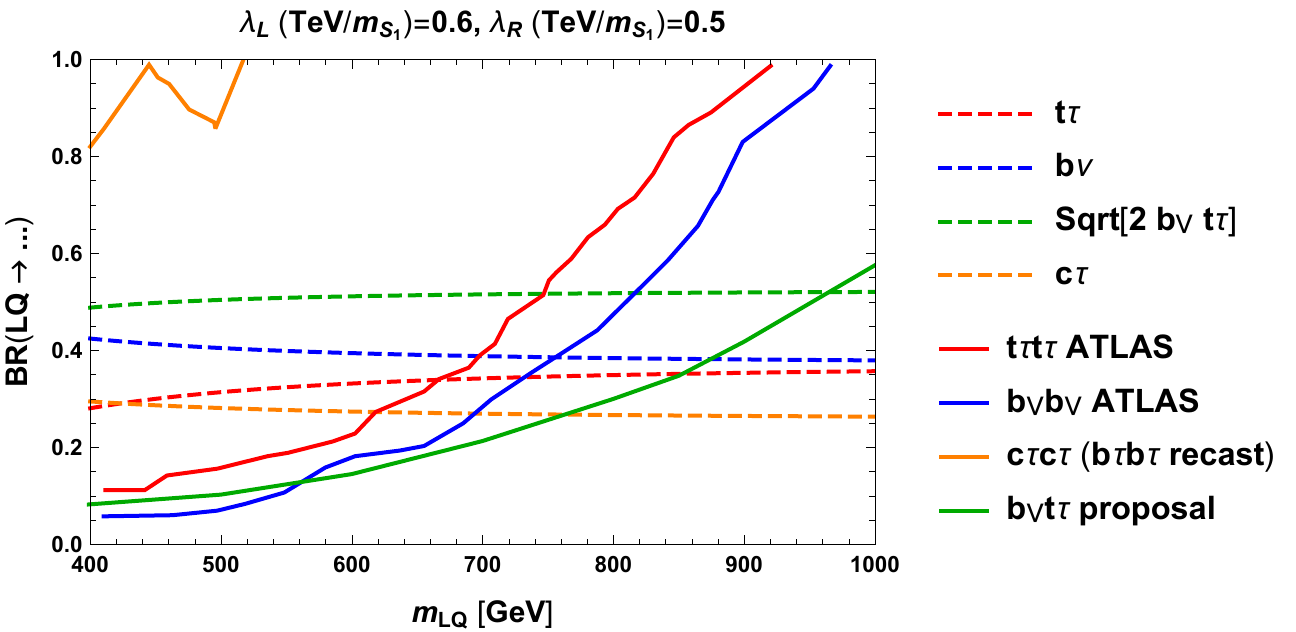} 
  \caption{\it Coverage of the different LQ pair searches with decays into third generation fermions. We show in dashed lines the branching ratios for the case where $x=0.5$ and $y=0.6$ (see main text for definition), while solid lines denote the branching ratio reach as a function of the leptoquark mass, for the $t \tau t \tau$ (red), $b \nu b \nu$ (blue), and $c \tau c \tau$ taken directly 
or rescaled from~\cite{Aaboud:2019bye}, as well as our proposed search in the $t \tau b \nu$ channel (green). For the mixed decay we plot the square of $2 BR(S_1 \to b \nu) BR(S_1 \to t \tau)$, see main text for details. }
\label{fig:RDM:LQpaircoverage}
\end{figure}

To summarize, we compare in figure~\ref{fig:RDM:LQpaircoverage} the
leptoquark mass coverage as a function of the BR for the four final states defined 
above for 36~fb$^{-1}$, where for the mixed decay 
$BR (S_1 \to t \tau) \approx BR (S_1 \to b \nu)$ is assumed. In the searches where both leptoquarks decay in the same manner what is usually reported is the leptoquark branching fraction into a given channel, but the event rate scales with the squared of this value. Hence, for the mixed decay we plot the square of $2 BR(S_1 \to b \nu) BR(S_1 \to t \tau)$. The solid lines correspond to the mass reach from the experiments and our two proposed searches, while the dashed lines correspond to our second benchmark\footnote{In our first benchmark we only can constrain the model using the $c \tau$ analysis, hence the mass reach is about 500 GeV.}. Over most of the leptoquark mass range considered, the mixed decay $b \nu t \tau$ (dashed green) 
search has a higher sensitivity than the $b \nu b \nu$ (dashed blue) and the $b \tau b \tau$ (dashed red) searches. The $c \tau$ final state is only relevant for the case where the branching fraction of $S_1$ to other states is below about 5 \%. As shown in our first benchmark, branching fractions even smaller into $b \nu$ and $t \tau$ do occur near the best fit point. 
\subsubsection{LQ associated production with a lepton}

A single leptoquark can also be produced in association with a SM fermion, as shown in the right panel of figure~\ref{fig:RDM:LQvisible}. For our particular choice of couplings and of leptoquark charge, 
two single leptoquark production processes will contribute:
$bg \rightarrow \nu_{\tau} S_1 $ and $cg \rightarrow \tau S_1$.
Combining with the three possible decays into SM particles shown in 
figure~\ref{fig:RDM:BRs}, one has a  final state with a single $b$+MET (mono-$b$), 
states with a single $\tau$ + heavy quark + MET, and states   
with two taus and a $c$ or a $t$ quark.

Searches exist for the mono-$b$ final state, but, due to the large backgrounds,
and the lack of kinematic handles,
we expect the sensitivity to be lower than the other channels. 

The final states with a single tau and a heavy quark are addressed elsewhere
in these proceedings, and will be relevant for the case where $\lambda_{R}\sim\lambda_{L}$.
Given the relatively low values of $\lambda_{L}$ compatible with the explanation 
of the flavour anomalies, we expect the reach in mass to be comparable
with the one from the mixed decay in pair production discussed in the
previous section.
 
\begin{figure}[!thp]
  \centering
  \includegraphics[width=0.46\textwidth]{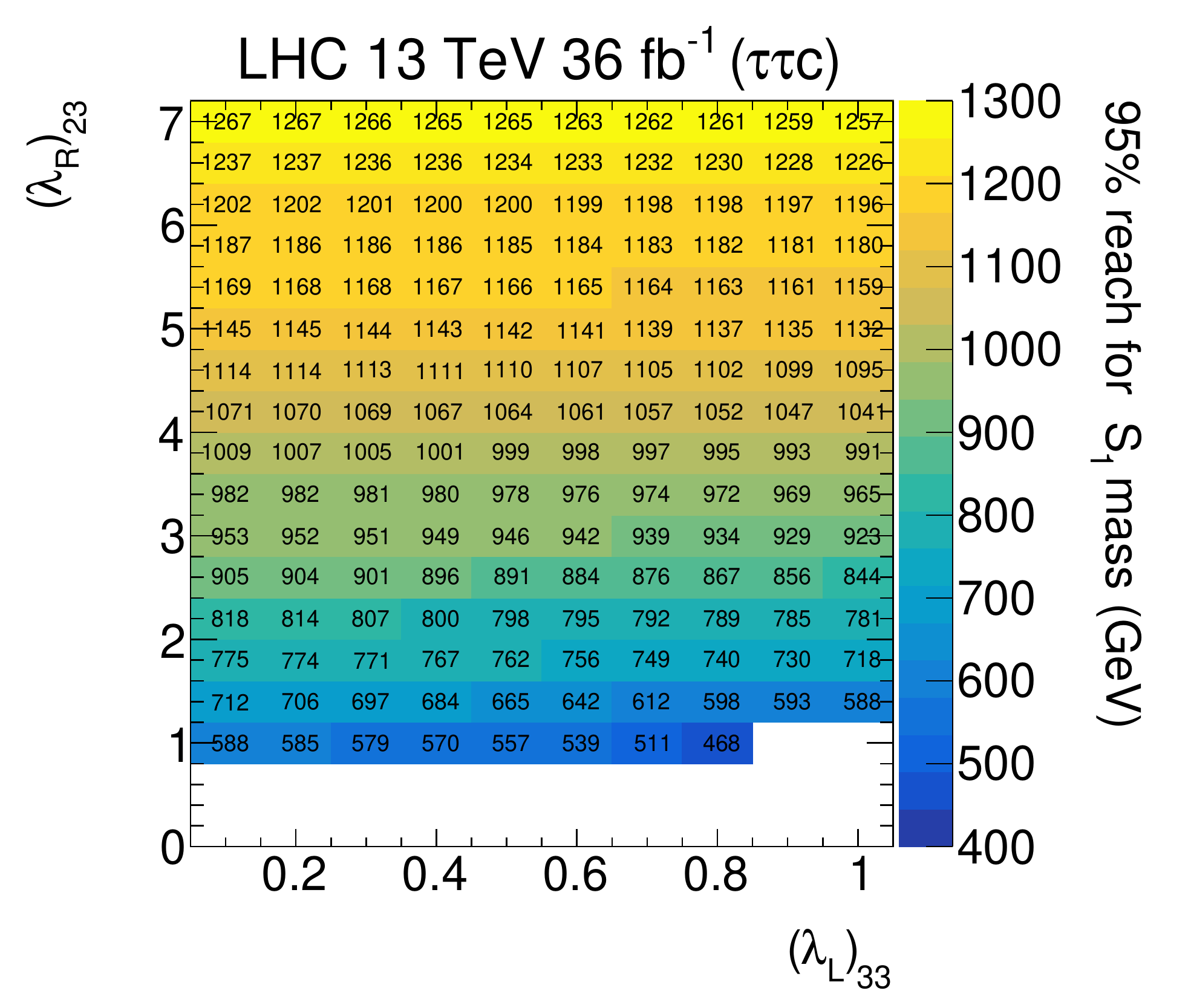}
  \caption{\it Excluded values of $\mLQ$ at 95\% CL in the plane defined by the
$(\lambda_{\sss L})_{33}-(\lambda_{\sss R})_{23}$ couplings for an integrated luminosity
of 36~fb$^{-1}$ for the CMS $\tau\tau c$ analysis as described in the text.}
\label{fig:RDM:recastcms}
\end{figure}

The $\tau\tau c$ final state will be the dominant channel for the case in which
$\lambda_{R}>\lambda_{L}$, as the event rate scales as 
$\lambda_{R}^2\times BR(S_1 \to c\tau)$. No search for this channel exists in the 
literature, but the CMS Collaboration has published a search for single 
leptoquark production in the $\tau\tau b$ channel \cite{Sirunyan:2018jdk}. 
As is the case for the ATLAS $b\tau b\tau$ analysis in the previous subsection, 
the algorithm used to tag jets from the fragmentation of $b$-quarks has
a certain sensitivity also for jets from the fragmentation of $c$-quarks.
Under the assumption that the kinematic acceptance of the analysis is
the same for the $\tau\tau b$ and for the $\tau \tau c$ final states, 
the results for the $\tau \tau c$ may be obtained by a simple rescaling
of the heavy-flavour tagging efficiency. In Ref.~\cite{Sirunyan:2018jdk}
they declare that they are using the combined secondary vertex algorithm at a working
point with an efficiency for $b$-jets of approximately 60\%.
For this algorithm they define in Ref.\cite{Sirunyan:2017ezt} a medium working point
with efficiency of 63\% for $b$-jets and 12\% for $c$-jets. We will
assume the ratio of these numbers for our rescaling.
The result is shown in figure~\ref{fig:RDM:recastcms}, where the value of $\mLQ$ excluded
at 95\% CL is shown in the plane defined by the 
$(\lambda_{\sss L})_{33}-(\lambda_{\sss R})_{23}$ couplings.
 
\begin{figure}[!thp]
  \centering
  \includegraphics[width=0.46\textwidth]{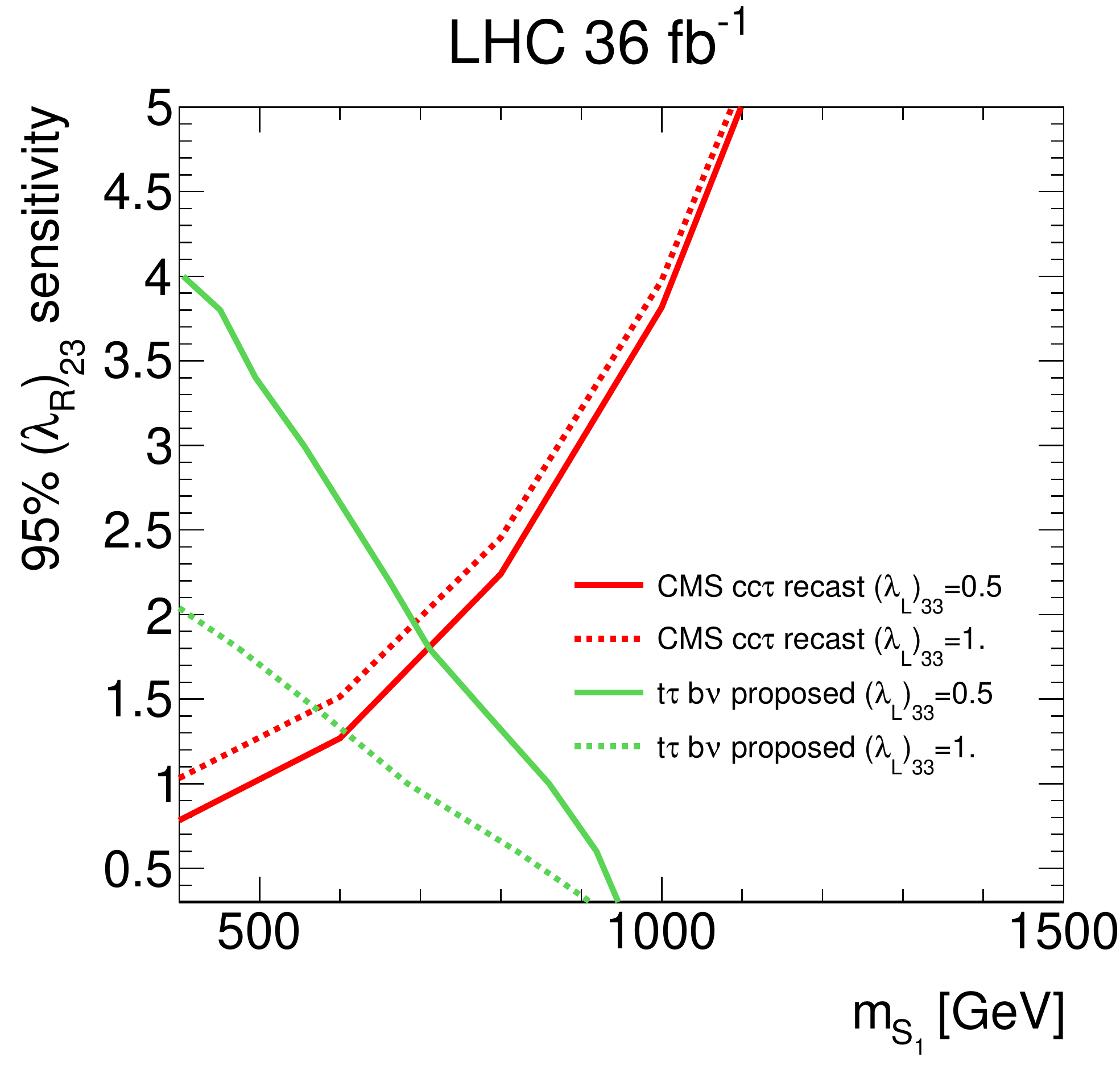}
  \caption{\it Comparison of the excluded $\lambda_R$ for the ``mixed" $b \nu t \tau$ proposed search (green) and the $c \tau \tau$ recasting of the CMS $b \tau \tau$ study. Solid (dashed) lines denote $\lambda_{L} = 0.5 (1.0)$.}
\label{fig:RDM:singlevsdouble}
\end{figure}

Finally we compare in figure~\ref{fig:RDM:singlevsdouble} the interplay of exclusions arising from pair and single leptoquark studies, where we show the 95 \% C.L exclusion on $\lambda_R$ as a function of \mLQ. For concreteness we included here only the proposed ``mixed" search $b \nu t \tau$ (in red) and the $c \tau \tau$ result obtained via a recasting of the CMS study (green), and present curves using two representative values of $\lambda_L$: 0.5 (solid) and 1.0 (dashed). As previously discussed, we clearly see that the mixed search is optimal for lower values of $\lambda_R$ (where the $c \tau$ final state has a comparable branching fraction with the other visible decays), while the single leptoquark search is optimal for high values of $\lambda_R$, where $c \tau$ is by far the dominant decay mode. As our benchmark points show that the flavor anomalies can be solved with either $\lambda_R \sim \lambda_L$ or $\lambda_R \gg \lambda_L$, both kind of searches are a necessary ingredient of a comprehensive campaign to probe the ``leptoquark solution" of the $\RD$ anomalies.
\subsection{Dark matter (missing energy) searches}
\label{subsec:RDM:mjet}
Since the $\chi_1$ particle is the lightest new particle with non-trivial SM quantum numbers, its production is strongly constrained by LHC searches, and it can only be ``hidden" if $\chi_1$ and $\chi_0$ are close in mass. The LHC production is depicted in figure~\ref{fig:RDM:MonoJet}. Since we always assume $m_{\chi_1} < m_{S_1}$, the decay of $\chi_1$ is entirely determined by the LQ branching ratios and the only decay channel is $\chi_1 \to S_1^*(\to lq) + \chi_0$.
As shown in figure~\ref{fig:RDM:BRs}, $\chi_1$ dominantly decays
to $\chi_0 + c\tau$ for $\lambda_R > \lambda_L$ (third and fourth benchmarks in Table~\ref{tab:RDM:benchmark_points}) or has similar BRs to $\chi_0 + c \tau$, $\chi_0 + b\nu$ and $\chi_0 + t \tau$ (first and second benchmarks) if $\lambda_R \lesssim \lambda_L$. Therefore searches for missing energy associated with jets, $b$-jets, tops or taus can be used to test our model.
\subsubsection{$b$-jets plus MET searches}

\begin{figure}[!h]
	\centering
	\includegraphics[width=0.5\textwidth]{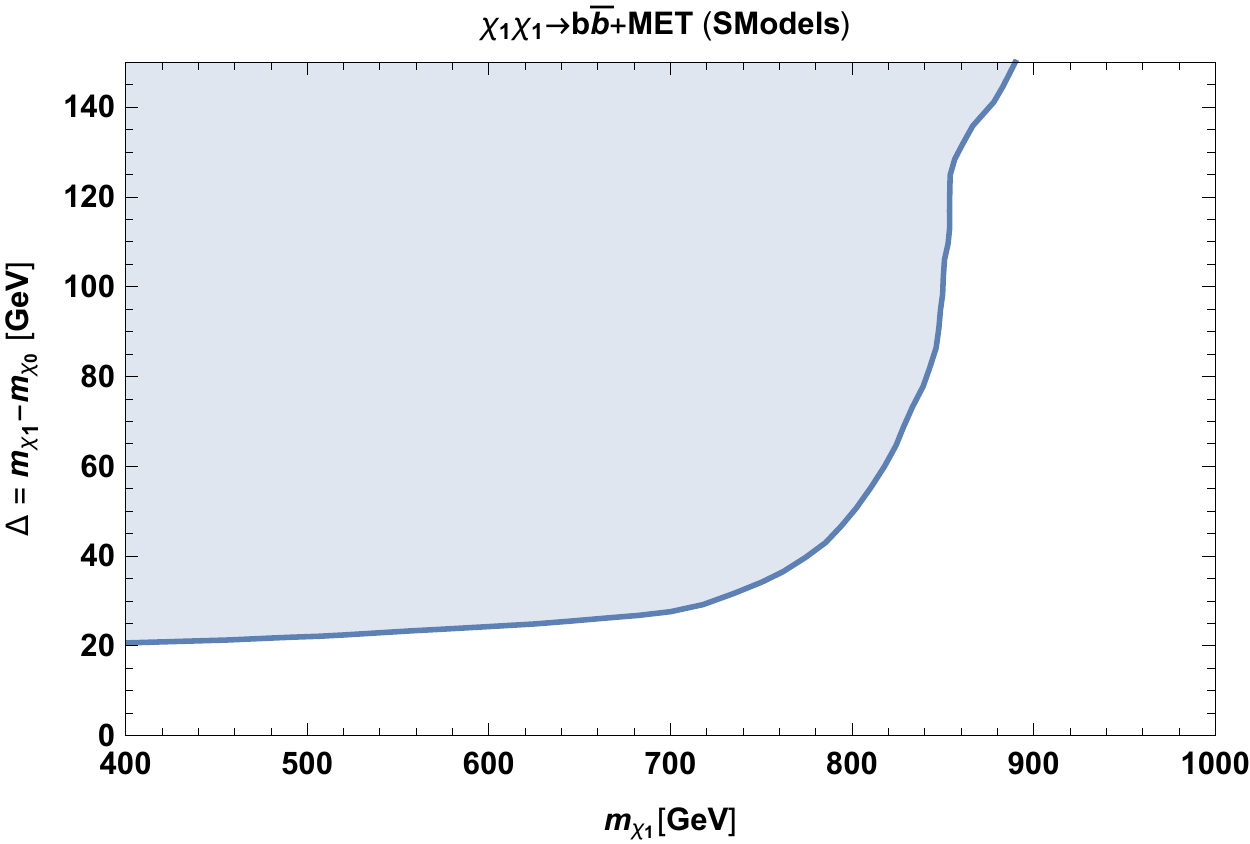}
	\caption{Region in the $\mxOne - \mxZero$ vs $\mxOne$ plane excluded by LHC searches for $b$-jets plus MET. The remaining model parameters are $y_{\chi} = 0.2$, $\lambda_L = 0.6, \lambda_R = 0.5$ and $\mLQ = 1$~TeV. The shaded region corresponds to the parameter space excluded by CMS searches for final states containing one or more $b$-jets and missing energy\cite{Sirunyan:2017kqq,Sirunyan:2017kiw}.
		\label{fig:RDM:T2bbexcl}
	}
\end{figure}

If $\lambda_L \gtrsim \lambda_R$ a sizeable fraction of the $\chi_1$s produced at the LHC decay to $b \nu$ or $t \tau$. In order to constrain this scenario we re-interpret the CMS results for the sbottom simplified model $p p \to \tilde{b} + \tilde{b} \to b \tilde{\chi}_1^0 + b \tilde{\chi}_1^0$. For simplicity we assume that the decay $\chi_1 \to \chi_0 b \nu$ has similar efficiencies and the same cross-section upper limits obtained by CMS for the sbottom scenario can be applied to our model.
We have used the {\sc SModelS}\cite{Ambrogi:2018ujg} implementation of the CMS searches at 13 TeV and 35.9~fb$^{-1}$ from Refs.\cite{Sirunyan:2017kqq,Sirunyan:2017kiw} and used it to recast the CMS bounds.
The cross-sections were computed using \mg\ and we have applied a $K$-factor of 1.5 to the $\chi_1$ pair production cross-section.  The results are shown in figure~\ref{fig:RDM:T2bbexcl}, where we have assumed the second benchmark of Table~\ref{tab:RDM:benchmark_points} and varied $m_{\chi_1}$ and $m_{\chi_0}$. As we can see, $
\chi_1$ masses up to 900~GeV can be excluded for sufficiently large mass gaps. However, once $m_{\chi_1}-m_{\chi_0}$ drops below 25~GeV the $b$-jets originating from the $\chi_1$ decay become too soft and the search looses its sensitivity.

\subsubsection{$\tau$ plus MET searches}

Once $\lambda_L \lesssim \lambda_R$, the $\chi_1$ decays are dominated by $\chi_1 \to \chi_0 + c \tau$.
Although there is no LHC search which targets this final state, searches for taus plus missing energy can still be sensitive to this decay mode. Here we consider the 13 TeV ATLAS search at 139~fb$^{-1}$ from Ref.\cite{Aad:2019byo}, which targets final states with two taus plus missing energy and has been used to constrain the $p p \to \tilde{\tau} + \tilde{\tau} \to \tau \tilde{\chi}_1^0 + \tau \tilde{\chi}_1^0$  simplified model.
Since in our model $\chi_1$ is colored, its production cross-section is much larger than for pair production of staus and much higher masses can be tested by the analysis.
The upper limits presented by ATLAS, however, are limited to $m_{\tilde{\tau}} < 450$~GeV and do not extend to the region of interest ($m_{\tilde{\chi}_1} \sim $ TeV). Therefore a full recasting was done using the tools provided by {\sc CheckMATE2}~\cite{Dercks:2016npn}.\footnote{The working point for the tau-tagging efficiency implemented in {\sc CheckMATE} seems to be slightly different from the one reported in Ref.\cite{Aad:2019byo}. Nonetheless the efficiencies obtained with our implementation are within 20\%-50\% of the ones obtained by ATLAS for the stau simplified model.} We also point out that since the final state considered here contains $c$-jets and the ATLAS analysis has a veto on $b$-jets, about 50\% of our signal will be lost due to mis-tagging of $c$-jets.

\begin{figure}[!h]
	\centering
	\includegraphics[width=0.5\textwidth]{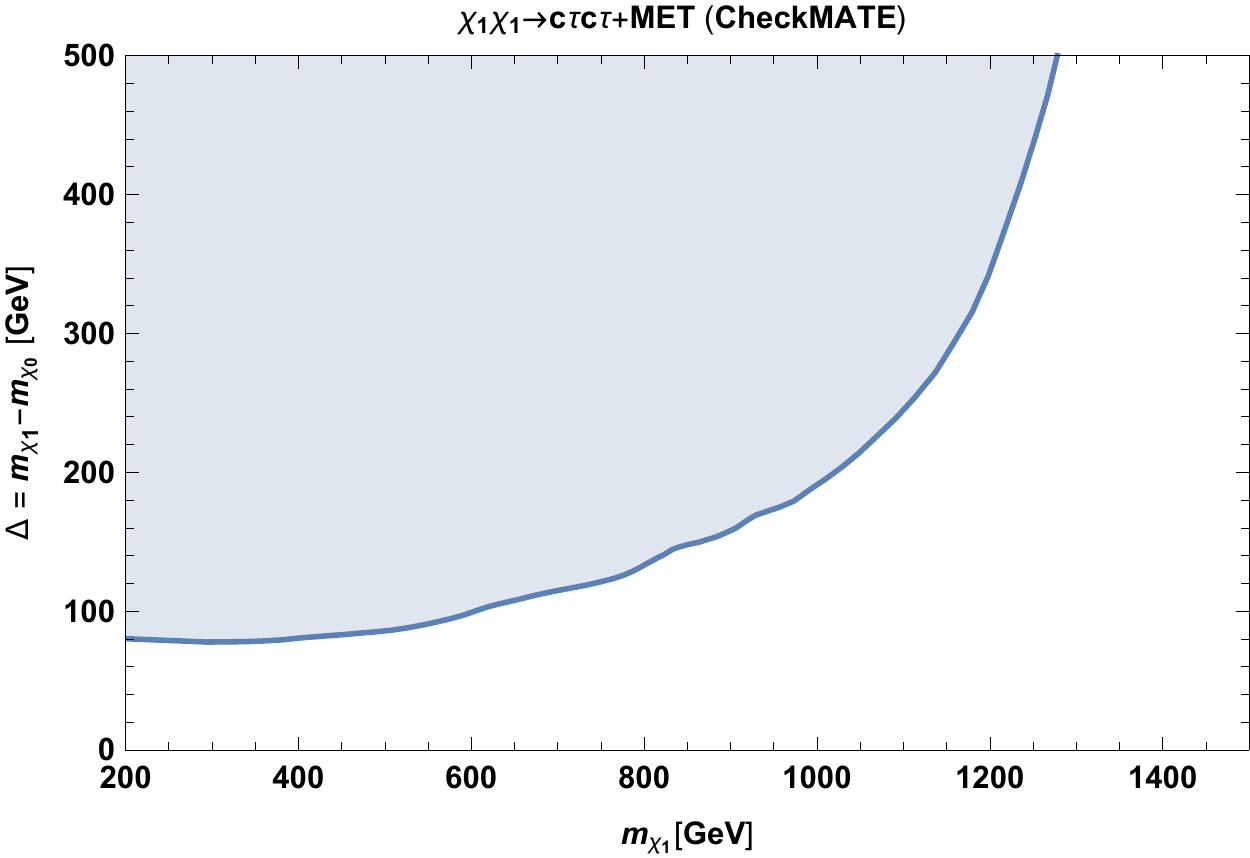}
	\caption{Region in the $\mxOne - \mxZero$ vs $\mxOne$ plane excluded by LHC searches for $\tau$ leptons plus MET. The remaining model parameters are $y_{\chi} = 0.2$, $\lambda_L = 0.16, \lambda_R = 2.0$ and $\mLQ = 1$~TeV. The shaded region corresponds to the parameter space excluded by ATLAS search for final states containing two hadronically decaying tau-jets and missing energy\cite{Aad:2019byo}. 
   \label{fig:RDM:TStauStauexcl}
	}
\end{figure}

Using \mg, {\sc Pythia 8.2}~\cite{Sjostrand:2014zea} and our analysis implementation in {\sc CheckMATE2 2.0.26} we compute the 95\% C.L. exclusion for the fourth benchmark model in Table~\ref{tab:RDM:benchmark_points}. Once again we apply a $K$-factor of 1.5 for the $\chi_1 \chi_1$ cross-section and vary $m_{\chi_1}$ and $m_{\chi_0}$. The result is shown in figure~\ref{fig:RDM:TStauStauexcl}, where we see that only points with a mass gap above 80 GeV can be excluded.

\subsubsection{jets plut MET searches}

\begin{figure}[!thp]
  \centering
  \includegraphics[width=0.4\textwidth]{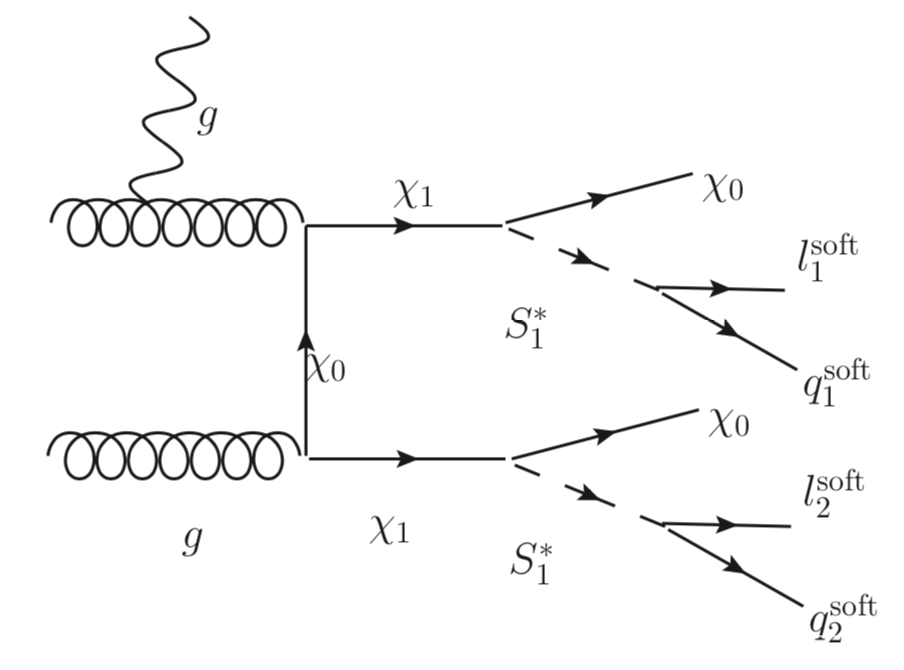} 
    \includegraphics[width=0.4\textwidth]{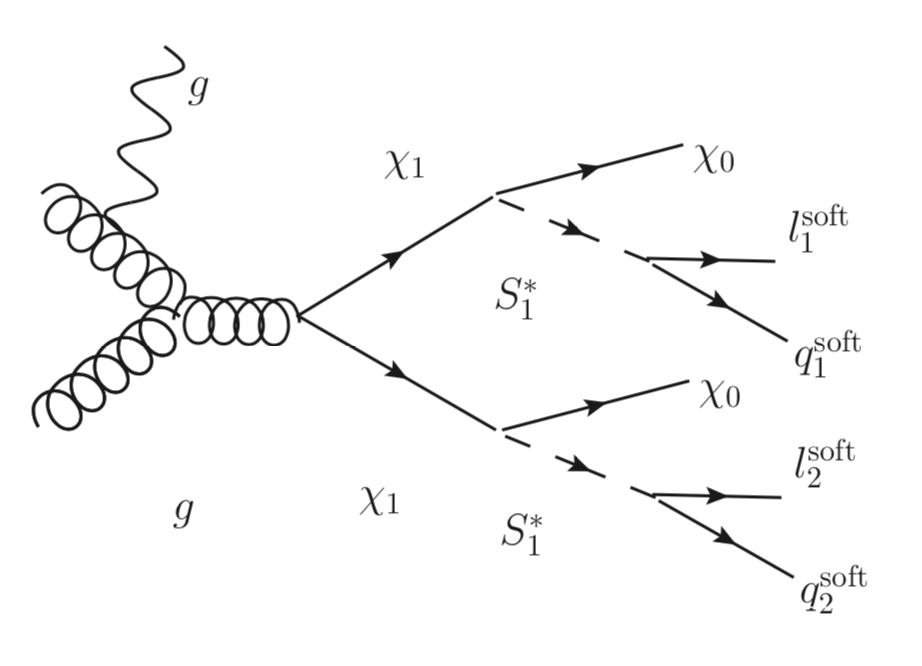} 
  \caption{\it Feynman diagram for monojet (plus soft leptons and quarks) signatures.}
\label{fig:RDM:MonoJet}
\end{figure}

As can be seen from figures~\ref{fig:RDM:T2bbexcl} and \ref{fig:RDM:TStauStauexcl}, these searches lose steam once the mass gap between the parent and the daughter particle becomes small (compressed spectra). The usual strategy against a compressed spectra is to boost the system with additional radiation, leading to the mono-X (typically jet) signals. However, if the SM decay products in the $\chi_1 \to \chi_0$ are hard enough to be reconstructed (but probably not to be triggered on) the LHC sensitivity gets greatly enhanced, as shown in e.g.~\cite{Schwaller:2013baa} for MSSM electroweakinos and soft-leptons. In our model this is shown in figure~\ref{fig:RDM:MonoJet}. However, the mass gap can be small enough such that the soft decay products fail to pass the reconstruction thresholds, in which case the pair production of $\chi$ particles can only be constrained by  jet + MET searches. We stress that this is an unavoidable bound, that, for a given $\Delta$, sets a lower mass on the dark sector masses. We will thus ignore here the MET searches that also include resolved (soft or hard) leptons. 

\begin{figure}[!h]
	\centering
	\includegraphics[width=0.5\textwidth]{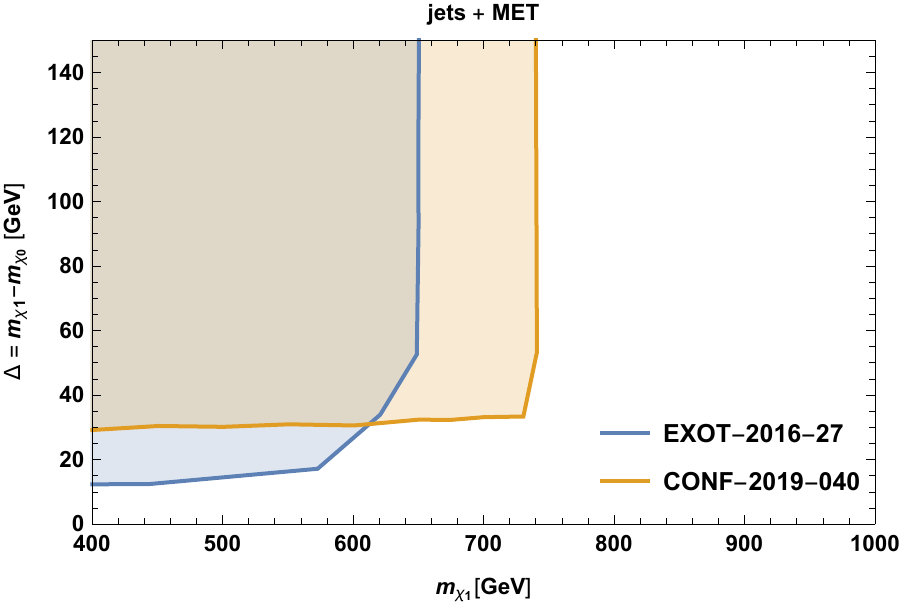}
	\caption{Region in the $\mxOne - \mxZero$ vs $\mxOne$ plane excluded by LHC studies for jets plus MET for a) more than one hard jet~\cite{ATLAS:2019vcq} (orange region) and b) a mono-jet search~ \cite{Aaboud:2017phn} (blue region) . Note that the additional model parameters ($y_\chi, \lambda_{L,R}, \mLQ$) do not play any relevant role here.}
   \label{fig:RDM:jetsplusmet}
\end{figure}

In figure~\ref{fig:RDM:jetsplusmet} we show the corresponding constraints in the $\mxOne - \Delta$ plane. In blue we show the region excluded by~\cite{Aaboud:2017phn} (obtained via the \ma~\cite{Dumont:2014tja,Conte:2018vmg} implementation of said study) and in orange we show the region covered by the ATLAS search of two (or more) hard jets and MET~\cite{ATLAS:2019vcq}. As anticipated the excluded cross section is independent of $\mxOne$ unless the spectra is compressed: the mono-jet (multi-jet) study can exclude up to 650 (750) GeV if $\Delta \gtrsim 30 (10)$ GeV.
\subsection{Resonant lepto-quark + MET search}

The model under consideration can also be constrained by reinterpreting the
results of the CMS analysis~\cite{Sirunyan:2018xtm} specifically
searching, in 77.4~fb$^{-1}$ of LHC data, signatures of dark matter that
originate from the decay of a heavy leptoquark. In this search, the signal is
assumed to arise from the production of a pair of heavy scalar leptoquarks that
decay differently. The first leptoquark is hence considered to decay into a muon
and a jet while the second one decays into dark matter and mostly soft Standard
Model states. Consequently, the searched for signal is comprised of a
significant amount of missing energy, jets and a high-$p_T$ muon.

More specifically, preselected events are required to feature at least 100~GeV
of missing transverse energy and one isolated central
muon candidate with a transverse momentum $p_T > 60$~GeV and a pseudorapidity
$|\eta| < 2.4$.~\footnote{We note that our interest lies in the search with one $\tau$ lepton instead of one muon. We can, however, expect similar acceptances in the case of a leptonically decaying $\tau$. A detailed study by the experimental collaborations including also electrons and $\tau$-leptons would be an important contribution to both the leptoquark and DM LHC search programmes. } The analysis then restricts the properties of the hadronic
activity in the events by constraining the centrally reconstructed jets ($|\eta|
< 2.4$). The leading jet is hence enforced to have a transverse momentum
$p_T > 100$~GeV and to be
well separated, in the transverse plane, from the leading muon (by $\Delta R >
0.5$). The leading jet and leading muon are further considered to arise from the
decay of a first leptoquark.

The selection then vetoes events featuring $b$-tagged jets, hadronic taus and
electrons, as well as those featuring a second muon of electric charge opposite
to the leading muon one, in which the muon pair made of these two muons is
compatible with a $Z$-boson. Finally, the missing transverse momentum, being
considered to stem from the decay of a second leptoquark, is required to be well
separated in azimuth ($\Delta\phi>0.5$) from the leading jet and muon, and the
transverse mass of the system comprised of the muon and the missing energy is
required to be larger than 500~GeV.

We have reimplemented this search in the {\sc MadAnalysis}~5
framework~\cite{Conte:2012fm,Conte:2014zja,Conte:2018vmg}, and are currently
validating our reimplementation by trying to reproduce detailed cutflows kindly
provided by the CMS collaboration.

With the help of our (unvalidated) reimplementation, we considering two benchmark points for which the LQ mass if fixed to 1~TeV,
\begin{eqnarray}
\textrm{BP1}: \ m_{\chi_1} = 400 \ \mathrm{GeV}, \ m_{\chi_0} = 380 \ \mathrm{GeV}, \nonumber \\
\textrm{BP2}: \ m_{\chi_1} = 600 \ \mathrm{GeV}, \ m_{\chi_0} = 200 \ \mathrm{GeV}, 
\end{eqnarray}
and where the new physics couplings are given by $\lambda_{R}=2$, and $\lambda_L = 0.16$. This choice affects the LQ pair branching ratios and is motivated as to maximize the sensitivity of the CMS search. For $m_{S_1} = 1$ TeV, the total production cross section 
is about $3.4$ fb. In table \ref{tab:xsig95}, we show the expected, and the observed limits on the visible cross  sections at $95\%$ CL along with the acceptance times the efficiency ($A\times \epsilon$)  of the selection in the signal region 
\begin{table}[!t]
\begin{center}
\begin{tabular}{c | c | c | c }
\toprule
Benchmark Point & $\sigma^{95\%}_\mathrm{exp}$ & $\sigma^{95\%}_\mathrm{obs}$ & $A \times \epsilon \ [\%]$ \\ \hline
\textrm{BP1} & $62.24$ &  $58.25$ & $0.00225$ \\ \hline
\textrm{BP2} & $112.01$ & $104.53$ & $0.00125$  \\ 
\bottomrule
\end{tabular}
\end{center}
\caption{The values of the expected, and observed visible cross sections at the $95\%$ CL along with the acceptance times the efficiency for the two benchmark points BP1 and BP2.}
\label{tab:xsig95}
\end{table}
The investigated scenarios are out of reach of the LHC, at least in in the channel under consideration and there is very little hope to catch up such a huge gain in sensitivity even at the HL-LHC. This is not surprising as tau-enriched final states are more relevant than muon-enriched ones in our benchmarks. This should motivate the design, by the ATLAS and/or CMS collaborations, of a similar analysis targeting the tau channel.

\section{Dark Matter constraints}
\label{sec:RDM:dm}

In this section we study how the measured relic density and the null results of direct detection impact our model. We start by noting that the direct detection rate is  one-loop suppressed (see figure~\ref{fig:RDM:dd_1loop}), however given the current $\sigma_{SI}$ exclusions this suppression might not be enough to discard this observable. For the relic density we consider two distinct scenarios. On one hand we study the (co)-annihilation case, as done in~\cite{Baker:2015qna}, while on the other hand we also consider the conversion-driven freeze-out (CDFO) scenario~\cite{Garny:2017rxs}, also known as co-scattering~\cite{DAgnolo:2017dbv}. The main idea behind this mechanism is that the self-annihilation of the DM candidate is negligible: the DM abundance arises instead from conversion processes within the dark sector. Hence, this scenario requires small/tiny interactions between the SM and the DM candidate ($y_{DM}$ in our case), thus rendering the $BR(S_1 \to \chi_1 \chi_0)$ negligible. This will thus affect the relevance of the collider signatures, that are discussed in detail in section~\ref{sec:RDM:lhc}.

\subsection{Direct Detection}

\begin{figure}[!t]
\center
\scalebox{0.80}{
\begin{tikzpicture}
\node at (0,0) {\includegraphics[scale=0.43]{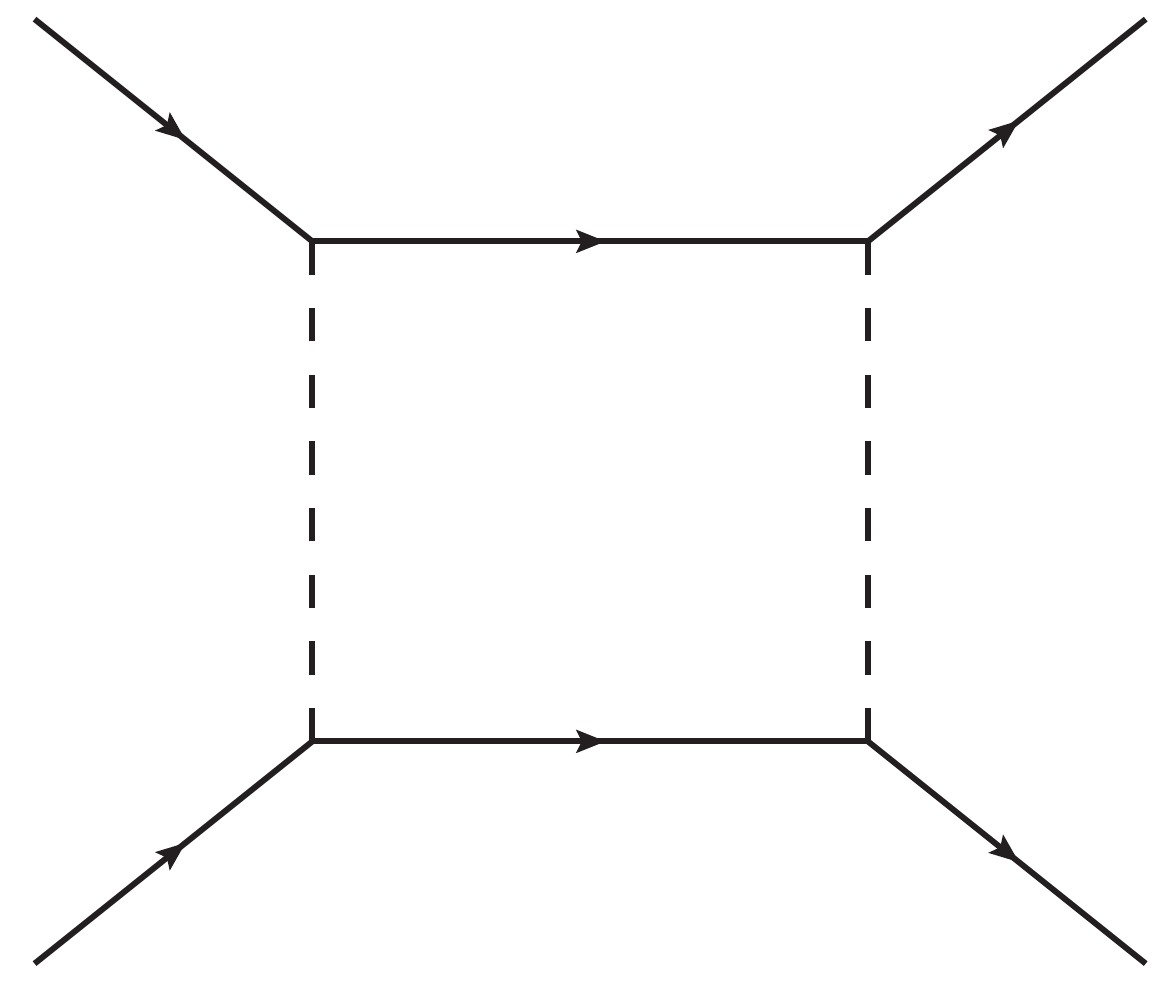} };
\node at (-2.6,2.3) {\scalebox{1.3}{$\chi_0$}};
\node at (2.6,2.3) {\scalebox{1.3}{$\chi_0$}};
\node at (0,1.4) {\scalebox{1.3}{$\chi_1$}};
\node at (-1.5,0) {\scalebox{1.3}{$S_1$}};
\node at (1.5,0) {\scalebox{1.3}{$S_1$}};
\node at (-2.4,-2.2) {\scalebox{1.3}{$q$}};
\node at (2.6,-2.2) {\scalebox{1.3}{$q$}};
\node at (0,-1.4) {\scalebox{1.3}{$\ell$}};
\end{tikzpicture}
}
\scalebox{0.80}{
\begin{tikzpicture}
\node at (0,0) {\includegraphics[scale=0.43]{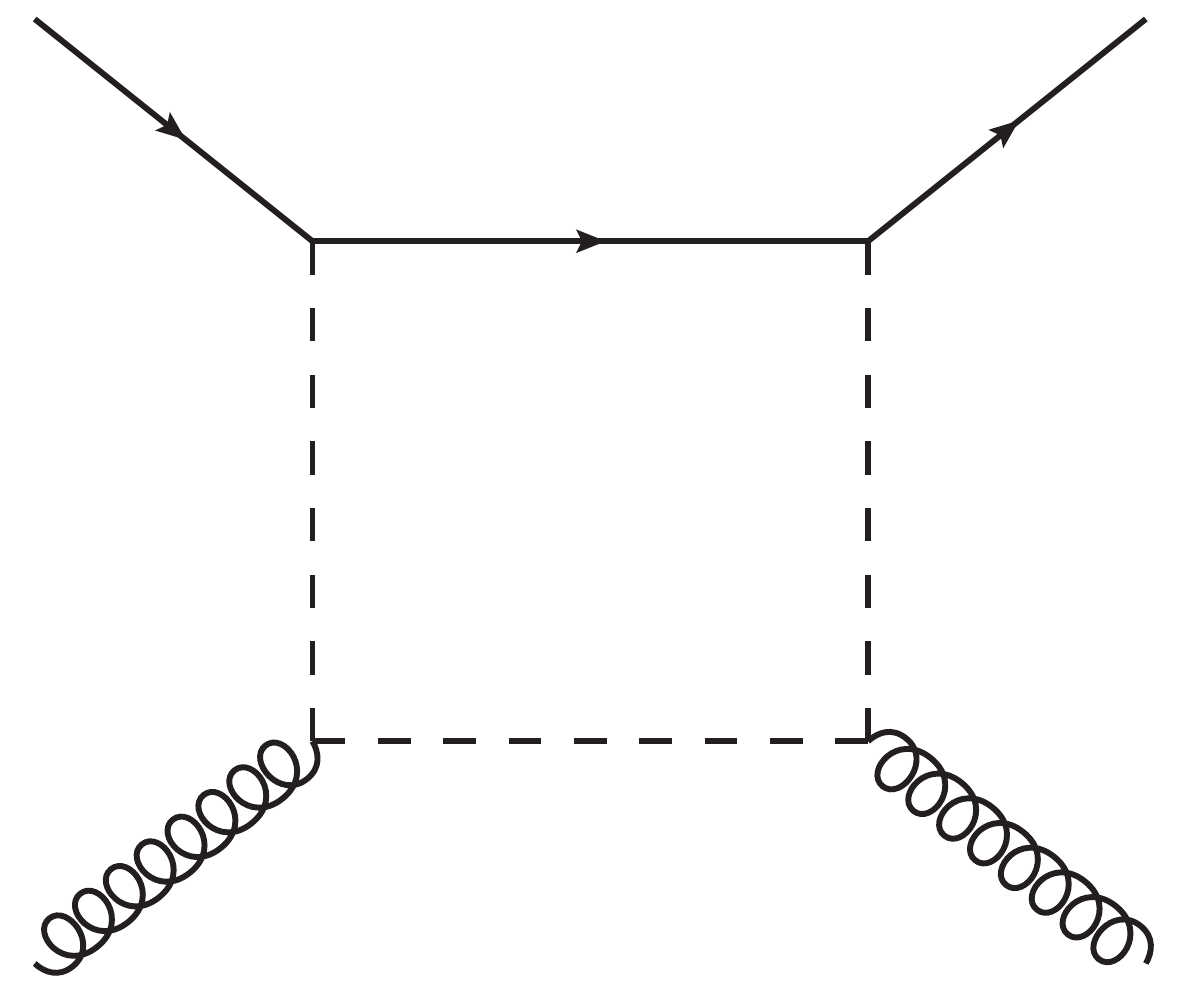} };
\node at (-2.6,2.3) {\scalebox{1.3}{$\chi_0$}};
\node at (2.6,2.3) {\scalebox{1.3}{$\chi_0$}};
\node at (0,1.4) {\scalebox{1.3}{$\chi_1$}};
\node at (-1.5,0) {\scalebox{1.3}{$S_1$}};
\node at (1.5,0) {\scalebox{1.3}{$S_1$}};
\node at (-2.4,-2.2) {\scalebox{1.3}{$g$}};
\node at (2.6,-2.2) {\scalebox{1.3}{$g$}};
\node at (0,-1.4) {\scalebox{1.3}{$S_1$}};
\end{tikzpicture}
}
\scalebox{0.80}{
\begin{tikzpicture}
\node at (0,0) {\includegraphics[scale=0.43]{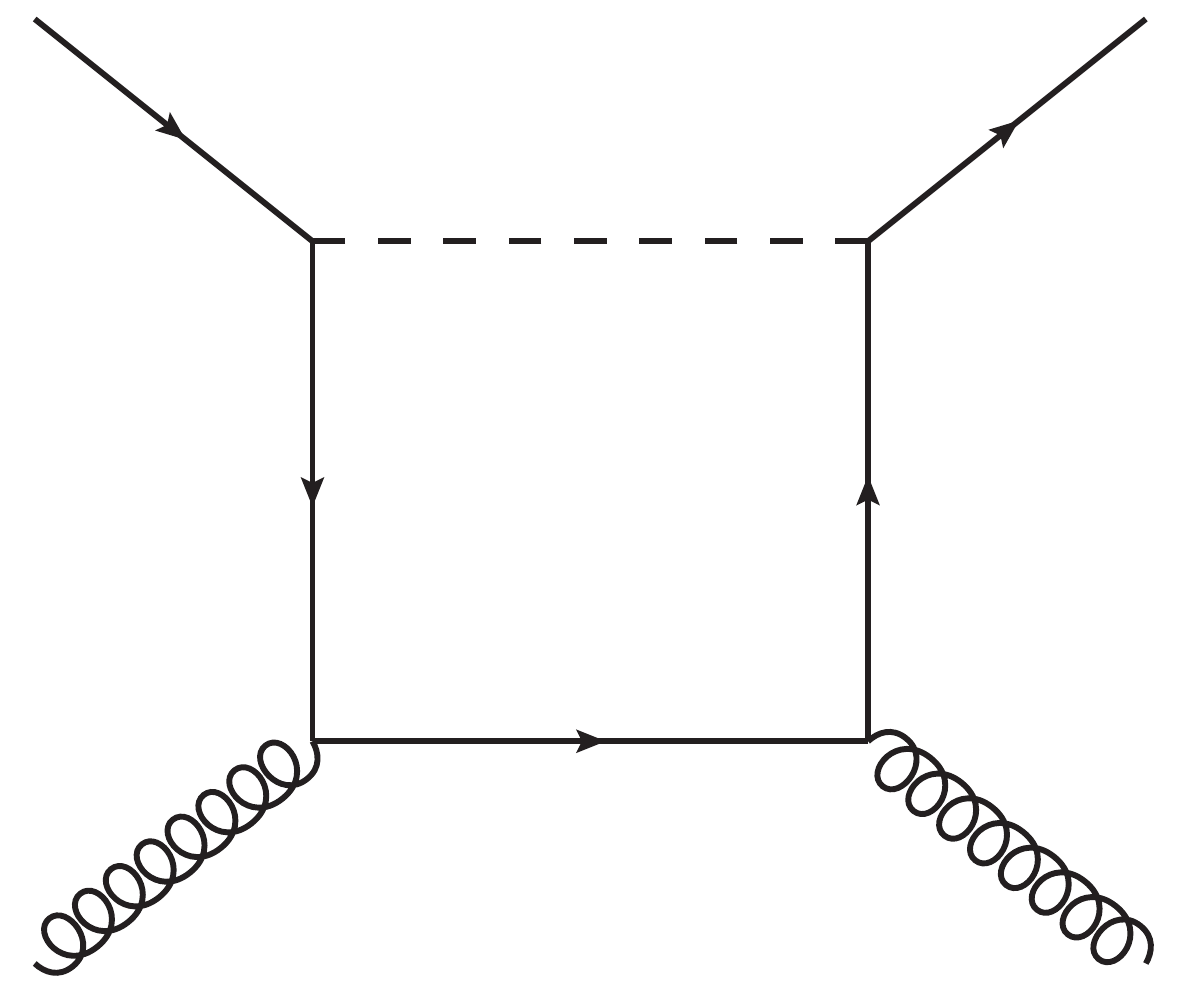} };
\node at (-2.6,2.3) {\scalebox{1.3}{$\chi_0$}};
\node at (2.6,2.3) {\scalebox{1.3}{$\chi_0$}};
\node at (0,1.4) {\scalebox{1.3}{$S_1$}};
\node at (-1.5,0) {\scalebox{1.3}{$\chi_1$}};
\node at (1.5,0) {\scalebox{1.3}{$\chi_1$}};
\node at (-2.4,-2.2) {\scalebox{1.3}{$g$}};
\node at (2.6,-2.2) {\scalebox{1.3}{$g$}};
\node at (0,-1.4) {\scalebox{1.3}{$\chi_1$}};
\end{tikzpicture}
}
\caption{Feynman diagrams for the direct detection of dark matter in our model at the lowest order in perturbation theory.}
\label{fig:RDM:dd_1loop}
\end{figure}

The scattering of $\chi_0$ on a nucleon proceeds via 1-loop diagrams, which we show in figure~\ref{fig:RDM:dd_1loop}. In the first diagram the loop include two leptoquark masses and the $\chi_1$ mass, which are already heavy enough compared with the typical momentum transfer, so that would naively induce a mass suppression. In the second diagram one would have three leptoquark masses and a $\chi_1$ mass, so that would be more suppressed than the diagram with three $\chi_1$ and one $S_1$ in the loop.

Integrating out the heavy degrees of freedom $\chi_1$ and $S_1$, we can obtain a rough estimate for the Wilson coefficient of the rightmost diagram, finding that $\ydm \lesssim 1$. This result implies that direct detection constraints have an impact on our parameter space, and hence a complete 1-loop calculation taking into account all possible diagrams and interference effects is incumbent and is left for future work~\footnote{Such a calculation was carried out for a similar model~\cite{Mohan:2019zrk}, however in that case $S_1$ is a $t$-channel mediator (instead of $s$-channel), and there is no $\chi_1$ particle.}.

\subsection{Relic Density}
 \begin{figure}[!htp]
  \centering
\setlength{\unitlength}{1\textwidth}
\begin{picture}(0.96,0.4)
 \put(0.0,-0.0){\includegraphics[width=0.5\textwidth, trim= {3.3cm 1.3cm 3cm 2cm}, clip]{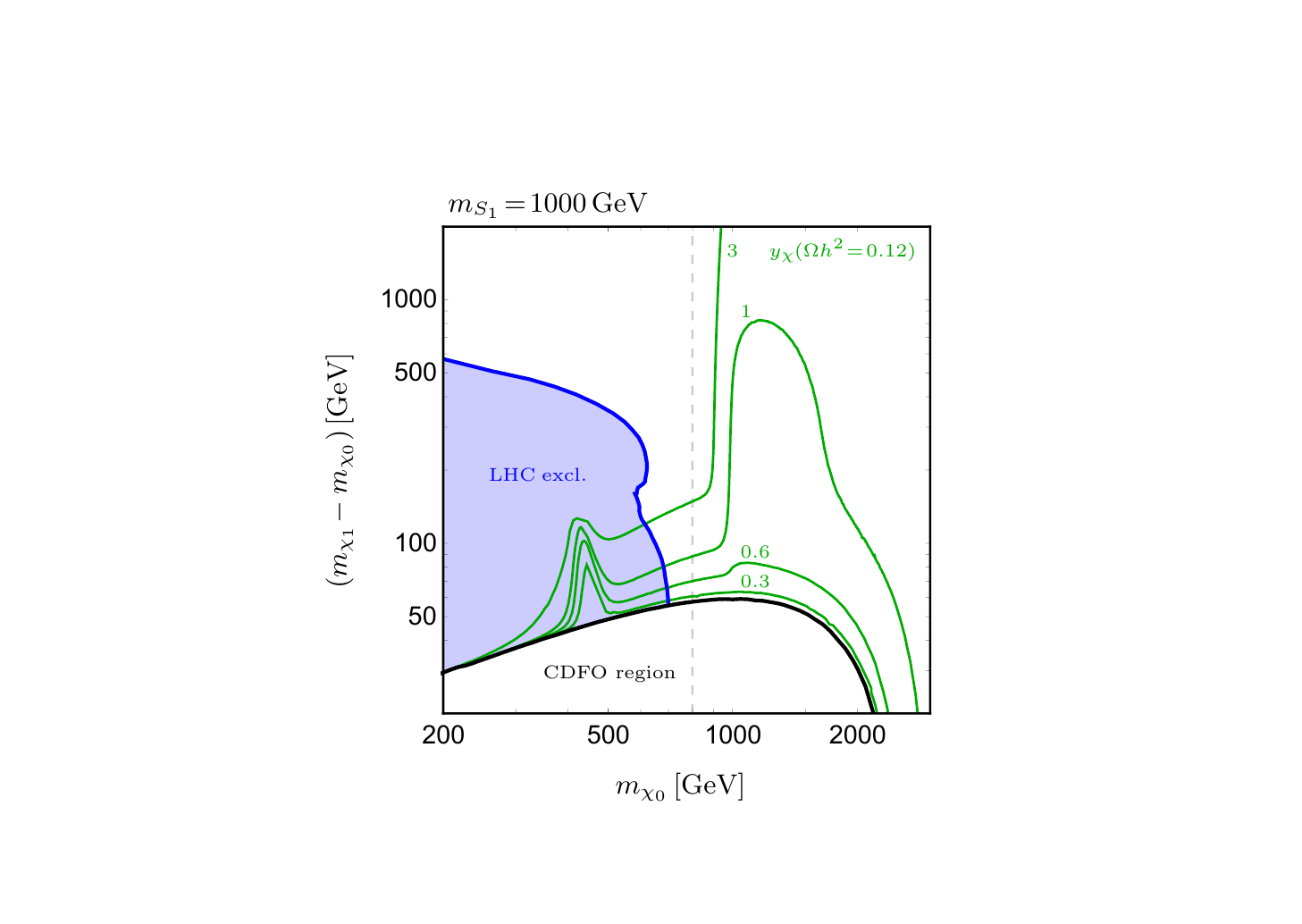}}
 \put(0.48,-0.0){\includegraphics[width=0.5\textwidth, trim= {3.3cm 1.3cm 3cm 2cm}, clip]{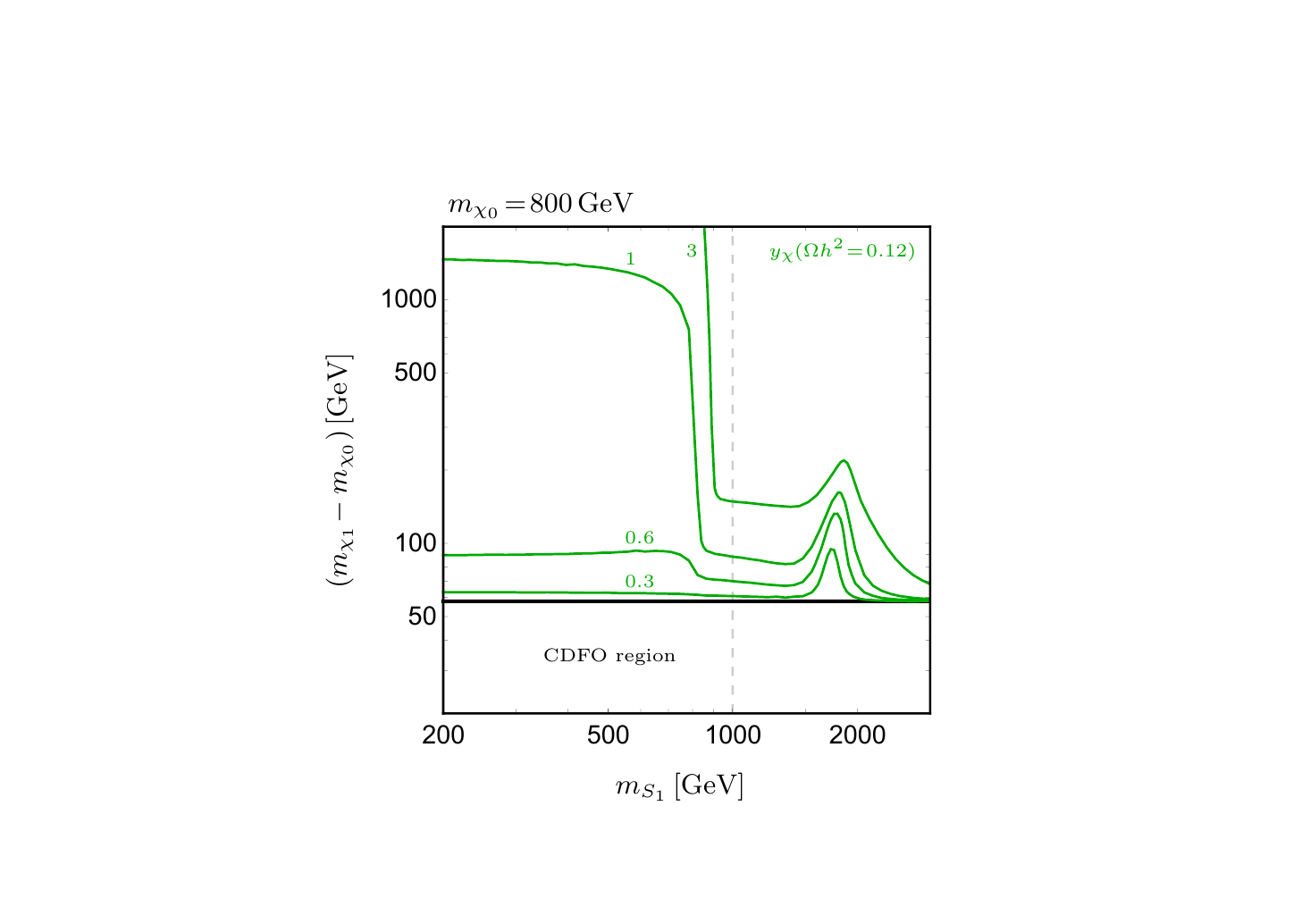}}
\end{picture}
  \caption{Parameter space satisfying $\Omega h^2 = 0.12$. The solid black line denotes the boundary between the standard freeze-out and conversion-driven freeze-out (CDFO) regime. The thin green curves denote contours of constant coupling $y_{\chi}$. We also overlay the envelope of the missing energy searches (blue region) discussed in the previous section.). \emph{Left panel:} Plane spanned by $\mxZero$ and $(\mxOne\! -\mxZero)$ while fixing $m_{S_1}=1000\,$GeV. \emph{Right panel:} Plane spanned by $m_{S_1}$ and $(\mxOne\! -\mxZero)$ while fixing $\mxZero=800\,$GeV. We choose $\lambda_L=0.6 \,m_{S_1}/\text{TeV}$ and $\lambda_L=0.5 \,m_{S_1}/\text{TeV}$. The vertical grey dashed line in the left panel indicates the fixed mass value in the right panel and vice versa.}
\label{fig:RDM:CDFO_WIMP}
\end{figure}

In figure~\ref{fig:RDM:CDFO_WIMP} we display contours of constant $y_{\chi}$ satisfying the correct relic abundance, $\Omega h^2 \simeq 0.12$~\cite{Aghanim:2018eyx}, computed with \micromegas~\cite{Belanger:2014vza}. We consider two slices of the parameter space. In the left panel we fix $m_{S_1}=1000\,$GeV and $\lambda_L=0.6, \lambda_L=0.5$ (second benchmark point in table~\ref{tab:RDM:benchmark_points}), which renders it safe from the leptoquark searches described in section~\ref{sec:RDM:lhc}. In the right panel we fix $\mxZero=800\,$GeV and $m_{S_1}$ while keeping $\lambda_L=0.6 \,m_{S_1}/\text{TeV}$ and $\lambda_L=0.5 \,m_{S_1}/\text{TeV}$
consistent with the flavour anomalies. 

As mentioned above, cosmologically allowed parameter space falls into two main classes: The standard thermal freeze-out regime with $y_\chi \sim {\cal O}(1)$ and the CDFO regime, residing above and below the thick black curve, respectively. For the latter the required couplings are small such that the decay of coannihilating partner is typically non-prompt~\cite{Garny:2017rxs}. In the standard freeze-out regime co-annihilations are important. The $y_\chi$ contours show two prominent features. The one at $\mxZero\sim m_{S_1}/2$ arises from resonantly enhanced coannihilation via an $s$-channel LQ, while the one at $\mxZero\sim m_{S_1}$ is due to the opening of DM pair-annihilation into a pair of LQs\@. Qualitatively, the phenomenology is very similar to the one of the top-philic charge parent model as studied in~\cite{Garny:2018icg}. Note that for $\mxZero< m_{S_1}/2$ loop-induced DM annihilation into gluons can become important~\cite{Garny:2018icg} which is, however, neglected here.

We overlay on top of these curves the ``envelope" of the missing energy searches, which clearly disfavors low masses except for small mass splittings, where solutions in the CDFO regime can be found. Here, the applicability of LHC searches depends on the lifetime of the metastable $\chi_1$ and is investigated elsewhere. For non-compressed mass spectra limits from the LHC require $\mxOne \gtrsim 750$ GeV. 

Note that in the CDFO region the $S_1$ branching fractions into the dark sector would be negligible, and the invisible decay of $S_1$ would not be seen at colliders. The only handle into the dark world are the $\chi_1$ searches (note that this is also true for the low $\mxOne$ region with compression). In the thermal freeze-out case one can have, besides the LQ, missing energy and direct detection searches, also information from the resonant leptoquark plus missing energy searches, thus allowing to firmly establish a connection between the flavor anomalies and the dark sector.

\subsection{Indirect Detection}

In general indirect detection constitutes an important probe of the self-annihilating nature of DM\@. 
In the considered model annihilation today dominantly takes place either via the $2\to2$ process 
$\chi_0 \chi_0 \to S_1 S_1$, if $\mLQ< m_{\chi_0}$, or 
via $2\to 4$ or loop-induced $2\to2$ processes otherwise. In the first case the leading contribution is $p$-wave suppressed and hence the corresponding cross sections are small. The same is true for the $2\to 4$ process which is suppressed by the off-shell $S_1$ propagator.
Points where annihilation dominantly proceeds via these processes are, hence, not expected to be challenged by current or upcoming indirect detection searches.\footnote{For instance, choosing $m_{\chi_0}=800\,$GeV, $m_{\chi_1}=1000\,$GeV, $y_\chi=0.71$ (obeying $\Omega h^2\simeq0.12$) and $\mLQ=500$ GeV, (according to benchmark point 1) yields a annihilation cross section $\langle \sigma v \rangle_0 \simeq 3\times 10^{31}\,\text{cm}^3/\text{s}$, way below current limits from gamma rays~\cite{Fermi-LAT:2016uux} or cosmic rays~\cite{Cuoco:2017iax}.
Note that the corresponding spectra (computed with {\sc\small MadDM}~\cite{Ambrogi:2018jqj}) are similar to the one for annihilation into $b\bar b$ considered as a benchmark in these analyses.}
If, in contrast, annihilation in the early Universe proceeds predominantly via the loop-induced $2\to2$ processes, requiring large $y_\chi$, the situation might change and the corresponding cross section today might be sizeable. In particular, the signature of mono-chromatic gamma lines, appearing at the same loop-level, can be a promising search strategy given higher sensitivity~\cite{Ackermann:2015lka,Abdallah:2018qtu}. However, as the region $\mLQ< m_{\chi_0}$ is not the prime focus of this study, we leave a detailed study of this case for future work.

\section{Conclusions and Outlook}
\label{sec:RDM:conclu}
The charged-current flavor anomalies hint at the present of leptoquarks near or at the TeV scale. Here we have studied the connection between these models and dark matter, which necessarily requires to add new particles and couplings. Employing a simplified model with only 6 parameters we have scanned the parameter space vis-a-vis the aforementioned flavor anomalies, relic density, direct detection and LHC constraints.

We have seen that when taken all these constraints simultaneously one is left with two possible scenarios: either the dark sector couples to the leptoquark mediator with a similar strength than the one required from taking the $\RD$ anomaly at face-value, which then implies a traditional freeze-out scenario to account for the current abundance of dark matter, or the leptoquark couples faintly to the dark sector, which in turn implies that the dark matter abundance arises from the conversion-driven freeze-out (co-scattering) mechanism.  In the second scenario the leptoquark branching fraction into the dark sector is negligible, and thus the phenomenology will consist of seemingly disconnected leptoquark ``anomaly-inspired" and traditional missing energy + X signatures. In the latter case, depending on the ``compression" between the dark sector particles, one can resolve the leptons and quarks from the $S_1 \to \chi_1 \chi_0 \to l q \chi_0 \chi_0$ decay chain, which would then allow to establish a firm connection between both new physics signals. In the case where the branching fraction into the dark sector is non-negligible, then the strongest indication of the leptoquark-dark matter connection is through the pair production of leptoquarks with one decaying into the dark sector and another one into hard leptons and quarks. Since this search is only being conducted currently by the CMS collaboration, we encourage the ATLAS collaboration to undertake this key study for a comprehensive dark matter program going beyond the mere annihilation into singlet final states.

Along this work we have pointed out the existence of three important holes in the campaign to optimally cover the leptoquark parameter space. On one side we have noted that for leptoquarks decaying into the third generation fermions (almost) exclusively, a search in the $b \nu t \tau$ should be added to the existing ones looking into $b \nu b \nu$ and $t \tau t \tau$, where one can benefit from a larger branching fraction, and devise cuts to keep the background under control, which we have demonstrated with a simple analysis. On the other side, noting that the solutions to the $\RD$ anomalies involve larger couplings connecting second generation quarks with third generation leptons, we advocate to study the $c \tau c \tau$ final state, and if possible and depending on the $c$-tagging efficiencies, also include the mixed cases, e.g $c \tau t \tau$ or $b \nu c \tau$ final states. Finally, it is also important to include single leptoquark searches decaying largely into $c \tau$, namely study the $c \tau \tau$ final state.

Our results show that the exploration of common solution to the $\RD$ anomalies and the dark matter puzzle is an interesting research avenue, and we thus plan to expand the preliminary results presented here into a comprehensive study of this model, vis-a-vis current constraints and also considering future projections of LHC and direct detection coverage.

\section*{Acknowledgements}
W would like to thank the organizers of the Les Houches ``Physics at TeV Colliders" 2019 Session for setting up an interesting workshop and providing an ideal atmosphere for scientific exchange. 
The work of A.L. was supported by the S\~ao Paulo Research Foundation (FAPESP), project
2015/20570-1.
J.H.~acknowledges support from the F.R.S.-FNRS, of which he is a postdoctoral researcher.
The work of D.S. is based upon work supported by the National Science Foundation under Grant No. PHY-1915147.

\let\Herwig\undefined
\let\Pythia\undefined
\let\Sherpa\undefined
\let\Rivet\undefined
\let\Professor\undefined
\let\eps\undefined
\let\mc\undefined
\let\mr\undefined
\let\mb\undefined
\let\tm\undefined
\let\be\undefined
\let\ee\undefined
\let\mLQ\undefined
\let\RD\undefined
\let\RK\undefined
\let\fr\undefined
\let\micromegas\undefined
\let\mg\undefined
\let\ma\undefined
\let\mxOne\undefined
\let\mXZero\undefined
\let\bsp\undefined
\let\esp\undefined
\let\lag\undefined
\let\sss\undefined
\let\Ll\undefined
\let\eR\undefined
\let\QL\undefined
\let\QLbar\undefined
\let\uR\undefined
\let\uRbar\undefined
\let\ydm\undefined
\let\tev\undefined

%% file: vqlqvl/vq_lq_vl.main.tex
\graphicspath{{vqlqvl/}}

\def\lag{{\cal L}}
\def\sss{\scriptscriptstyle}
\def\as{\alpha_{\sss s}}
\def\gs{g_{\sss s}}
\def\gL{g_{\sss w}}
\def\gY{g_{\sss Y}}
\def\gF{G_{\sss F}}
\def\tw{\theta_{\sss W}}
\def\dR{d_{\sss R}}
\def\eR{\ell_{\sss R}}
\def\Ll{L_{\sss L}}
\def\QL{Q_{\sss L}}
\def\QLbar{\bar Q_{\sss L}}
\def\uR{u_{\sss R}}
\def\uRbar{{\bar u}_{\sss R}}

\newcommand{\ch}{{\sc\small CalcHep}}
\newcommand{\co}{{\sc\small Contur}}
\newcommand{\fa}{{\sc\small FeynArts}}
\newcommand{\fr}{{\sc \small FeynRules}}
\newcommand{\ma}{{\sc\small MadAnalysis}~5}
\newcommand{\maddm}{{\sc\small MadDM}}
\newcommand{\mg}{{\sc\small MG5\_aMC}}
\newcommand{\micromegas}{{\sc\small MicrOMEGAs}}
\newcommand{\mthmtc}{{\sc\small Mathematica}}
\newcommand{\nloct}{{\sc\small NLOCT}}
\newcommand{\riv}{{\sc\small Rivet}}
\def\LQVL{{\tt LQVL}}
\newcommand{\be}{\begin{equation*}}
\newcommand{\ee}{\end{equation*}}
\def\bsp#1\esp{\begin{split}#1\end{split}}
\def\bpm{\begin{pmatrix}}
\def\epm{\end{pmatrix}}


\chapter{Non-standard VLQ and LQ signatures from cascade decays}

{\it J.~Butterworth, G.~Cacciapaglia, L.~Corpe, T.~Flacke, B.~Fuks, L.~Panizzi, W.~Porod, D.~Yallup}

\label{sec:VQ_LQ_VL}

\begin{abstract}
Many Standard Model extensions contain vector-like-quarks (VLQs) as well as potentially lighter additional particles, {\it e.g.} leptoquarks (LQs), vector-like leptons (VLLs), or additional scalars, through which the VLQs can cascade decay. Searches for vector-like quarks do not straight-forwardly apply to the resulting new final states, bounds on VLQ masses can thus be altered, and Standard Model mesurements can provide valuable constraints on exotic VLQ or LQ decays. We consider several sample models in which VLQ or LQ cascade decays are present, and determine bounds from Standard Model measurements, using the \co \ package.
\end{abstract}

\section{Introduction}
Searches for new physics at the LHC based on simplified models representing the essential features of classes of theoretical scenarios have so far considered minimal extensions of the SM, where only one new state is added and allowed to decay directly into SM states. Underlying models however often predict chain decays with several new states which can lead to final states with many SM particles with peculiar kinematics. A recasting of direct searches might thus not be sensitive to the different properties of final states originated by their decays, and can in some cases only yield weak bounds on BSM state masses. On the other hand, SM measurements contain a plethora of information due to the large range of final states and distributions which have been experimentally tested, and can thus contribute to fill gaps in searches.

This contribution focuses on extensions of the SM featuring vector-like quarks (VLQs) with exotic decays, either into a neutral new scalar and a SM quark, or through chains involving leptoquarks (LQs) and vector-like leptons (VLLs).
Current  VLQ searches consider processes of pair and single production and assume that VLQs  interact exclusively with SM quarks (mostly third generation) and electroweak bosons. Under such hypotheses the bounds on the masses of the VLQs have been already pushed  to $1 - 1.4$~TeV, with the precise bound depending on the VLQ charge and its branching ratios (see {\it e.g.} \cite{Aaboud:2018pii,Sirunyan:2019sza}). 
Scenarios with LQs have also been widely considered in experimental searches (see {\it e.g.} \cite{Sirunyan:2018kzh,Aaboud:2019bye}).
VLLs on the other hand have received less attention, due to the fact that in a hadron collider their production cross-section is generally lower and more dependent on their interaction with the SM particles~\cite{Sirunyan:2019ofn}. 

The two scenarios analysed in this contribution are motivated by underlying models ({\it c.f.} Sec. \ref{sec:models}), and serve as working examples to assess the potential of current SM searches in setting new limits on the masses of the new states when exotic decays are included. 
A generic and model-independent simplified model describing the interactions of VLQs, LQs and VLLs between them, with exotic scalars and with SM states has been produced and used (isolating the relevant interactions) to simulate their process of production and decay at the LHC. The simulated events have then been tested against various searches and SM measurements (available in \textsc{Rivet}~\cite{Bierlich:2019rhm}) made at the LHC at various beam energies by both ATLAS and CMS 
through the \co \ framework~\cite{Butterworth:2016sqg}.

\section{Models}\label{sec:models}

\subsection{Model for VLQ interacting with a colour-singlet, electrically  neutral scalar $S_1^0$ (VLQ-$S_1^0$)}\label{sec:s10model}

As a first simple extension with VLQ cascade decays, we consider a model with a charge $2/3$ vector like quark $T$ and a colour-singlet, electrically neutral scalar or pseudo-scalar $S_1^0$, which (for example) is commonly present in Standard Model extensions with VLQs and a composite Higgs  \cite{Bizot:2018tds}. Bounds from cascade decays $T\rightarrow t S_1^0$ from LHC BSM searches in various $S_1^0$ decay channels have been studied  \cite{Serra:2015xfa,Chala:2017xgc,Aguilar-Saavedra:2017giu,Cacciapaglia:2019zmj,Anandakrishnan:2015yfa,Benbrik:2019zdp,Han:2018hcu}. In particular the cascade decays to tops and (especially $b$) jets have been shown to be less severely constrained than canonically studied VLQ decays \cite{Cacciapaglia:2019zmj,Anandakrishnan:2015yfa}. Here, we therefore focus on QCD pair production of $T\bar{T}$ with the subsequent decay $T\rightarrow t S_1^0\rightarrow t b \bar{b}$ or $T\rightarrow t S_1^0\rightarrow t gg$.

The  interaction part of the effective Lagrangian relevant for the cascade decay reads\footnote{A much more general implementation of the model will be made publicly available on the {\sc \small FeynRules} Model website  \url{https://feynrules.irmp.ucl.ac.be/wiki}. }
\be\bsp
 {\cal L}_{\rm VLQS} =&  \left(S_1^0 \bar T \left(\Gamma^L_TP_L + \Gamma^R_TP_R \right) t + \mbox{h.c.}\right) + \frac{g^2_s}{16\pi^2 v} S^0_1 G^a_{\mu\nu}\left(\kappa^0_G G^a_{\mu\nu}+ \tilde\kappa^0_G \tilde G^a_{\mu\nu} \right)+
  S_1^0 \bar b \left(\Gamma_b + i \tilde \Gamma_b\right)b \,,
\esp\ee
where the parameters $\kappa^0_G$ ($\tilde\kappa^0_G$) and $\Gamma_b$ ($\tilde \Gamma_b$) correspond to the scalar (pseudo-scalar) case. The mass scale in the denominator of the effective interaction of $S_1^0$ with the gluons has been arbitrarily set to the vacuum expectation value $v$ of the Higgs boson to factorise all the new physics in the $\kappa$ parameters.

\subsection{Model including interactions between VLQs, LQs and VLLs (VLQ-LQ-VLL)} \label{sec:VLQLQVLLmodel}

As a second example model, we extend the Standard Model by three species of scalar leptoquarks $S_1$, $R_2$
and $S_3$ that are coloured weak singlet, doublet and triplet fields
whose representation under the SM gauge group is $({\bf 3}, {\bf 1})_{-1/3}$,
$({\bf 3}, {\bf 2})_{7/6}$ and $({\bf 3}, {\bf 3})_{-1/3}$ respectively\footnote{In principle there
is also $\tilde R_2$ ($({\bf 3}, {\bf 2})_{1/6}$) which however cannot be used to explain
the B-physics anomalies. It corresponds essentially to $\tilde Q$ in supersymmetric models
with broken R-parity. We omit it here for the moment being as we do not expect any additional features. Other
representations for the weak singlet leptoquarks are also possible, but will be ignored for similar reasons.}. These states can either arise as composite states or as remnants
of breaking a larger group, see e.g.\ \cite{Gripaios:2009dq,Faber:2018afz} and 
references therein for various possibilitities.

Expanding those gauge fields in terms of their component fields, we define
\be
  S_1 = \Delta^{(-1/3)} \ , \qquad
  R_2 = \bpm \Delta^{(5/3)} \\ \Delta^{(2/3)}\epm\ ,\qquad
  S_3 = \bpm \frac{1}{\sqrt{2}}\Delta^{'(-1/3)} & \Delta^{'(+2/3)}\\
              \Delta^{(-4/3)} & -\frac{1}{\sqrt{2}}\Delta^{'(-1/3)} \epm\ ,
\label{eq:LQspecies}
\ee
where the subscripts refer to the electric charges of the various component
fields. We consider a minimal set of interactions
whose Lagrangian is given by
\be\bsp
 \lag_{\rm LQ} = &\ 
    {\bf\lambda^{(1)}_{\sss R}} \uRbar^c\ \eR^{\phantom{c}} \ S_1^\dag
  + {\bf\lambda^{(1)}_{\sss L}} (\QLbar^c \!\cdot\! \Ll^{\phantom{c}})\ S_1^\dag
  + {\bf\lambda^{(2)}_{\sss R}} \QLbar^{\phantom{c}}\ \eR^{\phantom{c}} \ R_2
  + {\bf\lambda^{(2)}_{\sss L}} \uRbar^{\phantom{c}}\
       (\Ll^{\phantom{c}}\!\cdot\! R_2) 
   \\ &  \hspace*{12mm}     
  + {\bf\lambda^{(3)}_{\sss L}} \Big(\QLbar^c\!\cdot\!\frac{\sigma_a}{2}
       \Ll^{\phantom{c}}\Big) S_3^{a\dag}
   + {\rm h.c.} 
   \ .
\esp\label{eq:lag}
\ee
Here, all flavour indices are understood and the dot stands for
the $SU(2)$-invariant product of two $SU(2)_L$ doublets. In addition,
$\lag_{\rm kin}$ contains gauge-invariant kinetic and mass terms for all new
fields. The ${\bf \lambda_{\sss L}}$ and ${\bf \lambda_{\sss R}}$ couplings are
$3\times 3$ matrices in the flavour space, that we considered real for
simplicity. In our conventions, we associate the first index $i$ of any
$\lambda_{ij}$ element with the corresponding quark generation whilst the second
one $j$ refers to the lepton generation.

We then supplement the model by four species of physical extra quarks $X$, $T$,
$B$ and $Y$ of electric charges of 5/3, 2/3, $-1/3$ and $-4/3$ respectively. A
model-independent effective parameterisation suitable for the phenomenology
related to these quarks has been developed in \cite{Fuks:2016ftf,Cacciapaglia:2018qep}.

Next, we consider a pair of vector-like leptons $E$ and $N$. As for their
vector-like quark counterparts, we keep their couplings to the Higgs and gauge
bosons as free parameters, so that the corresponding 
interaction
Lagrangian reads\footnote{We constrain ourselves
here to leptons which are either singlets, doublets or triplets with respect to $SU(2)_L$. In the latter
case we set the hypercharge to 0 as in case of seesaw type III.}
\be\bsp
 & {\cal L}_{\rm VLL} =
 - h \bigg[
     \bar E \Big(\hat\kappa_{\sss L}^{\sss E} P_L \!+\!
         \hat\kappa_{\sss R}^{\sss E} P_R\Big) \ell
   + \bar N \Big(\hat\kappa_{\sss L}^{\sss N} P_L \Big) \nu_\ell
   \bigg] +
  \frac{g}{2 c_{\sss W}} \bigg[
     \bar E\slashed{Z} \Big(\tilde\kappa_{\sss L}^{\sss E}P_L \!+\!
         \tilde\kappa_{\sss R}^{\sss E} P_R\Big) \ell
   + \bar N\slashed{Z} \Big(\tilde\kappa_{\sss L}^{\sss N}P_L\Big) \nu_\ell
   \bigg] \\
 &\ 
 + \frac{g}{\sqrt{2}} \bigg[
   + \bar E\slashed{\bar W} \Big(\kappa_{\sss L}^{\sss E} P_L\Big) \nu_\ell
   + \bar N\slashed{W} \Big(\kappa_{\sss L}^{\sss N} P_L \!+\!
         \kappa_{\sss R}^{\sss N} P_R\Big) \ell
   \bigg]\ + {\rm h.c.} \ ,
\esp\ee
using the same conventions as for ${\cal L}_{\rm VLQ}$ of \cite{Fuks:2016ftf,Cacciapaglia:2018qep}.
We assume the couplings $W-N-L$ to be suppressed as otherwise there are strong contraints
on the masses of vectorlike leptons close to the TeV range \cite{Sirunyan:2019ofn}.  

Finally, all possible interactions
involving one leptoquark and one vector-like-fermion (and one SM fermion), are
\be\bsp
  & {\cal L}_{\rm LQVL} =
   \Delta^{(-4/3)\dag} \bigg[
       \bar B^c \Big(\tilde\Gamma_{\sss L}^{\sss B} P_L +
            \tilde\Gamma_{\sss R}^{\sss B} P_R\Big) \ell +
       \bar q_d^c \Big(\tilde\Gamma_{\sss L}^{\sss E} P_L +
            \tilde\Gamma_{\sss R}^{\sss E} P_R\Big) L +
       \bar Y^c \Big(\tilde\Gamma_{\sss L}^{\sss Y} P_L\Big) \nu_\ell
   \bigg]\\
   &\ + \Delta^{(-1/3)\dag} \bigg[
       \bar T^c \Big(\Gamma_{\sss L}^{\sss T} P_L \!+\!
           \Gamma_{\sss R}^{\sss T} P_R\Big) \ell
     + \bar q_u^c \Big(\Gamma_{\sss L}^{\sss E} P_R \!+\!
           \Gamma_{\sss R}^{\sss E} P_L\Big) L
     + \bar B^c \Big(\Gamma_{\sss L}^{\sss B} P_L\Big) \nu_\ell
     + \bar q_d^c \Big(\Gamma_{\sss L}^{\sss N} P_R + 
          \Gamma_{\sss R}^{\sss N} P_L\Big) N
    \bigg]\\
   &\ + \Delta^{(+2/3)} \bigg[
       \bar B \Big(\hat\Gamma_{\sss L}^{\sss B} P_L \!+\!
           \hat\Gamma_{\sss R}^{\sss B} P_R\Big) \ell +
       \bar q_d \Big(\hat\Gamma_{\sss L}^{\sss E} P_R \!+\!
           \hat\Gamma_{\sss R}^{\sss E} P_L\Big) L +
       \bar T \Big(\hat\Gamma_{\sss L}^{\sss T} P_L\Big) \nu_\ell +
     + \bar q_u \Big(\hat\Gamma_{\sss L}^{\sss N} P_R
         + \hat\Gamma_{\sss R}^{\sss N} P_L\Big) N
    \bigg]\\
   &\ + \Delta^{(+5/3)} \bigg[
       \bar T \Big(\bar\Gamma_{\sss L}^{\sss T} P_L \!+\!
           \bar\Gamma_{\sss R}^{\sss T} P_R\Big) \ell +
       \bar q_u \Big(\bar\Gamma_{\sss L}^{\sss E} P_R \!+\!
           \bar\Gamma_{\sss R}^{\sss E} P_L\Big) L +
       \bar X \Big(\bar\Gamma_{\sss L}^{\sss X} P_L\Big) \nu_\ell
    \bigg]\ 
    + {\rm h.c.}\ .
\esp\ee
The corresponding couplings, $\Gamma$, $\hat\Gamma$, $\tilde\Gamma$ and
$\bar\Gamma$ are all three-dimensional vectors in flavour space.

For simplicity we only take one VLQ, one VLL and one LQ at a time and evalulate the bounds in two steps:
\begin{enumerate}
\item We consider the production processes $pp\to \Delta\, \Delta^\dagger \to L^+ L^- q \bar{q}$ and $pp\to \Delta\, \Delta^\dagger \to N \bar{N} q \bar{q}$ 
where $\Delta$ is any
of the  leptoquarks listed in eq.~(\ref{eq:LQspecies}). In more detail we focus on the LQ decays
\begin{align*}
\Delta^{(5/3)} &\to l^+ \ t \,\,,\,\, E^+ \ t \\
\Delta^{(2/3)} &\to l^+ \ b \,\,,\,\, E^+ \ b \,\,,\,\, \nu \ t\,\,,\,\, N \ t \\
\Delta^{(-1/3)} &\to l^- \ t \,\,,\,\, E^- \ t  \,\,,\,\,  \nu \ b\,\,,\,\, N \ b \\
\Delta^{(-4/3)} &\to l^- \ b \,\,,\,\, E^- \ b 
\end{align*}
and the VLL decays
\begin{align*}
E&\to Z \ l \,\,,\,\, h \, l   \,\,,\,\, W \, \nu \\  
N&\to Z \ \nu \,\,,\,\, h \, \nu   \,\,,\,\, W \, l 
\end{align*}
\item We then move on with the production processes $pp\to Q\, \bar{Q} \to \Delta\, \Delta^\dagger l^+ l^-$ and $pp\to Q\, \bar{Q} \to \Delta\, \Delta^\dagger \nu \bar{\nu}$,
where $Q$ is any of the VLQ $X$, $T$, $B$ or $Y$. In more detail
\begin{align*}
X &\to \Delta^{(5/3)} \ \nu  \,\,,\,\, \Delta^{(2/3)} \ l^+  \\
T &\to \Delta^{(5/3)} \ l^-  \,\,,\,\, \Delta^{(2/3)} \ \nu  \,\,,\,\, \Delta^{(-1/3)} \ l^+  \\
B &\to \Delta^{(2/3)} \ l^-  \,\,,\,\, \Delta^{(-1/3)} \ \nu  \,\,,\,\, \Delta^{(-4/3)} \ l^+  \\
Y &\to \Delta^{(-1/3)} \ l^-  \,\,,\,\, \Delta^{(-4/3)} \ \nu  
\end{align*}
Moreover, the $\Delta$  decay to the final states above.
\end{enumerate}
For simplicity we assume here that the decays into third generation SM-fermions dominate (except maybe
the leptons). We use \co\ to
get a first idea of the underlying bounds due to existing LHC data. However, we expect that final states
containing at least one top quark which decays semileptonically will lead to sizable missing transverse energy
$E_T^{\rm miss}$.
Moreover, also the decays of the VLL contain final states with sizable $E_T^{\rm miss}$.
Consequently, we expect that part of the parameter space will be constrained by supersymmetry searches which we plan to 
evaluated using \ma~in the future.

\section{Results}

\subsection{Results for the VLQ-$S_1^0$ model} 

	\begin{figure}[b!]
	\begin{center}
		\begin{tabular}{c}
			\includegraphics[]{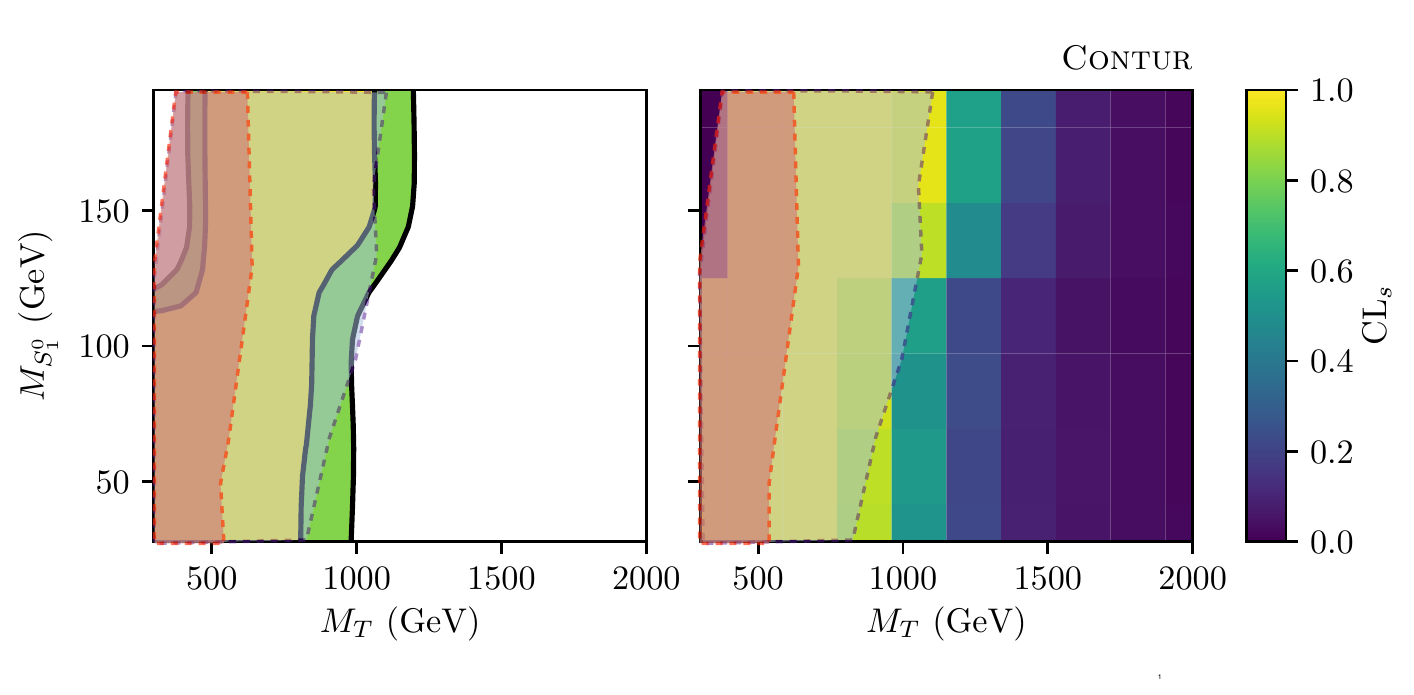}\\
			\includegraphics[]{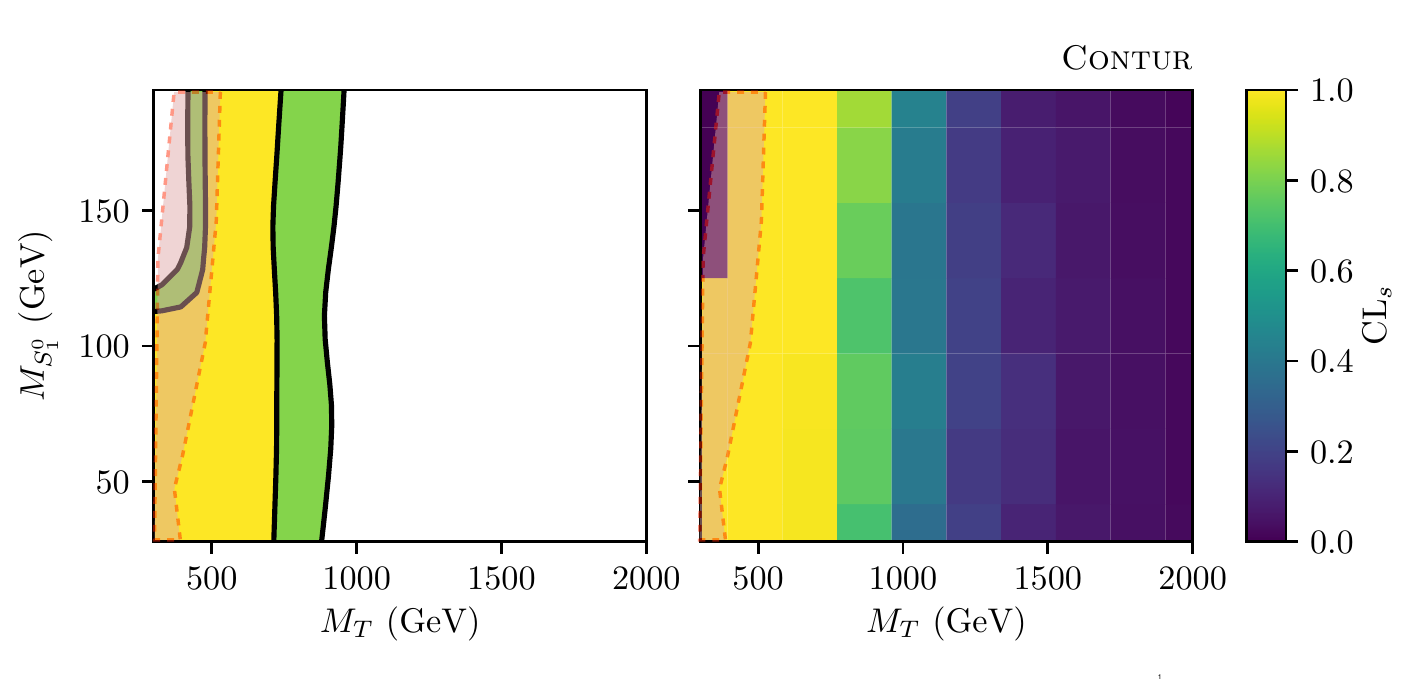}
		\end{tabular}
	\end{center}
	\caption{ Exclusion on the $M_{S_1^0}$ vs. $M_T$ mass plane by \co \ for $T\bar{T}$ production with subsequent cascade decay $T\rightarrow t S_1^0\rightarrow t b\bar{b}$ (top) or $T\rightarrow t S_1^0\rightarrow t jj$ (bottom). The right-hand figures show the heatmap of the obtained CL$_{s}$ values at each point. The left-hand figures show the areas which are excluded at 1 (2) $\sigma$ level in green (yellow). For comparison, red and blue areas show 2 $\sigma$ exclusions from a recast of several LHC searches \cite{Cacciapaglia:2019zmj}.}
	\protect\label{fig.s10}
\end{figure}

The model discussed in Sec.~\ref{sec:s10model} is implemented in {\sc \small FeynRules 2.3} \cite{Alloul:2013bka} at leading order. For simplicity we assume a pseudo-scalar $S_1^0$ ($\kappa_G^0 = 0 = \Gamma_b$) with 100 \% decay rate to either $b\bar{b}$ or $gg$ and narrow width. Events are generated with {\sc \small Herwig++ v.7.2} \cite{Bellm:2019zci} and compared to LHC measurements using \co \ v1.0.\footnote{See Sec.~\ref{sec:contur-update} of these proceedings for more details on the \co \ methodology.} The \co \ scan took into account a range of measurements from 7, 8 and 13~TeV data from both ATLAS and CMS.

Fig.\ref{fig.s10} shows the obtained constraints on $T\bar{T}$ pair production with subsequent decay $T\rightarrow t S_1^0\rightarrow t b\bar{b}$ (top) or $T\rightarrow t S_1^0\rightarrow t jj$ (bottom) in the $S_1^0$ vs. $T$ mass plane. As can be seen, the bounds from LHC search recasts are comparable to the bounds obtained with \co \ in the case of $S_1^0\rightarrow b\bar{b}$ decays, where  \verb+ATLAS_13_LMETJET+ ~\cite{Aaboud:2018uzf,Aaboud:2017fha,Aaboud:2018eki}, \verb+CMS_13_LMETJET+ ~\cite{Sirunyan:2018wem,Khachatryan:2016mnb,Sirunyan:2018ptc}, and \verb+ATLAS_13_TTHAD+ ~\cite{Aaboud:2018eqg} contribute most.
For $S_1^0\rightarrow jj$, \co \ yields stronger bounds than current recasts, with bounds arising mainly from \verb+ATLAS_13_LMETJET+ and \verb+CMS_13_LMETJET+.
This provides an example of the exclusion power of Standard Model measurements on BSM.

\subsection{Results for the VLQ-LQ-VLL model} 

\begin{figure}[ht!]
 \begin{center}
  \includegraphics[]{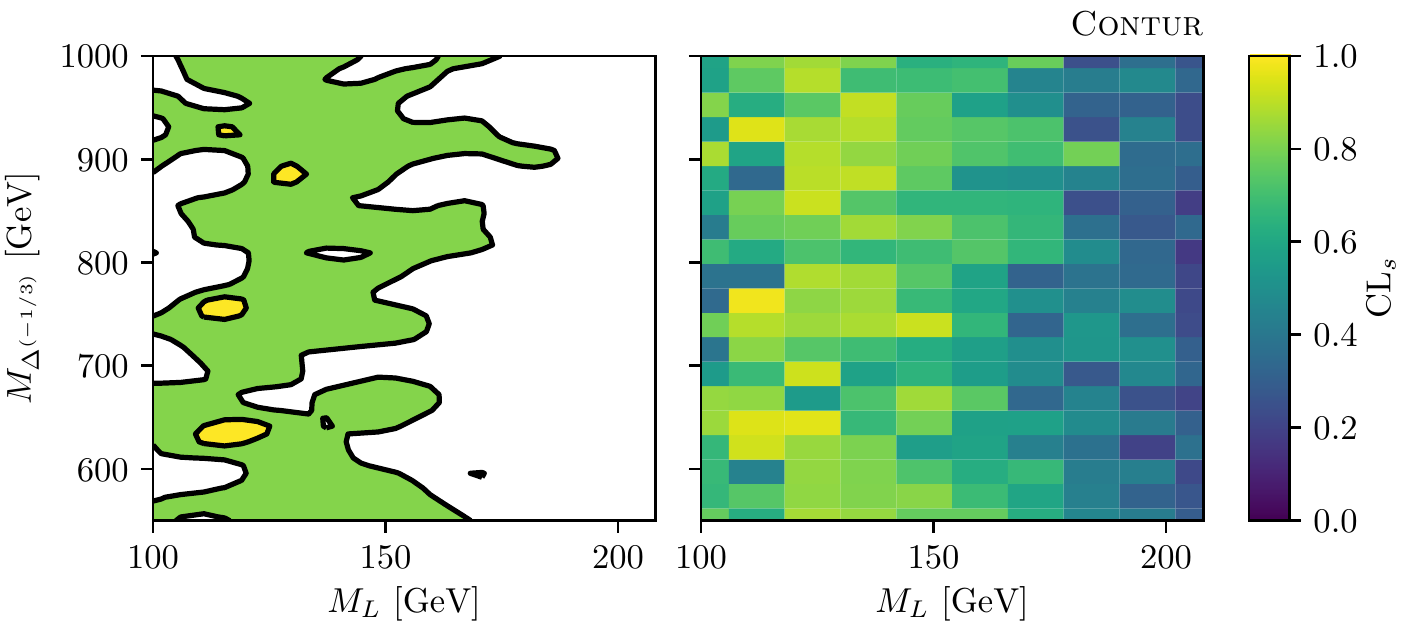}
 \end{center}
\caption{Exclusion on the $M_{L}$ vs.\ $M_{\Delta^{(-1/3)}}$ mass plane by \co\ for $L\bar{L}$ 
and $\Delta^{(-1/3)} \Delta^{(+1/3)}$ pair production with subsequent cascade decays. The underlying
parameters are given in the text.  }
\label{fig:MVLL_MLQ}
\end{figure}

\begin{figure}[ht!]
 \begin{center}
  \includegraphics[]{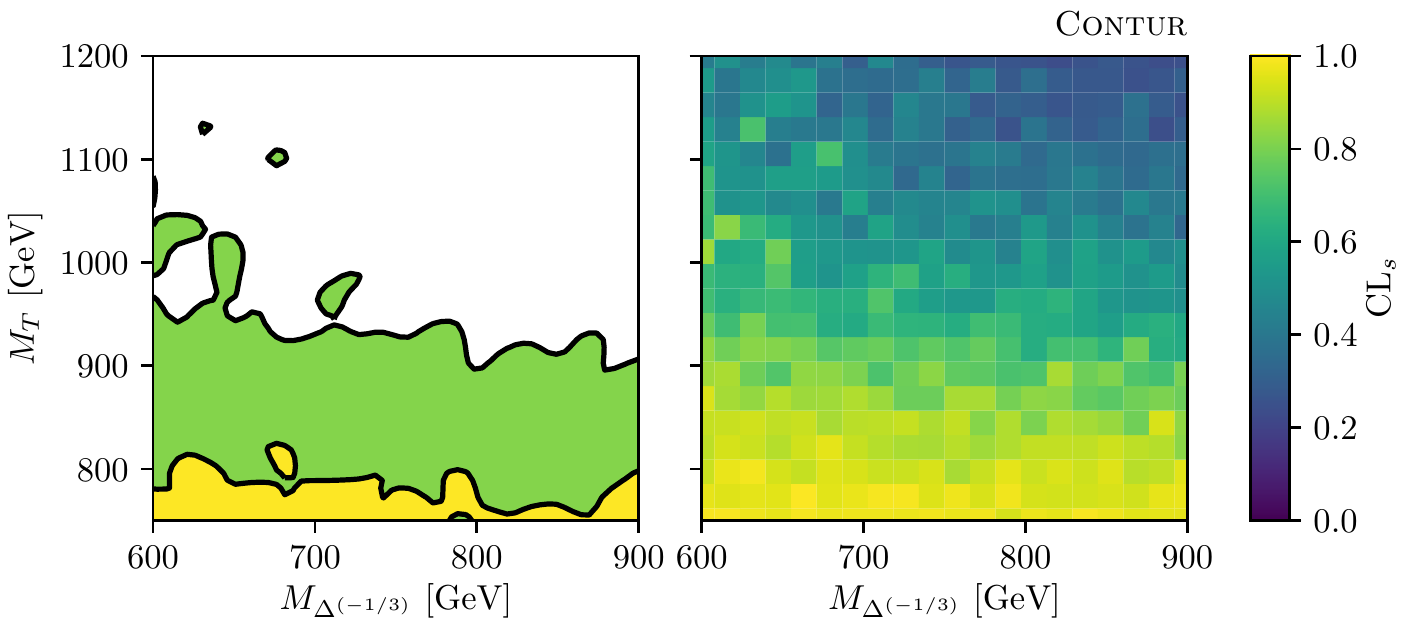}
 \end{center}
\caption{Exclusion on the $M_{\Delta^{(-1/3)}}$ vs.\ $M_{T}$ mass plane by \co\  for $T\bar{T}$ 
and $\Delta^{(-1/3)} \Delta^{(+1/3)}$ pair production with subsequent cascade decays. The underlying
parameters are given in the text and we have fixed $M_L=M_N=250$~GeV. }
\label{fig:MLQ_MT}
\end{figure}

In the spirit of simplified models we investigate a bench-mark scenario where $T$, 
$\Delta^{(-1/3)}$, $L$ and $N$ are all present at the same time.
Moreover, we here assumed that decays into final states containing
third generation fermions dominate.
To obtain a conservative bound we switch off
all couplings to fermions of the first two generations. For definiteness, 
we set $\hat\kappa_{\sss X}^{\sss F} = \tilde\kappa_{\sss X}^{\sss F} = \kappa_{\sss X}^{\sss F}=0.2$
with ${\sss X}={\sss L},{\sss R}$ and ${\sss F}={\sss E}, {\sss N}$. 
We have checked that we can reproduce the results of the ATLAS searches 
\cite{Aad:2015caa,Aaboud:2016qeg}
for leptoquarks within a few percents for several mass and branching ratio combinations
using the \co\ package. In a first step we have investigated to which extent the masses of
$L$ and $\Delta^{(-1/3)}$ are constrained. As can be seen in Fig.~\ref{fig:MVLL_MLQ} 
the bounds from the direct production of these particles are relatively mild. We have
checked that the cross section for the $L^+L^-$ pair production is below the bounds given
in \cite{Sirunyan:2019ofn}. As second step we have fixed $M_L=M_N=250$~GeV and varied $M_T$
and $M_{\Delta^{(-1/3)}}$. The corresponding exclusion-limits are presented in 
Fig.~\ref{fig:MLQ_MT}. Here we allowed for the simultaneous pair production of both
species, where the one for $T\bar{T}$ is significantly larger than the one for leptoquarks
if both have the same mass. We see that the bound on $M_T$ is well below a TeV with 
a mild dependence on  $M_{\Delta^{(-1/3)}}$. The analyses which contribute most are
\verb+ATLAS_13_JETS+ ~\cite{Aaboud:2017wsi,Aaboud:2019aii} followed by \verb+CMS_13_LMETJET+ ~\cite{Sirunyan:2018wem,Khachatryan:2016mnb,Sirunyan:2018ptc}
Similar to the example in the previous section this shows again the exclusion power of
SM measurements on BSM models.

\section{Conclusions }

We studied two theoretically motivated sample models in which VLQ or LQ cascade decays are present, and determined bounds from Standard Model measurements, using the \co \ package. 
	
The first sample model features the decay $T\rightarrow t S_1^0$ with $S_1^0\rightarrow jj \mbox{ or } b\bar{b}$. For top partner pair-production and $T$-decay into $tjj$ we find 95\% c.l. bounds of $M_T \gtrsim 700$~GeV with a very mild dependence on $M_{S_1^0}$, which exceeds bounds from previous search recasts. For the decay to $tb\bar{b}$ we find bounds $M_T \gtrsim 800 - 1050 $~GeV, depending on $M_{S_1^0}$, which are comparable to bounds from previous search recasts.
	
The second sample model features a more complex scenario, namely we allow in addition to the
standard decays of $T$ also for the decay
$T \to \tau^+ \Delta^{(-1/3)}$. The $\Delta^{(-1/3)}$ decays further into $\tau t$, $b\nu$,
$L t$ or $b N$. Fixing the mass of the heavy leptons to 250 GeV and including the pair
production of both, the top partners and the leptoquarks, we find  95\% c.l. bounds of 
$M_T \gtrsim 850$~GeV with a very mild dependence on $M_{\Delta^{(-1/3)}}$.
	
The above (theoretically motivated) physics cases and first benchmarks are examples for a much broader class of studies of ``visible cascade decays'' and the importance of SM measurements for constraining them. We provide examples of simplified models describing the interactions of VLQs, LQs, VLLs and exotic scalars, and plan to provide  publicly available model implementations in the near future.

\section*{Acknowledgements}

We thank the organizers of the Les Houches workshop for the stimulating and productive atmosphere.
J.B, L.C and D.Y.'s work has received funding from the European Union's Horizon 2020 research and innovation 
programme as part of the Marie Skłodowska-Curie Innovative Training Network MCnetITN3 (grant agreement no. 722104).
T.F.'s work has been supported by IBS under the project code, IBS-R018-D1. 
L.P. work is supported by the Knut and Alice Wallenberg foundation under the SHIFT project, grant KAW
2017.0100.
W.P.\ has been supported by DFG, project nr. PO 1337/7-1.

\let\Herwig\undefined
\let\Pythia\undefined
\let\Sherpa\undefined
\let\Rivet\undefined
\let\Professor\undefined
\let\eps\undefined
\let\mc\undefined
\let\mr\undefined
\let\mb\undefined
\let\tm\undefined
\let\be\undefined
\let\ee\undefined
\let\mLQ\undefined
\let\RD\undefined
\let\RK\undefined
\let\fr\undefined
\let\micromegas\undefined
\let\mg\undefined
\let\ma\undefined
\let\mxOne\undefined
\let\mXZero\undefined
\let\bsp\undefined
\let\esp\undefined
\let\lag\undefined
\let\sss\undefined
\let\Ll\undefined
\let\eR\undefined
\let\QL\undefined
\let\QLbar\undefined
\let\uR\undefined
\let\uRbar\undefined
\let\ydm\undefined
\let\tev\undefined

\let\lag\undefined
\let\sss\undefined
\let\as\undefined
\let\gs\undefined
\let\gL\undefined
\let\gY\undefined
\let\gF\undefined
\let\tw\undefined
\let\dR\undefined
\let\eR\undefined
\let\Ll\undefined
\let\QL\undefined
\let\QLbar\undefined
\let\uR\undefined
\let\uRbar\undefined
\let\ch\undefined
\let\co\undefined
\let\fa\undefined
\let\fr\undefined
\let\ma\undefined
\let\maddm\undefined
\let\mg\undefined
\let\micromegas\undefined
\let\mthmtc\undefined
\let\nloct\undefined
\let\riv\undefined
\let\LQVL\undefined
\let\be\undefined
\let\ee\undefined
\let\bpm\undefined
\let\epm\undefined

%% file: filc/FILC.main.tex
\graphicspath{{filc/}}
\newcommand{\Herwig}{H\protect\scalebox{0.8}{ERWIG}\xspace}
\newcommand{\Pythia}{P\protect\scalebox{0.8}{YTHIA}\xspace}
\newcommand{\Sherpa}{S\protect\scalebox{0.8}{HERPA}\xspace}
\newcommand{\Rivet}{R\protect\scalebox{0.8}{IVET}\xspace}
\newcommand{\Professor}{P\protect\scalebox{0.8}{ROFESSOR}\xspace}
\newcommand{\eps}{\varepsilon}
\newcommand{\mc}[1]{\mathcal{#1}}
\newcommand{\mr}[1]{\mathrm{#1}}
\newcommand{\mb}[1]{\mathbb{#1}}
\newcommand{\tm}[1]{\scalebox{0.95}{$#1$}}

\newcommand{\diff}{\mathrm{d}}
\newcommand{\sv}{\langle \sigma v \rangle}

\chapter{Freeze-in with large couplings at the LHC}
{\it   G.~B\'elanger,
  A.~Bharucha,
  N.~Desai, 
  B.~Fuks,
  A.~Goudelis,
  J.~Heisig,
  A.~Lessa,
  A.~Mjallal,
  S.~Sekmen,
  D.~Sengupta,
  J. Zurita}


\label{sec:FILC}

\begin{abstract}
We study freeze-in scenarios in which the interactions of dark matter with the Standard Model are mediated by a heavy vector-like fermion with a mass that is higher than the reheating temperature of the Universe. In such scenarios freeze-in requires sizeable interactions between the two sectors which, in turn, implies that prompt LHC dark matter searches become relevant, unlike in conventional scenarios of freeze-in. We examine the interplay of long-lived and prompt searches and show that their role is highly complementary in constraining different regions of the parameter space.
\end{abstract}
\section{Introduction}\label{sec:FILC:intro}

In order to establish a thermodynamical explanation for the observed dark matter (DM) abundance in the Universe an extension of the Standard Model (SM) by new particles, possibly with new interactions, appears to be necessary. While the freeze-out of a weakly interacting massive particle (WIMP) has been the most widely studied scenario of dark-matter genesis during recent decades, other possibilities exist, such as dark matter freeze-in\footnote{The production rates of non-thermalized dark matter through occasional scatterings or decays of thermal bath particles were first computed in~\cite{Ellis:1984eq,Kawasaki:1994af,Bolz:2000fu,Covi:2002vw,Pradler:2006hh,Rychkov:2007uq} (UV-dominated production) and \cite{McDonald:2001vt,Asaka:2005cn,Hall:2009bx} (IR-dominated production). The name `freeze-in' was introduced in~\cite{Hall:2009bx}.}. 
In most well-studied freeze-in scenarios DM interacts only extremely weakly, or \emph{feebly}~\cite{Hall:2009bx}, with the SM particles. This feature renders most DM searches targeting the WIMP scenario (direct, indirect detection and prompt LHC searches for missing energy events) inefficient, although a few exceptions do exist ({\it e.g.}~direct detection via light mediators~\cite{Hambye:2018dpi}). 

Among the most phenomenologically interesting freeze-in scenarios are those in which DM is one (typically the lightest) of several states belonging to some `dark' ${\cal{Z}}_2$-odd sector under which the SM particles are even. In such configurations, the heavier particles in the dark sector may even interact strongly with the SM\@. Then, they may be copiously pair-produced at high-energy colliders and subsequently decay into DM along with visible particles, typically with a macroscopic lifetime. In other words, the visible signals of most usual freeze-in scenarios fall into the realm of searches for long-lived particles. The situation becomes substantially more challenging if DM is the \textit{only} ${\cal{Z}}_2$-odd particle in the spectrum. In this case, DM is essentially inaccessible at colliders~\cite{Kahlhoefer:2018xxo}, and the only handle on the corresponding models are traditional (${\cal{Z}}_2$-even) mediator searches. 

There are, however, situations in which the previous picture breaks down. For instance, DM can be overproduced in the early Universe and subsequently diluted by injection of entropy in the primordial plasma~\cite{Co:2015pka}. In this work we investigate yet another possibility, in which DM production is suppressed due to relatively small reheating temperatures ($T_R$). We will assume that the interactions between DM and the SM are mediated by some heavy particle ($F$) and $m_{DM} < T_R \lesssim m_F$. In this case DM is mostly produced through higher-dimensional operators which scale with $T_R/m_F$\footnote{This scenario is similar to the production of gravitinos in supersymmetric models, except that in the gravitino case the suppression scale is related to the supersymmetry-breaking or the Planck scale.}.

Such a scenario is, arguably, less predictive than standard IR-dominated freeze-in (or freeze-out, for that matter) due to the presence of an additional free parameter entering the calculations: the reheating temperature. However, it is by no means excluded or disfavoured from a cosmological standpoint. Moreover, as we will see it allows for a fair amount of freedom in choosing the interaction strength between the DM and SM particles. In a sense, this scenario bridges the gap that exists between thermal freeze-out and freeze-in by requiring intermediate values for the DM-SM or DM-mediator coupling strength\footnote{Other such scenarios exist like conversion-driven freeze-out/co-scattering~\cite{Garny:2017rxs,DAgnolo:2017dbv}.}. This freedom has twofold consequences: first, microscopic models in which the thermal freeze-out or standard (\textit{i.e.} with a large $T_R$) freeze-in regimes are excluded may now become viable. Secondly, the scenario we consider here provides motivation to continue and intensify all existing or planned DM searches at the LHC and beyond since it constitutes a class of freeze-in configurations which can, indeed, be probed at DM detection experiments. 

The remainder of this contribution is structured as follows: in Sec.~\ref{sec:FILC:setup} we present some general features of the freeze-in mechanism and the concrete microscopic model we will study. In Sec.~\ref{sec:FILC:exp} we study the various constraints that our model is subject to, stemming from cosmology, flavour and collider observations. Finally, in Sec.~\ref{sec:FILC:summary} we summarise our main findings and conclude.
\section{Theoretical framework}\label{sec:FILC:setup}
\subsection{Freeze-in}\label{sec:FILC:fi}

Let us consider a simple ``dark'' (\textit{i.e.} ${\cal{Z}}_2$-odd) sector containing just the DM particle $s$ along with a charged parent $F$ (we will present a concrete such model in Sec.~\ref{sec:FILC:model}). Dark matter freeze-in can proceed either via decays of the heavier particle or from scattering processes involving bath particles in the initial state. Scattering processes can further fall into two classes: \emph{(i)} conversions, which (similarly to the case of freeze-in from decays) involve at least one charged parent in the initial state and \emph{(ii)} production of two DM particles from annihilation of two SM particles. In the standard freeze-in scenario where $ T_\text{R}\gg m_F > m_{s}$, class \emph{(ii)} is usually not considered as in many models it is of higher order in the feeble coupling and, hence, negligible. If, however, $m_F > T_\text{R}> m_{s}$ the situation can change as the DM yield of both the decays and the conversions is proportional to the charged parent abundance which can be very small after reheating, whereas pair-production of DM is proportional to the abundances of the -- potentially unsuppressed -- (light) SM degrees of freedom. Note that in everything that follows we will consider very light dark matter, namely $m_s \sim 12$ keV, the lowest value compatible with Lyman-$\alpha$ forrest bounds for the model considered here \cite{Belanger:2018sti}.

In order to quantify the abundance of $F$ we consider instantaneous reheating, assuming that the SM sector is reheated and thermalized at $T_\text{R}$ while the abundance of the parent is zero at this point in time. This is, of course, a rather unrealistic approximation. However, it constitutes a framework sufficient for the subsequent discussion. In this setup the (comoving) abundance of $F$, $Y_F$, is governed by the usual Boltzmann equation, which describes thermal freeze-out~\cite{Gondolo:1990dk,Griest:1990kh}:
\begin{equation}
\frac{ \diff Y_{F} }{\diff x} = \frac{1}{ 3 H}\frac{\diff s}{\diff x}\sv\left(Y_{F}^2-Y_{F,\text{eq}}^2\right)\,,
\end{equation}
with the initial condition $Y_{F}^2 (x_\text{R})=0$. Here $x=m_F/T$, $H$ is the Hubble parameter and $s$ the entropy density.

Let us define $T_\text{fo}$ to be the freeze-out temperature where $\sv Y_{F,\text{eq}}/s\simeq H$, \ie~where $Y_{F}$ departs from its equilibrium density in standard freeze-out. If the reheating temperature is above the freeze-out temperature of the mediator, $T_\text{R}>T_\text{fo}$, the production of $F$ from the thermal bath is efficient and it quickly thermalizes. In this case the DM abundance is essentially equal to the freeze-out abundance for $T_\text{R}\gg m_F$. If, however, $T_\text{R}\lesssim T_\text{fo}$ the mediator abundance becomes smaller than its (would-be) freeze-out abundance. In the limit $T_\text{R}\ll T_\text{fo}$ it is:
\begin{equation}
Y_{F,0}\simeq\int_{x_\text{R}}^{x_0} Y_{F,\text{eq}}^2\frac{s}{Hx}\langle\sigma v\rangle dx
\simeq 1.6\times10^{31}\,\frac{g_{*s}}{\sqrt{g_*}}\,\frac{ m_F}{\text{GeV}} \, \frac{\sv (x_\text{R})}{\text{cm}^3/\text{s}} \, x_\text{R}\, \text{e}^{-2x_\text{R}}
\end{equation}
provided that $\langle\sigma v\rangle$ is a sufficiently slowly varying function of $x$. This means that $Y_F$ decreases very fast with $x_\text{R}$ (faster than $Y_{F,\text{eq}}(x_\text{R})$). This consideration shows that decay and conversion processes become irrelevant relatively quickly if $T_\text{R}<T_\text{fo}$ is chosen. Once this happens, DM is mostly (pair-) produced through scattering of SM particles. In passing, let us point out that as the DM production process is mediated by the heavy charged parent $F$ appearing in the $t$-channel, the corresponding higher-dimensional effective operator leads to UV-dominated freeze-in and, hence, to a dependence of the DM abundance on $T_\text{R}$. 

\subsection{A simple charged parent model}\label{sec:FILC:model}

In order to render the previous discussion more concrete, we consider an extension of the SM by a real scalar DM candidate $s$ that is neutral under the SM gauge group along with a vector-like fermion $F$ which we take to be an $SU(3)_c \times SU(2)_L$ singlet but to carry a hypercharge of $-1$. Both $s$ and $F$ are taken to be odd under a $\mathbf{Z}_2$ symmetry which, provided $m_s < m_{F}$, renders $s$ completely stable. The SM particles are taken to be even under the same discrete symmetry.
 \\
 \\
The most general Lagrangian compatible with these symmetries reads
\begin{align}
\label{eq:FILC:lag}
{\cal{L}} & = {\cal{L}}_{\rm SM} + \partial_\mu s ~ \partial^\mu s  - \frac{\mu_s^2}{2} s^2 + \frac{\lambda_s}{4} s^4 + \lambda_{sh} s^2 \left(H^\dagger H\right) \\ \nonumber
& + i \bar{F} \slashed{D} F  - m_{F} \bar{F} F - 
\sum_{f} y_{s}^{f} \left(s \bar{F} \left( \frac{1+\gamma^5}{2} \right) f + {\rm{h.c.}} \right),
\end{align}
where $f = \lbrace e, \mu, \tau \rbrace$.
This model, as well as variants thereof assuming other gauge quantum number and spin assignments, has been studied in the context of IR-dominated freeze-in in\cite{Belanger:2018sti,Garny:2018ali}.
For simplicity, we assume flavor universal $y_s^f$ couplings, so that the model has a total of five free parameters,
\begin{equation}\label{eq:FILC:parameters}
m_s, m_F, \lambda_{sh}, \lambda_{s}, y_{s} \ ,
\end{equation}
{\it i.e.}~the DM mass, the vector-like fermion mass, the DM-Higgs quartic coupling, the DM quartic self-coupling and the universal Yukawa-type couplings betwen DM, the vector-like lepton and the right-handed SM leptons. The dark scalar mass is related to the $\mu_s$ parameter entering Eq.~(\ref{eq:FILC:lag}) through
\begin{equation}
\mu_s^2 = m_s^2 + \lambda_{sh} v^2 \ ,
\end{equation}
and the two $\lambda$ couplings are irrelevant for our purposes. We will thus set them to zero during the remainder of our analysis\footnote{We refer to, \textit{e.g.}, Refs.~\cite{McDonald:2001vt,Yaguna:2011qn} and \cite{Bernal:2015ova} for examples in which $\lambda_{sh}$ and $\lambda_s$ become relevant respectively.}.

The parent particle lifetime spans a wide range of values depending on the magnitude of $y_s, $which strongly affects the LHC phenomenology of our model as we will see in Sec.~\ref{sec:FILC:llpconstraints}. 
\begin{figure}[htbp]
	\centering
	\includegraphics[width=0.55\textwidth]{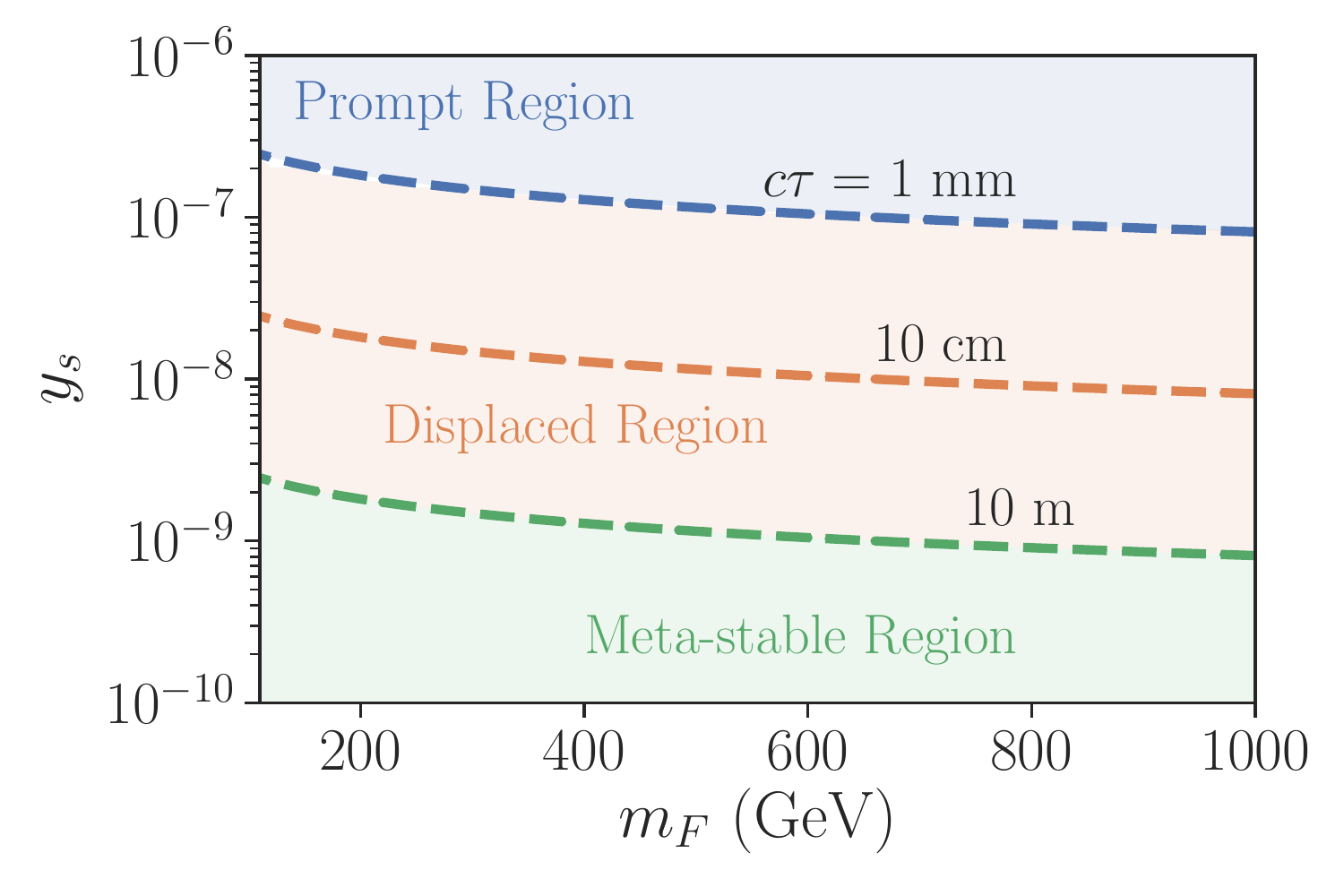}
	\caption{
		Contours of $c \tau$ for the mediator $F$ in the $y_s$- $m_F$ plane. The dark matter mass is assumed to be much smaller than $m_F$. The shaded regions approximately display the distinct regimes relevant for collider searches.}
	\label{fig:FILC:lifetimes}
\end{figure}    
In Fig.~\ref{fig:FILC:lifetimes} we illustrate this dependence in the $(m_F, y_s)$ plane assuming $m_s = 12$ keV and highlighting in different colours regions of parameter space in which $F$ decays mostly promptly ($c\tau \lesssim 1$ cm, blue region), through displaced vertices but within the LHC detectors ($c\tau \lesssim 10$ cm, red region) or outside the detector ($c\tau \lesssim 1$ m, green region). Needless to say that in reality, these regions are not as sharply separated as one might infer from Fig.~\ref{fig:FILC:lifetimes}. The decay length of $F$ is a statistically distributed quantity which also depends on the boost of $F$ which, in turn, varies from one event to another. In this respect, Fig.~\ref{fig:FILC:lifetimes} should be taken as a ballpark illustration of the proper lifetimes (and, hence, $(m_F, y_s)$ combinations) for which most of the decays occur promptly, displaced or outside the LHC detectors.

In order to perform our phenomenological analysis we have used the {\sc FeynRules} \cite{Alloul:2013bka} model implementation that was developed in the framework of \cite{Belanger:2018sti}, along with the corresponding {\sc UFO} \cite{Degrande:2011ua} and {\sc CalcHEP} \cite{Belyaev:2012qa} model files for use with {\sc MadGraph5\_aMC@NLO} \cite{Alwall:2014hca} (for the collider part of our analysis) and {\sc micrOMEGAs 5} \cite{Belanger:2018mqt} (for the computation of the freeze-in DM abundance). 
\section{Experimental constraints}\label{sec:FILC:exp}

We now move on to discuss the constraints that the model presented in Sec.~\ref{sec:FILC:model} is subject to. At first we discuss bounds from dark matter phenomenology, we then move on to flavour observables and, finally, LHC searches that affect our parameter space.

\subsection{Dark matter constraints}\label{sec:FILC:DMconstraints}

\begin{figure}[htbp]
\centering
\setlength{\unitlength}{1\textwidth}
\begin{picture}(0.96,0.33)
 \put(0.0,-0.02){\includegraphics[width=0.55\textwidth, trim= {3.3cm 2.2cm 3cm 2cm}, clip]{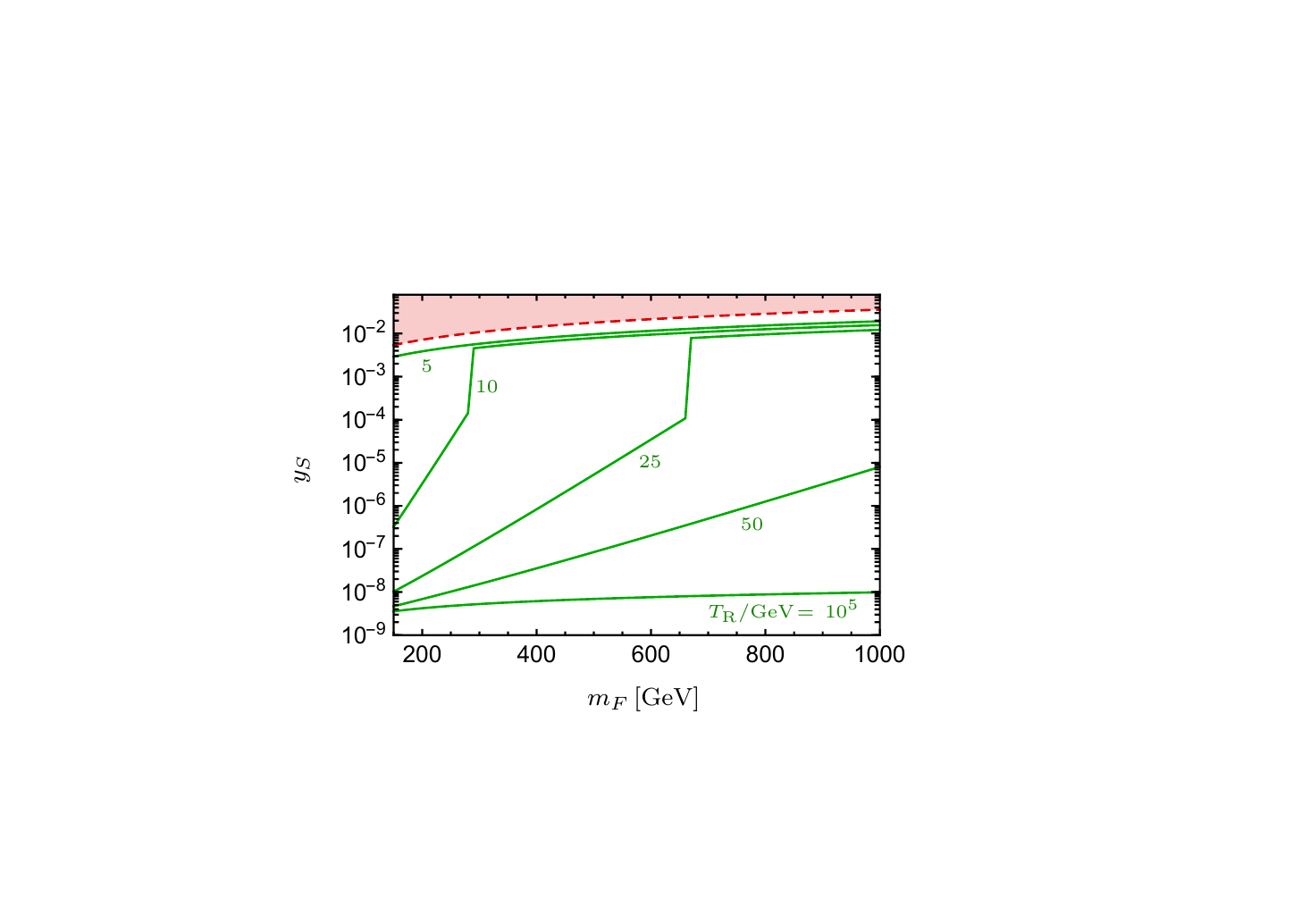}}
 \put(0.5,-0.021){\includegraphics[width=0.562\textwidth, trim= {3.3cm 2.2cm 3cm 2cm}, clip]{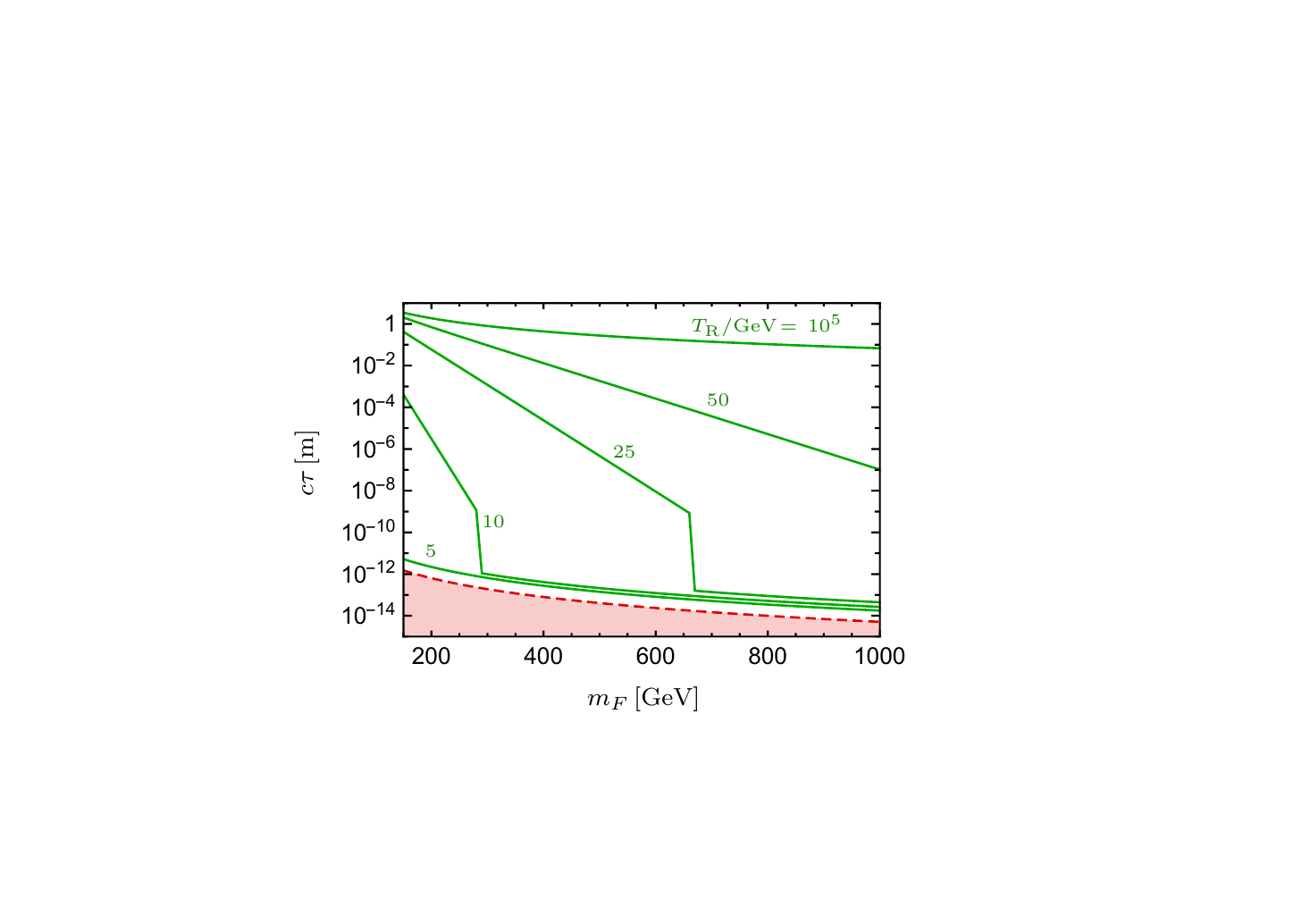}}
\end{picture}
\caption{
Contours of $\Omega h^2=0.118$ in the $y_s - m_F$ plane (left) and $c\tau - m_F$ plane (right) for different reheating temperatures and $m_s=12$ keV. The red shaded region represents the constraint from $\mu \rightarrow e\gamma$.}
\label{fig:FILC:relic}
\end{figure}

\paragraph{The dark matter abundance} 
In Fig.~\ref{fig:FILC:relic} we show contours of $\Omega h^2=0.118$ \cite{Aghanim:2018eyx} for four different (low) values of the reheating temperature, namely $T_\text{R} = 5, 10, 25$ and $50$ GeV (green lines) in the $(m_F, y_s)$ (left panel) and $(m_F, y_s)$ (right panel) planes. For comparison, we also include the result obtained assuming a high value of $T_\text{R} = 10^5$ GeV, which corresponds to conventional (IR-dominated) freeze-in. We observe that for decreasing $T_\text{R}$ increasingly large values of $y_s$ are required to reproduce the observed DM abundance in the Universe. This behaviour can be understood in light of the comments made in Sec.~\ref{sec:FILC:fi}. In conventional (\textit{i.e.} high $T_\text{R}$) freeze-in scenarios the relic density is usually dominated by DM production through $F$ decays, while the coupling $y_s$ required to reproduce the PLANCK observations is feeble. However, as the reheating temperature decreases the parent particle abundance becomes suppressed, DM production through scattering of SM particles can dominate, and larger values of $y_s$ are required.

Concretely, DM production through $F$ decays becomes subleading with respect to ($F$-mediated) annihilation processes $f\bar{f}\rightarrow ss$ once $T_\text{R}$ drops below the freeze-out temperature of $F$, $T_\text{fo}$, which is roughly given by $m_F/30$. This occurs for $m_F>680 (290)$ GeV for $T_\text{R}=25 (10)$ GeV and for all considered masses of $F$ when $T_\text{R}=5$ GeV. It is seen as a sharp transition in two of the curves in Fig.~\ref{fig:FILC:relic}. Note also that in all cases of scattering-dominated freeze-in, the couplings involved $y_s \approx 10^{-3}-10^{-2}$ are weak rather than feeble, thus clearly leading to prompt decays of $F$, whereas long-lived signatures (corresponding roughly to $c\tau > 10^{-3}$ m)  are relevant only for $T_\text{R}>50(10)$ GeV when $m_F=550 (150)$ GeV. If we considered higher values of the reheating temperature and/or the DM mass, long-lived signatures would be much more relevant, as \textit{e.g.} discussed in \cite{Belanger:2018sti}.

A few remarks are in order. In our calculations we assume a sharp transition of the initial conditions at $T_\text{fo}\approx T_\text{R}$ between the decay-dominated and the scattering-dominated regimes. More specifically, we assume that for $T_\text{R} > T_\text{fo}$ the parent particle $F$ follows its equilibrium distribution, whereas in the opposite regime we assume that its initial abundance is zero and it is produced through freeze-in. Moreover, for $T_\text{R} > T_\text{fo}$ we should also take into account the freeze-out of $F$, \textit{i.e.} the fact that below roughly $m_F/30$ the equilibrium distribution constitutes an underestimation of its true abundance. This component contributes to the DM density as
\begin{equation}
 \Omega_s h^2= \frac{m_s}{m_F} \Omega_F^\text{fo} h^2 
\end{equation}
where $\Omega_F^\text{fo}$ is the density of $F$ from the freeze-out mechanism. However, since $F$ is a charged particle, its relic abundance is rather small (and independent of $y_s$), while we are also considering a large mass hierarchy between $F$ and $s$. We have numerically checked that the combination of these two factors renders the post-freeze-out contribution of $F$ to the total DM density completely negligible. Lastly, note that although the sharp transition approximation is a rough one, it does not impact our conclusion that $F$ decay is prompt both before and after the transition. It simply means that for $m_F \simeq T_\text{fo}$, our estimates should be taken with a pinch of salt.

\paragraph{Direct detection through electron scattering}
The tree-level interaction between our DM candidate and the SM leptons opens up possibilities for direct detection through $s$ scattering off electrons. This interaction can be parameterized  in a model-independent way with respect to a reference cross section $\bar{\sigma}_{e}$, and a form factor $F_{DM}(q)$ \cite{Essig:2011nj}, 
\begin{eqnarray}
\overline{\sigma}_{e} &\equiv & \frac{\mu_{se}^{2}}{16\pi m_{s}^{2}m_{e}^{2}} |\overline{{M}_{se}(q)}|^{2}_{q^{2}=(\alpha m_{e})^{2}} \\
|\overline{{M}_{se}(q)}|^{2} &=& |\overline{{M}_{se}(q)}|^{2}_{q^{2}=(\alpha m_{e})^{2}} \times F_{DM}(q)^{2}
\end{eqnarray}
The reference cross section $\bar{\sigma}$, is the non-relativistic DM-electron scattering cross section with the momentum transfer $q$ set to the reference value $\alpha m_{e}$. $|\overline{{M}_{se}(q)}|^{2} $ is the spin-averaged squared matrix element for the dark matter-electron scattering.  Electron scattering experiments are typically sensitive to light dark matter interacting through ultra light mediators via either electron ionization or excitation.  For the model considered in this work, the reference cross section $\overline{\sigma}_{e}$ is proportional to $y_{s}^{4}/m_{F}^4$. Then, even in the case of relatively large, ${\cal{O}}(10^{-4})$ couplings an extremely light mediator ($\sim 1$ MeV) is required to give rise to observable signals. Since in what follows the heavy fermion mass will be taken to lie in the ${\cal{O}}(10^{2})$ GeV region, the resulting reference cross section is significantly below the threshold required to excite electrons. Hence, constraints from direct detection experiments relying on DM-electron scattering are negligible. 

\subsection{Constraints from flavour observables}\label{sec:FILC:flavourconstraints}

In the absence of any flavour protection mechanism, and for sufficiently low values of $m_s$, the model described by the Lagrangian of Eq.~(\ref{eq:FILC:lag}) gives rise to tree-level exotic contributions to the charged lepton decay widths, through processes of the type $l_1 \rightarrow l_2 ss$, as well as to contributions to radiative (three-body) decays such as $l_1 \rightarrow l_2 \gamma$ ($l_1 \rightarrow l_2 l_2 l_2$). For our choice $m_s = 12$ keV, the leading constraints stem from the very well-measured muon observables, in particular the measurement of the muon lifetime and, more importantly, the decay $\mu \rightarrow e\gamma$.

The muon lifetime has been measured extemely accurately, with a relative precision of $10^{-6}$ \cite{Tanabashi:2018oca}. In practice, in order to ensure that the decay mode $\mu \rightarrow e ss$ does not spoil the successful prediction of the Standard Model, we impose that the corresponding partial width should not exceed the $1\sigma$ uncertainty of the measured total muon width, $\Gamma_{\mu \rightarrow ess} \lesssim 3\times 10^{-25}$ GeV. For a heavy fermion mass of 100 GeV, we find that this condition can be satisfied as long as $y_s \lesssim 2\times 10^{-2}$, with the bound scaling as $1/m_F^4$.

The radiative muon decay $\mu \rightarrow e\gamma$ turns out to provide the most stringent constraint on our parameter space. Using the general expressions that can be found in \cite{Lavoura:2003xp}, neglecting the muon and electron masses in the loop integrals and keeping only the leading terms in $m_F^2/m_s^2$, the $\mu \rightarrow e\gamma$ partial width can be approximated as
\begin{equation}
\Gamma_{\mu \rightarrow e\gamma} \simeq \frac{16 m_\mu^5 y_s^4}{9 m_F^4 (16 \pi)^5} .
\end{equation}
At the same time, assuming that the muon decay width is dominated by the Standard Model contribution, we have as usual
\begin{equation}
\Gamma_{\mu \rightarrow e\gamma} = \frac{G_F^2 m_\mu^5}{192 \pi^3}
\end{equation}
where $G_F$ is the Fermi constant, the most precise determination of which comes exactly from the muon lifetime measurement. The branching ratio for $\mu \rightarrow e\gamma$, then, reads
\begin{equation}
Br(\mu \rightarrow e\gamma) \simeq \frac{y_s^4}{12 G_F^2 m_F^4 (16\pi)^2} .
\end{equation}
Given the latest constraint $Br(\mu \rightarrow e\gamma) < 4.2 \times 10^{-13}$ \cite{TheMEG:2016wtm}, we obtain the bound:
\begin{equation}
\frac{y_s}{m_F} \lesssim 3.6 \times 10^{-5} \ \ {\rm GeV}^{-1} 
\end{equation}
which, as we can see in Fig.~\ref{fig:FILC:relic} (red shaded region), approaches our phenomenologically interesting parameter space without affecting it substantially.
\subsection{Constraints from searches at the LHC}\label{sec:FILC:llpconstraints}

Our model contains a scalar dark matter field $s$ and a vector-like fermionic mediator $F$. Since $s$ is a singlet under the SM gauge symmetries and we have chosen to neglect the Higgs portal interaction appearing in the Lagrangian of Eq.~(\ref{eq:FILC:lag}), it can only be produced through the mediator $F$. 
\begin{figure}[htbp]
	\centering
	\begin{tikzpicture}
	\begin{feynman}
	\node [blob] (a);
	\vertex [above left=of a] (i1) {$p$};
	\vertex [below left=of a] (i2) {$p$};
	\vertex [right=of a] (b);
	\node [above right=of b,circle,fill,inner sep=1pt] (E1);
	\node [below right=of b,circle,fill,inner sep=1pt] (E2);
	\vertex [above right=of E1] (l1) {$e^+,\mu^+,\tau^+$};
	\vertex [below=of l1] (s1) {$s$};
	\vertex [below right=of E2] (l2) {$e^-,\mu^-,\tau^-$};
	\vertex [above=of l2] (s2) {$s$};
	
	\diagram*{
		(i1) -- [double] (a) -- [double] (i2),
		(E1) -- [fermion,edge label'={$\bar{F}$}] (b) -- [fermion,edge label'={$F$}] (E2),
		(s1) -- [scalar] (E1) -- [anti fermion] (l1), 
		(s2) -- [scalar] (E2) -- [fermion] (l2),
		(a) -- [photon,edge label'={$\gamma,Z$}] (b),
	};
	\end{feynman}
	\end{tikzpicture}
	\caption{\it Diagram for the main production mode of the mediator ($F$) and its decay to dark matter ($s$) at the LHC. Since we assume universal couplings to leptons, all decay modes (including mixed flavor) have the same branching ratio.}
	\label{fig:FILC:process}
\end{figure}
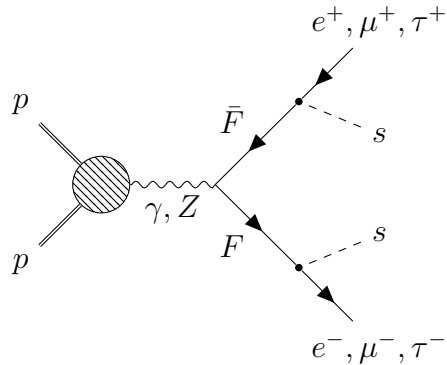
The main process relevant for production of $s$ at the LHC is illustrated in Fig.~\ref{fig:FILC:process}, which leads us to typically expect for a signal to emerge in searches for two (OS) leptons accompanied by missing energy (MET). However, as illustrated by Figs.\ref{fig:FILC:lifetimes} and \ref{fig:FILC:relic}, in most regions of parameter space the decay length of $F$ can vary over many orders of magnitude and $F$ can decay promptly, inside the detector with a macroscopic decay length or even outside the detector. The searches for $F$ can, therefore, be regrouped into three classes:
\begin{enumerate}
	\item Prompt decays ($c \tau \lesssim 1$~mm): the main signature in this case consists of 2 leptons and MET. In order to test this regime, we will use the corresponding ATLAS search from Ref.~\cite{Aad:2019vnb}.
	\item Displaced decays (10 mm $\lesssim c \tau \lesssim 1$~m): in this scenario $F$ can either decay inside the tracker or in the calorimeter. Hence, the signal can either consist of displaced leptons (which can be searched for with or without missing energy) or of disappearing tracks (DT) if the leptons from the $F$ decay are not associated to their parent track. In the former case we will consider the CMS displaced lepton + MET search of Ref.\cite{CMS:2016isf} whereas in the latter case we will rely on the ATLAS disappearing track search peformed in Ref.~\cite{Sirunyan:2018ldc}.
	\item Meta-stable decays ($c \tau \gtrsim 1$~m): if the $F$ decay length is of the order of a meter or larger, a fraction of the produced $F$'s can decay outside the detector. In this case the main signature is a charged track, corresponding to a heavy stable charged particle (HSCP). In order to constrain this scenario we will consider the 13~TeV ATLAS search for HSCPs~\cite{Aaboud:2019trc}.
\end{enumerate}

In Table~\ref{tab:FILC:searches} we present a summary of the searches we consider, their phenomenological signature and the approximate lifetime interval for which they apply.

\begin{table}[htbp]
	\begin{tabular}{|c|c|c|}
		\hline
		Lifetime range & Signature  & Reference \\ \hline
		Prompt Decays ($c \tau \lesssim 1$~mm) & 2 OS leptons + MET  & ATLAS-SUSY-2018-32~\cite{Aad:2019vnb} \\ \hline
		\multirow{2}{*}{Displaced Decays (10 mm $ \lesssim c \tau \lesssim 1$~m)} & displaced leptons plus MET & CMS-PAS-EXO-16-022 \cite{CMS:2016isf} \\
     & disappearing track & CMS-EXO-16-044 \cite{Sirunyan:2018ldc} \\ \hline
     		Meta-stable decays ($c \tau \gtrsim 1$~m) & HSCPs  & ATLAS-SUSY-2016-32~\cite{Aaboud:2019trc} \\ \hline
	\end{tabular}
	\caption{\it List of LHC searches used to test the final state illustrated in Fig.~\ref{fig:FILC:process}. The searches are split in three groups, corresponding to distinct lifetime ranges.}
	\label{tab:FILC:searches}
\end{table}

\paragraph{Searches for prompt leptons + MET}
 
If the mediator lifetime is of the order of a few mm or below, it will mostly decay promptly and prompt searches for leptons accompanied by MET apply. Since the topology in Fig.~\ref{fig:FILC:process} is similar to slepton pair-production and decay, we employ the 139~fb$ ^{-1}$ ATLAS search~\cite{Aad:2019vnb} which looked for an excess in the two leptons (electrons and muons) plus missing energy channel. This search requires the leptons to be produced within $|z_0 \sin \theta| = 0.5$~mm of the primary vertex,  where $\theta$ is the azimuthal angle of the lepton 3-momentum and $z_0$ its longitudinal impact parameter relative to the primary vertex. Therefore, this search is only relevant for lifetimes $ c \tau \lesssim \mathcal{O}(\mbox{1  mm})$. In order to apply the search results to constrain our model, we make use of the upper limits on the total production cross-section of the simplified slepton model ($p p \to \tilde{l}_{L,R} + \tilde{l}_{L,R} \to l^- l^+ \chi_1^0 \chi_1^0$) provided by ATLAS. Although $F$ is a vector-like fermion, we do not expect its spin to significantly change the signal efficiencies when compared to the slepton scenario. It is, therefore, a good approximation to apply the ATLAS cross-section upper limits to the process $ p p \to F + \bar{F} \to l^- l^+ s s $. In the near future we intend to properly re-implement this search in the {\sc MadAnalysis}~5 framework~\cite{Conte:2012fm,Conte:2014zja,Conte:2018vmg} for a more accurate recasting. However, even within our level of approximation there are two differences between the decay of $F$ and the simplified model considered by ATLAS that \textit{do} need to be accounted for:
\begin{itemize}
	\item first, since we assume universal couplings to leptons, about 55\% of the signal will contain at least one $\tau$-lepton. Here we make the conservative assumption that the search has zero sensitivity to these events. Therefore we rescale the $F$ production cross-section by 45\%.
	\item Second, for lifetimes larger than a few mm, some events will contain leptons with impact parameter above the value required by ATLAS. In order to take this effect into account we generate parton level events with  {\sc MadGraph5\_aMC@NLO} and apply the $|z_0 \sin \theta| < 0.5$~mm requirement to the leptons originating from $F$ decays in order to compute the fraction of events that contain prompt leptons. We then rescale the cross-section by this factor.
\end{itemize}

\begin{figure}[htbp]
 	\centering
\setlength{\unitlength}{1\textwidth}
\begin{picture}(0.98,0.74)
 \put(0.0,0.35){\includegraphics[width=0.55\textwidth, trim= {3.3cm 2.2cm 3cm 2cm},clip]{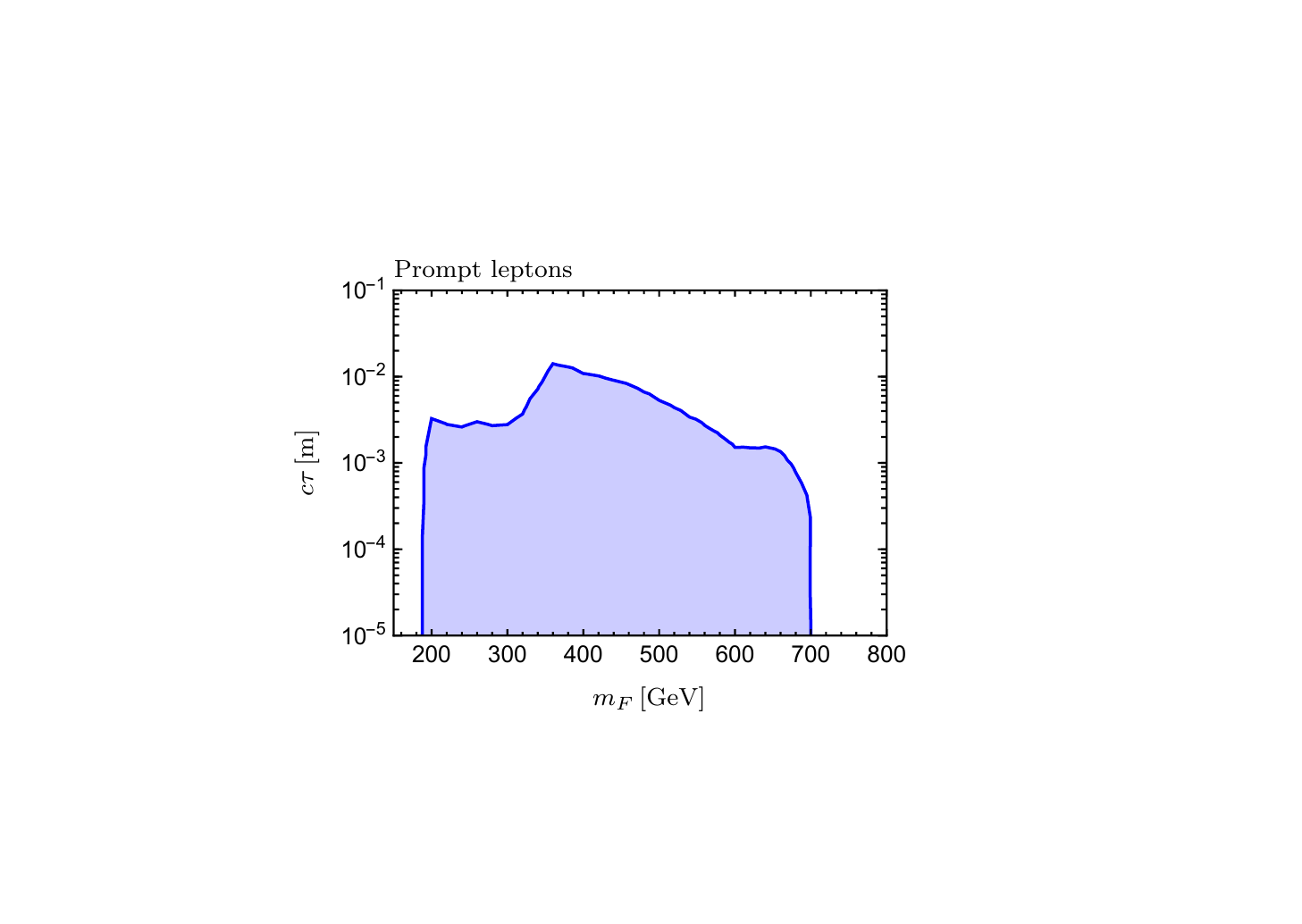}}
 \put(0.5,0.35){\includegraphics[width=0.55\textwidth, trim= {3.3cm 2.2cm 3cm 2cm},clip]{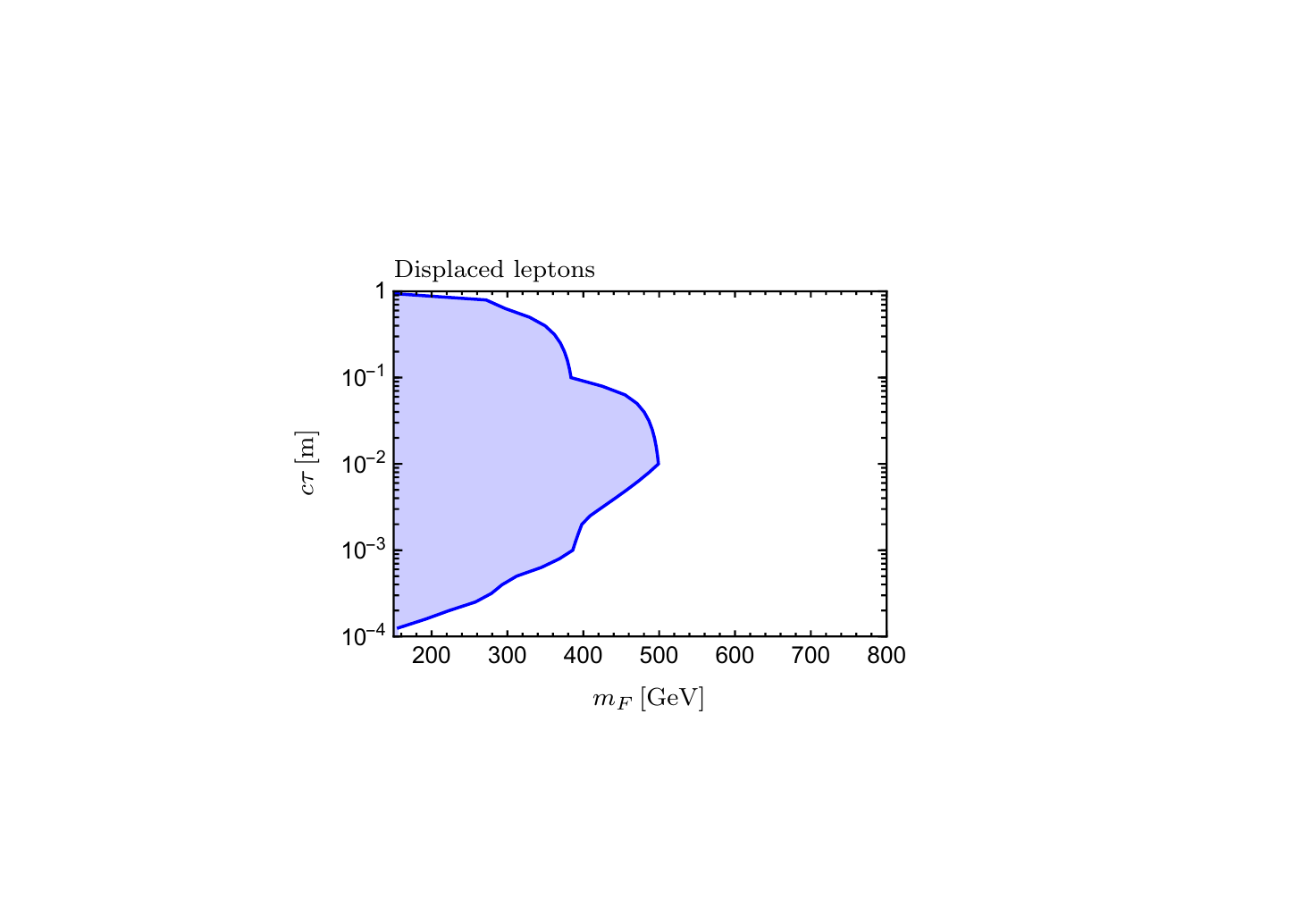}}
 \put(0.0,-0.02){\includegraphics[width=0.55\textwidth, trim= {3.3cm 2.2cm 3cm 2cm}, clip]{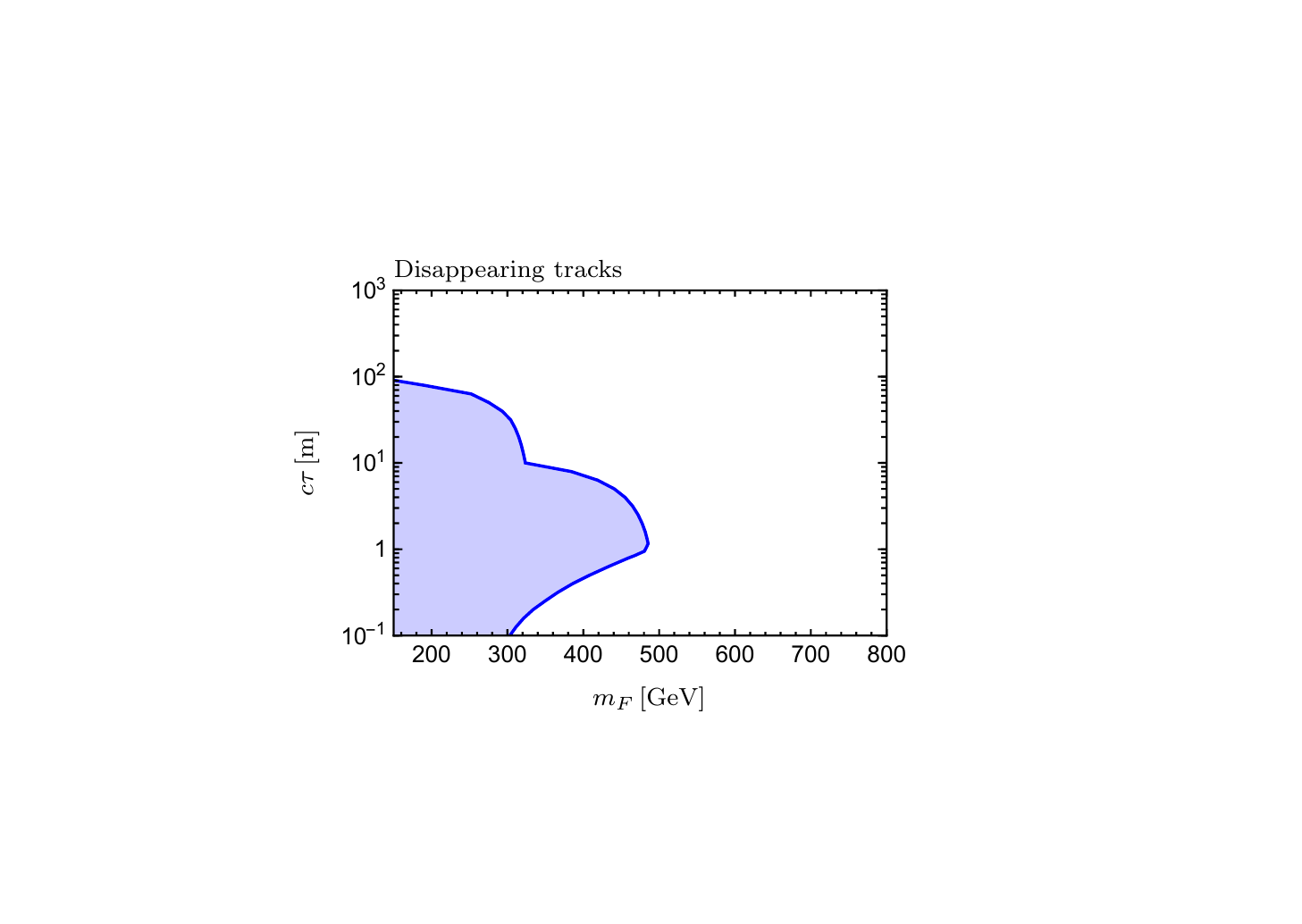}}
 \put(0.5,-0.02){\includegraphics[width=0.55\textwidth, trim= {3.3cm 2.2cm 3cm 2cm}, clip]{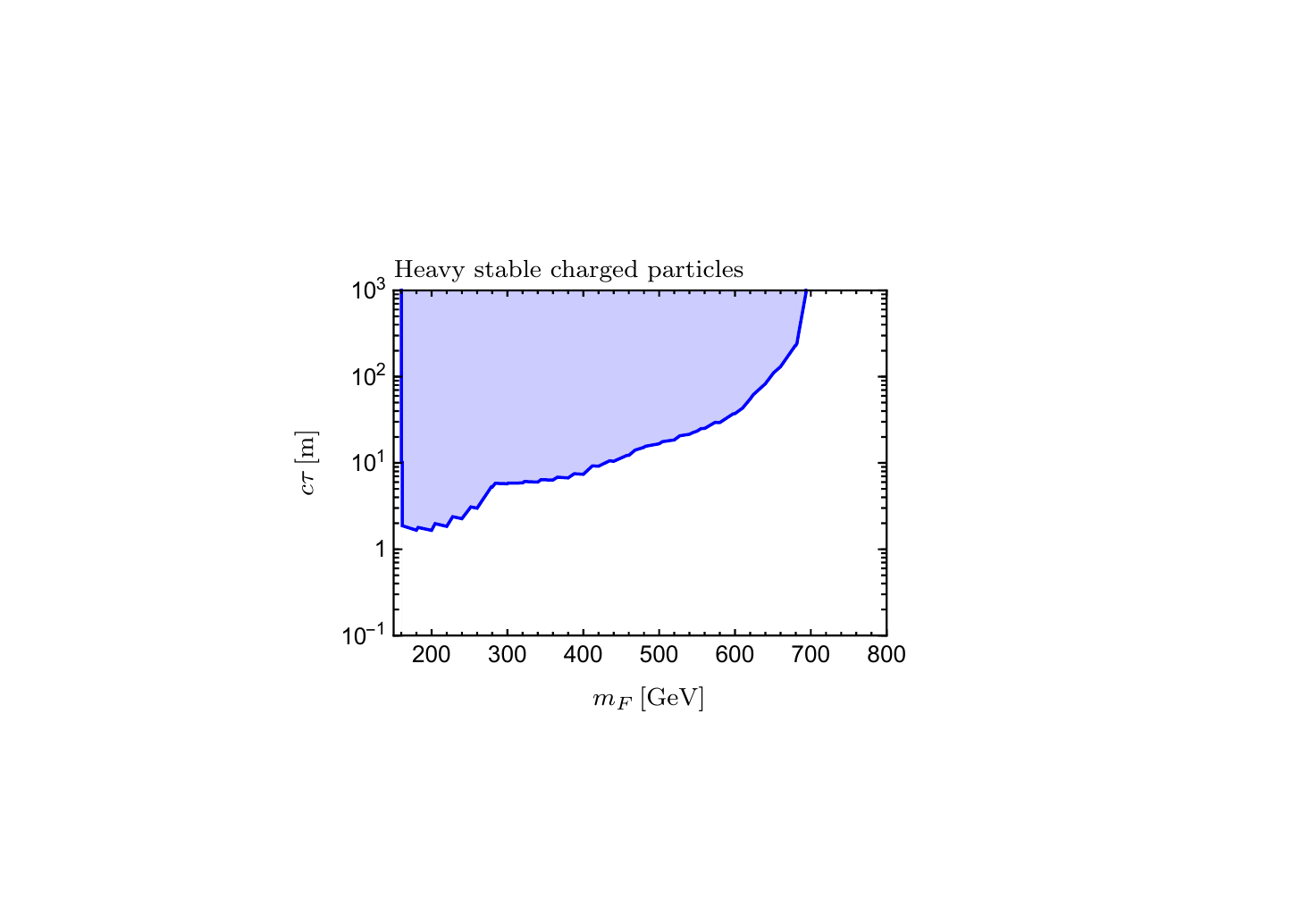}}
\end{picture}
 	\caption{%
 		95\% CL collider limits in the $m_F$-$c\tau$ plane. 
		\emph{Upper left:} Prompt ATLAS 2 leptons plus missing energy search from Ref.~\cite{Aad:2019vnb}.
		\emph{Upper right:} CMS displaced lepton search from Ref.\cite{CMS:2016isf}.
		\emph{Lower left:} Disappearing track search from Ref.~\cite{Sirunyan:2018ldc}.
		\emph{Lower right:} Heavy Stable Charged Particle (HSCP) search from Ref.~\cite{Aaboud:2019trc}.
		In all cases, the dark matter mass is assumed to be much smaller than $m_F$.
		}
 		\label{fig:FILC:collider}
 \end{figure}

Taking into account the above considerations, we compute the 95\% C.L. limit on $m_F$ as a function of lifetime, comparing the rescaled production cross-section for $F$ with the upper limit provided by \mbox{ATLAS} for the corresponding slepton mass and a massless neutralino. The $F$ production cross-section is computed at LO using {\sc MadGraph5\_aMC@NLO}, but a K-factor value of 1.2, typical of Drell-Yan-like processes at $13$ TeV, is applied. The results are shown in the top-left panel of Fig.~\ref{fig:FILC:collider}, in the $(m_F, c\tau)$ plane. As expected, the constraints are strongest for $c \tau < 1$~mm, excluding masses up to $m_F = 700$~GeV. However, once the lifetime is larger then 1~cm, the search loses all its sensitivity. Let us also point out that although masses below 200~GeV are not considered by the analysis, they are excluded by searches at 8 TeV.

\paragraph{Searches for displaced leptons + MET}

In the analysis performed in Ref.\cite{CMS:2016isf}, long-lived particles have been searched for using 2.6~fb$^{-1}$ of LHC data at a center-of-mass energy of 13~TeV. This search especially targeted long-lived particles that decay into leptons within the volume of the tracker, a signature which is typical of specific $R$-parity-violating supersymmetric scenarios in which the stop squark is long-lived and decays into a $b$-jet and an electron or a muon. The three search signal regions are populated by events featuring two isolated leptons of different flavors\footnote{In the following, the term `leptons' solely refers to electrons and muons.}. The properties of the two selected lepton candidates are such that they are hard (with a transverse momentum $p_T\gtrsim 40$ and 42~GeV for muons and electrons respectively), central (with a pseudorapidity $|\eta| < 2.4$) and with a transverse impact parameter $d_0$ lying between $0.2$~mm and $10$~cm. Moreover, the two leptons are constrained to be well separated in the transverse plane, with a $\Delta R$ distance of at least 0.5. The analysis then includes three signal regions relevant for important displacements ($d_0>1$~mm), moderate displacements ($d_0\in[0.5,1]$~mm) and smaller displacements ($d_0\in[0.2, 0.5]$~mm).

We compute exclusion limits by using the combined log likelihood ratio of all three signal regions. The corresponding results are shown in the top-right panel of Fig.~\ref{fig:FILC:collider}. We see that the constraints reach peak sensitivity for displacements of the order of a few cm, excluding heavy lepton masses up to $m_F \sim 500$ GeV.

\paragraph{Searches for disappearing tracks}

The CMS search for disappearing tracks at 13 TeV~\cite{Sirunyan:2018ldc} selects events with at least one track with $p_T > 55$~GeV and $|\eta| < 2.1$, having at least three missing hits in the outer calorimeter layer and small energy deposit in the electromagnetic calorimeter ($<10$~GeV).  While the full detector simulation is rather involved, the acceptances for a supersymmetric chargino model have been provided for the mass range between $100-900$~GeV and lifetimes between $\tau_0=0.3-333$~ns. Since we do not expect a significant difference in the angular distributions of Drell-Yan production between the chargino and our heavy lepton, we assume the same acceptances but weighted by the Drell-Yan production of $F$ instead.  

The caveat here is that we have assumed that the leptons from $F$ decay will not be correctly identified (which would have caused the event to be vetoed). Clearly, the leptons from $F$ decay are likely to be hard whereas the pions from chargino decay are too soft to be detected. However, given that $m_{e, \mu, s} \ll m_F$, we expect both the lepton and the DM candidate to carry equal $p_T$, and therefore form a large cone around the direction of the original track.  Thus any leptons from this decay will be seen as having a large impact parameter $d_0$ and not as coming from the primary vertex. We postpone a more detailed study of the lepton veto-induced loss of acceptance in our model for the future. 

The exclusion bounds obtained using these assumptions are shown in the bottom-left panel of Fig.~\ref{fig:FILC:collider}. The highest excluded mass is roughly $480$~GeV for a lifetime of $1$~m, whereas masses of $200$~GeV are excluded for lifetimes between $0.1-100$~m. It should also be mentioned that the published acceptance grid is not fine enough to allow for a smooth interpolation and the kinks in the exclusion curve are likely to constitute an artifact. 

\paragraph{Searches for Heavy Stable Charged Particles}

When the proper decay length of the mediator $F$ is of the order of a few meters or longer, a sizeable fraction of its decays will take place outside the detector. In this case searches for heavy stable charged particles are relevant. Here we consider the ATLAS search at 13~TeV with and integrated luminosity of 36.1~fb$^{-1}$, which looked for charged particles decaying inside the calorimeter or outside the detector\cite{Aaboud:2019trc}. In particular, colour-neutral HSCPs are required to decay outside the ATLAS detector: $R > 12$~m and $|z| > 23$~m, where $R$ ($z$) is the transverse (longitudinal) distance between the decay point and the primary vertex.  Limits on the signal yield are provided for 8 distinct signal regions which differ by the number of HSCP candidates reconstructed in each event (one or two) and the minimum value for the reconstructed HSCP mass.

Trigger and HSCP reconstruction efficiencies have been made publicly available by ATLAS, which allows us to recast the search using truth level events at the hadron level\footnote{Information about the recasting and its validation as well as a sample code can be found in \hyperlink{https://github.com/llprecasting/recastingCodes}{https://github.com/llprecasting/recastingCodes}.}. Using {\sc MadGraph5\_aMC@NLO} for the generation of parton level events and {\sc Pythia 8.2} \cite{Sjostrand:2014zea} for showering and hadronization, we compute the signal efficiencies for the model considered here as a function of the mediator mass and lifetime. Once again, we apply a constant K-factor of 1.2 to the $F$ production cross-section and compute the 95\% C.L. exclusion using the best signal region for the given mediator mass. The results are shown in the bottom-right panel of Fig.~\ref{fig:FILC:collider}. Since the search requires all $F$ decays to take place outside the ATLAS detector, it is only sensitive to decay lengths larger than a few meters and is not relevant for the small $T_\text{R}$ scenario discussed in Sec.~\ref{sec:FILC:DMconstraints}.

\section{Summary and discussion}\label{sec:FILC:summary}

\begin{figure}[htbp]
\centering
\setlength{\unitlength}{1\textwidth}
\begin{picture}(0.96,0.33)
 \put(0.0,-0.02){\includegraphics[width=0.55\textwidth, trim= {3.3cm 2.2cm 3cm 2cm}, clip]{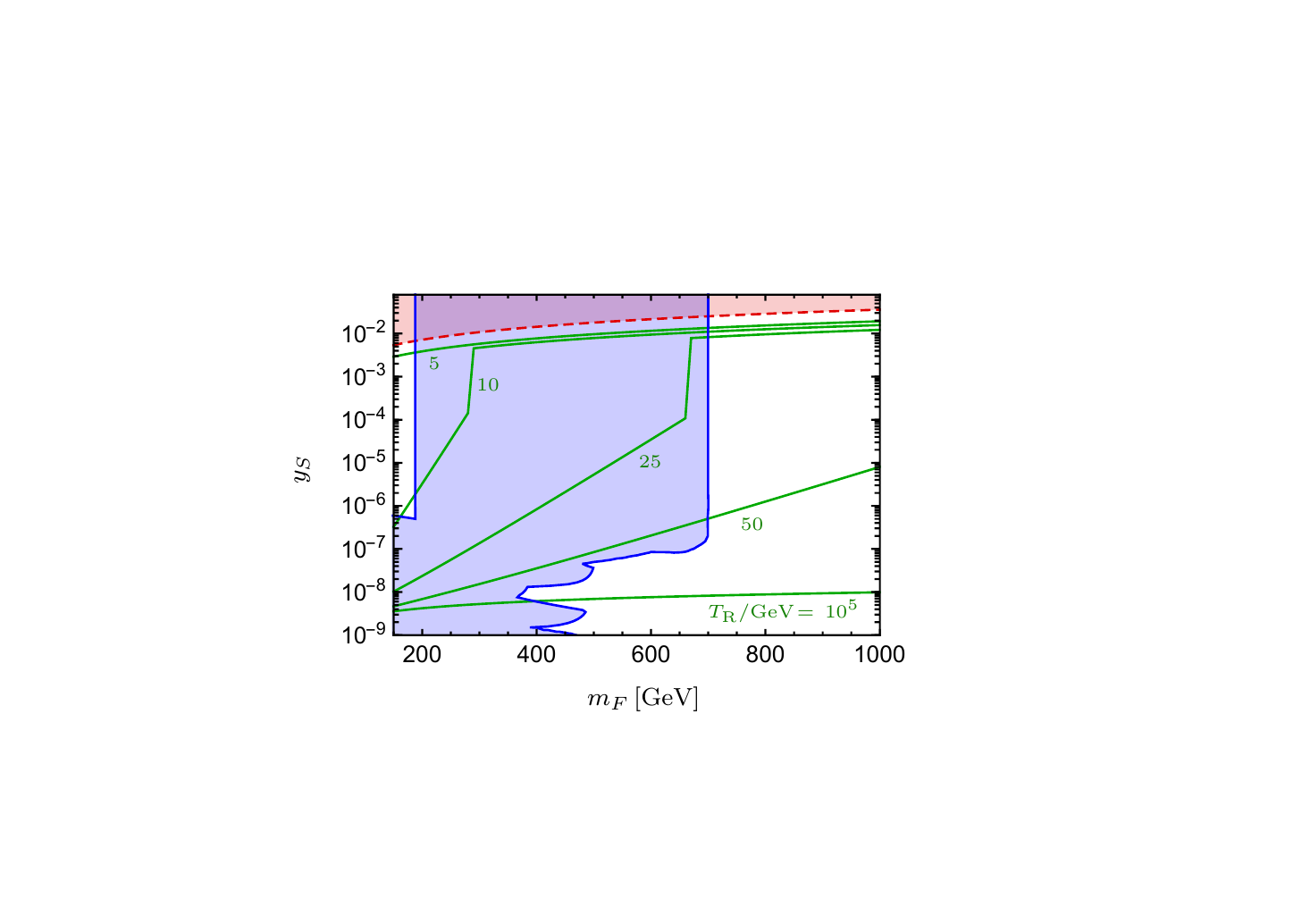}}
 \put(0.5,-0.021){\includegraphics[width=0.562\textwidth, trim= {3.3cm 2.2cm 3cm 2cm}, clip]{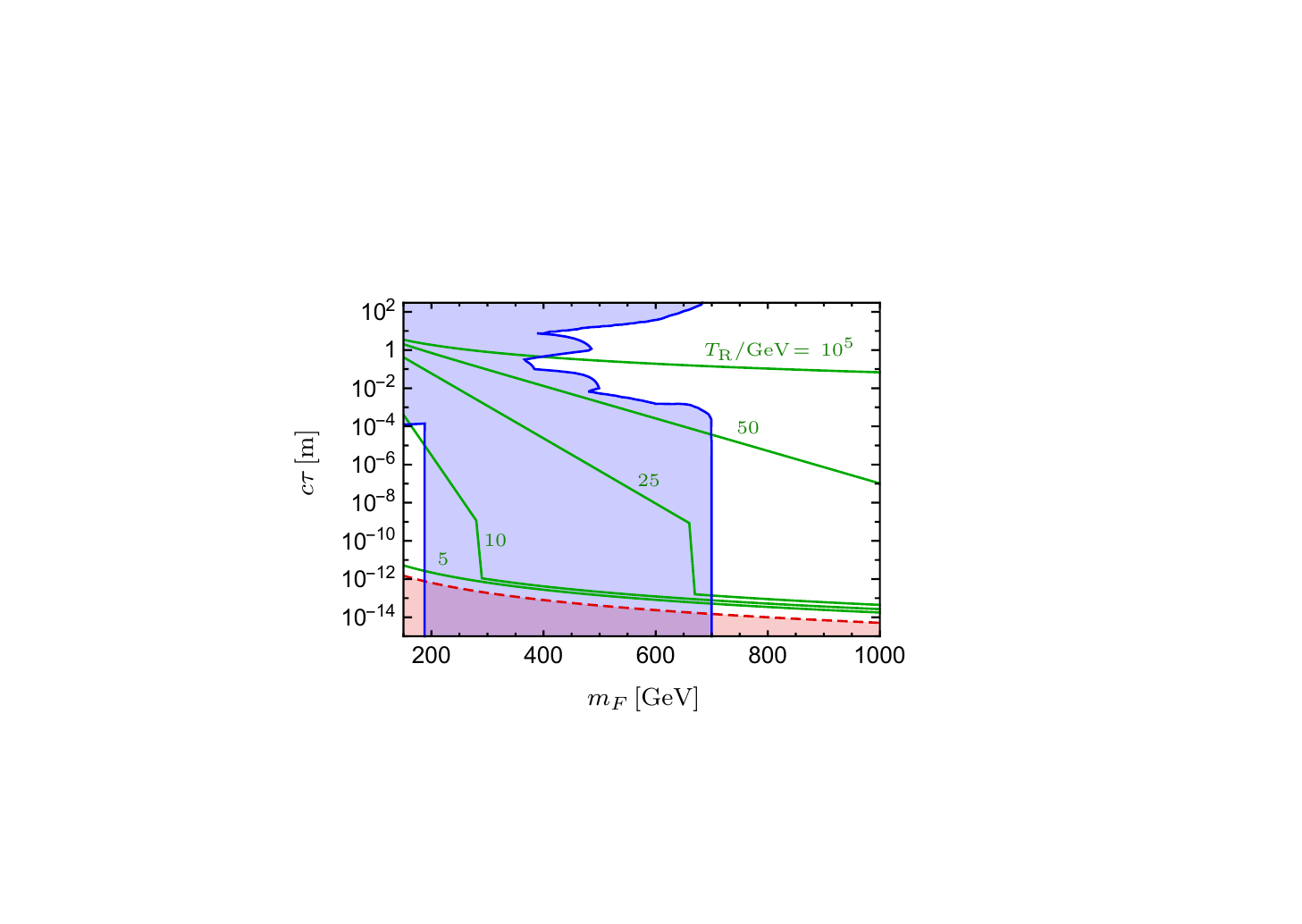}}
\end{picture}
\caption{
Summary of the results obtained in this work in the $(m_f, y_s)$ (left panel) and $(m_f, c\tau)$ (right panel) planes, assuming a dark matter mass $m_s = 12$ keV: Contours along which the cosmic DM abundance as measured by Planck can be explained through freeze-in of $s$ (green lines) along with global constraints from searches for prompt dileptons plus MET, displaced leptons plus MET, disappearing tracks and Heavy Stable Charged Particles (blue shaded region) and parameter space region disfavoured by $\mu \rightarrow e\gamma$ (red shaded region).}
\label{fig:FILC:summaryplot}
\end{figure}

The results obtained in the previous sections are summarised in Fig.~\ref{fig:FILC:summaryplot}, in which we overlay the $(m_f, y_s)$ (left panel) and $(m_f, c\tau)$ (right panel) values for which the model described in Sec.~\ref{sec:FILC:model} can explain the observed dark matter abundance in the Universe (green contours) for different values of the reheating temperature $T_\text{R}$ and assuming a DM mass $m_s = 12$ keV, along with the flavour (red shaded regions) and collider (blue shaded regions) constraints discussed in Sec.~\ref{sec:FILC:llpconstraints}.

We observe that the combination of the various constraints offers an almost complete parameter space coverage up to heavy fermion masses of roughly 700 GeV, and the role of prompt searches is crucial in order to complete this picture. The largest sensitivity is reached in the prompt and detector-stable limit, while the constraints for intermediate decay lengths of $0.1 - 1$ m turn out to be weaker. It is also interesting that the flavour constraints, and in particular $\mu \rightarrow e\gamma$ are only a factor of a few away from actually probing part of the cosmologically interesting parameter space. Limits on $\mu \rightarrow e\gamma$, which are expected to improve during the next few years, can provide an extremely interesting handle on some freeze-in scenarios with a heavier parent particle, which are kinematically inaccessible at the LHC.

Our results show that, as we already mentioned in the Introduction, conventional LHC dark matter searches are of vital importance, not only in order to assess the viability of traditional freeze-out scenarios, but also in order to probe more exotic ones such as freeze-in with a low reheating temperature. In this respect, they probe regions of the ``freeze-in parameter space'' which are inaccessible to long-lived particle searches. Our analysis illustrates, hence, the fact that there are indeed interesting freeze-in scenarios which can be probed at the LHC and that different LHC dark matter searches are complementary with each other. 
\section*{Acknowldgements}

J.H.~acknowledges support from the F.R.S.-FNRS, of which he is a postdoctoral researcher. The work of A.L. was supported by the S\~ao Paulo Research Foundation (FAPESP), project 2015/20570-1. The work of DS is  supported by the National Science Foundation under Grant No. PHY-1915147.

\let\Herwig\undefined
\let\Pythia\undefined
\let\Sherpa\undefined
\let\Rivet\undefined
\let\Professor\undefined
\let\eps\undefined
\let\mc\undefined
\let\mr\undefined
\let\mb\undefined
\let\tm\undefined
\let\diff\undefined
\let\sv\undefined

%% file: llp-dm/LLP_DM.main.tex
\graphicspath{{llp-dm/}}



\chapter{Long-lived signatures of conversion-driven freeze-out}
{\it B.~Fuks, J.~Heisig, A.~Lessa, J.\,M.~No, S.~Sekmen, D.~Sengupta, J.~Zurita}


\label{sec:LLP_DM}
  
\begin{abstract}
We consider a dark matter model where the relic density is set by conversion-driven feeze-out. 
This mechanism predicts long-lived particles at the LHC requiring small couplings and a relatively small mass splitting in
the dark sector. As such it falls somewhat in between the typical targets of currently performed long-lived particle searches
either considering extreme mass compressions or very small couplings, but lack their combinations.
We study various search strategies covering the whole range of possible decay lengths $\mathcal{O}(1\,\text{mm}\!-\!1\,\text{m})$,
pointing out blind spots in the experimental search programme and estimating the corresponding potential for improvement.
\end{abstract}

\section{Introduction} \label{sec:LLP_DM_intro}

The existence of dark matter (DM) is one of the prime motivations for physics beyond the standard model (SM) and hence
its detection consists in an important scientific goal of the LHC experiments. While the weakly interacting massive particle (WIMP)
paradigm has mainly guided our DM search strategies in the past,
other possibilities exist and have now come into focus. 
In this work we compare various search strategies for DM models containing long-lived particles with decay length 
$\mathcal{O}(1\,\text{mm}\!-\!1\,\text{m})$, and small mass splittings 
$\mathcal{O}(10)~\text{GeV}$, as predicted by conversion-driven freeze-out~\cite{Garny:2017rxs,DAgnolo:2017dbv,Garny:2019kua}. 
As a benchmark model we take 
the one studied in~\cite{Garny:2017rxs} supplementing the SM by a Majorana fermion DM candidate and 
a scalar bottom partner, both odd under a new $Z_2$ symmetry under which all SM
particles are even. 

This example is interesting as it reveals blind spots in the performed LHC analyses targeted to
other benchmark scenarios, where the long lifetime is typically considered to arise either from an extreme mass compression (long-lived chargino scenario) or a very small (effective) coupling (\emph{e.g.}~split supersymmetry or gravitino scenarios). 
However, the case in-between, as it occurs here, is often not explicitly considered. 

We study various different searches aiming at covering as much of the model parameter
space as possible. For very small decay lengths, common DM searches requiring missing energy plus jets can be
applied~\cite{Aaboud:2016tnv,ATLAS-CONF-2019-040}. These searches typically assume direct DM production, or its production via the prompt decays of a parent particle. 
Depending on the inclusiveness of the search, the latter could also be sensitive to intermediate decay lengths.
However, as the corresponding SM backgrounds are sizeable,
missing energy searches are typically outperformed by dedicated searches towards large decay lengths.

For very large decay lengths $\gtrsim1\,$m, searches for heavy stable charged particles (in this case $R$-hadrons)~\cite{Chatrchyan:2013oca,CMS-PAS-EXO-16-036,ATLAS:2014fka,Aaboud:2019trc}
can be applied and provide great prospects while being extremely inclusive. However, the sensitivity drops quickly for smaller decay length due to 
the exponentially suppressed number of $R$-hadrons passing through the whole detector. 
Therefore, intermediate lifetimes can only be sensitively searched for when exploiting the displaced decay
of the parent particle. In this respect we consider searches for disappearing tracks~\cite{Aaboud:2017mpt}, displaced jets~\cite{Aaboud:2017iio} and delayed jets~\cite{Sirunyan:2019gut}. 
However, these searches are less inclusive than the two cases mentioned above. Furthermore, displaced and delayed jets searches are not
targeted to small mass splittings, and hence soft visible decays of the long-lived parent particle.

\section{The dark matter model} \label{sec:LLP_DM_themodel}

We extend the SM by a singlet Majorana fermion DM candidate $\chi$ and 
a colored scalar bottom partner $\widetilde b$ both transforming odd under a new $Z_2$ symmetry (while SM
particles are even). 
The new physics interactions are described by
\begin{equation}
	\mathcal{L}_\text{int} \subset D_\mu \tilde b^\dag D^\mu\tilde b - \Big(\lambda_{\chi} \widetilde{b} \bar{b}\frac{1-\gamma_5}{2}\chi +\text{h.c.}\Big),
	\label{eq:LLP_DM_sbottommodel}
\end{equation}
where $b$ is the $b$-quark field, $D_\mu$ the covariant derivative and $\lambda_{\chi}$ the DM coupling.
The model resembles a supersymmetric simplified model with a bino-like neutralino and a right-handed sbottom, except for 
the coupling $\lambda_\chi$, being considered a free parameter here.
Within the model, conversion-driven freeze-out allows one to explain the measured relic density for small couplings, 
$\lambda_\chi = \mathcal{O}(10^{-7}\!\!-\!10^{-6})$ and relatively small mass splittings, 
$\Delta m_{\chi \widetilde b} = m_{\widetilde b} - m_\chi \lesssim 35\,$GeV~\cite{Garny:2017rxs}. The 
respective parameter space resides below the thick black curve in Fig.~\ref{fig:LLP_DM_param_space}.
The thin green and grey dotted curves denote contours of constant coupling and decay length, respectively, 
predicting the measured relic density, $\Omega h^2\simeq 0.12$~\cite{Aghanim:2018eyx}.
A  qualitatively similar phenomenology is obtained for a model with a top partner~\cite{Garny:2018icg} or a leptonic partner~\cite{Junius:2019dci}.

\section{Search strategies} \label{sec:LLP_DM_searches}

As the bottom partner $\widetilde b$ is strongly interacting, its
pair-production cross section at the LHC is large. In the regime of non-prompt decays
considered here, it hadronizes and forms $R$-hadron bound states
before decaying into a $b$-quark and DM\@. The corresponding process is
sketched in Fig.~\ref{fig:LLP_DM_prod_diag}. In the following we study and
discuss various search strategies in the context of this signature.

\subsection{Missing energy searches}

\begin{figure}[t]
\centering
\setlength{\unitlength}{1\textwidth}
\begin{picture}(0.64,0.5)
 \put(0.0,-0.02){\includegraphics[width=0.605\textwidth, trim= {3.3cm 0.7cm 3cm 2cm},clip]{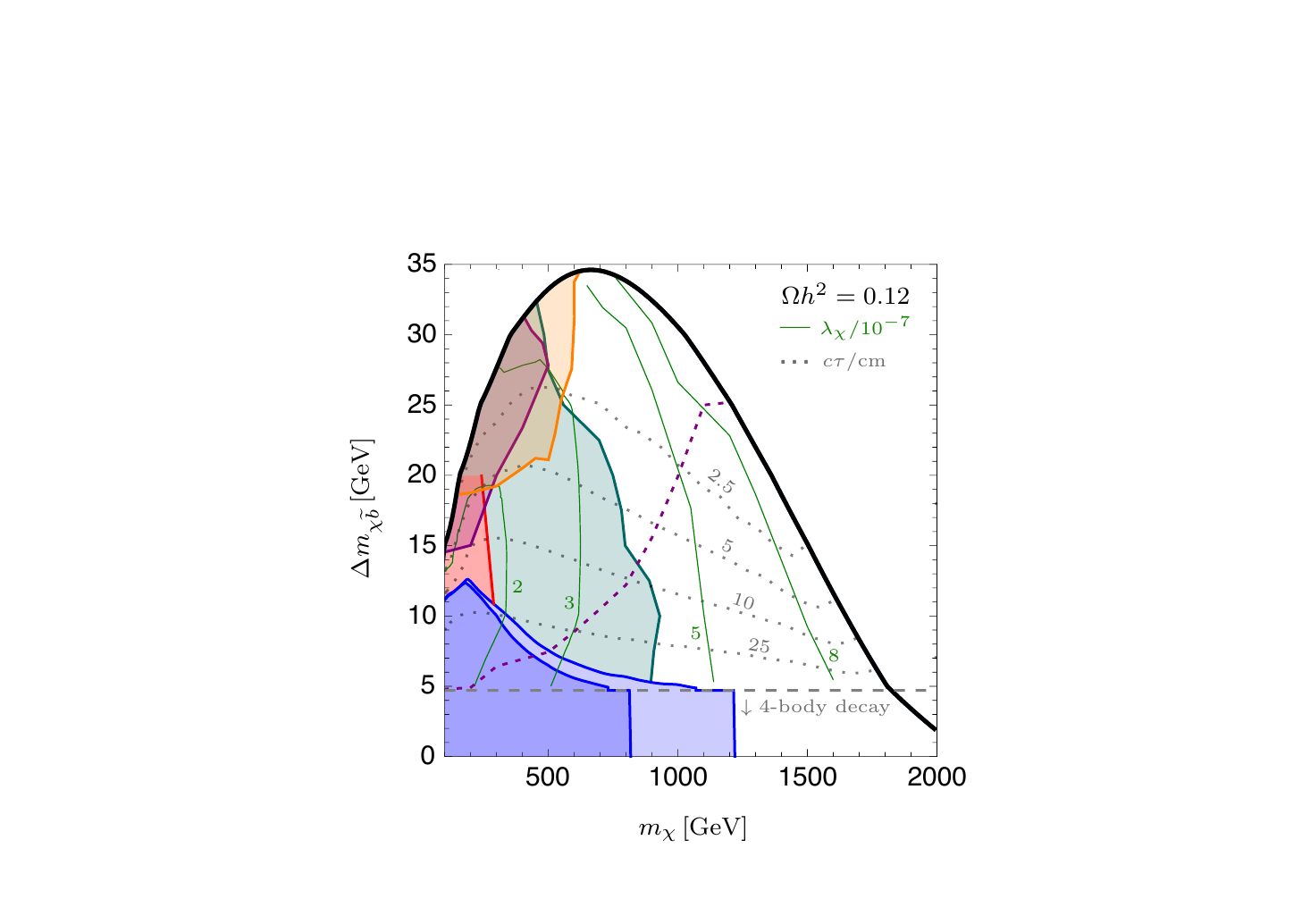}}
\end{picture}
\caption{
Allowed parameter space in the considered conversion-driven freeze-out scenario in the DM-mass vs. mass-splitting plane. The thin green and dotted grey curves denote contours of constant DM coupling and decay length, respectively (taken from~\cite{Garny:2017rxs}). We show 95\% CL exclusion limits derived from the following analyses. The dark and light blue regions are excluded from $R$-hadron searches at the 8~\cite{Chatrchyan:2013oca} and 13\,TeV LHC~\cite{CMS-PAS-EXO-16-036}, respectively, as reinterpreted in~\cite{Garny:2017rxs}. The red and orange regions are excluded by the monojet~\cite{Aaboud:2016tnv} and multijet plus missing energy~\cite{ATLAS-CONF-2019-040} analyses, respectively, while the teal and purple regions represent the limits obtained from the disappearing track~\cite{Aaboud:2017mpt} and displaced jet~\cite{Aaboud:2017iio} searches, respectively. Finally, the purple dotted curve illustrates the limit that would be obtained after dropping the invariant mass cut of the last search (see the text for details).
\label{fig:LLP_DM_param_space}
}
\end{figure}

\begin{figure}[t]
\centering
\setlength{\unitlength}{1\textwidth}
\begin{picture}(0.25,0.165)
 \put(0.0,0.00){\includegraphics[width=0.17\textwidth]{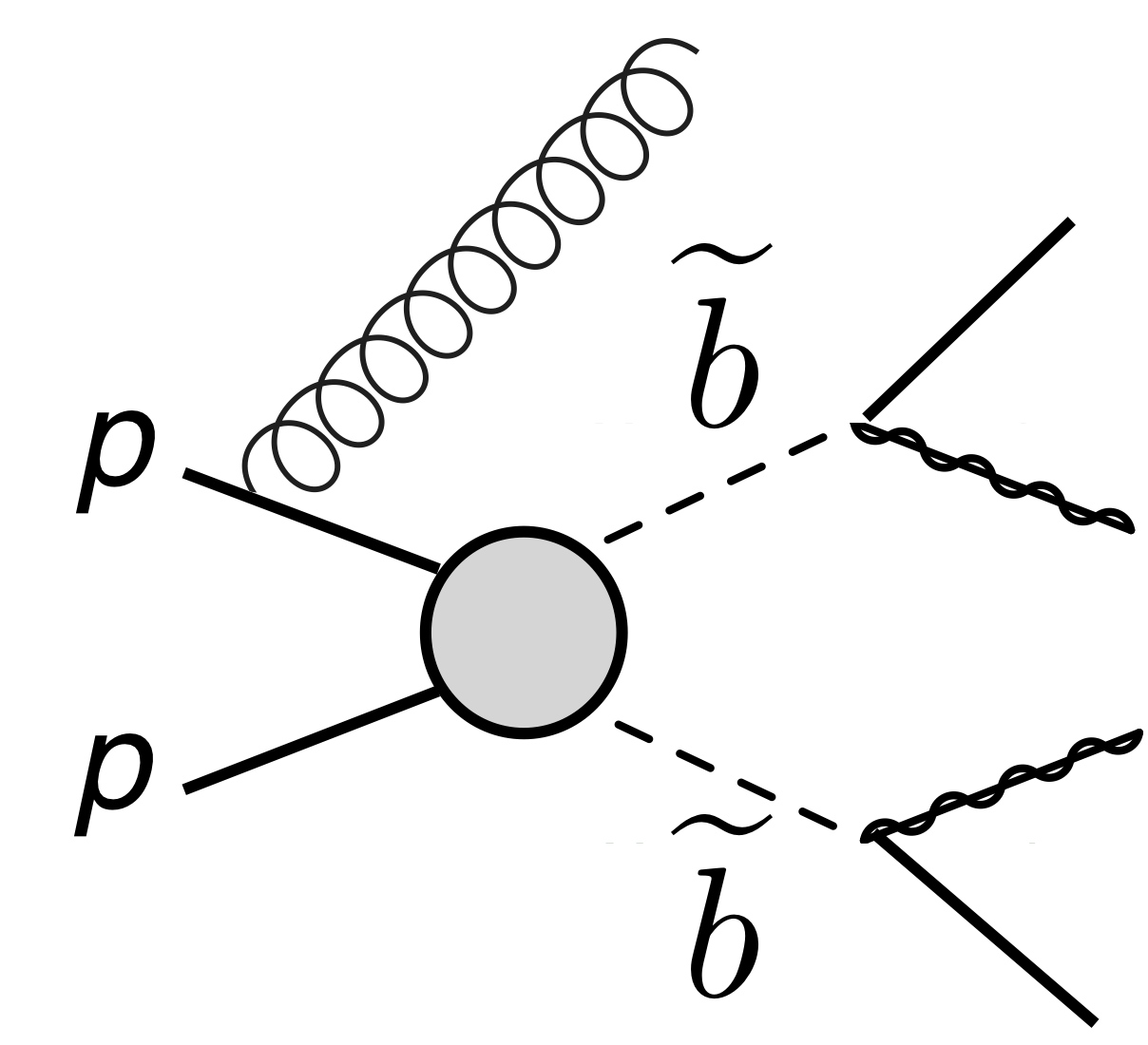}}
\end{picture}
\caption{%
Production and decay diagram at LHC (in this case with ISR).
\label{fig:LLP_DM_prod_diag}
}
\end{figure}

Among the most inclusive search strategies are missing energy searches.
In Ref.~\cite{Garny:2017rxs}, the results of an early Run~2 ATLAS monojet search for new physics in 3.2~fb$^{-1}$
of LHC data~\cite{Aaboud:2016tnv} have been reinterpreted. The red
curve in Fig.~\ref{fig:LLP_DM_param_space} shows the boundary of the correspoding 
95\% confidence level (CL) exclusion (the red shaded area being excluded), as taken from~\cite{Garny:2017rxs}.
In the experimental analysis, a signal is selected by
requiring events to feature a large amount of missing transverse energy
($E_T^{\rm miss} > 250$~GeV) and a hard central leading jet with a transverse
momentum $p_T$ larger than 250~GeV and pseudorapidity $|\eta|<2.4$. Moreover,
a charged lepton veto is applied and
the search strategy allows for a subleading jet activity provided it is well
separated from the missing energy, the number of jets with $p_T > 30$~GeV,
$|\eta|<2.8$ and with an angular separation from the missing transverse momentum
in azimuth $\Delta\phi$ of at least 0.4, being constrained to be at most 4.
The selection is then divided into several exclusive and inclusive signal regions
with different $E_T^{\rm miss}$ requirements.
Whereas not directly targetting long-lived new physics, this search is able to
constrain sbottom-neutralino configurations featuring a not too heavy neutralino
$m_\chi \lesssim 300$~GeV for moderate mass splittings of about 10--20~GeV.

Here, we additionally update those bounds by including the impact of a recent ATLAS
search for supersymmetry in a multijet plus missing energy final
state~\cite{ATLAS-CONF-2019-040}. As for the early Run~2 analysis, events
exhibiting a large amount of missing energy ($E_T^{\rm miss} > 300$~GeV), a
hard central leading jet ($p_T > 200$~GeV and $|\eta|<2.8$), and no
reconstructed electron or muon are preselected. However, in contrast, there is
no limit on the number of subleading jets of $p_T>50$~GeV, provided that the
first three jets are well separated from the missing transverse momentum in
azimuth ($\Delta\phi > 0.4$), and that the effective mass defined as the sum of
the $p_T$ of all reconstructed jets and the missing transverse energy satisfy
$m_{\rm eff} > 800$~GeV. Different signal regions are then defined depending on
the jet multiplicity and properties, the missing energy significance and the
exact value of the effective mass.  This analysis has been recast and validated 
within the{ \sc MadAnalysis}~5
framework~\cite{Conte:2012fm,Conte:2014zja,Dumont:2014tja,Conte:2018vmg}
with details of validation provided in the url \cite{ma5}, and the recast code available in 
\cite{multi}.

We reinterpret this search, within the context 
of the simplified model described in this work. To this end, we generate Monte Carlo samples
of 100000 parton-level events describing the production and decay of a pair of scalar bottom 
partners, possibly together with up to two additional QCD partons. Event generation is achieved with
 {\sc MadGraph5\_aMC@NLO}~\cite{Alwall:2014hca} and we use the MLM prescription to merge 
 events featuring different jet multiplicities at the hard-scattering level~\cite{Mangano:2006rw}. Parton showering and
 hadronization are achieved with {\sc Pythia 8.2}~\cite{Sjostrand:2014zea} and event reconstruction
 is performed using the anti-$k_T$ algorithm~\cite{Cacciari:2008gp}, as implemented in
{\sc FastJet}\cite{Cacciari:2011ma}. Detector simulation with  parameters set to match 
 the performance of the reinterpreted search has been performed with {\sc Delphes}~3~\cite{deFavereau:2013fsa}.
The results of the reinterpretation are presented
  in Fig.~\ref{fig:LLP_DM_param_space}, where the orange curve delineates 
  the constraint  on the parameter space in question, as an exclusion at 95 $\%$ using 
  the CLs prescription. The constraint is dominated by the 
  two jet bin with the lowest values of $m_{\rm eff} $ and missing transverse momentum 
  criteria. This is understandable as we predominantly rely on radiation jets to pass the analysis selection, the
new physics spectra being too compressed to lead to hard objects.
  The difference in the behaviour of
  the constrained region between the monojet analysis 
  and the multijet plus missing energy analysis can be understood as stemming from two factors. The first one
consists in the increased luminosity used in the multijet search. Secondly, the multijet search 
  prioritizes larger mass gaps, while the monojet targets more compressed regions. Overall, 
  we observe that up to $m_{\chi}\sim 500$ GeV is ruled for $\Delta m_{\tilde{\chi} b}\sim 35$
  GeV.

Those typical searches for supersymmetry through the production of a large
amount of missing energy in association with an important hadronic activity are
however unsensitive to more compressed new physics spectra. In
order to circumvert this issue, the CMS collaboration has performed a
traditional search~\cite{Sirunyan:2019xwh} for squarks and gluino using the
$M_{T2}$ kinematic variable~\cite{Lester:1999tx} and extended it by including a
signal region specifically focusing on events featuring a disappearing
track that could be interpreted as the impact of a long-lived particle. In the
standard search region, the selection vetoes the
presence of leptons and either requires the presence of one central jet ($p_T >
30$~GeV and $|\eta|<2.4$) or a low hadronic activity ($H_T < 1.2$~TeV) together
with $E_T^{\rm miss} > 250$~GeV, or of a larger hadronic activity and
$E_T^{\rm miss} > 30$~GeV. Once again, the missing transverse momentum is
imposed to be well separated from the four leading jets, the missing transverse
energy to be mainly originating from the jet activity and one finally requires
$M_{T2}>200$~GeV if $H_T < 1.5$~TeV, or $M_{T2}>400$~GeV otherwise. The $M_{T2}$
cut is lowered to 200~GeV in all cases if the event features a disappearing track.
The implementation of such an analysis in the {\sc MadAnalyss}~5
framework~\cite{Conte:2012fm,Conte:2014zja,Dumont:2014tja,Conte:2018vmg}
is ongoing.

\subsection{Heavy stable charged particles}

Towards large lifetimes, searches for heavy stable chareged particles become
sensitive (see \emph{e.g.}~\cite{Chatrchyan:2013oca,CMS-PAS-EXO-16-036,ATLAS:2014fka,Aaboud:2019trc}).
These searches make use of
observables related to large ionization energy loss and time of flight, typical
for heavy charged particles traveling at low velocity through the detector. In
this case, events are selected if they feature a well reconstructed track candidates with a large
transverse momentum and a large amount of missing
transverse momentum or a isolated muon candidate. 
In Fig.~\ref{fig:LLP_DM_param_space} we show the 95\% CL exclusion from an reinterpretation of the 8\,TeV~\cite{Chatrchyan:2013oca}
and early 13\,TeV~\cite{CMS-PAS-EXO-16-036} CMS analysis for finite lifetimes performed in~\cite{Garny:2017rxs}
as the dark and light blue curve (shaded area), respectively. The latter excludes the region of very small mass splittings 
$\Delta m_{\chi \widetilde b} \lesssim5\,$GeV up to around 1.2\,TeV. However, it quickly loses constraining power towards larger 
values of $\Delta m_{\chi \widetilde b}$.

\subsection{Disappearing tracks}

Searches for disappearing tracks by the ATLAS~\cite{Aaboud:2017mpt} and CMS~\cite{Sirunyan:2018ldc} collaborations
also have the potential to constrain the scenario considered here,
particularly for $\widetilde b$ decay lengths in the range $c \tau \in [0.1 - 1]$ m. In the following, we analyze the sensitivity of the disappearing track ATLAS analysis using 36.1~fb$^{-1}$ of 13~TeV LHC data~\cite{Aaboud:2017mpt}.
We use the electroweak production event selection from the ATLAS analysis (despite our signal being produced via strong interactions), 
since the ATLAS strong production selection requires, besides the leading-$p_T$ jet, several jets with a transverse momentum $p_T > 50$ GeV in the event. Such an option is usuaslly not featured by
our signal, due to the small mass splitting $\Delta m_{\chi\tilde{b}}$ between the new physics states.

The kinematic event selection for electroweak 
production requires the presence of at least one jet with $p_T > 140$~GeV,
a large amount of missing transvers energy ($E_T^{\rm miss} > 140$~GeV) that must be well separated in azimuth
(with $\Delta\phi > 1$) from the four
leading jets with $p_T>50$~GeV.
In addition, the selection vetoes the presence of reconstructed electron and muon, and events passing the kinematic selection so far are imposed to exhibit
{\it tracklets} (short isolated tracks reconstructed solely from hits in the pixel detector, with no requirements from the SCT and TRT detectors~\cite{Aaboud:2017mpt}), with four specific properties.
The corresponding requirements are:
\begin{enumerate}
\item {\it$p_T$ and isolation}: The $p_T$ of the candidate tracklet must be greater that 20 GeV, and the separation $\Delta R$ between the 
tracklet and any jet with $p_T > 50$~GeV must be greater than 0.4. Furthermore, the tracklet is required to be isolated (see~\cite{Aaboud:2017mpt} for details). 
\item {\it Geometrical acceptance}: The tracklet properties must satisfy $0.1 < |\eta| < 1.9$.
\item {\it Quality requirements}: The tracklet is required to have hits in all four pixel layers (located at a radial distance $R \in [33.25,\, 122.5]$\,mm 
from the interaction point). 
\item {\it Disappearance condition}: The number of SCT hits associated with the tracklet must be zero.
\end{enumerate}%
The last two requirements may be satisfied if the $\widetilde b$ decays outside the pixel detector and before the first SCT layer. Besides, the track hits on the first SCT layer from the  
jets coming out of the $\widetilde b$ decay are required to have a minimal distance ($d_{\min}$) 
from the would-be hit of the extrapolated tracklet in this layer. 
This minimal distance is such that, taken the resolution $\sigma_{\mathrm{SCT}} \simeq 17$ $\mu$m of the SCT
from~\cite{ATLAS-CONF-2014-047}, the $\chi^2$ for matching the tracks is at least 15. That is,
\begin{equation}
\label{eq:LLP_DM_dmin_DT}
d_{\min}> \sqrt{15} \,\sigma_{\mathrm{SCT}}\,.
\end{equation}
Among the candidate tracklets (if more than one) passing all the above requirements, the one with the largest $p_T$ is selected, and the analysis defines its signal region 
for tracklets with $p_T > 100$ GeV.

In order to estimate the sensitivity of this search to the present model, we generate $p p \to \tilde{b} \tilde{b} j$ (with $\tilde{b} \to \chi b$ decays) hard-scattering signal events
with {\sc MadGraph5\_aMC@NLO}, imposing that the additional hard-jet satisfies $p_T > 140$ GeV\footnote{This allows for a preliminary estimate of the 
sensitivity. A full recasting of the ATLAS analysis including multipartonic matrix element merging, parton shower and hadronization
is left for the future.\label{fn:LLP_DM_disappartonn}} and the two $\widetilde b$ states as yielding potential tracklets. We compute the fraction of events that pass the kinematic selection 
and the $\widetilde b$ tracklet selection at generator level. Furthermore, since $\widetilde b$ states form $R$-hadrons, we need to multiply the resulting efficiencies by the 
probability that the $R$-hadron ends up in a state with charge $\pm1$ (given approximately by $0.5$, which we obtain from~\cite{Aaboud:2019trc}). 
The {\it quality requirement} for the candidate tracklet is interpreted here as the $\widetilde b$ decaying outside the pixel detector volume (the ATLAS inner detector configuration 
measures are taken from Fig.~2 of~\cite{ATL-PHYS-PUB-2019-011}), and together with the tracklet {\it disappearance condition} ($\widetilde b$ decay before the first ATLAS SCT layer, 
together with its visible decay product satisfying the inequality~\eqref{eq:LLP_DM_dmin_DT}) mainly dictates the signal selection acceptance $\times$ efficiency. 
We also include in our analysis the efficiency for a generator level tracklet candidate to be identified as a tracklet at the reconstruction level, 
estimated approximately as $0.45$ from the auxiliary material from~\cite{Aaboud:2017mpt}.

The resulting excluded region is illustrated by the solid teal curve in Fig.~\ref{fig:LLP_DM_param_space}, that shows that the analysis is sensitive 
to bottom partner masses ranging up to $m_{\tilde{b}} \sim 900$\,GeV and to splittings $\Delta m_{\chi \widetilde b} \sim 10$\,GeV (corresponding to decay 
lengths $c\tau \sim 20$\,cm, in the ``sweet-spot'' of the ATLAS disappearing track search). Due to the high resolution of the ATLAS tracker,
the condition~\eqref{eq:LLP_DM_dmin_DT} on the visible decay product does not have a significant effect even though the mass splitting here ($\mathcal{O}(10)~\text{GeV}$) is much larger than in the long-lived chargino model ($\mathcal{O}(100)~\text{MeV}$)
targeted by the analysis.
We expect this to also hold in a refined analysis performed at hadron level which is, however, left for future work.  

\subsection{Displaced jets}

For decay lengths between a few millimeters and a few centimeters, the long-lived particle dominantly decays within the inner tracker, resulting in displaced $b$-jets and missing energy. 
A few ATLAS and CMS searches targeting this topology exist~\cite{Aaboud:2017iio,Aaboud:2019opc,Sirunyan:2018vlw}.

We consider here the ATLAS search for displaced vertices and missing energy using 32.8~fb$^{-1}$ of 13~TeV LHC data~\cite{Aaboud:2017iio}. This search 
is dedicated to final states exhibiting a large amount of missing energy and vertices containing at least five charged tracks displaced from the primary vertex by a distance between 4~mm and 30~cm.
The benchmark scenario considered by the ATLAS analysis is a long-lived gluino simplified model, where the gluino decays into two jets and a neutralino. In this scenario, gluino masses 
of 1.5--2.5~TeV are excluded, the precise exclusion depending on the neutralino mass and gluino lifetime. The signal region considered by ATLAS requires the presence of a large amount of missing transverse energy
($E_T^{\rm miss} > 250$~GeV), either one jet with $p_T > 70$~GeV or two jets with $p_T > 25$~GeV, and
one or more displaced vertices containing at least 5 charged tracks and of invariant mass ($m_{\rm DV}$) larger than 10~GeV.

\begin{figure}
	\centering
	\includegraphics[width=0.45\textwidth]{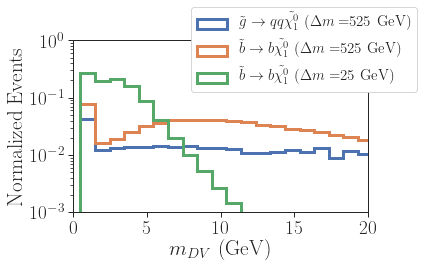}
	\caption{The invariant mass distribution for candidate displaced vertices. The blue histogram shows the distribution for a gluino simplified model with a large mass gap: $(m_{\tilde{g}},m_{\tilde{\chi}_1^0}) = (625 \mbox{ GeV}, 100 \mbox{ GeV})$. The orange histogram shows the distribution for the model considered here with masses $(m_{\tilde{b}},m_{\chi}) = (625 \mbox{ GeV}, 100 \mbox{ GeV})$, while the green histogram shows the same distribution but for the compressed scenario: $(m_{\tilde{b}},m_{\chi}) = (625 \mbox{ GeV}, 600 \mbox{ GeV})$.
		\label{fig:mDV}
	}
\end{figure}

Although the model considered here can produce events with large missing energy and sufficiently hard jets, the small mass gap between the long-lived parent and its daughter results in displaced vertices with invariant masses typically much smaller than 10\,GeV. In Fig.~\ref{fig:mDV}, we compare the $m_{\rm DV}$ distribution for the gluino simplified model considered by ATLAS with masses $m_{\tilde g} = 625$~GeV and $m_{\tilde\chi_1^0} = 100$~GeV, and the model considered here with and without a large mass gap. For cases with a large mass gap, a sizeable fraction of the events contains vertices with large invariant masses. However, in the compressed case almost all events fail the $m_{\rm DV} > 10$~GeV requirement.

In order to estimate the sensitivity of the search, we recast it by making use of the trigger and DV reconstruction efficiencies provided by ATLAS in the auxiliary material of Ref.~\cite{Aaboud:2017iio}.
We then use {\sc MadGraph5\_aMC@NLO} and {\sc Pythia 8.2} to generate hadron-level events and compute the signal efficiency for the model described in Sec.~\ref{sec:LLP_DM_themodel}. Events are normalized to cross sections matching next-to-leading order calculations with the resummation of the next-to-next-to-leading threshold logarithms, as obtained from {\sc NNLL-fast}~\cite{Beenakker:2016lwe,Beenakker:2010nq}.

As expected, most of the events fail the $m_{\rm DV}$ cut, suppressing the signal yield. The resulting 95\% CL exclusion is illustrated by the solid purple curve in Fig.~\ref{fig:LLP_DM_param_space} that shows that only points with very large cross sections (small $\widetilde b$ masses) and a mass gap larger than 15\,GeV are excluded.
Since the main loss in sensitivity is due to the invariant mass requirement for the displaced vertices, we try to estimate what could be the reach resulting from relaxing this cut. In order to achieve this, we assume that the SM background remains unchanged and the DV reconstruction efficiency for vertices with $m_{\rm DV} < 10$\,GeV is the same as the one for $m_{\rm DV} = 15$\,GeV. Although these certainly are optimistic assumptions, it allows us to use the efficiencies provided by the ATLAS collaboration when smaller mass cuts are used.
The result is shown by the purple dashed line in Fig.~\ref{fig:LLP_DM_param_space}, the excluded region being now significantly enhanced, extending up to 1\,TeV bottom partner masses for small lifetimes (large mass splittings within the considered scenario).

Once again we stress that this is an optimistic and probably unrealistic projection. Nevertheless, it illustrates the impact of the invariant mass cut on the sensitivity to models with small mass gaps and reveals the potential gain of relaxing this cut. To achieve this, the background might be reduced by other means, \emph{e.g.}~by requiring a larger displacement. In fact, Fig.~\ref{fig:LLP_DM_param_space} shows a significant region where the displaced jets without a $m_{\rm DV}$ cut would outperform the disappearing track search (\emph{e.g.}~for $c\tau>2.5\,$cm).

\subsection{Delayed jets}

Another option for distinguishing the long lifetime of some particles is to measure the timing information of their decay products, and search for delays with respect to the collision time.  This method was exploited in a recent CMS analysis~\cite{Sirunyan:2019gut}, where timing capabilities of the CMS electromagnetic calorimeter (ECAL) were used to identify non-prompt or ``delayed'' jets. The analysis is sensitive to long-lived particles decaying within the ECAL barrel volume extending up to 1.79 m and covering $|\eta| < 1.48$.  The analysis uses only calorimetric information to reconstruct jets and imposes a set of quality criteria on the ECAL cells and energy fractions.  Jet timing is calculated from the median of the times of ECAL cells associated with the jet, and is required to be at least 3~ns to discriminate from detector, beam, pileup and cosmic ray backgrounds.  Events are imposed to have at least one delayed central jet with $p_T > 30$~GeV and $|\eta| < 1.48$.  Additionally, one imposes that
$E_T^{\rm miss} > 300$~GeV to eliminate SM multijet backgrounds and other beam-related backgrounds.

The analysis was interpreted using a gauge-mediated supersymmetry breaking model, where a 1--3\,TeV long-lived gluino decays into a gluon and a 10 GeV\,gravitino. The large gluon-gravitino mass difference leads to high-$p_T$ jets and a high signal efficiency.  In the DM models considered here, the small $\widetilde{b}\,\,\!$-$\chi$ mass difference typically results in a much softer jet $p_T$ spectrum.  However, a preliminary study of events generated with {\sc MadGraph5\_aMC@NLO} (hard-scattering), {\sc Pythia 8.2} (parton showering and hadronization) and {\sc Delphes 3} (detector simulation) leads to a sizable fraction of events passing the jet $p_T$ and $E_T^{\rm miss}$ selection, especially for points with a relatively high mass difference. The larger cross sections for lower masses would further improve the signal significance in this very low background analysis.  Where the simulation of delayed jet selection criteria is non-trivial in existing public tools, delayed jet selection efficiencies depending on the jet transverse momentum and pseudorapidity would be a fundamental requirement for obtaining reliable estimates of the sensitivity of this analysis.

\section{Conclusions} \label{sec:LLP_DM_concl}

In this contribution, we considered a conversion-driven feeze-out DM scenario predicting 
long-lived particles with decay length $\mathcal{O}(1\,\text{mm}\!-\!1\,\text{m})$ at the LHC\@. 
The scenario is characterized by small DM couplings and
relatively small mass splittings ($\mathcal{O}(10)\,\text{GeV}$) in the dark sector, \emph{i.e.}~between the parent particle pair-produced at the LHC and the DM particle it decays into.
We considered searches for heavy stable charged particles ($R$-hadrons), disappearing tracks, displaced (and delayed) jets as well as prompt missing energy searches covering the whole range of possible decay lengths. The strongest constraint can be reached towards the upper edge of the occurring decay lengths within the model, $c\tau\sim1\,$m, by searches for heavy stable charged particles, reaching DM masses around 1.2\,TeV. For intermediate lifetimes, searches for disappearing tracks are most sensitive and constrain DM masses up to $\sim 900$\,GeV for $\Delta m_{\chi \widetilde b}\sim 10~$GeV (which corresponds to a decay length of around 20\,cm). Towards smaller lifetimes the search loses sensitivity. Finally, for $c\tau\lesssim2.5\,$cm ($\Delta m_{\chi \widetilde b}\gtrsim 25\,$GeV), the considered multijet plus missing energy search the most sensitive, constraining DM masses ranging up to around 600\,GeV.

The search for displaced jets is only sensitive to a very small region of the parameter space, with masses below $m_\chi=500\,$GeV -- a region being already excluded by the disappearing tracks and missing energy search. This is mainly due to the lower invariant mass cut on the tracks arising from the displaced vertex which is affordable in the scenario considered in the experimental analysis with a large mass gap, but that rejects most of the signal for the considered model with small mass splittings. Relaxing this cut potentially has a very large effect and would render the displaced jet search to be the most sensitive for $\Delta m_{\chi \widetilde b}\gtrsim 15\,$GeV ($c\tau\lesssim 10\,$cm).
Determining whether this is possible (above a certain minimal displacement) requires further studies considering the relevant backgrounds.

\section*{Acknowledgements}
We thank G.~B\'elanger, A.~Bharucha, B.~Bhattacherjee, N.~Desai, A.~Goudelis, P.~Pani and H.-S.~Shao
for fruitful discussions.
J.H.~acknowledges support from the F.R.S.-FNRS, of which he is a postdoctoral researcher.
The work of A.L.~was supported by the S\~ao Paulo Research Foundation (FAPESP), under the project
2015/20570-1\@. 
J.M.N.~acknowledges support from the Ram\'on y Cajal Fellowship contract RYC-2017-22986, from the 
Spanish MINECO's ``Centro de Excelencia Severo Ochoa'' programme under grant SEV-2016-0597, from the 
European Union's Horizon 2020 research and innovation programme under the Marie Sklodowska-Curie grant agreements 690575 (RISE InvisiblesPlus) and 
674896 (ITN ELUSIVES) as well as from the Spanish Proyectos de I$+$D de Generaci\'on de Conocimiento via the grant PGC2018-096646-A-I00.



%% file: gwc/GWC.main.tex
\graphicspath{{gwc/}}
\newcommand{\Contur}{C\protect\scalebox{0.8}{ONTUR}\xspace}
\newcommand{\Herwig}{H\protect\scalebox{0.8}{ERWIG}\xspace}
\newcommand{\Pythia}{P\protect\scalebox{0.8}{YTHIA}\xspace}
\newcommand{\Sherpa}{S\protect\scalebox{0.8}{HERPA}\xspace}
\newcommand{\Rivet}{R\protect\scalebox{0.8}{IVET}\xspace}
\newcommand{\Professor}{P\protect\scalebox{0.8}{ROFESSOR}\xspace}
\newcommand{\eps}{\varepsilon}
\newcommand{\mc}[1]{\mathcal{#1}}
\newcommand{\mr}[1]{\mathrm{#1}}
\newcommand{\mb}[1]{\mathbb{#1}}
\newcommand{\tm}[1]{\scalebox{0.95}{$#1$}}
%
\newcommand\simge{\mathrel{%
   \rlap{\raise 0.511ex \hbox{$>$}}{\lower 0.511ex \hbox{$\sim$}}}}
\newcommand\simle{\mathrel{
   \rlap{\raise 0.511ex \hbox{$<$}}{\lower 0.511ex \hbox{$\sim$}}}}

\newcommand{\slashchar}[1]%
        {\kern .25em\raise.18ex\hbox{$/$}\kern-.70em #1}
\def\lsim{\mathrel{\raise.3ex\hbox{$<$\kern-.75em\lower1ex\hbox{$\sim$}}}}
\def\gsim{\mathrel{\raise.3ex\hbox{$>$\kern-.75em\lower1ex\hbox{$\sim$}}}}
\newcommand\CA{{\cal A}}\newcommand\CCA{$\CA$}
\newcommand\CB{{\cal B}}\newcommand\CCB{$\CB$}
\newcommand\CC{{\cal C}}\newcommand\CCC{$\CC$}
\newcommand\CD{{\cal D}}\newcommand\CCD{$\CD$}
\newcommand\CE{{\cal E}}\newcommand\CCE{$\CE$}
\newcommand\CF{{\cal F}}\newcommand\CCF{$\CF$}
\newcommand\CG{{\cal G}}\newcommand\CCG{$\CG$}
\newcommand\CH{{\cal H}}\newcommand\CCH{$\CH$}
\newcommand\CI{{\cal I}}\newcommand\CCI{$\CI$}
\newcommand\CJ{{\cal J}}\newcommand\CCJ{$\CJ$}
\newcommand\CK{{\cal K}}\newcommand\CCK{$\CK$}
\newcommand\CL{{\cal L}}\newcommand\CCL{$\CL$}
\newcommand\CM{{\cal M}}\newcommand\CCM{$\CM$}
\newcommand\CN{{\cal N}}\newcommand\CCN{$\CN$}
\newcommand\CO{{\cal O}}\newcommand\CCO{$\CO$}
\newcommand\CP{{\cal P}}\newcommand\CCP{$\CP$}
\newcommand\CQ{{\cal Q}}\newcommand\CCQ{$\CQ$}
\newcommand\CR{{\cal R}}\newcommand\CCR{$\CR$}
\newcommand\CS{{\cal S}}\newcommand\CCS{$\CS$}
\newcommand\CT{{\cal T}}\newcommand\CCT{$\CT$}
\newcommand\CU{{\cal U}}\newcommand\CCU{$\CU$}
\newcommand\CV{{\cal V}}\newcommand\CCV{$\CV$}
\newcommand\CW{{\cal W}}\newcommand\CCW{$\CW$}
\newcommand\CX{{\cal X}}\newcommand\CCX{$\CX$}
\newcommand\CY{{\cal Y}}\newcommand\CCY{$\CY$}
\newcommand\CZ{{\cal Z}}\newcommand\CCZ{$\CZ$}
\newcommand\ub{\underbar}
\newcommand\ul{\underline}
\newcommand\be{\begin{equation}}
\newcommand\ee{\end{equation}}
\newcommand\bea{\begin{eqnarray}}
\newcommand\eea{\end{eqnarray}}
\newcommand\ba{\begin{array}}
\newcommand\ea{\end{array}}
\newcommand\nn{\nonumber}
\newcommand\tx{\textstyle}
\newcommand{\half}{\ensuremath{\frac{1}{2}}}
\newcommand{\third}{\ensuremath{\frac{1}{3}}}
\newcommand{\fourth}{\ensuremath{\frac{1}{4}}}
\newcommand{\fifth}{\ensuremath{\frac{1}{5}}}
\newcommand{\thalf}{\textstyle{\frac{1}{2}}}
\newcommand{\tthalf}{\textstyle{\frac{3}{2}}}
\newcommand{\fsixths}{\textstyle{\frac{5}{6}}}
\newcommand{\tthird}{\textstyle{\frac{1}{3}}}
\newcommand{\tfourth}{\textstyle{\frac{1}{4}}}
\newcommand{\teighth}{\textstyle{\frac{1}{8}}}
\newcommand{\tfifth}{\textstyle{\frac{1}{5}}}
\newcommand{\thhalf}{\ensuremath{\frac{3}{2}}}
\newcommand{\fourthirds}{\ensuremath{\frac{4}{3}}}
\newcommand\dagg{\dagger}
\newcommand\ts{\thinspace}
\newcommand\ra{\rightarrow}
\newcommand\Ra{\Rightarrow}
\newcommand\Lra{\Longrightarrow}
\newcommand\longra{\longrightarrow}
\newcommand\leftra{\leftrightarrow}
\newcommand\llra{\longleftrightarrow}
\newcommand\olra{\overleftrightarrow}
\newcommand\mev{{\rm MeV}}
\newcommand\gev{{\rm GeV}}
\newcommand\tev{{\rm TeV}}
\newcommand\MeV{{\rm MeV}}
\newcommand\GeV{{\rm GeV}}
\newcommand\TeV{{\rm TeV}}
\newcommand\nb{{\rm nb}}
\newcommand\pb{{\rm pb}}
\newcommand\fb{{\rm fb}}
\newcommand\ifb{{\rm fb}^{-1}}
\newcommand\ecm{\sqrt{s}}
\newcommand\rshat{\sqrt{\shat}}
\newcommand\shat{\hat s}
\newcommand\nin{\noindent}
\newcommand\et{E_T}
\newcommand\etmiss{\slashchar{E}_T}
\newcommand\emiss{\slashchar{E}}
\newcommand\jet{{\rm jet}}
\newcommand\jets{{\rm jets}}
\newcommand\cth{c_{\theta}}
\newcommand\sth{s_{\theta}}
\newcommand\cthst{c_{\theta^*}}
\newcommand\sthst{s_{\theta^*}}
\newcommand\cphst{c_{\phi^*}}
\newcommand\sphst{s_{\phi^*}}
\newcommand\bth{b_{\theta}}
\newcommand\bthst{b_{\theta^*}}
\newcommand\bphst{b_{\phi^*}}
\newcommand\cbeta{c_\beta}
\newcommand\sbeta{s_\beta}
\newcommand\cbetap{c_{\beta'}}
\newcommand\sbetap{s_{\beta'}}
\newcommand\cdelta{c_\delta}
\newcommand\sdelta{s_\delta}
\newcommand\cgamma{c_\gamma}
\newcommand\sgamma{s_\gamma}
\newcommand\cTh{c_\Theta}
\newcommand\sTh{s_\Theta}
\newcommand\cpsi{c_\psi}
\newcommand\spsi{s_\psi}
\hyphenation{Goldstone}
\hyphenation{longi-tudinal}
\hyphenation{coup-ling}
\hyphenation{coup-lings}
\hyphenation{ATLAS}

\chapter{Confronting Gildener-Weinberg Higgs bosons with LHC measurements}
{\it J.~M.~Butterworth, K.~Lane, D.~Sperka}



\label{sec:GWC}

\begin{abstract}
  Gildener-Weinberg (GW) models of electroweak symmetry breaking solve two
  fundamental problems of the Standard Model: (1)~What {\em{\ul{natural}}}
  mechanism makes the 125~GeV Higgs boson $H$ light? (2)~Given that most
  proposed answers to~(1) require multiple Higgs multiplets, generally
  doublets, what can {\em{\ul{naturally}}} explain the apparently SM
  couplings of $H$ to gauge bosons and fermions~\cite{Sirunyan:2018koj,
    Aad:2019mbh}? The GW mechanism of approximate scale invariance answers
  both questions, but at a price. The masses of extra Higgs bosons $\CH$ in
  these models satisfy a sum rule:
  $(\sum_{\CH} M_{\CH}^4)^{1/4} = 540\,\gev$. This is a powerful constraint
  which we probe with LHC data in this report. 
\end{abstract}

\section{Overview of the Gildener-Weinberg 2HDM}
\label{sec:GWC:section1}


In 1976, E.~Gildener and S.~Weinberg (GW)~\cite{Gildener:1976ih} proposed a
scheme, based on broken scale symmetry, to generate a light Higgs boson $H$
in multi-scalar models of electroweak symmetry breaking. What GW did not
appreciate then, for there was no reason for them to, was that that $H$ was
also aligned~\cite{Gunion:2002zf}. That is, of all the scalars, its couplings
to gauge bosons and fermions were exactly those of the single Higgs boson of
the Standard Model (SM)~\cite{Weinberg:1967tq}. Like the Higgs boson's mass,
its alignment is protected by the approximate scale
symmetry~\cite{Lane:2018ycs}.

Generalizing the work of S.~Coleman and E.~Weinberg~\cite{Coleman:1973jx}, GW
assumed a multi-Higgs multiplet potential $V_0$ with only quartic interaction
terms. Assuming that all gauge boson and fermion masses arise from their
couplings to Higgs bosons, such a theory is scale-invariant at the classical
level. This $V_0$ has a trivial minimum, the one at which all scalar fields
$\Phi = 0$ so that $(V_0)_{\rm min} = 0$. This Lagrangian also has a
nontrivial extremum. It occurs along a ray in scalar-field space and it is a
{\em flat minimum} if the quartic couplings satisfy certain positivity
conditions. This flat minimum corresponds to spontaneously broken scale
symmetry, so there is a massless Goldstone boson, a ``dilaton'', and it is
$H$. Higgs alignment is then a simple consequence of the linear combination
of fields composing $H$ having the {\em {\ul{same form}}} as the Goldstone
bosons $w^\pm$ and $z$ that become the longitudinal components of the $W^\pm$
and $Z$ bosons.

This classically-exact scale symmetry is {\ul{explicitly}} broken by the
first-order term $V_1$ in the Coleman-Weinberg loop expansion of the
effective scalar potential~\cite{Coleman:1973jx}: $V_0 + V_1$ has a deeper
minimum than the trivial one at zero fields. It occurs at a specific vacuum
expectation value $\langle H\rangle = v$, an explicit breaking of scale
invariance. Then $M_H$ and all other masses in the theory are proportional
to~$v$.

In 2012, Lee and Pilaftsis~\cite{Lee:2012jn} proposed a simple model of the
GW mechanism using the two Higgs doublets
\be\label{eq:GWC:Phii}
\Phi_i = \frac{1}{\sqrt{2}}\left(\ba{c}\sqrt{2} \phi_i^+ \\ \rho_i + i
  a_i \ea\right), \quad i = 1,2.
\ee
Here, $\rho_i$ and $a_i$ are neutral CP-even and odd fields,
respectively. The potential $V_0$ is
\bea\label{eq:GWC:Vzero}
V_0(\Phi_1,\Phi_2) &=&\lambda_1 (\Phi_1^\dagg \Phi_1)^2 +
\lambda_2 (\Phi_2^\dagg \Phi_2)^2 +
\lambda_3(\Phi_1^\dagg \Phi_1)(\Phi_2^\dagg \Phi_2)\nn \\
&+& \lambda_4(\Phi_1^\dagg \Phi_2)(\Phi_2^\dagg \Phi_1)+
\thalf\lambda_5\left((\Phi_1^\dagg \Phi_2)^2 + (\Phi_2^\dagg
  \Phi_1)^2\right).
\eea
All five quartic couplings are real so that $V_0$ is CP-invariant.

In 2018, Lane and Shepherd studied this model and, {\em inter alia}, updated
its consistency with LHC data of the previous six
years~\cite{Lane:2018ycs}. For this, the following $\CZ_2$ symmetry was
imposed to prevent tree-level flavor-changing interactions among fermions,
$\psi$, induced by neutral scalar exchange~\cite{Glashow:1976nt}:
\be\label{eq:GWC:Z2}
\Phi_1 \to -\Phi_1,\,\, \Phi_2 \to \Phi_2, \quad
\psi_L \to -\psi_L,\,\, \psi_{uR} \to \psi_{uR},\,\,
\psi_{dR} \to \psi_{dR}.
\ee
This is the usual type-I 2HDM~\cite{Branco:2011iw}, but with $\Phi_1$ and
$\Phi_2$ interchanged. In type-I, all Yukawa couplings of the new, heavier
Higgs bosons are proportional to $\tan\beta$ in the alignment limit. It was
chosen to remain consistent with limits from CMS~\cite{Khachatryan:2015qxa}
and ATLAS~\cite{Aaboud:2018cwk} on charged Higgs decay into $t\bar b$; also
see Ref.~\cite{Sirunyan:2019arl}.The limits from these papers are consistent
with $\tan\beta \simle 0.5$ for $M_{H^\pm} \simeq 200$--$500\,\gev$. Type-I
couplings also suppress gluon fusion of $A,H'$ by $\tan^2\beta$, where $A$
and $H'$ are the massive CP-odd and CP-even Higgs of the 2HDM. We refer to
this version of the model as GW-2HDM.\footnote{{\bf Experimentalists beware!}
  Because of the $\Phi_1$--$\Phi_2$ interchange relative to the type-I
  definition in Ref.~\cite{Branco:2011iw}, experimental limits on $\tan\beta$
  for the usual type-I are limits on $\cot\beta$ for the GW-2HDM.}

The flat minimum of the potential $V_0$ occurs along the ray
\be\label{eq:GWC:theray}
\Phi_{1\beta} = \frac{1}{\sqrt{2}} \left(\ba{c} 0\\ \phi\,\cos\beta
  \ea\right),\quad
\Phi_{2\beta} = \frac{1}{\sqrt{2}} \left(\ba{c} 0\\ \phi\,\sin\beta \ea\right).
\ee
Here $\phi > 0$ is any real mass scale and $0 < \beta < \pi/2$. The massless
CP-even dilaton $H$ and its massive orthogonal combination $H'$ are
\bea\label{eq:GWC:Hcombo}
H &=& \rho_1\cos\beta + \rho_2\sin\beta,\nn\\
H' &=& -\rho_1\sin\beta + \rho_2\cos\beta.
\eea
This $H$ is the {\ul{same}} linear combination of $\rho_1$ and $\rho_2$ as
the electroweak Goldstone bosons $z = a_1\cos\beta + a_2\sin\beta$ and
$w^\pm = \phi_1^\pm \cos\beta + \phi_2^\pm \sin\beta$ that are eaten by $Z^0$
and $W^\pm$. {\ul{This}} is the origin of $H$ being perfectly aligned.

The one-loop potential, $V_1$, explicitly breaks scale symmetry but, of
course, not electroweak symmetry. This picks out a particular value
$v = \langle H \rangle$ of~$\phi$ at which $V_0 + V_1$ has a deeper minimum
than the trivial one for $V_0$, and it mixes $H$ and $H'$ to become the mass
eigenstates
\bea\label{eq:GWC:H1H2}
H_1 &=& \cos\delta H - \sin\delta H' = \cos\beta'\rho_1 + \sin\beta'\rho_2,\nn\\
H_2 &=& \sin\delta H + \cos\delta H' = -\sin\beta'\rho_1 + \cos\beta'\rho_2,
\eea
where $\beta' = \beta -\delta$.\footnote{Because $H^\pm$ and $A$ must remain
  orthogonal to~the combinations defining $w^\pm$ and~$z$, their masses are
  unchanged by $V_1$.} The angle~$\delta$ measures the departure of the Higgs
boson $H_1$ from perfect alignment. Being one-loop corrections, $\delta$ and
$\delta/\beta$ are only a few percent and, so, $H_1 \cong H$ is very nearly
an aligned Higgs boson~\cite{Lane:2018ycs}. The decays
$A,H_2 \to W^+W^-,\, ZZ,\, ZH_1$ and $H^\pm \to W^\pm Z, \, W^\pm H_1$ are
highly suppressed by $\sin^2\delta$; observing them from a new, heavier
{\ul{spinless}} boson would be significant, if not fatal, blows to GW
models. As noted earlier, all couplings of $H^\pm$, $A$ and $H_2$ are
proportional to $\tan\beta$ up to a correction of
$\CO(\delta^2)$~\cite{Lane:2018ycs}. From now on, we refer interchangeably to
the 125~GeV Higgs boson as $H_1$ or $H$, as clarity requires.

The nonzero one-loop mass of {\ul{the}} Higgs boson~$H$~is
given by~\cite{Gildener:1976ih},\cite{Lee:2012jn},\cite{Lane:2018ycs}
\be\label{eq:GWC:MHsq}
M_H^2 = \frac{1}{8\pi^2 v^2}\left(6M_W^4 + 3M_Z^4
  + M_{H'}^4 + M_A^4 + 2M_{H^\pm}^4 - 12m_t^4\right).
\ee
In accord with first-order perturbation theory, all the masses on the right
side of this formula are obtained from zeroth-order perturbation theory,
i.e., from $V_0$ plus gauge and Yukawa interactions; see
Refs.~\cite{Gildener:1976ih,Lane:2018ycs}. As $v$ is the only mass scale in
the theory, all masses in Eq.~(\ref{eq:GWC:MHsq}) are proportional to
$v$; e.g, $M_W = \thalf g v = M_Z/\cos\theta_W$, so that $v = 246\,\gev$ and
$\tan\beta = v_2/v_1$.

Eq.~(\ref{eq:GWC:MHsq}) implies a remarkable sum rule for the masses of
the new Higgs bosons in the GW-2HDM~\cite{Lee:2012jn,Hashino:2015nxa,
  Lane:2018ycs}:
\be\label{eq:GWC:MHsum}
\left(M_{H'}^4 + M_A^4 + 2M_{H^\pm}^4\right)^{1/4} = 540\,\gev.
\ee
With appropriate modification of the left side to allow for different Higgs
sectors, this sum rule holds in {\em{\ul{any}}} GW model of electroweak
breaking in which the only weak bosons are $W$ and $Z$ and the only heavy
fermion is the top quark. The second-order loop correction will modify the
right side, but we would be surprised if it were more than
$\CO(100\,\gev)$. Thus, in GW models, new Higgs bosons should be found at
surprisingly low masses. Furthermore, the larger the Higgs sector in a GW
model, the lighter will be the masses of at least some of the new Higgs
bosons.

\begin{figure}[h!]
\includegraphics[width=0.9\textwidth]{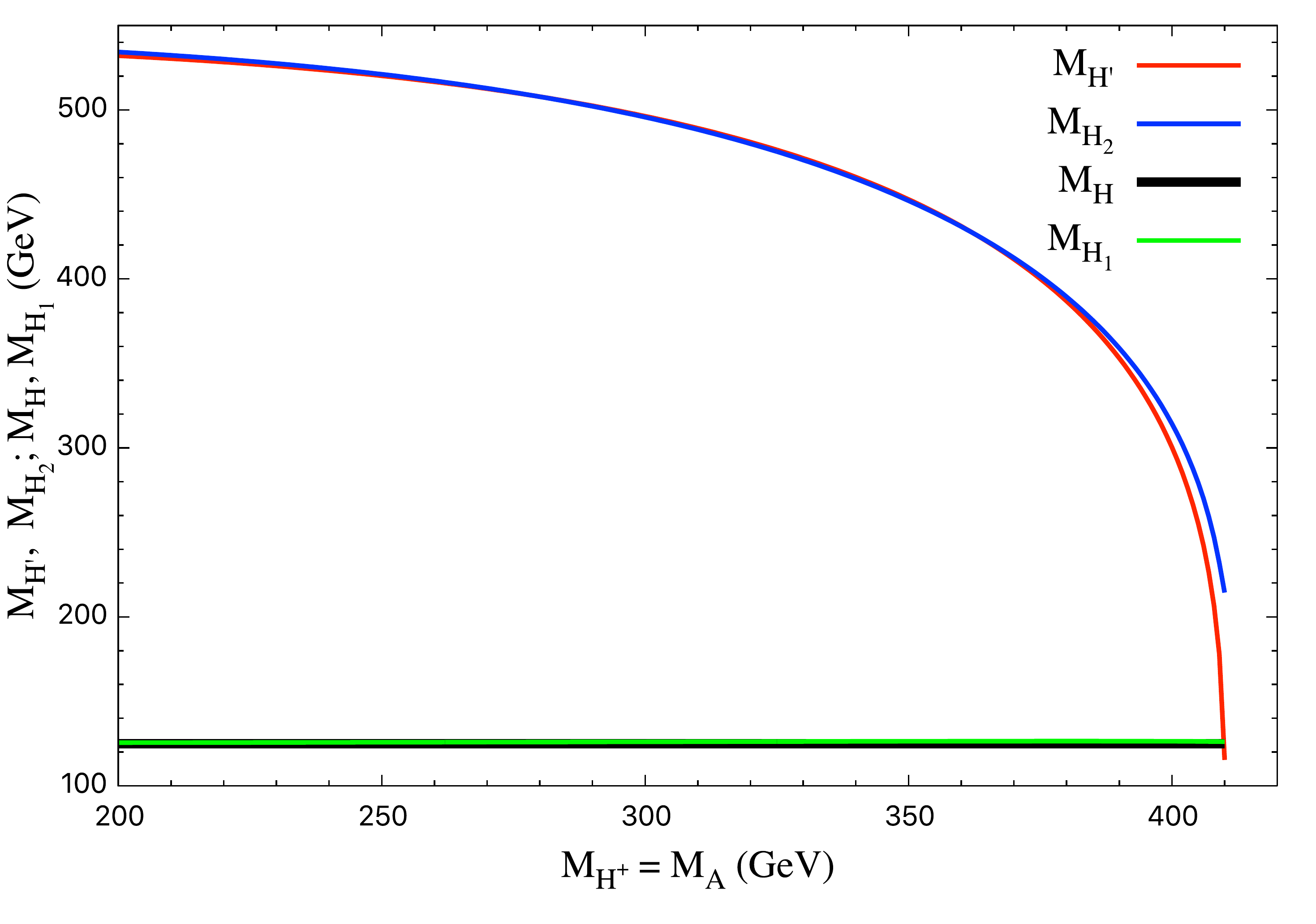}
\caption{The CP-even masses in tree ($M_H$ and $M_{H'}$) and 1-loop
  ($M_{H_1}$ and $M_{H_2}$) approximations, calculated as a function of
  $M_{H^\pm} = M_A$ and using the sum rule~(\ref{eq:GWC:MHsum}).}
\label{fig:GWC:CPeven_all}
\end{figure}

Beyond this observation, Eq.~(\ref{eq:GWC:MHsum}) has a significant
impact on searching for the new GW
Higgses. Figure~\ref{fig:GWC:CPeven_all} shows the CP-even masses $M_H$
(from Eq.~(\ref{eq:GWC:MHsq}), with initial input Higgs mass
$M_H = 125\,\gev$), $M_{H'}$ (from Eq.~(\ref{eq:GWC:MHsum})), and the
corresponding eigenvalues $M_{H_1}$ and $M_{H_2}$ of the CP-even mass matrix
at one-loop order. They are calculated as a function of
$M_{H^\pm} = M_A = 200$--$410\,\gev$. (The constraint $M_{H^\pm} = M_A$ is
important; it makes the contribution to the $T$-parameter from the new
scalars vanish identically~\cite{Battye:2011jj, Pilaftsis:2011ed}.) Three
observations:

\begin{itemize}

\item 1) $M_H$ and $M_{H_1}$ are very close along the entire range of
$M_{H^\pm}$; their ratio is 1.004--1.012. This is not surprising since
Eq.~(\ref{eq:GWC:MHsq}) and diagonalizing the CP-even mass matrix should
be equivalent.

\item 2) $M_{H'} \to 0$ at $M_{H^\pm} = 410.22\,\gev$, the boundary of the sum
rule for $M_{H^\pm} = M_A$. This is both unphysical and in conflict with LHC
(but not LEP) limits on light Higgses decaying to $b\bar b$ and
$\gamma\gamma$.

\item 3) $M_{H'} \cong M_{H_2}$ up to $M_{H^\pm} \simeq 370\,\gev$; beyond
  that $M_{H'}$ starts to dive rapidly to zero while $M_{H_2} \to 214\,\gev$
  at $M_{H^\pm} = 410\,\gev$.

\end{itemize}

To remedy this, we propose using $M_{H_2}$ over the entire range of
$M_{H^\pm} = M_A$ for estimates of production cross sections and decay
branching ratios of the GW Higgs bosons. We recommend this approach for
searches by ATLAS and CMS. For example, in a search involving all three GW
Higgs bosons (say, $pp \to A \to Z H_2$ and $pp \to H^\pm \to W^\pm H_2$,
with $H_2 \to b \bar b$), one could use ellipsoidal search regions in
($M_{H^\pm},M_A, M_{H_2}$)-space, roughly consistent with $M_{H^\pm} = M_A$
and the sum rule, and then calculate the model's predicted
$\sigma\cdot {\rm BR}$'s accordingly.

The sum rule~(\ref{eq:GWC:MHsum}) makes searches for the new Higgs bosons
lying below about $500\,\gev$ the most promising tests of GW models. In the
GW-2HDM, the new scalars are just $H^\pm$, $A$ and $H_2$. Their main
production cross sections at the LHC are $gg \to A,\, H_2$ and
$g\bar b \to H^+ \bar t(b)$. Assuming as we have that $M_{H^\pm} = M_A$, the
principal search modes are via the decays~\cite{Lane:2019dbc}:
\bea
\label{eq:GWC:Hpdecays}H^\pm &\to& t\bar b\,(b\bar t)\,\,{\rm and}\,\,W^\pm
H_2;\\
\label{eq:GWC:Adecays}A &\to& b\bar b,\,\,t\bar t \,\,{\rm and}\,\,ZH_2;\\
\label{eq:GWC:H2decays}H_2 &\to& b\bar b,\,\,t\bar t \,\,{\rm
  and}\,\,ZA,\,W^\pm H^\mp. 
\eea
Searches by CMS~\cite{Khachatryan:2015qxa,Sirunyan:2019arl} and
ATLAS~\cite{Aaboud:2018cwk} for $H^\pm \to t\bar b$ were noted above. CMS has
posted a search for $A$~or~$H_2 \to t \bar t$ that actually showed a
$1.9\,\sigma$ global ($3.5\,\sigma$ local) excess in $A \to t\bar t$ at
$M_A \simeq 400\,\gev$~\cite{Sirunyan:2019wph}. Searches for
$A(H_2) \to ZH_2(A) \to \ell^+\ell^- b \bar b$ have been reported by
ATLAS~\cite{Aaboud:2018eoy} and CMS~\cite{Sirunyan:2019wrn}. Their
95\%~C.L.~limits cross sections are well above GW-2HDM
expectations~\cite{Lane:2019dbc}.

From Fig.~\ref{fig:GWC:CPeven_all}, the decays $H^\pm \to W^\pm H_2$ and
$A \to Z H_2$ require $M_{A,H^\pm} \ge 400\,\gev$. Then, the rapid decrease
of $M_{H_2}$ to $214\,\gev$ at $M_{A,H^\pm} = 410\,\gev$ implies that these
decays are quickly dominated by the emission of longitudinally-polarized
$W,Z$. Thus, their rates grow as $p_{W,Z}^3$, and they quickly overcome
$A \to t\bar t$ and $H^\pm \to t\bar b$. This behavior is illustrated in
Table~\ref{tab:GWC:peecubed} where \Contur was used to compare the GW-2HDM
prediction for
$\sigma(pp \to A \to Z H_2) B(Z \to \ell^+ \ell^-) B(H_2 \to b \bar b)$ to
$\ell^+ \ell^- + \jets$ production at the 8-TeV~LHC; here
$\ell = e$~or~$\mu$.  For $M_A = M_{H^\pm}$, similar behavior of $H_2 \to ZA$
and $W^\pm H^\mp$ occurs above $M_{H_2} = 450\,\gev$.\footnote{The specific
  masses in these statements should be taken {\em cum grano salis}
  because of the distinct possibility of some modification of the right side
  of Eq.~(\ref{eq:GWC:MHsum}) by higher-order
  corrections.}$^{,}$\footnote{Given the large $b \bar b$ SM production rate,
  $gg \to H_2, \,A \to b\bar b$ is likely to be background limited for
  $\tan\beta \simle 0.5$; e.g., see Ref.~\cite{Sirunyan:2018ikr}.}

Confronting these signals with LHC data is the subject of
Sec.~\ref{sec:GWC:section2}.

\begin{table}[!h]
     \begin{center}{
  \begin{tabular}{|c|c|c|c|}
  \hline
$M_A$ (GeV) & $M_{H_2}$ (GeV) & $B(A \to ZH_2)$ & 8-TeV data exclusion\\
\hline
    400 & 314.3  & 0.0\%  & 26.6\% \\
    403 & 295.1  & 9.8\%  & 55.7\% \\
    407 & 259.4  &  42\%  & 97.9\% \\
    410 & 214.4  &  63\%  & 99.9\% \\
\hline
\end{tabular}}
\caption{The rapid onset of the branching ratio for $A \to Z H_2$ and its
  exclusion at the LHC for $\sqrt{s} = 8\,\tev$ and
  $Z H_2 \to \ell^+ \ell^- b \bar b$ $(\ell = e,\mu)$. As discussed in the
  text, this due to the growing dominance of the decay by emission
  longitudinally polarized $Z$-bosons and the corresponding growth of the
  decay rate as $p_Z^3$. The 26.6\% exclusion at $M_A = 400\,\gev$ is due to
  decay of top quarks produced in $A \to t\bar t$ and $H^\pm \to t \bar b$.}
\label{tab:GWC:peecubed}
\end{center}
\end{table}

\section{Confronting GW with \Contur}
\label{sec:GWC:section2}

\begin{figure}[h]
\includegraphics[width=1.0\textwidth]{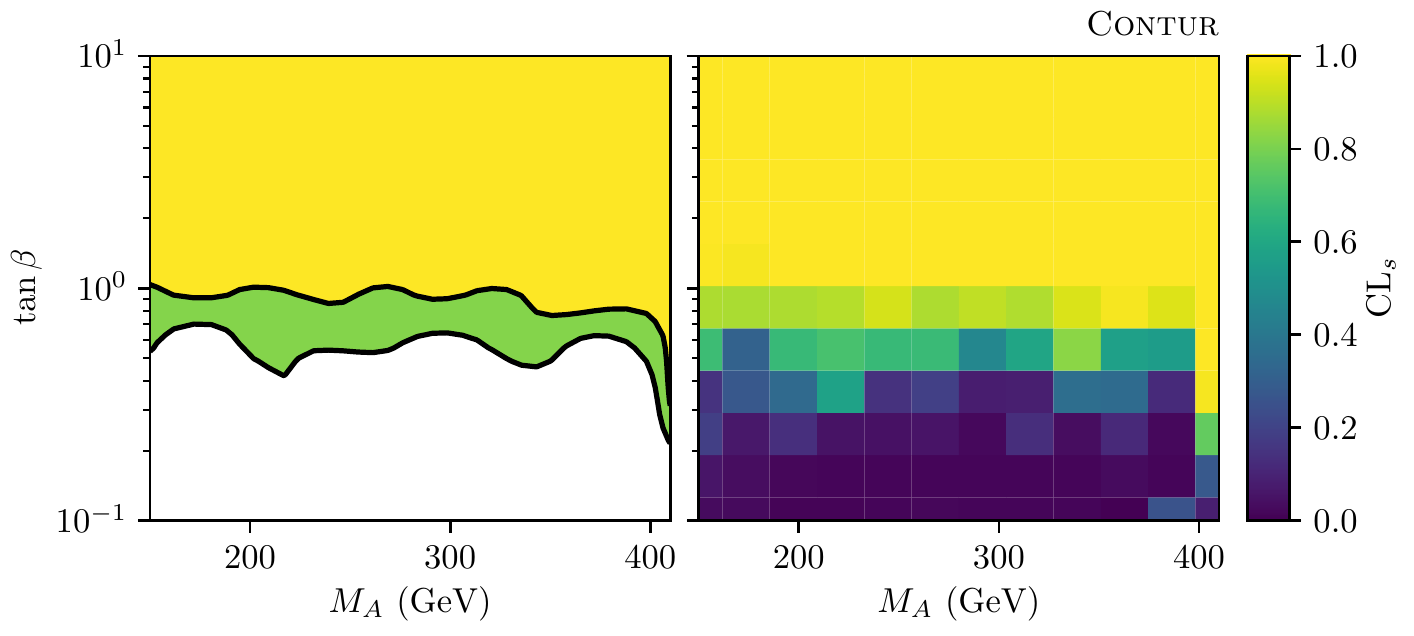}
\caption{Left: Sensitivity plot for the GW-2HDM of $\tan\beta$
  vs.~$M_A = M_{H^\pm}$ over the range allowed by the sum rule
  Eq.~(\ref{eq:GWC:MHsum}). Yellow areas signify exclusion at
  $\ge 95\%$~C.L.; green areas are 68--95\%. Right: The same exclusion shown
  on a continuous scale by the bar on the right, with 1.0~being fully
  excluded and 0.0~being zero sensitivity.}
\label{fig:GWC:combo_1}
\end{figure}

In this section we use \Contur~\cite{Butterworth:2016sqg} to determine
whether and where the GW-2HDM model is in conflict from LHC measurements. In
a nutshell, BSM processes are generated by
\Herwig~\cite{Bellm:2019zci} and \Contur compares these
particle-level simulations with existing measurements contained in the \Rivet
library of analyses~\cite{Bierlich:2019rhm}. Since, with the exception of
relatively few isolated excursions, all these measurements agree with SM
predictions, these comparisons provide a ``health check'' on new physics
models. Currently, most analyses in \Rivet 
use data from the 8-TeV
run of the LHC. This may not be much of a disadvantage for probing the
GW-2HDM. The sensitivity to its signals --- the new Higgs bosons below about
500~GeV --- is generally not better with $36\,\ifb$ at $13\,\tev$ than with
$25\,\ifb$ at~7 and $8\,\tev$ because SM background rates often increase
faster than signal rates, especially for low-mass signals. See, e.g., the
expected significances for observing $H_1 \to b \bar b$ with the Run~1 and
$36\,\ifb$ Run~2 datasets in Ref.~\cite{Sirunyan:2017elk}. For more
information on \Contur in this document, see~\ref{sec:contur-update}.

Figure~\ref{fig:GWC:combo_1} shows the sensitivity of LHC data to the decay
signals in Eqs.~(\ref{eq:GWC:Hpdecays}--\ref{eq:GWC:H2decays}) as a function
of $\tan\beta$. Signal events were generated assuming $M_A = M_{H^\pm}$ as
explained in Sec.~\ref{sec:GWC:section1}.\footnote{The endpoint
  $M_A = 410\,\gev$ is the boundary of the sum rule~(\ref{eq:GWC:MHsum}) when
  $M_A = M_{H^\pm}$.}  Generally, the GW-2HDM is fully excluded for
$\tan\beta > 1$. The 1-$\sigma$ and 2-$\sigma$ excluded regions are
consistent with the finding in Ref.~\cite{Lane:2018ycs} that limits on
$\sigma(pp \to H^\pm \to t\bar b)$ from CMS~\cite{Khachatryan:2015qxa} and
ATLAS~\cite{Aaboud:2018cwk} require $\tan\beta \simle 0.5$. The exception to
this in Fig.~\ref{fig:GWC:combo_1} (right) is at $M_A \simge 403\,\gev$ where
the 95\%~C.L.~upper limit is $\tan\beta = 0.3$; also see
Table~\ref{tab:GWC:peecubed}. This sensitivity comes from data on
$\ell^+\ell^- + \jets$ and $\ell^\pm \etmiss + \jets$ and it is due to the
rapid turn-on of the decays $A\to Z H_2$ and $H^\pm \to W^\pm H_2$ discussed
above. The production cross sections for both processes are proportional to
$\tan^2\beta$.

The sensitivity plot Fig.~\ref{fig:GWC:combo} of $M_{H_2}$
vs.~$M_A = M_{H^\pm}$ treats them as independent variables, but (1) they are
restricted to their ranges allowed by the sum rule~Eq.~(\ref{eq:GWC:MHsum})
and (2) $\tan\beta$ is fixed at~0.5, approximately its largest value allowed
by data. So, this figure allows a much wider range of masses to be probed by
LHC data than does Fig.~\ref{fig:GWC:combo_1}. There is a sizable fully
excluded region in the lower right corner. Again, it is due to the conflict
between the model and data on $\ell^+\ell^- + \jets$ and
$\ell^\pm + \etmiss + \jets$. A sliver of this contains the masses consistent
with $M_A = M_{H^\pm}$ and the sum rule, depicted in
Fig.~\ref{fig:GWC:CPeven_all}. It is shown as the purple band in this
figure, whose lower edge is $M_{H'}$ and upper edge is $M_{H_2}$. The
separation between the two masses above $M_A = 370\,\gev$ accounts for the
band's thickening at the high end. The excluded portion of the band
corresponds exactly to the one in Fig.~\ref{fig:GWC:combo_1}

\begin{figure}[h]
\includegraphics[width=1.0\textwidth]{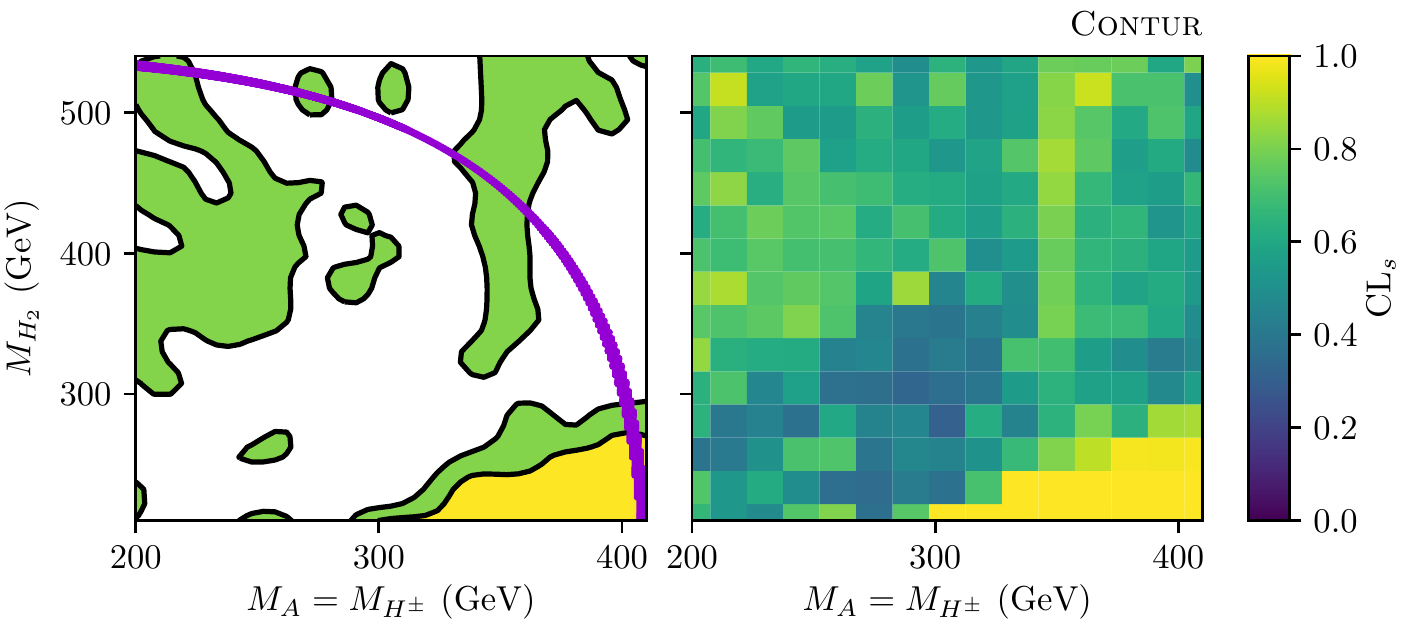}
\caption{Left: Excluded regions in a plot of $M_{H_2}$ vs.~$M_A = M_{H\pm}$,
  treated as independent variables, for fixed $\tan\beta = 0.50$. The purple
  band shows the masses $M_{H'}$ (lower edge) and $M_{H_2}$ (upper edge)
  determined by the sum rule~(\ref{eq:GWC:MHsum}) and the one-loop CP-even
  mass matrix in Fig.~\ref{fig:GWC:CPeven_all}. The color scheme is as in
  Fig.~\ref{fig:GWC:combo_1}. Right: The corresponding continuous scale of
  exclusions with 1.0~being fully excluded and 0.0~being zero sensitivity.}
\label{fig:GWC:combo}
\end{figure}

\section{Summary}
\label{sec:GWC:section3} 

In this report we described the Gildener-Weinberg mechanism which naturally
accounts for and stabilizes the low mass of the Higgs boson $H$ and its
apparently SM couplings to fermions and gauge bosons. We illustrated the
mechanism in the simple 2HDM proposed by Lee and Pilaftsis in 2012, just
before the announcement of the discovery of $125\,\gev$ Higgs at the LHC. In
2018, Lane and Shepherd modified this model with type-I Higgs couplings to
fermions to make it consistent CMS and ATLAS searches for
$H^\pm \to t \bar b$; they determined that $\tan\beta = v_2/v_1 \simle 0.5$
at $\le 95\%$~C.L.

A remarkable feature of this GW-2HDM is the sum rule~(\ref{eq:GWC:MHsum})
limiting the masses of the new Higgs bosons $H^\pm$, $A$, and $H_2$ to be
less than about $500\,\gev$. Thus, searching for and finding --- or excluding
--- these new Higgses at the LHC is far and away the best way to test GW
models {\em \ul{in this decade}}.

This paper is a start on that program. We outlined a simplified scheme for
such searches, based on the sum rule and the equality, $M_{H^\pm} = M_A$,
that makes the contribution to the $T$-parameter from the new scalars
vanish. And we put the GW-2HDM to a wide-ranging test using \Contur to
compare its major signals,
Eqs.~(\ref{eq:GWC:Hpdecays}--\ref{eq:GWC:H2decays}), with LHC
measurements. This powerful tool did its job. It turned up the one instance
that the GW-2HDM with $\tan\beta \simeq 0.5$ is in conflict with LHC
data: For $M_A = M_{H^\pm} \simge 403\,\gev$ and $M_{H_2} \simle 295\,\gev$,
the rates for $pp \to A \to Z H_2 \to \ell^+\ell^- b\bar b$ and
$pp \to H^\pm \to W^\pm H_2 \to \ell^\pm \etmiss \, b\bar b$ exceed the
measured rates of these final states at greater than the 95\%~C.L.~for
$\tan\beta > 0.3$.


\section*{Acknowledgments}

We thank the organizers and conveners of the Les Houches workshop, ``Physics
at TeV Colliders'', for a stimulating meeting.  We also thank Gustaaf
Brooijmans, Bill Murray and Eric Pilon for stimulating discussions, comments
and collaboration. K.~Lane thanks the Laboratoire d'Annecy-le-Vieux de
Physique Th\'eorique (LAPTh) and the CERN Theory Group for their support and
hospitality during the initial stage of this research. This work has received
funding from the European Union's Horizon 2020 research and innovation
programme as part of the Marie Sklodowska-Curie Innovative Training Network
MCnetITN3 (grant agreement no. 722104).








\let\Herwig\undefined
\let\Pythia\undefined
\let\Sherpa\undefined
\let\Rivet\undefined
\let\Professor\undefined
\let\eps\undefined
\let\mc\undefined
\let\mr\undefined
\let\mb\undefined
\let\tm\undefined

%% file: 2hdm/2hdma.main.tex
\graphicspath{{2hdm/}}

\newcommand{\Herwig}{H\protect\scalebox{0.8}{ERWIG}\xspace}
\newcommand{\Pythia}{P\protect\scalebox{0.8}{YTHIA}\xspace}
\newcommand{\Sherpa}{S\protect\scalebox{0.8}{HERPA}\xspace}
\newcommand{\Rivet}{R\protect\scalebox{0.8}{IVET}\xspace}
\newcommand{\Yoda}{Y\protect\scalebox{0.8}{ODA}\xspace}
\newcommand{\hepdata}{HEPData\xspace}
\newcommand{\contur}{\textsc{Contur}\xspace}
\newcommand{\eps}{\varepsilon}
\newcommand{\mc}[1]{\mathcal{#1}}
\newcommand{\mr}[1]{\mathrm{#1}}
\newcommand{\mb}[1]{\mathbb{#1}}
\newcommand{\tm}[1]{\scalebox{0.95}{$#1$}}

\chapter{Sensitivity of LHC measurements to a two-Higgs-doublet plus pseudoscalar DM model}
{\it J.~M.~Butterworth, M.~Habedank,  P.~Pani}


\label{sec:2hdma}


\section{Introduction}
\label{sec:2hdma:intro}

In this short contribution we use \contur~\cite{Butterworth:2016sqg} to examine the
sensitivity of ATLAS, CMS and LHCb measurements available in \Rivet 3.1~\cite{Bierlich:2019rhm} to
a Dark Matter model involving two Higgs doublets and an additional pseudoscalar mediator~\cite{Bauer:2017ota}.
This model has been the subject of several searches at the LHC \cite{Sirunyan:2018gdw,Aaboud:2019yqu} and in particular was studied
using a combination of ATLAS searches, which we will compare with in the following.

\begin{figure}[htbp]
  \begin{center}
    \includegraphics[width=0.48\textwidth]{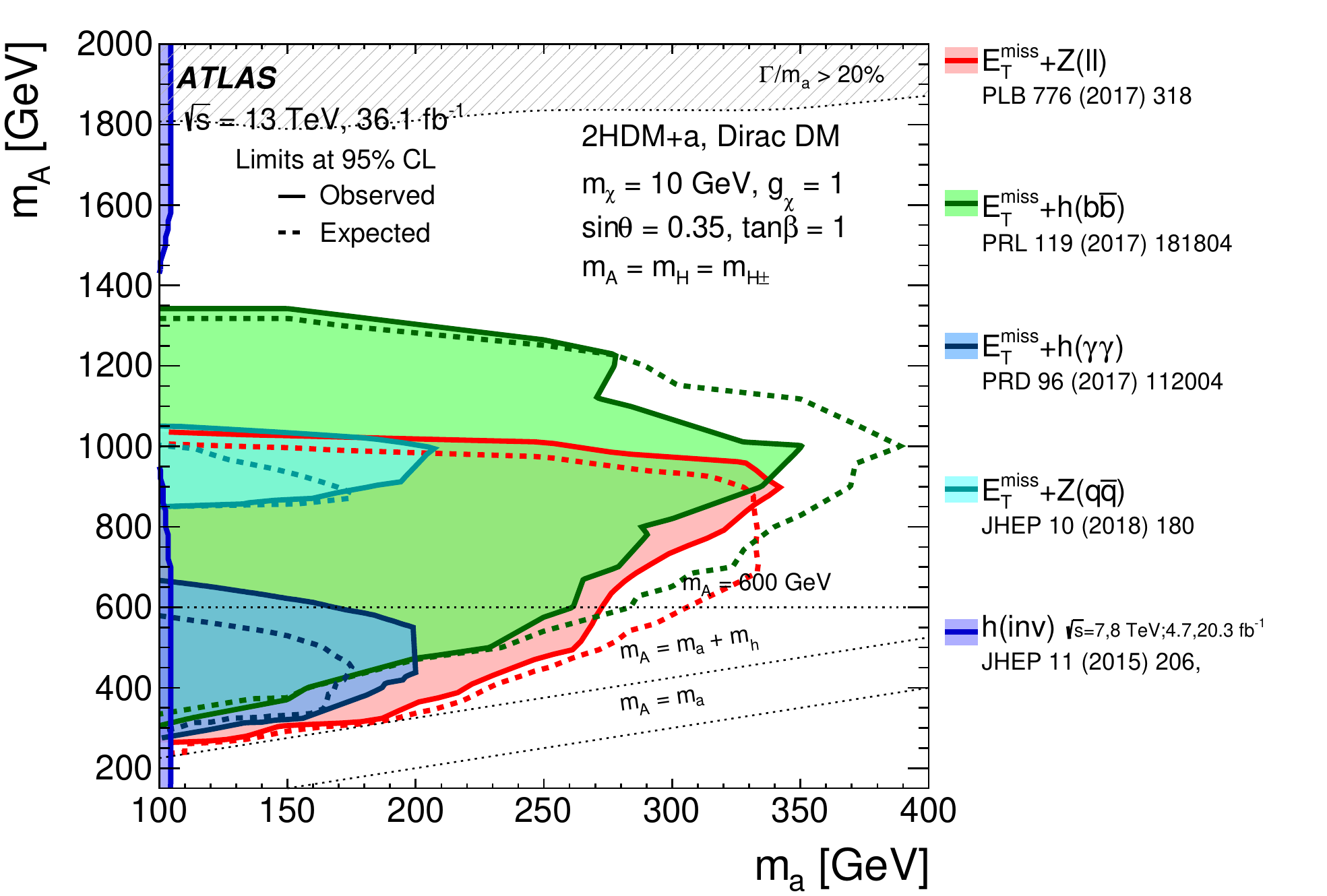}
    \includegraphics[width=0.48\textwidth]{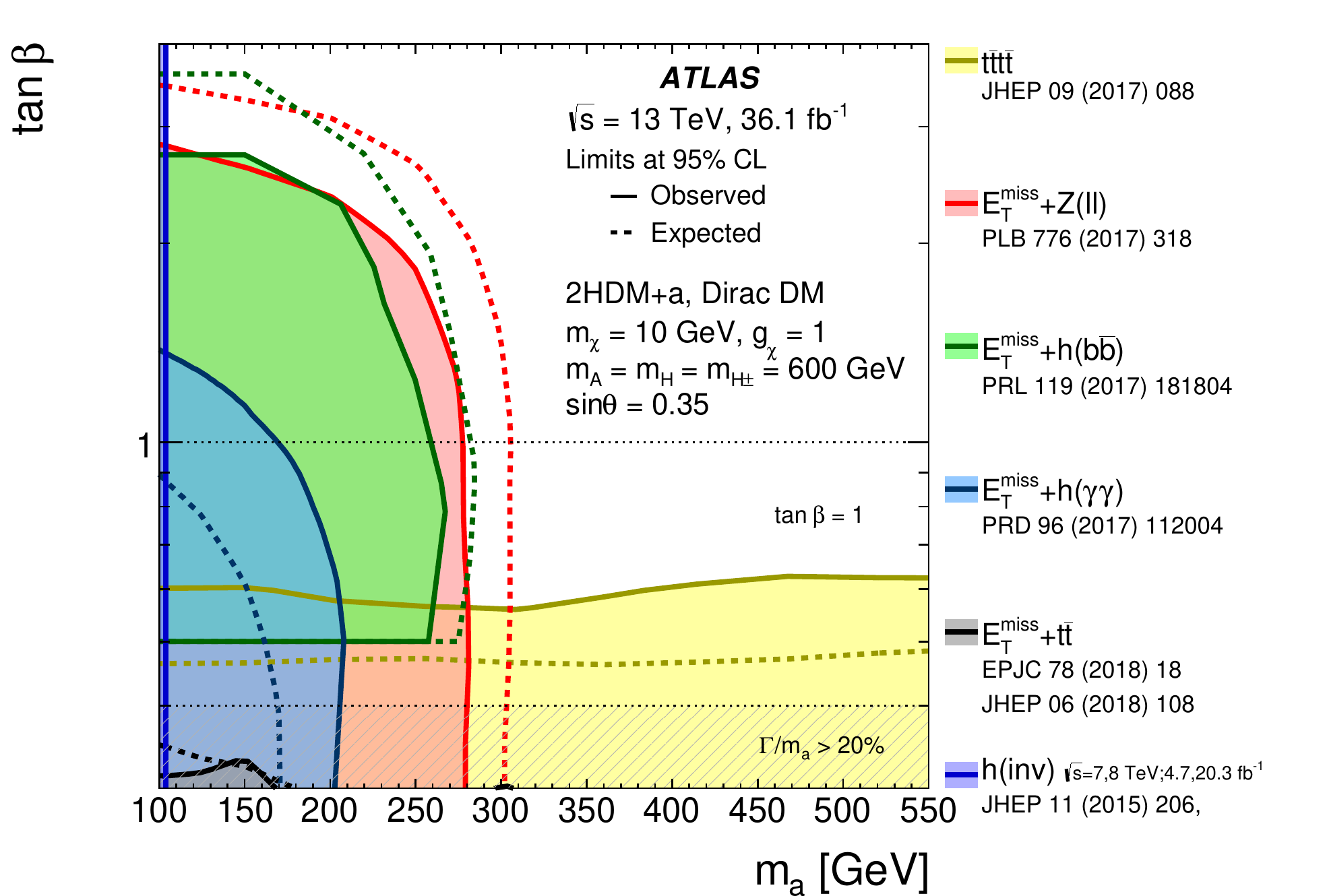}
    \caption{Figure 19 from Ref.\cite{Aaboud:2019yqu}, showing ATLAS exclusion limits
    from a combination of searches.}
    \label{fig:2hdma:atlas}
  \end{center}
\end{figure}

We focus on two parameter scans, those of Fig.\,19 in Reference~\cite{Aaboud:2019yqu}, shown for convenience in
Fig.~\ref{fig:2hdma:atlas}. In these scans, which follow the recommendations of the LPCC DMWG \cite{Abe:2018bpo}, 
the masses of all the exotic Higgs bosons ($A, H, H^\pm$) are set to be degenerate, the mass of the DM candidate $\chi = 10$~GeV, the coupling of the pseudoscalar mediator $a$
to $\chi$ is unity, $\sin\theta = 0.35$ where $\theta$ is the mixing between the two neutral CP-odd 
weak eigenstates, and we set $\sin(\beta - \alpha)$, the sine of the difference of the mixing angles 
in the scalar potential containing only the Higgs doublets, to unity, meaning we are in the 
aligned limit so that the lightest mass eigenstate has SM Higgs couplings.

This model is characterised by a particularly rich phenomenology, 
dominated by the production of the lightest pseudoscalar or the heavier Higgs boson partners, 
via loop-induced gluon fusion, associated production with heavy-flavour quarks or associated
production with a SM Higgs or $Z$ boson.
Very diverse characteristic signatures can be produced by various decay chains of these bosons.
Many of these signatures remain largely unexplored.

For this reason, the use of \contur with its wide range of analysis signatures, coupled with the inclusive
approach of Herwig in generating all processes leading to BSM particle production, is of particular interest, as a contribution to a systematic identification of all signatures that can potentially constrain the model.

\section{Results}

\begin{figure}[htbp]
  \begin{center}
    \includegraphics[width=1.0\textwidth]{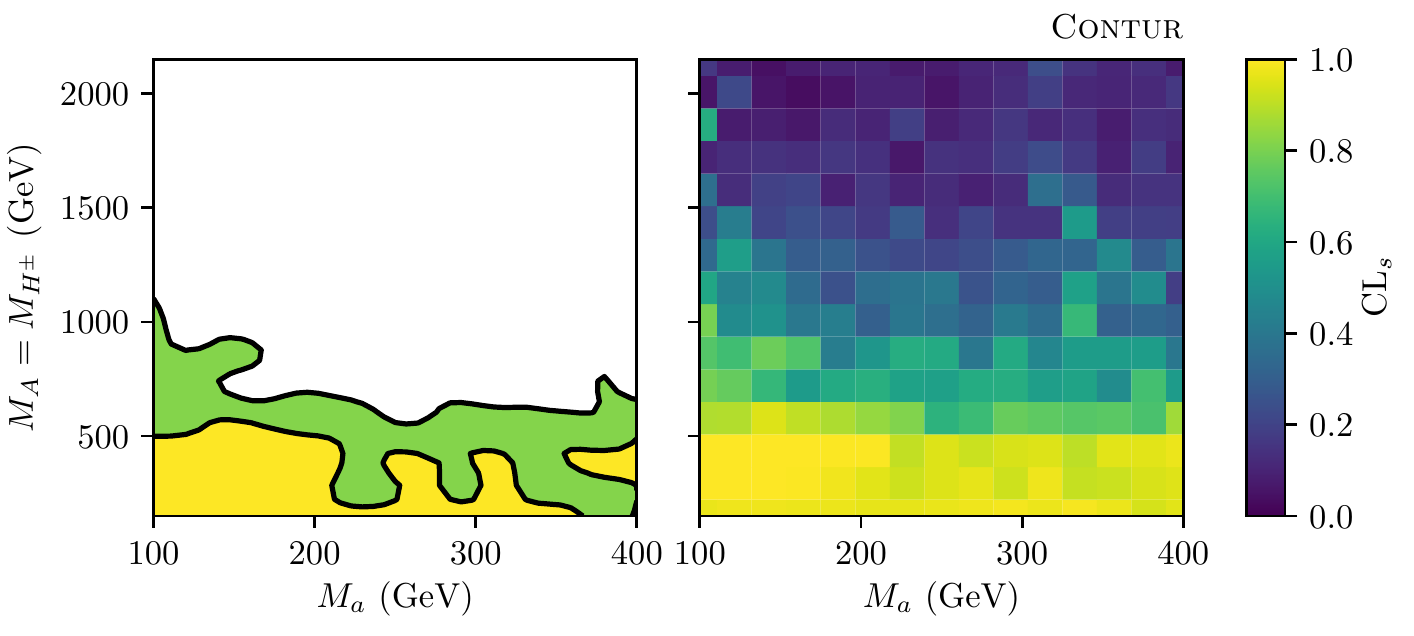}
    \includegraphics[width=1.0\textwidth]{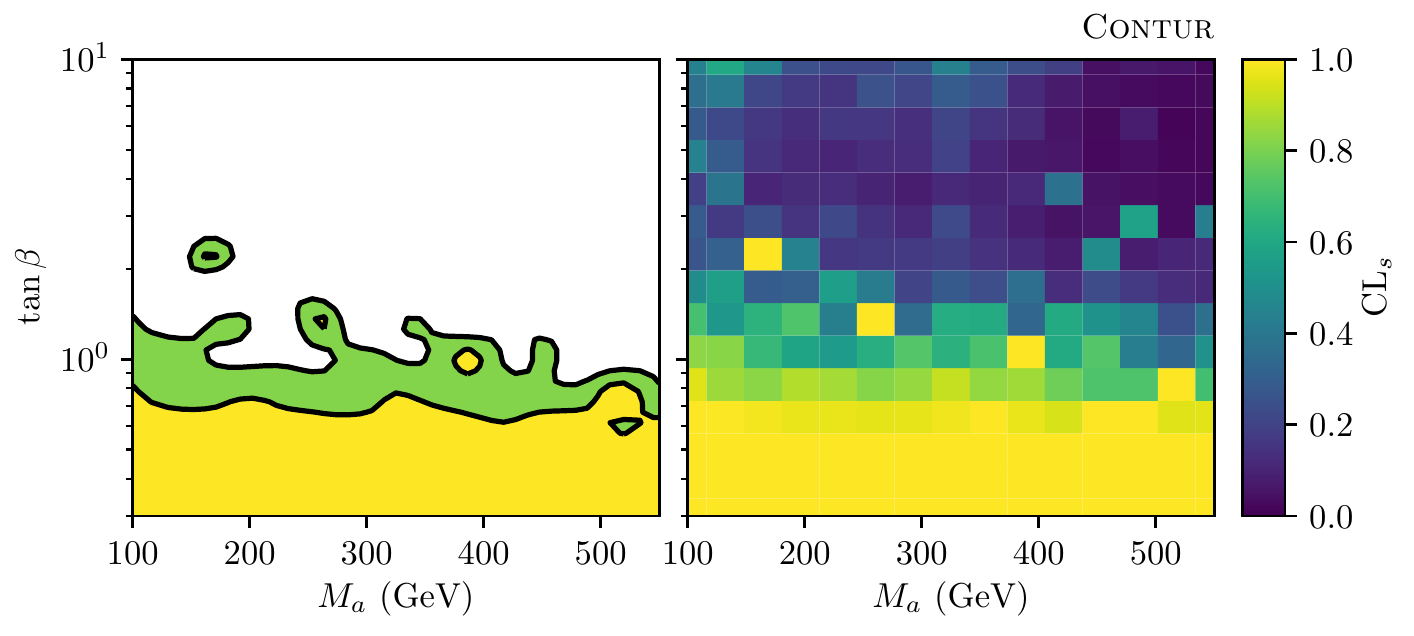}
    \caption{\contur scans over the parameter planes shown in Fig.~\ref{fig:2hdma:atlas} in the $M_a$--$M_A$ (top) and $M_a$--$\tan\beta$ plane (bottom). Left: 95\% (68\%) excluded contours in yellow (green). Right: Underlying heatmap of exclusion at each investigated parameter space point.}
    \label{fig:2hdma:contur}
  \end{center}
\end{figure}

We use \contur to scan over the same parameter plane as ATLAS, generating the BSM events with 
\Herwig~\cite{Bellm:2019zci} and looking for parameter points where an observably significant number of events would have entered the fiducial phase space of the measurements. The results are shown in Fig.~\ref{fig:2hdma:contur} (see Section~\ref{sec:contur-update} of these proceedings for
more detail on the \contur methodology).

\begin{figure}[htbp]
	\begin{flushleft}
		\parbox{.39\textwidth}{\flushright \contur\\ATLAS, 7\,TeV, $ZZ$}
		\hspace{0.05\textwidth}
		\parbox{.4\textwidth}{\flushright \contur\\ ATLAS \& CMS, lepton/W + jets}
	\end{flushleft}
	\vspace{-25pt}
	\begin{center}	
		\includegraphics[width=.4\textwidth, valign=t]{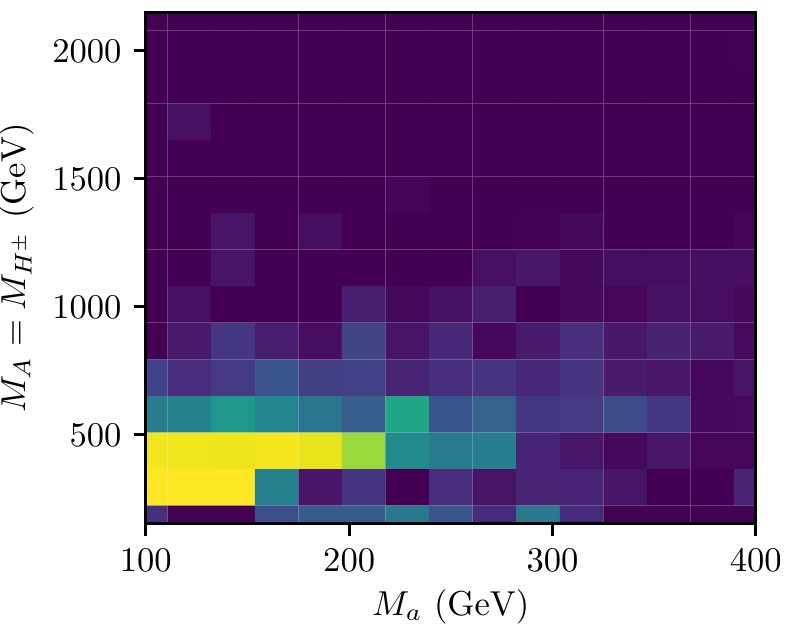}
		\hspace{0.05\textwidth}
		\includegraphics[width=.4\textwidth, valign=t]{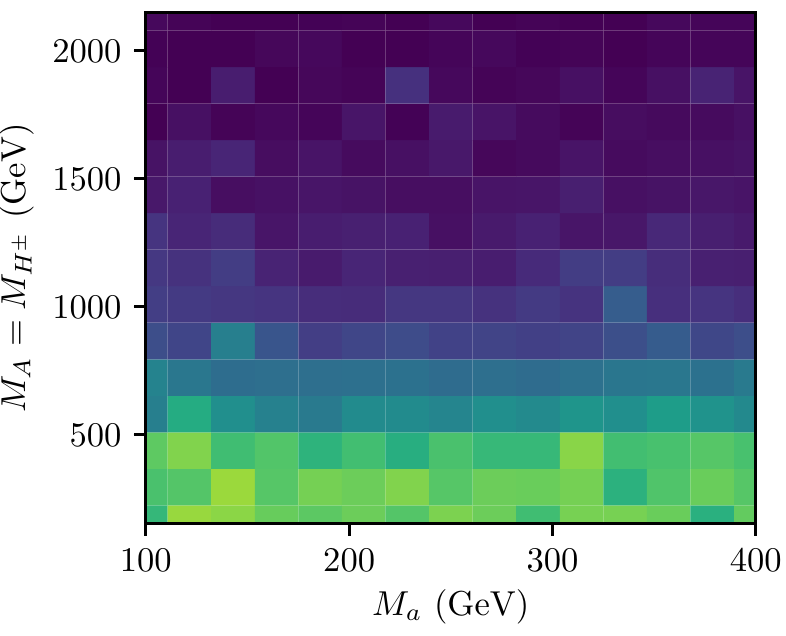}
		\hspace{0.05\textwidth}
		\includegraphics[width=.069\textwidth, valign=t]{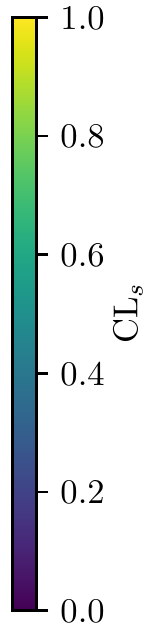}
                \caption{Heatmaps of exclusion at each considered parameter space point in the $M_a$--$M_A$ plane for the $ZZ$ measurement at $\sqrt{s}=7$\,TeV \cite{Aad:2012awa} by ATLAS (left) and the combination of various ATLAS and CMS analyses with signatures comprising a lepton or vector boson and jets \cite{Sirunyan:2018wem, Aaboud:2018eki, Chatrchyan:2013vbb, Aaboud:2017fye} (right) as used by \contur.}

		\label{fig:2hdma:analyses}
	\end{center}
\end{figure}

The overall sensitivity is mostly the combined result of what would be 1-2 $\sigma$ contributions to a wide range of cross-section measurements. Fig.\,\ref{fig:2hdma:analyses} shows exemplary heatmaps of analyses contributing to the sensitivity in the $M_a$--$M_A$ scan. Especially analyses involving $W$ boson production and leptonic decay (i.e. lepton-plus-missing-energy final states) contribute to the sensitivity. It generally arises from the production of all the exotic
Higgs bosons, with subsequent decays either to top or directly to $W$ bosons. For example for $m_A = m_H = m_{H^{\pm}} = 435$ GeV and $m_a = 250$ GeV, 
the production cross-section for the CP-odd Higgs boson $A$ is about $6$ pb and its dominant decay channels are in
$t\bar t$ (85\%) and $\chi\bar\chi$ (15\%). In addition, the production cross section for the CP-even Higgs boson $H$ is
$3$ pb, and it decays into a top-pair with a branching ratio of 88\%, while the second dominant decay mode is
$H\rightarrow aZ$ (10\%). For lower pseudoscalar mediator masses as $m_a = 100$ GeV, the $H\rightarrow aZ$ branching ratio increases to
30\% and this final state dominates the sensitivity to this model (Fig~\ref{fig:2hdma:analyses} left). All cross sections have been calculated with Herwig 7 \cite{Bahr:2008pv,Bellm:2015jjp}
and with Madgraph5\_aMC@NLO \cite{Alwall:2014hca} and were found in good agreement.
\\

Because \contur takes the SM as being identical to the data, the absolute 
sensitivity should be treated with caution. In addition, correlated systematic uncertainties between different measurements are not accounted for,
as the information is not provided by the experiments. Nevertheless, the conclusion that there is sensitivity in these
final states, even with current data, is robust. 

In passing we note that naively, there is apparent strong sensitivity in the $H \rightarrow WW$ fiducial cross section
measurements. However, these measurements typically apply a $b$-jet veto, or use a $b$-jet control region, to suppress the
large background from $t\bar{t}$. In the case of CMS~\cite{Khachatryan:2016vnn}, this is applied at detector level and is not 
included as part of the fiducial 
cross section definition (and thus not applied in the Rivet routine). Therefore, the analysis cannot be used here, as in reality
many of our signal events would fail the veto, but we cannot evaluate how many.  The same applies to the CMS $W$+jet measurements.
In the case of ATLAS~\cite{Aad:2016lvc}, the cross section
is measured as a function of jet multiplicity and is consistently implemented in Rivet, so in principle the analysis could be used. However,
there is a large data-driven background subtraction, and our signal would most likely contribute in both the control region and signal region
of the analysis, so simply counting the contribution to the signal region is likely to overestimate our sensitivity. For these reasons we
also exclude the ATLAS measurement from the current study.

\section{Conclusion}

There is interesting sensitivity to the two-Higgs-doublet plus pseudoscalar DM model, across several measured final states, due to the quite complex
phenomenology of the model. This phenomenology can change a lot when the parameters change, and
a wider set of parameter scans would be of interest, relaxing some of the current assumptions imposed on the parameters.

Future searches for this model should also consider final states involving top and/or $W$ production, even in the absence of
a large missing energy signature. 

Of course, the full run 2, coming run 3, and HL-LHC measurements can be expected to have a substantial impact. 
If future $H \rightarrow WW$ measurements
can be made less model-dependent, they may also make a significant contribution.

\let\contur\undefined
\let\Herwig\undefined
\let\Pythia\undefined
\let\Sherpa\undefined
\let\Rivet\undefined
\let\Professor\undefined
\let\eps\undefined
\let\mc\undefined
\let\mr\undefined
\let\mb\undefined
\let\tm\undefined

%% file: HiggsZa/HiggsZa.main.tex
\graphicspath{{HiggsZa/}}


\newcommand{\mg}{{\sc MadGraph\_MC@NLO}}
\newcommand{\pythia}{{\sc Pythia 8.2}}
\newcommand{\delphes}{{\sc Delphes 3}}

\chapter{Looking for exotic Higgs decays via $h \to Z a$}

{\it
A. Bharucha, J. M. Butterworth, N. Desai, S. Gascon-Shotkin, S. Jain, A. Lesauvage, 
G. Moreau, S. Mutzel, J. M. No, J. Quevillon, C. Smith, K. Tobioka, N. Vignaroli, S. L. Williamson, J. Zurita}


\label{sec:HiggsZa}

\begin{abstract}
We study the exotic decay of the 125 GeV Higgs boson into a $Z$ boson and a light pseudoscalar $a$ beyond the Standard Model, $h \to Z a$. 
Such decay is well-motivated in a variety of new physics scenarios, ranging from axion-like particles to extended Higgs sectors.
We analyse the LHC sensitivity to this Higgs decay mode in several decay modes $a \to xx$ of the light pseudoscalar, namely 
$xx = \gamma \gamma$, $\mu^+ \mu^-$, $\tau^+ \tau^-$. 
Besides deriving model-independent LHC sensitivity estimates on 
BR($h \to Z a$) $\times$ BR($a \to x x$), we interpret our results in terms of axion-like particles with masses $m_a < 35$ GeV, showing that this search would provide a powerful window to discover such states at the LHC.
\end{abstract}

\section{Introduction}
\label{sec:HiggsZa:intro}

Exploring the possible existence of exotic decays of the 125 GeV Higgs boson $h$ constitutes a primary avenue to search for new physics beyond the Standard Model (SM) at the LHC (see~\cite{Curtin:2013fra} for a review). 
Current data from ATLAS and CMS measurements of Higgs signal strengths are compatible with the existence of a 125 GeV Higgs boson
branching fraction into beyond the SM (BSM) states as large as $\mathcal{O}$(20\%)~\cite{Aad:2019mbh}. Direct searches targeting specific BSM decay modes of the 125 GeV Higgs boson have the potential to probe much smaller branching fractions, and thus they may provide a unique window into new BSM light states. 

In this work we focus on exotic decays of $h$ into light pseudoscalars via $h \to Z a$. Such decays arise naturally in well-motivated extensions of the SM, e.g.~via the presence of axion-like particles (ALPs)~\cite{Georgi:1986df,Brivio:2017ije,Bauer:2017ris,Bauer:2017nlg,Quevillon:2019zrd} or in the context of non-minimal Higgs sectors featuring a light scalar/pseudoscalar~\cite{Chisholm:2016fzg,Baum:2018zhf,No:2015xqa}. Decays of the 125 GeV Higgs boson into a $Z$ boson plus a BSM state may also occur in other SM extensions (e.g. $h \to Z Z_D$ with $Z_D$ a hidden gauge boson, see~\cite{Gopalakrishna:2008dv,Davoudiasl:2012ag,Davoudiasl:2013aya}). 
The LHC potential to probe the decay $h \to Z a$ has been recently highlighted by Bauer et al. in~\cite{Bauer:2017ris}, considering $a$ to be an ALP decaying to di-photon or di-electron final states. In this work we perform for the first time a 
detailed analysis of the LHC sensitivity to the exotic decay $h \to Z a$ in the $Z \to \mu\mu$, $a \to \gamma\gamma$ final state, for which no present ATLAS or CMS search exists (although we note that
this final state was measured in 8~TeV collisions~\cite{Aad:2016sau}, without a Higgs mass constraint). In addition, we consider the leptonic decays $a \to \mu\mu$ and $a \to \tau\tau$ and derive approximate limits on the corresponding branching fractions by reinterpreting existing ATLAS and CMS analyses. In all cases, we restrict ourselves to the mass range $m_a < 34$ GeV, for which the Higgs boson decays into two on-shell states. 
We derive limits on the branching fractions BR($h \to Z a$) $\times$ BR($a \to \gamma \gamma$), BR($h \to Z a$) $\times$ BR($a \to \mu \mu$) and BR($h \to Z a$) $\times$ BR($a \to \tau \tau$) respectively in sections~\ref{sec:HiggsZa:gammagamma},~\ref{sec:HiggsZa:mumu} and \ref{sec:HiggsZa:tautau}, and in section~\ref{sec:HiggsZa:ALPs} we interpret our results in terms of the ALP effective field theory (EFT) extension of the SM (see~\cite{Brivio:2017ije,Bauer:2017ris} and references therein).

\section{LHC searches for $h \to Z a$}
\label{sec:HiggsZa:LHC}

\subsection{$a \to \gamma\gamma$}
\label{sec:HiggsZa:gammagamma}

The decay of the state $a$ into two photons provides a very clean final state in combination with the leptonic decay of the $Z$ boson coming from the Higgs decay. Depending on the mass $m_a$ of the light pseudoscalar state and on the event kinematics, the two photons may be resolved in the detector or appear as one reconstructed object (as noted in~\cite{Bauer:2017ris}), the latter occuring for strongly boosted ALPs. The minimal angular separation $\Delta R$ the 
two photons need to have to be resolved in the detector is dictated by the resolution of the Electromagnetic Calorimeter (ECAL) barrel and endcap: for CMS, individual crystals in the ECAL span $d \eta / d \phi \sim 0.0175$ for the barrel (for the endcap, this can be up to $0.05$ depending on $\eta$), with most of the photon energy ($\sim 97\%$) captured by a $5 \times 5$ array of crystals~\cite{CERN-LHCC-97-033, Bayatian:922757}, spanning a $\Delta R \sim 5 \times 0.0175 \simeq 0.09$. 
As a result, for $\Delta R_{\gamma\gamma} < 0.1$ photon energy deposits in the CMS ECAL start overlapping and are likely to be reconstructed as one single photon. For converted photons, this $\Delta R_{\gamma\gamma}$ may be slightly larger. A similar minimal angular separation to resolve the two photons is found for ATLAS (see e.g.~\cite{ATLAS:2012soa}). Below this angular separation, it is still viable to differentiate between single and multiple photons by means of shower-shape analyses~\cite{ATLAS:2012soa}, a possibility we do not discuss further here.

\begin{figure}[htbp]
  \begin{center}
    \includegraphics[width=0.55\textwidth]{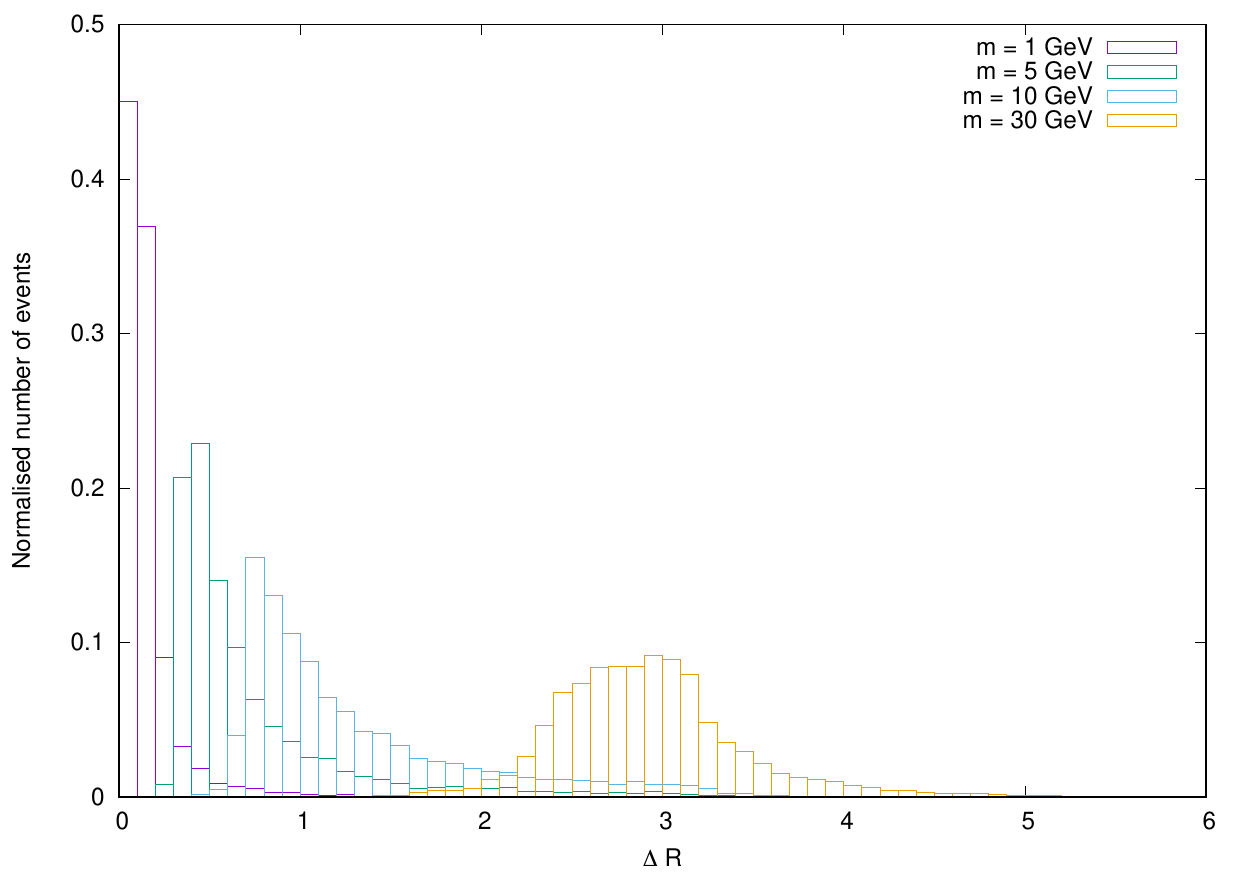}
    \caption{Kinematic event distribution for the $\Delta R_{\gamma\gamma}$ separation between the two photons in $p p \to h \to Z a$, $a \to \gamma\gamma$ for various values of $m_a$.}
    \label{fig:HiggsZa:DRaa}
    
    \vspace{-2mm}
    
  \end{center}
\end{figure}

We thus begin our analysis by studying the $\gamma\gamma$ angular separation for the process $p p \to h \to Z a$ (the Higgs boson produced via gluon-fusion) at LHC 13 TeV, followed by the decay $a \to \gamma\gamma$, as a function of $m_a$. The $\Delta R_{\gamma\gamma}$ distribution is shown in Fig.~\ref{fig:HiggsZa:DRaa} for masses in the range $m_a \in [1,\, 30]$ GeV. For $m_a \lesssim 1$ GeV, a very large fraction of the events has $\Delta R_{\gamma\gamma} < 0.1$, and thus would fail a resolved analysis. For $m_a \ll 1$ GeV it is safe to assume the two photons are reconstructed as one, and thus possible to reinterpret the current LHC 13 TeV upper bounds on the $h \to Z \gamma$ branching fraction~\cite{Aaboud:2017uhw,Sirunyan:2018tbk} to yield a constraint ${\rm BR}(h\to Z a) \times {\rm BR}(a\to \gamma\gamma) < 0.01$ at 95\% 
C.L.\footnote{We use here the ATLAS analysis~\cite{Aaboud:2017uhw} with 36.1 fb$^{-1}$ of integrated luminosity.}. Whether this limit can be extrapolated to masses close to and above 1 GeV strongly depends on the details of the ATLAS and CMS photon identification and reconstruction algorithms (see e.g.~\cite{Aaboud:2016yuq}). Here we simply consider the fraction of signal events that satisfy $\Delta R_{\gamma\gamma} < 0.1$ as a measure of the degrading of the limit for increasing $m_a$ (e.g.~for $m_a = 500$ MeV this fraction is $0.83$, while it drops to $0.008$ for $m_a = 2$ GeV).   

Turning now to the analysis of $p p \to h \to Z a$, $a \to \gamma\gamma$ with two resolved photons, we simulate the signal\footnote{We use an in-house tune of the UFO model developed in~\cite{Bauer:2017ris}.} at LHC 13 TeV with \mg~\cite{Alwall:2014hca}, \pythia~(parton showering and hadronization)~\cite{Sjostrand:2014zea} and \delphes~(detector simulation)~\cite{deFavereau:2013fsa}. We adopt the N$^3$LO results from~\cite{Anastasiou:2016cez} for the single Higgs production cross section, which quotes a value of 48.58 pb for $\sigma (p p \to h)$ in gluon-fusion.
For the photon isolation in \delphes~we use a standard isolation cone $\Delta R = 0.5$, which we note causes a strong degrading of the signal efficiency for $m_a \lesssim 5$ GeV (as is apparent from Fig.~\ref{fig:HiggsZa:DRaa} and the discussion above). We thus consider pseudoscalar masses in the range $m_a \in [5,\,34]$ GeV. The event selection is performed as follows:  
\begin{itemize}
    \item We require two isolated photons with $p_T^{\gamma} > 10$ GeV and $|\eta_{\gamma}| < 2.5$, and two oppositely charged muons with $p_T^{\mu} > 10$ GeV and $|\eta_{\mu}| < 2.4$.
    
    \item The two muons must satisfy $m_{\mu\mu} \in [75,\,105]$ GeV. Similarly, the combination of the two muons and two photons must reconstruct the Higgs boson mass, 
    $m_{\mu\mu\gamma\gamma} \in [115,\,135]$ GeV.
    
    \item The two photons must satisfy $m_{\gamma\gamma} + p^{\gamma\gamma}_T < m_h/2$. This condition is effective in suppressing the reducible $Z +$ jets SM background (see below), while maintaining a high efficiency for the signal for all $m_a$.
\end{itemize}
For events that pass the event selection, a signal region is defined in the $m_{\gamma\gamma}$ variable, where we look for an excess over the background in a $\pm5$ GeV mass window centered in the $m_a$ mass hypothesis.

\begin{figure}[htbp]
  \begin{center}
    \includegraphics[width=0.7\textwidth]{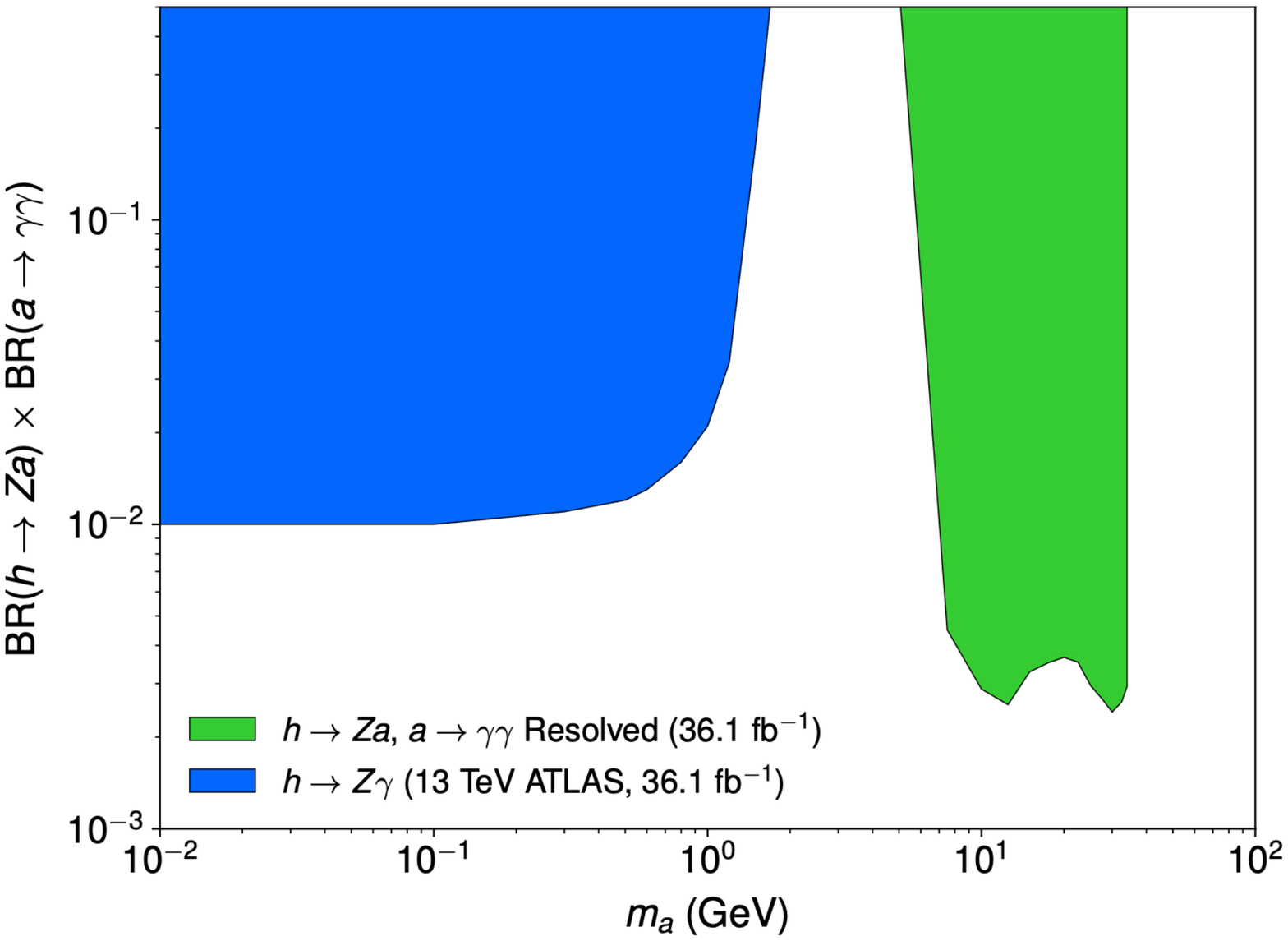}
    
    \vspace{-3mm}
    
    \caption{95\% C.L. upper limits for ${\rm BR}(h\to Z a) \times {\rm BR}(a\to \gamma\gamma)$ in our resolved di-photon analysis (green) and via $h \to Z \gamma$ when the two photons are too collimated to be resolved by the detector (blue) at LHC 13 TeV with $36.1$ fb$^{-1}$.}
    \label{fig:HiggsZa:UL_gammagamma}
    
    \vspace{-4mm}
    
  \end{center}
\end{figure}

The main SM backgrounds are $p p \to \mu \mu \gamma \gamma$ (with an inclusive cross section at LHC 13 TeV of 0.109 pb) and the reducible, but very large $p p \to j j \mu \mu$ background coming dominantly from $Z +$ jets, with the two jets faking photons. We obtain the approximate 
$j\to \gamma$~fake rates in the detector barrel and endcap as a function of the jet $p_T$ from~\cite{Jet_Photon_SummerStudent} (see also~\cite{CMS:2012ad})\footnote{We stress that for 
low jet transverse momentum, $p_T \lesssim 30$ GeV, the uncertainty on the $j\to \gamma$~fake rate becomes rather large. Since this is precisely the most important kinematical region in our analysis (the two photons from the $a \to \gamma\gamma$ decay have fairly low $p_T$ in general), the aforementioned uncertainty may somewhat degrade the sensitivity of the analysis. We leave a detailed investigation of this issue for future work.}. 
We then compute the expected number of 13 TeV LHC SM background and signal (for ${\rm BR}(h\to Z a) \times {\rm BR}(a\to \gamma\gamma) = 1$) events, with an integrated luminosity of $36.1$ fb$^{-1}$, in the $m_{\gamma\gamma}$ signal region as a function of the $m_a$ hypothesis. From these expectations, we derive the 95\% C.L. expected exclusion sensitivity for ${\rm BR}(h\to Z a) \times {\rm BR}(a\to \gamma\gamma)$ as a function of $m_a$ from the present resolved di-photon analysis, using a simple likelihood ratio of signal + background and background only as the test statistic. The CLs method~\cite{Read_2002} is used to quote the 95\% C.L. upper limits. The resulting (expected) limits are shown in Fig.~\ref{fig:HiggsZa:UL_gammagamma}. These limits however do not take into account any potential systematic uncertainties, particularly from our simulation of the reducible $Z +$ jets background and the lack of sufficient statistics in our Monte Carlo event sample given the very large cross section for this process, as well as from the $j\to \gamma$~fake rate (see above). As a result, our derived 95\% C.L. exclusion sensitivity may be regarded as a first approximation, until a more detailed analysis of the reducible SM backgrounds for this process is granted.

Fig.~\ref{fig:HiggsZa:UL_gammagamma} also shows the expected 95\% C.L. exclusion sensitivity on ${\rm BR}(h\to Z a) \times {\rm BR}(a\to \gamma\gamma)$ as a function of $m_a$ from the reinterpretation of the ATLAS $h \to Z \gamma$ analysis~\cite{Aaboud:2017uhw} (when the two photons are too collimated to be resolved by the detector). 
Altogether, we find that for ${\rm BR}(a\to \gamma\gamma) = 1$, values of the exotic Higgs branching fraction ${\rm BR}(h\to Z a) \sim 0.01$ are probed with present LHC data for $m_a < 1$ GeV, 
whereas in the region of the resolved analysis, $10\,{\rm GeV} \lesssim m_a \lesssim 35$ GeV, the sensitivity approaches the per-mille level, reaching values ${\rm BR}(h\to Z a) \sim 0.003$. We also emphasize that it would be possible to significantly close the gap between the $h \to Z \gamma$ limits and our resolved analysis, corresponding to masses $m_a \in [1,\,10]$ GeV, by lowering the $\Delta R$ photon isolation cone used in our analysis down to $\Delta R \sim 0.2$, a study we leave for the future.

\subsection{$a \to \mu\mu$}
\label{sec:HiggsZa:mumu}

Contrary to the case of the previous section, for leptonic decays $a \to \ell \ell$
($\ell = e, \mu$) there exist ATLAS and CMS searches that target the corresponding final state, and so it becomes possible to reinterpret existing LHC searches to obtain limits on ${\rm BR}(h\to Z a) \times {\rm BR}(a\to \ell\ell)$. In this section we concentrate on the decay $a \to \mu \mu$.  
Among the suite of LHC searches 
targeting final states with four leptons, the ATLAS 13 TeV search for exotic Higgs decays via $h \to Z Z_D \to 4 \ell$ (with $Z_{D}$ a BSM gauge boson)~\cite{Aaboud:2018fvk}
provides the most comprehensive coverage of our signal, having two key advantages 
w.r.t.~other analyses: {\it (i)} the analysis considers both electrons and muons (as opposed to~\cite{Sirunyan:2018mgs,ATLAS:2018bsg} which employ only muons); {\it (ii)} the analysis requires a reconstructed $Z$ boson together with a varying mass resonance, as opposed to other analyses which typically consider the decay of the Higgs boson to two resonances of the same mass (see e.g.~\cite{Sirunyan:2018mgs}) and whose di-lepton invariant mass selection requirements discard our signal. 

\begin{figure}[!htp]
  \centering
  \includegraphics[width=0.5\textwidth]{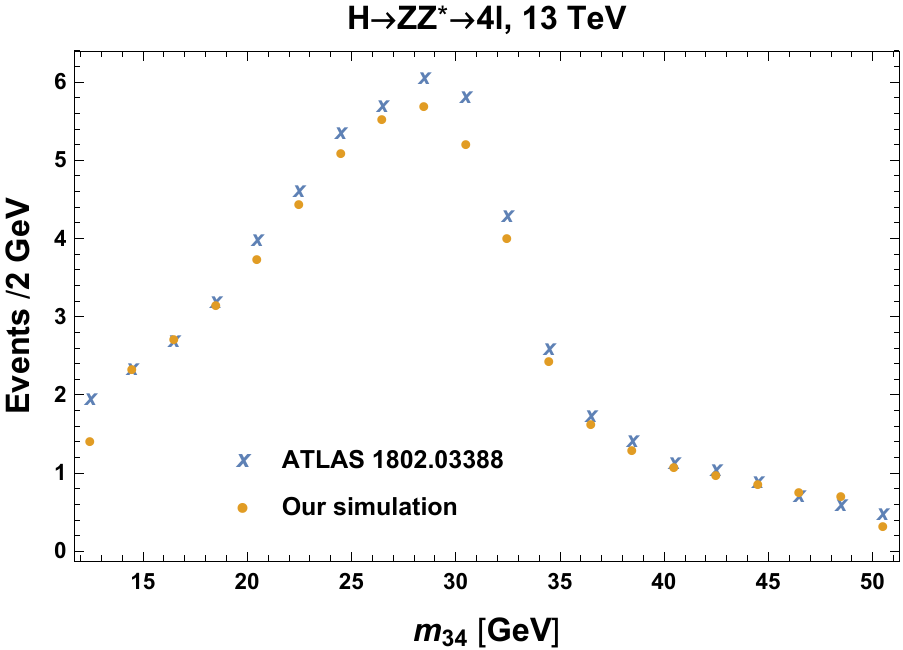}
  \hspace{2mm}
  \includegraphics[width=0.47\textwidth]{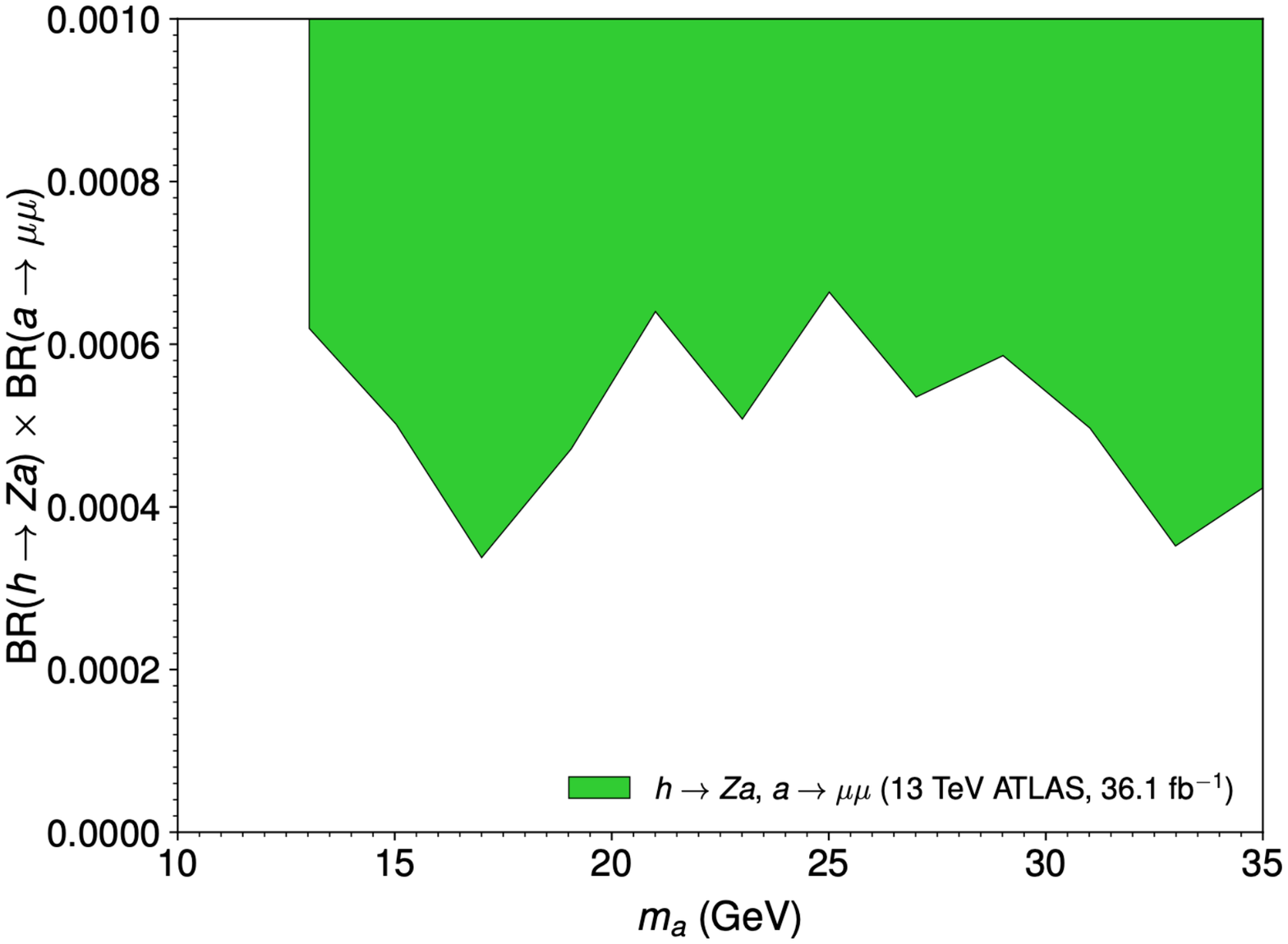} 
  \caption{Left: Comparison of the reported $H \to Z Z^* \to 4\ell$ SM background distribution in the $m_{34}$ variables of the ATLAS 
  collaboration (orange circles) versus our simulation with a flat $0.9$ lepton reconstruction efficiency (blue crosses), for the 13 TeV ATLAS analysis~\cite{Aaboud:2018fvk}. Right: 
  95\% C.L. excluded values of ${\rm BR}(h\to Z a) \times {\rm BR}(a\to \mu\mu)$ as a function of $m_a$.}
\label{fig:HiggsZa:ATLAS4l}
\end{figure}

We then reinterpret in this work the ATLAS analysis~\cite{Aaboud:2018fvk} to derive the present LHC sensitivity to $p p \to h \to Z a$ ($Z \to \ell\ell$, $a \to \mu\mu$).  
We have validated our analysis by first considering a parton level\footnote{A full-fledged analysis including parton shower, hadronization and detector simulation is warranted, but outside of the scope of this work and will be presented elsewhere.} sample of the largest SM background, $h \to Z Z^* \to 4\ell$, and comparing the acceptance of our Monte Carlo sample 
generated with \mg~with the $m_{34}$ kinematical distribution published by ATLAS (Fig.~2 
of~\cite{Aaboud:2018fvk}). The event selection requires
\begin{itemize}
    \item Two pairs of same-flavour (SF), opposite-sign (OS) leptons, $\ell_{i}$.
    \item $p_T(\ell_1,\,\ell_2,\,\ell_3) > 20,\,15,\,10$ GeV.
    \item $\Delta R (\ell, \ell^{'}) > 0.1\, (0.2)$ for SF (OF).
    \item 115 GeV $ < m_{4\ell} < $ 130 GeV.
    \item $m_{i,j} > 5 $ GeV.
    \item 50 GeV $< m_{1,2} < 106$ GeV and 12 GeV $< m_{3,4} <$~115 GeV.
\end{itemize}
with $i, j = 1,2,3,4$ indexing the leptons.
After these cuts the ATLAS collaboration displays the $m_{34}$ mass distribution (namely, the pair of leptons that do not reconstruct the $Z$ mass) in the range 
$m_{34} \in [11,\,59]$ GeV. In order to account for the detector effects not included in the parton level generation we have assumed a flat lepton reconstruction efficiency of $\sim0.9$. The results of our validation are shown in the left panel of Fig.~\ref{fig:HiggsZa:ATLAS4l}, and are in reasonable agreement with each other. 

After validation, we proceed to derive the limits on 
${\rm BR}(h\to Z a) \times {\rm BR}(a\to \mu\mu)$ as a function of the pseudoscalar mass $m_a$. We use \mg~for the signal event generation (with the same UFO model as in section~\ref{sec:HiggsZa:gammagamma}, and the same N$^{3}$LO value for the single Higgs production cross section from~\cite{Anastasiou:2016cez}). 
In order to derive a 95\% C.L. exclusion for our model, we compare the $2\sigma$ upper limits quoted by ATLAS with our expected signal rate as a function of 
${\rm BR}(h\to Z a) \times {\rm BR}(a\to \mu\mu)$. 
For each mass point, we consider for this purpose 
the most constraining individual bin of the $m_{34}$ distribution. 
The resulting LHC 13 TeV 95\% C.L. exclusion for ${\rm BR}(h\to Z a) \times {\rm BR}(a\to \mu\mu)$ with 36.1 fb$^{-1}$ is shown in the right panel of Fig.~\ref{fig:HiggsZa:ATLAS4l} as a function of $m_a$. 
While this search can in principle provide sensitivity to masses $m_a > 35$ GeV (up to around $50$ GeV)
for an $h \to Z^* a$ decay, we concentrate here in the region $m_a < 35$ GeV for which the Higgs boson can decay into on-shell $Z$ and $a$ states.

\subsection{$a \to \tau\tau$}
\label{sec:HiggsZa:tautau}

As in section~\ref{sec:HiggsZa:mumu} for the $a \to \mu\mu$ decay mode, existing 8 TeV and 13 TeV ATLAS and CMS searches may be used to probe the $p p \to h \to Z a$ ($Z \to \mu\mu$, $a \to \tau\tau$) signature at the LHC.  
However, for the $\mu\mu\tau\tau$ final state all the available searches target the decay $h \to a a \to \mu\mu\tau\tau$~\cite{CMS:2016cqw,Sirunyan:2018mbx,Aad:2015oqa}, whereas in our case the invariant masses for the di-muon and the di-tau systems will be rather different for on-shell decays $h \to Z a$. In particular, the 8 TeV ATLAS analysis~\cite{Aad:2015oqa} requires a reconstructed di-muon mass $m_{\mu\mu} \in [2.8,\,70]$ GeV, and the 13 TeV CMS analysis~\cite{Sirunyan:2018mbx} requires a reconstructed di-muon mass $m_{\mu\mu} \in [14,\,64]$ GeV. Thus, both analyses 
discard\footnote{These searches may nevertheless be sensitive to 
$h \to Z^* (\to \mu\mu) \,a$, for $m_a \gtrsim 50$ GeV. We defer a study of this region of parameter space for the future.} our signal events $h \to Z (\to \mu\mu)\, a$. 

Fortunately, the signal event selection from the 8 TeV CMS search~\cite{CMS:2016cqw} for $h \to a a \to \mu\mu\tau\tau$ events provides coverage to pseudoscalar masses in the approximate range $m_a \in [18,\,34]$ GeV for our $h \to Z a$ signal.
The CMS analysis is divided according to the various decay modes of the $\tau$ leptons: leptonic (into an electron, $\tau_e$, or a muon, $\tau_\mu$) or hadronic ($\tau_h$), with the following final states considered: $\mu\mu\,\tau_e\tau_e$, $\mu\mu\,\tau_e\tau_\mu$, $\mu\mu\,\tau_e\tau_h$, $\mu\mu\,\tau_\mu\tau_h $ and $\mu\mu\,\tau_h\tau_h$. The two reconstructed muons are required to satisfy $|\eta| < 2.4$ and $p_T > 18$ GeV ($p_T > 9$ GeV)\footnote{For the $\mu\mu\,\tau_\mu\tau_e$ and $\mu\mu\,\tau_\mu\tau_h$ final states, the subleading muon from the $a \to \mu\mu$ decay may be selected with $p_T > 5$ GeV provided that the muon from the $\tau$ decay satisfies $p_T > 9$ GeV, in order for the event to be triggered upon.} for the leading (subleading) muon, together with certain isolation criteria.
For the tau leptons, a reconstructed $\tau_e$ is required to satisfy $|\eta| < 2.5$ and $p_T > 7$ GeV, a reconstructed $\tau_\mu$ is required to satisfy $|\eta| < 2.4$ and $p_T > 5$ GeV (see however footnote 8) and a reconstructed $\tau_h$ is required to satisfy $|\eta| < 2.3$ and $p_T > 15$ GeV. In all cases, there are further isolation and reconstruction criteria which must be satisfied (see~\cite{CMS:2016cqw} for details). In addition, the distance between any pair of leptons in the event is required to be $\Delta R_{\ell\ell} > 0.4$, and the reconstructed invariant masses for the $\mu\mu$, $\tau\tau$ and $\mu\mu\tau\tau$ systems need to satisfy $|m_{\mu\mu\tau\tau} - 125 \,\mathrm{GeV}| < 25$ GeV and $|m_{\mu\mu} - m_{\tau\tau}|/m_{\mu\mu} < 0.8$. Under the assumption of perfect $m_{\mu\mu}$ and $m_{\tau\tau}$ reconstruction for our $h \to Z a \to \mu\mu\tau\tau$ signal, satisfying the latter cut would require $m_{\tau\tau} = m_a > 18$ GeV (since $m_{\mu\mu} = m_Z = 91$ GeV). Due to the presence of neutrinos in the decay of the $\tau$-leptons, the invariant mass reconstruction for di-tau resonances generally involves the use of likelihood algorithms~\cite{Elagin:2010aw,Bianchini:2014vza}. From the public information of the performance of these methods in the reconstruction of SM Higgs and $Z$ bosons in $h \to \tau\tau$ and $Z \to \tau\tau$ decays~\cite{Elagin:2010aw}, we assume here that $m_{\tau\tau}$ and $m_{\ell\ell\tau\tau}$ are smeared compared to their true values with a Gaussian with RMS width~$\sigma = 0.15\, m_a$ (incidentally, this makes the analysis sensitive to pseudoscalar masses slightly below 18 GeV).


We perform a validation of our analysis by generating $p p \to h \to a a \to \mu\mu\tau\tau$ signal events with \mg~and \pythia~for the two mass points $m_a = 20$ GeV and $m_a = 60$ GeV detailed in the CMS analysis~\cite{CMS:2016cqw}, obtaining the expected number of signal events in each of the possible di-tau decay modes considered: $\tau_e\tau_e $, $\tau_e\tau_\mu$, $\tau_e\tau_h $ $\tau_\mu\tau_h$ and $\tau_h\tau_h$. Our results tend to overestimate the number of expected events for categories including $\tau_e$ (by a factor $\sim 1.5-2$), and particularly $\tau_h$, while for categories with $\tau_\mu$ we agree better with the CMS expected number of signal events (or slightly underestimate it). We then define a fudge factor (which we define for each category from a linear interpolation between our validation results for $m_a = 20$ GeV and $m_a = 60$ GeV) to approximately account for the mismatch between the CMS expected number of signal events and ours (typically larger) due to features of the CMS selection which we do not capture, such as particle isolation and the precise performance of the $m_{\tau\tau}$ reconstruction algorithm at low invariant masses.

\begin{figure}[!htp]
  \centering
  \includegraphics[width=0.70\textwidth]{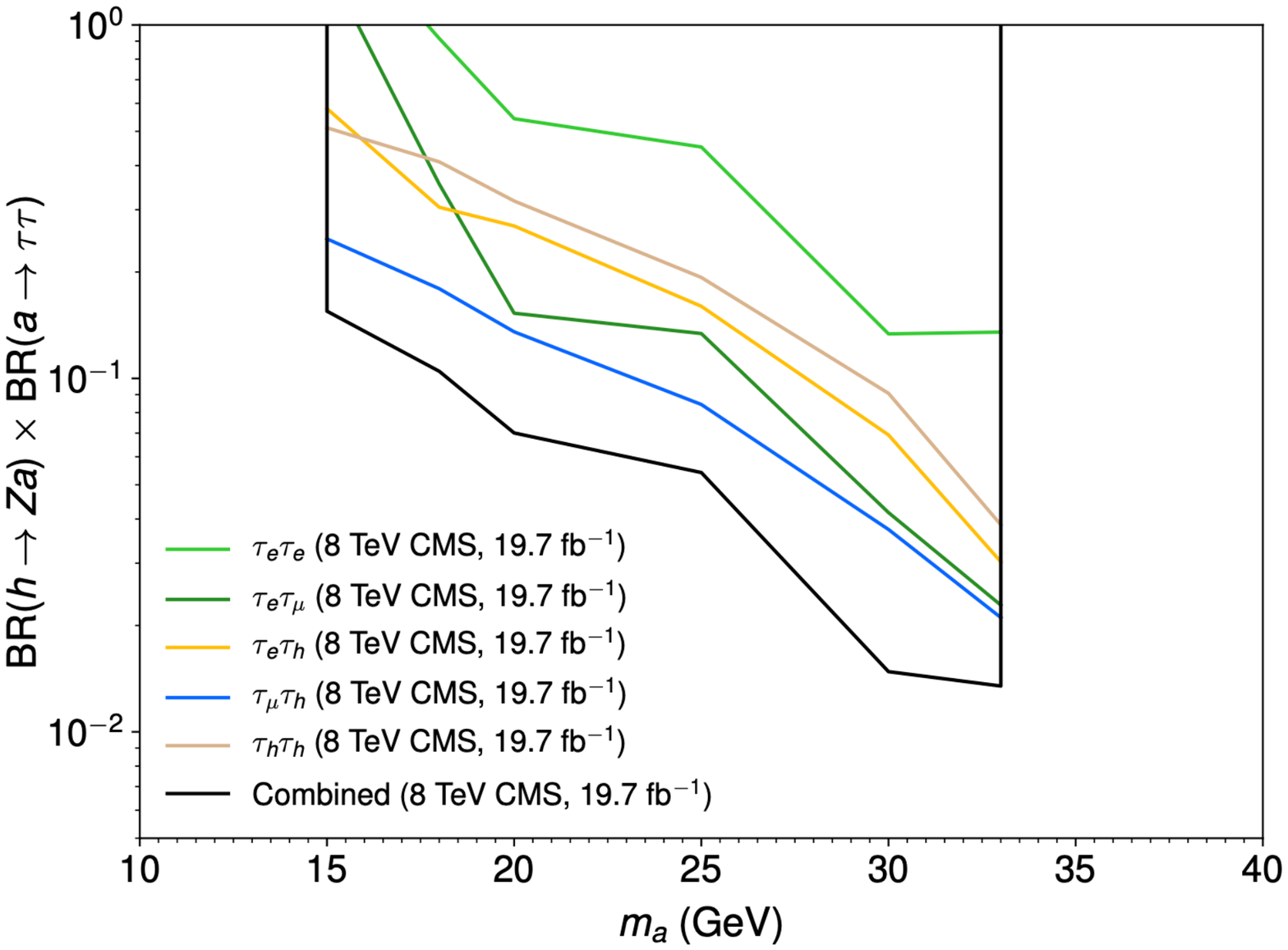} 
  \caption{95\% C.L. excluded values of ${\rm BR}(h\to Z a) \times {\rm BR}(a\to \tau\tau)$ as a function of $m_a$ from the reinterpretation of the 8 TeV CMS analysis~\cite{CMS:2016cqw}. We show separately the limits from the $\mu\mu\,\tau_e\tau_e$ (light green), $\mu\mu\,\tau_e\tau_\mu$ (dark green), $\mu\mu\,\tau_e\tau_h$ (yellow),  $\mu\mu\,\tau_\mu\tau_h$ (blue) and $\mu\mu\,\tau_h\tau_h$ (brown), as well as the combination (black).}
\label{fig:HiggsZa:tautau}
\end{figure}

We then generate 8 TeV LHC signal events for $p p \to h \to Z(\to \mu\mu) a$, $a \to \tau\tau$ with \mg~and \pythia~in the mass range $m_a \in [15,\, 33]$ GeV, and compute the expected number of events after selection (for ${\rm BR}(h\to Z a) \times {\rm BR}(a\to \tau\tau) = 1$) with an integrated luminosity of $19.7$ fb$^{-1}$. From these expectations, we derive the 95\% C.L. exclusion sensitivity on ${\rm BR}(h\to Z a) \times {\rm BR}(a\to \tau\tau)$ as a function of $m_a$ using the same likelihood method as in section~\ref{sec:HiggsZa:gammagamma}. The resulting limits are shown in Fig.~\ref{fig:HiggsZa:tautau} for each of the five search categories, as well as the combined limit.

\section{Model interpretation: Axion-like particles} 
\label{sec:HiggsZa:ALPs}

While the previous LHC sensitivity studies of the exotic Higgs boson decay $h \to Z a$ ($a \to \gamma\gamma, \ell\ell$) are model independent, it is also useful to cast the results into 
specific, well-motivated BSM theories. This also allows to explore the complementarity with other probes of the parameter space of such theories, an aspect we leave for a future study.
In this work we consider the EFT for the SM extended by an axion-like particle, $a$~\cite{Georgi:1986df,Brivio:2017ije,Bauer:2017ris,Quevillon:2019zrd}. 
This ALP extension is inspired by axion models which aim, amongst other potential motivations, to solve the strong CP problem, and at the same time provides a flexible phenomenological framework to interpret the constraints from the LHC. 

In full generality, ALPs are assumed to couple both to SM gauge bosons and fermions. Only including terms up to dimension-5 in a linear EFT expansion, we therefore modify the SM Lagrangian as follows:
\begin{eqnarray}
  {\cal L}^{d\le 5}_{\rm eff}\supset \mathcal{L} _{\rm SM}+\frac{1}{2} \partial_\mu a \partial^\mu a - \frac{1}{2}m^{2}_{a} a^2+ \sum_f \frac{m_f}{\Lambda}C_f\, a \bar f \gamma_5 f  -\frac{e^2}{4 \Lambda} \,C_{\gamma}\, a \, F_{\mu\nu}\tilde{F}_{\mu\nu} \label{eq:HiggsZa:lagrangian}.
\end{eqnarray}
where the coupling  of the ALP to the photon is given by $C_{\gamma}$ and the coupling to the fermion $f$ (of mass $m_f$) is given by $C_f$. The ALP could couple to the other gauge 
bosons (see e.g. the discussion in~\cite{Alonso-Alvarez:2018irt}) but this is not relevant to our analysis. Note that in this Lagrangian the ALP couplings to the Higgs boson are not present. In the linear EFT expansion, the dimension-5 coupling of the Higgs boson $h$ to a $Z$ boson and the ALP vanishes~\cite{Georgi:1986df,Brivio:2017ije,Bauer:2017ris}, as the operator can be rewritten in terms of the ALP-fermion operators using the equations of motion. There are two possible sources\footnote{This is so in the context of a linear EFT that we are considering here. For a {\sl chiral} EFT expansion the ALP-$h$-$Z$ coupling is present at the lowest (leading) order in the expansion, see~\cite{Brivio:2017ije} for details.} for the decay $h\to Z a$. The first is from higher-dimensional terms, which can be incorporated in a straightforward way by including the following effective interaction in our Lagrangian:
\begin{equation}
\label{eq:HiggsZa:Leff6}
   {\cal L}_{\rm eff}^{d= 7}
  \supset \frac{1}{\Lambda^3} C_{Zh}^{(7)} \left( \partial^\mu a\right) 
    \left( \phi^\dagger\,i \overset{\text{\tiny$\leftrightarrow$}}{D}_\mu\,\phi + \mbox{h.c.} \right) \phi^\dagger\phi + \dots \,,
\end{equation}
where $C_{Zh}^{(7)}$ is the effective coupling between the ALP and the Higgs and $Z$ bosons~\cite{Bauer:2017nlg}. The second source of this coupling is loop effects, notably from the top quark, as detailed in Ref.~\cite{Bauer:2017nlg}. The full effective coupling between the Higgs, the $Z$ boson and the ALP can then be written as:
\begin{equation}
  C_{Zh}^{\rm eff} = - \frac{N_c\,y_t^2}{8\pi^2}\,T_3^t\,C_{t}\,F
     + \frac{v^2}{2\Lambda^2}\,C_{Zh}^{(7)} \,.
\end{equation}
where $F$ is defined in Ref.~\cite{Bauer:2017nlg} via
\begin{equation}
   F = \int_0^1\!d[xyz]\,\frac{2m_t^2-x m_h^2-z m_Z^2}{m_t^2-xy m_h^2-yz m_Z^2-xz m_a^2} \,,
\end{equation}
using the abbreviation $d[xyz]\equiv dx\,dy\,dz\,\delta(1-x-y-z)$.
Note that the numerical size of the coefficients of $C_t$ and $C_{Zh}^{(7)}$ is comparable for an EFT cut-off scale $\Lambda=1$ TeV.
In terms of this effective coupling $C_{Zh}^{\rm eff}$, the Higgs partial decay width into $Z a$, for $m_a < m_h - m_Z$, can be expressed as
\begin{equation}
   \Gamma_{h\to Za} = \frac{m_h^3}{16\pi\Lambda^2} \left| C_{Zh}^{\rm eff} \right|^2 
     \lambda^{3/2}\bigg(\frac{m_Z^2}{m_h^2},\frac{m_a^2}{m_h^2}\bigg) \,.
\end{equation}
with $\lambda(x, y) = (1 - x - y)^2 - 4 x y$. 

Having established the expression for the decay width of $h\to Z a$, we now determine the different possibilities for the decay of the ALP in this model. As we are focusing on ALPs below the electroweak scale (in fact, lighter than the $Z$ boson), clearly ALP decays into (on-shell) massive electroweak bosons are not possible, leaving fermion pairs, photons and gluons as potential decay signatures.
In the following we define two model benchmarks for which we analyse the LHC sensitivity discussed in the previous sections. In the first, we neglect the ALP-fermion couplings, setting $C_f=0$. 
In the second, we choose the opposite situation where all the ALP coupling to 
photons vanishes ($C_\gamma=0$), and consider {\it i)} all $C_f$ to be equal and {\it ii)} only $C_\mu$ or $C_\tau$ to be non-zero. 

\subsection{$C_f=0$} 
\label{sec:HiggsZa:ALPs_BP1}

\begin{figure}[htbp]
  \begin{center}
    \includegraphics[width=0.7\textwidth]{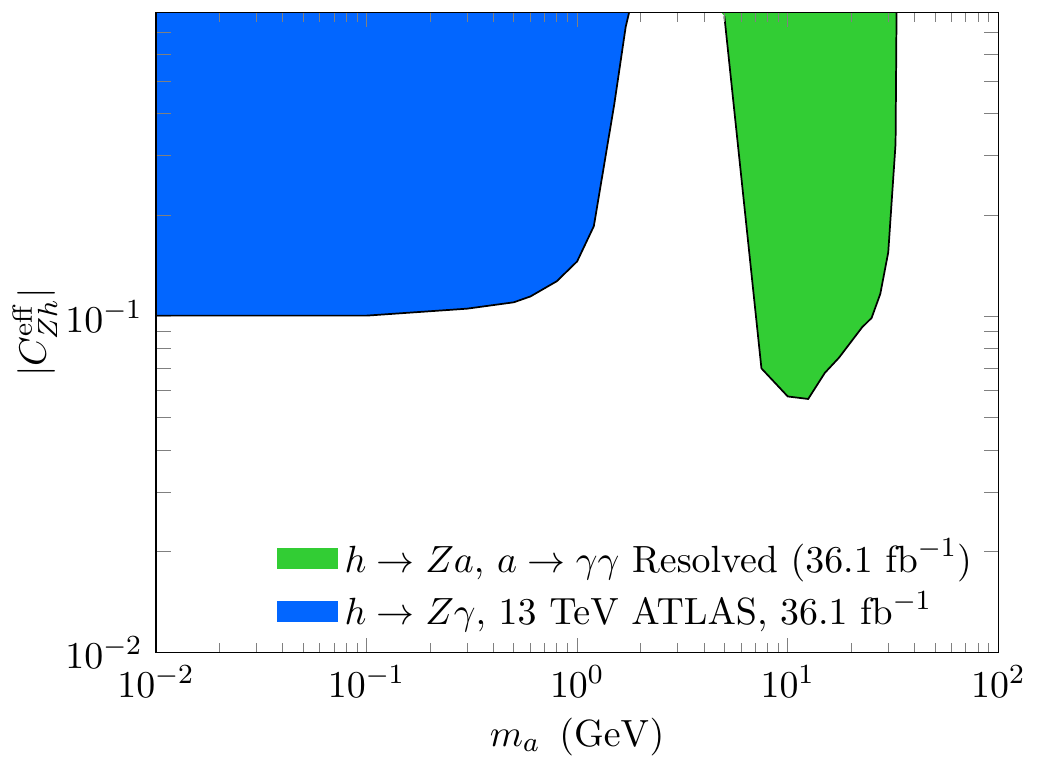}
    \caption{95\% C.L. upper limits on the Wilson coefficient $|C_{Zh}^{\mathrm{eff}}|$ for $C_f=0$ from bounds on ${\rm BR}(h\to Z a) \times {\rm BR}(a\to \gamma\gamma)$ where the photons are resolved (green) and from bounds on $h \to Z \gamma$ when the two photons are too collimated to be resolved by the detector (blue) at LHC 13 TeV with $36.1$ fb$^{-1}$. We have assumed that the ALP does not couple to fermions, and $\Lambda=1$ TeV.}
    \label{fig:HiggsZa:Cgamma}
   \end{center}
\end{figure}

In such a case, the decay rate of the ALP into photons is given by~\cite{Bauer:2017nlg} 
\begin{equation}
      \Gamma(a\to\gamma\gamma) 
   = \frac{4\pi\alpha^2 m_a^3}{\Lambda^2}\,\big| C_{\gamma} \big|^2 \,.
\end{equation}
For this benchmark point, in the kinematic range that we are considering, the branching fraction to photons will be 100\%. Therefore limits on the $h\to Z a\to\gamma\gamma$ final state will provide a bound on the Wilson coefficient $C_{Zh}^{\mathrm{eff}}(= v^2/(2\Lambda^2)\,C_{Zh}^{(7)})$. Note that we do not make any assumption about $C_{\gamma}$ other than that it is sufficiently large that the ALP decays within the detector. The results are presented in Fig.~\ref{fig:HiggsZa:Cgamma}. We find that the exclusion on $C_{Zh}^{\mathrm{eff}}$ goes down to $\sim 10^{-1}$ for small values of $m_a$ where the photons are unresolved, and down to $\sim 5\cdot 10^{-2}$ at $m_a=10$ GeV for the case that the photons are resolved, assuming that $\Lambda=1$ TeV. Note that the difference from Ref.~\cite{Bauer:2017nlg} in the blue region corresponding to the limits from $h\to Z\gamma$ arises due to the fact that we allow $C_\gamma$ to vary such that the ALP always decays within the detector.

\vspace{-3mm}

\subsection{$C_\gamma=0$} 
\label{sec:HiggsZa:ALPs_BP2}
The second benchmark point is the case where all couplings to gauge bosons vanish. In this case, the decay rate into charged leptons is given by ~\cite{Bauer:2017nlg}
\begin{equation}
   \Gamma(a\to\ell^+\ell^-) 
   = \frac{m_a m_\ell^2}{8\pi\Lambda^2} \left| C_{\ell}^{\rm eff} \right|^2
    \sqrt{1-\frac{4m_\ell^2}{m_a^2}} \,.
\end{equation}
\begin{figure}[t]
  \begin{center}
    \includegraphics[width=0.48\textwidth]{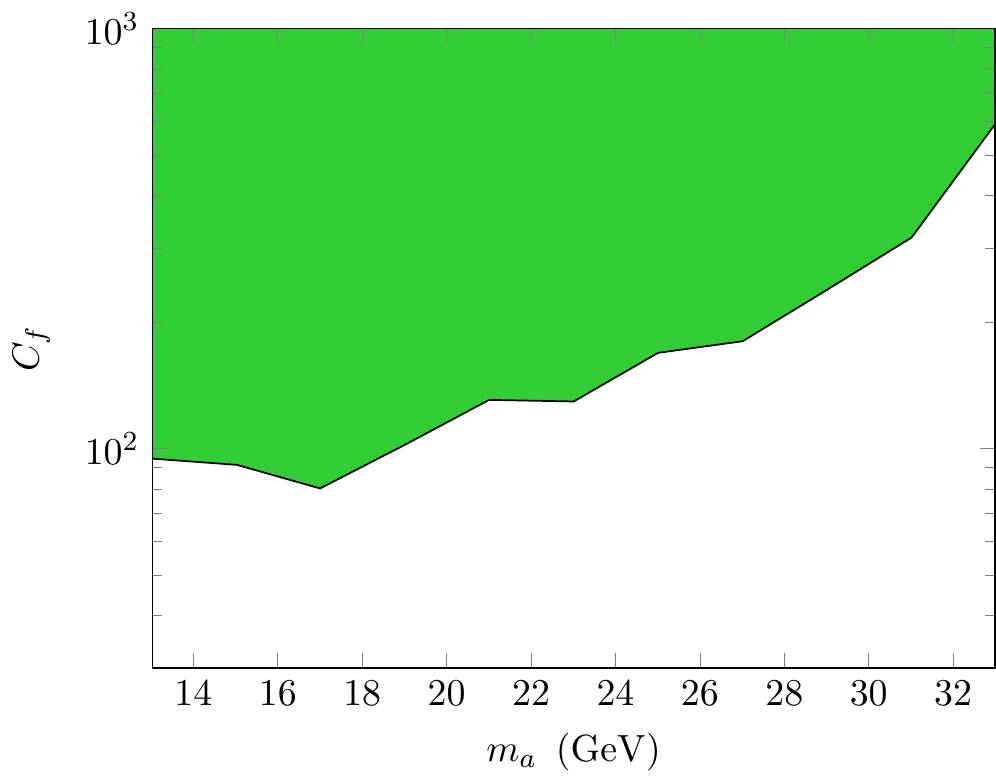}\hspace{.5cm}\includegraphics[width=0.48\textwidth]{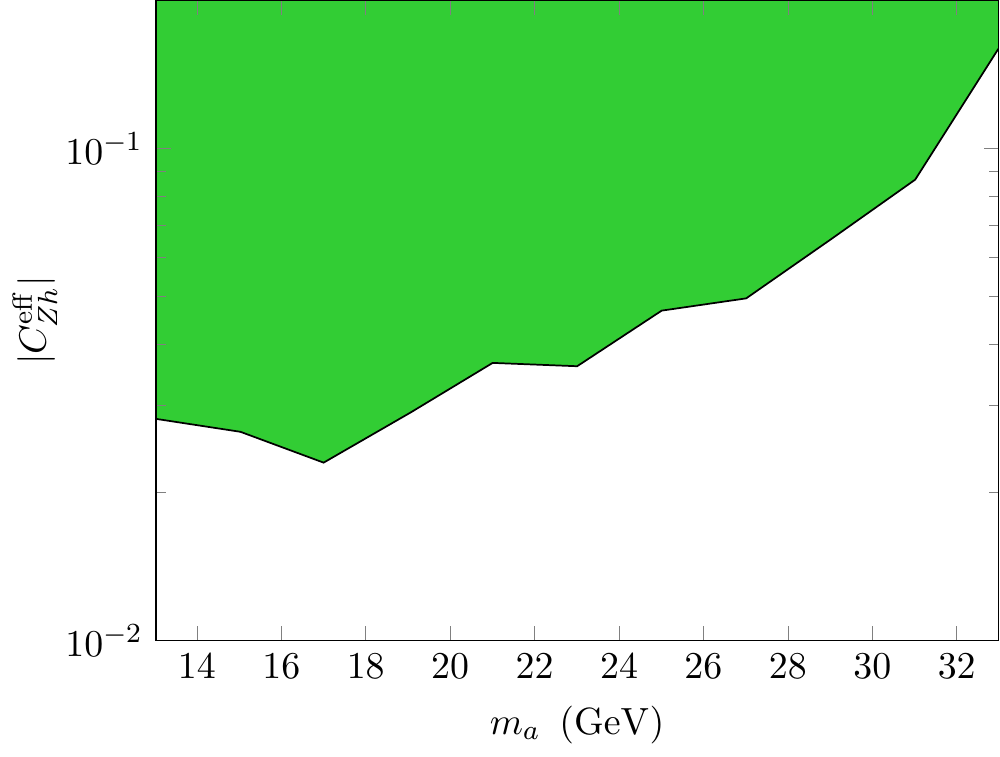}
    \caption{95\% C.L. upper limits on: the Wilson coefficient $C_{f}$ assuming a universal coupling of the ALP to fermions and that $C_{Zh}^{(7)}=0$ (left); the Wilson coefficient $|C_{Zh}^{\mathrm{eff}}|$ assuming that the ALP decays 100\% to muons (right). These limits are taken from the bounds obtained in Sec.~\ref{sec:HiggsZa:mumu} on ${\rm BR}(h\to Z a) \times {\rm BR}(a\to \mu\mu)$ in the case that $C_\gamma=0$, assuming $\Lambda=1$ TeV.}
    \label{fig:HiggsZa:Cmu}
   \end{center}
   
   \vspace{-4mm}
   
\end{figure}
\begin{figure}[t]
  \begin{center}
    \includegraphics[width=0.48\textwidth]{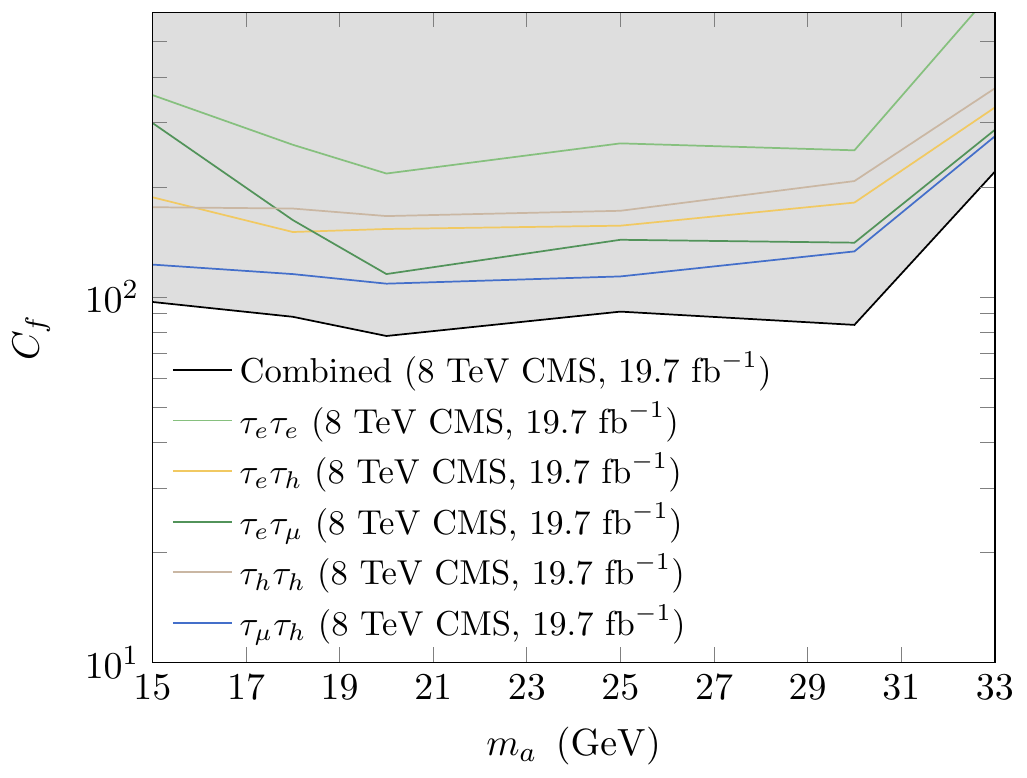}\hspace{.5cm}\includegraphics[width=0.48\textwidth]{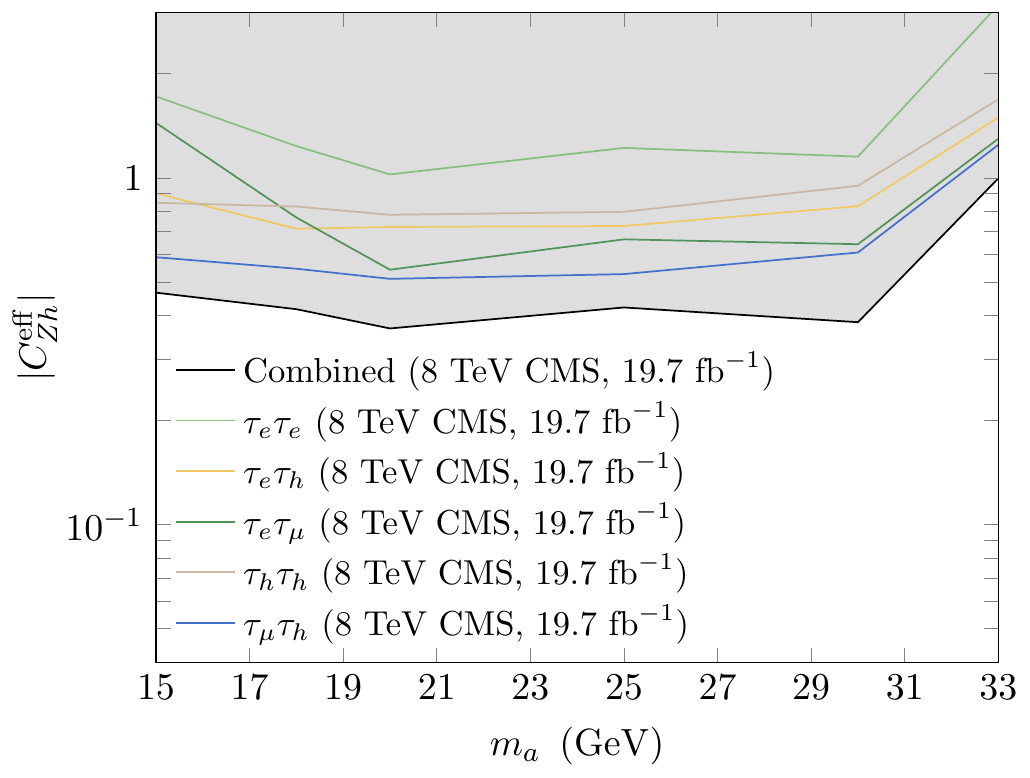}
    \caption{95\% C.L. upper limits on: the Wilson coefficient $C_{f}$ assuming a universal coupling of the ALP to fermions and that $C_{Zh}^{(7)}=0$ (left); the Wilson coefficient |$C_{Zh}^{\mathrm{eff}}|$ assuming that the ALP decays 100\% to tau leptons (right). These limits are taken from the bounds obtained in Sec.~\ref{sec:HiggsZa:tautau} on ${\rm BR}(h\to Z a) \times {\rm BR}(a\to \tau\tau)$ in the case that $C_\gamma=0$, assuming $\Lambda=1$ TeV.}
    \label{fig:HiggsZa:Ctau}
   \end{center}
   
   \vspace{-6mm}
   
\end{figure}
We consider two possibilities: a) the case where all $C_f$ are equal and b) the case where only the muon or tau lepton coupling is present, and all other fermion couplings to the ALP vanish. In scenario a), for the $h\to Za$ decay we assume that $C_{Zh}^{(7)}=0$, and that this decay is mediated by the top loop. For the ALP decay, in order to calculate the branching fraction several decay modes need to be accounted for, i.e.~decays into hadrons and into heavy quarks. The necessary decay rates are~\cite{Bauer:2017nlg}:
\begin{align}
   \Gamma(a\to\mbox{hadrons}) 
   \,=&\, \frac{\alpha_s^2(m_a) n_q^2\,m_a^3}{{32\pi^3}\Lambda^2}
    \left[ 1 + \left( \frac{97}{4} - \frac{7n_q}{6} \right) \frac{\alpha_s(m_a)}{\pi} \right]
  |C_{f}|^2, \\
       \Gamma(a\to Q\bar Q) 
   \,=&\,\frac{3m_a\,\overline{m}_Q^2(m_a)}{8\pi\Lambda^2}
    \sqrt{1-\frac{4m_Q^2}{m_a^2}} \,\left| C_{f} \right|^2\,,
\end{align}
where the number of light quarks, $n_q$, is equal to 3 and $\overline{m}_Q$ is the MS-bar mass of the heavy quark $Q$. In scenario b), we assume that the contribution of $C_{Zh}^{\mathrm{eff}}$ to the $h\to Za$ decay width dominates, and that the branching ratio of the ALP to muons or taus is 100\% respectively. Again we do not make any assumption about the size of $C_{\mu}$ or $C_\tau$ beyond that it is large enough such that the ALP decays within the detector. The results are presented in Fig.~\ref{fig:HiggsZa:Cmu} and in Fig.~\ref{fig:HiggsZa:Ctau}. We see that for the case a) the lowest bound on $C_f$ in the $\sim 15-30$ GeV region is $\sim 80$ for both the muon and tau lepton searches, when $\Lambda=1$ TeV. In the case b) the constraint on $C_{Zh}^{\mathrm{eff}}$ from the muon final state ($\sim 2\cdot 10^{-2}$) is stronger than that from tau leptons ($\sim 4\cdot 10^{-1}$). Clearly very large values of the coupling $C_f$ or conversely low scales are still allowed.

\section{Conclusions}
\label{sec:HiggsZa:conclusions}

In this contribution, we have considered the exotic decay of the 125 GeV Higgs boson $h$ into a $Z$ boson and a BSM pseudoscalar state $a$. This Higgs decay mode may occur in many BSM scenarios, such as axion-like particle extensions of the SM or theories with non-minimal Higgs sectors. The corresponding branching fraction of the Higgs is currently allowed at the $\mathcal{O}(20\%)$ level by LHC measurements of Higgs signal strengths, and could be probed much more efficiently via dedicated searches. Here we study specific final states where the pseudoscalar $a$ decays into photons, muons or $\tau$-leptons. In the former, we perform an analysis of the 13 TeV LHC sensitivity with 36.1 fb$^{-1}$ integrated luminosity, in the mass range $m_a \in [5, 34]$ GeV, where the two photons from the $a \to \gamma\gamma$ decay are isolated/resolved. In this region we reach a sensitivity to BR$(h \to Z a)\times$ BR$(a \to \gamma\gamma) \sim 0.003$. Using ATLAS searches for the $h \to Z \gamma$ Higgs decay mode we also derive limits for pseudoscalar states lighter that $1$ GeV, when the two photons are very collimated and appear as one in the detector, constraining in this case BR$(h \to Z a)\times$ BR$(a \to \gamma\gamma) \sim 0.01$. 

For the leptonic decays of $a$ we reinterpret existing ATLAS and CMS searches: to probe the $a \to \mu\mu$ decay mode, we use the 13 TeV ATLAS search (with 36.1 fb$^{-1}$ integrated luminosity) for exotic Higgs decays into four leptons via 
$h \to Z Z_{D}$~\cite{Aaboud:2018fvk}, constraining values of BR$(h \to Z a)\times$ BR$(a \to \mu\mu)$ in the range $\sim (4 - 6)\times 10^{-4}$ for $m_a \in [10,\,35]$ GeV. For the $a \to \tau\tau$ decay mode, despite the existence of 13 TeV LHC searches targeting the decay mode $h \to \mu\mu \, \tau\tau$, these are not sensitive to the kinematic region of our $h \to Z a $, $Z \to \mu\mu$ $a \to \tau\tau$ signal. However, we find the 8 TeV CMS search for exotic Higgs decays $h \to a a \to \mu\mu \tau\tau$ with 19.7 fb$^{-1}$ integrated luminosity~\cite{CMS:2016cqw} provides sensitivity to our signal in the mass range $m_a \in [15, \,33]$ GeV, reaching values BR$(h \to Z a)\times$ BR$(a \to \tau\tau) \sim 0.02$ in the upper end of that mass range.        

Finally, we cast our limits into the linear EFT extension of the SM by an axion-like 
particle, considering two simple benchmarks where respectively the coupling of the ALP to SM fermions $C_{f}$ or to photons $C_{\gamma}$ vanishes.
In the former, the $a \to \gamma\gamma$ search from section~\ref{sec:HiggsZa:gammagamma} would allow values of $C_{Zh}^{\mathrm{eff}} \lesssim 0.1$ TeV$^{-1}$ to be probed with $36.1$ fb$^{-1}$ of LHC 13 TeV data. In the latter, existing ATLAS and CMS analyses are sensitive to 
$C_{Zh}^{\mathrm{eff}} \lesssim 0.03$ TeV$^{-1}$ ($0.4$ TeV$^{-1}$) when the decay $a \to \mu\mu$ ($a \to \tau \tau$) is the dominant ALP decay mode. These searches also allow the ALP-fermion coupling $C_{f}$ to be probed when it is universal (all SM fermions having the same $C_{f}$) and responsible for the decay $h \to Z a$ via loop effects.

\section*{Acknowledgements}

We would like to thank the organisers of the Les Houches "Physics at TeV  Colliders" 2019 workshop for their warm hospitality and for the very friendly and productive environment. We also thank Martin Bauer and Andrea Thamm for providing us with the ALP UFO model used in this work, as well as Satyaki Bhattacharya for his help in setting the limit tool. J.M.N. was supported by Ram\'on y Cajal Fellowship contract RYC-2017-22986, and acknowledges further support from the Spanish MINECO's ``Centro de Excelencia Severo Ochoa" Programme under grant SEV-2016-0597, from the European Union's Horizon 2020 research and innovation programme under the Marie Sklodowska-Curie grant agreements 690575  (RISE InvisiblesPlus) and 674896 (ITN ELUSIVES) and from the Spanish Proyectos de I$+$D de Generaci\'on de Conocimiento via grant PGC2018-096646-A-I00.
K.T. is supported by the US Department of Energy grant DE-SC0010102 and also by his startup fund at Florida State University (Project id: 084011- 550- 042584).

\let\mg\undefined
\let\pythia\undefined
\let\delphes\undefined

%% file: vbf/VBF.main.tex
\graphicspath{{vbf/}}

\newcommand{\Herwig}{H\protect\scalebox{0.8}{ERWIG}\xspace}
\newcommand{\Pythia}{P\protect\scalebox{0.8}{YTHIA}\xspace}
\newcommand{\Sherpa}{S\protect\scalebox{0.8}{HERPA}\xspace}
\newcommand{\Rivet}{R\protect\scalebox{0.8}{IVET}\xspace}
\newcommand{\Professor}{P\protect\scalebox{0.8}{ROFESSOR}\xspace}
\newcommand{\eps}{\varepsilon}
\newcommand{\mc}[1]{\mathcal{#1}}
\newcommand{\mr}[1]{\mathrm{#1}}
\newcommand{\mb}[1]{\mathbb{#1}}
\newcommand{\tm}[1]{\scalebox{0.95}{$#1$}}


\chapter{Extra Scalar Boson Searches at the LHC through Vector Boson Fusion}
{\it S.~Fichet, S.~Gascon-Shotkin, A.~Lesauvage, G.~Moreau}


\label{sec:VBF}

\begin{abstract}
Extra scalar bosons predicted in several scenarios beyond the Standard Model (SM) might have been missed so far by the present experimental 
searches based on the Large Hadron Collider (LHC) data. We study the Vector Boson Fusion (VBF) mode of production of a such a scalar which has the generic possibility to be comparable in rate 
with the main gluon-gluon Fusion (ggF) mechanism. The VBF presents a different final state from the ggF, with two additional jets that can optimise the sensitivity reach of the signal
and prove to be particularly helpful for the tagging of final state when the diphotons from the decays of a too light scalar field do not allow an efficient triggering.
We show that the VBF productions of generic neutral CP-odd or CP-even scalar fields exhibit specific kinematical distributions which could indeed allow to
distinguish between their signatures at LHC and the SM background. In order to reach this conclusion, we have used a theoretical framework based on an effective field theory 
including various types of interactions, and, run Monte Carlo simulations for the signal and background in order to confront them with each other. 
\end{abstract}

\section{Introduction}

Many extensions of the Standard Model (SM) of elementary particle physics include an electrically 
neutral scalar boson in their field content with a mass below a hundred GeV: light Higgs bosons,
the radion, the dilaton, the axion\dots Although in 2012 the Large Hadron Collider (LHC) discovered a particle compatible with the SM Higgs boson as a $125$~GeV resonance, one should thus consider the 
possibility that some lighter or heavier scalar particles might have been missed so far in the LHC data analyses.

Regarding the SM Higgs boson rates, the second production mechanism at a proton beam machine of $14$~TeV -- after the gluon-gluon Fusion (ggF) -- is the Vector Boson Fusion (VBF),
over an hypothetical mass range between $100$~GeV and $1$~TeV. Hence one could expect that the VBF is crucial as well for an extra (pseudo)scalar field production. 
Furthermore, depending on the scenario where the scalar field is arising, the photon contribution to the VBF might be competitive with the $Z^0$ and $W^\pm$ boson exchanges since 
both the extra scalar couplings to the photons and to the ElectroWeak (EW) fields [$Z^0$, $W^\pm$] might appear at the tree-level~\footnote{See for instance Ref.~\cite{Ahmed:2015uqt,Angelescu:2017jyj} for the radion field of warped
extra dimensions.} -- in contrast with the Higgs boson interaction configuration. Such a new contribution tends to increase the VBF cross section relatively to the ggF one. Moreover, there is the generic possibility 
that the new scalar coupling to the top quark is reduced compared to the corresponding Yukawa interaction so that the VBF amplitude would be enhanced with respect to the ggF one.

For establishing the LHC potential reach for VBF productions of generic neutral CP-odd/even (under the combined Charge Parity symmetry)
scalar fields in an effective parameter space, one has to find selective kinematical cuts possibly taking possibly benefit of
the two characteristic VBF jets of the final state, in particular for scalar boson masses below typically $\sim 65$~GeV where the golden scalar decay into a pair of on-shell Z bosons is kinematically closed while their 
decay into diphoton is extremely difficult to detect experimentally (as the LHC diphoton triggers are bandwidth-limited for such soft photon productions)~\footnote{At low diphoton invariant masses, boosted diphoton 
events with high $p_T$ transverse momentum can still be triggered, but at the price of a weaker selection efficiency.}.

For this purpose we have performed Monte Carlo simulations of the events for the signal, using the {\tt Feynrules} code~\cite{Degrande:2011ua}, and the SM Background, 
with {\tt MadGraph5\_aMC@NLO}~\cite{Alwall:2011uj} interfaced with {\tt Pythia8}~\cite{Sjostrand:2007gs} for hadronisation as well as {\tt FastJet}~\cite{Cacciari:2006sm,Cacciari:2011ma} 
and {\tt Delphes}~\cite{deFavereau:2013fsa} for detector reconstruction.

\section{The effective model}

Diagrammatically, the studied scalar boson production via VBF occurs through the double radiation off quarks of photons or electroweak gauge bosons then merging to create the new scalar field.
Hence, these processes involve only the new scalar coupling to two photons $\gamma$, two $Z^0$ bosons or two $W^\pm$ bosons. 
We use an effective theory approach to describe the scalar interaction with SM EW gauge bosons. The scalar mass can be smaller than the EW symmetry breaking scale.
When it is the case, we make the extra assumption that the scalar has large tree-level $SU(2)_L\times U(1)_Y$ couplings, so that the loop-induced EW breaking contributions 
are subleading. Under this condition, the interactions of a neutral CP-even or CP-odd scalar $\phi$ with the EW gauge bosons are respectively described by the following 
dimension-5 effective gauge invariant Lagrangians 
\begin{equation}
{\cal L}_{\rm eff}\supset\phi \left( \frac{1}{f_G}G^{\mu\nu\,a}G_{\mu\nu}^a+ \frac{1}{f_W}W^{\mu\nu\,b}W_{\mu\nu}^b
+\frac{1}{f_B}B^{\mu\nu}B_{\mu\nu}+\frac{1}{f_H}|D^\mu H|^2
\right)
\label{SM_VBF:EFTop1}
\end{equation}
\begin{equation}
{\cal L}_{\rm eff}\supset\phi \left( \frac{1}{\tilde f_G}G^{\mu\nu\,a}\tilde G_{\mu\nu}^a+ \frac{1}{\tilde f_W}W^{\mu\nu\,b}\tilde W_{\mu\nu}^b
+\frac{1}{\tilde f_B}B^{\mu\nu}\tilde B_{\mu\nu}
\right)
\label{SM_VBF:EFTop2}
\end{equation}
where $\tilde V^{\mu\nu}=\frac{1}{2}\epsilon^{\mu\nu\rho\sigma}V_{\rho \sigma}$, $H$ represents the SM Higgs doublet, $D^\mu$ the covariant derivative,
the $f$'s denote high-energy scales of new physics, $a,b$ are summed group generator indices whereas 
$\mu$, $\nu$ stand for summed Lorentz indices and the rank-2 tensors are the field strengths for all the SM
gauge bosons before EW symmetry breaking (using standard notations). 
After this breaking, the effective Lagrangian, for example in the CP-even case, contains the $\phi Z Z$ interactions 
\begin{equation} 
{\cal L}_{\rm eff}\supset \frac{1}{f_Z} \phi  (Z_{\mu\nu})^2 +\frac{m_Z^2}{ 2f_H}\phi (Z_\mu)^2\,, 
\end{equation} where $f_Z^{-1}=s_w^2 f_B^{-1}+c_w^2 f_W^{-1}$ and $s_w^2\equiv{\sin^2(\theta_w)}\approx 0.23$. 
The effective theory is valid as long as the $f$'s (related to Kaluza-Klein mass scales for instance) are larger than the typical energies 
going through the vertices. 
The $\phi$ scalar mixing with the SM Higgs boson is assumed to be small to ensure that the SM Higgs field has SM-like couplings compatible with the 
LHC signal strength measurements. 
The scalar fields entering Eq.~(\ref{SM_VBF:EFTop1})-(\ref{SM_VBF:EFTop2}) are taken to be the mass eigenstates.

The CP-even couplings might be those of a radion in a model with a warped extra dimension along which matter is propagating.
Notice that if EW brane kinetic terms are negligible in such models, one has $f_W=f_B$ \cite{Fichet:2013ola, Fichet:2013gsa} which implies that the $\phi F^{\mu\nu}Z_{\mu\nu}$ 
coupling vanishes, a property which can be used for model discrimination \cite{Baldenegro:2017aen}.

The CP-odd scalar field can be a pseudo Nambu-Goldstone boson from an approximate global symmetry, just like those appearing in composite Higgs models. The couplings to gauge fields are then induced by the many fermion resonances populating the TeV scale (see e.g Ref.~\cite{Belyaev:2016ftv} or Ref.~\cite{Fichet:2016xvs}).

\section{Numerical analyses}

\subsection{Cross sections}

\begin{figure}[h]\begin{center}
\includegraphics[width=10cm]{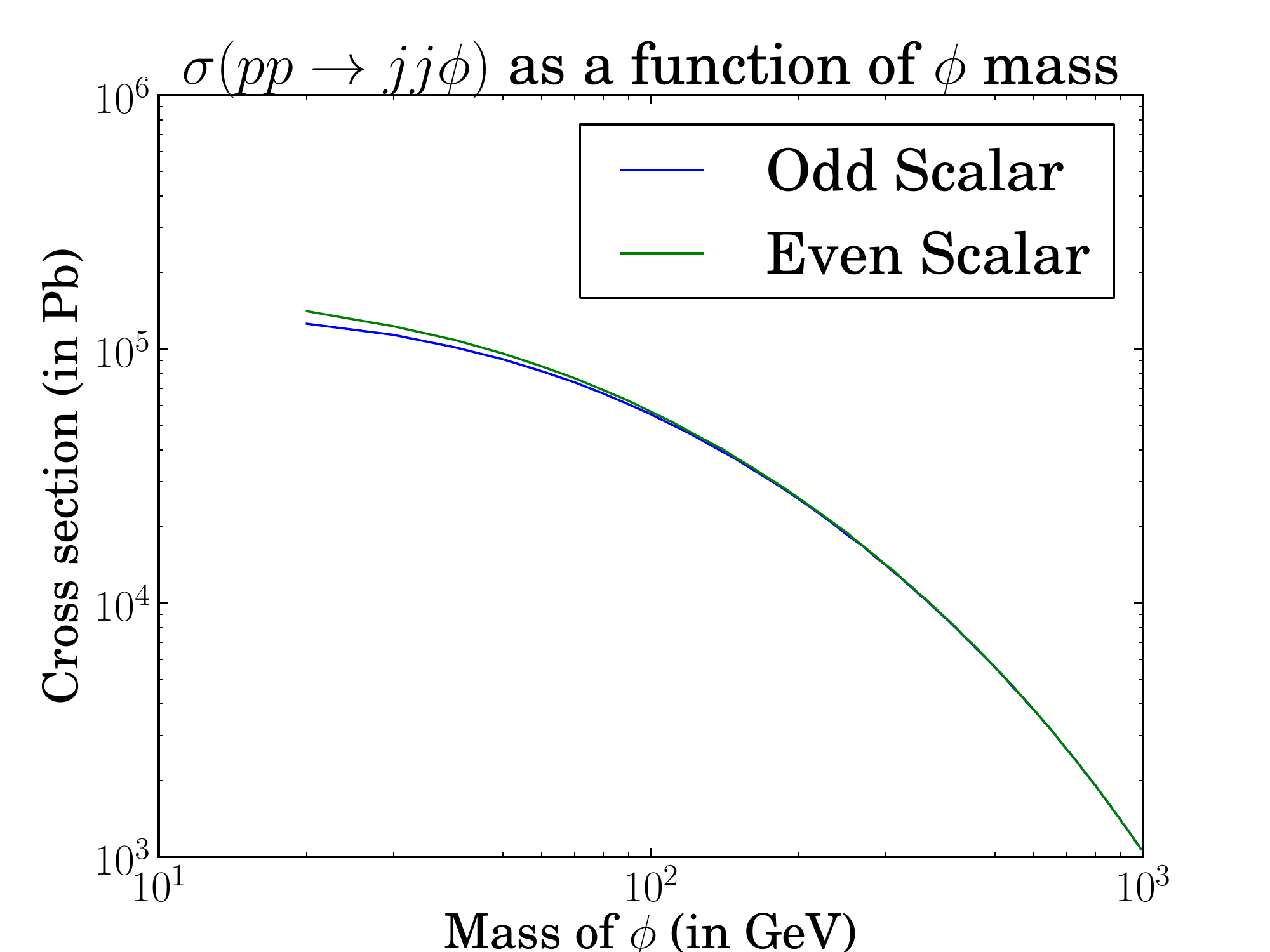}
\caption{Cross section (in Pb) for the VBF production of a generic light scalar, $pp\to \phi qq$, at the $13$~TeV LHC as a function of the scalar mass $m_\phi$ (in GeV).
Both odd and even scalar field cases are considered. A typical scale value $f_H=f_B=f_W=1$~TeV ($\tilde f_B=\tilde f_W=1$~TeV) is used in case of a CP even (odd) scalar field.}
\label{fig:Xsection}
\end{center}\end{figure}

First we present the cross sections, obtained with the {\tt Feynrules} code, for the VBF production of a light scalar, $pp\to \phi qq$ ($q$ denoting light quarks), at the $13$~TeV LHC. 
The results are shown in Fig.~\ref{fig:Xsection} for a typical scale value $f_H=f_B=f_W=1$~TeV ($\tilde f_B=\tilde f_W=1$~TeV) in case of a CP even (odd) scalar field.
For instance, we see that with $m_\phi=200$~GeV, the number of events with an integrated luminosity of $36.1$~fb$^{-1}$ ($100$~fb$^{-1}$) would be around $7\ 10^8$ ($2\ 10^9$) 
which could be promising for detection even after a signal suppression by a small branching ratio of the scalar field decay. 
Any model builder can use the result presented in Fig.~\ref{fig:Xsection} to deduce the rate values in a specific scenario using the present generic parametrisation.

\vspace{1cm}

\subsection{Kinematical distributions}

Following the preliminary analysis of Ref.~\cite{Brooijmans:2018xbu},
in order to investigate the capability of such an analysis to be able to distinguish the production of a CP-even from a CP-odd scalar particle as defined in the above effective model, and from SM background processes, we constructed both CP-even and CP-odd instances of the model with the {\tt Feynrules} code in the form of Universal FeynRules Output (UFO) files.  These were then propagated to the {\tt MadGraph5\_aMC@NLO\_v2\_5\_5} program for generation of parton-level events then to {\tt Pythia8} for showering and to {\tt Delphes} (using 'CMS flavor') for a detector reconstruction at $\sqrt{s}=13$ TeV of the process $pp\rightarrow\phi qq$ , $\phi\rightarrow\gamma\gamma$, for each of $m_{\phi}=20$ and $70$ GeV, as well as for generation of events from the process $pp\rightarrow\gamma\gamma + 2$ jets  within the SM. The event generations were performed for the following three choices of parameters.
\begin{itemize}
\item Two different cases of a CP-even scalar boson:
\begin{itemize}
\item $f_B=f_W=1$~TeV and $f_{H}\to \infty$,  corresponding to the case of a CP-even scalar boson coupling to  two $Z^0$ bosons via the $(Z_{\mu\nu})^2$ Lorentz structure, 
to two photons $\gamma$ via $(F_{\mu\nu})^2$ and to two $W^\pm$ bosons via $(W_{\mu\nu})^2$; case called CP-even$_{1/f_V}$,
\item $f_H=1$~TeV and $f_{B,W}\to \infty$,  corresponding to the case of a CP-even scalar boson coupling to  two $Z^0$ bosons via the $(Z_{\mu})^2$ Lorentz structure, 
to two photons $\gamma$ via $(A_{\mu})^2$ and to two $W^\pm$ bosons via $(W_{\mu})^2$; case called CP-even$_{1/f_H}$.
\end{itemize}
\item  The case of a CP-odd$_{1/\tilde f_V}$ scalar field with $\tilde f_B=\tilde f_W=1$~TeV, in which the coupling to two $Z^0$ bosons always occurs via the $Z_{\mu\nu}\tilde Z^{\mu\nu}$ Lorentz structure,
to two photons $\gamma$ via $F_{\mu\nu}\tilde F^{\mu\nu}$ and to two $W^\pm$ bosons via $W_{\mu\nu}\tilde W^{\mu\nu}$.
\end{itemize}

\begin{figure}[h]\begin{center}
\includegraphics[width=7.2cm]{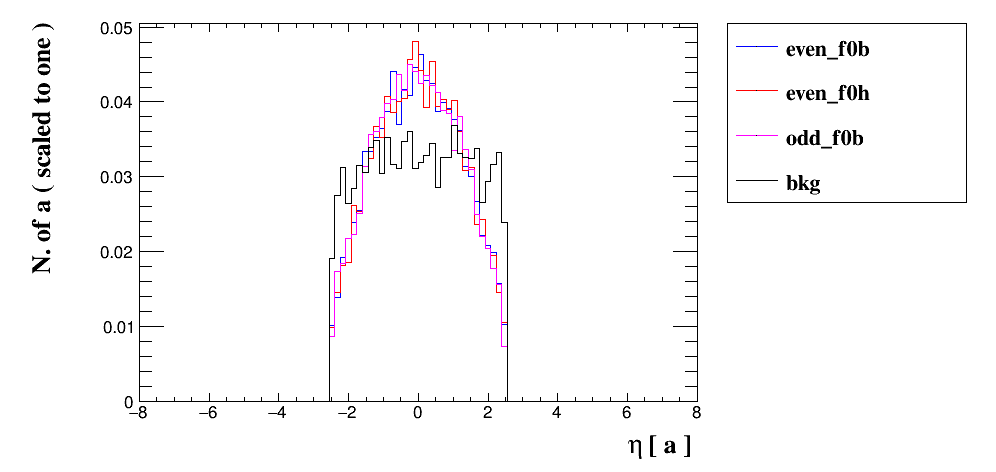}    \hspace{1cm}   \includegraphics[width=7.2cm]{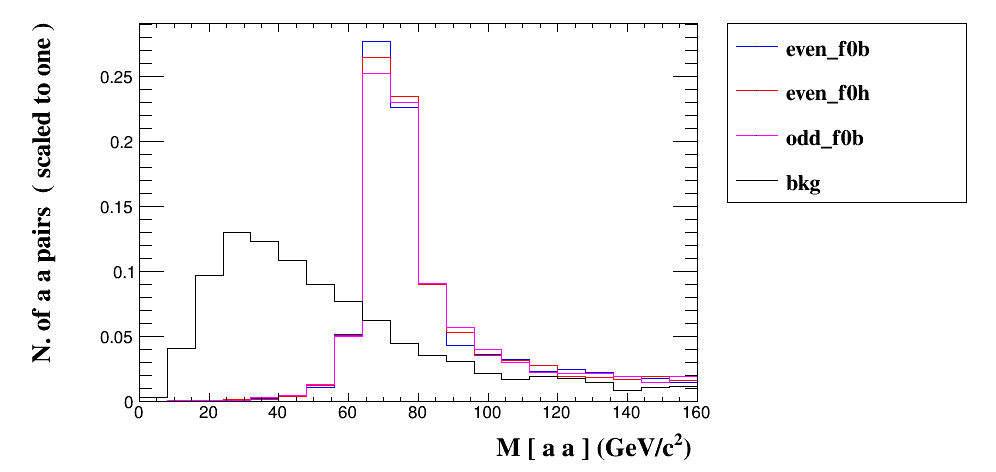}
\caption{Left-hand side: Normalised distribution of $\eta$ for a photon (denoted $a$ on the figure) which can be approximated by $-ln\tan(\theta /2)$, $\theta$ and $\phi$ here being the polar and azimuthal angles.
Right-hand side: Normalised distribution of the invariant mass (in GeV) of the diphoton system. We consider a scalar mass $m_\phi=70$~GeV. The signal and SM background (black) are superimposed over each other.
A typical scale value $f_B=f_W=1$~TeV, $f_H=1$~TeV and $\tilde f_B=\tilde f_W=1$~TeV is used respectively in the case CP-even$_{1/f_V}$ (blue), CP-even$_{1/f_H}$ (red) and CP-odd$_{1/\tilde f_V}$ (purple).}
\label{SM_VBF:fig:XseriesA}
\end{center}\end{figure}

\begin{figure}[h]\begin{center}
\includegraphics[width=7.2cm]{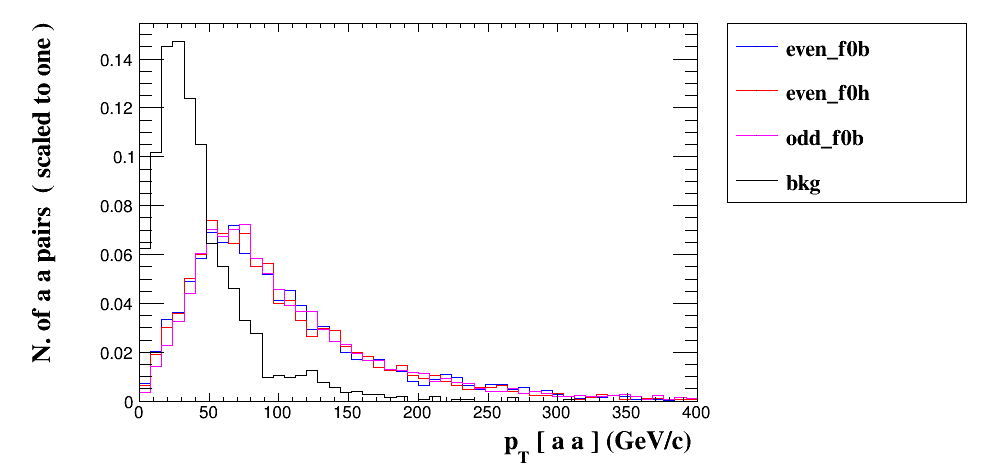}    \hspace{1cm}   \includegraphics[width=7.2cm]{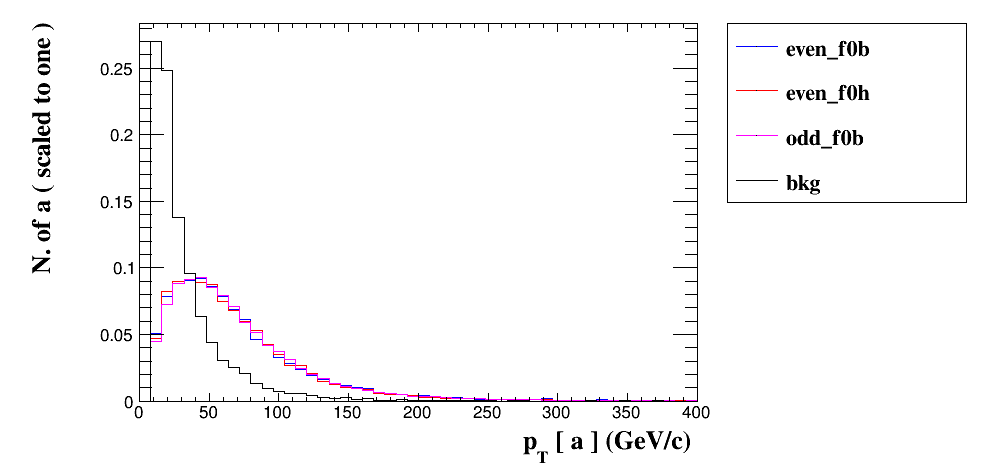}
\caption{Left-hand side: Normalised distribution of the diphoton system transverse momentum, $p_{T_{\gamma\gamma}}$ (in GeV).
Right-hand side: Normalised distribution of the photon system transverse momentum, $p_{T_{\gamma}}$ (in GeV). 
We consider a scalar mass $m_\phi=20$~GeV. The signal and SM background (black) are superimposed over each other.
A typical scale value $f_B=f_W=1$~TeV, $f_H=1$~TeV and $\tilde f_B=\tilde f_W=1$~TeV is used respectively in the case CP-even$_{1/f_V}$ (blue), CP-even$_{1/f_H}$ (red) and CP-odd$_{1/\tilde f_V}$ (purple).}
\label{SM_VBF:fig:XseriesB}
\end{center}\end{figure}

\begin{figure}[h]\begin{center}
\includegraphics[width=7.2cm]{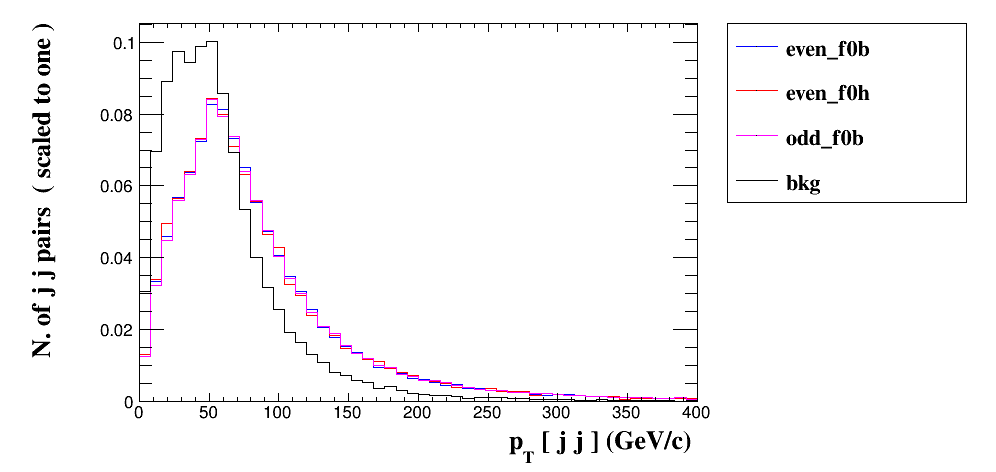}    \hspace{1cm}   \includegraphics[width=7.2cm]{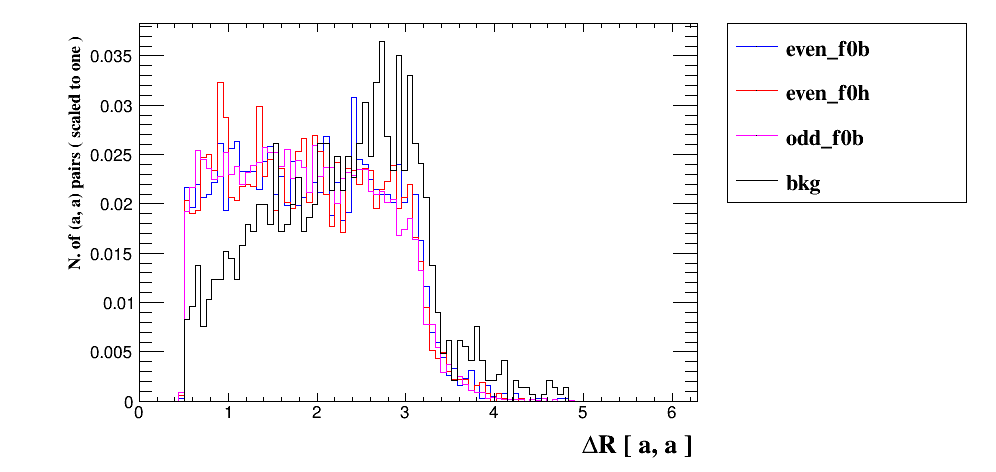}
\caption{Left-hand side: Normalised distribution of the dijet system transverse momentum, $p_{T_{jj}}$ (in GeV).
Right-hand side: Normalised distribution of the diphoton system quantity: $\Delta R^2=\sqrt{\Delta\theta^2+\Delta\phi^2}$ ($\theta$ and $\phi$ denoting the polar and azimuthal angles, respectively). 
We consider a scalar mass $m_\phi=20$~GeV. The signal and SM background (black) are superimposed over each other.
A typical scale value $f_B=f_W=1$~TeV, $f_H=1$~TeV and $\tilde f_B=\tilde f_W=1$~TeV is used respectively in the case CP-even$_{1/f_V}$ (blue), CP-even$_{1/f_H}$ (red) and CP-odd$_{1/\tilde f_V}$ (purple).}
\label{SM_VBF:fig:XseriesC}
\end{center}\end{figure}

\begin{figure}[h]\begin{center}
\includegraphics[width=7.2cm]{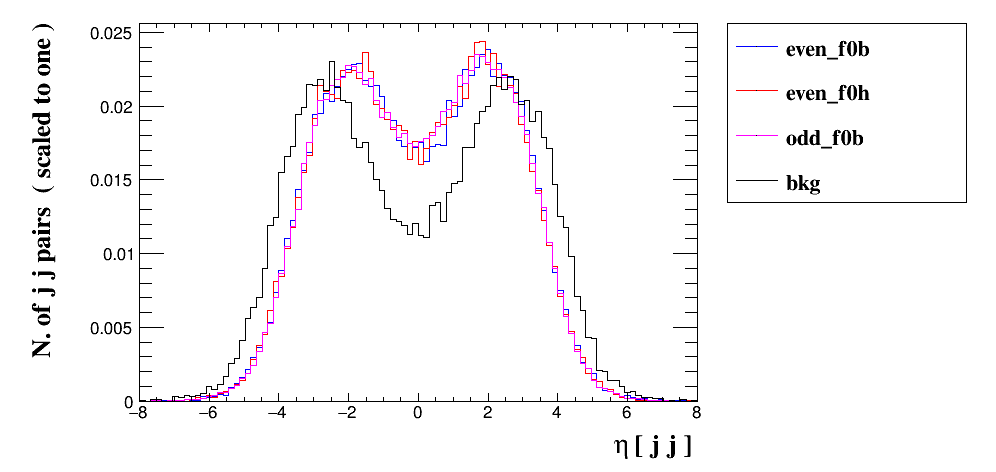}    
\caption{Normalised distribution of the same quantity $\eta$ as above, but now for the dijet system.
We consider a scalar mass $m_\phi=20$~GeV. The signal and SM background (black) are superimposed over each other.
A typical scale value $f_B=f_W=1$~TeV, $f_H=1$~TeV and $\tilde f_B=\tilde f_W=1$~TeV is used respectively in the case CP-even$_{1/f_V}$ (blue), CP-even$_{1/f_H}$ (red) and CP-odd$_{1/\tilde f_V}$ (purple).}
\label{SM_VBF:fig:XseriesD}
\end{center}\end{figure}

\noindent
Fig.~\ref{SM_VBF:fig:XseriesA}-\ref{SM_VBF:fig:XseriesD} 
show kinematical distributions for $m_{\phi}=20$ and $70$ GeV without selection or acceptance criteria and for each of the above three cases of light scalar:  CP-even$_{1/f_V}$, CP-even$_{1/f_H}$
and CP-odd$_{1/\tilde f_V}$, as well as for the SM background.  The areas of all distributions have been normalized to unity. We see that it is difficult to discriminate between the CP odd and CP even scalar fields or among the two different
CP even types of coupling. In contrast, there exist kinematical distributions allowing to reduce significantly the SM background, using some relevant cuts. Since the distributions shown are at parton level, further precise studies must be
undertaken in order to determine whether the differences in shape are as marked after parton showering/hadronisation, reflecting only partly the influence of the underlying event.

\vspace{1cm}

\vspace{1cm}

\vspace{1cm}

\vspace{1cm}

\section{Conclusion}

Hence, for two examples of low scalar masses at $20$~GeV and $70$~GeV, we have shown that some kinematical distribution shapes should allow for a reasonable discrimination 
of the extra scalar boson VBF production against the SM background.
This result combined with the obtained potentially large number of VBF events makes the VBF production process at LHC a promising channel of search for physics beyond the SM. 
It also turns out to be more challenging to distinguish among the CP even and CP odd scalar bosons so those two kinds of production would sum up to increase the signal.

\section*{ACKNOWLEDGEMENTS}
S.G.-S. would like to acknowledge the support of HiggsTools, an Initial Training Network (ITN) supported by the 7th Framework Programme of the European Commission.
All the authors acknowledge the organisers of the ``Les Houches'' Workshop for the whole organisation and the friendly atmosphere during the workshop.

\let\Herwig\undefined
\let\Pythia\undefined
\let\Sherpa\undefined
\let\Rivet\undefined
\let\Professor\undefined
\let\eps\undefined
\let\mc\undefined
\let\mr\undefined
\let\mb\undefined
\let\tm\undefined

%% file: wwtj/wwtj.main.tex
\graphicspath{{wwtj/}}


\newcommand{\mg}{{\sc MadGraph\_MC@NLO}}
\newcommand{\pythia}{{\sc Pythia 8.2}}
\newcommand{\delphes}{{\sc Delphes 3}}

\chapter{Probing the Top and Charm Yukawa Couplings in $p p \to VV+t/c+j$}

{\it A. Falkowski, S. Ganguly, P. Gras, J. M. No, K. Tobioka, E. Venturini, N. Vignaroli, T. You}

\label{sec:wwtj}

\begin{abstract}
The top and charm Yukawa couplings of the Higgs boson are important measurements to 
target in the Standard Model. They can be probed in complementary ways to on-shell Higgs decays through processes 
involving an off-shell Higgs. We estimate the sensitivity of the HL-LHC in the $p p \to VV + q + j$ channel to modifications of the 
Yukawa $\delta y_q$, where $q=t, c$ and $V=W^\pm, Z$. We find a rough $95\%$ CL sensitivity of around $|\delta y_t| \sim 0.2-0.3$ and $|\delta y_c| \sim 1-3$ 
assuming an amount of reducible background a factor 1 to 10 of the irreducible SM signal. This preliminary study can serve as a baseline for more refined future projections.
\end{abstract}

\section{Introduction}
\label{sec:wwtj:intro}

Almost a decade after its discovery, the Higgs boson remains a mysterious lynchpin of the Standard Model (SM). Its 
apparent simplicity as an elementary scalar belies an undesired arbitrariness in its properties -- from the unexplained pattern of 
Yukawa couplings to the precarious nature of the electroweak symmetry-breaking potential. This makes it all the more important to probe the Higgs 
experimentally in every way possible. Such a program of measurements of the Higgs sector will benefit from the upcoming high-luminosity LHC (HL-LHC) 
upgrade enabling new types of analyses to become competitive~\cite{Cepeda:2019klc}.  

Recently, it has been suggested by Ref.~\cite{Henning:2018kys} to measure the Higgs couplings without the Higgs; in other words, to consider processes 
with an off-shell Higgs. While many such studies have already been performed and are being pursued, others remain to be investigated. The relation of 
processes involving longitudinal gauge bosons to those with an on-shell Higgs can be categorised systematically by considering higher-dimensional operators 
in the SM effective field theory (SMEFT). The conclusions also follow more generally from the energy-growing behaviour that arises when breaking delicate cancellations 
in the electroweak sector. 

\begin{figure}[h!]
\centering
\includegraphics[scale=0.7]{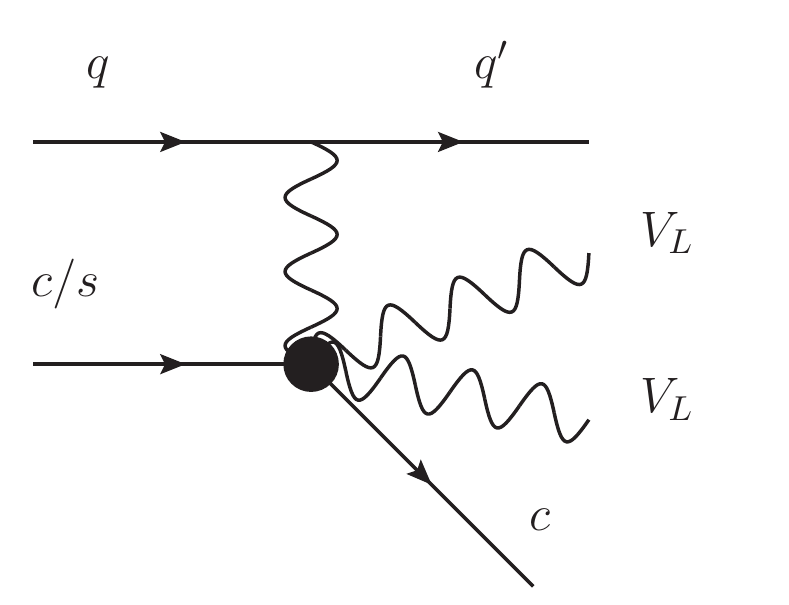} 
\includegraphics[scale=0.6]{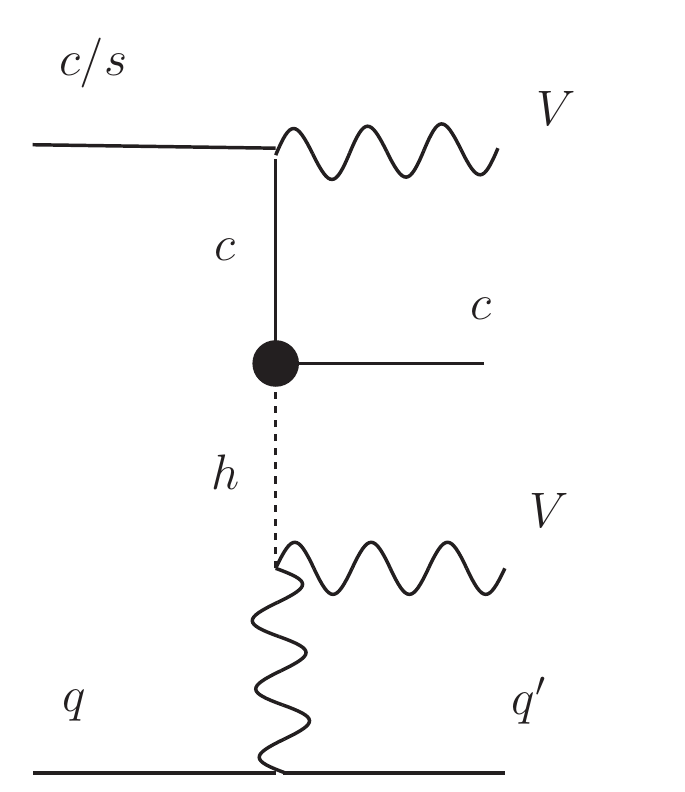} 
\caption{ \small Feynman diagrams for the signal of modification to $y_c$ as seen in the Feynman gauge (left) or in the unitary gauge (right).}
\label{fig:wwtj:diagram}
\end{figure}

Here we follow up on the top Yukawa study of the $p p \to VV + t + j$ channel in Ref.~\cite{Henning:2018kys} and make first projections also for the charm Yukawa. 
In Fig.~\ref{fig:wwtj:diagram} we show the Feynman diagram for our target process that is modified by the SMEFT dimension 6 operator
\begin{equation}
\mathcal{L}_\text{SMEFT} \supset \frac{Y_q}{v^2}|H|^2\bar{Q}_LHq_R \, ,
\label{eq:wwtj:op}
\end{equation}
where $q=t,c$. On the left of Fig.~\ref{fig:wwtj:diagram} is the diagram "without the Higgs" in a non-unitary gauge in which the longitudinal 
degrees of freedom of the vector bosons appear explicitly; on the right is the corresponding unitary gauge picture with a virtual off-shell Higgs. 
By setting $|H|^2 \sim v^2$ in the effective operator (\ref{eq:wwtj:op}), we see that this induces a modification of the top Yukawa coupling that can be probed 
through precision measurements of the Higgs couplings in on-shell decays or constrained in global SMEFT fits (see e.g. Refs.~\cite{Ellis:2018gqa, vanBeek:2019evb, Biekotter:2018rhp, Almeida:2018cld, Brivio:2019ius,deBlas:2019wgy, Falkowski:2019hvp} for some recent such fits). However, this operator also contains $|H|^2 \sim \phi^+ \phi^-$ 
corresponding to the longitudinal gauge boson components in the non-unitary gauge. The operator will therefore necessarily modify correlated processes such as the one 
in Fig.~\ref{fig:wwtj:diagram}. It benefits moreover from an energy-growing behaviour $\sim E^2/\Lambda^2$ as one probes higher energies closer to 
the new physics threshold $\Lambda$. This may compensate somewhat for the limited precision of these measurements.

In the next Section we study first the top Yukawa's signal sensitivity and estimate the top background. In Section~\ref{sec:wwtj:charm} we consider 
the charm Yukawa, before concluding in Section~\ref{sec:wwtj:conclusion}.

\section{Top Yukawa}
\label{sec:wwtj:top}

As the heaviest particle in the SM, the top quark and its coupling to the Higgs may be particularly sensitive to new physics. The projected reach of the HL-LHC from on-shell Higgs measurements is reviewed in Ref.~\cite{Cepeda:2019klc}. Here we estimate the HL-LHC sensitivity of the complementary off-shell process following the analysis of Ref.~\cite{Henning:2018kys} (see also the related process with linear growth in Refs.~\cite{Degrande:2018fog, Barger:2019ccj}) starting 
with a calculation of the signal expectation with a naive sensitivity estimate before describing our $t\bar{t}$ background simulations. 

\subsection{Signal}

\begin{figure}[h!]
\begin{center}
\includegraphics[scale=.75]{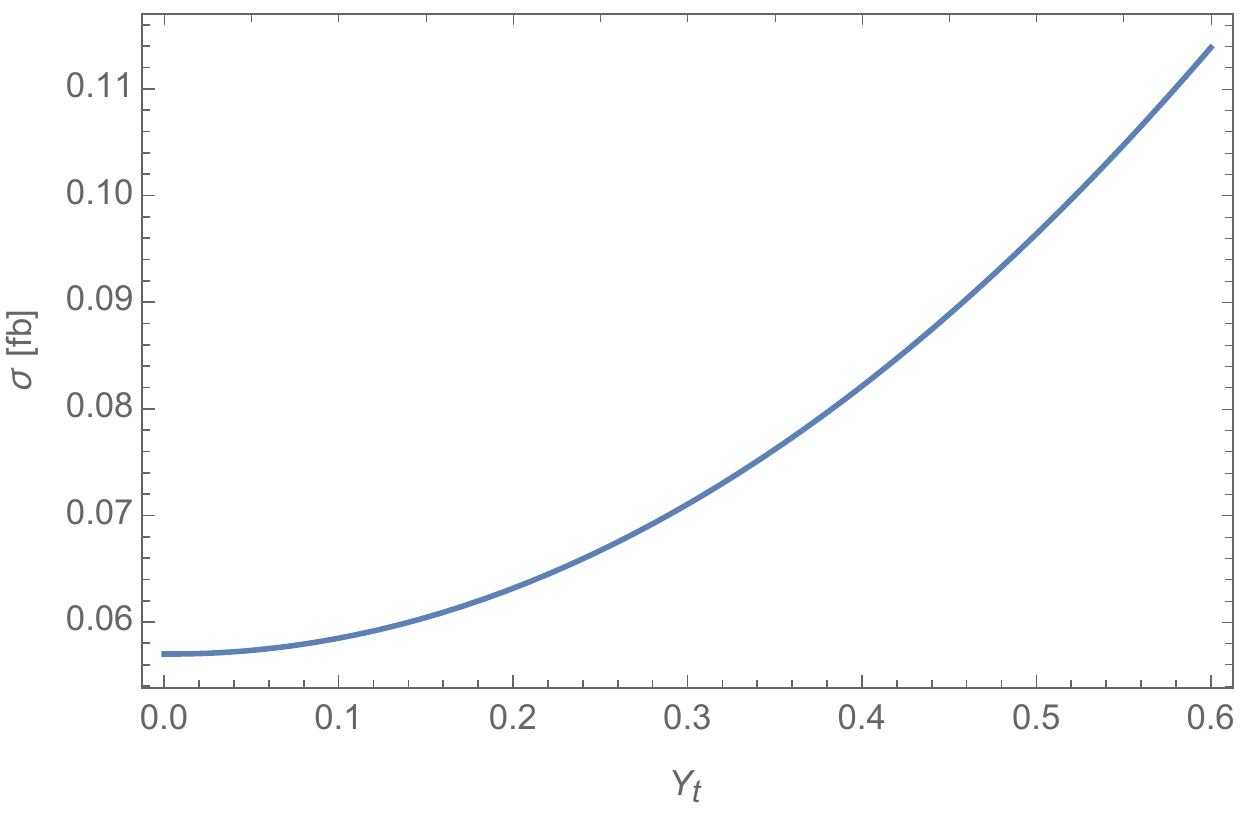}\\
\caption{Cross section (in fb) for $W^\pm W^\pm$ production, with \emph{cuts 2} applied, as a function of $Y_t$. 
\label{fig:wwtj:WpmWpm_yt_cut1}}
\end{center}
\end{figure}

In the presence of the dimension 6 operator (\ref{eq:wwtj:op}) the cross-section for $p p \to V V + t + j$ can be expanded as
\begin{equation}
\sigma \sim \sigma_{SM} + Y_t \frac{1}{v^2} \sigma_{int} + Y_t^2 \frac{1}{v^4} \sigma_{BSM^2} \, .
\end{equation}
We illustrate the quadratic dependence on the Wilson coefficient, for example, in the $W^\pm W^\pm$ channel in Fig.~\ref{fig:wwtj:WpmWpm_yt_cut1}, which is 
non-negligible. We apply the following cuts: 
\begin{equation}
\text{\emph{cut 1 (2)}:} \quad p_T^t > 250 \, (500) \text{ GeV} \, , \, |\eta_j| > 2.5 \, , \, p_T^j > 30 \text{ GeV} \, , \, E_j > 300 \text{ GeV} \, .
\end{equation}
For reference we report here the dependence of each channel on both the looser and tighter cuts 1 and 2 respectively. We use MadGraph~\cite{Alwall:2014hca} with a model implementing the 
operator~\eqref{eq:wwtj:op} to simulate events and calculate the dependence on the Yukawa modification. 

The number of SM signal events with the cuts above applied are given in Table~\ref{tab:wwtj:yt0}. For the $WW$ channels with cut 1 we find good agreement 
with Ref.~\cite{Henning:2018kys} in each decay sub-channel (we include leptonic decays of the $\tau$'s in the leptonic categories), though we find discrepancies 
in the $WZ$ and $ZZ$ estimates for cut 1 and obtain more suppression when applying cut 2, which warrants further study.

In Tables~\ref{tab:wwtj:yta2},~\ref{tab:wwtj:ytb2} and~\ref{tab:wwtj:ytc2}, the cross sections and their dependence on $Y_t$ in each decay channel are 
indicated, with \textit{cut 2} applied (as used to estimate the sensitivity of our search). 

\begin{table}[h!]
\begin{center}
\begin{tabular}{|c|c|c|c|c|c|}
\hline
& 0$\ell$ & 1$\ell$ & $\ell^\pm \ell^\mp$ & $\ell^\pm \ell^\pm$ & 3$\ell$  (4$\ell$) \\ %
\hline
\hline
$W^\pm W^\mp$ & 1.277/0.0682 & 0.9034/0.04826 & 0.1598/0.008538 & 0 & 0 \\
\hline
 $W^\pm W^\pm$ &1.102/0.03656 & 0.7799/0.02587 & 0 & 0.1380/0.004576 & 0  \\
 \hline
$WZ$ &  6.329/0.4799 & 2.239/0.1698 & 0.7089/0.05375 & 0 & 0.2508/0.01902   \\
\hline
 $ZZ$ &   0.1810/0.01387 & 0 & 0.04054/0.003106 & 0 & 0.002270/0.0001740  \\
 \hline
 \end{tabular}
 \caption{$Y_t=0$ (SM) and cut 1/cut 2 on $p_T^t$. Cross sections are in fb}
 \label{tab:wwtj:yt0}
\end{center}
\end{table}

\begin{table}[h!]
\begin{center}
\begin{tabular}{|c|c|c|}
\hline
& 0$\ell$ & 1$\ell$   \\ 
\hline
\hline
$W^\pm W^\mp$ & $0.06820 + 0.007311 \, Y_t + 0.2783 \, Y_t^2$ &  $0.04826 + 0.005174\, Y_t + 0.1969\, Y_t^2$   \\
\hline
 $W^\pm W^\pm$ & $0.03656 - 0.0004092 \, Y_t + 0.1233 \,Y_t^2$ & $0.02587 - 0.0002896 \, Y_t + 0.08726 \,Y_t^2$ \\
 \hline
$WZ$ & $0.4800 + 0.001319 \,Y_t + 0.05116 \,Y_t^2$ & $0.1698 + 0.0004665\, Y_t + 0.01810\, Y_t^2$  \\
\hline
$ZZ$ & $0.01387 + 0.002090 \, Y_t + 0.04232 \, Y_t^2$& 0  \\
 \hline
 \end{tabular}
 \caption{Channels without leptons and with a single lepton; cut 2 on $p_T^t$. Cross sections are in fb.}
 \label{tab:wwtj:yta2}
\end{center}
\end{table}

\begin{table}[h!]
\begin{center}
\begin{tabular}{|c|c|c|}
\hline
& $\ell^\pm \ell^\mp$ & $\ell^\pm \ell^\pm$    \\
\hline
\hline
$W^\pm W^\mp$ & $0.008538 + 0.0009152 \,  Y_t + 0.03483 \, Y_t^2 $& 0 \\
\hline
 $W^\pm W^\pm$ &  0 & $0.004576 - 0.00005123\, Y_t + 0.01544 \, Y_t^2$  \\
 \hline
$WZ$ &  $0.05375 + 0.0001477 \, Y_t + 0.005730 \, Y_t^2 $ & 0  \\
\hline
$ZZ$  & $0.003106 + 0.0004682 \, Y_t + 0.009479\, Y_t^2$ & 0 \\
 \hline
 \end{tabular}
 \caption{Channels with 2 leptons; cut 2 on $p_T^t$. Cross sections are in fb.}
 \label{tab:wwtj:ytb2}
\end{center}
\end{table}

\begin{table}[h!]
\begin{center}
\begin{tabular}{|c|c|}
\hline
&  3$\ell$ (4$\ell$)   \\
\hline
\hline
$W^\pm W^\mp$ & 0\\
\hline
 $W^\pm W^\pm$ &   0 \\
 \hline
$WZ$ & $0.01902 + 0.00005225 \, Y_t + 0.002027 \, Y_t^2$ \\
\hline
$ZZ$   & $0.0001740 + 0.00002622 \, Y_t + 0.0005308 \, Y_t^2$\\
 \hline
 \end{tabular}
 \caption{Channels with  3 and 4 leptons; cut 2 on $p_T^t$. Cross sections are in fb. }
 \label{tab:wwtj:ytc2}
\end{center}
\end{table}

\begin{figure}[h!]
\centering
\includegraphics[scale=0.7]{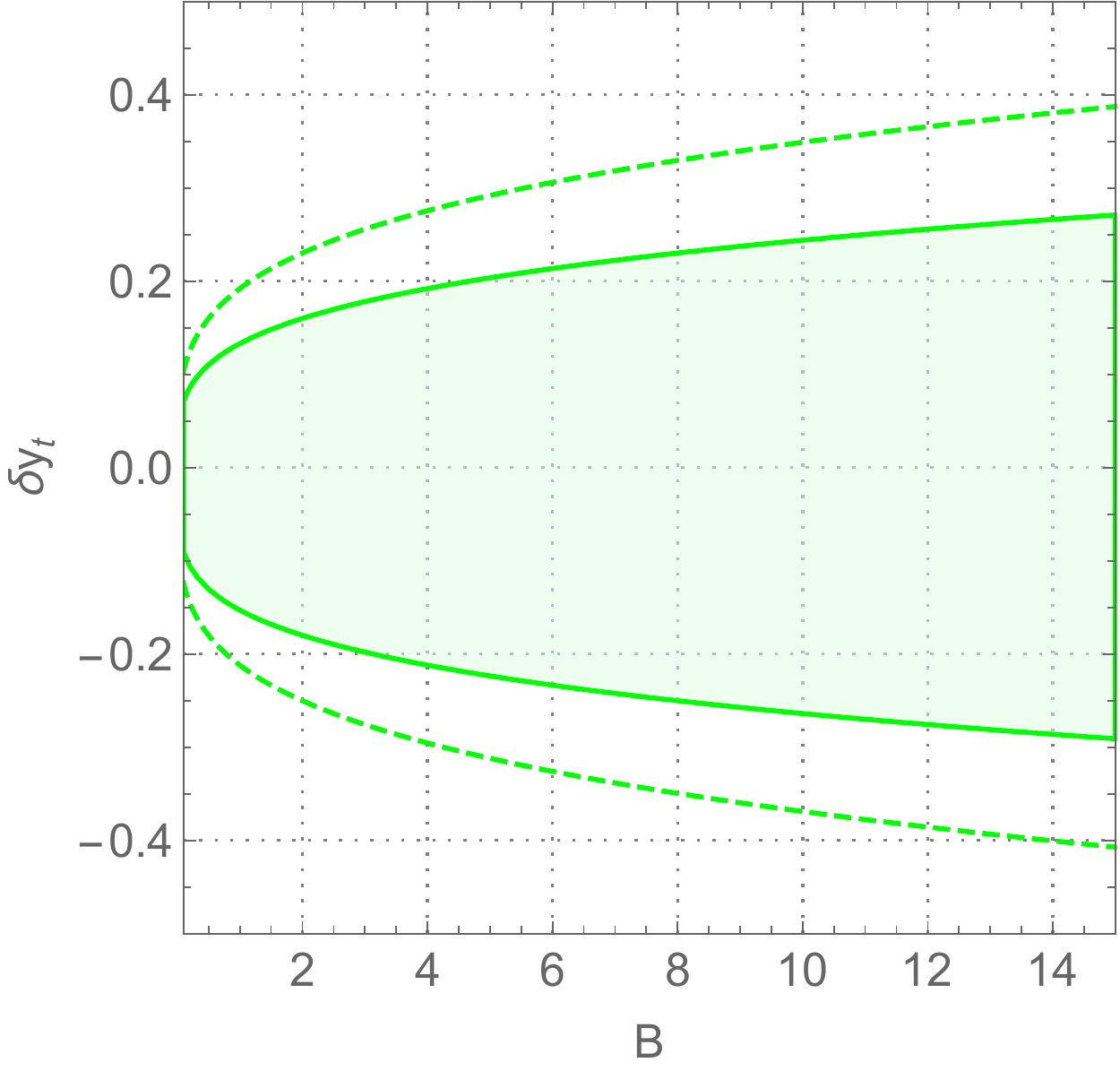} 
\caption{\small Projection of the 1- and 2-$\sigma$ sensitivity, in solid and dashed lines respectively, at the $3 \text{ab}^{-1}$ HL-LHC 
to modifications of the top Yukawa $\delta y_t$.  }
\label{fig:wwtj:yt}
\end{figure}

We may make a rough sensitivity projections based on the signal dependence on the Yukawa modification due to $Y_t$ relative to the SM value $y_t$, 
\begin{equation}
\delta y_t \equiv \frac{3}{2}\frac{Y_t}{y_t} \, .
\end{equation}
The significance $\sigma$ is estimated by applying a naive background rescaling factor $B$ that parametrises the number of background events as a ratio of the SM signal events: 
\begin{equation}
\sigma = \frac{N_{SM+BSM} - N_{SM}}{\sqrt{B \times N_{SM}}} \, .
\label{eq:wwtj:sigma}
\end{equation}
The resulting sensitivity at HL-LHC with $3 \text{ab}^{-1}$ is shown in Fig.~\ref{fig:wwtj:yt}. Cut 2 has been applied as this led to increased sensitivity, 
though we note that optimal choices of cuts will rely on a detailed study of individual sub-channels together with their background estimates that we leave to future work. 
In the next Section we describe in detail our analysis to estimate the dominant $t \bar{t}$ background. 

\subsection{Backgrounds}

	\subsubsection{Introduction}
	
	We restrict the study to the final state with one boosted hadronic top ($p_{T} > 250\,$GeV) that can be identified with a top tagging 
	technique~\cite{Plehn:2011tg,Anders:2013oga,Aaboud:2018psm,CMS-PAS-JME-16-003,CMS:2019gpd}. A rejection factor of 
	$\gtrsim 10$ is obtained for working point with a 80\% top quark selection efficiency.
	We now review background processes with one top quark that decays hadronically. 
	
	\subsubsection{Event selection and classification}

	Vector bosons that decay hadronically can be identified if they are boosted enough, $p_{\text{T}} \gtrsim 200\,$GeV. 
	The boson is reconstructed as a jet with a large radius parameter, and a discriminator as the jet mass is used for the identification~\cite{Aaboud:2018psm,CMS-PAS-JME-16-003}.
	Events are classified according to the number of leptons as in the previous 
	sections (only electrons and muons are considered). 
	The category with two opposite-sign leptons is split into two categories, with and without a tagged vector boson.
	The leptons are required to fulfill the following isolation criteria: the sum of the transverse momentum of particle within a $R$ distance, $R=\sqrt{\Delta\eta^2+\Delta\varphi^2}$ of 0.4 must be smaller than 15\% of the lepton transverse momentum.
	The vector boson tagging is applied for all the categories covering a signal final state with one or two hadronic vector boson. 
	Table~\ref{tab:wwtj:cat} lists the categories with the  respective targeted signal processes.
	When vector boson tagging is applied, a threshold of 200$\,$GeV is applied on the transverse momentum of the jet. 
	The $\text{t}\bar{\text{t}}$ background is further suppressed by rejecting events with one b-tagged jet or more whose 
	$R$ distance from the reconstructed top quark is larger than 0.8 are rejected. If several top quarks are reconstructed, 
	the one with the highest transverse momentum is considered. 
	For the category $\ell^\pm\ell^\mp$ + 1 V$_h$ it is also required that the two reconstructed leptons have the same flavour 
	and that their four-momentum sum has a mass of $\pm 20\,$GeV around 91$\,$GeV. 
	
	\begin{table}	
		\caption{Event categories. Here leptons refer to muons, electrons and their antipaticles. The letter V$_h$ refers to a vector boson, indifferently 
		W or Z.}\label{tab:wwtj:cat}
		\begin{tabular}{l|p{15em}|l} 		
			Category name & Selection & Selected signal\\
			\hline 
			0 $\ell$ + 2 V$_h$ & No isolated lepton and two tagged vector bosons & $\overset{\scriptscriptstyle(-)}{\text{t}}$V($\rightarrow$jj)V($\rightarrow$jj)\\
			1 $\ell$ + 1 V$_h$ & One isolated lepton and one tagged vector boson & $\overset{\scriptscriptstyle(-)}{\text{t}}$W($\rightarrow\ell\nu$)V($\rightarrow$jj)\\
			$\ell^\pm\ell^\pm$ & Two same-sign isolated leptons & $\overset{\scriptscriptstyle(-)}{\text{t}}$W$^\pm$($\rightarrow\ell^\pm\overset{\scriptscriptstyle(-)}{\nu}$)W$^\pm$($\rightarrow\ell^\pm\overset{\scriptscriptstyle(-)}{\nu}$)\\
			$\ell^\pm\ell^\mp$      & Two opposite-sign leptons & 		$\overset{\scriptscriptstyle(-)}{\text{t}}$W$^\pm$($\rightarrow\ell^\pm\overset{\scriptscriptstyle(-)}{\nu}$)W$^\mp$($\rightarrow\ell^\mp\overset{\scriptscriptstyle(-)}{\nu}$)\\
			$\ell^\pm\ell^\mp$ + 1 V$_h$ & Two opposite-sign isolated leptons and one tagged vector boson & $\overset{\scriptscriptstyle(-)}{\text{t}}$Z($\rightarrow\ell^\pm\ell^\mp$)V($\rightarrow$jj))\\
			3 $\ell$      & Three isolated leptons & $\overset{\scriptscriptstyle(-)}{\text{t}}$Z($\rightarrow\ell^\pm\ell^\mp$)W$^\pm$($\rightarrow\ell^\pm\overset{\scriptscriptstyle(-)}{\nu}$)\\
			4 $\ell$      & At least four isolated leptons  & $\overset{\scriptscriptstyle(-)}{\text{t}}$Z($\rightarrow\ell^{\pm}\ell^{\mp}$)Z($\rightarrow\ell^{\pm}\ell^{\mp}$)\\
		\end{tabular}
	\end{table}

	\subsubsection{Event yield estimation}
	
	To estimate the background process event yields, samples are generated at leading order with Madraph5\_aMC@NLO~\cite{Alwall:2014hca}. 
	Madspin~\cite{Artoisenet:2012st} is used for the top quarks and the vector boson decays. The parton distribution function (PDF) NNPDF~2.3 NLO~\cite{Ball:2018iqk} 
	is used and the strong coupling is set to $\alpha_s(m_{\text{Z}})=0.119$ at the Z boson mass. Parton showering is 
	performed with {\sc Pythia}~8. A 250$\,$GeV threshold is applied on the top quark (at parton level). Particles are clustered in jets using 
	the anti-$k_{\text{T}}$ algorithm implemented in fastjet with a distance parameter of 0.4. One jet with energy above 300$\,$GeV and 
	pseudorapidity $|\eta| > 2.5$ is required. Quark and gluons are required to be within the pseudorapidity range $|\eta|<5$ to account for the calorimeter acceptance.  
	
	A factor of 80\% is applied on the event yields to account for the top tagging efficiency. For the b-jet veto the efficiency parametrization 
	of the loose working point given in~\cite{Sirunyan:2017ezt} is used, assuming similar performance for HL-LHC. The acceptance for 
	b-tagging is extended to $|\eta| = 4$ to account for the larger coverage of the trackers of the experiment Phase~II 
	upgrade~\cite{Collaboration:2017mtb,CMSCollaboration:2015zni}.
	The vector boson tagging is emulated by reconstructing anti-$k_{\text{T}}$ jets with 
	a distance parameter $R=0.8$ and applying an efficiency of 80\% for a jet from a vector boson decay and 0.04\% for a jet from a prompt quark or gluon.
	The event yield estimate from the $\text{p}\text{p}\rightarrow\text{t}\bar{\text{t}}\text{j}\text{j}$ process is provided in Table~\ref{tab:wwtj:ttjj}.
	
	The cross section of the $\text{pp}\rightarrow\text{t}\bar{\text{t}}\text{Wj}$ process, that will contribute to the categories with one or more leptons, 
	has been estimated at leading order. After requiring one top or antitop quark with $p_{\text{T}}>250\,$GeV and a forward jet, 
	with $|\eta| > 2.5$ and $E>300\,GeV$, the cross section is, leading to less than one event for 3000$\,\text{fb}^{-1}$. The contribution 
	from $\text{pp}\rightarrow\text{t}\bar{\text{t}}\text{W}$ with one hadronic top quark with $p_{\text{T}}>250\,$GeV and one leptonic top quark from which the b-jet passes the forward jet requirement ($|\eta| > 2.5$ and $E>300\,$GeV) is also smaller 
	than one event before the b-jet veto.
	
	The background from the processes with a top pair in the final state is substantial for the hadronic and semi-leptonic channels, 
	while it is negligible for the fully leptonic channels. We expect that the background for leptonic channels will mainly come from 
	events where a QCD jet was misidentified as a top quark, e.g. from the production of two W bosons with two jets. Such contribution remains to be estimated. 

\begin{table}
	\caption{Contribution from the $\text{t}\bar{\text{t}}\text{j}\text{j}$ process. The event yields are given for an integrated 
	luminosity $3000\,\text{fb}^{-1}$ at $\sqrt{s} = 14\,\text{TeV}$. The quoted uncertainties are the statistical uncertainty} \label{tab:wwtj:ttjj}
	\centering
	
\begin{tabular}{l|r}	
	Category                      & t$\bar{\text{t}}$jj event yield \\
	\hline 
	0 $\ell$ 2 V$_{h}$   	      &  10$\,$200. $\pm$ 4.5e+03   \\
	1 $\ell$ 1 V$_{h}$   	      &   175$\,$000. $\pm$ 1.9e+04 \\
	$\ell^\pm\ell^\pm$   	      &            -                \\
	$\ell^\pm\ell^\mp$            &  42$\,$600. $\pm$ 9.3e+03   \\
	$\ell^\pm\ell^\mp$  1 V$_{h}$ &  12$\,$200. $\pm$ 5.0e+03   \\
	3 $\ell$                      &                -            \\
	4 $\ell$                      &                -            \\

\end{tabular} 
\end{table}


\section{Charm Yukawa}

\label{sec:wwtj:charm}

We now study the process with two gauge bosons plus a jet plus a tagged charm in the final state~\footnote{See e.g. Refs~\cite{Perez:2015aoa, Han:2018juw, ATL-PHYS-PUB-2018-016} for some recent studies on probing the charm Yukawa.}. In this 
way we can directly probe the modification to the Yukawa coupling of the charm due to the operator (\ref{eq:wwtj:op}), 
\begin{equation}
\delta y_c \equiv \frac{y^{BSM}_c}{y_c}=\frac{3}{2}\frac{Y_c}{y_c} , .
\end{equation}

\begin{figure}[h!]
\centering
\includegraphics[scale=0.5]{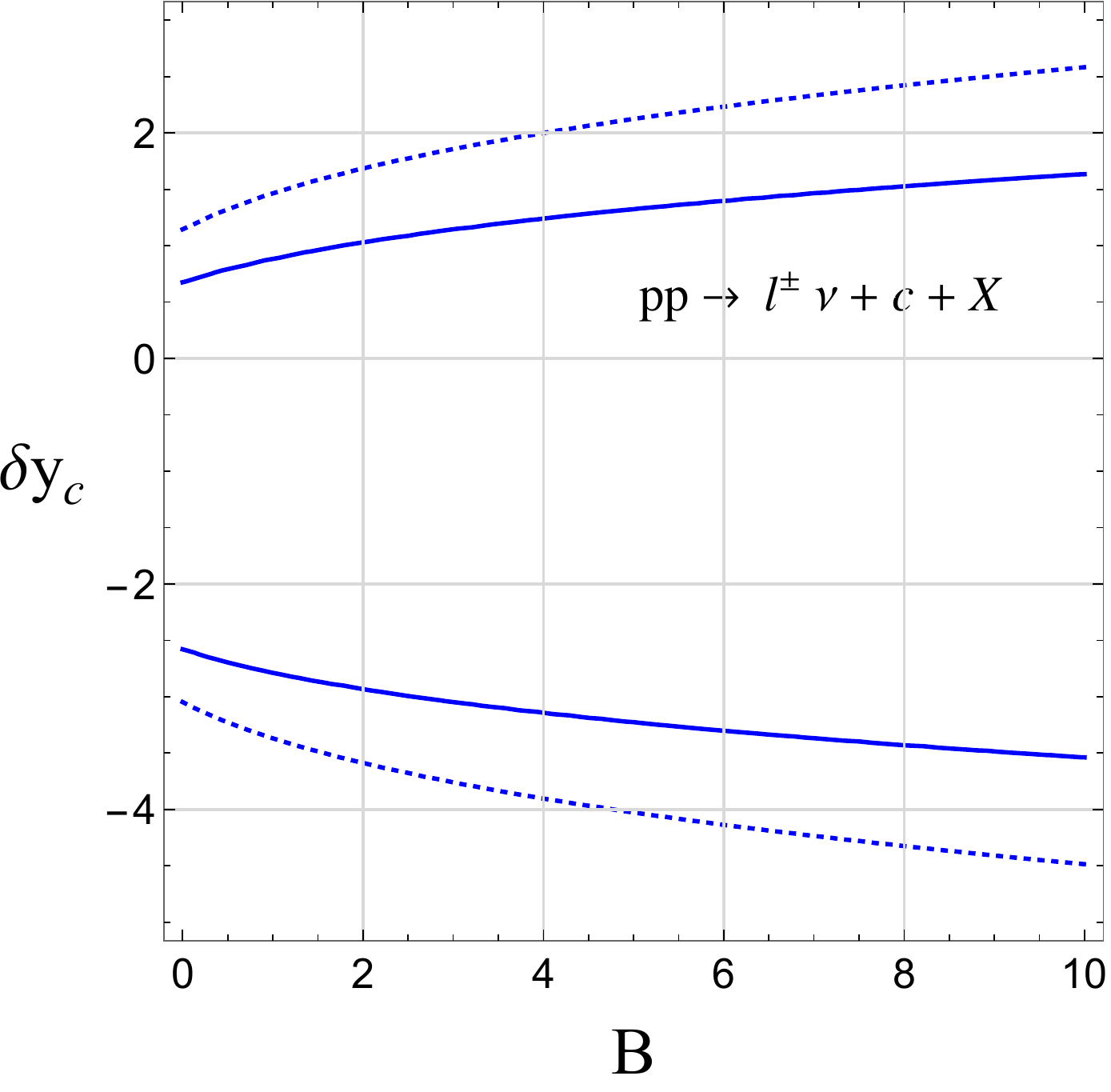} 
\caption{\small  1$\sigma$ (continuous curve) and 2$\sigma$ (dotted curve) HL-LHC (14 TeV, 3 ab$^{-1}$) sensitivity on 
$\delta y_c$, as a function of an arbitrary amount of reducible background, calculated as a factor $B$ of the irreducible background given by the SM signal.}
\label{fig:wwtj:Yc}
\end{figure}

We can estimate the BSM contribution to the inclusive cross section by expanding it in the SM ($Y_c=0$) plus the SM-BSM interference 
term ($\propto Y_c$) plus the pure BSM term ($\propto Y^2_c$):

 \begin{equation}
 \sigma \approx \sigma^{SM}(Y_c=0) + Y_c \, \sigma^{INT}(Y_c=1) + Y^2_c \, \sigma^{BSM}(Y_c=1) \ .
 \end{equation}\\
\noindent
We find the following cross sections in the different $VVcj$ channels:\\
\begin{center}
\begin{tabular}{|c|c|c|c|}
\hline
& SM ($Y_c=0$) & INT ($Y_c=1$)  & BSM ($Y_c=1$) \\ \hline
$W^+ W^- cj$ & 2.3 pb & 0.58 pb & 63 pb \\ \hline
$W^+ Z cj$ & 0.86 pb & 0.17 pb & 17 pb \\ \hline
$W^- Z cj$ & 0.79 pb & 0.09 pb & 9.1 pb \\ \hline
$Z Z cj$ & 0.19 pb & 0.14 pb & 15 pb \\ \hline
$W^+ W^+ cj$ & 29 fb & 0.42 fb & 94 fb \\ \hline
$W^- W^- cj$ & 23 fb & 0.31 fb & 90 fb \\ \hline
\end{tabular}
\end{center}

The significance can be estimated using the rescaling factor $B$ as for the top case in Eq.~\ref{eq:wwtj:sigma}. By considering the results in the 
semileptonic channel, $\ell^{\pm}\nu$+$c$+$X$, and considering a 25$\%$ c-tagging efficiency (see e.g. Refs.~\cite{ATL-PHYS-PUB-2018-016, Han:2018juw}), we find a 
rough estimate of the background-free sensitivity as

\begin{equation}
\delta y_c \lesssim \quad 0.7 \,  (1\sigma)\,  -\,1.1  \,  (2\sigma) \qquad \text{(HL-LHC, 3 ab$^{-1}$)} 
\end{equation}

Note that in this estimate we do not include the contribution of reducible backgrounds, for example the $t \bar{t}$ background estimated 
in the previous section. It is important to evaluate the contribution of this background to obtain a realistic estimate, though 
we consider an idealised estimate of the sensitivity in the absence of backgrounds to be a useful starting point for more realistic studies to aim towards. 
Fig. \ref{fig:wwtj:Yc} shows the 1$\sigma$ (continuous curve) and the 2$\sigma$ (dotted curve) HL-LHC sensitivity on $\delta y_c$, as a function of an arbitrary 
amount of reducible background, calculated as a factor $B$ of the irreducible background given by the SM signal.

\section{Conclusion}

\label{sec:wwtj:conclusion}

In this Proceedings note we made sensitivity projections for the top Yukawa at the HL-LHC and analysed its $t\bar{t}$ background, 
also considering for the first time the charm Yukawa, in the off-shell Higgs process $p p \to V V + t/c + j$. For the top Yukawa we find a comparable sensitivity 
to Ref.~\cite{Henning:2018kys} of $|\delta y_t| \sim 0.2 - 0.3$. The charm Yukawa sensitivity is $\delta y_c \sim \mathcal{O}(1)$ in the semileptonic channel, 
though this will depend in large part on improvements in charm tagging and controlling backgrounds sufficiently. 

In future work we aim to incorporate detailed background estimates in each sub-channel and identify optimal cuts and 
channels for a more realistic analysis. Furthermore several other related processes remain to be investigated in the off-shell Higgs program; 
the development of these measurements could complement the on-shell Higgs ones and provide a useful input to global fits of the SMEFT. 
As the HL-LHC will enable more sensitive EFT measurements in the next decade, we have focused on those projections here. In the longer 
term off-shell Higgs measurements could form an important part of Higgs studies at the 100 TeV FCC-hh.

\section*{Acknowledgements}

We would like to thank the organisers of the Les Houches "Physics at TeV  Colliders" 2019 workshop for their warm hospitality and for the friendly environment. 
J.M.N. was supported by Ram\'on y Cajal Fellowship contract RYC-2017-22986, and acknowledges further 
support from the Spanish MINECO's ``Centro de Excelencia Severo Ochoa" Programme under grant SEV-2016-0597, from the European Union's Horizon 2020 research and 
innovation programme under the Marie Sklodowska-Curie grant agreements 690575  (RISE InvisiblesPlus) and 674896 (ITN ELUSIVES) and from the Spanish Proyectos de I$+$D 
de Generaci\'on de Conocimiento via grant PGC2018-096646-A-I00.
K.T. is supported by the US Department of Energy grant DE-SC0010102 and also by his startup fund at Florida State University (Project id: 084011- 550- 042584).
N.V. is supported by the ERC grant NEO-NAT.  TY is supported by a Branco Weiss Society in Science Fellowship and partially supported by STFC consolidated grant ST/P000681/1.

\let\be\undefined
\let\ee\undefined
\let\bea\undefined
\let\eea\undefined
\let\mg\undefined
\let\pythia\undefined
\let\delphes\undefined

%% file: ttHZ/tthz.main.tex
\graphicspath{{ttHZ/}}






\chapter{Constraining the $t \bar{t} hZ$ vertex from $t\bar{t}hZ$ and $t\bar{t}hj$ production channels}
{\it S.~Banerjee, R.~S.~Gupta, S.~Jain,  E.~Venturini}

\label{sec:ttHZ}



\begin{abstract}
In this chapter, we consider two processes through which  operator generating the $Zt_R\bar{t}_R$ and $hZt_R\bar{t}_R$  interactions can be probed. These interaction is  uncorrelated to $Z$-decays to bottom quarks and thus unconstrained by LEP data. With the objective of constraining these couplings we specifically study the energy growth of the cross-sections in the processes $p p \to t \bar{t} h Z$ and $p p \to t \bar{t} h j$. A first analysis for the future 27 TeV and 100 TeV machines yield a percent to permille level level bound on this coupling.
\end{abstract}


\section{Introduction}

Colliders such as the Large Hadron Collider and other proposed future machines would study the interactions of elementary particles at the multi-TeV scale for the first time ever.  As this will correspond to the fundamental scale ever probed,  the data collected by these machines would be valuable whether or not any new physics beyond the Standard Model (SM) is directly discovered.  
 
Of particular interest are the interactions of the top quark with the electroweak  and Higgs  bosons. From a purely experimental point of view many of the interactions of the top quark are still unconstrained. For example, the $Zt_R \bar{t}_R$ and $hZt_R \bar{t}_R$ couplings, the focus of this chapter, are still poorly constrained compared to the analogous couplings to lighter quarks which were precisely measured by LEP in $Z$-boson decays.
 
Studying the top couplings to electroweak bosons is also very well motivated theoretically. This is because  the top quark and its BSM partners often play a crucial role in electroweak symmetry breaking, for instance by radiatively contributing to the Higgs potential. The mass scale of the top partners is  not expected to be much larger than the TeV scale due to naturalness considerations and thus this sector might lead to sizeable indirect effects. In composite models~\cite{Agashe:2004rs, Pomarol:2008bh} for example, integrating out the top partners generates all the dimension-6 operators that can lead to deviations of the coupling of the $Z$-boson to the top ,  
\begin{equation}
\mathcal{O}_{HQ_L}^{(1)}=i H^\dagger \overleftrightarrow{D}_\mu H \bar{Q}_L\gamma^\mu Q_L~~~~ \mathcal{O}^{(3)}_{HQ_L}= i H^\dagger \sigma^a \overleftrightarrow{D}_\mu H \bar{Q}_L \sigma^a \gamma^\mu Q_L
~~~~\mathcal{O}_{Ht_R}= i H^\dagger \overleftrightarrow{D}_\mu H \bar{t}_R \gamma^\mu t_R
\end{equation}
While the operators involving the left-chiral fermions deform both the SM $Zb\bar{b}$ and $Zt\bar{t}$ couplings and are thus constrained by LEP measurements (specifically $Z$-pole and anomalous triple gauge coupling  measurements),  the operator $\mathcal{O}_{Ht_R}$ is completely unconstrained.\footnote{Note that there is a linear combination of $\mathcal{O}_{HQ_L}^{(1)}$ and $\mathcal{O}_{HQ_L}^{(3)}$ that generates only a  $Z t_L \bar{t}_L$ deformation but not a $Z b_L \bar{b}_L$ deformation. This linear combination is also not constrained by LEP. As in the case of $\mathcal{O}_{Ht_R}$, to constrain this direction in EFT space we need to  analyse the other processes mentioned in the discussion.} This operator when expanded in the unitary gauge yields,
\begin{equation}
\frac{\mathcal{O}_{Ht_R}}{\Lambda^2}=-\frac{g v^2}{2 c_{\theta_W} \Lambda^2}(1+ \frac{2 h}{v}+\frac{h^2}{v^2})Z_\mu \bar{t}_R \gamma^\mu t_R,
\end{equation}
where $\Lambda$ is the scale of new physics. In this work we will compute the bounds on the coefficient of $(h/v) Z_\mu \bar{t}_R \gamma^\mu t_R$,
 \begin{equation}
 g_{hZt_R} = -\frac{g}{c_{\theta_W}}\frac{v^2}{\Lambda^2} c_{Ht_R}
 \label{cutoff}
 \end{equation}
where $c_{Ht_R}$ is the Wilson coefficient of the operator $\mathcal{O}_{Ht_R}$.  In composite models we expect $c_{Ht_R}\sim g_*^2$~\cite{Pomarol:2008bh} where, $1\lesssim g_*\lesssim 4 \pi$, is the coupling characterising the strong sector.

At the LHC, the $Zt\bar{t}$ coupling can be constrained in $t\bar{t}Z$ production. Contribution of $\mathcal{O}_{Ht_R}$  to $t\bar{t}Z$ production however does not grow with energy and thus only $\mathcal{O}$(1)  constraints on deviations of the $Z$ coupling from its SM value  are expected from this process at the LHC~\cite{Rontsch:2014cca, Dror:2015nkp}. In this note, we explore the feasibility of probing the  contribution of this operator to the $t\bar{t} \to hZ$ and $tZ \to th$ processes in $pp$ collisions. These contributions grow quadratically with energy with respect to the SM (energy growth in the analogous process for light quarks was studied in ~\cite{Banerjee:2018bio, Banerjee:2019pks, Banerjee:2019twi}) and hence stronger constraints may be possible. This operator also generates contributions  to $tW \to tW$ process that grow quadratically in energy with respect to the SM but this has already been studied in Ref.~\cite{Dror:2015nkp}. An analogous channel involving $t\bar{t} \to W^+ W^-$ in $pp$ collisions can also be interesting to look into.

This chapter is divided as follows. In Section~\ref{sec:tthZ}, we discuss the $t\bar{t} \to hZ$ scattering and show the energy growth of the interference piece as a function of the $Zh$ invariant mass. In Section~\ref{sec:tthj}, we show the energy growth in the $tZ \to th$ scattering as a function of the $th$ invariant mass. Finally we conclude in Section~\ref{sec:conclusions}.

\section{$p p \to t \bar{t} h Z$}
\label{sec:tthZ}

\begin{figure}[htb!]
\begin{center}
\includegraphics[scale=1.0]{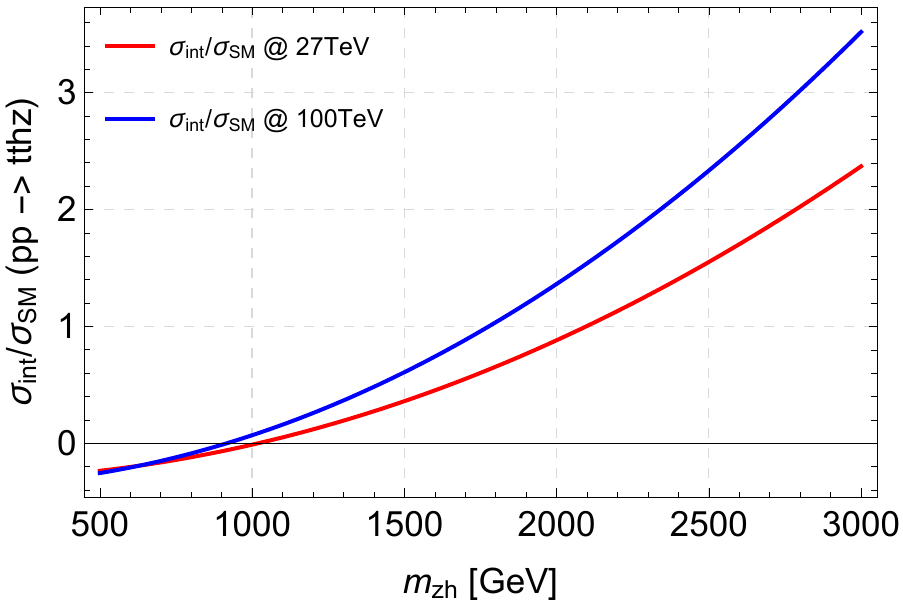}\\
\caption{Results from \texttt{MG5\_aMC\@NLO} simulation of the $pp \to t\bar{t} h Z$ process: ratio $\sigma_{int}^{(g_{hZt_R} = 1)}/\sigma_{SM}$ for $g_{hZt_R} = 0.1$. The \texttt{UFO}~\cite{Degrande:2011ua} model implementation of the Standard Model Effective Field Theory (SMEFT) is done within the \texttt{FeynRules}~\cite{Alloul:2013bka} framework.
\label{fig:intSM_tthZ}}
\end{center}
\end{figure}

In this first analysis, we consider the process $p p \to t \bar{t} hZ$. Apart from the SM processes, we encounter the $gg$ initiated diagram with the effective $t\bar{t}hZ$ vertex. At 14 TeV, the SM cross-section is $\sim 1.5$ fb without including the branching ratios of the various particles. Hence, even upon considering a semi-leptonic channel comprising $h/Z \to b\bar{b}$, we end up with $110$ events at 3 ab$^{-1}$. However, various detector effects will need to be folded in including $b$-tagging efficiencies. Upon demanding a flat 70\% $b$-tagging efficiency, one obtains $\sim 15$ events. Owing to this reason, we reconsider this channel for the planned future hadron colliders at 27 TeV (HE-LHC) and 100 TeV (FCC-hh). The respective SM cross-sections for these two colliders with the default scale choices in \texttt{MG5\_aMC\@NLO}~\cite{Alwall:2014hca} are $~8$ fb and $130$ fb without the decays. Hence, we can see that going to higher energy colliders have strong advantages in probing this coupling. Because we want to look into the higher energy effects, it is important to mention the cross-sections with cuts on $m_{Zh}$, \textit{i.e.}, on the $Zh$ invariant mass. With $m_{Zh} > 500$ GeV, 1000 GeV and 1500 GeV, the respective SM cross-sections at 27 TeV (100 TeV) colliders are 2.2 fb (47 fb), 0.3 fb (10.7 fb) and 0.07 fb (3.6 fb) without including the branching ratios. It is thus more feasible to consider the semi-leptonic channel for the 27 TeV collider (yielding $\sim 70$ events) and the fully leptonic channel (yielding $\sim 160$ events) for the 100 TeV collider. Now, upon considering $g_{hZt_R} = 0.1$, one obtains the respective interference cross-section values as -0.2 fb (3.2 fb), 0.08 fb (8.2 fb) and 0.06 fb (6.4 pb) for $m_{Zh} > 500$ GeV, 1000 GeV and 1500 GeV for the 27 TeV (100 TeV) machine. Thus, we can clearly see the interference term growing quadratically in energy with respect to the SM cross-section in Figure~\ref{fig:intSM_tthZ}. The dominant backgrounds (that we do not consider in this proceeding) ensue from $t\bar{t}hh, t\bar{t}ZZ, t\bar{t}hb\bar{b}, t\bar{t}Zb\bar{b}$ and fake backgrounds involving $c$-jets and light jets~\cite{Banerjee:2019jys}. We leave the full analysis for a future work.

\begin{figure}[htb!]
\begin{center}
\includegraphics[scale=0.4]{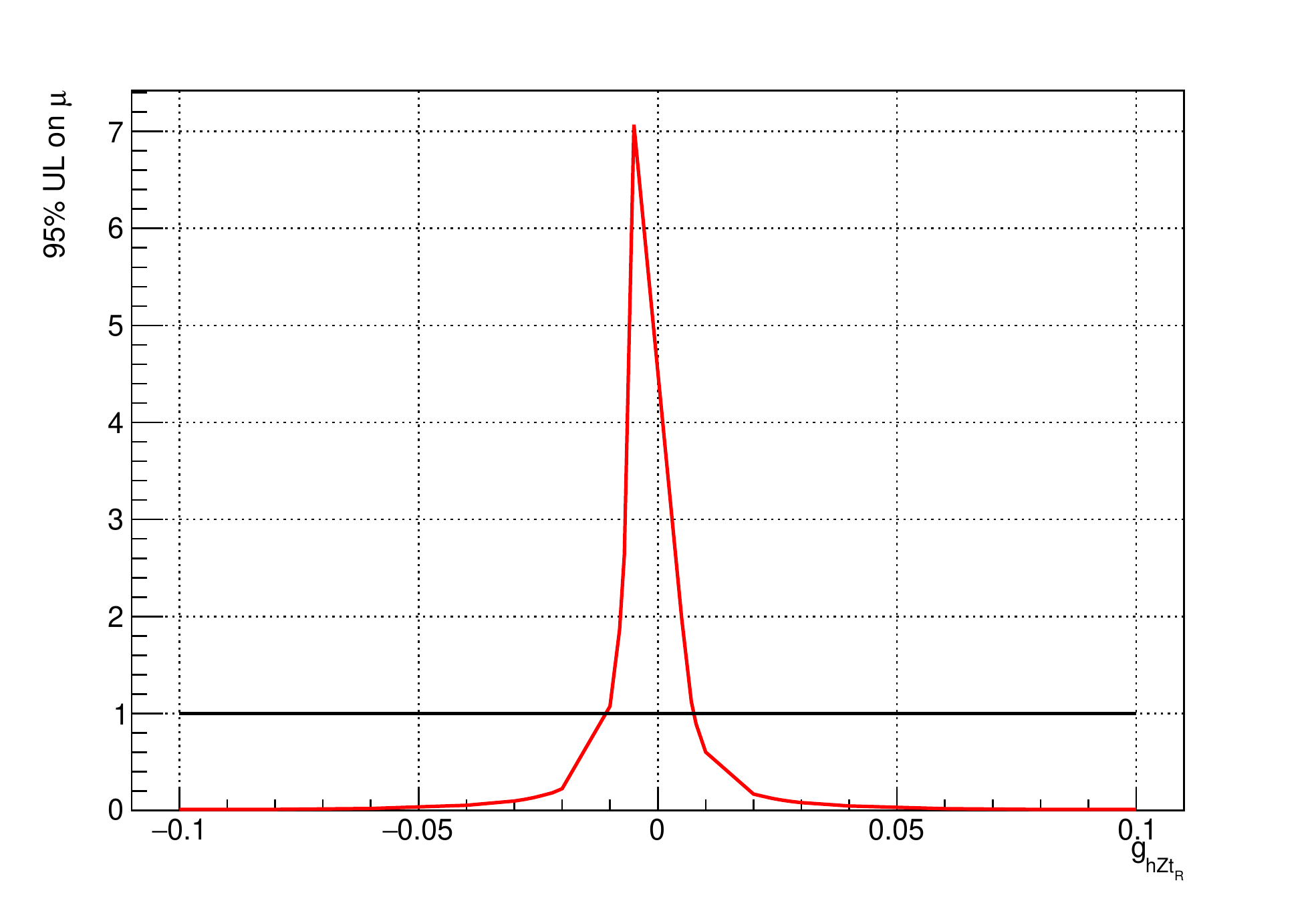}~~~~~~~\includegraphics[scale=0.4]{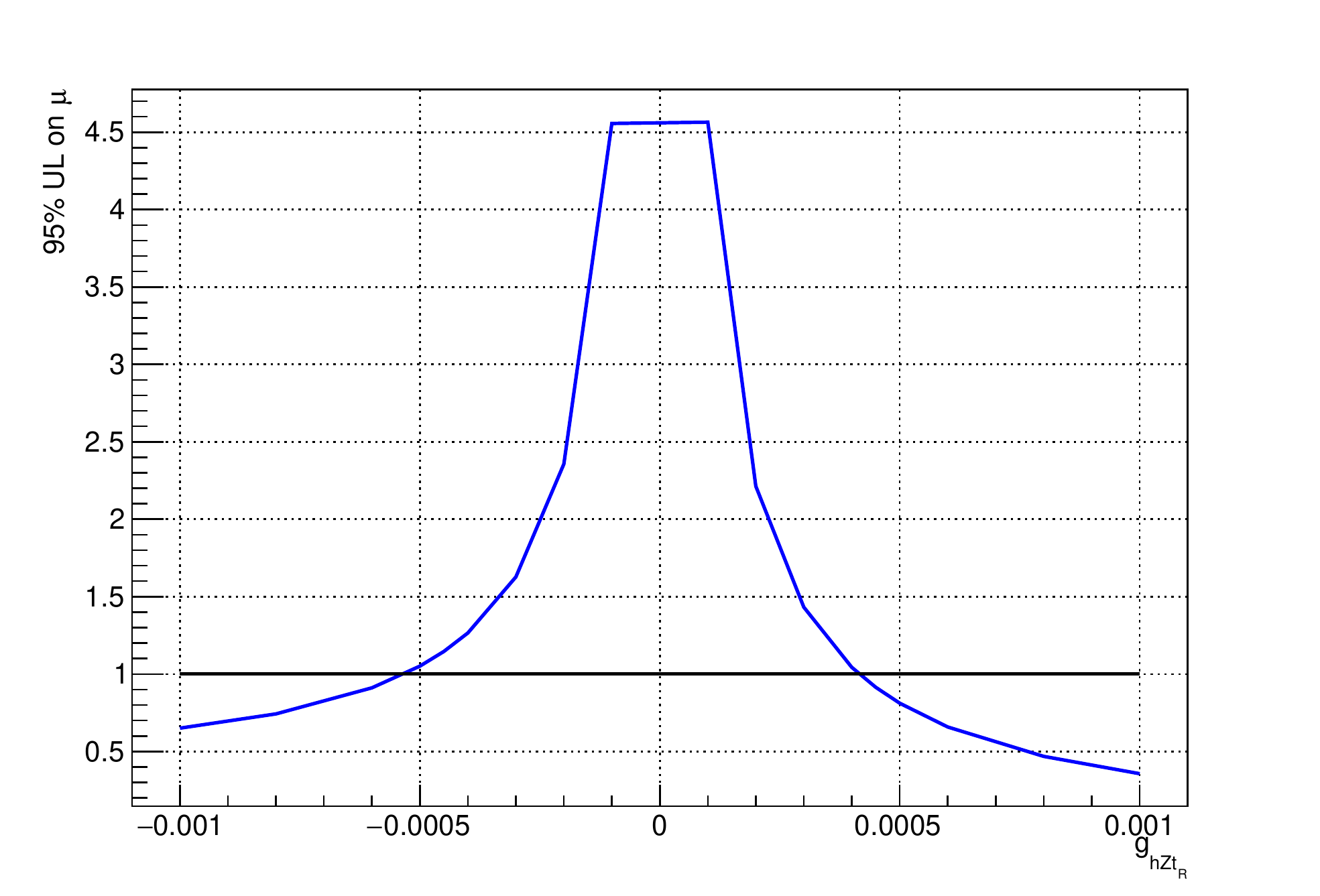}
\caption{$\mu$ as a function of $g_{hZt_R}$ for the 27 TeV (100 TeV) machine in the left (right) panel. All the values below the black line are excluded. 
\label{fig:UL}}
\end{center}
\end{figure}

Upon treating  the SM $t\bar{t}hZ$ production as the background, our first estimates for the allowed coupling range at 95\% C.L. are
\begin{equation}
 -0.007 \; (-0.0006) \lesssim \; g_{hZt_R} \lesssim 0.007 \; (0.00045)  
\label{bounds}
\end{equation}
for the 27 TeV @ 3 ab$^{-1}$ (100 TeV @ 30 ab$^{-1}$) machine.  Using $c_{Ht_R}\sim g_*^2$ and Eq.~\ref{cutoff} this corresponds to a scale $\Lambda\gtrsim 4$ TeV  for any  $g_* \gtrsim 1.6$ ($g_* \gtrsim 0.5$). As there are no events beyond this energy in  both our 27 TeV and 100 TeV studies, our bounds respect EFT validity requirements.  A plot showing the exclusion ($\mu$) as a function $g_{hZt_R}$ is shown in Figure~\ref{fig:UL}. The limits are quoted using a simple likelihood ratio of signal + background and a background only hypothesis as the test statistic. We employ the CLs method~\cite{Read_2002} to quote the 95\% C.L. upper limits.

\section{$p p \to t \bar{t} h j$}
\label{sec:tthj}

An alternative process that can be used to constrain $\mathcal{O}_{Ht_R}$ is $pp\to t\bar{t} hj$. In particular, for the quark-gluon initiated scenario, one encounters the $tZ\to th$ scattering which gives a contribution to the amplitude that grows with $\hat{s}$, with respect to the SM process. Upon exploiting this energy growth and analysing the $m_{th}$ distribution, one can improve the sensitivity to the $g_{hZt_R}$ parameter.

The cross-section for $pp\to t\bar{t} hj$ is more than two order of magnitudes larger than the process studied in Section~\ref{sec:tthZ}. For this channel, the cross-section is $\sim 0.64$ pb at 14 TeV. The cross-section is enhanced to $\sim 3.5$ pb for a 27 TeV machine. Hence it is feasible to study the fully leptonic channel and obtain constraints on $g_{hZt_R}$, even without going to the FCC-hh.

However, the cross-section is dominated by gluon-gluon induced processes, which do not receive contributions from $\mathcal{O}_{Ht_R}$ and thus do not interfere with the diagrams involving this effective vertex. At 14 TeV (27 TeV) the $gg$ initiated channels constitute $\sim 68$\% (74\%) of the total $\sigma_{SM}$. The $gq$ initiated subprocesses, which are relevant for this analysis, comprise $\sim 74\%$ ($\sim 80\%$) of the remaining cross-section. The $qq$ scattering provides an even smaller contribution; in this case there can be amplitudes involving the aforementioned effective vertex. However, these diagrams do not grow in energy with respect to SM.

As a consequence, despite the huge difference in the total cross-section with respect to the $pp\to t\bar{t}hZ$ process, the interference term is not as much enhanced in comparison to what we found in Section~\ref{sec:tthZ}. Upon expanding in powers of $g_{hZt_R}$,
\begin{equation}
\label{eq:SigmaEFT}
\sigma=\sigma_{SM} + g_{tht_R} \sigma_{int} +g_{tht_R}^2 \sigma_{BSM^2}
\end{equation}
one has $\sigma_{int}\sim -4$ fb ($\sim -7$fb) at 14 TeV (27 TeV) and, therefore, for $g_{hZt_R} = 0.1$, $\sigma_{int}^{(g_{hZt_R} = 1)}\sim -0.4$ fb ($\sim -0.7$fb) at 14 TeV (27 TeV). This is of the same order of magnitude larger as the interference cross-section for the process $pp\to t\bar{t}hZ$. Besides, the interference for the present process only corresponds to $< 1\%$ of the total cross-section. Furthermore, only a fraction of the $gq\to t\bar{t} h q$ interference cross-section is affected by $\mathcal{O}_{Ht_R}$ and shows energy growth. Thus, it is difficult to observe the energy growing information from the full $p p \to t \bar{t} h j$ process. In order to see the energy growth, we first rely on the $tZ\to th$ scattering process.

\begin{figure}[htb!]
\begin{center}
\includegraphics[scale=.45]{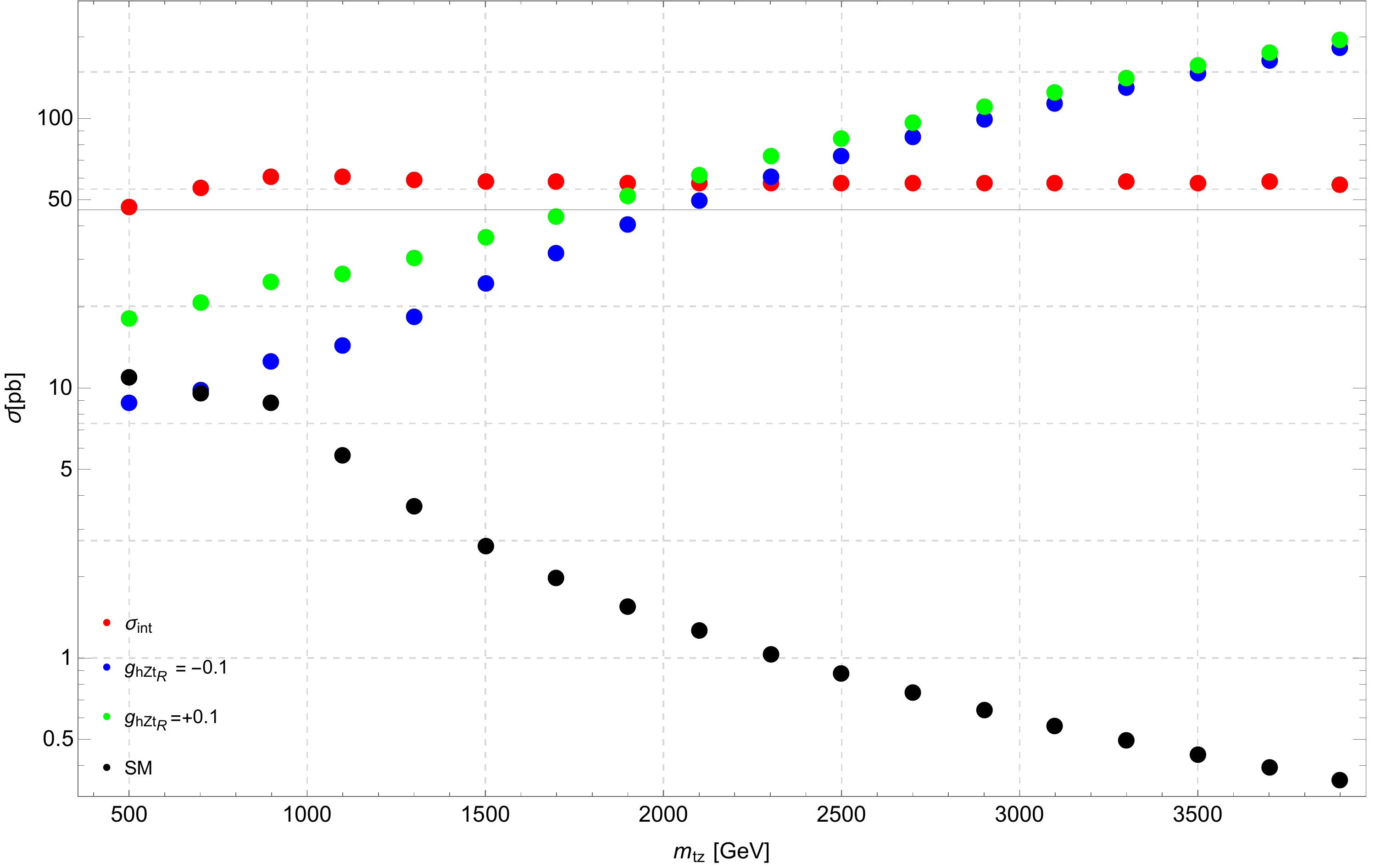}\\
\caption{Results from a \texttt{MG5\_aMC\@NLO} simulation for the $tZ\to th$ process: cross sections within SM, with $g_{hz{t_R}}=\pm 0.1$ and interference term $\sigma_{int}$. 
\label{fig:tZ}}
\end{center}
\end{figure}

\begin{figure}[htb!]
\begin{center}
\includegraphics[scale=.75]{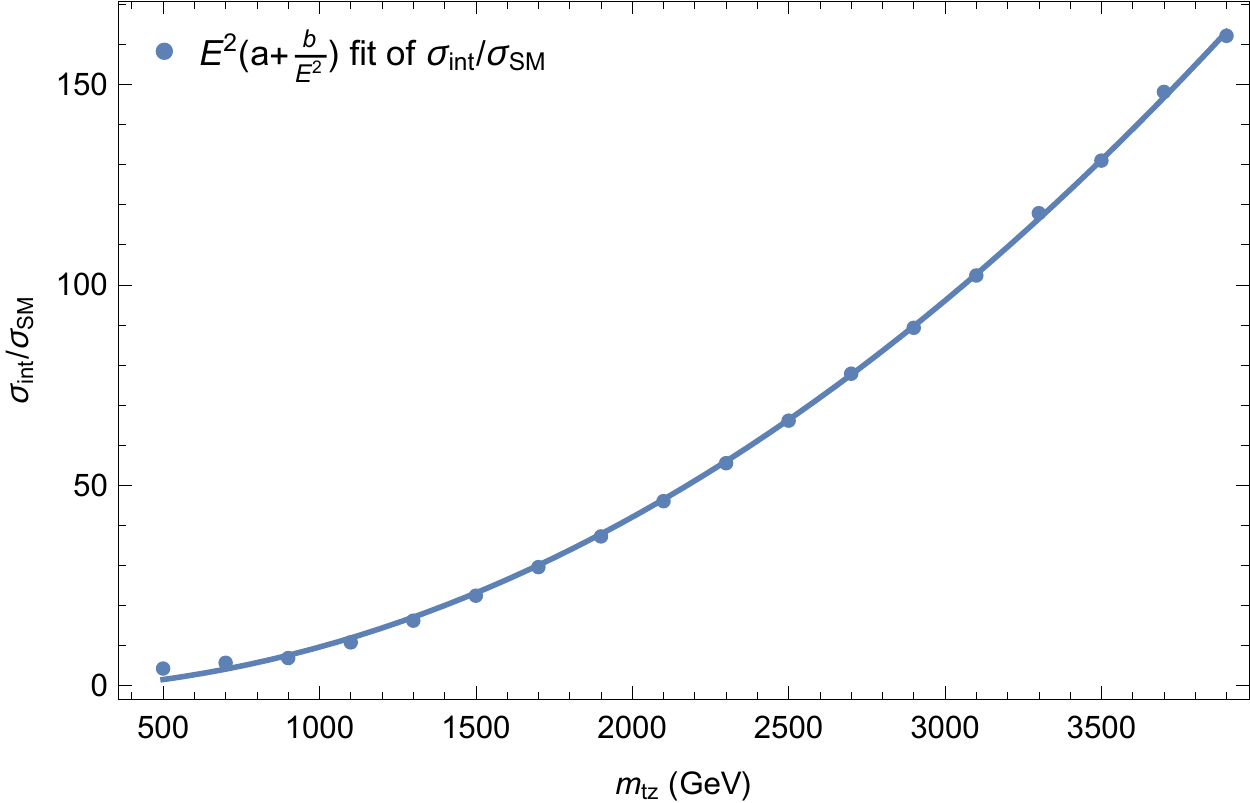}\\
\caption{Results from a \texttt{MG5\_aMC\@NLO} simulation of the $tZ\to th$ process: ratio $\sigma_{int}/\sigma_{SM}$. 
\label{fig:intSM_tZ}}
\end{center}
\end{figure}

The 4-point amplitude for $tZ\to th$, $A^{BSM}_{tZ\to th}$, can receive contributions from the contact interactions induced by $\mathcal{O}_{Ht_R}$. This scales as $ A^{BSM}_{tZ\to th}/A^{SM}_{tZ\to th} \sim m_{th}^2/\Lambda^2$. Therefore, since the SM cross-section falls off as $1/\hat{s}$, we expect a leading $\sim \hat{s}$ growth for the cross-section in presence of non-zero values for $g_{hZt_R}$ owing to the fact that in the large energy regime it is dominated by the $\sigma_{BSM^2}$ term (see Eq.~\ref{eq:SigmaEFT}). On the other hand, the interference $\sigma_{int}$ should behave as a constant in $\hat{s}$. These expectations are indeed consistent with the results of our simulations for $tZ\to th$ as shown in Figure~\ref{fig:tZ}. A pseudorapidity cut $|\eta |<2$ has been applied on the final state objects in order to remove the forward singularity. In Figure~\ref{fig:intSM_tZ}, the ratio between the interference and the SM terms of the cross-section is shown. It is evident that the $m_{th}$ distribution is consistent with the expected quadratic growth. We must also mention here that the $tZ\to th$ process is complementary to the $tW\to tW$ scattering studied in Ref.~\cite{Dror:2015nkp}, which yields similar results.

\section{Conclusions and Outlook}

In this work we have explored the possibility of constraining the $hZt_R \bar{t}_R$ coupling by studying the  $p p \to t \bar{t} h Z$ and $p p \to t \bar{t} h j$ processes where the EFT contributions grow quadratically in energy with respect to the SM.  Probing this coupling is important as it is unconstrained by LEP unlike the corresponding  couplings involving lighter quarks. Furthermore measuring this coupling can be a probe of naturalness in many interesting BSM theories, such as composite Higgs models, where they are generated upon integrating out the top partners. We find  $p p \to t \bar{t} h Z$ to be the more sensitive of the two processes.  Our preliminary estimates indicate that the $(h/v)Zt_R \bar{t}_R$ coupling can be measured at the percent to permille level (see Eq.~\ref{bounds}) which translates to a  bound $\Lambda >$ few TeV for the new physics scale.

\label{sec:conclusions}
\section*{Acknowledgements}

We would like to thank Satyaki Bhattacharya for his help in setting the limit tool. EV, SJ and SB would like to thank the organisers of the Les Houches Workshop Series: Physics at TeV Colliders for hospitality.

%% file: higgseftobs/higgseftobs.main.tex
\graphicspath{{higgseftobs/}}

\newcommand{\RIVET}{{\textsc{rivet}}\xspace}
\newcommand{\PYTHIA}{{\textsc{pythia}}\xspace}
\newcommand{\MADGRAPH}{\textsc{MadGraph}\xspace}
\newcommand{\MCATNLO}{\textsc{mc@nlo}\xspace}
\newcommand{\MGvATNLO}{\MADGRAPH{}5\_a\MCATNLO}

\chapter{EFT parametrization of Higgs observables}

{\it A.~Gilbert, J.~Langford, N.~Wardle}

\label{sec:higgseftobs}


\section{Introduction}

Effective field theories (EFTs) provide a consistent set of perturbations of the standard model (SM) Lagrangian~\cite{Patrignani:2016xqp,Brivio:2017vri} in which new physics enters at some higher energy scale $\Lambda$, assumed to be much larger than the electroweak scale and beyond our current direct detection reach. The SM EFT is constructed by the addition of higher-dimensional operators to the SM Lagrangian,

\begin{equation}
\mathcal{L}_{\mathrm{EFT}}=\mathcal{L}_{\mathrm{SM}}+\sum_{i}\frac{C_{i}^{(6)}}{\Lambda^{2}}\mathcal{O}_{i}^{(6)}+\sum_{i}\frac{C_{i}^{(8)}}{\Lambda^{4}}\mathcal{O}_{i}^{(8)} + \cdots,
\end{equation}

where $(D)$ denotes an operator of dimension $D$ and $C_{i}^{(D)} / \Lambda^{D-4}$ are Wilson coefficients which take non-zero values in the presence of new physics. The leading deviations from the SM are generally expected to occur at $D=6$.

The study of EFT effects in Higgs boson production and decay is of growing interest due to the large $\sqrt{s} = 13\,\mathrm{TeV}$ data set accumulated during LHC Run~2, and the increased sensitivity to kinematic information this provides. The ATLAS and CMS Collaborations have produced EFT constraints via interpretations of Simplified Template Cross Section (STXS) measurements~\cite{ATL-PHYS-PUB-2017-018,CMS-PAS-HIG-19-005}. The STXS framework~\cite{deFlorian:2016spz} defines a set of fiducial region for each Higgs boson production mode that are chosen to cover a broad range in relevant observables, for example the Higgs boson $\pT$ and additional jet multiplicity for gluon fusion production, and the $\pT$ of the vector boson in $\PW$ and $\PZ$ associated production.

One approach to constrain EFT coefficients is to start from an analysis which measures a set of unfolded fiducial cross sections of sensitive observables, such as the STXS.\@ Each cross section is then parametrized by the coefficients, and the best-fit values and associated confidence intervals can be determined. The parametrization of the cross sections could come from known analytic expressions, or via Monte Carlo simulation which incorporates an EFT Lagrangian model. This approach has the advantage of not requiring a full detector simulation of the events, and means new models can readily be applied to existing analyses. Conversely, a limitation is that it assumes there is no change of experimental acceptance due to the non-SM kinematics. Methods based on full simulation, for example Ref.~\cite{Sirunyan:2019twz}, can incorporate this information and utilise more optimal observables. This report focusses on the former approach, and is based on the procedure outlined in Refs.~\cite{Hays:2673969,ATL-PHYS-PUB-2019-042}.

 A software package has been developed, EFT2Obs~\footnote{\texttt{https://github.com/ajgilbert/EFT2Obs}}, that aims to facilitate the process of deriving an EFT parametrization. It utilises \MADGRAPH{}~v2.6.6~\cite{Alwall:2014hca} for simulation, interfaced to \PYTHIA~v8.2~\cite{Sjostrand:2014zea} for hadronisation and showering. The fiducial selection and histograms of observables are defined in the \RIVET{}~v3.0.1~\cite{Bierlich:2019rhm} framework. While this report focusses on the EFT parametrization of Higgs boson production, the tool is generic and can be applied to arbitrary observables for any model with new couplings implemented in \MADGRAPH{} via \textsc{FeynRules}~\cite{Alloul:2013bka}, subject to the limitations described in Section~\ref{sec:higgseftobs:summary}.

\section{Method}

This section outlines the main features of EFT2Obs. As input, the user must define the process to simulate using the standard \MADGRAPH{} syntax, as well as the \textsc{FeynRules} model to import which defines the EFT Lagrangian. The subset of $N$ model parameters that will be included in the cross section parametrization, denoted $c_{j}$, where $j=1 \ldots N$ are also specified. Finally, one or more \RIVET{} routines should be provided, defining histograms of observables to be parametrized, with each fiducial bin labelled with an index $i$ in the following.

The cross section for bin~$i$ can be expressed as the sum shown in Eq.~\ref{eq.higgseftobs.xs_eft}, where $\sigma_i^{\rm{int}}$ is the leading term in the EFT expansion ($\propto~1/\Lambda^2$), and accounts for the interference with the SM amplitude. The SM-independent term is labelled as $\sigma_i^{\rm{BSM}}$ ($\propto~1/\Lambda^4$).

\begin{equation}\label{eq.higgseftobs.xs_eft}
    \sigma_i^{\rm{EFT}}(c_{j}) = \sigma_i^{\rm{SM}} + \sigma_i^{\rm{int}}(c_{j}) + \sigma_i^{\rm{BSM}}(c_{j})
\end{equation}

Dividing through by the SM cross section, $\sigma_i^{\rm{SM}}$, provides a scaling function for each bin, $\mu_i(c_j)$, which parametrizes deviations in the cross section in terms of the $c_j$ parameters.  The general form of the scaling function is given in Eq.~\ref{eq.higgseftobs.mu_eft}.

\begin{equation}\label{eq.higgseftobs.mu_eft}
    \mu_i(c_j) = 1 + \sum_j A_j^{i} c_j + \sum_{jk} B_{jk}^{i} c_j c_k
\end{equation}

The tool then automates running \MADGRAPH{} to generate events, passing them through \PYTHIA{} before running \RIVET{} on the output. It handles splitting these steps into several independent parallel jobs, supporting both multi-process running on a single machine as well as a number of commonly used batch computing systems.

In order to determine the $A_{j}$ and $B_{jk}$ coefficients the cross sections in each bin, $\sigma_i^{\rm{EFT}}(c_{j})$, is sampled at a series of points in $c_{j}$. One approach is to generate multiple samples, each with varied $c_{j}$ values. There are however some drawbacks to this approach: the minimum number of independent samples needed is $2N + (N^{2}-N)/2$, which may become computationally prohibitive if $N$ is large. In addition, the statistical uncertainties in each sample must be propagated to the extracted coefficients, which may have a large impact if the $C_{j}$ for each sample are not well-chosen. For example, the precision on $\sigma_{\mathrm{int}}$, calculated from the difference between two samples giving $\sigma_{\mathrm{SM}}$ and $\sigma_{\mathrm{SM}} + \sigma_{\mathrm{int}}$ will be limited if the $C_{j}$ are chosen such that $\sigma_{\mathrm{int}} << \sigma_{\mathrm{SM}}$.

A more efficient approach, and the one utilised in EFT2Obs, is to generate a single sample in which a set of weights is calculated for each event, using the reweighting module in \MADGRAPH{}~\cite{Mattelaer:2016gcx}. Each weight $W$ corresponds to the original event weight $W_{\mathrm{orig}}$ multiplied by the ratio of LO squared matrix elements: ${|M_{\mathrm{new}}|}^{2} / {|M_{\mathrm{orig}}|}^{2}$, with the numerator evaluated at the specified values of $c_{j}$ and the denominator giving the matrix element used for the event generation. The tool will automatically generate the reweighting configuration for \MADGRAPH{}. For this it uses the set of parameters requested by the user and corresponding set of non-zero values, denoted $D_{i}$. Considering a simple case where only two operators are considered, $c_{1}$ and $c_{2}$, Table~\ref{tab:higgseftobs:weights} gives the  set of parameter configurations that will be evaluated. The generalisation to any number of operators is straightforward: two weights are needs to calculate the linear and squared terms for each parameter in turn, and an additional weight is needed for each $c_{j}c_{k}$ combination where $j \neq k$.

\begin{table}[hbtp]
\centering
\caption{The set of weights for the two parameter case. The values of $c_{1}$ and $c_{2}$ for each weight are given, along with expressions for the weights in terms of the $A_{j}$ and $B_{jk}$ factors which are sampled in each event. The final column gives the expression for the weights after a transformation is applied.}
\begin{tabular}{@{} c c c c c @{}}
\hline
Label & $c_{1}$ & $c_{2}$ & Weight & Transformed weight \\
\hline
$W_{1}$ & $0$ & $0$ & $1$ & $1$ \\
$W_{2}$ & $0.5 D_{1}$ & $0$ & $1 + 0.5 D_{1} A_{1} + 0.25 {D_{1}}^{2} B_{11}$ & $D_{1} A_{1}$ \\
$W_{3}$ & $D_{1}$ & $0$ & $1 + D_{1} A_{1} + {D_{1}}^{2} B_{11}$ & ${D_{1}}^{2} B_{11}$ \\
$W_{4}$ & $0$ & $0.5 D_{2}$ & $1 + 0.5 D_{2} A_{2} + 0.25 {D_{2}}^{2} B_{22}$ & $D_{2} A_{2}$ \\
$W_{5}$ & $0$ & $D_{2}$ & $1 + D_{2} A_{2} + {D_{2}}^{2} B_{22}$ & ${D_{2}}^{2} B_{22}$ \\
$W_{6}$ & $D_{1}$ & $D_{2}$ & $1 + D_{1} A_{1} + D_{2} A_{2} + {D_{1}}^{2} B_{11} + {D_{2}}^{2} B_{22} + D_{1}D_{2}B_{12}$ & $D_{1}D_{2}B_{12}$  \\
\hline
\end{tabular}\label{tab:higgseftobs:weights}
\vspace{0.5cm}
\end{table}

The final output from \RIVET{} is a set of histograms for each observable, one per event weight configuration. In each histogram the sum of the weights and sum of squared weights is stored, allowing the mean weight and associated standard error to be determined. From these the averages $\overline{A_{j}}$ and $\overline{B_{jk}}$ can be extracted. However it is important to note that as shown in Table~\ref{tab:higgseftobs:weights} the $A_{j}$ and $B_{jk}$ distributions are not sampled directly. For example in $W_{3}$ the distribution of $1 + D_{1}A_{1} + D_{(1)}^{2}B_{11}$ is sampled when $c_{1} = D_{1}$, since both the interference and SM-independent contributions are included by default. While the $\overline{A_{j}}$ and $\overline{B_{jk}}$ can be determined via linear algebra from the set of $\overline{W_{i}}$, it is generally not possible to isolate their standard errors given only the standard errors on the $\overline{W_{i}}$. As a solution, an event-by-event transformation of the weights is performed before the events are passed through $\RIVET$, in order to isolate the $A_{j}$ and $B_{jk}$ distributions. The transformed weights are given in the final column of Table~\ref{tab:higgseftobs:weights}.

\section{Example}

This section gives an example of parametrizing the $\PZ\PH$ production process using the Higgs Effective Lagrangian (HEL) model~\cite{Contino:2013kra,Alloul:2013naa}, and using the STXS \RIVET{} routine to study the effect on the $\pT(\PZ)$ distribution. The HEL model introduces 39 flavour independent dimension-6 operators, $\mathcal{O}_{j}$. Table~\ref{tab:higgseftobs:eft_vertices} summarises the relevant operators considered in this example, and the normalisation convention used for each of the corresponding parameters $c_{j}$.

\begin{table}[h!]
    \centering
    \caption{The HEL operators and the corresponding parameters considered in this example. Also given are the relevant vertices which are modified by these operators.}
    \begin{tabular}{p{2cm} p{5cm} p{3cm} p{3cm}}
    \hline
        Operator & Expression & Parameter & Relevant vertices \\ \hline
        $\mathcal{O}_A$ & $|H^2|B_{\mu\nu}B^{\mu\nu}$ & $c_A=\frac{m_W^2}{g'^2}\frac{f_A}{\Lambda^2}$ & \ensuremath{\PH\PGg\PGg} \\[6pt]
        $\mathcal{O}_{HW}$ & $i(D^\mu H)^\dagger\sigma^a(D^\nu H)W^a_{\mu\nu}$ & $c_{HW}=\frac{m_W^2}{2g}\frac{f_{HW}}{\Lambda^2}$ & \ensuremath{\PH\PW\PW}, \ensuremath{\PH\PZ\PZ}  \\[6pt]
        $\mathcal{O}_{HB}$ & $i(D^\mu H)^\dagger(D^\nu H)B_{\mu\nu}$ & $c_{HB}=\frac{m_W^2}{g'}\frac{f_{HB}}{\Lambda^2}$ & \ensuremath{\PH\PZ\PZ}  \\[6pt]
        $\mathcal{O}_{W}$ & $i(H^\dagger \sigma^a \overleftrightarrow{D}^\mu H)D^\nu W^a_{\mu\nu}$ & $c_{WW}=\frac{m_W^2}{g}\frac{f_{WW}}{\Lambda^2}$ & \ensuremath{\PH\PW\PW}, \ensuremath{\PH\PZ\PZ} \\[6pt]
        $\mathcal{O}_B$ & $i(H^\dagger \overleftrightarrow{D}^\mu H)\partial^\nu B_{\mu\nu}$ & $c_{B}=\frac{2m_W^2}{g'}\frac{f_{B}}{\Lambda^2}$ & \ensuremath{\PH\PZ\PZ} \\[6pt]
    \hline
    \end{tabular}\label{tab:higgseftobs:eft_vertices}
\end{table}

Within EFT2Obs a \texttt{proc\_card.dat} file is specified which imports the HEL model and defines the leading order ZH process, limited to at most one vertex that depends on an EFT parameter using the \texttt{NP<=1} syntax.

\begin{Verbatim}[frame=single,fontsize=\small]
import model HEL_UFO
define ell+ = e+ mu+ ta+
define ell- = e- mu- ta-
define j2 = g u c d s u~ c~ d~ s~
generate p p > h ell+ ell- / j2 ell+ ell- vl vl~ a h w+ w- NP<=1 @0
output zh
\end{Verbatim}

Using this file as input, the \MADGRAPH{} directory structure for the process is then created. A configuration file in JSON format is also needed, which in this case specifies that $c_{A}$, $c_{HW}$, $c_{HB}$, $c_{WW}$ and $c_{B}$ are the parameters of interest, that $D_{i} = 0.01$ should be used in the reweighting, and that all other parameters should be set to zero throughout. It should be noted that the choices of $D_{i}$ are in principle arbitrary, but should be chosen such that the resulting weights are not so large or small as to cause issues with numerical precision. From this configuration file a  reweighting card is created automatically which specifies the 20 points in $c_{i}$ that will be evaluated.

A sample of events is then generated and the reweighting and weight transformation described in the previous section are performed. The events are hadronised and showered with \PYTHIA{} and passed through the \RIVET{} routine. From the output histograms the scaling terms for each bin of the $\pT(\PZ)$ distribution are calculated, along with the associated statistical uncertainty. An example of the output for a single bin and just two parameters is given below.

\begin{Verbatim}[frame=single,fontsize=\small]
-----------------------------------------------------------------
Bin 0    numEntries: 77325      mean: 1.25e-10   stderr: 2.95e-15
-----------------------------------------------------------------
Term                 |        Value |       Uncert | Rel. uncert.
-----------------------------------------------------------------
chw                  |       4.7348 |       0.0083 |       0.0018
chb                  |       1.4179 |       0.0025 |       0.0018
chw^2                |       7.5364 |       0.1640 |       0.0218
chb^2                |       0.6758 |       0.0147 |       0.0218
chw * chb            |       4.5137 |       0.0982 |       0.0218
\end{Verbatim}

\begin{figure}[htbp]
  \begin{center}
    \includegraphics[width=0.65\textwidth]{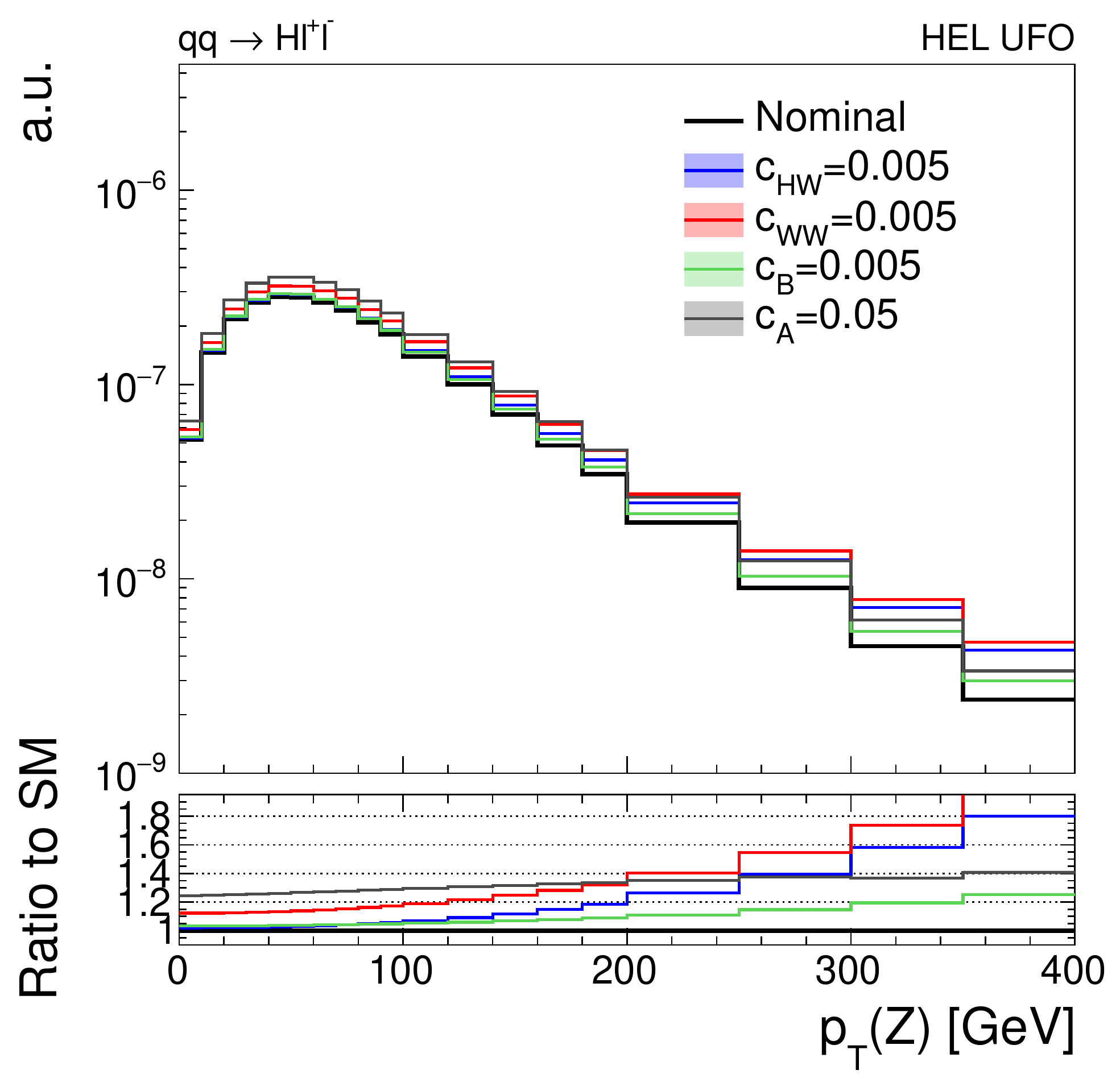}
    \caption{Distribution of $\pT(\PZ)$ comparing the SM expectation (black) to the SM with the inclusion of non-zero EFT coefficients in the Higgs Effective Lagrangian.}
    \label{fig:higgseftobs:plot1}
  \end{center}
\end{figure}

The uncertainty on each term can be useful in estimating the additional number of events needed to reduce the uncertainty to a particular level. The full parametrisation in four coarse bins of $\pT(\PZ)$ is given in Table~\ref{tab:scaling_table}. Figure~\ref{fig:higgseftobs:plot1} shows the distribution expected in the SM along with a few examples for non-zero values of the $c_{j}$. The statistical uncertainty on the $A_{j}$ and $B_{jk}$ is propagated to these distributions.

\begin{table}[htb]
    \centering
    \caption{The scaling functions produced by the EFT2Obs tool for four bins of $\pT(\PZ)$ in $\PZ\PH$ production.}
    \setlength\tabcolsep{10pt}
    \begin{tabular}{|c|c|}
    \hline
    $\pT(\PZ)$ bin [GeV] & Scaling \\
    \hline
$0$--$50$ & \parbox{0.6\columnwidth}{$1  + 24.9\,\,c_{WW} + 6.9\,\,c_{B} + 4.7\,\,c_{HW} + 1.4\,\,c_{HB} + 4.7\,\,c_{A} + 160.2\,\,{c_{WW}}^{2} + 12.9\,\,{c_{B}}^{2} + 7.5\,\,{c_{HW}}^{2} + 0.7\,\,{c_{HB}}^{2} + 7.2\,\,{c_{A}}^{2} + 88.8\,\,c_{WW}\,c_{B} + 60.9\,\,c_{WW}\,c_{HW} + 18.2\,\,c_{WW}\,c_{HB} + 59.9\,\,c_{WW}\,c_{A} + 17.1\,\,c_{B}\,c_{HW} + 5.1\,\,c_{B}\,c_{HB} + 18.5\,\,c_{B}\,c_{A} + 4.5\,\,c_{HW}\,c_{HB} + 10.7\,\,c_{HW}\,c_{A} + 3.2\,\,c_{HB}\,c_{A}$} \\
\hline
$50$--$100$ & \parbox{0.6\columnwidth}{$1  + 29.4\,\,c_{WW} + 8.3\,\,c_{B} + 8.1\,\,c_{HW} + 2.4\,\,c_{HB} + 5.1\,\,c_{A} + 227.6\,\,{c_{WW}}^{2} + 18.6\,\,{c_{B}}^{2} + 25.2\,\,{c_{HW}}^{2} + 2.3\,\,{c_{HB}}^{2} + 8.6\,\,{c_{A}}^{2} + 127.9\,\,c_{WW}\,c_{B} + 126.5\,\,c_{WW}\,c_{HW} + 37.9\,\,c_{WW}\,c_{HB} + 77.1\,\,c_{WW}\,c_{A} + 35.9\,\,c_{B}\,c_{HW} + 10.7\,\,c_{B}\,c_{HB} + 23.6\,\,c_{B}\,c_{A} + 15.1\,\,c_{HW}\,c_{HB} + 17.7\,\,c_{HW}\,c_{A} + 5.3\,\,c_{HB}\,c_{A}$} \\
\hline
$100$--$200$ & \parbox{0.6\columnwidth}{$1  + 43.3\,\,c_{WW} + 12.5\,\,c_{B} + 20.0\,\,c_{HW} + 6.0\,\,c_{HB} + 5.6\,\,c_{A} + 518.6\,\,{c_{WW}}^{2} + 43.5\,\,{c_{B}}^{2} + 160.6\,\,{c_{HW}}^{2} + 14.4\,\,{c_{HB}}^{2} + 12.1\,\,{c_{A}}^{2} + 298.2\,\,c_{WW}\,c_{B} + 503.6\,\,c_{WW}\,c_{HW} + 150.8\,\,c_{WW}\,c_{HB} + 127.1\,\,c_{WW}\,c_{A} + 145.6\,\,c_{B}\,c_{HW} + 43.6\,\,c_{B}\,c_{HB} + 38.5\,\,c_{B}\,c_{A} + 96.2\,\,c_{HW}\,c_{HB} + 41.9\,\,c_{HW}\,c_{A} + 12.6\,\,c_{HB}\,c_{A}$} \\
\hline
$200$--$400$ & \parbox{0.6\columnwidth}{$1  + 91.4\,\,c_{WW} + 26.9\,\,c_{B} + 66.4\,\,c_{HW} + 19.9\,\,c_{HB} + 6.1\,\,c_{A} + 2521.2\,\,{c_{WW}}^{2} + 219.5\,\,{c_{B}}^{2} + 1630.0\,\,{c_{HW}}^{2} + 146.2\,\,{c_{HB}}^{2} + 23.6\,\,{c_{A}}^{2} + 1485.2\,\,c_{WW}\,c_{B} + 3825.9\,\,c_{WW}\,c_{HW} + 1145.7\,\,c_{WW}\,c_{HB} + 295.3\,\,c_{WW}\,c_{A} + 1128.2\,\,c_{B}\,c_{HW} + 337.8\,\,c_{B}\,c_{HB} + 89.0\,\,c_{B}\,c_{A} + 976.2\,\,c_{HW}\,c_{HB} + 127.6\,\,c_{HW}\,c_{A} + 38.2\,\,c_{HB}\,c_{A}$} \\
\hline
\end{tabular}\label{tab:scaling_table}
    \end{table}

\section{Investigating acceptance corrections}
As previously mentioned, the approach documented in this section is limited by assuming the introduction of non-SM kinematics has a negligible effect on the experimental acceptance. By including acceptance cuts in the \RIVET{} routine, it becomes possible to study EFT corrections to this acceptance. 

\begin{figure}[htbp]
  \begin{center}
    \includegraphics[width=0.65\textwidth]{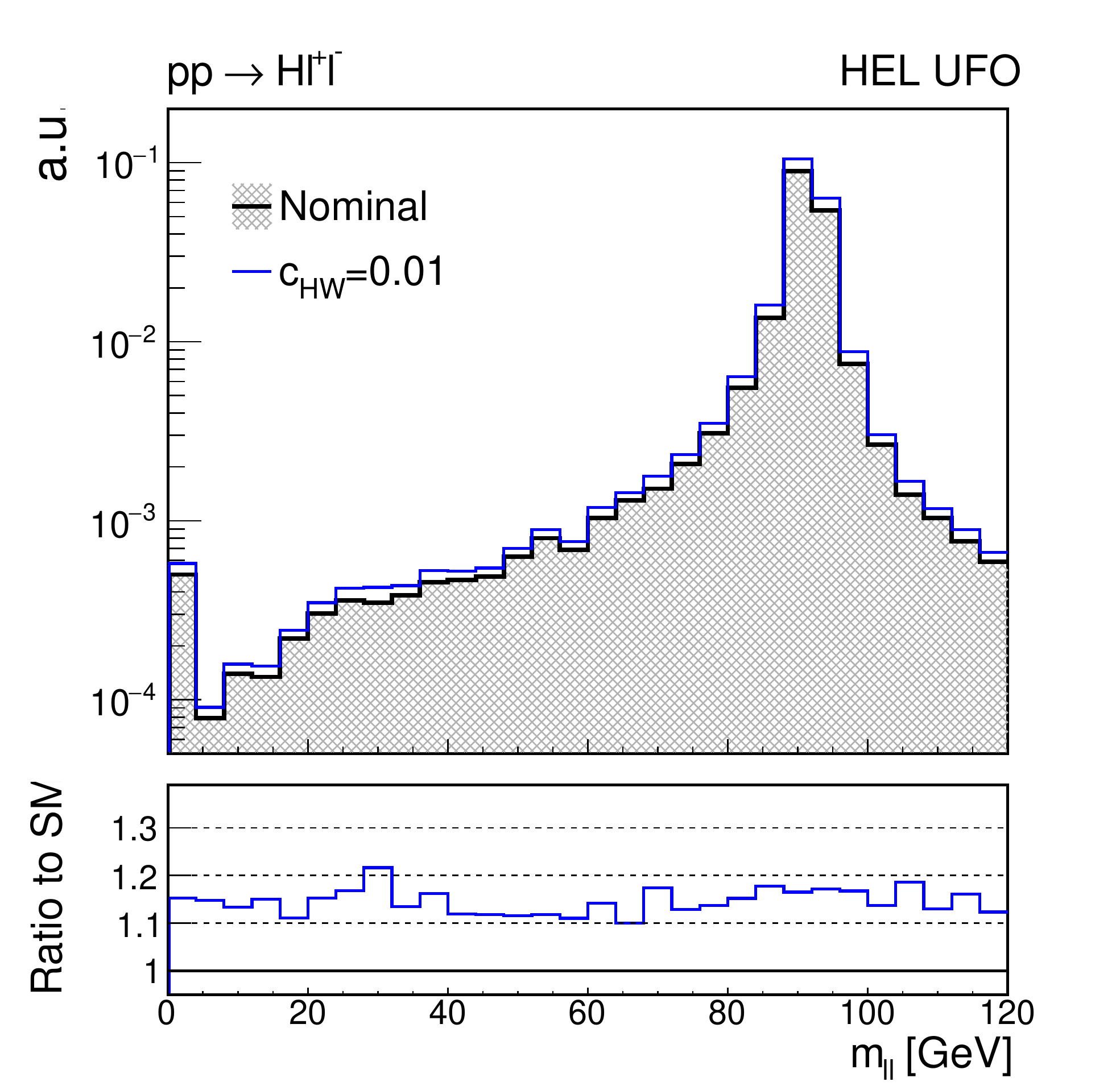}
    \caption{Distribution of $m_{\ell\ell}$ comparing the SM expectation (black hatched) to the SM with the inclusion of contributions from the HEL parameter, $c_{HW}$ (0.01).}
    \label{fig:higgseftobs:plot2}
  \end{center}
\end{figure}

Figure \ref{fig:higgseftobs:plot2} shows the dilepton invariant mass, $m_{\ell\ell}$, distribution for the $\PZ\PH$ process in the SM and for a non-zero value of $c_{HW}$. Experimental acceptance ensures events at low $m_{\ell\ell}$ do not enter current analyses and the effects of such selections can be studied with the EFT2Obs tool. 
In this particular case, the $c_{HW}$ dependency varies by at most a few per cent across the entire $m_{\ell\ell}$ spectrum.

\section{Summary}\label{sec:higgseftobs:summary}

This report has described the main features of a software tool, EFT2Obs, to assist in the parametrization of observables by EFT parameters. The current implementation features several limitations that will be addressed in future developments. In particular, the tool requires there is at most one vertex containing the new operators in each diagrams produced. In the example of the Higgs processes given here this means that new physics contributions can be studied in either Higgs boson production or decay, but not both simultaneously. This restriction can be lifted by sampling an expanded set of points in the reweighting step, such that terms at more than second order in the $c_{i}$ can be evaluated. In addition, the parametrization of processes at next-to-leading order in QCD is not currently supported, but can implemented following a conceptually similar approach to the leading-order case presented here.

\let\PYTHIA\undefined
\let\MADGRAPH\undefined
\let\MCATNLO\undefined
\let\MGvATNLO\undefined

%% file: likelihood/likelihoodinfo.main.tex
\graphicspath{{likelihood/}}

\newcommand{\ma}{{\sc MadAnalysis\,5}\xspace}
\newcommand{\smodels}{{\sc SModelS}\xspace}
\newcommand{\met}{E_T^{\rm miss}}

\newcommand{\sk}[1]{{\color{blue}[{\bf Sabine:} #1]}}


\chapter{On the use of likelihood information for reinterpreting LHC results}
{\it G.~Alguero, J.~Araz, B.~Fuks, S.~Kraml, W.~Waltenberger}


\label{sec:likelihoodinfo}

\begin{abstract}
Both the ATLAS and CMS collaborations have started to provide information on
background correlations for some of their searches.
This is extremely useful for re-interpretation purposes, as it
allows one to construct a proper likelihood from the combination of signal regions (SRs) of a given search.
The two collaborations follow quasi-ortogonal approaches: 
while CMS provides the linear correlations between the background contributions as simple covariance matrices, ATLAS recently released the full likelihood model for two analyses. 
We discuss the usage of this (simplified or full) likelihood information in two public re-interpretation tools, 
\ma~\cite{Conte:2018vmg} and \smodels~\cite{Ambrogi:2018ujg}.
\end{abstract}


\section{CMS simplified likelihoods} \label{sec:likelihoodinfo:cms-covariances}

For most of its Run~2 analyses of $36\Ufb^{-1}$ of data, the CMS SUSY group made available approximate covariance matrices for the background estimates in various search regions, or a reduced set thereof. These covariance matrices can be used to build a simplified likelihood~\cite{CMS:2242860} for any signal model.
The underlying assumptions are that systematic uncertainties in the signal modeling can be neglected, and that uncertainties on the background contributions are Gaussian such that the distribution of the number of background events is symmetric around the expectation.
Although approximate, the combination of SRs through this simplified likelihood scheme greatly improves the constraining power of analysis recasts.

The combination of SRs via the simplified likelihood approach has been available 
in \smodels since v1.2.0 (Aug 2018)~\cite{Ambrogi:2018ujg}. 
We adopt this implementation in \smodels, {\it i.e.}\ the relevant python module
 for use in \ma (v1.8 onwards).
In order to comply with the \ma framework \cite{Conte:2018vmg}, the covariance information has to be included in the \verb|.info| file associated with the studied analysis. For each region, the covariance with every other region must be supplied. There is thus a list of covariances with two entries: the paired region and the value of the covariance. 
The standard syntax of the \verb|.info| file 
\begin{verbatim}
<analysis id="analysis name">
    <region type="signal" id="region name">
        <nobs> ... </nobs>
        <nb> ... </nb>
        <deltanb> ... </deltanb>
    </region>
    ...
</analysis>
\end{verbatim}
specifying, for each SR, the number of observed events \verb|<nobs>|,
expected background events \verb|<nb>| and their uncertainty \verb|<deltanb>|, 
is therefore extended as follows: 
\begin{verbatim}
<analysis id="analysis name" cov_subset="combined SRs">
    <region type="signal" id="region name">
        <nobs> ... </nobs>
        <nb> ... </nb>
        <deltanb> ... </deltanb>
        <covariance region="first SR name">...</covariance>
        <covariance region="second SR name">...</covariance>
        ...
        <covariance region="last SR name">...</covariance>
    </region>
    ...
</analysis>
\end{verbatim}

If a covariance is not supplied, it will be considered as zero in the covariance matrix. However, if a region does not contain any covariance field, it will not be included in the SR combination.
This feature can be useful if the covariance matrix is available only for a subset of SRs, like for example in the CMS multilepton plus missing transverse energy ($\met$) search \verb|CMS-SUS-16-039| \cite{Sirunyan:2017lae} which provides covariances only for SRs of type~A.
In order to keep trace of which subset of SRs is combined, we introduce a \verb|cov_subset| attribute in the \verb|<analysis>| tag, in which one can give a brief description of the subset of SRs for which the covariance is available.
For the \verb|CMS-SUS-16-039| example, this is 
\begin{verbatim}
  <analysis id="cms_sus_16_039" cov_subset="SRs_A">
\end{verbatim}
This description will be printed in the output file (\verb|CLs_output_summary.dat|) with the result from the simplified likelihood combination, just after the usual
exclusion information under the format
\begin{verbatim}
  <set> <tag> <cov_subset> <exp> <obs> <CLs> ||
\end{verbatim}
The successive elements consist of the dataset name, the analysis name,  the description of the subset of combined SRs, the expected and observed cross section upper limits at 95\% confidence level (CL), and finally the exclusion level, $1-\textrm{CL}_s$. A concrete example reads
\begin{verbatim}
  defaultset  cms_sus_16_039  [SL]-SRs_A  10.4851515  11.1534040  0.9997  ||
\end{verbatim}
where `SL' stands for `simplified likelihood'. The statistical error, usually provided after the double bar, is not printed here as it is already encoded in the simplified likelihood calculation.

To illustrate the gain in constraining power, we stay with the example of the \verb|CMS-SUS-16-039| analysis, which is available as recast code \cite{CMS-SUS-16-039-recast} in the \ma PAD (Public Analysis Database), and whose \verb|.info| file we extended by the covariance information provided by CMS for the SRs of class~A.
The analysis is searching for supersymmetry through multilepton plus $\met$ final states, and the set of SRs class A specifically concerns final states with three electrons or muons comprising at least one opposite-sign-same-flavour pair.

The targeted process is chargino/neutralino associated production, $pp\to \tilde\chi^\pm_1\tilde\chi^0_2$, with $\tilde\chi^\pm_1\to W^\pm\tilde\chi^0_1$ and $\tilde\chi^0_2\to Z^0\tilde\chi^0_1$ decays. 
Focusing on such a signal, we simulate events with 
{\sc MadGraph\,5\_aMC@NLO}~\cite{Alwall:2014hca} and 
{\sc Pythia\,8}~\cite{Sjostrand:2014zea}, assuming wino-like $\tilde\chi^\pm_1$, 
$\tilde\chi^0_2$ with $m_{\tilde{\chi}^{\pm}_1} = m_{\tilde{\chi}^0_2}$ and 100\% branching ratio into the $WZ+\met$
final state (so-called TChiWZ simplified model).
We normalise the event samples to cross sections evaluated at the
next-to-leading order matched with next-to-leading-logarithmic threshold
resummation (NLO+NLL)~\cite{Fuks:2012qx,Fuks:2013vua}, using the CTEQ6.6 PDF
set~\cite{Nadolsky:2008zw}.\footnote{See
\url{https://twiki.cern.ch/twiki/bin/view/LHCPhysics/SUSYCrossSections}.}
We then reinterpret the results of the CMS-SUS-16-039 search in \ma.

\begin{figure}[t]
        \centering
        \includegraphics[width=0.495\textwidth]{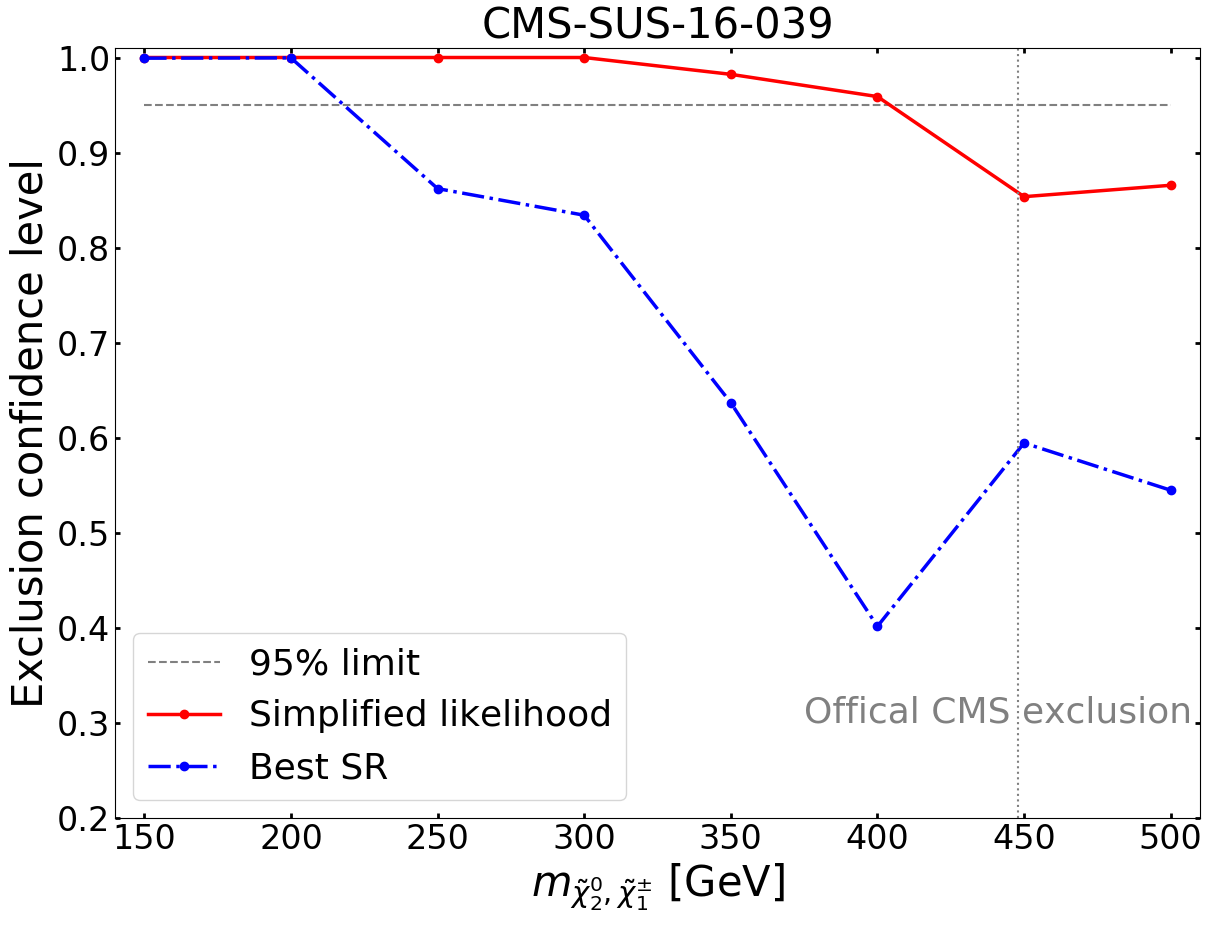}~
        \includegraphics[width=0.505\textwidth]{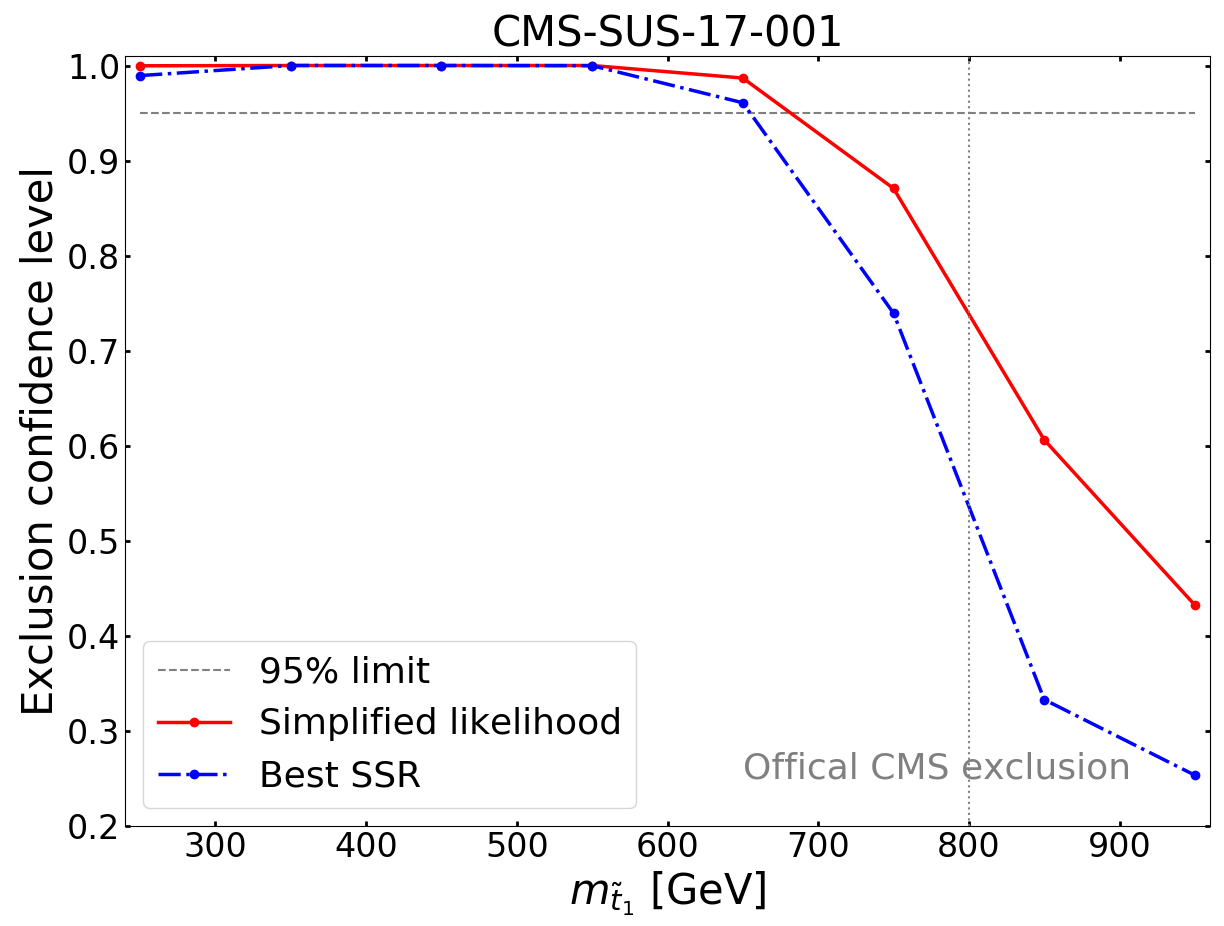}
    \caption{Exclusion CL from recasts of the CMS multilepton analysis \cite{Sirunyan:2017lae} for the TChiWZ simplified model (left) and of the CMS dilepton stop search \cite{Sirunyan:2017leh} for the T2tt simplified model (right).
    In both figures, $m_{\tilde{\chi}^0_1}=50\UGeV$. The official CMS bounds are
    indicated as vertical dotted lines.}
    \label{fig:likelihoodinfo:simplike}
\end{figure}

Figure~\ref{fig:likelihoodinfo:simplike}\,(left) displays the resulting exclusion CL as a function of the common chargino/neutralino mass $m_{\tilde{\chi}_1^\pm,\tilde{\chi}_2^0}$ for a fixed lightest neutralino mass $m_{\tilde{\chi}^0_1}=50\UGeV$. It turns out that in the conventional approach of using only the `best'
({\it i.e.}\ the most sensitive) SR for limit setting (blue), only values up to $m_{\tilde{\chi}_1^\pm,\tilde{\chi}_2^0}\approx 220\UGeV$ can be excluded at the 95\%~CL. The combination of SRs in the simplified likelihood approach allows to extend this range to $m_{\tilde{\chi}_1^\pm,\tilde{\chi}_2^0}\approx 400\UGeV$ (red), which is only slightly below the official CMS bound (vertical dotted line).

Given the covariance matrix, one may also define new `aggregated' search regions tailored to the new physics model of interest, provided they are defined as the union of existing analysis search regions. In this case the covariance matrix can be used to extract  the total uncertainty on the background expectation in the aggregated regions~\cite{CMS:2242860}. The implementation of the simplified likelihood in \smodels\ makes use of this option in order to reduce the computing time~\cite{Ambrogi:2018ujg}. As some information (and thus precision) is
lost in the aggregation process, care must be taken to aggregate SRs without too much loss in constraining power 
(see {\it e.g.}\ Fig.~1 in ref.~\cite{Ambrogi:2018ujg} and the related discussion in there).

Because of the huge number of SRs in some analyses, CMS sometimes introduces `super signal regions' (SSRs) in order to facilitate recasting. These SSRs are usually aggregated SRs, that should allow for the reproduction of official CMS bounds with good precision without the need of combination.
One example available in the \ma\ PAD is the recast code \cite{CMS-SUS-17-001-recast} emulating the CMS search for stops in the dilepton final state \verb|CMS-SUS-17-001|, which includes three SSRs~\cite{Sirunyan:2017leh}. 
Figure~\ref{fig:likelihoodinfo:simplike}\,(right) compares the constraints 
on stop pair production, $pp\to\tilde t_1^{}\tilde t_1^*$, 
followed by $\tilde t_1\to\tilde\chi_1^0$ with 100\% BR (T2tt simplified model) using only the best SSR (blue), or combining the three SSRs in the simplified likelihood approach (red). As above, we simulated events with {\sc MadGraph\,5\_aMC@NLO} and {\sc Pythia\,8}, 
and applied NLO-NLL cross sections \cite{Borschensky:2014cia}
as quoted by the SUSY cross section working group. 
We observe that the combination gives only a slight, though still noticable, improvement. 

Super signal regions are also defined in \verb|CMS-SUS-16-039|. 
In this case they are, however, {\it not} appropriately aggregated regions from the full analysis, and neither of the mass points would be excluded
in Figure~\ref{fig:likelihoodinfo:simplike}\,(left), as the exclusion CL would never reach the 95\% level.

Last but not least, the covariances published by CMS encode only the linear correlation between the background contributions, {\it i.e.}\ they assume symmetric Gaussian uncertainties. To correctly account for asymmetries, ref.~\cite{Buckley:2018vdr} proposes a convenient and easy-to-use simplified likelihood scheme that keeps the next-to-leading term in the large $N$ expansion. We hope that this will be adopted in the future.

\section{ATLAS full likelihoods} \label{sec:likelihoodinfo:atlas-likelihood}

The ATLAS collaboration recently started to provide complete likelihoods associated with statistical fits used in searches for new physics on
{\sc HEPData}~\cite{ATL-PHYS-PUB-2019-029}. This is based on
\textsc{HistFactory}~\cite{Cranmer:1456844} and embedded in a JSON scheme,
which enables the usage of the \textsc{HistFactory} structure outside the
\textsc{Root} framework. 
For a given likelihood, the \textsc{Pyhf} package~\cite{lukas_2019_3334365} allows for, {\it e.g.}, the calculation of the CL$_s$ value from the corresponding statistical model.
So far, such JSON likelihood data is available for two analyses, the ATLAS 2019 sbottom multi-$b$ search~\cite{Aad:2019pfy} and the search for direct
stau production~\cite{Aad:2019byo}, both using the full Run~2 dataset ($139\Ufb^{-1}$).

For the usage of this full likelihood information in \ma, 
the relevant JSON files must be located in the same analysis folder as the
recast code (done automatically when downloading from the \ma\ online repository at the time of the PAD installation). 
Moreover, the \verb|.info| file  
must include new \verb|<pyhf>| elements specifying the names of the JSON files
together with the corresponding channels (ensembles of SRs) and the regions they
include, as defined in the JSON files. In the
example of the sbottom search~\cite{Aad:2019pfy} ({\tt ATLAS-SUSY-2018-031}), assuming that the 
signal does not leak into validation/control regions (VRs/CRs) as only SRs
are 
currently being recasted, this gives for regions~A
\begin{verbatim}
<pyhf id="RegionA">
  <name>atlas_susy_2018_031_SRA.json</name>
    <regions>
      <channel name="SR_meff">SRA_L SRA_M SRA_H</channel>
      <channel name="VRtt_meff"> </channel>
      <channel name="CRtt_meff"> </channel>
    </regions>
</pyhf>
\end{verbatim}
and analogous for regions B and C. 
Whenever the exclusion is computed by means of the \textsc{Pyhf} package, the
results are reported in the \verb|CLs_output_summary.dat| output file in the
form
\begin{verbatim}
  <set> <tag> <SR> <best?> <exp> <obs> <CLs> ||
\end{verbatim}
No statistical error information is printed (to the right of the double bars),
as it is already accounted for in the likelihood calculation.

For these proceedings, we implemented the {\tt ATLAS-SUSY-2018-031} analysis in the \ma\ framework. Focusing on the benchmark point
$m_{\tilde b_1}=1100\UGeV$, 
$m_{\tilde\chi^0_2}=330\UGeV$ and  
$m_{\tilde\chi^0_1}=200\UGeV$ from the ATLAS study, 
we simulated sbottom pair production followed by $\tilde b_1\to b\tilde\chi^0_2$, $\tilde\chi^0_2\to h\tilde\chi^0_1$  
with
{\sc MadGraph5\_aMC@NLO} and {\sc Pythia~8}; 
we used the LO cross sections from {\sc MadGraph} and merged matrix elements featuring
up to two additional jets following the MLM prescription~\cite{Mangano:2006rw}. 
The result in the \verb|CLs_output_summary.dat| file reads 
\begin{verbatim}
defaultset atlas_susy_2018_031 [pyhf]-RegionA-profile 1 0.0020761 0.0013821 0.9959  ||
defaultset atlas_susy_2018_031 [pyhf]-RegionB-profile 0 -1        -1        0.0000  ||
defaultset atlas_susy_2018_031 [pyhf]-RegionC-profile 0 0.1139845 0.1846410 0.0293  ||
\end{verbatim}
and includes expected (\verb+<exp>+) and observed (\verb+<obs>+) cross sections
excluded at the 95\% CL, as well as the exclusion level obtained for each
likelihood (\verb+<CLs>+).
Further validation and
tests are ongoing, both for the \textsc{Pyhf} interface and the
implementation of the ATLAS-SUSY-2018-031 analysis within \ma. The
integration of theoretical uncertainties into the likelihood is also planned.

Work on an interface of the \textsc{Pyhf} package with \smodels\ is also underway. This is much helped by the fact that both the sbottom and stau ATLAS analyses provide signal acceptance and efficiency maps for all SRs in addition to cross section upper limits, which is exactly what is needed by \smodels. 
The implementation of these maps in the \smodels\ database is currently being validated, with a release foreseen for the v1.2.3 database. 
Last but not least we note that the communication with the \textsc{Pyhf} team on questions regarding the usage of the JSON scheme is handled via github’s issue tracking system and is thus transparent and open to all, see {\it e.g.}~\url{https://github.com/scikit-hep/pyhf/issues/620}.

\section{Conclusions} \label{sec:likelihoodinfo:conclusions}

After \smodels\,v1.2, \ma\ v1.8 will also be able to exploit the
covariance information provided by the CMS collaboration in the simplified
likelihood scheme. 
We have demonstrated the gain in constraining power, and hope
CMS will continue to make background correlation data available in digitised form for the final Run~2 results.
We also advocate to use the scheme of ref.~\cite{Buckley:2018vdr} whenever assymetric uncertainties are relevant for simplified likelihoods.

Moreover, the
integration of the full likelihood information via the JSON format, as recently released by the
ATLAS collaboration for two analyses, is on the way in both \ma\ and \smodels.
We highly welcome this important move by ATLAS to publish full likelihoods.

\let\ma\undefined
\let\smodels\undefined

%% file: anacor/BSM_AnalysisCorrelations.main.tex
\graphicspath{{anacor/}}


\chapter{Determination of Independent Signal Regions in LHC Searches for New Physics}
{\it A.~Buckley, B.~Fuks, H.~Reyes-Gonz\'alez, W.~Waltenberger, S.~L.~Williamson}


\label{sec:BSM_AnalysisCorrelations}

\begin{abstract}
The first two runs of the LHC have aggregated an astonishing number of published searches for new physics -- nearly 600 from ATLAS and CMS alone at the time of writing. Since no ``smoking gun'' signatures have been found, attention has turned to less canonical Beyond the Standard Model (BSM) models than the (typically simplified) ones originally used by the experiments to contextualise their results. Reinterpretation of BSM searches in the contexts of more complex or subtle models is hence now a very important topic, but full exploitation of the rich trove of experimental results requires care with analysis combination: with so many experimental results, manual cross-referencing of $\mathcal{O}(600)$ publications to avoid double-counting is not a realistic option. Indeed, even with the time to compare all pairs of released studies, it is non-trivial, to say the least, for a person outside the experimental collaborations to answer the question of which analyses can be considered to be approximately uncorrelated. This Les Houches contribution aims at filling this gap, using a probabilistic approach to identify pairs of analyses that can safely be considered uncorrelated in global BSM reinterpretation fits.
\end{abstract}


\section{Introduction}
\label{sec:BSM_AnalysisCorrelations:Sec_Introduction}


At the time of this writing, the LHC search programmes for new physics has produced hundreds of publications with an even larger number of individual simplified model space (SMS) results.
The results are typically presented as upper limits on production cross sections or signal efficiency ``maps'' for a small number
of simplified models. It is therefore of utmost importance that the wider community is able to re-derive these results and to determine the viability of less explored models under the current collider constraints. To obtain an accurate prediction, a recasting of multiple analyses exploring different parts of the parameter space must often be carried out, but this raises the question: which analyses can be combined? In this manner, we want to determine subsets of analyses whose event selection criteria do not overlap.

Generally, analyses implemented by different collaborations ({\it e.g.}~ATLAS {\it vs.}~CMS) can safely be considered to be uncorrelated, as are those carried out at different centre of mass energies ({\it e.g.}~13 TeV
{\it vs.}~7 TeV). In addition, within this work we look only at signal regions, and do not consider possible correlations between control regions, as we assume such correlations to be second order effects and negligible for reinterpretation purposes. Our considerations thus focus on the signal regions, and we attempt to answer the question of whether any given pair of signal regions overlap in the space of all measured quantities or not. Using the analyses contained both in  {\tt{SModelS}}~\cite{Kraml:2013mwa, Ambrogi:2017neo, Dutta:2018ioj, Heisig:2018kfq, Ambrogi:2018ujg} and {\tt{MadAnalysis}~5}~\cite{Conte:2012fm, Conte:2014zja, Dumont:2014tja,Conte:2018vmg}, we want to determine the parameter space covered by the topologies that populate these signal regions and generate events to cover all the signal regions, enabling us to infer correlations between the analyses.

To determine which pairs of analyses are uncorrelated, we pursue the following strategy.
Restricting ourselves to analyses implemented both in the {\tt{SModelS}} v1.2.2 and {\tt{MadAnalysis}~5} v1.8.20 frameworks, we use the {\tt{SModelS}} database to extract the
simplified models that a given analysis is sensitive to, alongside with the mass ranges
for the BSM particles. We then sample this space of simplified model mass parameters and create random, realistic events from these simplified models using {\tt{MadGraph5\_aMC@NLO}}~\cite{Alwall:2014hca}. These events are then passed to a modified version of {\tt{MadAnalysis}~5}, which as usual checks whether an event passes the cuts in each signal region (SR), but in addition records the corresponding set of SR cuts passed for every event. Using this information, a statistical bootstrap procedure is used to extract a correlation matrix relating every signal region through the events that co-populate them.





\section{Implementation}
\label{sec:BSM_AnalysisCorrelations:Sec_Implementation}

To investigate the analysis correlations, we chose to use the overlap of analyses implemented both in {\tt{SModelS}} and {\tt{MadAnalysis}~5}: at the time of publication,  these are ATLAS-SUSY-2015-06~\cite{Aaboud:2016zdn}, CMS-SUS-16-039 \cite{Sirunyan:2017lae}, CMS-SUS-16-033~\cite{Sirunyan:2017cwe} and CMS-SUS-17-001~\cite{Sirunyan:2017leh}, all being searches for supersymmetry (SUSY). A description of the signal regions under consideration is shown in \cref{tab:BSM_AnalysisCorrelations:Table_A0ASR,tab:BSM_AnalysisCorrelations:Table_A0BSR,tab:BSM_AnalysisCorrelations:Table_A0CSR,tab:BSM_AnalysisCorrelations:Table_A0DSR,tab:BSM_AnalysisCorrelations:Table_A0ESR,tab:BSM_AnalysisCorrelations:Table_A0FSR,tab:BSM_AnalysisCorrelations:Table_A0G-KSR,tab:BSM_AnalysisCorrelations:Table_A0SS,tab:BSM_AnalysisCorrelations:Table_A1SR,tab:BSM_AnalysisCorrelations:Table_A2SR,tab:BSM_AnalysisCorrelations:Table_A3SR} in the appendix. 

\subsection{Topology identification:}
Using the {\tt{SModelS}} database, we extract the topologies that a certain analysis is sensitive to, as denoted in table \ref{BSM_AnalysisCorrelations:Table_Analysis_Topologies}, and consequently the mass parameter space the analysis can access. A topology describes a specific cascade decay, which, in the case of simplified models, is reduced to a 2- or 3- body decay with symmetric branches. A full list of the topologies contained in {\tt{SModelS}} is given in \url{https://smodels.github.io/}. In the case of 3- body decays, the topology is reduced further so that the mass of the intermediate particle is fixed as a function of that of the mother and granddaughter. 

Two analyses may both be sensitive to a given topology, but in different yet overlapping regions of the simplified model space that depends on the masses of the mother and daughter particles, denoted by $m_0$ and $m_1$ respectively. We want to generate events that randomly populate the union of the two regions without doubly populating the regions that are in common. We deduce this convex mass hull using the efficiency maps in {\tt{SModelS}}: the efficiency maps provide upper limits on the production cross sections as a function of the masses of the simplified model particles. If the efficiency map of the point chosen in the parameter space is 0, then we move on to probe a different part of the parameter space until eventually a contour can be interpolated around the range of mass points that can be touched by the different analyses. For topologies involving two-body decays, this curve is obtained by sampling values of $m_0$, and determining the minimum and maximum $m_1$ values for which the regions are populated by the corresponding signal. When three-body decays are involved, the scan is expanded tri-dimensionally to determine minimum and maximum values of $m_2$ (the mass of the granddaughter particle) for a given $m_1$ and $m_0$ for which the SRs are populated.

\subsection{Event generation and analysis:}
We produce 50000 events evenly distributed over the corresponding convex mass hulls for each topology. These events have been generated with {\tt{MadGraph5\_aMC@NLO}}, using the standard MSSM UFO implementation shipped with the package~\cite{Christensen:2009jx}. Hard-scattering events
have been matched with parton showers, as described with {\tt Pythia}~8.2~\cite{Sjostrand:2014zea} that has also handled hadronisation. As parameter inputs, we used template SLHA files~\cite{Skands:2003cj}, specific for each topology.

Hadron-level events are passed to {\tt{MadAnalysis}~5}, whose expert mode (from v1.8.20 onwards) has been augmented by two new functions ({\tt DumpSR} and {\tt HeadSR}) dedicated to the writing to a file of the information on how a given event populates the various signal regions of the recasted analyses under consideration. In practice, the main executable of the recasting module (located in {\tt tools/PAD/Build/Main/main.cpp}) has to be modified so that {\tt HeadSR} is called prior to the event loop (in the {\tt Execute} function), and that {\tt DumpSR} is called within the event loop ({\it i.e.}~for each event). The former function writes, as the first line of the output file, the signal regions as they are ordered internally in the code. The latter function writes, for each event, a set of 0 and 1 indicated whether the event populates each signal region (following the internal ordering). Both functions take an {\tt ostream} object as argument.

\subsection{Statistical framework:}
The augmented {\tt MadAnalysis~5} output consists in a grid of binary flags indicating the pass/fail status of each of $N_\mathrm{evt}$ event for each of the $N_\mathrm{SR}$ signal-region selections. In general, each event can have more than one ``true'' SR flag. This defines the correlation or co-population which makes analysis combination problematic, as a single unusual event could produce multiple observed analysis excesses. It would not make sense to count all such measures of the same event as if they were independent.
\begin{table}
\begin{center}
\begin{tabular}{|c|c|c|}
\hline
Tag & Analysis & Topologies \\ 
\hline
A0(A-K,SS) & CMS-SUS-16-039 & TChiWZ,TChiWZ(off), TChiWH, TChipmSlepL, \\ & & TChiChipmStauStau, TChiChipmSlepStau  \\

\hline
A1 & CMS-SUS-16-033 & T1, T1bbbb, T1tttt(off), T2, T2bb, T2tt(off)    \\
\hline
A2 & CMS-SUS-17-001 & T2tt(off), T6bbWW  \\
\hline
A3 & ATLAS-SUSY-2015-06 & T1, T2 \\ \hline
\end{tabular}
\caption{Table showing the sample of analyses investigated, with the names of the topologies reached and the tag-name used to refer to the analysis in the figures that follow.}
\label{BSM_AnalysisCorrelations:Table_Analysis_Topologies}
\end{center}
\end{table}
To reduce this large matrix of binary flags to a more useful figure, we first make it much \textit{larger}: by use of bootstrap sampling from a unit Poisson distribution, every event is multiplied by a set of $N_\mathrm{boot}$ ``bootstrap samples'', with sampled weights $w \sim \mathrm{Pois}(\mu = 1)$ which replace the 0/1 binary SR weights. The total data structure is now a $N_\mathrm{evt} \times N_\mathrm{SR} \times N_\mathrm{boot}$ tri-dimensional array of SR event weights - the SRs in each bootstrap history correlated through their common weight fluctuation. The event axis is then summed over, reducing to a two-dimensional $N_\mathrm{SR} \times N_\mathrm{boot}$ array of SR weight sums $W = \sum_\mathrm{evt} w$, over the $N_\mathrm{boot}$ ``histories''. This two-dimensional array allows us to compute the correlation between two SRs $i$ and $j$ as usual via the covariance:
\begin{eqnarray}
\mathrm{cov}_{ij} = \langle W_i W_j \rangle - \langle W_i \rangle \langle W_j \rangle\,,
\end{eqnarray}
and
\begin{eqnarray}
\mathrm{\rho}_{ij} = \mathrm{cov}_{ij} / \sqrt{ \mathrm{cov}_{ii} \mathrm{cov}_{jj}}\,.
\end{eqnarray}
The averages here are computed over the bootstrap histories.

The correlations matrix $\rho_{ij}$ acts as a sliding scale of event sharing between $-1$ and $1$, and whether the combination of SRs $i$ and $j$ is acceptable can be decided by applying a cutoff $|\rho_{ij}| < \rho_\mathrm{max}$. With asymptotic statistics, $\rho_{ij}$ should be positive, as there is no mechanism (such as normalisation by the sum of SR yields) to generate true negative correlations where an increase in yield for one SR due to weight fluctuations leads to the active depopulation of another. But with finite statistics random negative correlations can occur, and empirical study of their distribution can help in setting the $\rho_\mathrm{max}$ cutoff.

It is useful to note that this measure of correlation reflects where a pair of SRs are significantly correlated due to shared events, which is not the same as the logical existence of sharing. For example, naively one might imagine that an exclusive, low-statistics SR entirely contained within a much more inclusive, high-population one would be fully correlated with its parent. But this is not so: the fractional contribution of the events from the exclusive bin to the more inclusive one's weight-sum $W$ may be negligibly small, such that a significant excess in the exclusive bin would not register as a significant change in the inclusive one. This normalised measure of sensitivity on the total yields of both SRs (which are the inputs to {\it e.g.}~a Poisson profile likelihood calculation) automatically captures this nuance in a way that studies of cut overlaps cannot.
\begin{figure}
    \centering
    \includegraphics[width=75mm]{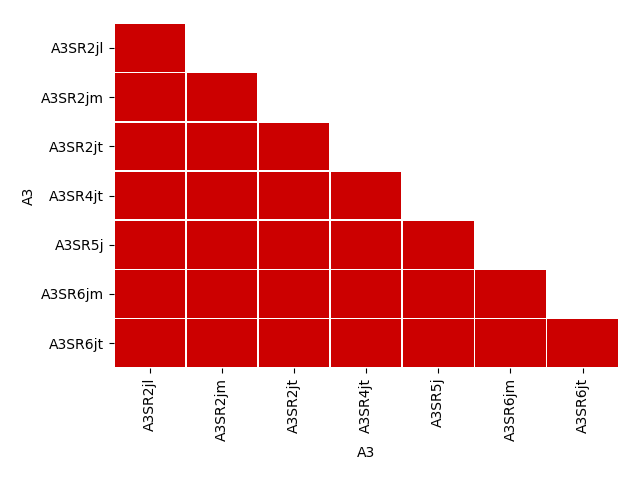}
    \caption{Matrix showing the correlation between the SRs of the A3 ATLAS analysis ~\cite{Aaboud:2016zdn}. As expected for ``inclusive'' SRs, all the blocks in the matrix are red, denoting correlation between the SRs.}
    \label{fig:BSM_AnalysisCorrelations:Fig_corrA3A3}
\end{figure}

\begin{figure}
\centering     
\subfigure[A0(A-K,SS) vs. A0(A-K,SS)]{\label{fig:BSM_AnalysisCorrelations:Fig_corrA0A0}\includegraphics[width=75mm]{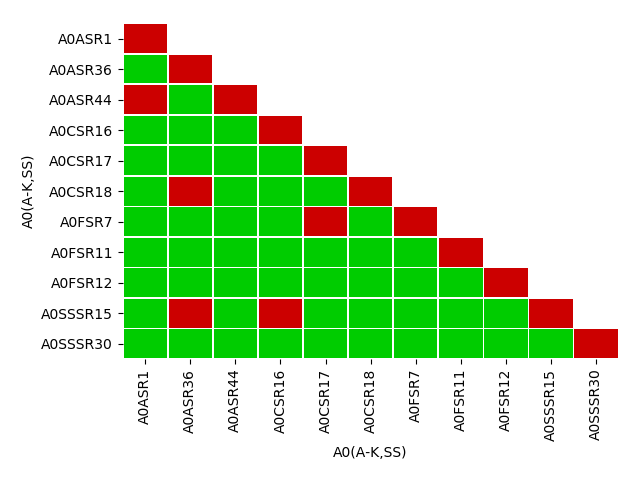}}
\quad
\subfigure[A1 vs. A1]{\label{fig:BSM_AnalysisCorrelations:Fig_corrA1A1}\includegraphics[width=75mm]{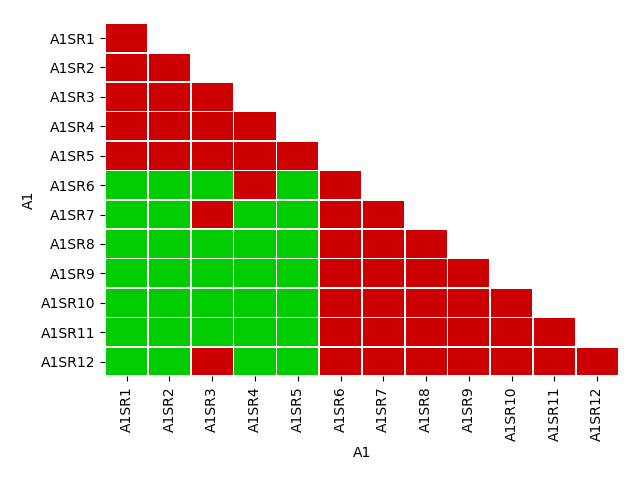}}
\quad
\subfigure[A2 vs. A2]{\label{fig:BSM_AnalysisCorrelations:Fig_corrA2A2}\includegraphics[width=75mm]{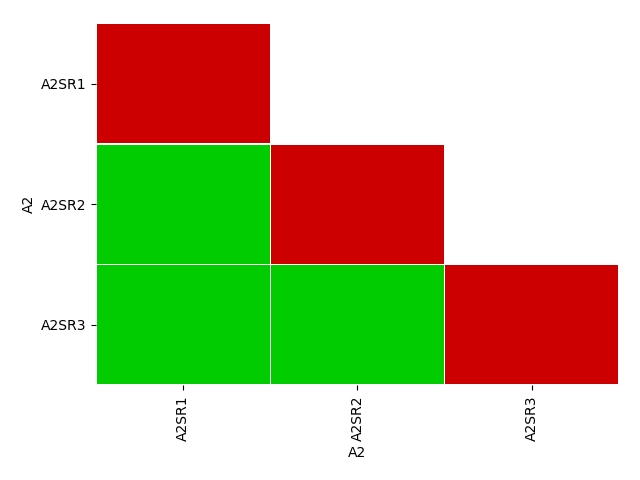}}
\quad
\subfigure[A1 vs. A0(A-K,SS)]{\label{fig:BSM_AnalysisCorrelations:Fig_corrA1A0}\includegraphics[width=75mm]{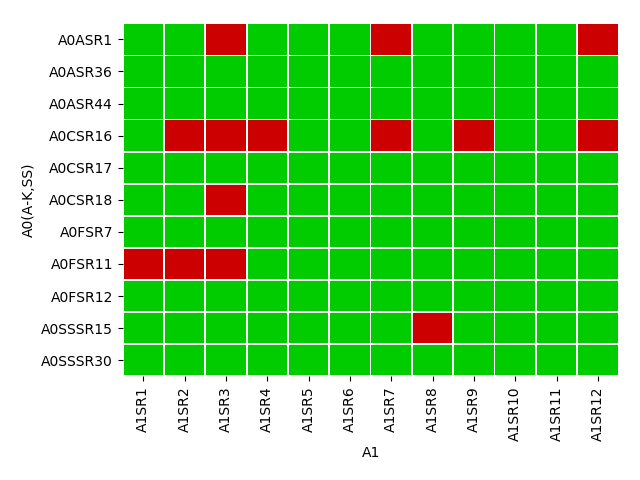}}
\quad
\subfigure[A0 vs. A2]{\label{fig:BSM_AnalysisCorrelations:Fig_corrA0A2}\includegraphics[width=75mm]{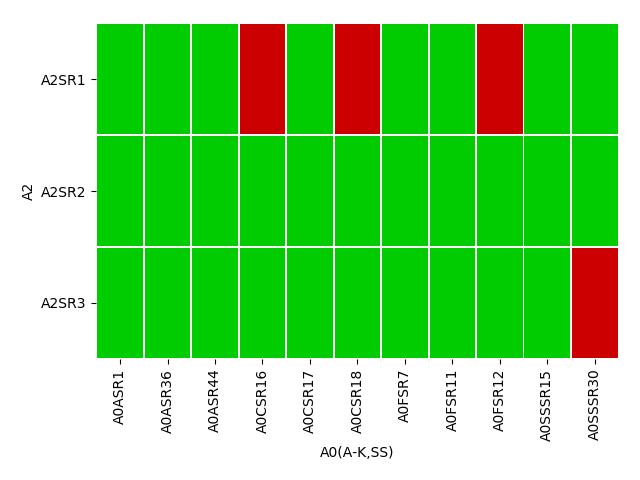}}
\quad
\subfigure[A1 vs. A2]{\label{fig:BSM_AnalysisCorrelations:Fig_corrA1A2}\includegraphics[width=75mm]{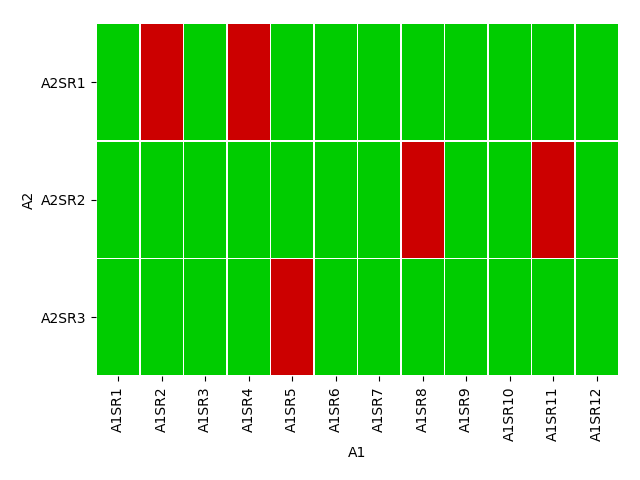}}
\caption{Matrices showing the correlations between the SRs of the CMS analyses under consideration that are populated with at least 300 events: A0(A-K,SS), A1 and A2. The green blocks denote that two SRs can be treated as approximately independent from each other, whereas the red blocks indicate that they are correlated.}\label{fig:BSM_AnalysisCorrelations:Fig_CMScorrs}
\end{figure}

We have implemented this bootstrap procedure in a {\tt Python} program called TACO ({\it Testing Analyses' COrrelations}), available at \url{https://github.com/hreyes91/TACO}. TACO takes as input a data frame in which each column corresponds to a SR, each row to an event, and which is filled with 1 and 0 depending on whether the event passes the cuts of the corresponding signal region. The program then generates a user-defined number of $N_{\rm boot}$ histories (here taken as 1200) and computes the correlations between the considered SRs from the produced $N_\mathrm{SR} \times N_\mathrm{boot}$ matrix. Afterwards, TACO determines if the SRs are approximately uncorrelated based on a user-defined $\rho_{\rm max}$ cut-off (here 0.05) and substitutes the $\rho_{ij}$ coefficients with a 0 (if $|\rho_{ij}| < \rho_\mathrm{\rm max}$) or a 1 (if $|\rho_{ij}| > \rho_\mathrm{max}$) to produce an ``independence matrix'', which is the final value of the calculation.

The implementation of the bootstrap procedure was applied to the signal regions populated with at least 300 events; this includes those in the A1, A2, A3 analyses and 11 SRs from the  A0(A-K,SS) analysis (see table~\ref{BSM_AnalysisCorrelations:Table_Analysis_Topologies} for the naming scheme). In our results given in Figs~\ref{fig:BSM_AnalysisCorrelations:Fig_corrA3A3} and \ref{fig:BSM_AnalysisCorrelations:Fig_CMScorrs}, green blocks denote SRs which are independent, whereas red blocks are not. In all figures exhibiting correlations between signal regions within a sole analysis (Figs.~\ref{fig:BSM_AnalysisCorrelations:Fig_corrA3A3}, \ref{fig:BSM_AnalysisCorrelations:Fig_corrA0A0}, \ref{fig:BSM_AnalysisCorrelations:Fig_corrA1A1}, \ref{fig:BSM_AnalysisCorrelations:Fig_corrA2A2}), the anti-diagonal elements are trivially correlated. Fig.~\ref{fig:BSM_AnalysisCorrelations:Fig_corrA3A3} depicts the results originating from the A3 ATLAS analysis. As expected for ``inclusive'' SRs, these are tagged as correlated. For the considered CMS analyses, results are shown in Fig.~\ref{fig:BSM_AnalysisCorrelations:Fig_CMScorrs} where we can see that ``exclusive'' SRs are normally estimated as uncorrelated. For instance, Fig.~\ref{fig:BSM_AnalysisCorrelations:Fig_corrA1A1} shows that most of the A1SR(1-5) regions, where the number of $b$-jets is $N_b=0$, are uncorrelated with the set of A1SR(6-12) regions in which $N_b\geq 1$. In addition, the the pairs of SRs that seem correlated feature a $\rho$ parameter very close to the $\rho_\mathrm{max} = 0.05$ threshold, with $\rho=~0.06$. They could thus be expected to be found uncorrelated as soon as more events are included. In this way, the results we present are on the conservative side in the sense that correlated SRs will surely be flagged as correlated, but independent SRs can erroneously be flagged correlated if the number of bootstrap iterations is too low. Moreover, in Figs.~\ref{fig:BSM_AnalysisCorrelations:Fig_corrA1A0}, \ref{fig:BSM_AnalysisCorrelations:Fig_corrA0A2} and \ref{fig:BSM_AnalysisCorrelations:Fig_corrA1A2}  the determination of the independency between SRs of \textit{different} analyses is presented. As a quick sanity check, we see for example that the SRs A0CSR017 and ASCSR018 in fig \ref{fig:BSM_AnalysisCorrelations:Fig_corrA0A0} are independent; this is because despite looking in the same dilepton invariant and stansverse mass ranges, the missing transverse energy $p_T^{\rm miss}$ ranges do not overlap.
For the most part, the SRs from the A0(A-K,SS), A1 and A2 analyses were determined as independent from each other, while the rest of these SRs are expected to be determined as such once the number of populating events is increased.

\section{Conclusion}
\label{sec:BSM_AnalysisCorrelations:Sec_Conclusion}
In this contribution, a probabilistic method for the determination of approximately uncorrelated pairs of analyses' SRs was developed. It is based on simplified models, and first results were presented using the intersection of analyses present in both {\tt{SModelS}} and {\tt{MadAnalysis}~5}. It is our pronounced hope that the effort will be repeated with a wider set of analyses with the aim to compile a complete detailed list of analyses that can be treated as independent from each other. Specifically, {\tt{CheckMATE}}~\cite{Dercks:2016npn,Drees:2013wra} and its database is a potential future target for this effort. Given such a database, a potential future extension of this work is to identify the subset of independently populated SRs that maximises limit-setting power within a BSM parameter scan.
The source code developed for this work can be accessed on Zenodo, see \url{http://doi.org/10.5281/zenodo.3634740}.
In this document, only events originating from simplified model topologies were produced. 
In a future project, we wish to introduce also more complicated events to cover potential correlations that might be missed by too simplistic  simplified models-based events.

\section*{Acknowledgements}
\label{sec:BSM_AnalysisCorrelations:Sec_Acknowledgements}
The authors thank Sabine Kraml and Sezen Sekmen for insightful discussions.  
AB thanks the European Union for funding of the Marie Sklodowska-Curie Innovative Training Network MCnetITN3 (grant agreement no. 722104), and The Royal Society for University Research Fellowship grant UF160548. HRG
is funded by the Consejo Nacional de Ciencia y Tecnología, CONACyT, scholarship no.
291169. BF thanks his mother's cat. 
This work
also has been partly supported by French state funds managed by the Agence Nationale de la Recherche (ANR)
in the context of the LABEX ILP (ANR-11-IDEX-0004-
02, ANR-10-LABX-63), which in particular funded the
scholarship of SLW. 

\section*{Appendix: Description of the considered signal regions}\label{sec:BSM_AnalysisCorrelations:sec_appendix_a}

\begin{table}[h]\begin{center}
\begin{tabular}{|c|c|c|c|c|}
\hline
$M_{T}~[\mathrm{GeV}] $&$ p^\text{miss}_{T}~[\mathrm{GeV}] $&$ M_{ll}<75~\mathrm{GeV} $&$ 75 \geq M_{ll}<105~\mathrm{GeV} $&$ M_{ll}\geq 105~\mathrm{GeV} $ \\
\hline
& 50--100 & A0ASR1 & A0ASR15 & A0ASR32 \\\cline{2-5}
& 100--150 & A0ARA2 & A0ASR16 & A0ASR33 \\\cline{2-5}
& 150--200 & A0SRA3 & A0ASR17 & A0ASR34 \\\cline{2-5}
0--100& 200-250 & A0ASR4 & A0ASR18 & A0ASR35\\ \cline{2-5}
& 250--400 &  & A0ASR19 &  \\\cline{2-2}\cline{4-4}
& 400--500 & A0ASR5 & A0ASR20 & A0SR36  \\\cline{2-2}\cline{4-4}
& $\geq$ 550 &  & A0ASR21 &   \\
\hline
&50--100 &A0ASR6 & A0ASR22 & A0ASR37 \\\cline{2-5}
100--160 & 100-150 & A0ASR7 & A0ASR23 & A0ASR38 \\\cline{2-5}
 & 150--200 & A0ASR8 & A0ASR24 & A0ASR39 \\\cline{2-5}
& $\geq$ 200 & A0ASR9 & A0ASR25 & A0ASR40 \\\cline{2-5}
\hline
& 50--100 & A0ASR10 & A0ASR26 & A0ASR41 \\ \cline{2-5}
& 100--150 & A0ASR11 & A0ASR27 & A0ASR42 \\\cline{2-5}
$\geq$ 160 & 150-200 & A0ASR12 & A0ASR28 & A0ASR43 \\ \cline{2-5}
& 200--250 & A0ASR13 & A0ASR29 &  \\\cline{2-4}
& 250--400 & A0ASR14 & A0ASR30 & A0ASR44  \\ \cline{2-2} \cline{4-4}
& $\geq$ 400 &  & A0SAR30 &   \\ \cline{2-2} \cline{4-4} 
\hline
\end{tabular}
  \caption{\label{tab:BSM_AnalysisCorrelations:Table_A0ASR} Summary of the A0A signal regions in the A0 analysis, cf.~Ref.~\cite{Sirunyan:2017lae}.}
  \end{center}
 \end{table}

 \begin{table}[h]
  \begin{center}
 \begin{tabular}{|c|c|c|c|}
 \hline
 $M_{T} (\mathrm{GeV})$ & $p^\text{miss}_{T} (\mathrm{GeV})$ & $M_{ll}< 100 \mathrm{GeV}$ & $M_{ll}\geq 100 \mathrm{GeV}$  \\\hline
 0-120 & 50-100 & A0BSR1 & A0BSR4 \\\cline{2-4}
  & $>100$ & A0BSR2 & A0BSR5 \\\hline
 $>120$ & $> 50$ & A0BSR3 & A0BSR6 \\ \hline

 \end{tabular}
    \caption{\label{tab:BSM_AnalysisCorrelations:Table_A0BSR} Summary of the A0B signal regions in the A0 analysis, cf.~Ref.~\cite{Sirunyan:2017lae}.}
    \end{center}
  \end{table} 

\begin{table}[h]
\begin{center}
\begin{tabular}{|c|c|c|c|c|}
\hline
$p^\text{miss}_{T}~[\mathrm{GeV}]$ & $75\geq M_{ll}<105~\mathrm{GeV}$ & $M_{T2}(l_{1}l_{2}) [\mathrm{GeV}]$ & $M_{ll}<75~\mathrm{GeV}$ & $M_{ll}\geq 105~\mathrm{GeV}$\\
\hline
50--100 & A0CSR06 & &  A0CSR01  &  A0CSR012 \\\cline{1-2}\cline{4-5}
100--150 & A0CSR07 & &  A0CSR02  &  A0CSR013 \\\cline{1-2}\cline{4-5}
150--200 & A0CSR08 & &  A0CSR03  &  A0CSR014 \\\cline{1-2}\cline{4-5}
200--250 & A0CSR09 & 0-100 &  A0CSR04  &  A0CSR015\\\cline{1-1}\cline{4-5}
250--300 &  &  &   &  \\\cline{1-2}
300--400 & A0CSR10 &  &  A0CSR05  &  A0CSR016\\\cline{1-2}
$\geq$ 400 & A0CSR11 &  &   &  \\\hline
50--200 & & $\geq$ 100 & \multicolumn{2}{c|}{A0CSR017} \\\cline{1-1}\cline{4-5}
$\geq$200 & & & \multicolumn{2}{c|}{A0CSR018} \\\hline
\end{tabular}
  \caption{\label{tab:BSM_AnalysisCorrelations:Table_A0CSR} Summary of the A0C signal regions in the A0 analysis, cf.~Ref.~\cite{Sirunyan:2017lae}.}
  \end{center}
 \end{table}

\begin{table}[h]

\begin{center}
\begin{tabular}{|c|c|c|c|c|}
\hline
$M_{T}~[\mathrm{GeV}]$ & $p^\text{miss}_{T}~[\mathrm{GeV}]$ & $M_{ll}<75~\mathrm{GeV}$ & $75\geq M_{ll}<105~\mathrm{GeV}$ &$ M_{ll}\geq 105~\mathrm{GeV}$ \\ 
\hline
& 50--100 & A0DSR1 & A0DSR6 & A0DSR11 \\\cline{2-5}
& 100--150 & A0DSR2 & A0DSR7 & A0DSR12 \\\cline{2-5}
0--100 & 150--200 & A0DSR3 & A0DSR8 & A0DSR13 \\\cline{2-5}
 & 200--250 & A0DSR4 & A0DSR09 & A0DSR14 \\\cline{2-4}
 & $\geq$250 & A0DSR5 & A0DSR10 &  \\\hline
$\geq$100 & 50--200 & \multicolumn{3}{c|}{A0DSR15} \\\cline{2-5}  

& $\geq$200 & \multicolumn{3}{c|}{A0DS016} \\ 
\hline

\end{tabular}
  \caption{\label{tab:BSM_AnalysisCorrelations:Table_A0DSR} Summary of the A0D signal regions in the A0 analysis, cf.~Ref.~\cite{Sirunyan:2017lae}.}
  \end{center}
 \end{table}

\begin{table}[h]
\begin{center}
\begin{tabular}{|c|c|c|c|c|}
\hline
$M_{T2}(l_{1},\tau)~[\mathrm{GeV}]$ &$p^\text{miss}_{T}~[\mathrm{GeV}]$ & $M_{ll}<60~\mathrm{GeV}$ & $60 \geq M_{ll}<100~\mathrm{GeV}$& $M_{ll}\geq 100~\mathrm{GeV}$ \\
\hline
& 50--100 & A0ESR1 & A0ESR6 & \\ \cline{2-4} 
& 100--150 & A0ESR2 & A0ESR7 & \\ \cline{2-4}
0--100 & 150--200 & A0ESR3 & A0ESR8 & A0ESR11 \\ \cline{2-4} 
& 200--250 & A0ESR4 & A0ESR9 &  \\ \cline{2-4}
& $\geq$ 250 & A0ESR5 & A0ESR10 &  \\ \hline
$\geq 100$&$\geq$50 &\multicolumn{3}{c|}{A0ESR15} \\ \hline

\end{tabular}
  \caption{\label{tab:BSM_AnalysisCorrelations:Table_A0ESR} Summary of the A0E signal regions in the A0 analysis, cf.~Ref.~\cite{Sirunyan:2017lae}.}
  \end{center}
 \end{table}

\begin{table}[h]
\begin{center}
\begin{tabular}{|c|c|c|c|}
\hline
$M_{T2}(l_{1},\tau)~[\mathrm{GeV}]$ &$p^\text{miss}_{T}~[\mathrm{GeV}]$ & $ M_{ll}<100~\mathrm{GeV}$& $M_{ll}\geq 100~\mathrm{GeV}$ \\
\hline
& 50--100 & A0FSR1 & A0FSR7 \\ \cline{2-4} 
& 100--150 & A0FSR2 & A0FSR8 \\\cline{2-4} 
0--100& 150--200 & A0FSR3 & A0FSR9 \\\cline{2-4} 
& 200--250 & A0FSR4 &  \\\cline{2-3} 
& 250--300 & A0FSR5 & A0FSR10 \\ \cline{2-3} 
& $\geq$ 300 & A0FSR6 &  \\ \hline
$\geq$ 100 & 50--200 & \multicolumn{2}{c|}{A0FSR11} \\ \cline{2-4} 
& $\geq$ 200 & \multicolumn{2}{c|}{A0FSR12} \\  \hline

\end{tabular}
  \caption{\label{tab:BSM_AnalysisCorrelations:Table_A0FSR} Summary of the A0F signal regions in the A0 analysis, cf.~Ref.~\cite{Sirunyan:2017lae}.}
  \end{center}
 \end{table}

\begin{table}[h]
\begin{center}
\begin{tabular}{|c|c|c|c|c|c|}
\hline
$p_{T}^\text{miss}~[\mathrm{GeV}]$ & \multicolumn{2}{c}{0~$\tau_{h}$} &1~$\tau_{h}$ & \multicolumn{2}{c|}{2~$\tau_{h}$}  \\ \cline{2-6}
& nOSSF$\geq$2 & nOSSF$<$ 2 & nOSSF$\geq$ 0 & nOSSF$\geq$ 2 & nOSSF$<$ 2 \\   
\hline
0--50 & A0GSR1 &  A0HSR1 & A0ISR1 & A0JSR1 & A0KSR1 \\ \hline
50--100 & A0GSR2 &  A0HSR2 & A0ISR2 & A0JSR2 & A0KSR2 \\\hline
100--150 & A0GSR3 &  A0HSR3 & A0ISR3 & A0JSR3 &  \\\cline{1-5}
150--200 & A0GSR4 &  A0HSR4 & A0ISR4 & A0JSR4 & A0KSR3  \\ \cline{1-2}
$\geq$200 & A0GSR5 &   &  &  &  \\ \hline

\end{tabular}
  \caption{\label{tab:BSM_AnalysisCorrelations:Table_A0G-KSR} Summary of the A0(G-K) signal regions in the A0 analysis, cf.~Ref.~\cite{Sirunyan:2017lae}.}
  \end{center}
 \end{table}
 
 \begin{table}[h]
     \begin{center}
     \resizebox{\textwidth}{!}{
     \begin{tabular}{|c|c|c|c|c|c|c|c|c|c|c|c|c|c|}
     \hline
         $N_\text{jets}$ & \multicolumn{7}{c|}{0} & \multicolumn{6}{c|}{1}  \\ \hline
         $M_{T}$~[GeV] & \multicolumn{5}{c|}{$<$100} & \multicolumn{2}{c|}{$>$100} & \multicolumn{4}{c|}{$<$100} & \multicolumn{2}{c|}{$>$100} \\ \hline
         $p_{T}^{ll}$~[GeV] & \multicolumn{3}{c|}{$<$50} & \multicolumn{2}{c|}{$>$50} & \multicolumn{2}{c|}{} & \multicolumn{2}{c|}{$<$50} & \multicolumn{2}{c|}{$>$50} & \multicolumn{2}{c|}{} \\ \hline  
         $p_{T}^\text{miss}<100$ GeV & \multicolumn{3}{c|}{A0SSSR1} &\multicolumn{2}{c|}{A0SSSR6} &\multicolumn{2}{c|}{A0SSSR11} &\multicolumn{2}{c|}{A0SSSR16} &\multicolumn{2}{c|}{A0SSSR21} &\multicolumn{2}{c|}{A0SSSR26} \\ \hline 
         $ 100 \leq p_{T}^\text{miss}<150$ GeV &++& \multicolumn{2}{c|}{A0SSSR2} &\multicolumn{2}{c|}{A0SSSR7} &\multicolumn{2}{c|}{A0SSSR12} &\multicolumn{2}{c|}{A0SSSR17} &\multicolumn{2}{c|}{A0SSSR22} &\multicolumn{2}{c|}{A0SSSR27} \\ \cline{2-14}
         &- -& \multicolumn{2}{c|}{A0SSSR3} &\multicolumn{2}{c|}{A0SSSR8} &\multicolumn{2}{c|}{A0SSSR13} &\multicolumn{2}{c|}{A0SSSR18} &\multicolumn{2}{c|}{A0SSSR23} &\multicolumn{2}{c|}{A0SSSR28} \\ \hline
         $ 150 \leq p_{T}^\text{miss}<200$ GeV& \multicolumn{3}{c|}{A0SSSR4} &\multicolumn{2}{c|}{A0SSSR9} &\multicolumn{2}{c|}{A0SSSR14} &\multicolumn{2}{c|}{A0SSSR19} &\multicolumn{2}{c|}{A0SSSR24} &\multicolumn{2}{c|}{A0SSSR29} \\ \hline
         $p_{T}^\text{miss}\geq 200$ GeV& \multicolumn{3}{c|}{A0SSSR5} &\multicolumn{2}{c|}{A0SSSR10} &\multicolumn{2}{c|}{A0SSSR15} &\multicolumn{2}{c|}{A0SSSR20} &\multicolumn{2}{c|}{A0SSSR25} &\multicolumn{2}{c|}{A0SSSR30} \\ \hline
     \end{tabular}}
     \caption{\label{tab:BSM_AnalysisCorrelations:Table_A0SS} Summary of the A0SS signal regions in the A0 analysis, cf.~Ref.~\cite{Sirunyan:2017lae}.}
     
     \end{center}
 \end{table}
  \begin{table}[h]
\begin{center}
\begin{tabular}{|c|c|c|c|c|}

\hline
& $N_\text{jet}$ & $N_\text{$b$-jet}$ & $H_{T}~[\text{GeV}] $ & $H_{T}^\text{miss}~[\text{GeV}]$ \\ 
\hline
A1SR1 & $\geq$ 2 & 0 & $\geq$ 500 & $\geq$ 500 \\
A1SR2 & $\geq$ 3 & 0 & $\geq$1500 &$\geq$  750  \\
  A1SR3 & $\geq$ 5 & 0 &$\geq$ 500 & $\geq$500 \\
 A1SR4 & $\geq$  5 & 0 &$\geq$ 1500 &$\geq$ 750 \\
  A1SR5 & $\geq$  9 & 0 &$\geq$ 1500 &$\geq$ 750  \\
 A1SR6 & $\geq$ 2& $\geq$2&$\geq$500&$\geq$500 \\
 A1SR7 & $\geq$ 3& $\geq$1& $\geq$750 & $\geq$750  \\
  A1SR8 & $\geq$  5 & $\geq$ 3&$\geq$500 &$\geq$ 500  \\
  A1SR9 & $\geq$ 5&$\geq$2&$\geq$ 1500 &$\geq$ 750 \\
  A1SR10 & $\geq$  5&$\geq$ 3& $\geq$750 &$\geq$ 750  \\
  A1SR11 & $\geq$  7&$\geq$ 1&$\geq$300& $\geq$ 300 \\ 
  A1SR12 & $\geq$  5& $\geq$1& $\geq$750 & $\geq $ 750 \\  
 \hline

  \end{tabular}
  \caption{\label{tab:BSM_AnalysisCorrelations:Table_A1SR} Summary of the signal regions in the A1 analysis, cf.~Ref.~\cite{Sirunyan:2017cwe}.}
  \end{center}
 \end{table} 
\begin{table}[h]\begin{center}
\begin{tabular}{|c|c|c|c|}
\hline
$m_{T2}(l_1 l_2)~[\mathrm{GeV}]$ & $100- 140  $ & $140 - 240  $ &  $> 240$ \\ \hline
$E_{T}^\text{miss}>200~\mathrm{GeV}$ & A2SR1 & A2SR2 & A2SR3 \\ 
\hline
\end{tabular}
  \caption{\label{tab:BSM_AnalysisCorrelations:Table_A2SR} Summary of the signal regions in the A2 analysis, cf.~Ref.~\cite{Sirunyan:2017leh}.}
  \end{center}
 \end{table}
 
\begin{table}[h]
\begin{center}
\resizebox{\textwidth}{!}{
\begin{tabular}{|c|c|c|c|c|c|c|c|}
\hline
& A3SR2jl & A3SR2jm & A3SR2jt & A3SR4jt & A3SR5j & A3SR6jm & A3SR6jt \\ \hline
$E_{T}^\text{miss}[\mathrm{GeV}]>$ & \multicolumn{7}{c|}{200}\\ \hline 
$p_{T}(j_{1})[\mathrm{GeV}]>$ & 200 & 300 & \multicolumn{5}{c|}{200} \\ \hline
$p_{T}(j_{2})[\mathrm{GeV}]>$ & 200 & 50 & 200 & \multicolumn{4}{c|}{100} \\\hline
$p_{T}(j_{3})[\mathrm{GeV}]>$ & \multicolumn{3}{c|}{-} & \multicolumn{4}{c|}{100} \\\hline
$p_{T}(j_{4})[\mathrm{GeV}]>$ & \multicolumn{3}{c|}{-} & \multicolumn{4}{c|}{100} \\\hline
$p_{T}(j_{5})[\mathrm{GeV}]>$ & \multicolumn{4}{c|}{-} & \multicolumn{3}{c|}{50} \\\hline
$p_{T}(j_{6})[\mathrm{GeV}]>$ & \multicolumn{5}{c|}{-} & \multicolumn{2}{c|}{50} \\\hline
$\Delta\phi(\mathrm{jet}_{1,2,(3)},E_{T}^\text{miss})_\text{min}>$ & 0.8 & 0.4 & 0.8 & \multicolumn{4}{c|}{0.4} \\ \hline 
$\Delta\phi(\mathrm{jet}_{i>3},E_{T}^\text{miss})_\text{min}>$ & \multicolumn{3}{c|}{-} & \multicolumn{4}{c|}{0.2} \\  \hline
$E_{T}^\text{miss}/\sqrt{H_{T}}~[\mathrm{GeV}^{1/2}]>$ & \multicolumn{2}{c|}{15} & 20 & \multicolumn{4}{c|}{-} \\ \hline
$\mathrm{Aplanarity}> $ & \multicolumn{3}{c|}{-} & \multicolumn{4}{c|}{0.04} \\ \hline
$E_{T}^\text{miss}/m_\text{eff}(N_{j})>$ & \multicolumn{3}{c|}{-} & 0.2 & \multicolumn{2}{c|}{0.25} & 0.2 \\ \hline
$m_\text{eff}(\mathrm{incl.})~[\mathrm{GeV}]>$&1200 & 1600 & 2000 & 2200 & 1600 & 1600 & 2000 \\ \hline
\end{tabular}}
 \caption{\label{tab:BSM_AnalysisCorrelations:Table_A3SR} Summary of the signal regions in the A3 analysis, cf.~Ref.~\cite{Aaboud:2016zdn}.}
\end{center}
 \end{table}


%% file: adl/LH19ADLoverlap.main.tex
\graphicspath{{adl/}}

\newcommand{\Herwig}{H\protect\scalebox{0.8}{ERWIG}\xspace}
\newcommand{\Pythia}{P\protect\scalebox{0.8}{YTHIA}\xspace}
\newcommand{\Sherpa}{S\protect\scalebox{0.8}{HERPA}\xspace}
\newcommand{\Rivet}{R\protect\scalebox{0.8}{IVET}\xspace}
\newcommand{\Professor}{P\protect\scalebox{0.8}{ROFESSOR}\xspace}
\newcommand{\eps}{\varepsilon}
\newcommand{\mc}[1]{\mathcal{#1}}
\newcommand{\mr}[1]{\mathrm{#1}}
\newcommand{\mb}[1]{\mathbb{#1}}
\newcommand{\tm}[1]{\scalebox{0.95}{$#1$}}
\newcommand{\adl}{\textsc{ADL}\xspace}
\newcommand{\adltnm}{\textsc{adl2tnm}\xspace}
\newcommand{\cutlang}{\textsc{CutLang}\xspace}
\newcommand{\smodels}{\textsc{SModelS}\xspace}

\chapter{Using ADL for analysis comparison}

{\it
  H.~B.~Prosper, 
  S.~Sekmen,
  W.~Waltenberger}

\label{sec:LH19ADLoverlap}

\begin{abstract}
In this study, we explore the potential of analysis description languages to assist, and even automate, the comparison of multiple analyses, and report on recent progress towards these goals.  
\end{abstract}

\section{Introduction}
\label{sec:LH19ADLcomparison:introduction}

The unprecedented amount of data collected by the LHC is inspiring physicists to continue designing an ever growing number and diversity of analyses. These analyses focus on different final states involving different types of objects with varying properties, different kinematic variables, upon which varying selections are applied.  As a result, each analysis covers a subset of the multi-dimensional space of event properties.  In some cases, it is readily apparent that these subsets are  disjoint, as they are based on distinct object types. But, sometimes, the subsets overlap, especially when there are multiple analyses exploring similar final states.

This diversity is a tremendous physics resource because it has the potential to increase the amount of information that can be extracted from a given data-set. However, analysis diversity can also cause confusion when attempting to get a complete view of which signatures are covered, which ones are not, which analyses have disjoint subsets, and which ones have overlapping subsets.  This information can be very helpful when designing a new analysis. For example, the information could suggest which sub-space of event properties could benefit from greater attention.  This information is also crucial for interpretation studies whether conducted within or outside the experiments, where a typical task is choosing the set of analyses that are optimal for testing a given model of new physics. In this study, we explore the use of the analysis description language  approach to develop methods and tools for comparing analyses in the space of event properties.   

\section{Analysis description languages}
\label{sec:LH19ADLcomparison:ADL}

An analysis description language is a domain specific language (DSL) capable of describing the physics contents of an LHC  analysis in a standard and unambiguous way.  In this approach, the description of the analysis components is decoupled from the software frameworks that run the analysis.  Analysis description languages, which can be used both by the experimental and phenomenological communities, would bring numerous benefits to the LHC community, ranging from analysis preservation that goes beyond the lifetimes of experiments or analysis software to facilitating the abstraction, design, visualization, validation, combination, reproduction, interpretation and overall communication of the contents of LHC analyses. 

Several attempts have been made to build domain specific languages for analyses at particle colliders. These include  the well-established ROOT TTreeFormula and TTree::Draw, YADL (YAML ADL in the F.A.S.T. framework), LINQtoROOT, PartiQL and NAIL (NAtural Analysis Implementation Language).  Here, we focus on \adl~\cite{Brooijmans:2016vro}, whose development has continued well beyond its genesis at the Les Houches PhysTeV 2015 workshop.  

\adl consists of a plain text file containing blocks with a keyword-value structure. The blocks make clear the separation of analysis components such as object definitions, variable definitions, and event selections. The keywords specify domain specific analysis concepts and operations, such as object, region, select, weight, etc.  The syntax rules include standard mathematical and logical operations, comparison and optimization operators, reducers (e.g. size, max, min), 4-vector algebra and common HEP-specific functions (e.g. angular separations).  The \adl file is  complemented by self-contained functions encapsulating variables with complex algorithms (e.g. $M_{T2}$, aplanarity) or non-analytic variables (e.g. efficiency tables or variables resulting from a machine learning model.  The \adl syntax is capable of referring to these functions.

Apart from documenting the flow of an analysis, \adl can also be used for running the analysis on events with the help of dedicated tools.  For example, \adltnm~\cite{Brooijmans:2018xbu} is a Python transpiler that converts an analysis written in \adl into C++ or Python event-by-event analysis code.  \cutlang~\cite{Sekmen:2018ehb, Unel:2019reo} is a runtime interpreter that directly runs an analysis written in \adl on events without the need for compilation.

\section{ADL and analysis comparison}
\label{sec:LH19ADLcomparison:ADLcomp}

The domain specific and declarative nature of \adl makes it an ideal construct for comparing analyses.  Having the analysis described in a single human-readable text file decoupled from framework code already makes comparison "by eye" considerably simpler than comparisons of descriptions in publications, analysis notes, and certainly simpler than comparing multiple C++-based analysis codes. Moreover, comparisons can be performed in more advanced, automated, ways by building tools of increasing sophistication as the need arises.

Such tools can be used standalone for comparative studies of the rich analysis spectrum available at the LHC.  They can also be incorporated into interpretation tools such as \smodels~\cite{Kraml:2013mwa, Ambrogi:2017neo} to identify  analyses that yield uncorrelated results, which we refer to as "uncorrelated analyses".  Since \smodels does not run analyses on events, but only directly uses interpretation results from the experiments, an additional tool is required to provide information on analysis comparisons.  Within the context of \smodels, comparison tools are also intended to be used, for example, to systematically scout for interesting combinations of analysis results or large-scale analysis combinations, as has been outlined in a recent talk~\cite{icl}.

The development of methods to compare analyses and their implementation in comparison tool prototypes are currently in progress. Indeed, both the prototype of \adltnm, examples of \adl files, as well as preliminary versions of the "by eye" and "by sampling" comparison methods, described in the subsections below, can be found at the repository

{\tt https://github.com/hbprosper/adl2tnm}.

\noindent The Python programs in this repository have been tested with Python 3.6.7, but should work with newer versions. The programs may work with Python 2, but this is not supported.

In order to help test the development of these methods, 10 CMS SUSY searches at 13\,TeV with 35.9\,fb$^{-1}$ of data--already included in the SModelS analysis database--have been implemented in ADL.  These analyses are searches in the bb/cc final state~\cite{Sirunyan:2017kiw} (for direct sbottom/stop production); jets$+E_T^{miss}$ with $H_T^{miss}$~\cite{Sirunyan:2017cwe};  same-sign dileptons~\cite{Sirunyan:2017uyt}; one lepton final states with $M_J$~\cite{Sirunyan:2017fsj};  multileptons~\cite{Sirunyan:2017hvp}; one lepton final states with $\delta\phi$~\cite{Sirunyan:2017mrs}; WH~\cite{Sirunyan:2017zss} (for EWK SUSY); $\gamma + E_T^{miss}$~\cite{Sirunyan:2017nyt}; $\gamma + H_T$~\cite{Sirunyan:2017yse}, and the fully hadronic final state~\cite{Sirunyan:2017wif} (for direct stop pair production).  The \adl implementations of these analyses can also be found in the above repository.

In the following, we describe the methods being developed and how, concretely, \adl can be used for this purpose.
We are developing analysis comparisons at four different, but complementary, levels of sophistication as explained in the following subsections.

\subsection{Analysis comparison "by eye"}
\label{sec:LH19ADLoverlap:byeye}

```The simplest way of comparing analyses is obviously to view their \adl files.  However, given the well-defined block structure and standard domain specific syntax in the \adl, this process can be made easier and more systematic.  \adl file contents can be queried with the help of a script.  For a given set of analyses, the user can list the complete set of objects attributes and event variables featured in any number of analyses, and search which selection was applied in quantities of interest.  The user can query event level quantities like jet multiplicity, hardest jet $p_T$, $E_T^{miss}$, $H_T$, or any other variable listed in the analyses, and can receive information on which analysis and which selection region uses that quantity and what selection criteria are applied on it.  Object level selection criteria, such as those on jet $p_T$, electron $\eta$, muon $d_0$, etc. can also be queried and listed for all analyses.  As a result, the Python program (a simple prototype of which exists in the repository mentioned above) can give a collective view of the ranges covered by quantities of interest in multiple analyses. 
`
\subsection{Analysis comparison through static analysis}
\label{sec:LH19ADLoverlap:simple}

The comparison by eye can be taken one step further by automating the comparison of selection criteria through a static analysis of \adl files.  We are investigating three levels of comparison: 
\begin{itemize}
\item The user can filter the list of analyses and/or selection regions whose selection criteria on a given variable is in a certain range.  For example, one can ask for the list of analyses whose jet multiplicity is $>6$ or which yields a discrete sub-space that overlaps the one defined by this criterion.  An analysis or region requiring at least 4 jets would fall into this range while an analysis or region requiring 3 to 5 jets would not. 
\item The user can choose a set of selection variables and find out if their selection ranges overlap.  For example, one can select lepton multiplicity and $E_T^{miss}$ and find out if the selection criteria for these quantities overlap in a given set of analyses.
\item A more extensive analysis comparison can be made using a default set of predefined quantities consisting of simple variables such as object multiplicities, $E_T^{miss}$, object momenta or pseudorapidities, etc.  
\end{itemize}
In each case above, if an analysis does not impose a selection on a given quantity, the output explicitly states this.  

The static analysis of \adl files is a very fast and practical approach, however, it is better suited for qualitative studies.  One point to note is that this approach treats every quantity in an analysis as independent.  For example, a high level variable like transverse mass $M_T$ is considered independently of the lepton and $E_T^{miss}$ from which it is calculated.  However, the static analysis of \adl files can find and state all dependencies that are explicitly stated as functions in the \adl files.  For example, to guide the user, it can state that $M_T$ is dependent on lepton and $E_T^{miss}$, or that both $M_{T2}$ and the razor variables are dependent on jets and $E_T^{miss}$.  In the latter example, the user can then perform a subsequent comparison on jets and $E_T^{miss}$.

\subsection{Analysis comparison using physics events}
\label{sec:LH19ADLoverlap:events}

Another way for analysis comparison is to run the set of analyses on real or simulated events and find out which subset of events, if any, simultaneously survive selection criteria from multiple analyses or selection regions.  \adltnm and \cutlang, the tools currently available to run, event by event, analyses written in \adl,  can readily accomplish this task.  
In order to facilitate analysis comparison, for every event, \adltnm notes which regions accepted the event and writes the results to a flat ROOT tree with one branch per region. The \adltnm tree is provided to \smodels for analysis of overlaps.

This approach is conceptually simple, leads to quantitative information about overlaps, and can be used to estimate the degree of correlation between regions or between analyses.  However, one should note that having a broadly inclusive set of events with physics processes relevant for the analyses under study is a necessity for this approach to work.  

\subsection{Analysis comparison using random sampling}
\label{sec:LH19ADLoverlap:full}

If the goal is simply overlap detection, there is an alternative option for quantitative comparison of analyses without the need for an inclusive set of physics events.  This approach involves determining the union of multi-dimensional sets of event properties covered by a given set of analyses and populating the space with with a sufficiently dense set of randomly sampled points.  The \adl construct is ideal for this approach, since its domain specific syntax and well-defined structure make it possible to systematically parse an \adl file and determine the space of event properties and setting boundaries on these.

The procedure begins by constructing the complete list of all event level variables (i.e., those with a single instance per event) upon which a selection is applied.  Examples of such variables include jet multiplicity, the hardest lepton $p_T$, $H_T$, dilepton invariant mass, etc.  Then, all the objects used to compute these variables are traced, and the set of properties with which these objects were selected are determined.  These objects, their properties, and their multiplicities define the multi-dimensional space of event properties of an analysis.  The union of the set of properties over multiple analyses defines the super-set of event properties that define the space to be populated with points. 

In order to populate the space with points, the boundaries of the event properties need to be known.  These boundaries can be defined once and for all using a suitable sample of physics events and be made available as a default.  The selections on properties in the \adl files also impose boundaries, which will further limit the space to be populated.  Once the space and boundaries are defined, random numbers are generated event by event to populate the space with a single point. However, we can reduce the amount of sampling to be done, event by event, by first sampling the object multiplicities, then sampling the objects' properties.  The generated random numbers are then used for calculating higher order variables.  The analysis proceeds exactly the same way as for real or simulated events.

Populating high-dimensional spaces can suffer from the curse of dimensionality. But, if we find ourselves similarly cursed, the curse can be lifted or at least mitigated with some degree of importance sampling. One advantage of internally generated pseudo-events is that the generation is considerably faster than generating events with the correct physics-motivated multi-dimensional distributions. Therefore, it would be entirely feasible to generate billions of pseudo-events in order to populate the space with sufficient density.

\section{Outlook}

The analysis description language approach offers possibilities that were unanticipated at its inception. We describe one below. But, first, we make a few general remarks about the future of \adl.

In the near term, the grammar of \adl will be formally defined so that modern parsing tools can be deployed to render the current tools considerably more robust and to increase the level of automatic syntax validation with useful error reporting. Concurrently, the prototype tool to transpile from \adl to Python will be completed and brought to the same level of functionality as \adltnm. The major work to be done is moving all tools from the prototype stage to the production stage, which will require engaging the expertise of computer scientists to build the \adl abstract syntax trees (AST) that describe LHC analyses. Work in this direction is in progress.

One advantage of the analysis description language approach that was unanticipated, is that it lends itself to the creation of tools that could automate the determination of all the disjoint sub-spaces of the full bounded space of event properties such that the union of the former equals the latter. This offers the tantalizing possibility of running multiple, possibly correlated, analyses and automatically partitioning their results on real and simulated data into disjoint sets. If this could be done, it would go a long way towards automating the combination of analyses regardless of their degree of correlation. The point is that since each sub-space is disjoint by constructions, the count in each is statistically independent of the counts in the other sub-spaces. This would greatly simplify the construction of likelihood functions that describe the combination of the results from multiple analyses because the results that would be combined would be statistically independent yet contain the same information as the correlated results.

In the longer term, we envisage much more sophisticated static analysis of multiple \adl files using natural language processing combined with artificial intelligence. We believe this to be a realistic possibility. The first step would be to use the same technology that permits the automatic conversion of speech to text to make it possible to use natural speech to query a future ADL file database. The underlying analysis of the returned boolean statements would be done by an AI-enabled analyzer.

\let\Herwig\undefined
\let\Pythia\undefined
\let\Sherpa\undefined
\let\Rivet\undefined
\let\Professor\undefined
\let\eps\undefined
\let\mc\undefined
\let\mr\undefined
\let\mb\undefined
\let\tm\undefined

%% file: contur/contur-update.main.tex
\graphicspath{{contur/}}

\let\Yoda\undefined
\let\hepdata\undefined

\newcommand{\Herwig}{H\protect\scalebox{0.8}{ERWIG}\xspace}
\newcommand{\Pythia}{P\protect\scalebox{0.8}{YTHIA}\xspace}
\newcommand{\Sherpa}{S\protect\scalebox{0.8}{HERPA}\xspace}
\newcommand{\Rivet}{R\protect\scalebox{0.8}{IVET}\xspace}
\newcommand{\Yoda}{Y\protect\scalebox{0.8}{ODA}\xspace}
\newcommand{\hepdata}{HEPData\xspace}
\newcommand{\contur}{\textsc{Contur}\xspace}
\newcommand{\eps}{\varepsilon}
\newcommand{\mc}[1]{\mathcal{#1}}
\newcommand{\mr}[1]{\mathrm{#1}}
\newcommand{\mb}[1]{\mathbb{#1}}
\newcommand{\tm}[1]{\scalebox{0.95}{$#1$}}

\chapter{\contur Update: Correlations and SM Theory}

{\it A.~Buckley,
  J.~M.~Butterworth,
  L.~Corpe,
  G.~Watt,
  D.~Yallup,
  }

\label{sec:contur-update}


\section{Introduction}
\label{sec:contur-update:intro}

\contur is a software package written in Python which allows
new physics models to be compared to existing measurements, using analyses archived
in \Rivet~\cite{Buckley:2010ar,Bierlich:2019rhm}. It was first presented in Ref.~\cite{Butterworth:2016sqg},
where the concept was illustrated using a simplified Dark Matter model.
Further physics results have been presented in the previous Les Houches proceedings~\cite{Brooijmans:2018xbu}(contributions 14 \& 20)
and subsequent publications~\cite{Amrith:2018yfb,Butterworth:2019wnt,Allanach:2019mfl,Butterworth:2019iff}, 
as well as in these proceedings\footnote{See sections 
\ref{sec:SearchMeasurementComplementarity}, \ref{sec:projname}, \ref{sec:VQ_LQ_VL} and \ref{sec:GWC}.}.
During this time, the code has evolved and several new features have been implemented.
In this contribution we give an overview of these functional and technical updates, and, for those not 
covered in other contributions, we also present some first results. Further details on some of the statistical treatment can be found in Ref.~\cite{yallup}.

In addition to the functional developments described below, we note that the code
now has a recommended stable release (1.0.0), is recommended for use with version 3 of \Rivet~\cite{Bierlich:2019rhm},
and is now distributed and developed
via gitlab at \texttt{https://gitlab.com/hepcedar/contur}. The webpage with
documentation is still accessible via hepforge \texttt{https://contur.hepforge.org/}.
The command-line options to use the new features described below are all visible by typing \texttt{contur -\--help} as usual.

\section{Use of correlation information on uncertainties}
\label{sec:contur-update:use-of-corr-info-on-uncertainties}
In looking for potential BSM contributions to the fiducial phase space of existing measurements, 
\contur scans multiple final states in the same run. Care must therefore be taken not to double-count
the same BSM events multiple times, should the same events contribute to several different measurements,
since to do some would overestimate the statistical power of the sensitivity. \contur avoids this by
classifying the measurements into non-overlapping ``analysis pools'', either based on the experiment, the LHC run period,
or the final state. For example, ATLAS 7~TeV and 8~TeV inclusive jet measurements are statistically independent and a combined 
sensitivity may be derived using both. They will be placed in separate pools. 
Likewise, the sensitivity from 8~TeV jet measurements in non-overlapping rapidity regions may be
combined, as may measurements from ATLAS and CMS even if measured in an identical phase space. Sensitivity from  a CMS dilepton measurement may also
be combined with that from a CMS dijet measurement. However, an ATLAS 8~TeV multijet measurement \textit{may not} be combined
with an ATLAS 8~TeV inclusive jet measurement, since the same events may contribute to both. In this case, \contur will evaluate
the sensitivity of both, but use only the most sensitive in deriving its final result\footnote{The classification of measurements into
these non-overlapping analysis pools of statistically correlated measurements is done in an SQLite (\texttt{sqlite.org}) 
database bundled with \contur, called \texttt{analyses.sql}.}. 

This deals well with statistical correlations. Correlated systematic uncertainties are trickier to handle.
A particularly worrying possibility from the point of view of limit setting is that a more-or-less constant increase in a differential
cross section due to some BSM contribution might be counted as an excess in each bin of the measurement, giving a strong exclusion, 
when in fact a single relatively insignificant excursion of a correlated systematic uncertainty (for example, in the integrated luminosity) 
could absorb the entire effect. In its initial, and still default, incarnation, \contur avoids this possibility by only allowing
the most significantly discrepant bin in any given differential measurement to contribute to the final sensitivity. While
being conservative, this approach has the obvious disadvantage of throwing away information and, for example, reducing sensitivity 
when a BSM resonant peak is split over more than one bin of a measurement. 
Increasingly, the experimental results uploaded to \hepdata~\cite{Maguire:2017ypu} include detailed information about the
bin-to-bin correlations, and the latest version of \contur can now (optionally) make use of this.

\subsection{Method}

The measurements come with statistical uncertainties, and sometimes a systematic uncertainty, which are uncorrelated between
bins, and potentially a series of systematic uncertainties which are correlated between bins. Within a single distribution,
each of these correlated components is characterised (i.e. scaled) by a single nuisance parameter. A fit is then performed, 
varying these parameters to maximise the likelihood for measurement and theory to be consistent, assuming
a Gaussian probability distribution for these scaling parameters centred on zero with a width of one. 
The covariance matrix is constructed
and the likelihood of a given BSM signal injection being consistent with the measurement (and thus, by inference, an exclusion) 
is calculated with the data shifted according to the fitted nuisance parameters,
and including the uncorrelated systematic uncertainties, the data statistical uncertainty and the Monte Carlo statistical
uncertainty. 

\subsection{Implementation Details}

The use of a bin-by-bin correlation information in \contur{} was made possible by updates to the \Yoda{} format and the back-end of the \hepdata website.
Previously, \Yoda{} histograms only had the capacity to store a total uncertainty for each bin. In response to the fact that LHC measurements are increasingly 
dominated by systematics uncertainties (which are often correlated between bins), the format was augmented to include correlation information.
Starting from \Yoda{} 1.7, an \texttt{ErrorBreakdown} annotation in the header of \Yoda{} histograms was permitted, codifying the breakdown of uncertainty sources/nuisance parameters and their impacts on each bin.
The annotation contents are in the YAML format to allow a quick and efficient read-in of the information. Internal \Yoda{} functions and methods were written so that if an 
uncertainty breakdown was present, it could be easily accessed and manipulated. In particular, it could be converted into a covariance matrix using standard linear algebra,
assuming that statistical uncertainties (or uncertainties specifically marked as ``uncorrelated'') are treated as uncorrelated between bins (and affect only the diagonal of the covariance matrix). 
Other uncertainties are assumed to be 100\% correlated between bins, and the outer product of the vector of 
impacts in each bin is taken to produce the covariance matrix for a given nuisance parameter or uncertainty source. An example of the format is shown below.

{\footnotesize
\begin{verbatim}
BEGIN YODA_SCATTER2D_V2 /REF/RIVET_ANALYSIS_NAME/d01-x01-y01
Variations: [""]
ErrorBreakdown: {
0: {'stat,Data_statistics': {dn: -2.5347e-06, up: 2.5347e-06}, 
'sys,DY_TH_Scale_uncertainty': {dn: -9.6354e-08, up: 9.6354e-08}, 
'sys,JET_GroupedNP_1': {dn: -9.6549e-07, up: 9.6549e-07}}, 
1: {'stat,Data_statistics': {dn: -8.6171e-06, up: 8.6171e-06}, 
'sys,DY_TH_Scale_uncertainty': {dn: -8.4579e-07, up: 8.4579e-07}, 
'sys,JET_GroupedNP_1': {dn: -8.6178e-06, up: 8.6178e-06}}, 
2: {'stat,Data_statistics': {dn: -2.6374e-06, up: 2.6374e-06}, 
'sys,DY_TH_Scale_uncertainty': {dn: -4.6791e-07, up: 4.6791e-07}, 
'sys,JET_GroupedNP_1': {dn: -6.7696e-06, up: 6.7696e-06}}, 
3: {'stat,Data_statistics': {dn: -2.0552e-06, up: 2.0552e-06}, 
'sys,DY_TH_Scale_uncertainty': {dn: -1.3018e-07, up: 1.3018e-07}, 
'sys,JET_GroupedNP_1': {dn: -4.9286e-07, up: 4.9286e-07}}}

IsRef: 1
Path: /REF/RIVET_ANALYSIS_NAME/d01-x01-y01
Title: Inclusive cross-section measurements
Type: Scatter2D
---
# xval        xerr-         xerr+         yval          yerr-         yerr+  
1.500000e+02  1.500000e+02  1.500000e+02  3.953700e-05  2.714067e-06  2.714067e-06
4.000000e+02  1.000000e+02  1.000000e+02  3.694700e-04  1.221623e-05  1.221623e-05
6.500000e+02  1.500000e+02  1.500000e+02  1.746400e-04  7.280268e-06  7.280268e-06
9.500000e+02  1.500000e+02  1.500000e+02  2.680900e-05  2.117476e-06  2.117476e-06
END YODA_SCATTER2D_V2

\end{verbatim}
}

There already exist dozens of \hepdata entries where the uncertainty breakdown was provided when the results of the search or measurement were uploaded.
The \hepdata upload format allows for these as additional labels in each bin which specify the uncertainty name, as well as the signed or symmetrized, relative or absolute effect of that uncertainty on a given bin.
Previously, there was no way to propagate this information when converting the \hepdata entry into a \Yoda{} file, and the various sources were summed in quadrature to give a total uncertainty for the \Yoda{} histograms.
The back-end of the \hepdata website was therefore updated so that the additional uncertainty labels were converted into \texttt{ErrorBreakdown} annotations by default when propagating data into the \Yoda{} format.
This \texttt{ErrorBreakdown} information was then propagated into the latest \Rivet{} reference data during the usual synchronisation between \Rivet and \hepdata at the time of new \Rivet{} releases.
As a result of this change, the \texttt{ErrorBreakdown} information became available for the reference data of dozens of \Rivet routines, and could then be used in \contur{}.

The \contur{} code needed to be updated in several ways to be able to take account of the uncertainty breakdowns available in many analyses. The current workflow is as follows.
All reference data are first checked for error breakdowns. If the error breakdowns are available, covariance matrices are built as described above. 
To speed up the procedure, only nuisance parameters which have an effect of at least 1\% anywhere on the range of a histogram are processed by default\footnote{This materiality cutoff is adjustable via a command-line option.}.
If theory predictions are being used (see Section~\ref{sec:contur-update:theory}), theory uncertainties on SM background predictions 
are treated as 100\% correlated between bins by default, 
but may also be treated as
100\% uncorrelated if desired.
The matrices are checked to ensure they are invertible.
If no error breakdowns are available or the matrix is not invertible, the code falls back to the single-bin approximation described at the start of this section, and all bins are treated as a separate test.
If the diagonalised covariance matrix has determinant zero, the entry is treated as pathological and discarded.

When evaluating the agreement between the reference data and the prediction(s) to determine the CL$_{s}$ exclusion, a likelihood function $\mathcal{L}$ is constructed.
This likelihood is the product of the Poisson probability to see an observed number of events in each bin, given the prediction (with or without signal).
The nuisance parameters are included in the prediction, and each nuisance parameter is Gaussian-constrained.
The nuisances are individually profiled by finding the best-fit value of the nuisance $\nu_{i}$ across all bins, using the error breakdown provided.
The best-fit value of each nuisance parameter is either obtained analytically by solving $\partial \log \mathcal{L} / \partial \nu_{i} =0$, or using a numerical maximization of the log-likelihood.
The analytical solution is faster for histograms with less than 5 bins, but the numerical solution is faster for 5 bins or more. By default, the choice of method is made automatically.

The final agreement between data and prediction is obtained from the $\chi^2$ by taking the dot product of the residuals in each bin (after profiling of each nuisance parameter) with the inverted total covariance matrix in the usual way.
The $p$-values for the null and alternative hypotheses are calculated from the $\chi^2$, and used to evaluate the CL$_{s}$ exclusion.
The  CL$_{s}$ is truncated at zero if a bin if found to agree better with BSM+SM than SM alone (something which can never happen unless
theory predictions are being used, as described in Section~\ref{sec:contur-update:theory}.

\subsection{Results}

\begin{figure}[htbp]
  \begin{center}
    \includegraphics[width=0.6\textwidth]{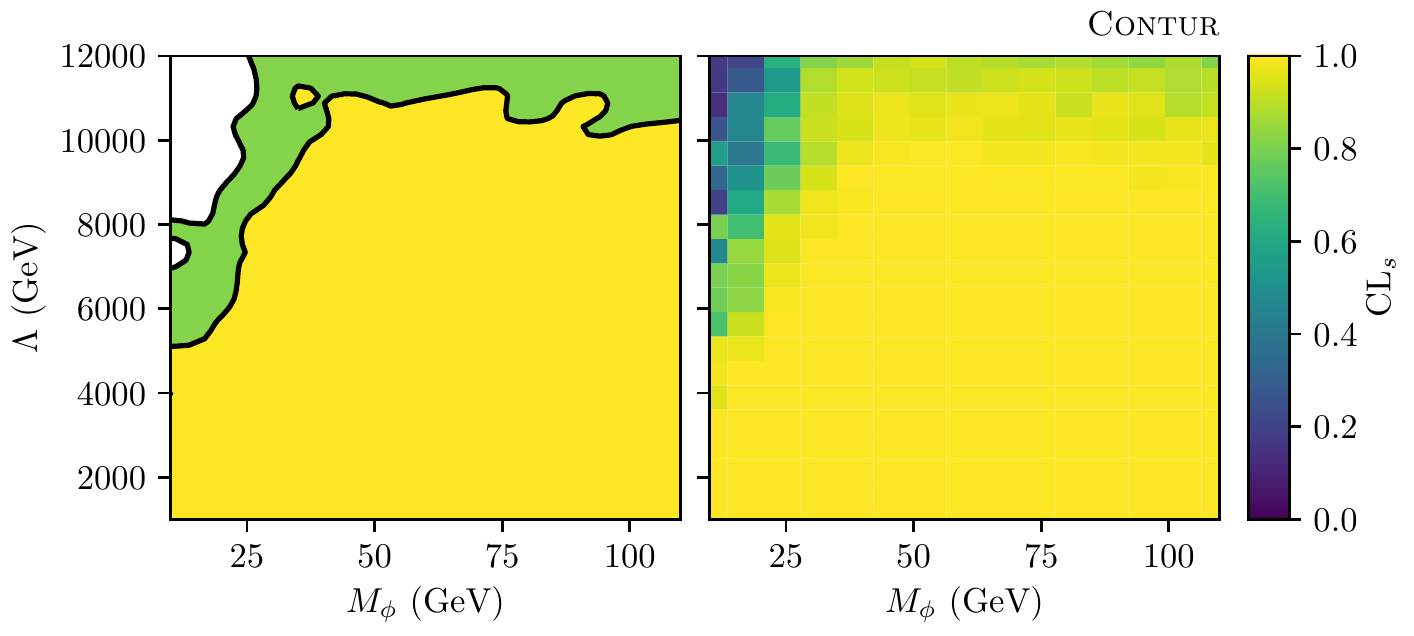}
    \includegraphics[width=0.6\textwidth]{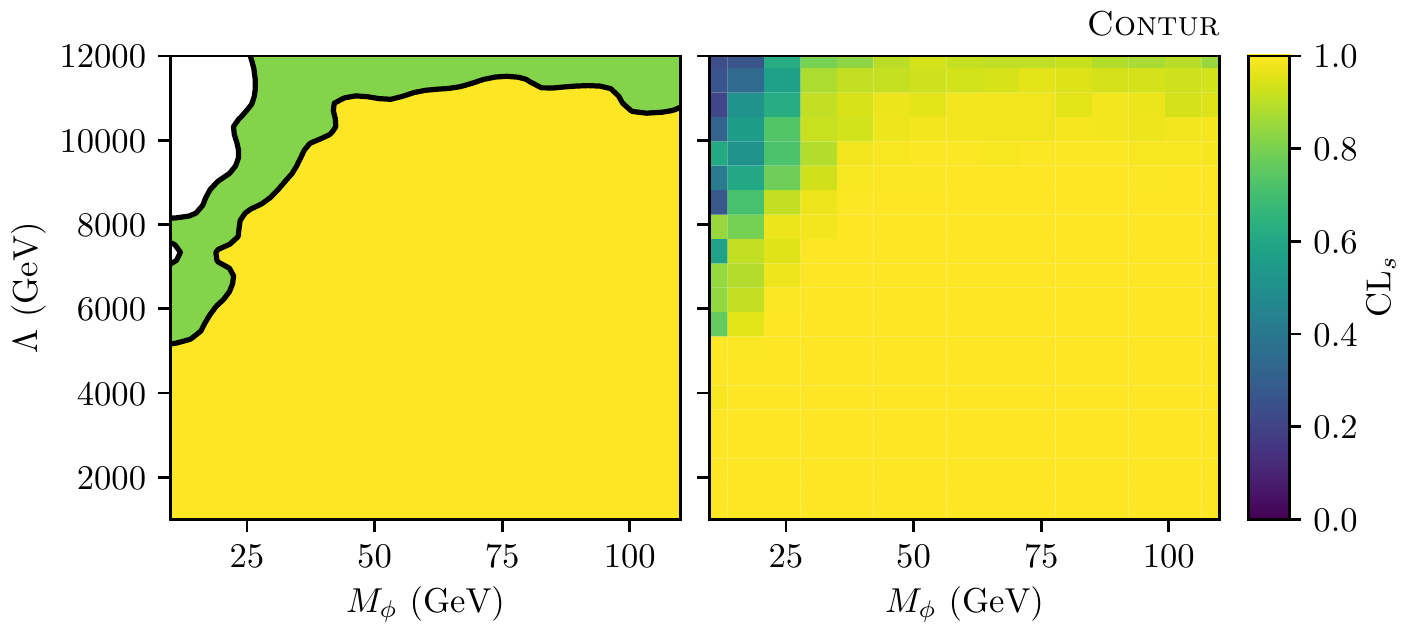}
    \caption{Exclusion with the default \contur approach (upper) and using correlation information (lower) for
    the light scalar model described in Ref.\cite{Brooijmans:2018xbu}(20). The left hand plots show the 68\% and 95\%
    confidence levels, derived from the right hand plots which show the full heatmap on the right hand scale.}
    \label{fig:contur-update:cpescan}
  \end{center}
\end{figure}

The correlation treatment was tested on a number of different models and final states. As an example, we
focus here on a model which introduces a light scalar $\phi$ which decays to diphotons. Production and decay
take place via effective couplings to gauge bosons which are set to one and suppressed by a scale $\Lambda$. Thus,
as well as photons, the final states which are produce will include jets, leptons and missing energy, since 
the $\phi$ is generally produced in association with a $W$ or $Z$. Correlation information for many 
relevant measurements is available, making this model a good exercise ground for their treatment. 
Figure~\ref{fig:contur-update:cpescan} shows the sensitivity (scanned across $\Lambda$ and $M_\phi$, 
with and without the correlation information. It can be seen that the sensitivity without correlation
information is very similar to that shown in \cite{Butterworth:2019wnt}, indicating 
both code stability over several \contur, \Rivet, and \Herwig versions (the version used here 
is 7.2~\cite{Bahr:2008pv,Bellm:2019zci}) and a minor improvement due to a change in 
the treatment of the ATLAS $Z+\gamma$ measurement~\cite{Aad:2016sau}, where the electron and muon data 
are now treated separately rather then being assigned to the same analysis pool.
When correlation information is used, the sensitivity is again very similar, 
with very modest improvement visible at the edges at high scales and low masses, where the limits
also become somewhat smoother. The latter effect is expected since correlations should reduce the
impact of the signal moving between bins of a measurement as the mass of the $\phi$ changes.

\begin{figure}[htbp]
  \begin{center}
    \includegraphics[width=0.3\textwidth]{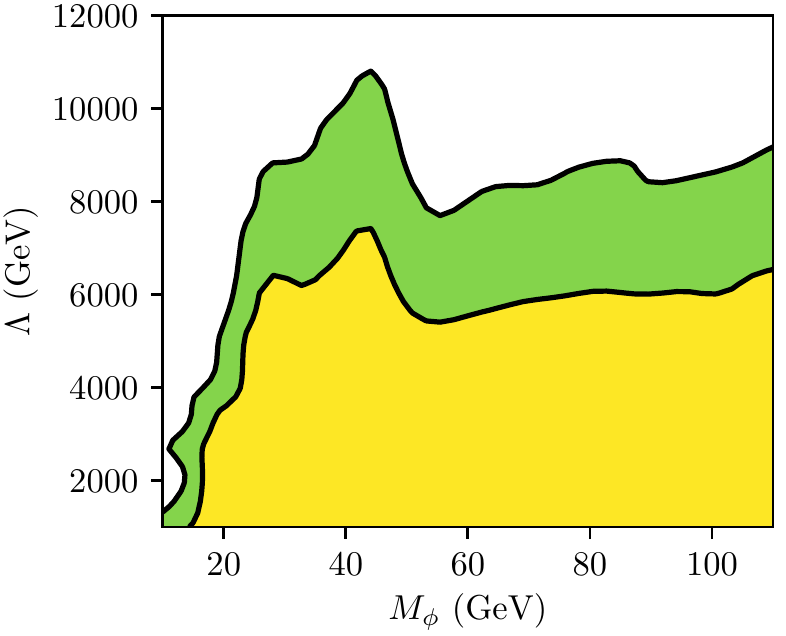}
    \includegraphics[width=0.3\textwidth]{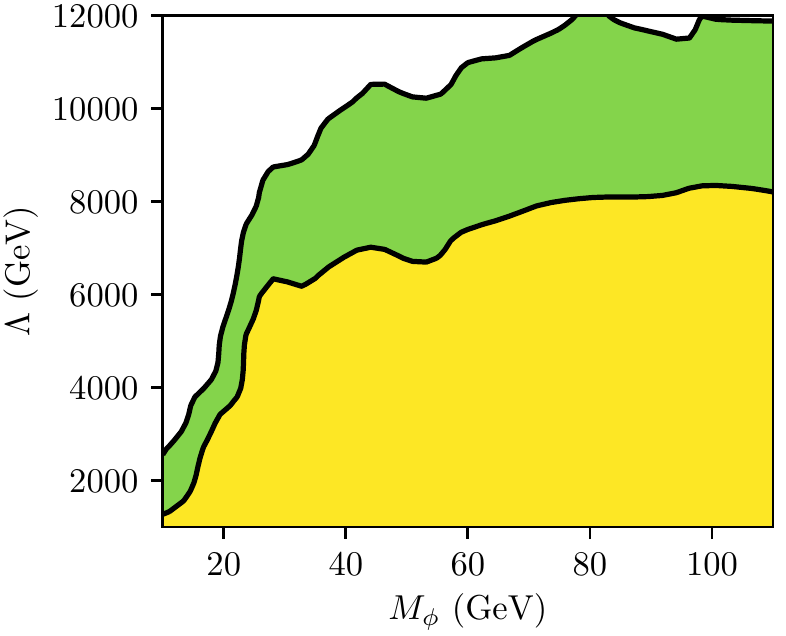}
    \caption{Exclusion from inclusive (di)photon measurements (see text) with the default \contur approach (left) 
    and using correlation information (right) for
    the light scalar model described in Ref.\cite{Brooijmans:2018xbu}(20). The 68\% and 95\%
    confidence levels are shown.}
    \label{fig:contur-update:cpescangg}
  \end{center}
\end{figure}

The overall exclusion comes from a mix of measurements, some of which do not have correlation information
available. The impact of using correlations is seen more clearly on individual measurements where the information
is provided. This is shown in Fig.~\ref{fig:contur-update:cpescangg}, where the exclusion derived from the ATLAS 8~TeV inclusive photon and diphoton
measurements is shown, with and without using correlation information. There is a marked increase in the sensitivity 
for $M_\phi > 60$~GeV or so. In the uncorrelated case, the sensitivity in this region is driven by the diphoton mass distribution of
\cite{Aaboud:2017vol}, where since the $\phi$ is a narrow resonance, the signal appears generally in a single bin. 
In the correlated case, 
the strongest sensitivity comes from the transverse momentum of the diphoton system (from the same paper), where a systematic
deviation builds up over several bins, increasing towards high $p_T^{\gamma\gamma} = p_T^\phi$.

\section{Use of SM theory predictions}
\label{sec:contur-update:theory}

By default \contur treats the data as being identically equal to the SM, an approximation justified to some extent
by the fact that all the measurements used have been shown to agree with the SM, and at some level therefore equivalent 
to treating the whole data set as a data-driven control region. However, if we wish to take SM predictions and their uncertainties 
at face value, this assumption is not good enough. In particular it may overestimate they exclusions in cases where the theory 
uncertainties on the SM are similar to or larger than those on the measurement, since the theory uncertainties are not used. Also, in
cases where there is an excess of data over the SM (which my definition is not ``significant'', since all the measurement papers
state consistency with the SM), stacking a BSM contribution on top of the SM may even improve the agreement with the measurement.
\contur has the capacity to consider any theory calculation as the SM ``background'', if the prediction is provided as
a \Yoda{} object and is visible via the usual \Rivet search path\footnote{Technically, instead of a path name beginning 
with \texttt{/REF},
the path name of the \Yoda{} histogram should begin with \texttt{/THY}.}. A library of such predictions is being assembled 
gradually. We note that if experiments upload the theory predictions used in the paper to \hepdata, 
they can be exported as \Yoda{} files, 
greatly facilitating this process. 

\begin{figure}[htbp]
  \begin{center}
    \includegraphics[width=0.6\textwidth]{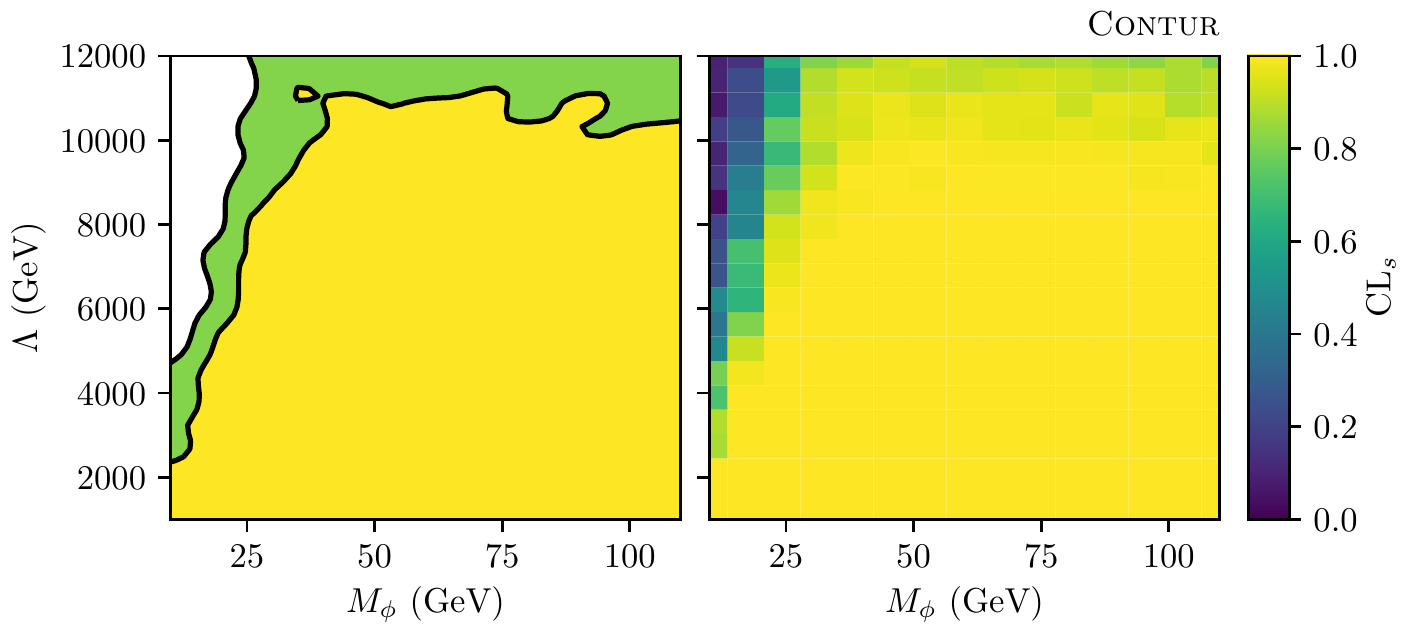}
    \includegraphics[width=0.6\textwidth]{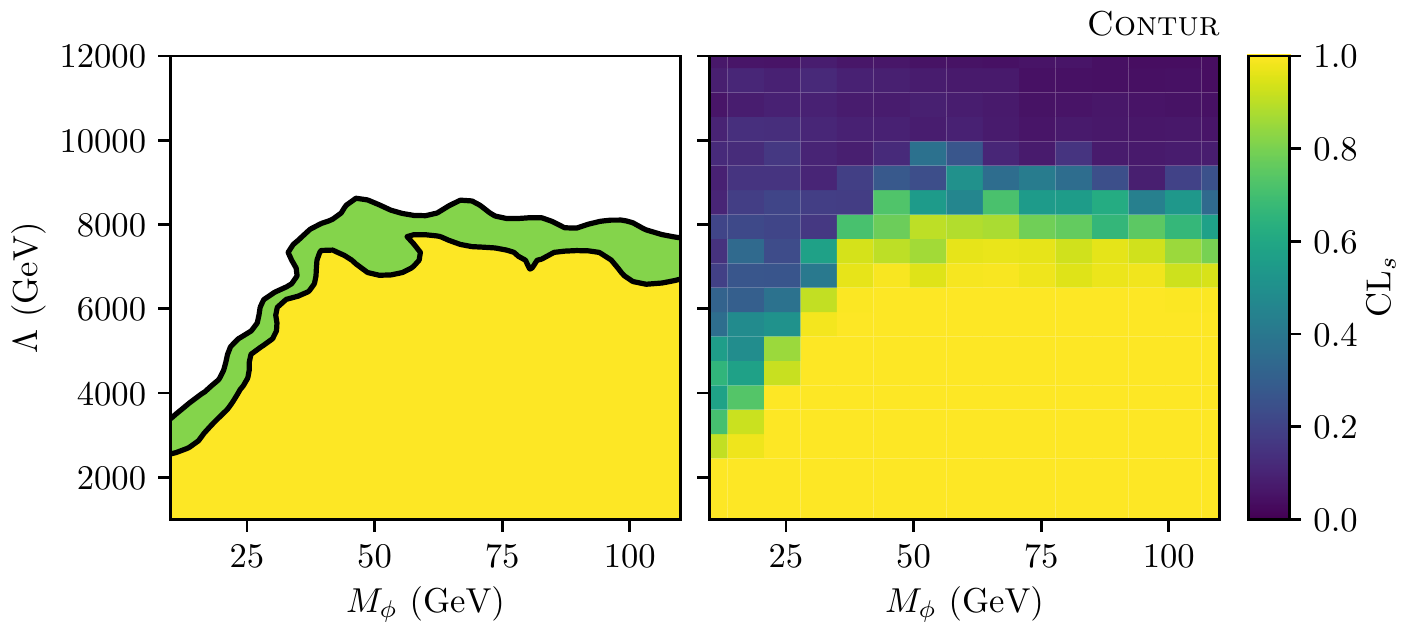}
    \caption{Exclusion with the default \contur approach (upper) and using SM theory information (lower) for
    the light scalar model described in Ref.\cite{Brooijmans:2018xbu}(20). The left hand plots show the 68\% and 95\%
    confidence levels, derived from the right hand plots which show the full heatmap on the right hand scale.}
    \label{fig:contur-update:cpescan-th}
  \end{center}
\end{figure}

Since the majority of measurements relevant to the light scalar model discussed in the previous
section have theory predictions already available in \contur, (from~\cite{Chen:2019zmr,Boussaha:2018egy,Catani:2011qz}),
we present some examples of their use here.
To begin with, show the heatmap comparable to the upper half of Fig.~\ref{fig:contur-update:cpescan}, not using the theory 
predictions yet, but only using to measurements for which they are available. The only significant difference is a reduction
in sensitivity at low $M_\phi$, which is due to the removal of the ATLAS 7~TeV $Z/W + \gamma$ measurements\cite{Aad:2013izg}, for which
the SM prediction is not yet in \contur. This is shown in the upper half of Fig.~\ref{fig:contur-update:cpescan-th}.

The lower portion of Fig.~\ref{fig:contur-update:cpescan-th} shows the exclusion coming from the same measurements, but now with
SM theory predictions used as the null signal hypothesis, including the appropriate uncertainties\footnote{Theory uncertainties
are treated as uncorrelated, since several have a significant statistical component, and treating them as 100\% correlated
gives a very poor agreement between measurements and the SM. The new correlation treatment for data uncertanties described in the
previous section is not used.}. There is a reduction
in sensitivity. This is due to both factors discussed above. Since the theory uncertainties are generally not negligible compared
to the experimental uncertainties for isolated photon measurements, all measurements show some reduction in sensitivity when 
the theory uncertainty is taken into account at face value, on the assumption it unconstrained by other data. 
For example, the SM prediction for the diphoton results~\cite{Aaboud:2017vol} discussed previously, taken from 
Refs.~\cite{Boussaha:2018egy,Catani:2011qz}, have uncertainties similar to the measurement (in fact, larger at lower $p_T^{\gamma\gamma}$
and smaller at higher $p_T^{\gamma\gamma}$), reducing the sensitivity of this measurement to scales below about 3~TeV (compared to 6~TeV
when the data are used as the null hypothesis).

This is not the dominant 
effect, however. The most sensitive measurement of high $\Lambda$ for most $M_\phi$ is the 
ATLAS $\gamma$+missing $E_T$ measurement
from \cite{Aad:2016sau}. This measurement in fact shows a modest excess over the SM prediction, particularly
for the $\gamma\gamma+$ missing $E_T$ zero jet channel
($1.18^{+0.52}_{-0.44} (\textrm{stat.}) ^{+0.48}_{-0.49}(\textrm{syst.}) \pm 0.02(\textrm{lumi.})$ 
fb compared to $0.395^{+0.049}_{-0.037}$ fb from MCFM version 7.0 as used in \cite{Aad:2016sau}
\footnote{See \cite{Campbell:2019dru} for the most recent version of MCFM.}.) 
Thus, adding a small BSM contribution
in fact improves the consistency with the data, and the exclusion limit is therefore reduced.

\section{Use of search analyses}

\Rivet can now implement detector smearing and efficiencies, so that search analyses which
have not been unfolded for detector effects may be implemented~\cite{Buckley:2019stt}. \contur is now
able to make use of this feature. More details, and some studies, 
are presented in Section~\ref{sec:SearchMeasurementComplementarity}.

\section*{Acknowledgements}

We thank the organizers of the Les Houches workshop for the stimulating and productive atmosphere.
A.B, J.B, L.C and D.Y.'s work has received funding from the European Union's Horizon 2020 research and innovation
programme as part of the Marie Skłodowska-Curie Innovative Training Network MCnetITN3 (grant agreement no. 722104). 

\let\Yoda\undefined
\let\contur\undefined
\let\Herwig\undefined
\let\Pythia\undefined
\let\Sherpa\undefined
\let\Rivet\undefined
\let\Professor\undefined
\let\eps\undefined
\let\mc\undefined
\let\mr\undefined
\let\mb\undefined
\let\tm\undefined

%% file: recastcomp/recastcomp.main.tex
\graphicspath{{recastcomp/}}
\let\adl\undefined
\let\adltnm\undefined
\let\cutlang\undefined
\let\smodels\undefined

\newcommand{\Herwig}{H\protect\scalebox{0.8}{ERWIG}\xspace}
\newcommand{\Pythia}{P\protect\scalebox{0.8}{YTHIA}\xspace}
\newcommand{\Sherpa}{S\protect\scalebox{0.8}{HERPA}\xspace}
\newcommand{\Rivet}{R\protect\scalebox{0.8}{IVET}\xspace}
\newcommand{\Professor}{P\protect\scalebox{0.8}{ROFESSOR}\xspace}
\newcommand{\eps}{\varepsilon}
\newcommand{\mc}[1]{\mathcal{#1}}
\newcommand{\mr}[1]{\mathrm{#1}}
\newcommand{\mb}[1]{\mathbb{#1}}
\newcommand{\tm}[1]{\scalebox{0.95}{$#1$}}

\newcommand{\smodels}{\textsc{SModelS}\xspace}
\newcommand{\rivet}{\textsc{Rivet}\xspace}
\newcommand{\gambit}{\textsc{GAMBIT}\xspace}
\newcommand{\colliderbit}{\textsc{ColliderBit}\xspace}
\newcommand{\buckfast}{\textsc{BuckFast}\xspace}
\newcommand{\checkmate}{\textsc{CheckMATE}\xspace}
\newcommand{\madanalysis}{\textsc{MadAnalysis}\xspace}
\newcommand{\adl}{\textsc{ADL}\xspace}
\newcommand{\adltnm}{\textsc{adl2tnm}\xspace}
\newcommand{\cutlang}{\textsc{CutLang}\xspace}
\newcommand{\pythia}{\textsc{Pythia~8}\xspace}
\newcommand{\madgraph}{\textsc{MadGraph5\_aMC@NLO}\xspace}
\newcommand{\geant}{\textsc{Geant}\xspace}
\newcommand{\delphes}{\textsc{Delphes}\xspace}
\newcommand{\fastsim}{\textsc{SuperFastSim}\xspace}

\newcommand{\anders}[1]{{\bf \color{purple} Anders: #1}}
\newcommand{\tomas}[1]{{\bf \color{red} Tomas: #1}}
\newcommand{\are}[1]{{\bf \color{pink} Are: #1}}

\newcommand{\toremove}[1]{{\color{gray} #1}}

\chapter{Comparison of recasting tools}
{\it
  J.~Y.~Araz,
  A.~Buckley,
  N.~Desai,
  B.~Fuks,
  T.~Gonzalo,
  P.~Gras,
  A.~Kvellestad,
  A.~Raklev,
  R.~Ruiz de Austri,
  S.~Sekmen}

\label{sec:recastcomp}

The plethora of LHC searches for new particles target an impressive range of scenarios of physics beyond the Standard Model (BSM). In addition to providing the raw search results, that consist in the expected Standard Model (SM) background rates and the observed number of events in each search region, the LHC experiments usually interpret their results in terms of a few specific BSM scenarios. In this context, often so-called \textit{simplified models}, with only a few free parameters, are used due to the inherent difficulty of exploring many-dimensional parameter spaces. The phenomenology predicted by these models is generally much more restricted than that of the complete BSM models they originate from. Furthermore, many fundamentally different BSM models can give rise to similar signals in LHC searches. 

In other words, the scientific potential of a given LHC search extends far beyond the BSM interpretations performed directly by the experiment. In order to realise this potential, the phenomenology community must be able to reinterpret the search results within BSM models that are different from, and often more general than, the particular BSM scenarios studied by the experiments. 
The success and reliability of such reinterpretation (or \textit{recasting}) efforts depend on dedicated work from both experiment and theory: The experiments must provide sufficiently detailed information -- preferably in standardised electronic formats -- regarding how each search was performed. The phenomenology community must develop public tools that can utilise this information to reproduce and recast the LHC results with acceptable accuracy. In this section we focus on the contribution from the phenomenology side. We briefly describe some of the major public tools for LHC recasting, and check how closely they match each other in reproducing the experiment cutflow for a rather challenging BSM signal region.


\section{Recasting tools}
\label{sec:recastcomp:section1}

The computationally fastest way to reinterpret an LHC search in a new BSM model is to map the model to an appropriate set of simplified models for which the LHC experiments have already provided interpretations. This is the approach taken by the tool \smodels~\cite{Ambrogi:2017neo}. The low computational cost makes this a well-suited approach for large BSM parameter scans. However, for complicated BSM models, the derived limits can often be overly conservative. 

A more general, but computationally much more expensive, approach is to reproduce the experimental search directly on simulated events for the new BSM model. This is the method we focus on here. The software tools \checkmate~\cite{Dercks:2016npn,Drees:2013wra}, \rivet~\cite{Buckley:2010ar,Bierlich:2019rhm}, \madanalysis~\cite{Conte:2012fm,Conte:2014zja,Dumont:2014tja,Conte:2018vmg} and \colliderbit~\cite{Balazs:2017moi}, and the two interpreters \adltnm~\cite{Brooijmans:2018xbu} and \cutlang~\cite{Sekmen:2018ehb, Unel:2019reo}, both working on an analysis description language~\cite{Brooijmans:2016vro}, are all based on this full-simulation approach, but differ in terms of implementation and the targeted use-case.

\vspace{5pt}
\noindent \textbf{\checkmate}

\noindent CheckMATE2 (Check Models At Terascale Energies) is a programme package which accepts simulated event files in many formats for any given model, and if necessary, generates events on-the-fly using the Monte Carlo generators \madgraph~\cite{Alwall:2014hca} or \pythia~\cite{Sjostrand:2014zea}. The programme then determines whether the model is excluded at the 95\% confidence level (CL) by comparing predictions to published experimental results. The programme structure allows for easy addition of new analyses.

\vspace{5pt}
\noindent \textbf{\rivet}

\noindent The \rivet toolkit (Robust Independent Validation of Experiment and Theory) is a system developed for the validation and tuning of Monte Carlo event generators. A large (and ever growing) set of experimental analyses are supplied with \rivet\ ($\sim 950$ analyses as of version 3.1.0, released in Dec 2019).
\rivet has become the LHC standard for archiving measurement analysis logic in SM, top and Higgs physics, which can provide indirect constraints on BSM models. From \rivet~2.5 onward, a detector-emulation machinery is also available, based on analysis-specific efficiency and smearing functions. This feature permits the reproduction also of reconstruction-level direct BSM searches.

\vspace{5pt}
\noindent \textbf{\madanalysis}~5

\noindent The \madanalysis~5 package~\cite{Conte:2012fm,Conte:2014zja} is a general framework for new physics phenomenology at particle colliders, which ease the design and the recasting of any collider analysis. Its recasting module relies on \delphes~\cite{deFavereau:2013fsa} for simulating the response of the detector, allows for the automatical recast of the results of about 40 ATLAS and CMS analyses (in its to-be-released version 1.8.0), and estimate how they constrain any BSM signal by the evaluation of confidence level exclusions. Details on the re-implementations and their validation can be found on the \madanalysis Public Analysis Database (PAD)~\cite{Dumont:2014tja,Conte:2018vmg}\footnote{https://madanalysis.irmp.ucl.ac.be/wiki/PublicAnalysisDatabase}. In its upcoming v1.8 release, \madanalysis~5 will allow the user to use efficiency and smearing functions to parameterise the detector effects.

\vspace{5pt}
\noindent \textbf{\colliderbit}

\noindent The Global and Modular BSM Inference Tool (\gambit)~\cite{Athron:2017ard} is a multi-purpose inference tool capable of performing computationally intensive global fits of BSM models. Amongst the different physics modules that constitute \gambit, \colliderbit~\cite{Balazs:2017moi} is responsible for the computation of collider observables and likelihoods. \colliderbit has been designed with an emphasis on speed, due to its intended use in large parameter scans and global fits. In particular, \colliderbit performs a fast simulation of LHC signal events using a parallelised version of \pythia. The events are passed through a homebrew detector simulation, \buckfast~\cite{Balazs:2017moi}, which performs fast four-vector smearing, before they are subjected to efficiency functions published by the experiments. \colliderbit is shipped with a large, and steadily growing, collection of implemented LHC searches, with a focus on searches for supersymmetry.\footnote{A light-weight, standalone version of \colliderbit, \textsc{ColliderBit Solo}, will be released in the near future. This allows the user to apply \colliderbit's fast detector simulation and collection of LHC analyses to HepMC events provided as input. For the recast comparisons discussed here, the \colliderbit results were obtained using \textsc{ColliderBit Solo} with HepMC input.}

\vspace{5pt}
\noindent \textbf{\adl}, \textbf{\adltnm} and \textbf{\cutlang} 

\noindent An Analysis Description Language (\adl) is a domain specific, declarative language describing the physics contents of a collider  analysis (\textit{e.g.}\ object and variable definitions, event selections) in a standard and unambiguous way, independent of analysis frameworks. \adltnm is a {\sc Python} transpiler that converts an analysis written in \adl into its C++ or {\sc Python} version.  \cutlang is a runtime \adl interpreter that directly runs an analysis in \adl form on events without the need for compilation.  \adl, \adltnm and \cutlang are usable for both experimental and phenomenological analyses.

\section{Comparison setup}
\label{sec:recastcomp:section2}

Due to their differences in philosophy and implementation, it is worthwhile to investigate to what extent the results from the above tools differ when emulating the same LHC search. The spread in results may also speak in a quantitative manner to the general reliability of current recasting tools. For this purpose we compare how the different tools reproduce the cutflow for a particular signal region of a specific analysis, when provided with the same set of simulated signal events before detector simulation.

\begin{figure}[t]
  \begin{center}
    \includegraphics[width=6cm]{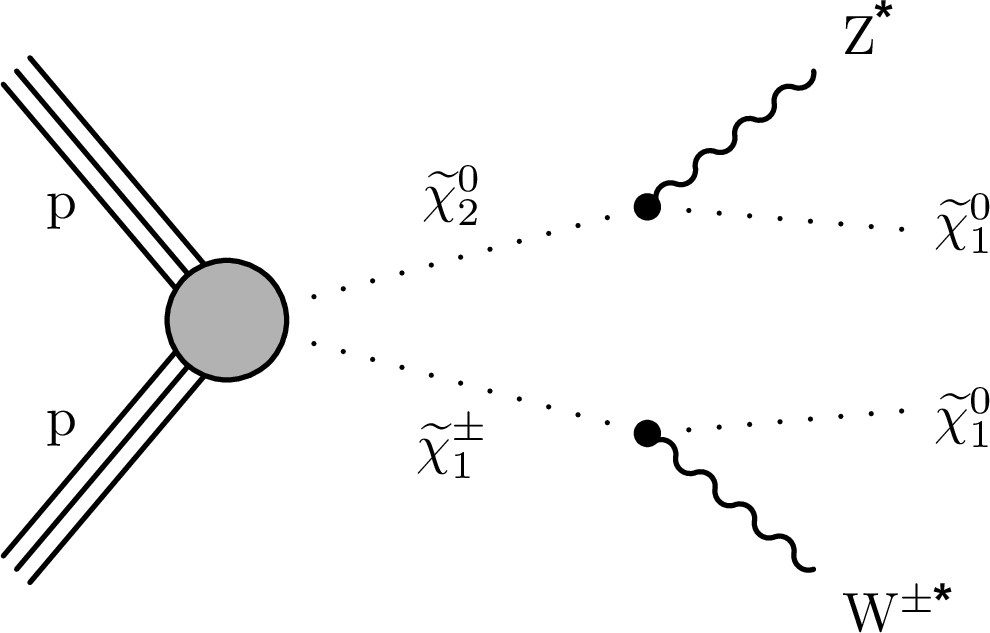}
    \caption{Representative Feynman diagram illustrating the signal process targeted by the considered CMS search. Figure taken from Ref.~\cite{Sirunyan:2018iwl}}
    \label{fig:recastcomp:CMSSignalDiagram}
  \end{center}
\end{figure}

The starting point for our comparison is a CMS search at 13 TeV using final states with two low-$p_T$ oppositely charged leptons and missing transverse momentum~\cite{Sirunyan:2018iwl}. This search is designed to target production of a chargino and neutralino pair ($\tilde{\chi}_{1}^{\pm} \tilde{\chi}_2^0$), with subsequent decays down to the lightest stable neutralino ($\tilde{\chi}_1^0$), when the mass splittings between the supersymmetric particles are small. Such a specific search is by constrution sensitive to details in the implementation as focusing on non-usual soft objects. We illustrate the signal process considered in this CMS search by the Feynman diagram in Fig.~\ref{fig:recastcomp:CMSSignalDiagram}. The low-$p_T$ leptons in the signal arise from leptonic decays of the off-shell $Z$ and $W$, while the two stable $\tilde{\chi}_1^0$ produce the missing transverse momentum.

The baseline for our recast comparison is the ``electroweakino cutflow'' provided by CMS in the auxiliary material for Ref.~\cite{Sirunyan:2018iwl}.\footnote{\url{http://cms-results.web.cern.ch/cms-results/public-results/publications/SUS-16-048/index.html}}  We summarise the cuts here for clarity:
\begin{enumerate}
    \item two muons, each with a transverse momentum $p_T$ lying in [5, 30]~GeV;
    \item opposite electric charges for the two muons;
    \item transverse momentum of the dimuon pair $p_T(\mu\mu)$ larger than 3 GeV;
    \item dimuon invariant mass $M(\mu\mu)$ between 4 and 50 GeV;
    \item veto of events with a dimuon invariant mass $M(\mu\mu)$ between 9 and 10.5 GeV;
    \item missing transverse momentum $p_T^\textrm{miss}$ between 125 and 200 GeV;
    \item $\mu$ and $p_T^\textrm{miss}$ trigger requirement;
    \item at least one ISR jet;
    \item transverse hadronic energy $H_T$ larger than 100 GeV;
    \item $p_T^\textrm{miss}/H_T$ between 0.6 and 1.4;
    \item veto of events featuring $b$-tagged jets;
    \item veto of events with a ditau invariant mass $M(\tau\tau)$ between 0 and 160 GeV;
    \item transverse mass of each muon/missing energy system $M_T(\mu_i,p_T^\textrm{miss})$ less than 70 GeV.
\end{enumerate}
We use the $(150,142.5)$\,GeV benchmark point from Ref.~\cite{Sirunyan:2018iwl}, meaning that we set $m_{\tilde{\chi}_{1}^{\pm}} = m_{\tilde{\chi}_{2}^{0}} = 150$~GeV and $m_{\tilde{\chi}_{1}^{0}} = 142.5$~GeV. For such a compressed electroweakino spectrum, the CMS search relies on a dedicated trigger for dimuon events with low missing $p_T$, which explains the focus on muon pairs in the CMS electroweakino cutflow. The above event selection cuts and the corresponding CMS cutflow numbers are reproduced in the two first columns of Table~\ref{tab:recastcomp:recastcutflow}.

To clearly show the cutflow comparison across the different recast tools, we generate a common set of 250000 Monte Carlo $\tilde{\chi}_{1}^{\pm} \tilde{\chi}_{2}^{0}$ signal events using \madgraph~\cite{Alwall:2014hca}, with \pythia~\cite{Sjostrand:2014zea} handling the electroweakino decays, the parton showering and hadronisation. The recast tools cutflows are all scaled according to the CMS numbers for cross-section times luminosity, \textit{i.e.}\ to match the CMS starting point of 172004 signal events before any cuts. Following the CMS analysis, we require the produced $\tilde{\chi}_{1}^{\pm}$ and $\tilde{\chi}_{2}^{0}$ to decay exclusively into the lightest neutralino $\tilde{\chi}_1^0$ through an off-shell $W$ and $Z$ boson respectively. The CMS signal simulation assumes SM branching ratios for the virtual $Z$ and $W$ boson, but taking into account the Breit-Wigner shapes. We attempt to reproduce this using the \pythia settings \texttt{23:mMin = 24:mMin = 1}, which sets the lower limit of the allowed $Z$ and $W^\pm$ Breit-Wigner mass ranges to $1$ GeV.\footnote{Since performing this comparison we have been informed that CMS used \texttt{23:mMin = 24:mMin = 0.1} for their signal simulation. While this will have caused a systematic difference between the CMS signal events and our signal events, we expect the effect to be small compared to statistical uncertainties and the differences stemmig from the detector simulation.} The simulated events are stored in HepMC format, and later provided as input for the different recast tools.

\section{Cutflow comparison}
\label{sec:recastcomp:section3}

In order to obtain realistic signal rate estimates from a set of simulated events, detector effects must be taken into account. CMS uses a proprietary detector simulation based on the \geant package~\cite{Agostinelli:2002hh}, as well as faster version~\cite{Abdullin:2011zz}. Recasting projects outside the experiments have to rely on available open-source detector simulators, or write their own. Of the recasting tools used in this comparison, \checkmate, \madanalysis and \adl use the fast detector simulation \delphes~\cite{deFavereau:2013fsa}. \madanalysis can also use its own detector simulation, known as \fastsim\ and relying on object reconstruction efficiency and smearing function~\cite{Araz:inprep}, with two different reconstruction methods, one is clustered jet based (Jet) and one is substructure based (Cons). We include \madanalysis cutflows for all three detector simulation options in our comparison. Lastly, \rivet and \colliderbit use their own home-brewed fast detector smearing. Although a detailed comparison of all detector simulations is beyond the scope of this work\footnote{See Refs.~\cite{Balazs:2017moi} and~\cite{Buckley:2019stt} for a comparison between the fast detector simulators in \colliderbit and \rivet, respectively, with \delphes.}, we expect that most of the difference in the cutflows will arise from the different simulations. The last source of difference across all recasting tools lies in their specific implementation of the given CMS search, with efficiency functions and selection cuts.

\begin{figure}[h]
  \begin{center}
    \includegraphics[width=0.95\textwidth]{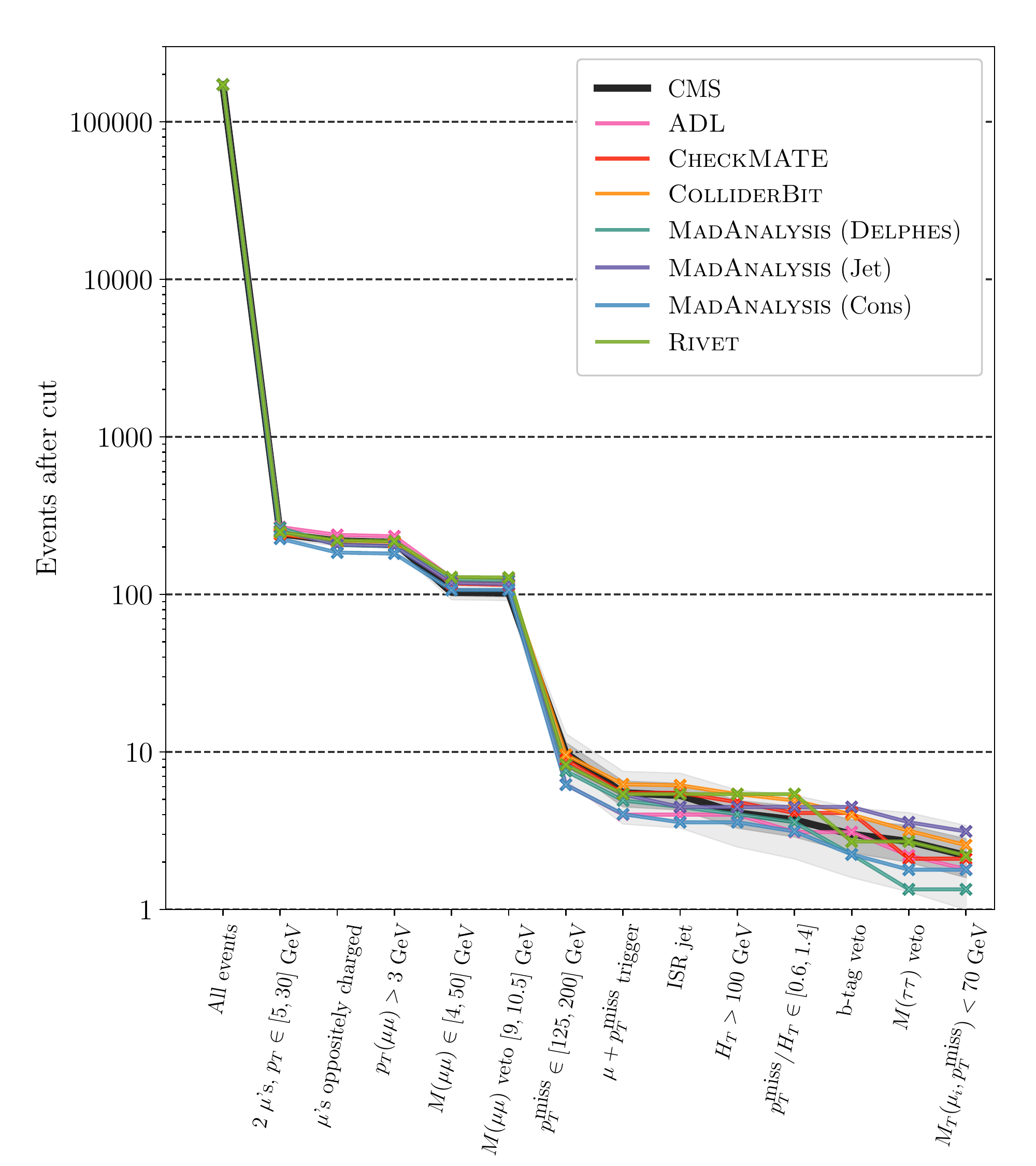}
    \caption{Visualisation of the cutflow comparison in Table~\ref{tab:recastcomp:recastcutflow}. The inner and outer grey bands indicate the $1\sigma$ and $2\sigma$ statistical uncertainty for the CMS cutflow, respectively, as given in Ref.~\cite{Sirunyan:2018iwl} (auxiliary material).}
    \label{fig:recastcomp:plot1}
  \end{center}
\end{figure}

The cutflows obtained with the different tools for the $(150,142.5)$ GeV benchmark point can be seen in Table~\ref{tab:recastcomp:recastcutflow}.
\begin{table}
\renewcommand{\arraystretch}{1.4}
  \begin{center}
    {\scriptsize
      \begin{tabular}{lrrrrrrrr}
      \toprule\\
      & CMS & \checkmate & \rivet & \textsc{MA} \delphes & \textsc{MA} Jet & \textsc{MA} Cons & \adl & \colliderbit\\
      \midrule
      All events & 172004 & 172004 & 172004 & 172004 & 172004 & 172004 & 172004 & 172004\\
      2 $\mu$'s, $p_T \in [5,30]$ & 242.7 & 240.8 & 249.1 & 264.9 & 253.2 & 225.7 & 267.6 & 245.8\\
      $\mu$'s oppositely charged & 218.5 & 217.4 & 220.2 & 205.7 & 207.1 & 184.4 & 238.7 & 214.6\\
      $p_T(\mu\mu) > 3$ & 213.8 & 212.6 & 217.4 & 201.6 & 203.0 & 181.6 & 233.9 & 209.9\\
      $M(\mu\mu) \in [4,50]$ & 103.3 & 117.0 & 128.7  & 124.5 & 118.3 & 106.6 & 127.3 & 122.8\\
      $M(\mu\mu)$ veto $[9,10.5]$ & 102.2 & 114.9  & 128.0 & 122.5 & 116.3 & 106.6 & 125.2 & 121.7\\
      $p_T^\textrm{miss} \in [125,200]$ & 9.8 & 8.9 & 8.3 & 7.6 & 8.3 & 6.2 & 6.2 & 9.6\\
      $\mu + p_T^\textrm{miss}$ trigger & 5.5 & 5.5 & 5.4 & 4.9 & 5.4 & 4.0 & 4.0 & 6.2\\
      ISR jet & 5.3 & 5.5 & 5.4 & 4.5 & 4.5 & 3.6 & 4.0 & 6.1\\
      $H_T > 100$ & 4.1 & 4.8 & 5.4 & 4.0 & 4.5 & 3.6 & 4.0 & 5.4\\
      $p_T^\textrm{miss}/H_T \in [0.6,1.4]$ & 3.7 & 4.1 & 5.4 & 3.6 & 4.5 & 3.1 & 3.1 & 4.9\\
      b-tag veto & 3.0 & 4.1 & 2.7 & 2.2 & 4.5 & 2.2 & 3.1 & 4.0\\
      $M(\tau\tau)$ veto & 2.7 & 2.1 & 2.7 & 1.3 & 3.6 & 1.8 & 2.2 & 3.2\\
      $M_T(\mu_i,p_T^\textrm{miss}) < 70$ & 2.2 & 2.1 & 2.2 & 1.3 & 3.1 & 1.8 & 1.8 & 2.6\\
      \bottomrule
      \end{tabular}
    }
  \end{center}
  \caption{Comparison of cutflows for all recasting tools, including three cutflows using \madanalysis, one with \delphes and two with \fastsim with different reconstruction methods (Jet and Cons). All dimensionful kinematical variables in the first column are given in units of GeV.}
  \label{tab:recastcomp:recastcutflow}
\end{table}
A visual representation of the cutflow comparison is given in Fig.~\ref{fig:recastcomp:plot1}. This figure also includes grey bands indicating the statistical uncertainty on the CMS cutflow numbers, as provided in Ref.~\cite{Sirunyan:2018iwl} (auxiliary material). 

All of the recasting tools reproduce the CMS cutflow reasonably well, with the \rivet and \checkmate cutflows matching very well the CMS expectation after the last cut.
\madanalysis (with \fastsim-Cons), \adl and \colliderbit all give a final signal prediction within 20\% of the CMS result, while \madanalysis with \fastsim-Jets and with \delphes end up $\sim 40$\% above and below the CMS number, respectively. However, when interpreting these results it should be kept in mind that statistical uncertainties grow relatively large towards the end of the cutflow. We note that five out of seven tools lie in the CMS statistical $1\sigma$ band after the final cut.

In the early stages of the cutflow there are some notable discrepancies -- in particular following the cut $M(\mu\mu) \in [4,50]$\ GeV, after which the recast cutflows significantly overshoots the CMS cutflow. The only exception is the \madanalysis Cons cutflow, but this cutflow undershoots the CMS one in the preceding steps. However, after the requirement that $p_T^\textrm{miss} \in [125,200]$\ GeV, all the recast cutflows stay within the $2\sigma$ uncertainty band of the CMS cutflow.

The three \madanalysis cutflows serve as a useful illustration of the impact different choices of fast detector simulation and reconstruction algorithms can have. Since all cutflows in our comparison start from the same set of truth-level signal events, and the three \madanalysis cutflows further share a common implementation of the cutflow logic, the discrepancies between these three cutflows are likely due to the differences in the detector simulation and reconstruction step.

\section{Summary}
\label{sec:recastcomp:section4}

The recasting of LHC searches in the context of complete BSM models gives a more powerful understanding of the results of the experiments. In this work we have performed a detailed comparison of the reinterpretation of a particular signal region of a CMS analysis with many of the publicly available recasting tools: \checkmate, \rivet, \madanalysis, \adl and \colliderbit. We have shown that all of them reproduce fairly well the cutflow provided by CMS in a specific benchmark scenario, but there are minor discrepancies mainly due to using different detector simulation and reconstruction algorithms. 

In the early stages of this comparison work we identified and corrected some clear discrepancies in the implementations of the CMS analysis across the different recast tools. The comparison presented here therefore illustrates the residual level of systematic discrepancy one can expect between the different recast tools and their detector simulation and reconstruction inputs after any obvious analysis logic differences have been dealt with. Such cross-checks highlight one of the benefits of having a handful of different recast tools with independent analysis implementations and detector simulation and reconstruction choices.  

However, having multiple implementations of the same LHC search in different tools does represent a significant amount of duplicated effort. Developing a more unified software framework for implementation and preservation of BSM searches has the potential long-term benefit of freeing up significant resources in the recasting community. This could lead to a much larger combined set of implemented analyses available for reinterpretation studies, and also allow for even more in-depth validations of the analyses against the simulation results provided by the LHC experiments.


\let\Herwig\undefined
\let\Pythia\undefined
\let\Sherpa\undefined
\let\Rivet\undefined
\let\Professor\undefined
\let\eps\undefined
\let\mc\undefined
\let\mr\undefined
\let\mb\undefined
\let\tm\undefined

\let\smodels\undefined
\let\rivet\undefined
\let\gambit\undefined
\let\colliderbit\undefined
\let\buckfast\undefined
\let\checkmate\undefined
\let\madanalysis\undefined
\let\adl\undefined
\let\madgraph\undefined
\let\geant\undefined
\let\delphes\undefined
\let\fastsim\undefined

\let\anders\undefined
\let\addanders\undefined
\let\tomas\undefined
\let\are\undefined
\let\toremove\undefined

%% file: searchmeasurement/SearchMeasurementComplementarity.main.tex
\graphicspath{{searchmeasurement/}}

\newcommand{\Herwig}{H\protect\scalebox{0.8}{ERWIG}\xspace}
\newcommand{\Pythia}{P\protect\scalebox{0.8}{YTHIA}\xspace}
\newcommand{\Powheg}{P\protect\scalebox{0.8}{OWHEG}\xspace}
\newcommand{\Sherpa}{S\protect\scalebox{0.8}{HERPA}\xspace}
\newcommand{\Rivet}{R\protect\scalebox{0.8}{IVET}\xspace}
\newcommand{\Gambit}{G\protect\scalebox{0.8}{AMBIT}\xspace}
\newcommand{\Professor}{P\protect\scalebox{0.8}{ROFESSOR}\xspace}
\newcommand{\eps}{\varepsilon}
\newcommand{\mc}[1]{\mathcal{#1}}
\newcommand{\mr}[1]{\mathrm{#1}}
\newcommand{\mb}[1]{\mathbb{#1}}
\newcommand{\tm}[1]{\scalebox{0.95}{$#1$}}
\definecolor{darkmagenta}{rgb}{0.55, 0.0, 0.55}
\newcommand{\benj}[1]{\textbf{\color{darkmagenta}Benj: #1}}
\newcommand{\addbenj}[1]{\textbf{\color{darkmagenta}#1}}

\chapter{Complementary of searches and measurements in re-interpretation}

{\it  T.~Berger-Hryn'ova,
  A.~Buckley,
  J.~M. Butterworth,
  L.~Corpe,
  D.~Yallup,
  M.~van Beekveld,
  B.~Fuks,
  T.~Gonzalo,
  A.~Kvellestad}

\label{sec:SearchMeasurementComplementarity}

Re-interpretation is an increasingly important topic in the high energy physics community.
As no new particles have been discovered since the Higgs boson, the field is pivoting from a theory-driven approach to a data-driven approach.
It is therefore crucial that Large Hadron Collider (LHC) measurements/searches are preserved, both in terms of analysis logic and results. Furthermore, it is necessary that infrastructure is developed to allow for the rapid and reliable re-interpretation of these LHC results. It could then easily be checked whether new proposed models are ruled out by existing results, avoiding the need to use resources on a dedicated search.
Conversely, gaps in the existing analysis programme at the LHC could be identified, leading to the proposal of new analyses dedicated to these specific cases.

Searches are typically optimised towards one set of models, or a family of models, but could have sensitivity to new, as-yet-unproposed models.
Measurements are typically preserved via the process of ``unfolding'' to particle-level ({\it i.e.}~correcting for detector effects), and storing the analysis logic in a \Rivet{}~\cite{Buckley:2010ar,Bierlich:2019rhm} routine. It is then straightforward to check how any potential new signal could have showed up in a measurement preserved in this way: one only needs to run simulated signal events through the routine. Indeed, a framework called \Contur~\cite{Butterworth:2016sqg} has been built to automatically scan the bank of LHC measurements to verify whether a proposed new model is already excluded. Examples of how \Contur{} can be used to set constraints on new models can be found in Refs.~\cite{Allanach:2019mfl,Butterworth:2019wnt,Amrith:2018yfb} and the procedure is also discussed in Section~\ref{sec:contur-update} of these proceedings. 

Searches, however, are typically not unfolded. In such cases detailed re-interpretation requires some form of detector simulation, in addition to the signal event generation and implementation of the analysis logic. As different use cases require different levels of speed and accuracy in these simulations, several public tools have been developed in the phenomenology community for LHC search re-interpretation, such as \textsc{CheckMATE} \cite{Dercks:2016npn}, \textsc{MadAnalysis}~5 \cite{Conte:2014zja,Dumont:2014tja,Conte:2018vmg} and \textsc{ColliderBit} (part of \Gambit) \cite{Balazs:2017moi,Athron:2017ard}. In addition, the package \textsc{SModelS} \cite{Ambrogi:2017neo} performs a fast re-interpretation without Monte Carlo (MC) simulations, by mapping predictions for total production cross sections for all signals of a given beyond-the-Standard-Model (BSM) model to the efficiencies associated with a set of searches targeting specific simplified model signatures. Conclusive statements about the viability of the initial model can then be drawn after a comparison with published excluded rates at the 95\% confidence level. Currently each of these tools comes with its own database of implemented LHC searches.


In this contribution, we investigate two questions. Firstly, we examine whether, using the smearing functionality recently added to \Rivet{}~\cite{Buckley:2019stt}, it is possible to preserve searches in the same way as measurements are preserved, and thus automate the re-interpretation process for searches to be as fast and efficient as it is for measurements in tools such as \Contur{}. 
Secondly we examine a series of MSSM parameter points which have been identified as high-likelihood scenarios in a \Gambit fit to results of searches for supersymmetry (SUSY), and see whether the measurements available to \Contur{} have any complementary impact.


\section{Preserving searches in \Rivet{}}
\label{sec:SearchMeasurementComplementarity:section1}

To test whether a \Rivet{} routine for a search can be used in an automated framework like \Contur{}, one should check whether detector-level distributions of the Standard Model (SM) background and signal in the analysis signal region can be reproduced within the uncertainties, using particle-level inputs only and a smearing approach to convert truth-level distributions into approximate detector-level ones. Two examples of how this can be achieved are shown below.

The first example consists of an ATLAS search~\cite{Aaboud:2016zdn} for supersymmetry in a final state featuring  jets, missing energy and vetoing the presence of leptons, using $3.2\textrm{~fb}^{-1}$ of 13~TeV LHC data. The routine for this SUSY search uses \Rivet{}'s pre-loaded jet, $b$-tagging, missing energy and lepton efficiency and energy smearing functions to emulate the detector-level distributions of signal and background based on particle-level inputs.
The second is a more involved example. We consider  an ATLAS search for high-mass di-lepton resonances with $139\textrm{~fb}^{-1}$ of 13~TeV LHC data~\cite{Aad:2019fac}, where the resolution function provided in the experimental publication is integrated as a custom smearing function in the \Rivet{} routine. 

In both cases, it has been possible to produce a \Rivet{} routine which is able to give a good representation of the background distribution(s) presented in the original ATLAS publication, so that both searches are therefore usable in the \Contur{} framework for re-interpretations.

\subsection{Example 1: a SUSY search preserved in \Rivet{}}
\label{sec:SearchMeasurementComplementarity:subsection1}

\begin{figure}[t]
  \begin{center}
    \includegraphics[width=5cm]{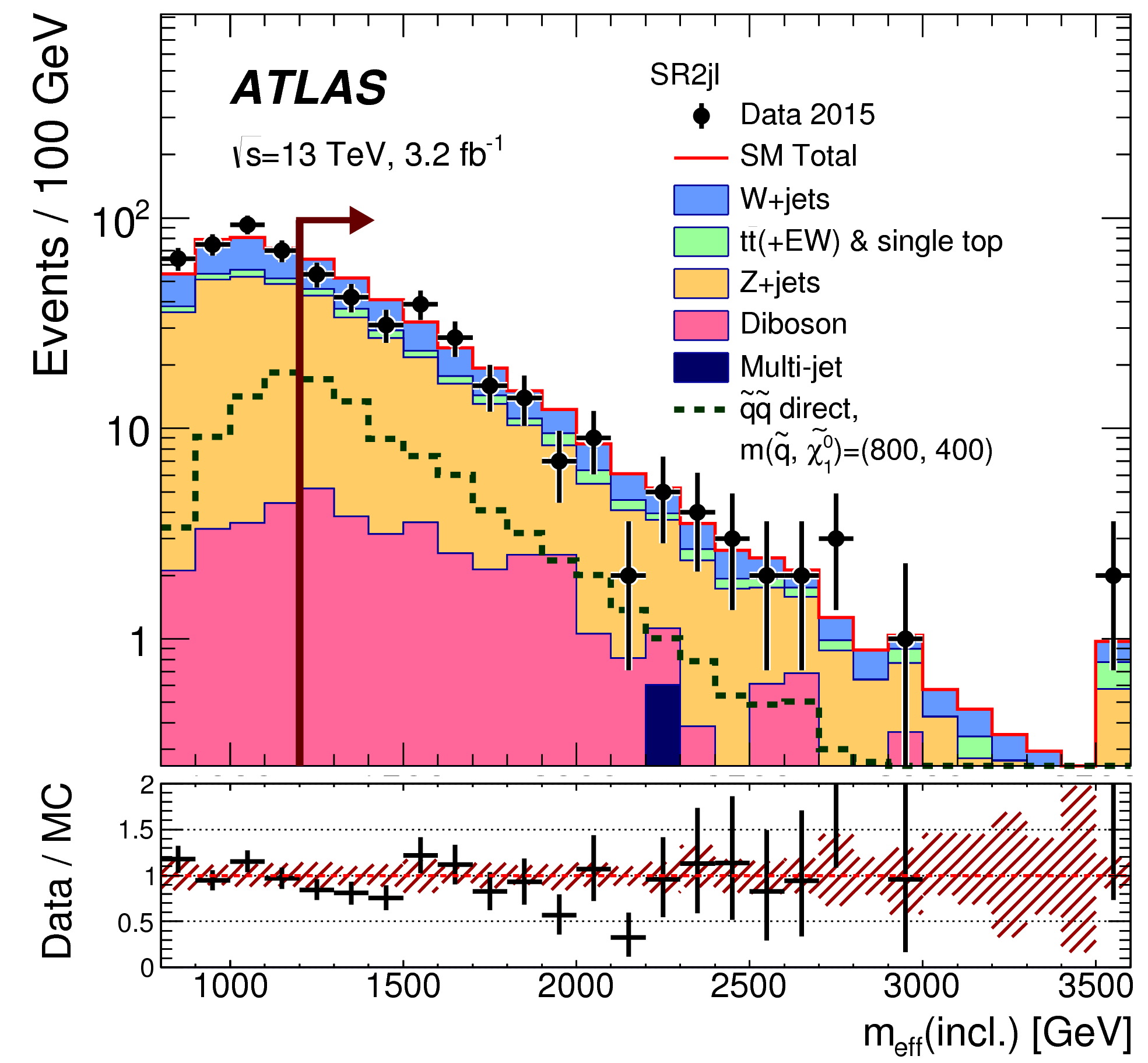}
    \includegraphics[width=5cm]{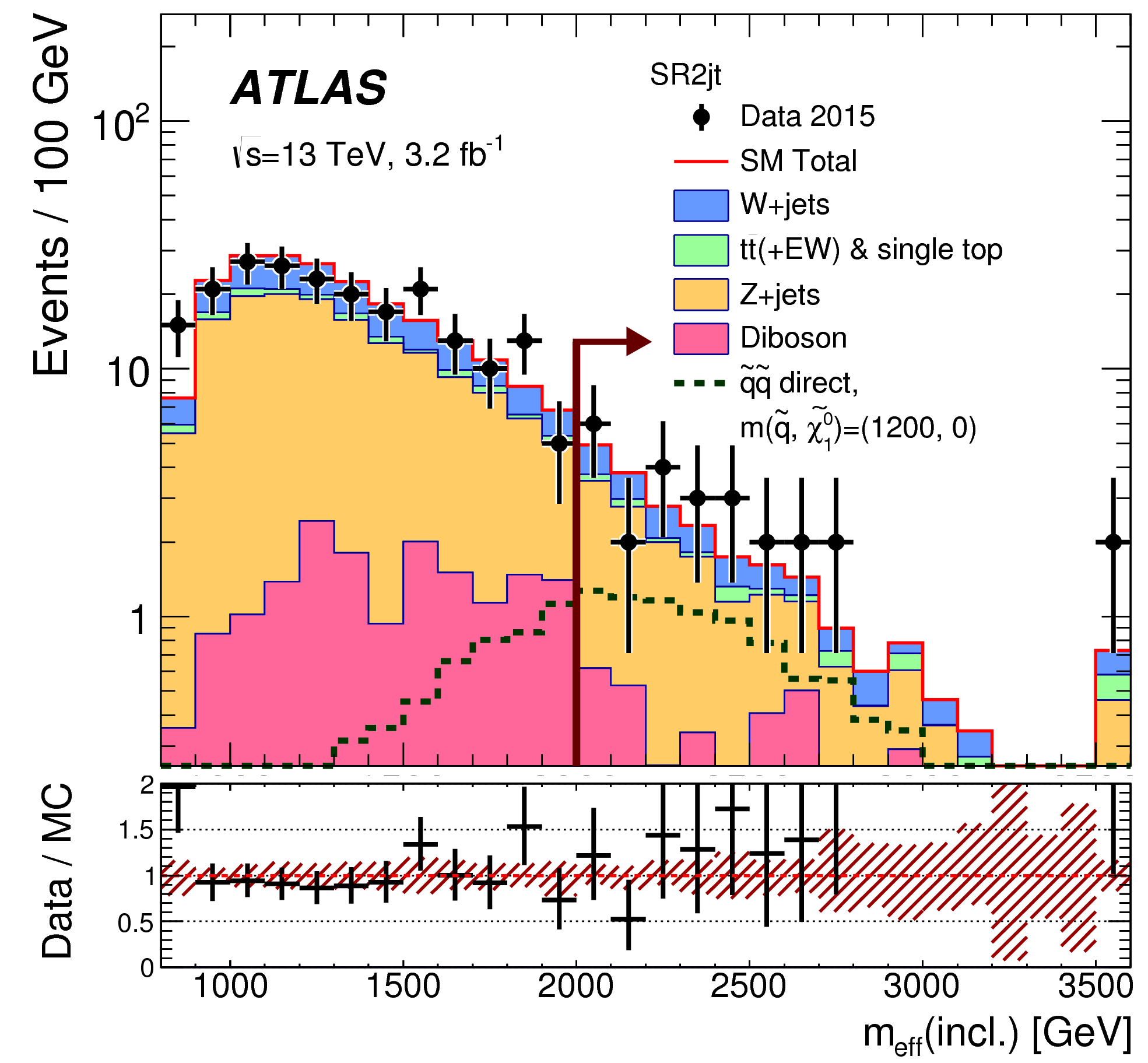}
    \includegraphics[width=5cm]{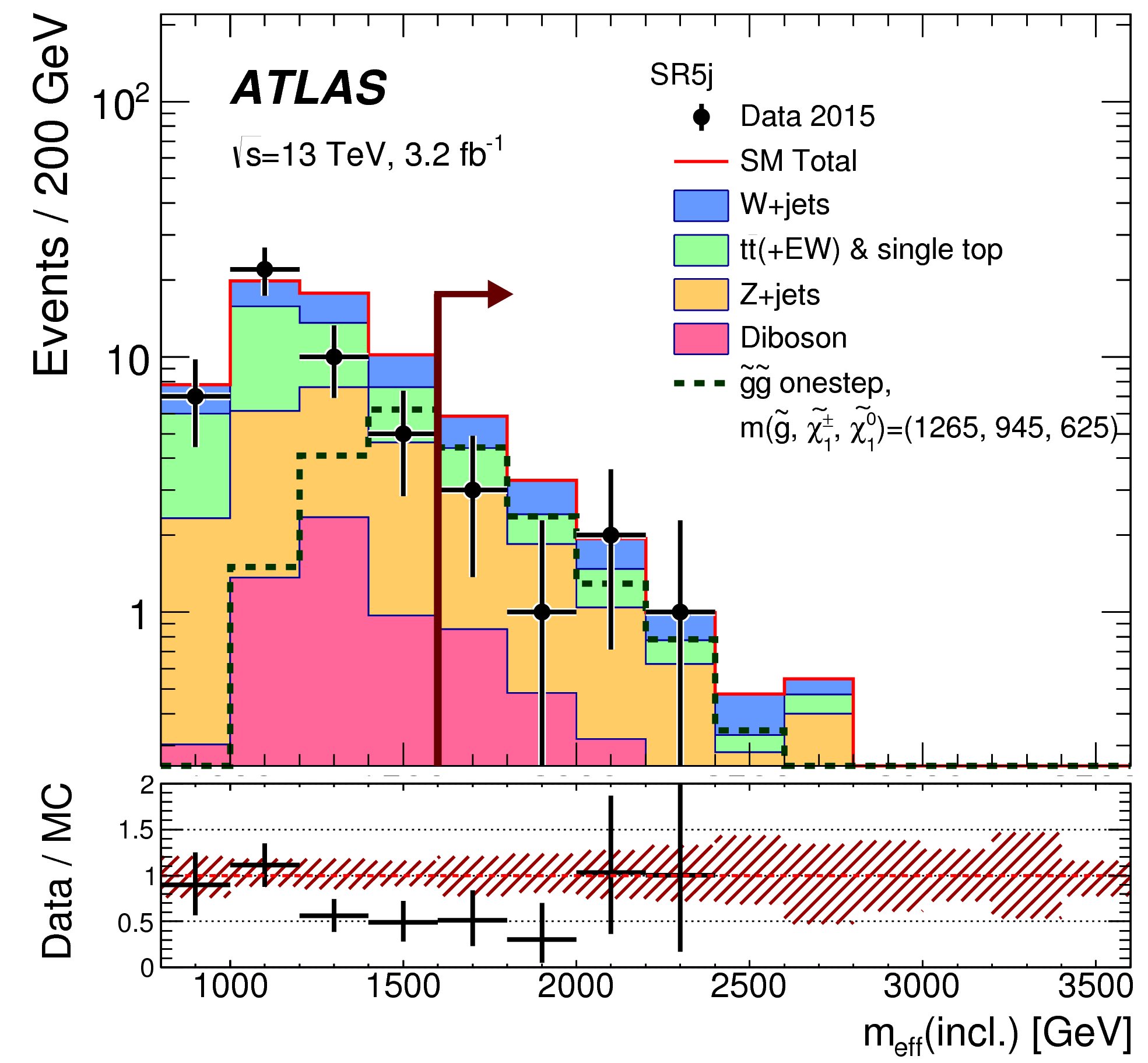}
    \includegraphics[width=5cm]{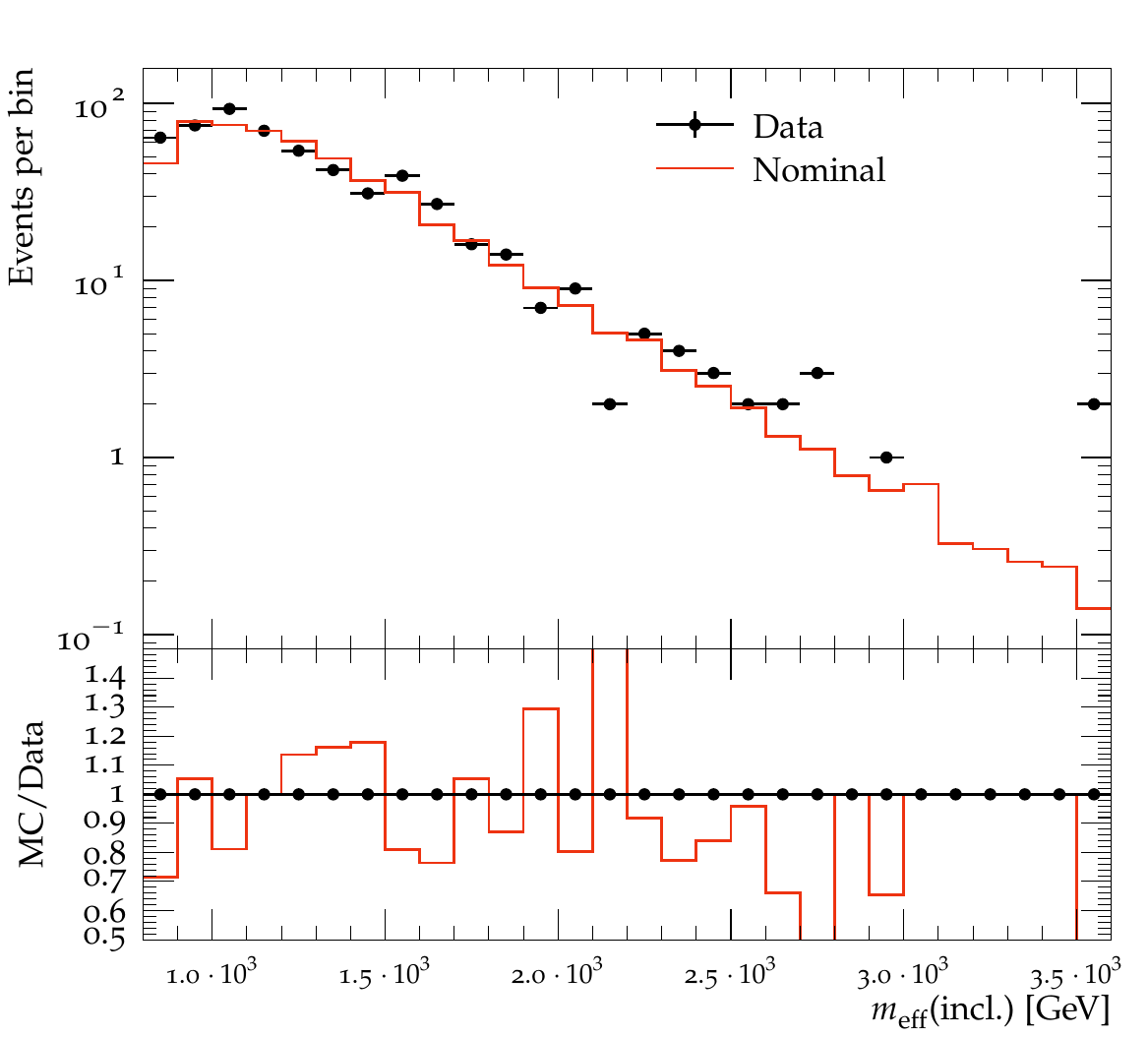}
    \includegraphics[width=5cm]{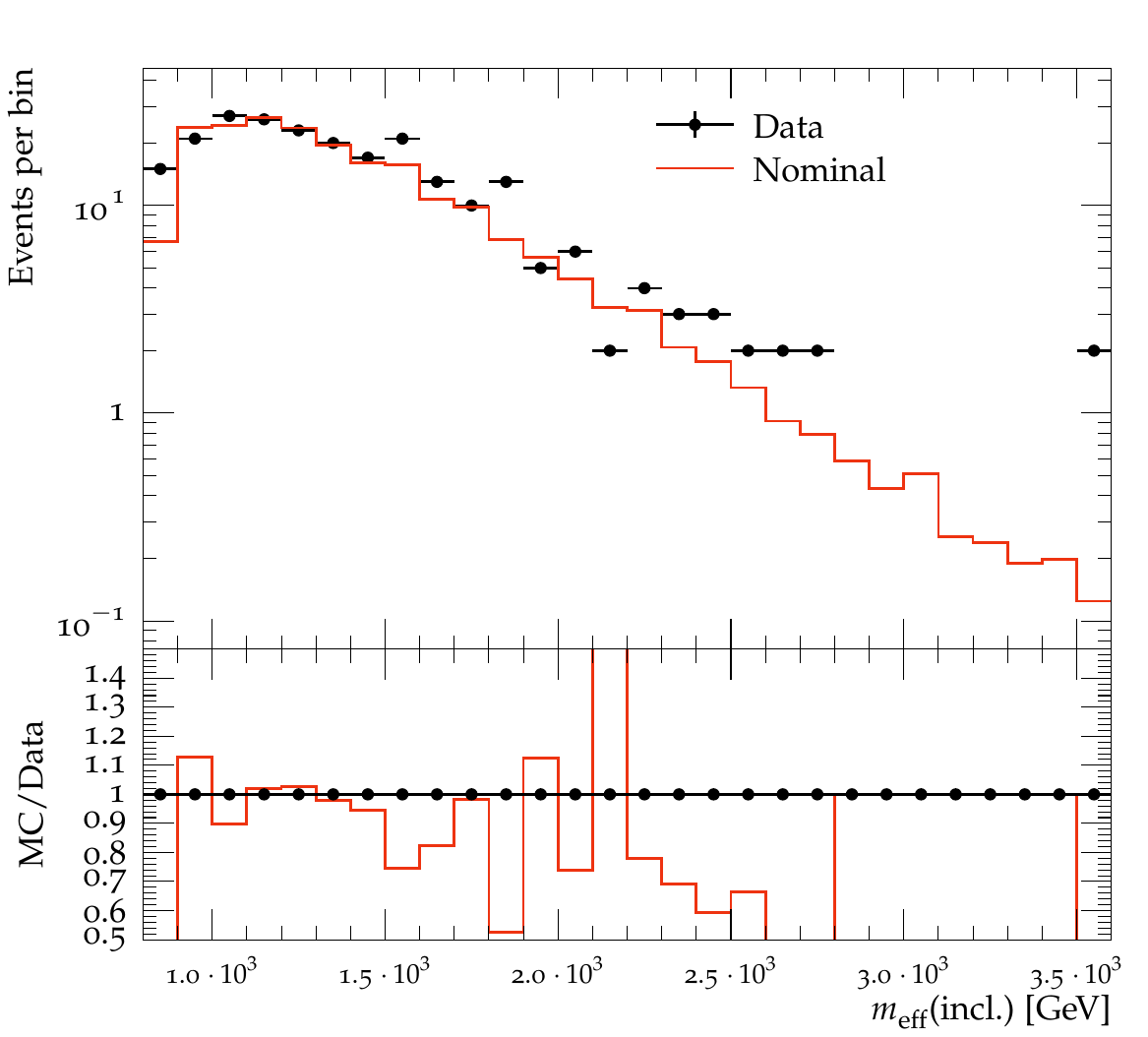}
    \includegraphics[width=5cm]{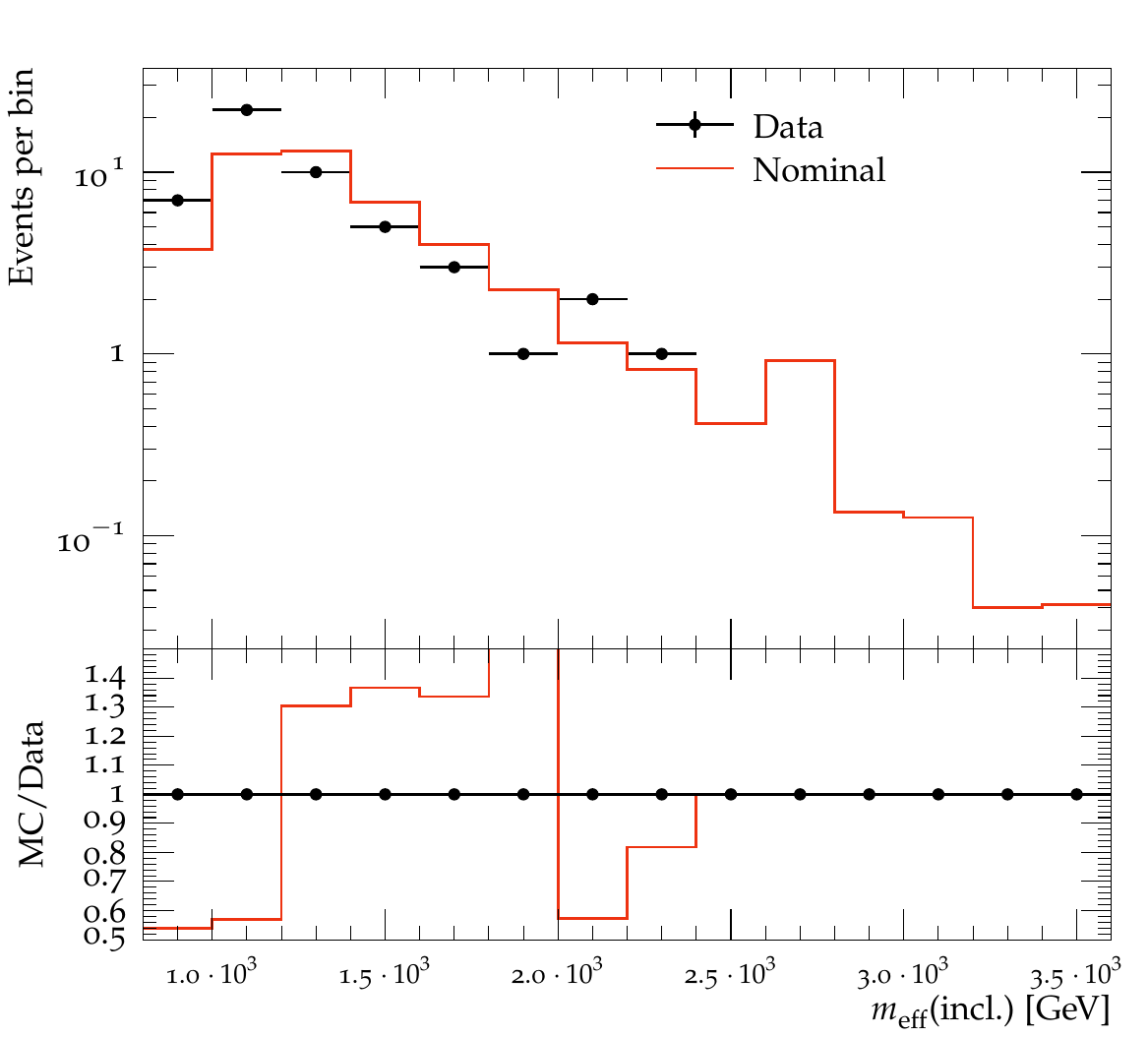}
    \caption{Examples of some of the $m_\textrm{eff}$ distributions presented in the ATLAS zero-lepton SUSY search~\cite{Aaboud:2016zdn} (top row), and comparisons of matching data and smeared MC simulation predictions obtained from the corresponding \Rivet{} routine (bottom row). ``Nominal'' in the second row subfigures refers to the sum of the dominant SM processes which contribute, as described in the main body of the text.}
    \label{fig:SearchMeasurementComplementarity:SUSYPaperMEffPlots}
  \end{center}
\end{figure}

The $3.2 \textrm{~fb}^{-1}$ zero-lepton SUSY search described in Ref.~\cite{Aaboud:2016zdn} has seven signal regions (SRs), which were designed to target three different possibilities for the all-hadronic decay of squarks and gluinos in simplified SUSY models that assume $R$-parity conservation. All the SRs require exactly zero reconstructed electrons or muons with  momentum above 10~GeV, and at least 200~GeV of missing transverse energy. The selection defining the different SRs then imposes the presence of 2 to 6 jets with varying transverse momentum requirements and imposes a region-specific threshold on the inclusive effective mass, $m_\textrm{eff}$, defined as the scalar sum of the transverse momentum of all reconstructed jets and the missing transverse energy. The distributions of the SM background as a function of $m_\textrm{eff}$ are included the corresponding HEPData entry\footnote{Tables 4--10 of \url{https://www.hepdata.net/record/ins1458270}} for the search.
A few examples are shown in the top row of Figure~\ref{fig:SearchMeasurementComplementarity:SUSYPaperMEffPlots}. These background distributions are precisely what is needed as an output from a \Rivet{} routine in order to use them in an automated framework like \Contur{}.

Replicating the analysis selection at the detector-level requires that the truth-level energies of these objects is smeared in a way that mimics the ATLAS detector reconstruction and object identification performance. Thankfully, since the objects used in this search are ``standard'' ATLAS objects, the built-in jet, missing energy and lepton momentum smearing efficiencies, which have recently been introduced in \Rivet{}~\cite{Buckley:2019stt}, can be used for this purpose. 

It was therefore possible to construct a \Rivet{} routine dedicated for this search\footnote{\url{https://rivet.hepforge.org/analyses/ATLAS_2016_I1458270}} using the smearing functions like:
\begin{lstlisting}
// Initialise and register projections
FinalState calofs(Cuts::abseta < 4.8);
FastJets fj(calofs, FastJets::ANTIKT, 0.4);
declare(fj, "TruthJets");
declare(SmearedJets(fj, JET_SMEAR_ATLAS_RUN2,
          JET_BTAG_ATLAS_RUN2_MV2C20), "RecoJets");

MissingMomentum mm(calofs);
declare(mm, "TruthMET");
declare(SmearedMET(mm, MET_SMEAR_ATLAS_RUN2), "RecoMET");

PromptFinalState es(Cuts::abseta < 2.47 &&
                    Cuts::abspid == PID::ELECTRON, true, true);
declare(es, "TruthElectrons");
declare(SmearedParticles(es, ELECTRON_RECOEFF_ATLAS_RUN2,
          ELECTRON_SMEAR_ATLAS_RUN2), "RecoElectrons");

// + similar for muons
\end{lstlisting}
This \Rivet routine has since been incorporated into the main \Rivet{} distribution.
Its performance was validated separately for the signal and SM background.
For the SUSY signal, it was possible to check that the analysis logic and smearing was working correctly using the sequential selections included in the auxiliary material attached to the original publication\footnote{https://atlas.web.cern.ch/Atlas/GROUPS/PHYSICS/PAPERS/SUSY-2015-06/}, in auxiliary Tables~1-7, or equivalently, in the HEPData entry.
For the SM background, the performance of the routine was checked by investigating the properties of $V$ plus jets events generated with \Sherpa{}~\cite{Bothmann:2019yzt} and top-antitop events generated with \Powheg\ and \Pythia\ \cite{Frixione:2007vw,Sjostrand:2014zea}. The comparison of the resulting distributions to those observed in data are shown in the bottom row of Figure~\ref{fig:SearchMeasurementComplementarity:SUSYPaperMEffPlots}.

This exercise demonstrates that approximate detector-level distributions from LHC searches can be reproduced using truth-level inputs and smearing functions in a \Rivet{} routine.

\subsection{Example 2: a dilepton ``bump-hunt'' search preserved in \Rivet{}}
\label{sec:SearchMeasurementComplementarity:subsection2}

In the high-mass di-lepton resonance search using $139 \textrm{~fb}^{-1}$ of 13~TeV data collected with the ATLAS detector described in Ref.~\cite{Aad:2019fac}, events are required to contain two same-flavour leptons, with an invariant mass $m_{ee}$ larger than 225~GeV. The background estimation is achieved using a fit to the observed data passing the analysis selection. 
Unlike the previous example, it is not possible to simply use default smearing functions to construct a \Rivet{} routine, since the objects used in the search are ``non-standard'', and the final background estimation uses a data-driven technique. However, the analysis team used simulated events to extract a resolution function from the fit. This resolution function was provided in the auxiliary material for the publication\footnote{https://atlas.web.cern.ch/Atlas/GROUPS/PHYSICS/PAPERS/EXOT-2018-08/} in Figures 10 and 11, and in the associated HEPData entry\footnote{https://www.hepdata.net/record/ins1725190?version=1} in Tables 23 and 24.

\begin{figure}[t]
  \begin{center}
    \includegraphics[width=7cm]{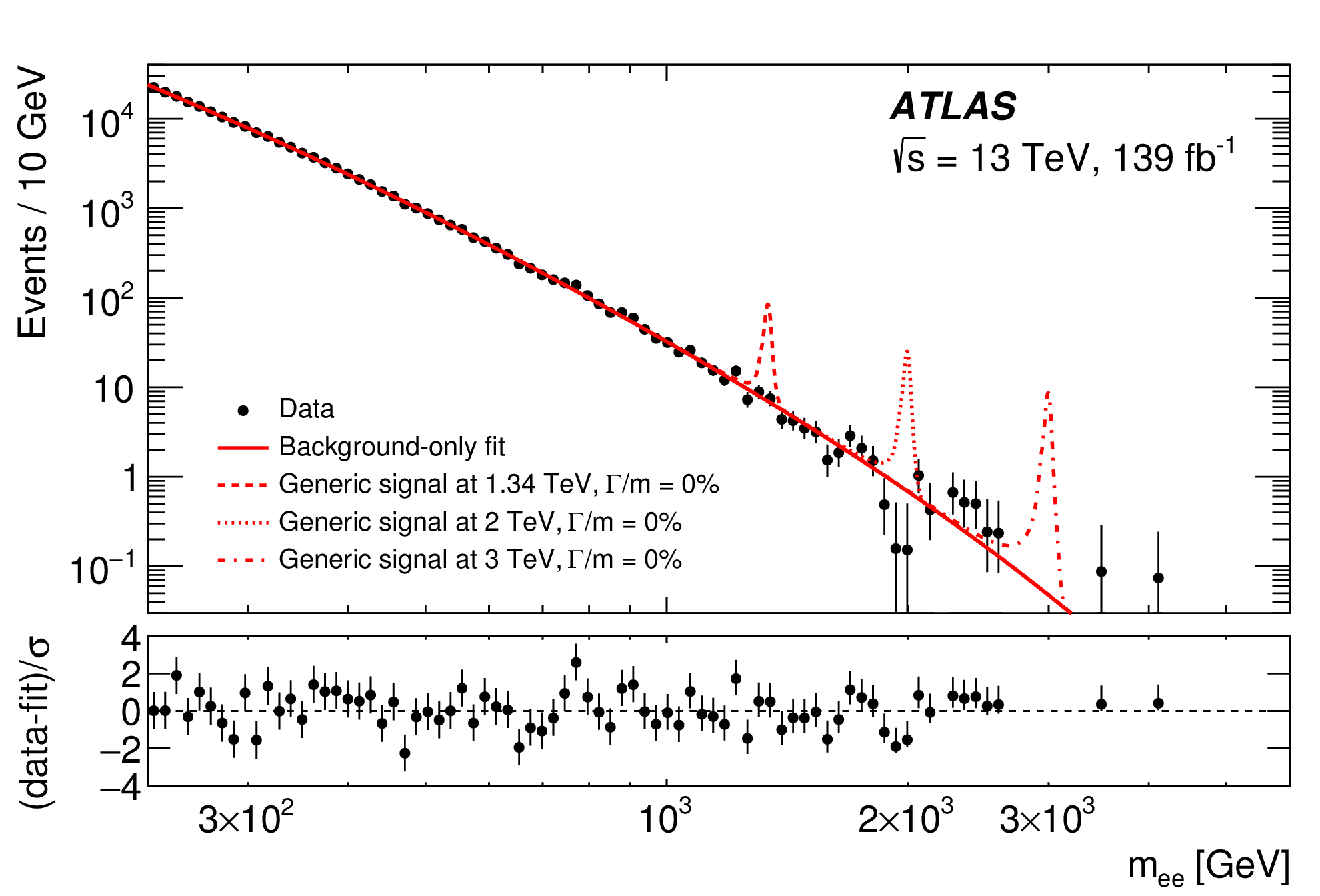}
    \includegraphics[width=7cm]{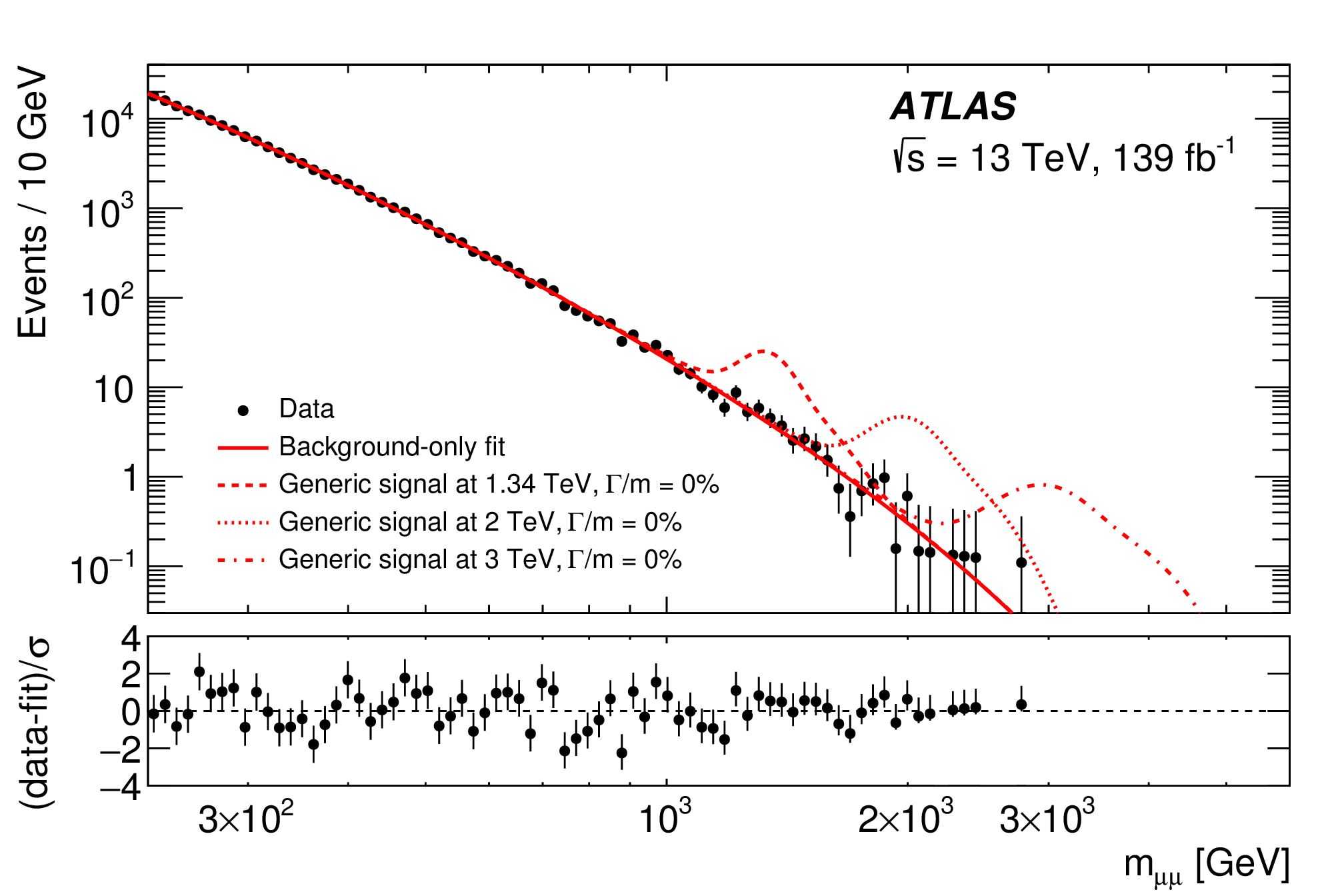}\\
    \includegraphics[width=7cm]{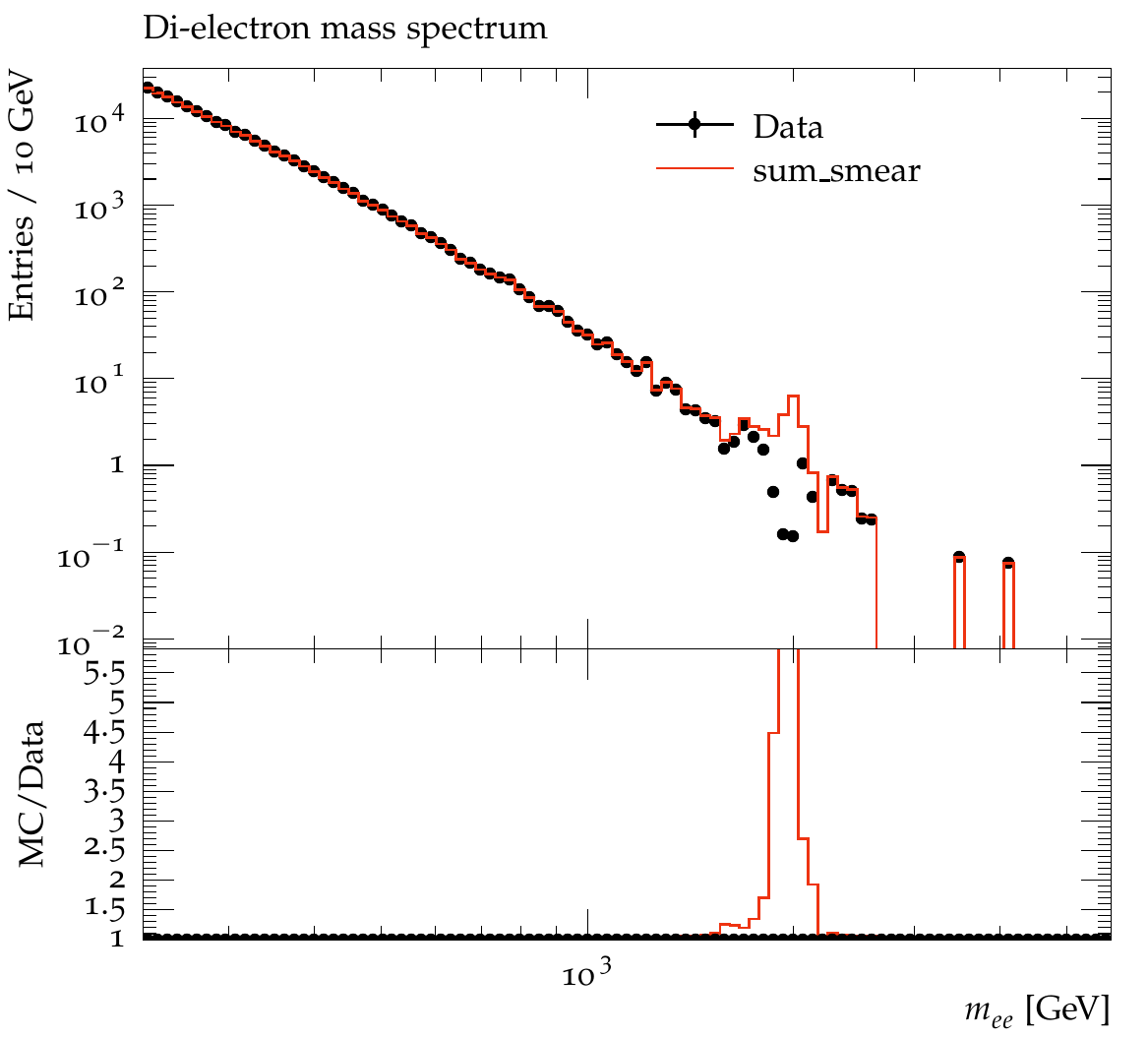}
    \includegraphics[width=7cm]{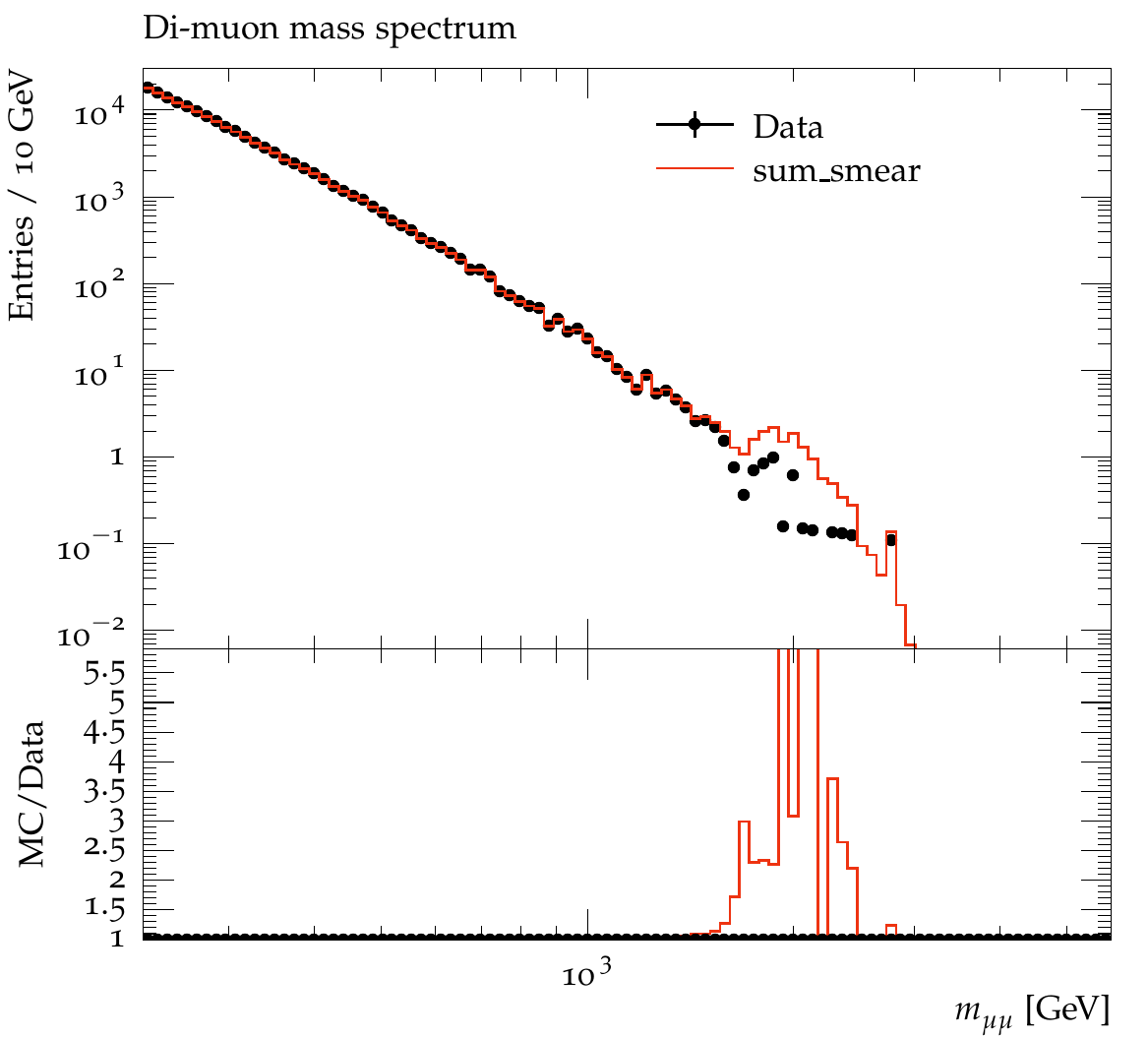}
    \caption{The dilepton invariant mass distributions in the SRs of the ATLAS dilepton resonance search of Ref.~\cite{Aad:2019fac} (top row), and examples of a simple $Z'$ signal generated with \Pythia{} ($m_{Z'} = 2$~TeV) after smearing in the \Rivet{} routine (bottom row). The ``sum\_smear'' tag in the bottom row subfigures refers to the smeared $Z'$ signal super-imposed on the observed background.}
    \label{fig:SearchMeasurementComplementarity:DYPaperMllPlots}
  \end{center}
\end{figure}

The background distributions in the di-electron and di-muon channels from the original experimental article can be seen in Figure~\ref{fig:SearchMeasurementComplementarity:DYPaperMllPlots}. The \Rivet{} routine needs to be able to replicate these detector-level background shapes using truth-level inputs. In order to do this, the resolution functions from the original publication are integrated into the \Rivet{} routine\footnote{https://rivet.hepforge.org/analyses/ATLAS\_2019\_I1725190.  Thanks to Peter Falke for assistance with ATLAS dilepton resolution parametrisations.} in the following way,

{\scriptsize%
\begin{lstlisting}
// Smear the dilepton mass with a CB + Gauss function
double muCB, sigCB, alpCB, nCB, muG, sigG, kappa;

// for electrons (similar for muons) 
const double lnm = log(mll);

static const vector<double> pmuCB = {0.13287, -0.410663, -0.0126743, 2.9547e-6};
muCB = pmuCB[0] + pmuCB[1]/lnm + pmuCB[2]*lnm + pmuCB[3]*pow(lnm, 4);

static const vector<double> psigCB = {0.0136624, 0.230678, 1.73254};
sigCB = sqrt(pow(psigCB[0],2) + pow(psigCB[1],2)/mll + pow(psigCB[2]/mll, 2));

alpCB = 1.59112;

static const vector<double> pnCB = {1.13055, 0.76705, 0.00298312};
nCB = pnCB[0] + pnCB[1]*exp(-pnCB[2]*mll);

static const vector<double> pmuG = {-0.00402708, 0.814172, -3.94281e-7, 7.97076e-6, -87.6397,
   -1.64806e-11};
muG = pmuG[0] + pmuG[1]/mll + pmuG[2]*mll + pmuG[3]*pow(lnm,3) + pmuG[4]/sqr(mll) +
    pmuG[5]*sqr(mll);

static const vector<double> psigG = {0.00553858, 0.140909, 0.644418};
sigG = sqrt(sqr(psigG[0]) + sqr(psigG[1])/mll + sqr(psigG[2]/mll));

static const vector<double> pkappa = {0.347003, 0.135768, 0.00372497, -2.2822e-5,
    5.06351e-13};
kappa = pkappa[0] + pkappa[1]*exp(-pkappa[2]*mll) + pkappa[3]*mll +  pkappa[4]*pow(mll,3);

static const vector<double> pkappa =  {0.347003, 0.135768, 0.00372497, -2.2822e-5, 5.06351e-13};
kappa = pkappa[0] + pkappa[1]*exp(-pkappa[2]*mll)  + pkappa[3]*mll + pkappa[4]*pow(mll,3);

// Smearing using calculated params(mll)
double mll_scale = -1;
do {
  mll_scale = (rand01() > kappa) ? randnorm(muG, sigG) : randcrystalball(alpCB,  nCB, muCB,
     sigCB);
} while (fabs(mll_scale) > 0.5);
const double mll_reco = (1 + mll_scale) * mll;
\end{lstlisting}
}%

The performance of the implemented smearing was checked by comparing the number of events passing the selection for a simple BSM model to those reported in the original article. 
An example of a 2~TeV $Z'$ signal after smearing can be seen in Figure~\ref{fig:SearchMeasurementComplementarity:DYPaperMllPlots}.
This exercise shows that even for non-standard objects, it is possible to include detector-level results into \Rivet{} and get good agreement with the fully-reconstructed results. 


\section{Comparison of performance of searches and measurements using \Contur}
\label{sec:SearchMeasurementComplementarity:section4}

The SUSY search mentioned in Section~\ref{sec:SearchMeasurementComplementarity:subsection1} probes final states featuring jets, missing energy and no lepton using $3.2 \textrm{~fb}^{-1}$ of 13~TeV data.
ATLAS has also published a measurement that is sensitive to BSM physics in the same final state. Specifically, this is a detector-corrected measurement of the ratio of fiducial cross sections for missing energy and jets, and dilepton pairs and jets~\cite{Aaboud:2017buf}. This measurement was also made using the same dataset ($3.2 \textrm{~fb}^{-1}$ at 13~TeV) as the SUSY search. Since the measurement was unfolded to particle level, it already had a \Rivet{} routine that was included within \Contur{}. This measurement will be refered to as the ``jetMET'' measurement hereafter.
Since the two analyses probe the same final state and use the same dataset, one would expect that they give similar sensitivity to new models when used for re-interpretation.

\begin{figure}[t]
\begin{center}
\includegraphics[width=0.8\textwidth]{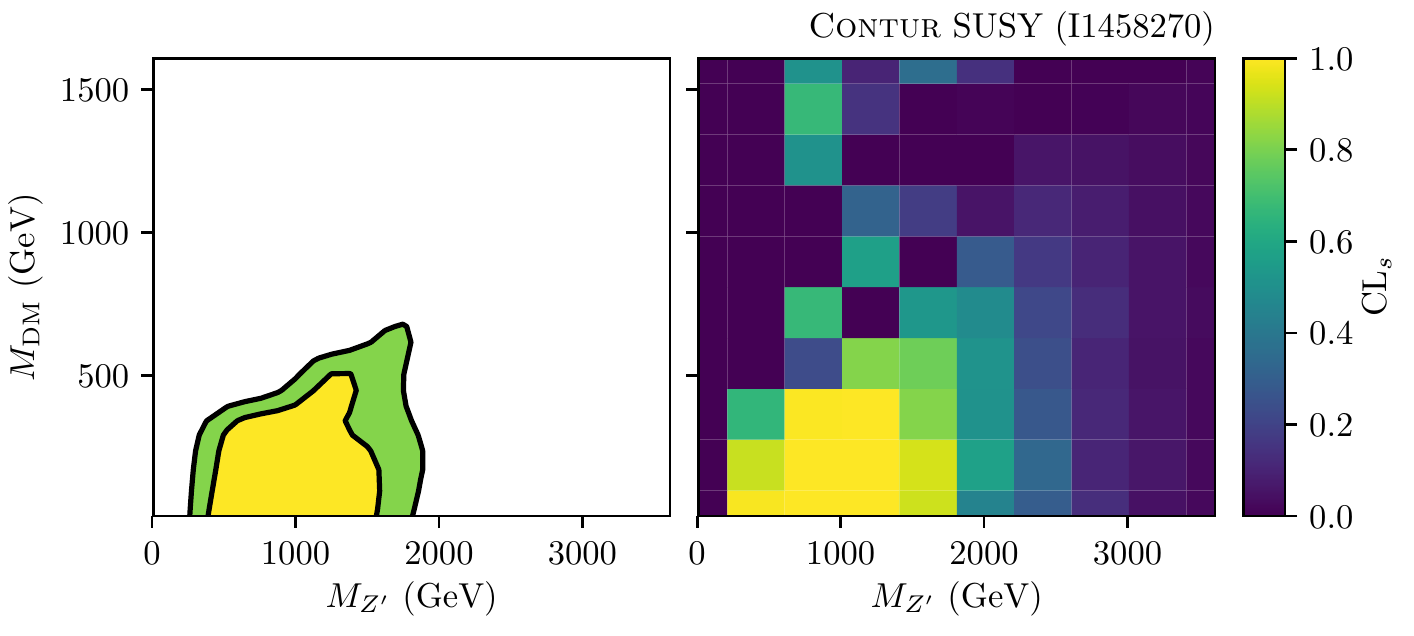}
\includegraphics[width=0.8\textwidth]{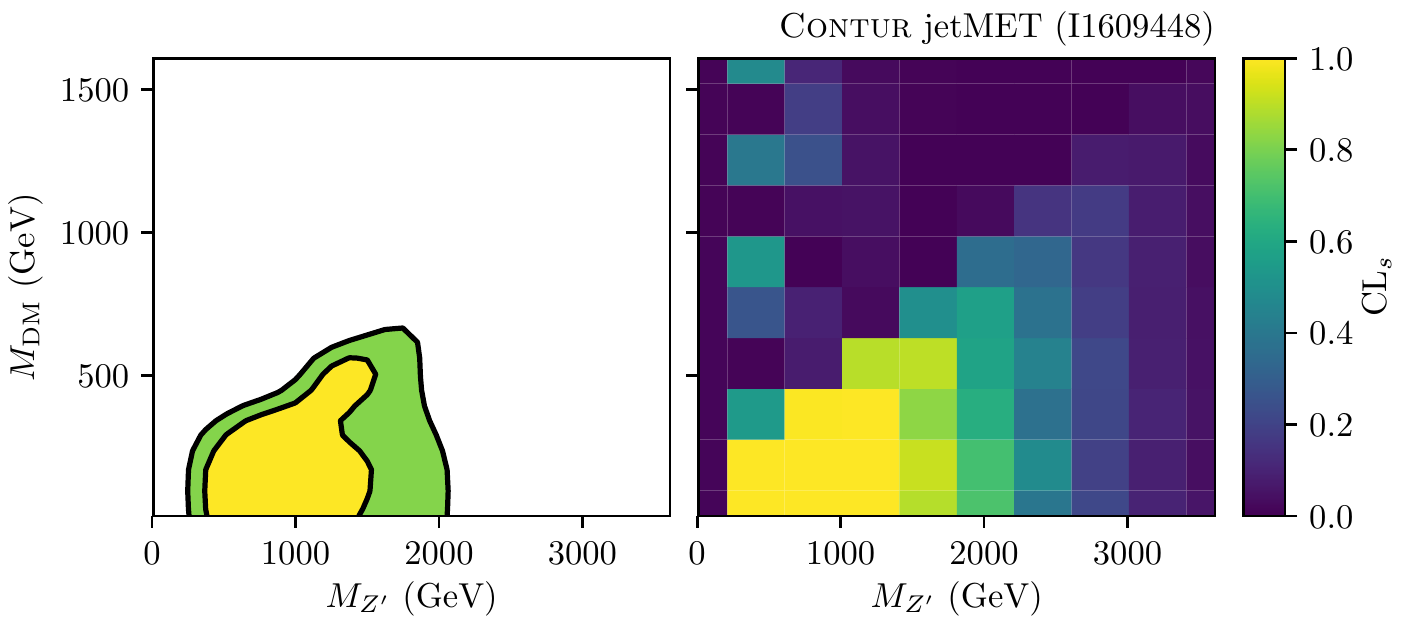}
\caption{Comparison of the \Contur{} exclusions for a benchmark dark matter model from an existing routine for a SM measurement of the jet+MET final state (lower panel), and for the SUSY routine described above (upper panel). The two analyses probe the same final state and use the same $3.2 \textrm{~fb}^{-1}$ of 13~TeV data. They are found to provide similar sensitivity to the benchmark DM model.}
\label{fig:SearchMeasurementComplementarity:SUSY_vs_jetMet}
\end{center}
\end{figure}

A benchmark dark matter model was chosen to compare the performance of the search and measurement of the jets+MET final state. The parameters were chosen as in Ref.~\cite{Kahlhoefer:2015bea}, and a scan of the exclusion from each routine as a functions of the mass of the dark matter (DM) candidate ($M_{DM}$) and the mass of the $Z'$ particle ($M_{Z'}$) predicted by the model was performed. The limits are calculated using the CL$_{s}$ method~\cite{Read:2002hq}. The results of both exclusion scans are shown in Figure~\ref{fig:SearchMeasurementComplementarity:SUSY_vs_jetMet}.
They show that the SUSY search, which was performed at detector-level, and whose \Rivet{} routine was written using \Rivet{}'s built in smearing and efficiency tables, performs roughly as well as the jetMET measurement which was unfolded the particle-level.
This demonstrates that searches and measurements of similar final states can both be used in machinery like \Contur{} to provide fast and reliable re-interpretation of results when compared to new physics models.

\section{Using search routines in \Contur to probe new models}
\label{sec:SearchMeasurementComplementarity:section3}

\begin{figure}[t]
\begin{center}
\includegraphics[width=0.8\textwidth]{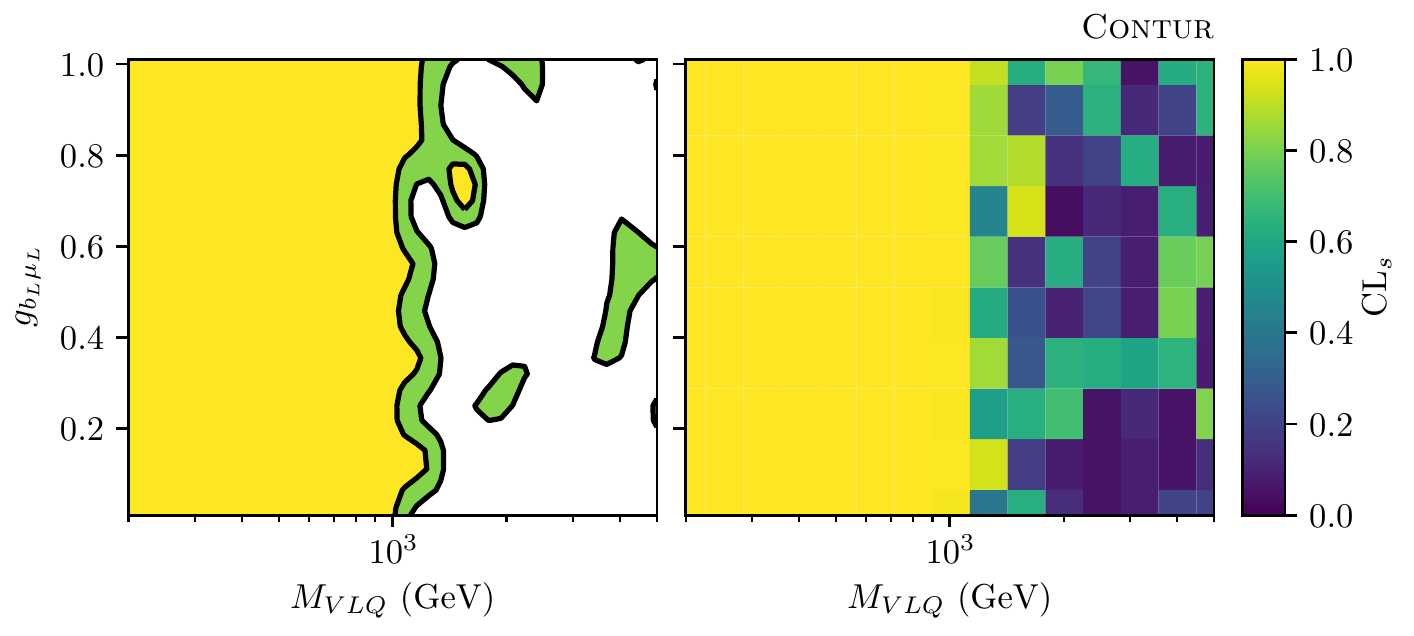}
\includegraphics[width=0.8\textwidth]{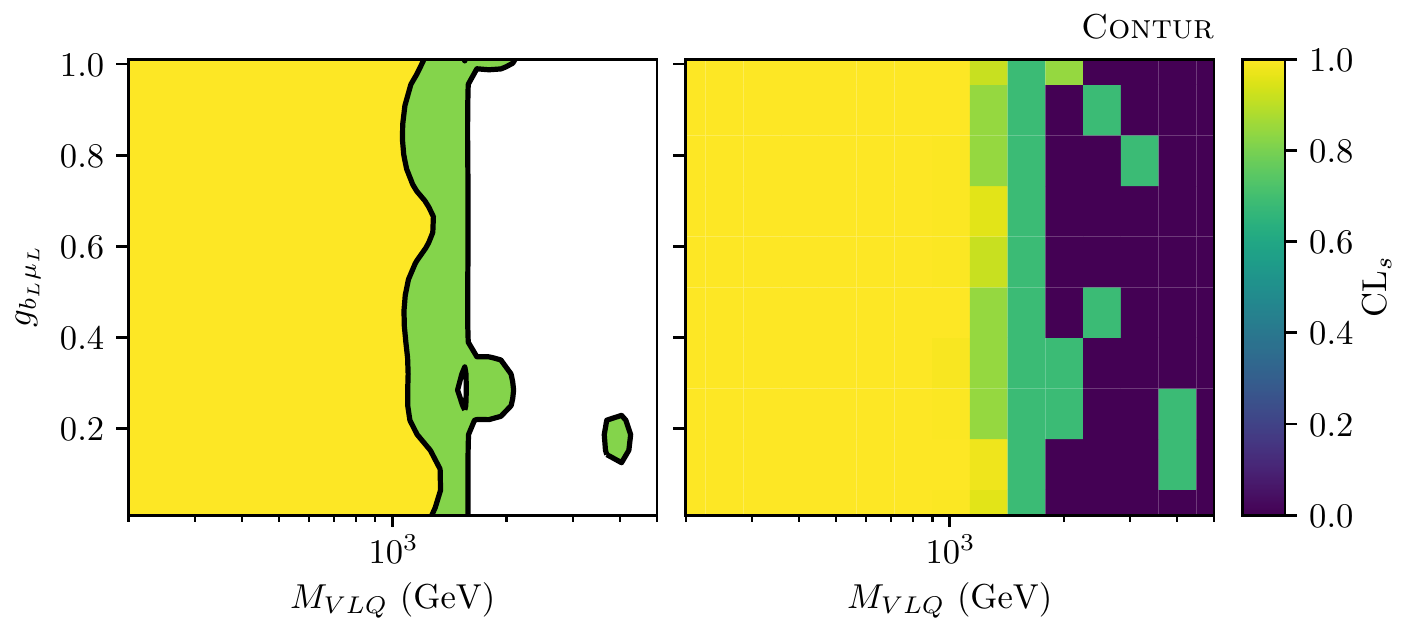}
\caption{Comparison of the \Contur{} exclusions for a LQ model from the 13~TeV measurements already in \Contur{} (top row) and the 13~TeV high mass di-lepton resonance search (bottom row).}
\label{fig:SearchMeasurementComplementarity:HMDY_VLQ}
\end{center}
\end{figure}

The high-mass di-lepton resonance routine was used to check the exclusion for a model which includes a vector leptoquark (LQ) singlet~\cite{Dorsner:2018ynv}. This type of model has been invoked to explain the observed anomalies in the $B$-sector~\cite{Buttazzo:2017ixm}.
The set of parameters where the leptoquark decays to second-generation fermions (with coupling $g_{b_{L}\mu_{L}}$) was tested.
The \Contur{} scan was performed as a function of the LQ mass ($M_{VLQ}$), and the results are shown in Figure~\ref{fig:SearchMeasurementComplementarity:HMDY_VLQ} for for all 13~TeV measurements (top row) and just the high-mass di-lepton resonance search (bottom row). The sensitivity of the measurements to this model comes from ATLAS and CMS measurements of $V$+jets and $t\bar{t}$ processes, as well as differential measurements of the jet production cross-sections. The search is found to be slightly more constraining than the measurements at high LQ masses, and is therefore largely complementary with the measurements to help exclude new models in \Contur{}.

\section{Complementarity in SUSY parameter scans}
\label{sec:SearchMeasurementComplementarity:gambit}

As with many BSM searches at the LHC, searches for electroweakinos (neutralinos and charginos) are usually based on \textit{simplified models}. These models use a reduced set of phenomenological parameters to simplify the complex task of optimizing the search and interpreting the results. However, this leaves open the question of what implications the search results have for the parameter spaces of more complete SUSY models. In \cite{Athron:2018vxy}, the \Gambit global fitting framework, and in particular the \textsc{ColliderBit} module, was used to re-interpret several ATLAS and CMS searches~\cite{Aaboud:2018jiw,Aaboud:2018sua,Aaboud:2018htj,Aaboud:2018zeb,CMS:2017fth,Sirunyan:2018iwl,Sirunyan:2017qaj,CMS-PAS-SUS-16-039} in the context of the full electroweakino sector of the Minimal Supersymmetric Standard Model (MSSM). This model, called the EWMSSM, is defined by four free parameters: $M_1$, $M_2$, $\mu$ and $\tan\beta$, which control the tree-level masses and mixings of the four neutralinos ($\tilde{\chi}^0_1$, $\tilde{\chi}^0_2$, $\tilde{\chi}^0_3$, $\tilde{\chi}^0_4$) and two charginos ($\tilde{\chi}^\pm_1$, $\tilde{\chi}^\pm_2$) of the MSSM. The gluino, sfermions and additional Higgs bosons are decoupled by setting the relevant MSSM mass parameters to several TeV.



\begin{figure}[t]
\begin{center}
\includegraphics[width=0.32\textwidth]{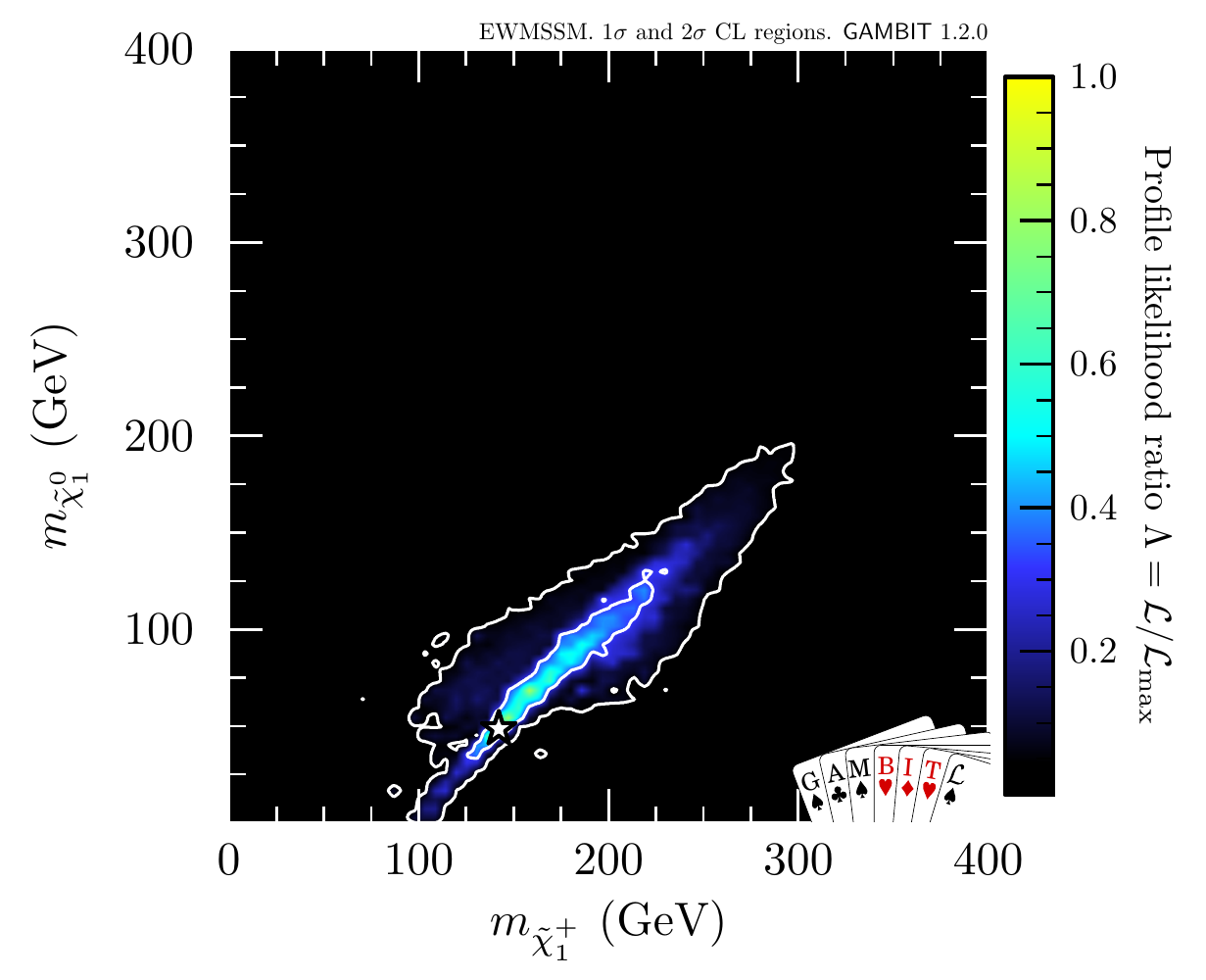}
\includegraphics[width=0.32\textwidth]{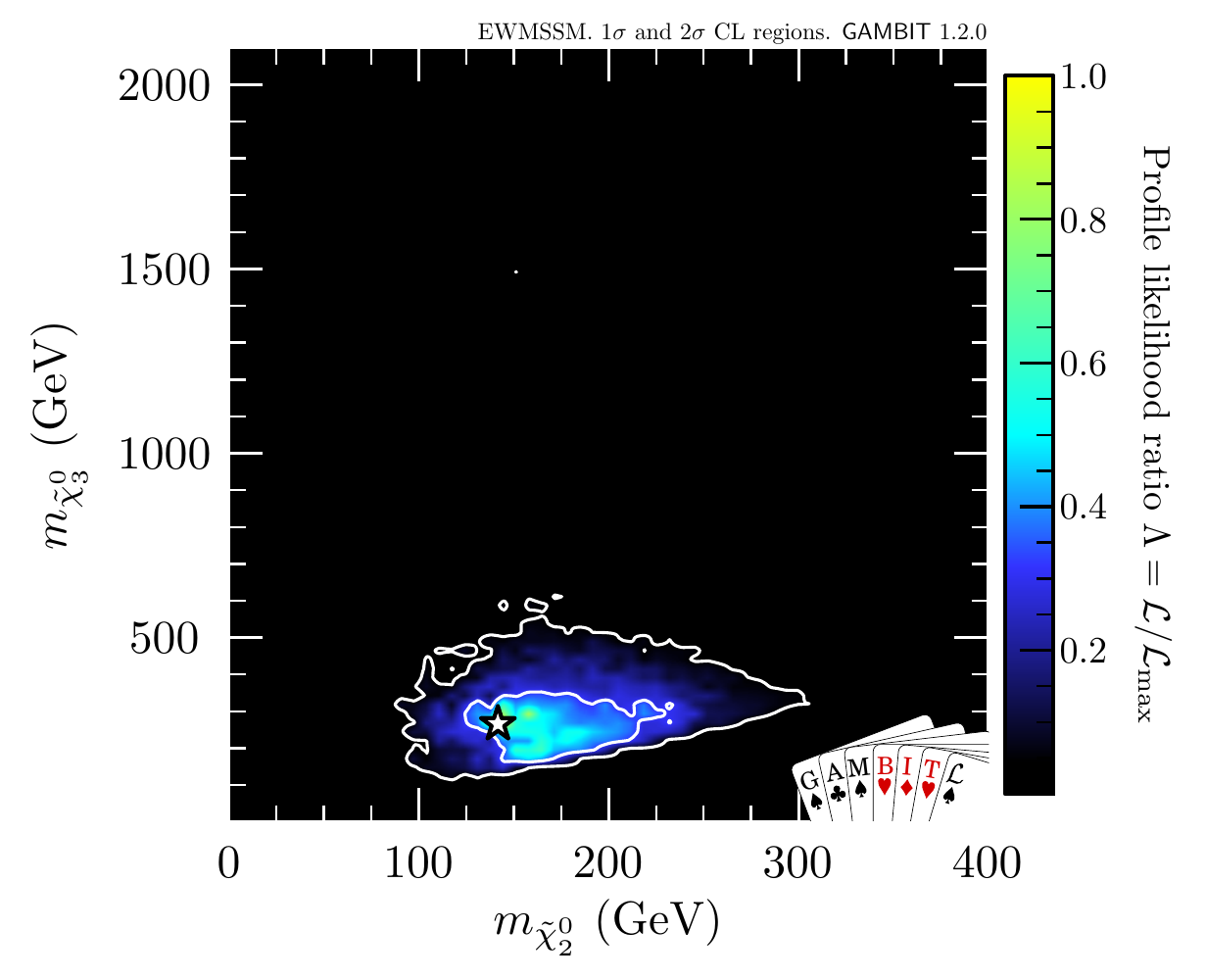}
\includegraphics[width=0.32\textwidth]{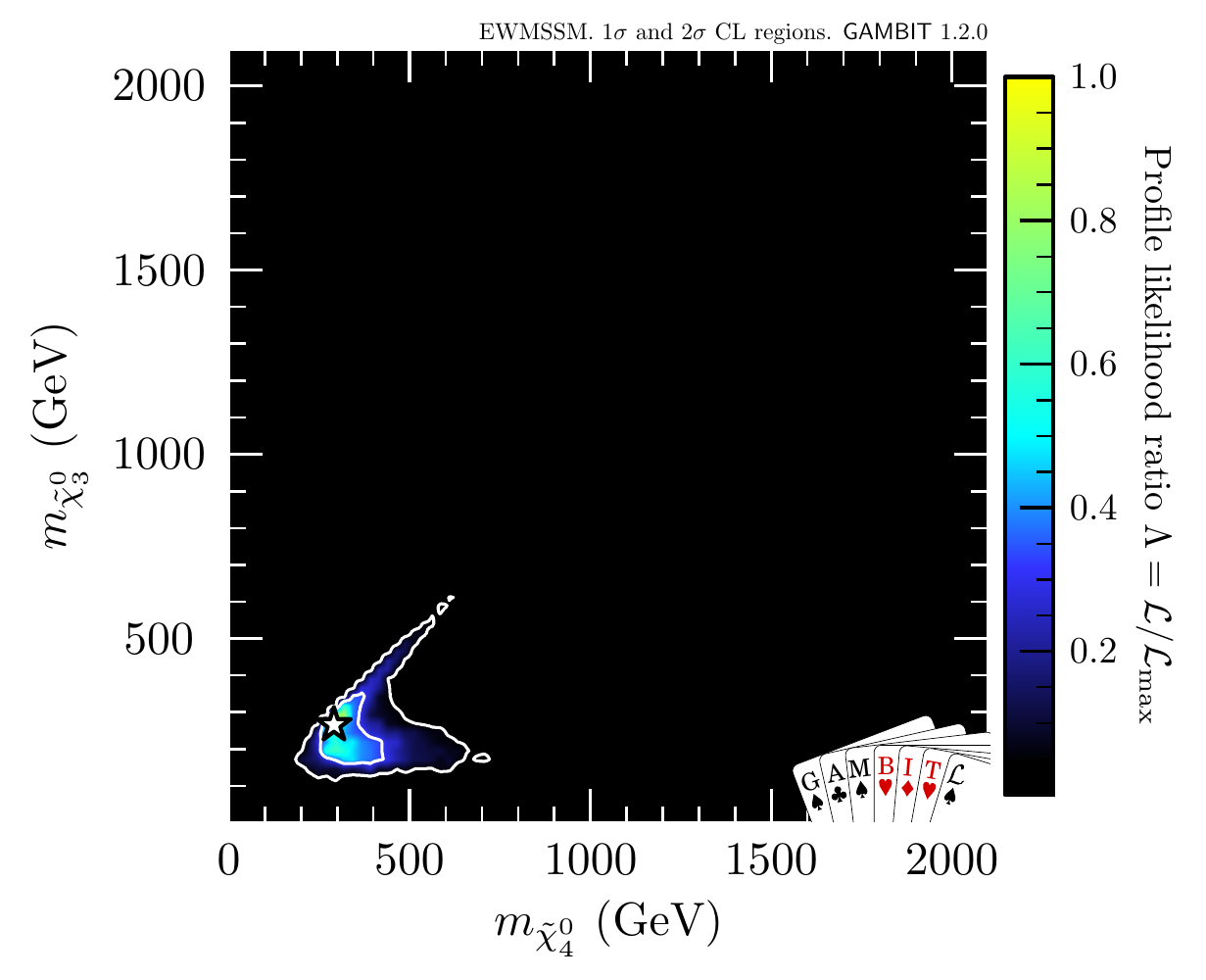}
\caption{Results from the combination and re-interpretation of LHC 13 TeV searches in a global fit of the EWMSSM, from \cite{Athron:2018vxy}. The figures show the profile likelihood ratio $\mathcal{L}/\mathcal{L}_\textrm{max}$ in the ($m_{\tilde{\chi}^{\pm}_1},m_{\tilde{\chi}^0_1}$) plane (left), the ($m_{\tilde{\chi}^0_2},m_{\tilde{\chi}^0_3}$) plane (middle), and the ($m_{\tilde{\chi}^0_4},m_{\tilde{\chi}^0_3}$) plane (right). The likelihood function used in the fit combines the likelihood contributions from all the re-interpreted ATLAS and CMS searches, as well as likelihood contributions from invisible Higgs and Z boson decays. $\mathcal{L}_\textrm{max}$ is the likelihood of the global EWMSSM best-fit point, while $\mathcal{L}$ is the highest EWMSSM likelihood for the given point in mass space. The white star marks the location of the best-fit point and the white contours show the $1\sigma$ and $2\sigma$ preferred regions relative to this point.
}
\label{fig:SearchMeasurementComplementarity:gambitresults}
\end{center}
\end{figure}

The analysis in \cite{Athron:2018vxy} shows that, in the MSSM, the masses of the charginos and neutralinos are only weakly constrained by the combination of electroweakino searches with $36$\textrm{~fb}$^{-1}$ of data at $\sqrt{s}=13$\ TeV. The loose constraints are in part a consequence of the parametric freedom of the EWMSSM compared to the simplified models, and partly due to a series of small data excesses that can offset tensions in other analyses. In fact, these small excesses conspire to produce a preferred subregion of the EWMSSM parameter space, at the $3\sigma$ level, compared to the SM-only expectation. The resulting best-fit EWMSSM subspace is shown in Figure~\ref{fig:SearchMeasurementComplementarity:gambitresults}, in terms of profile likelihood figures for different electroweakino mass planes. The $2\sigma$ region (relative to the best-fit point) predicts mass spectra with all the electroweakinos lighter than $\sim700$\ GeV. Further, the predicted spectra feature two mass gaps larger than $m_Z$, between the dominantly bino $\tilde{\chi}^0_1$ and the mostly wino (Higgsino) $\tilde{\chi}^0_2$/$\tilde{\chi}^\pm_1$, and between $\tilde{\chi}^0_2$/$\tilde{\chi}^\pm_1$ and the dominantly Higgsino (wino) $\tilde{\chi}^0_4$/$\tilde{\chi}^\pm_2$. The first of these mass gaps is evident in the first panel of Figure~\ref{fig:SearchMeasurementComplementarity:gambitresults}, where the $1\sigma$ region picks out the $m_{\tilde{\chi}^{\pm}_1} - m_{\tilde{\chi}^0_1} \sim m_Z$ region. The preference for these mass gaps comes from the fact that the fitted excesses are seen in searches that target SUSY signals with on-shell $Z$ and $W$ bosons.



\begin{figure}[t]
\begin{center}
\includegraphics[width=0.32\textwidth]{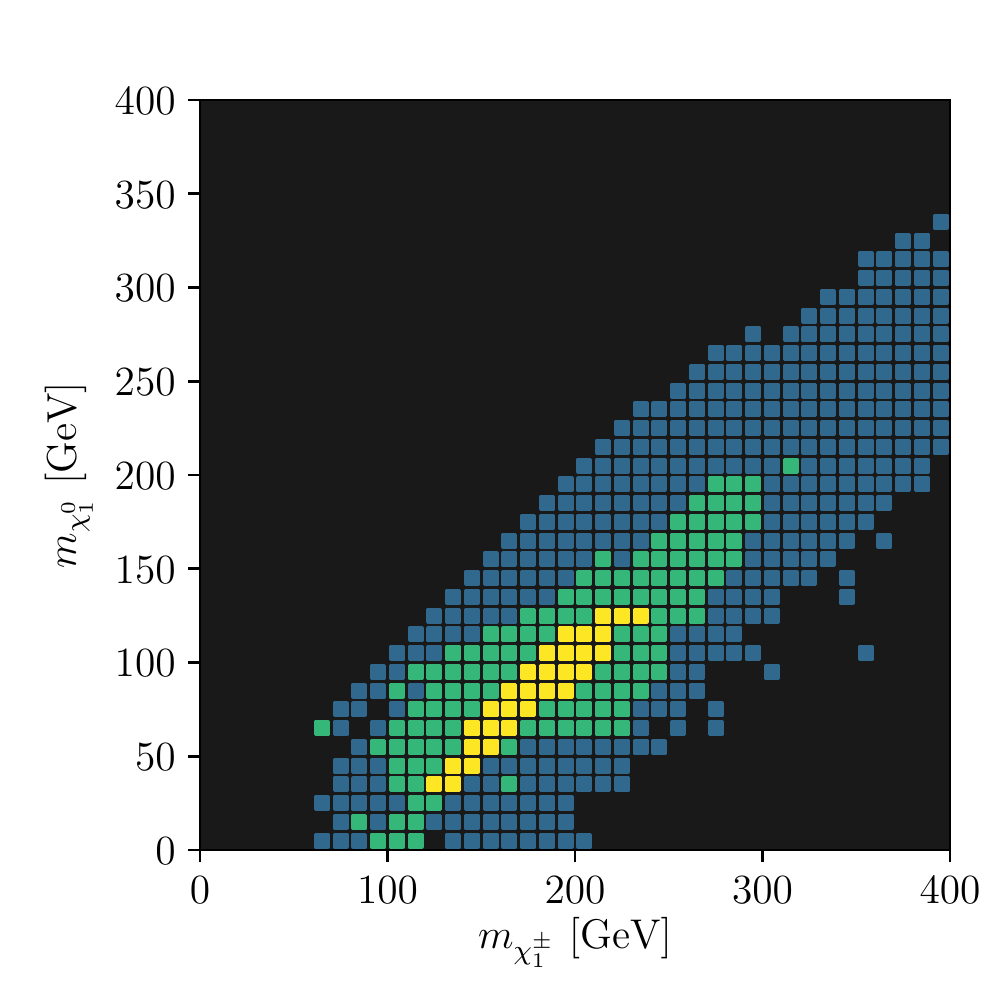}
\includegraphics[width=0.32\textwidth]{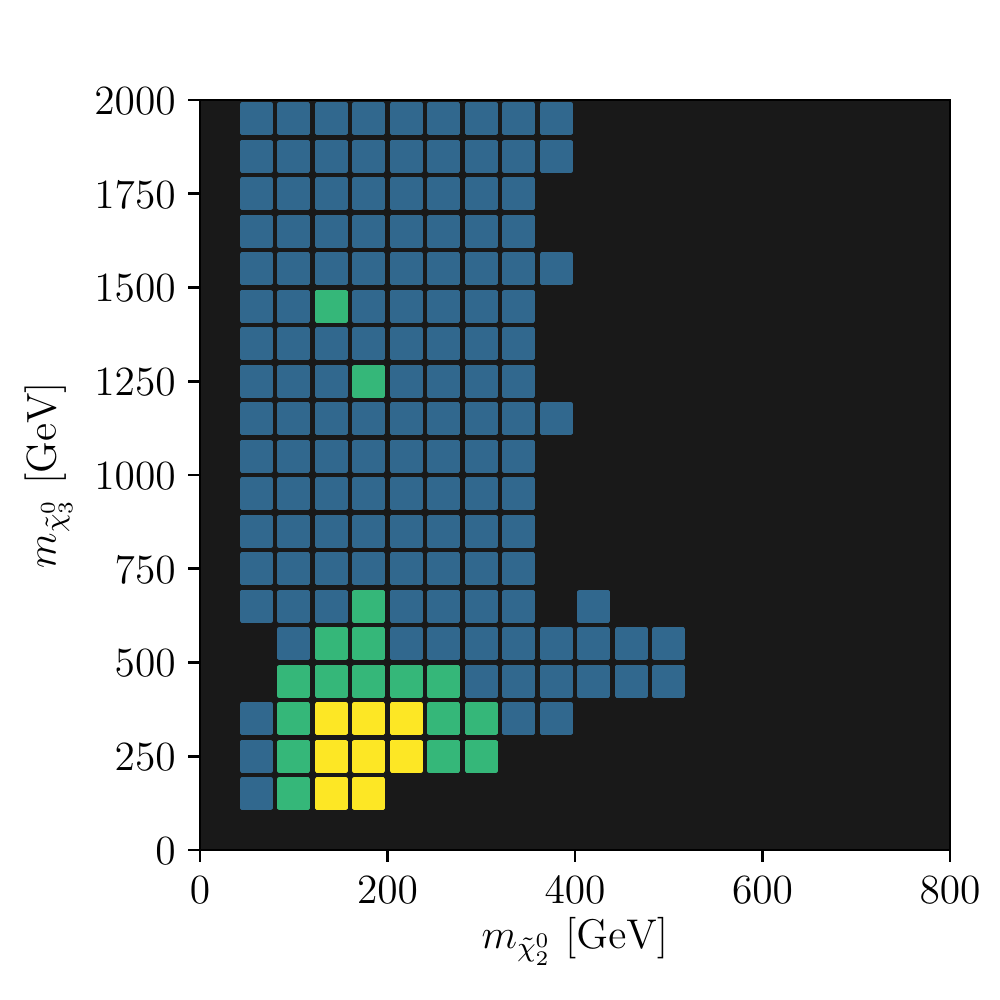}
\includegraphics[width=0.32\textwidth]{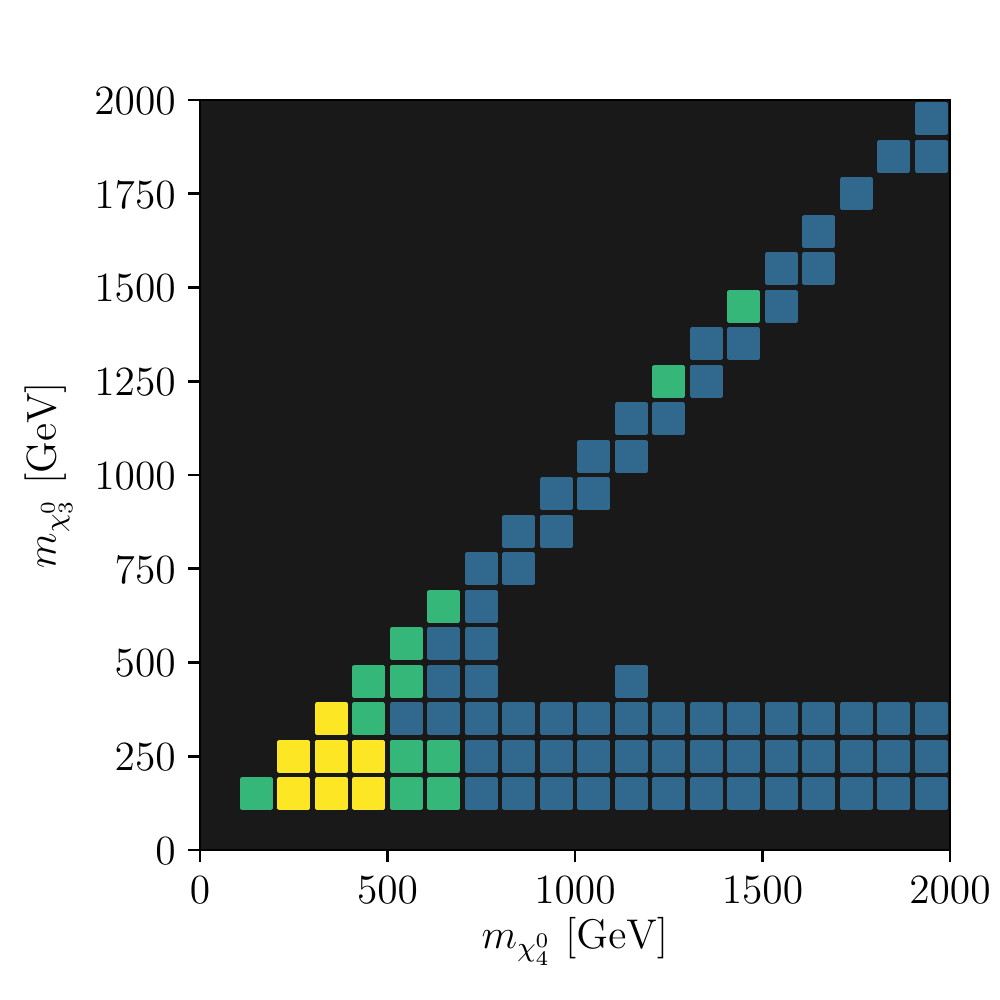}
\caption{Grids of EWMSSM points extracted from the best-fit EWMSSM parameter space surface identified by the \Gambit global fit in \cite{Athron:2018vxy}. The yellow, green and blue points are within the $1\sigma$, $2\sigma$ and $3\sigma$ preferred regions in the \Gambit fit, respectively.}
\label{fig:SearchMeasurementComplementarity:gambitgrids}
\end{center}
\end{figure}

Here we investigate to what extent the collection of LHC measurements implemented in \Contur can constrain these particular EWMSSM scenarios. To run \Contur on the full set of EWMSSM parameter samples from the \Gambit scan in \cite{Athron:2018vxy} is beyond the scope of this work, due to the significant computational cost. Instead we extract three grids of EWMSSM points from the profile-likelihood surfaces shown in Figure~\ref{fig:SearchMeasurementComplementarity:gambitresults} by binning the \Gambit samples in each mass plane and extracting the best-fit point within each bin.\footnote{The extracted points are not necessarily unique to a single grid. For instance, the global best-fit point from the \Gambit fit will appear in all three grids.}  The resulting grids of EWMSSM points are illustrated in Figure~\ref{fig:SearchMeasurementComplementarity:gambitgrids}. We include points that are within the $3\sigma$ region in the \Gambit fit result, as indicated by the colouring in Figure~\ref{fig:SearchMeasurementComplementarity:gambitgrids}. These parameter points are then passed to \Herwig~\cite{Bellm:2019zci} in the form of SLHA files~\cite{Skands:2003cj}. \Contur is used to scan over these points, as usual requesting \Herwig to generate all implied final states with one or more BSM 
particles outgoing from the matrix element, and looking for the impact on fiducial cross section measurements available in \Rivet.


\begin{figure}[t]
\begin{center}
\includegraphics[width=0.3\textwidth]{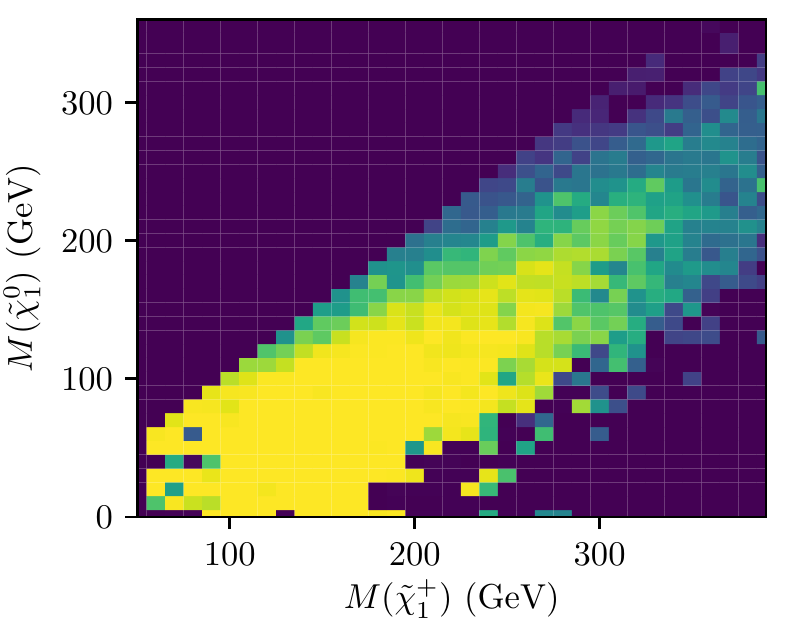}
\includegraphics[width=0.3\textwidth]{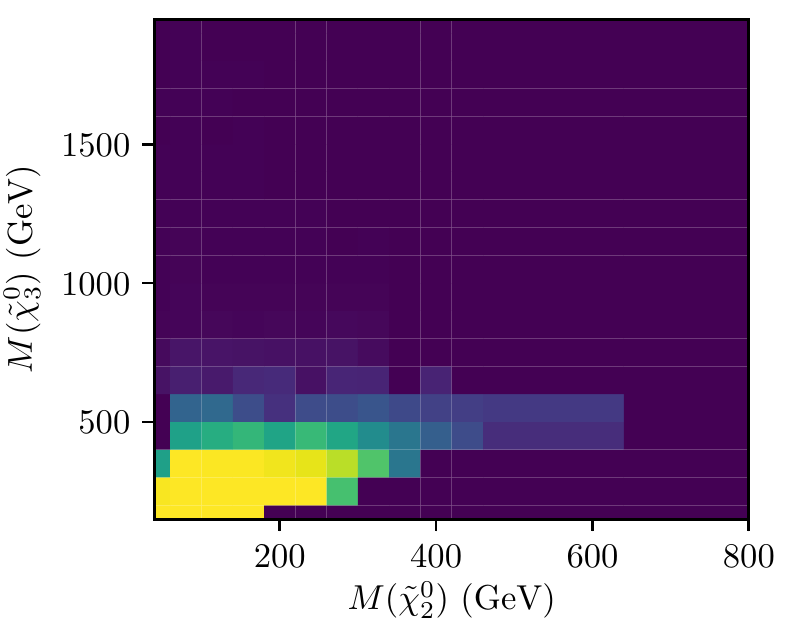}
\includegraphics[width=0.3\textwidth]{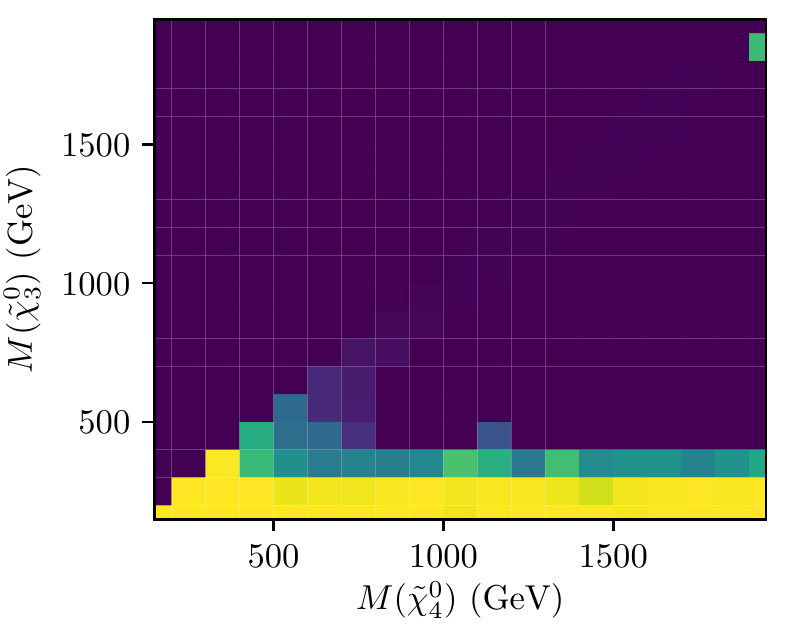}
\includegraphics[width=0.055\textwidth,raise=0.4cm]{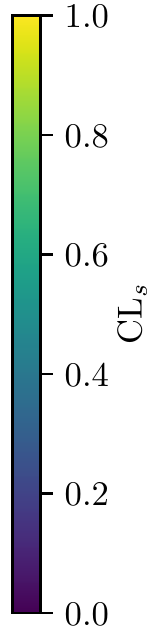}
\caption{\Contur exclusion sensitivity to the \Gambit points shown in Fig.~\ref{fig:SearchMeasurementComplementarity:gambitgrids}.}
\label{fig:SearchMeasurementComplementarity:gambit_contur}
\end{center}
\end{figure}

The resultant heatmaps of CL$_s$ values from \Contur are shown in Figure~\ref{fig:SearchMeasurementComplementarity:gambit_contur}. 
It can be seen that many of the parameter points preferred in the \Gambit combination of SUSY searches display tension with measurement results. For instance, in the ($m_{\tilde{\chi}^\pm_1},m_{\tilde{\chi}^0_1}$) plane (left), the 30 points extracted from the $1\sigma$ region of the \Gambit fit all end up with CL$_s > 0.95$ in the \Contur analysis. Similarly, in the ($m_{\tilde{\chi}^0_2},m_{\tilde{\chi}^0_3}$) and ($m_{\tilde{\chi}^0_4},m_{\tilde{\chi}^0_3}$) planes (middle and right), the measurements checked by \Contur show exclusion sensitivity to many of the EWMSSM points with $m_{\tilde{\chi}^0_3} \lesssim 400$\ GeV.
%
%
%
%
%
%
%
This comes principally from the ATLAS
$WWW$ measurement~\cite{Aaboud:2016ftt} in the channel with two same-sign leptons and two jets. 
There is also sensitivity from the ATLAS four-lepton line shape measurement~\cite{Aaboud:2019lxo} for parameter points where the heavier electroweakinos ($\tilde{\chi}^0_3$, $\tilde{\chi}^0_4$ and $\tilde{\chi}^\pm_2$) are not completely decoupled. This is not surprising, as what mainly drives the preference for having these states relatively light in the \Gambit fit is an excess seen in an ATLAS search for final states with four leptons plus missing energy~\cite{Aaboud:2018zeb}. This excess can more easily be accommodated when at least $\tilde{\chi}^0_3$, and possibly also $\tilde{\chi}^0_4$/$\tilde{\chi}^\pm_2$\footnote{Due to the dominantly wino or Higgsino nature of the $\tilde{\chi}^0_4$ state in these EWMSSM scenarios, the $\tilde{\chi}^\pm_2$ chargino is always close in mass to the $\tilde{\chi}^0_4$ neutralino.}, are not decoupled, since the decays $\tilde{\chi}^0_3 (\tilde{\chi}^0_4) \rightarrow Z \tilde{\chi}^0_2$ and $\tilde{\chi}^\pm_2 \rightarrow Z \tilde{\chi}^\pm_1$ increase the production rate for di-$Z$-boson events. Similarly, the production and decay of the heavier electroweakinos can increase the expected rates for multi-$W$-boson events, \textit{e.g.}\ through decay processes such as $\tilde{\chi}^0_3 (\tilde{\chi}^0_4) \rightarrow W^\pm \tilde{\chi}^\pm_1$, $\tilde{\chi}^\pm_2 \rightarrow W^\pm \tilde{\chi}^0_2$ and $\tilde{\chi}^0_3 \rightarrow h \tilde{\chi}^0_1 \rightarrow W^\pm W^{(*)\pm} \tilde{\chi}^0_1$.

\begin{figure}[t]
\begin{center}
\includegraphics[width=0.45\textwidth]{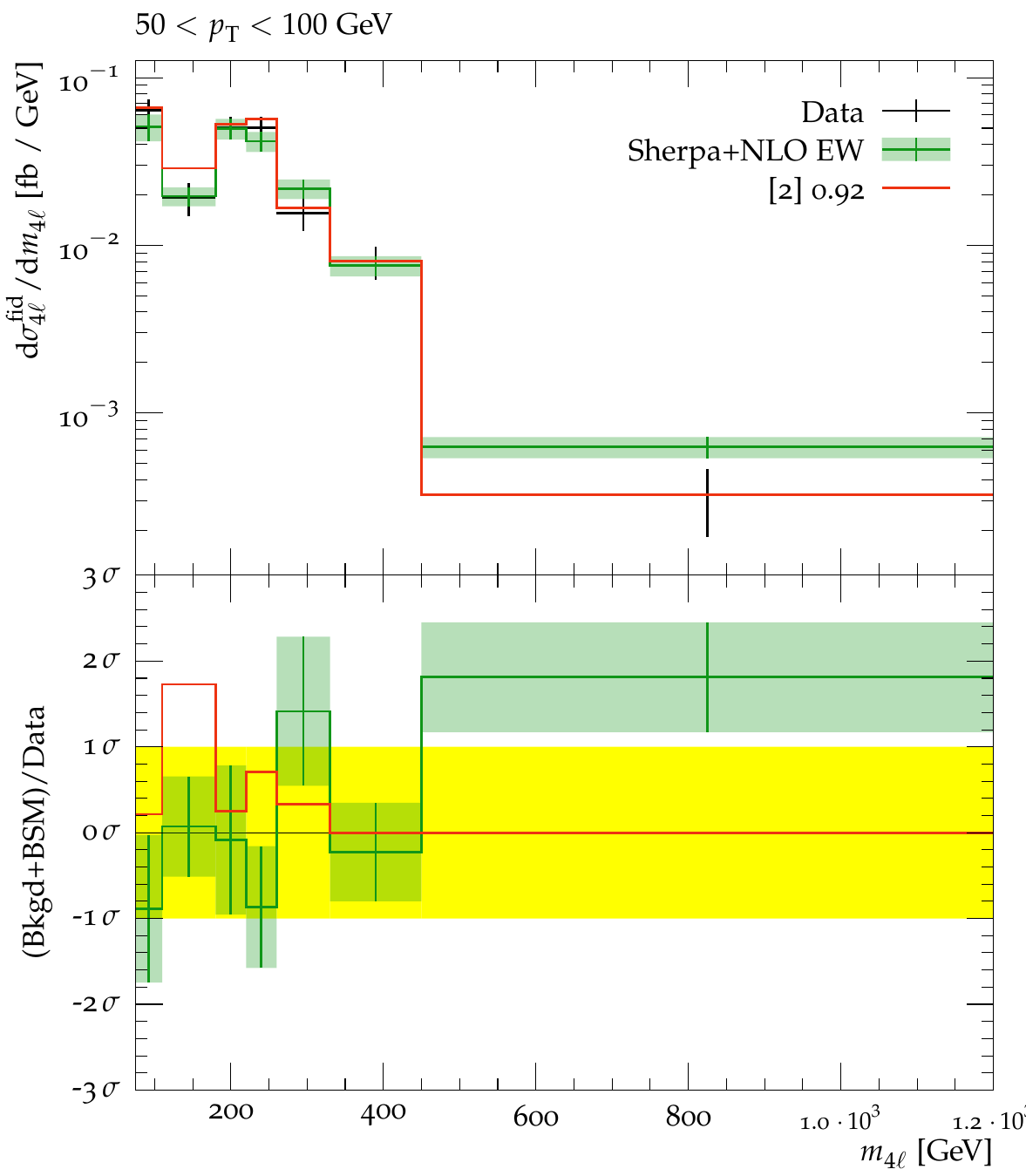}
\includegraphics[width=0.45\textwidth]{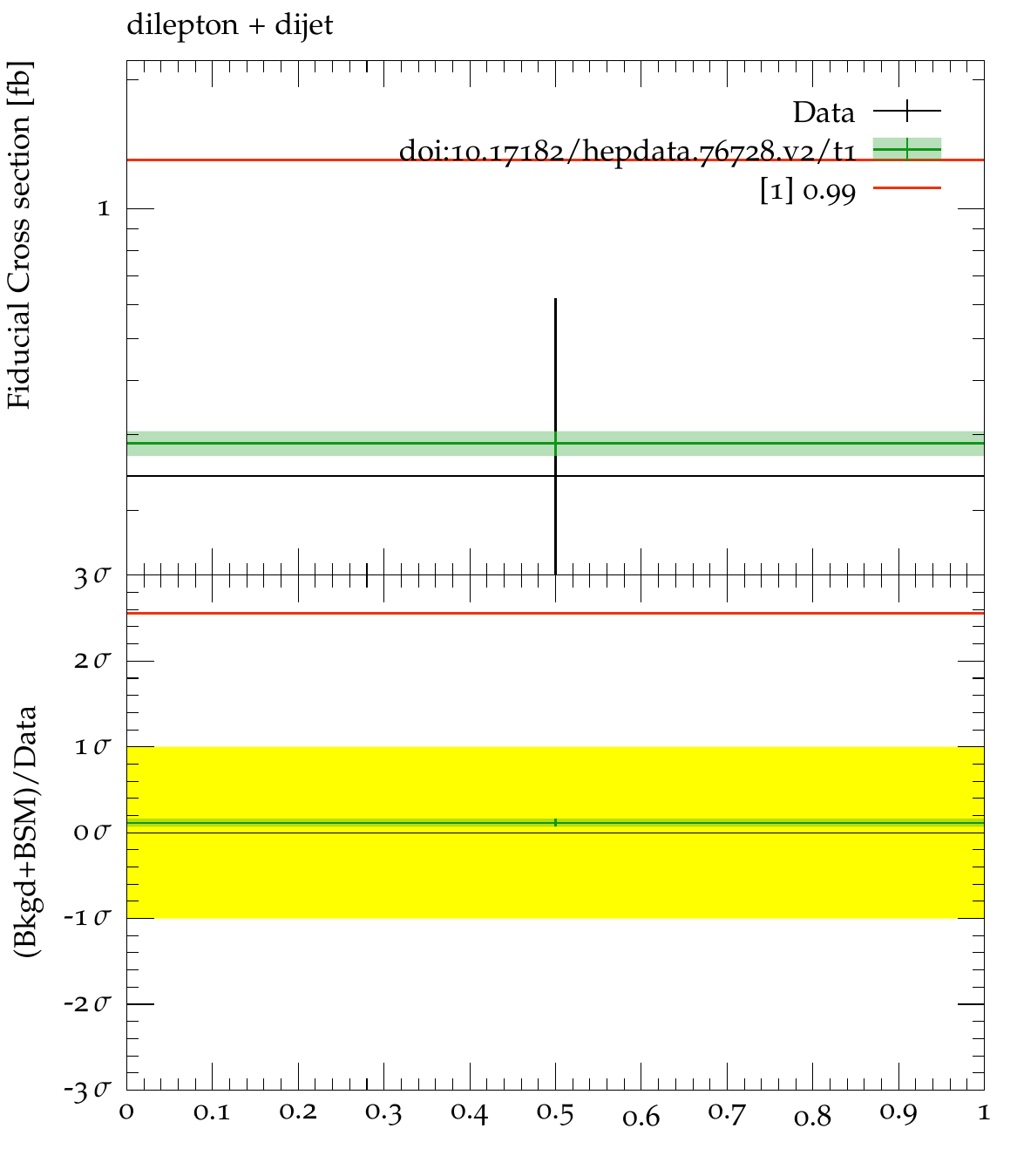}
\caption{Example \Rivet{} figures contributing to the \Contur{} exclusion for the third \Gambit hypersurface. On the left, the four-lepton lineshape
at intermediate $p_T^{4l}$ from \cite{Aaboud:2019lxo}, for an EWMSSM point with $(m_{\tilde{\chi}^0_1},m_{\tilde{\chi}^0_2},m_{\tilde{\chi}^0_3},m_{\tilde{\chi}^0_4},m_{\tilde{\chi}^\pm_1},m_{\tilde{\chi}^\pm_2}) = (55,94,169,199,98,205)$\ GeV.
On the right, the like-sign-dilepton plus dijet cross section from \cite{Aaboud:2016ftt} 
for a point with $(m_{\tilde{\chi}^0_1},m_{\tilde{\chi}^0_2},m_{\tilde{\chi}^0_3},m_{\tilde{\chi}^0_4},m_{\tilde{\chi}^\pm_1},m_{\tilde{\chi}^\pm_2}) = (74,201,202,1686,196,1686)$~GeV.
The upper figures show the cross section comparison between data (black), data and new physics (red) and the SM (green). The lower figures
show the ratios expressed as a significance.
}
\label{fig:SearchMeasurementComplementarity:gambit_rivet}
\end{center}
\end{figure}

The best-fit EWMSSM points from the \Gambit fit were preferred due their ability to fit excesses in some searches, while simultaneously avoiding strong tension with other searches. Since \Contur in this mode is treating the signal as background, such excesses would not be taken into account. However, neither of these measurements shows an excess of data over the SM. Example figures, with
the SM theory prediction indicated, are shown in Figure~\ref{fig:SearchMeasurementComplementarity:gambit_rivet}. 
Events from parameter points at higher masses also enter these and several other the fiducial cross sections -- typically those involving (multiple) gauge boson production -- indicating that with higher luminosity,
with more precise and/or exclusive measurements, some of these scenarios may be accessible to this approach. 



\section{Summary}
\label{sec:SearchMeasurementComplementarity:summary}

The analysis logic and reference data for two searches were preserved as \Rivet{} routines using smearing techniques to obtain approximate detector-level distributions from particle-level inputs. These routines could then be used in an automated re-interpretation tool, in this case, \Contur{}. The performance of these search routines was compared to that of measurements of similar final states for benchmark BSM models, and were found to be similar. This study shows that it is possible to use searches alongside measurements in automated re-interpretation workflows. This could speed up the re-interpretation of searches which are preserved in \Rivet{} routines, since it bypasses the need for an ad-hoc re-interpretation on a case-by-case basis. For
cases where lack of statistics precludes unfolding (for example in tails of distributions), these approaches are expected to be complementary in the
context of \Contur{}.

A selection of SUSY parameter points which are either favoured, or not excluded, by a combination of SUSY searches considered in \Gambit, were run through \Contur{} to check
for sensitivity in the existing LHC measurements. Several of the parameter points at low $m_{\tilde{\chi}^0_3}$ are disfavoured by the measurements,
particularly the four-lepton and same-sign-dilepton plus jet channels. There seems to be significant potential, and complementarity between
measurements, in regions where the SM contribution is large but well-known, and searches where the goal is usually to find regions where
the SM contribution is small. This is an area of interesting further study in the HL-LHC period.

\let\Contur\undefined
\let\Herwig\undefined
\let\Powheg\undefined
\let\Professor\undefined
\let\Pythia\undefined
\let\Rivet\undefined
\let\Sherpa\undefined
\let\eps\undefined
\let\mb\undefined
\let\mc\undefined
\let\mr\undefined
\let\tm\undefined

%% file: 4top/4top.main.tex
\graphicspath{{4top/}}

\newcommand{\Herwig}{H\protect\scalebox{0.8}{ERWIG}\xspace}
\newcommand{\Pythia}{P\protect\scalebox{0.8}{YTHIA}\xspace}
\newcommand{\Sherpa}{S\protect\scalebox{0.8}{HERPA}\xspace}
\newcommand{\Rivet}{R\protect\scalebox{0.8}{IVET}\xspace}
\newcommand{\Contur}{C\protect\scalebox{0.8}{ONTUR}\xspace}
\newcommand{\Professor}{P\protect\scalebox{0.8}{ROFESSOR}\xspace}
\newcommand{\eps}{\varepsilon}
\newcommand{\mc}[1]{\mathcal{#1}}
\newcommand{\mr}[1]{\mathrm{#1}}
\newcommand{\mb}[1]{\mathbb{#1}}
\newcommand{\tm}[1]{\scalebox{0.95}{$#1$}}

\chapter{Determining sensitivity of future measurements to new physics signals}

{\it D.~Kar, L.~Corpe}

\label{sec:projname}



No definite signs of new physics have been observed in LHC data since the discover of the Higgs boson in 2012. This has brought into focus a few alternative strategies for the search for new physics.
\Contur~\cite{Butterworth:2016sqg} is one such approach, where existing unfolded measurements are used to set bounds on different BSM models (the procedure is also discussed in Section~\ref{sec:contur-update} of these proceedings). 
In this exercise, we ask the opposite question: if we expect a certain measurement to be performed at the LHC, what precision needs to be achieved in order to exclude a certain region of parameter space of a specific BSM model. This approach can be generalised: one could produce  a list of topologies and phase spaces which,
when measured with specified precision, can constrain unexplored region of parameter space of models. We leave that as a future 
exercise.

\begin{figure}[h!]
  \centering
  \includegraphics[width=0.9\textwidth]{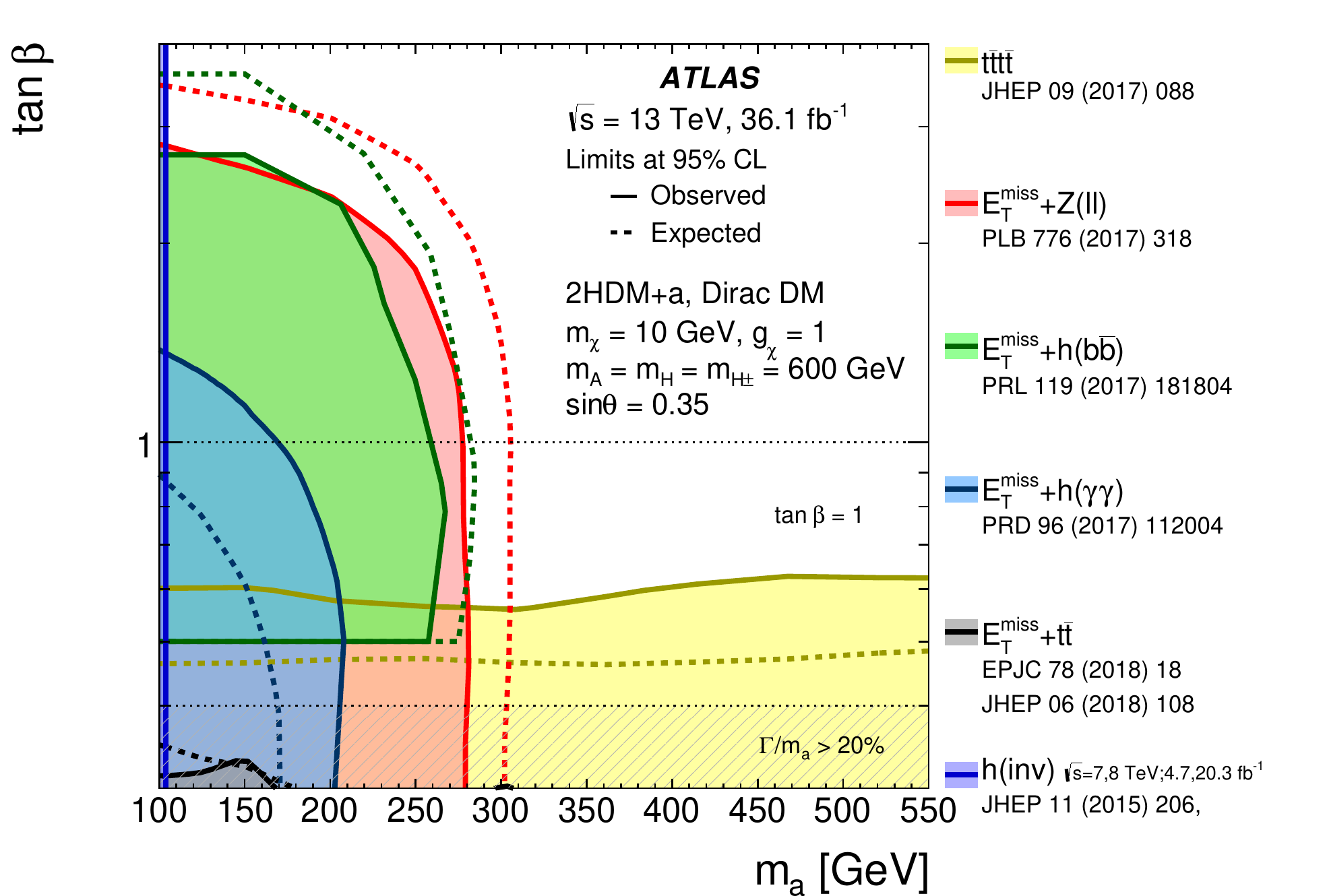}\qquad
  \caption{Regions in ($m_a$, $\tan\beta$) planes from the model considered excluded by data from $X$+MET and
  $tttt$ measurements~\cite{Aaboud:2019yqu}. More more details, please refer to the cited article.}
  \label{fig:atlaslimit}
\end{figure}

Since this is a proof-of-principle demonstration, the details and the viability of the model and parameter space is not of critical importance.
A benchmark two-Higgs doublet model (``2HDM+a'') is used, inspired by the combination of some of the ATLAS dark matter sear\-ches~\cite{Aaboud:2019yqu}. 
Figure~\ref{fig:atlaslimit} shows
that for low values of the $\tan \beta$ parameter (which is related to the ratio of vacuum expectation values of the two Higgs doublets), this model can be sensitive to the four-top-quark production process. To test if a future four top cross-section
measurement can constrain the model, we generate pseudodata with the \Pythia~\cite{Sjostrand:2014zea} event generator. 
Since the cross-section is extremely small,
we generate the largest background, semileptonic $t\overline{t}$ events, as proxy for the four top process. 
We only select events with at least 6 jets, 
2 of which must be tagged as containing $b$ quarks. We consider the $H_T$ distribution, which is the scalar sum of $p_T$ of all selected jets, constructed
using the \Rivet~\cite{Bierlich:2019rhm} analysis framework. All samples are normalised to 300 fb$^{-1}$ of integrated luminosity.

\begin{figure}[h!]
  \centering
  \includegraphics[width=0.9\textwidth]{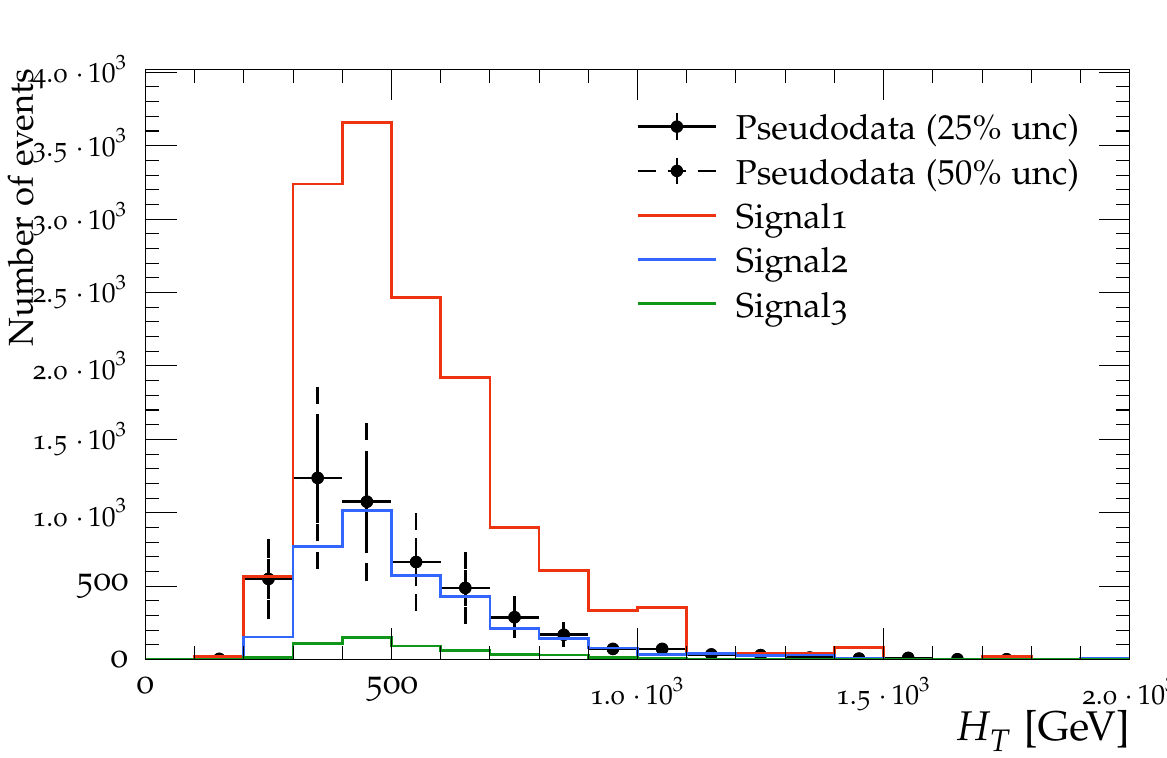}\qquad
  \caption{Distribution of $H_T$, normalised to 300 fb$^{-1}$ of integrated luminosity, for pseudodata shown
  with $25\%$ and $50\%$ flat uncertainties, and three signal samples}
  \label{fig:result}
\end{figure}

We considered two scenarios for the uncertainties associated with the pseudo-mea\-su\-re\-ment: $25\%$ and $50\%$ flat uncertainties, shown by solid and dashed error bars in Figure~\ref{fig:result}.
Three different signal scenarios are overlaid, corresponding to $\tan\beta$ values between $0.1$ to $0.5$, keeping $m_a=500$~GeV
and $M_A=600$~GeV, but the specific values are not the important aspect here. The signal samples where generated using
\Herwig~\cite{Bellm:2019zci}. As can be seen from Figure~\ref{fig:result}, the ``Signal1'', corresponding to the lowest $\tan\beta$ value will be excluded by such a 
measurement alone. Intuitively, this is due to the fact that the additional signal, on top of the measured data, would be well beyond the uncertainties of the measurement. ``Signal3'', which corresponds to the highest $\tan\beta$ value, can't be excluded just by this measurement, since the signal would be within the uncertainties of the measurement if stacked on top of the data. It is hard 
to say anything about the intermediate ``Signal2''. This approach can be quantified more rigorously with \Contur{}, which includes a set of available measurements at 7, 8 and 13~TeV center of mass energies from LHC (and we added this \textit{fake} measurement). For ``Signal2'', the combined
exclusion is about $98\%$ and $84\%$ for $25\%$ and $50\%$ flat uncertainties on all bins of our distribution. 
This indicates that if we can make the measurement with $\approx25\%$ uncertainty, we can exclude this signal at $95\%$ confidence level, as opposed
to making the measurement with $\approx50\%$ uncertainty, which would not definitely exclude the model at this parameter point. For ``Signal1'', the exclusion is $100\%$ for both the cases, while for ``Signal3'', the exclusion is about $60\%$ and $45\%$ respectively.

In this exercise, it is demonstrated that with the help of tools like \Rivet and \Contur, we can predict measurements
at what precision can be sufficient exclude certain parameter space of certain models.

\let\Herwig\undefined
\let\Pythia\undefined
\let\Sherpa\undefined
\let\Rivet\undefined
\let\Contur\undefined
\let\Professor\undefined
\let\eps\undefined
\let\mc\undefined
\let\mr\undefined
\let\mb\undefined
\let\tm\undefined

%% file: ml-1/PhenoAI.main.tex
\graphicspath{{ml-1/}}









\chapter{Practical Machine Learning for regression and classification and applications in HEP phenomenology}
{\it S.~Caron, A.~Coccaro, S.~Ganguly, S.~Kraml, A.~Lessa, S.~Otten, R.~Ruiz, H.~Reyes-Gonz\'{a}lez, R.~Ruiz de Austri, B.~Stienen, R.~Torre}


\label{sec:PhenoAI}

\begin{abstract}
This contribution explains practical applications of Machine Learning (ML) in HEP phenomenology and here in particular the possibility to examine HEP models of new physics in their full dimensionality.
ML models can be trained to provide likelihood evaluations, cross-sections and exclusion boundaries given the parameters of the HEP model. We discuss ways to accelerate the applicability of ML, to evaluate uncertainties in the ML model and to provide a sustainable reuse of training data and  ML model.
Various examples are discussed, such as the ML-based regression of cross sections in left-right symmetric models and in the inert doublet model (IDM) and the regression of likelihoods for the minimal supersymmetric standard model (MSSM7) as provided by the GAMBIT collaboration.
Links to the training data can be found at \url{https://www.phenoMLdata.org}. 
\end{abstract}

\section{Introduction}



In recent years machine learning (ML) has gained increasingly more traction in the particle physics community. Where  machine learning was first used 
predominantly only in HEP experiments for the creation of metrics in a cut-and-count analysis, since a few years it also finds applications in various other branches of particle physics. This article is aimed at theorists and experimenters who want to understand how machine learning can be used for regression and classification in HEP phenomenology.

Many recent examples of the application of ML in phenomenology come from simulations and the exploration of theoretical models.
Application range from simulations of particle physics events via so-called {\it generative} ML models ~\cite{Otten:2019hhl,Hashemi:2019fkn,DiSipio:2019imz} to proposals to search for new physics with the help of ML based {\it anomaly detection}, which is discussed elsewhere in these proceedings.
Early ML applications in particle physics phenomenology and theory include the estimation of parton density functions~\cite{Ball:2014uwa} and various applications to jet physics as summarised in ~\cite{Larkoski:2017jix}.
Furthermore it has been demonstrated that high-parameteric model exclusion boundaries~\cite{Caron:2016hib}
or likelihood (ratios) can be learned, explored and provided for further use via the use of ML models, see e.g. \cite{Brehmer:2018kdj}.
The ML model is made by learning numbers based on theoretical calculations or experimental measurements 
(e.g. cross-sections, likelihoods, posteriors, confidence level for exclusion) from training data given the parameters of the physical model and including even the experimental nuisance parameters \cite{Coccaro:2019lgs}.
Typically training data is provided by simulators (e.g. event generators) or
by evaluating the likelihood of a given parameter set and theory with the help of experimental data.
Examples could be a cross section of a new particle or an experimentally derived likelihood as a function of the parameters of a 20-dimensional Beyond the Standard Model (BSM) model. Here, an effective high-dimensional sampling of the full model parameter space is needed. 

Even though ML allows to explore physical parameters in their full dimensionality, the application of such an approach has so far been limited. The keen interest of the experimental and theoretical community at Les Houches led to this note describing ways to accelerate progress in researching ML models for phenomenology.

First, we wish to encourage the experimental and theoretical communities to study and release model data in its full dimensionality. The publication of high-dimensional data should become a standard practice. 
Furthermore, we propose in this contribution  
a standardised way to publish high-dimensional training data, 
a way to publish the trained ML model, and 
a code database to provide examples to quickly (re-)train ML models on such data.
We also present various examples, worked out within the Les Houches workshop, 
that follow this approach. 




\section{Project description}
\label{sec:PhenoAI:project}

The idea of this project is to simplify the creation and reuse of ML models made for HEP phenomenology.
For this we like to encourage to save the ML model, the HEP  data set and the code to train the ML models.
If the construction of ML models can be simplified, any publication on HEP phenomenology could use ML models to store probabilities, cross sections, and limits also for multidimensional HEP models.

\subsection{Classification and Regression with  uncertainties}
\label{sec:PhenoAI:uncertainties}

The majority of machine learning in high energy physics uses supervised learning. Within supervised learning there are two main branches: classification and regression. In classification the  ML algorithm learns to classify new data into a discrete number of classes (e.g. excluded or not excluded parameter configurations), whereas in regression the ML algorithms learns to predict  a continuous quantity (e.g. a cross section). Both these branches can be seen as a prediction of an output variable $y$ (or a set of output variables $\vec{y}$) given a set of input variables $\vec{x}$. The machine learning algorithm created with the data estimates the mapping function $\hat{f}(x)=y$ between the two. This estimator of the true mapping function $f(x)$ is learned from $n$ examples where $x_i$ and the true function values $f(x_i)$ are known.

Determining the best parameters of a particular ML model (e.g. a neural network) is called training. The core of machine learning is that this determination of parameters is automated based on data (hence called training data). As this data has a big influence on the performance of the algorithm, having a data set that is both large and information dense enough is often key to training a good machine learning algorithm.
In low dimensional problems (i.e. problems with few input variables) with a computationally cheap data generation procedure data can easily be generated with random sampling or grid sampling. 
In HEP phenomenology, however, data generation typically requires
the execution of a chain of numerical integration programs, e.g. event generators or simulations and often even a (time-consuming) comparison to experimental data.
HEP more often than not deals with high dimensional functions for which the truth values are computationally costly to acquire. Techniques like active learning can predominantly sample parameter regions that are difficult to learn for the ML algorithm and can help to mitigate these problems e.g. to learn exclusion boundaries of HEP  models \cite{Caron:2019xkx}.
 
Since there is freedom in the choice of ML models/algorithms, a limited amount of training data and uncertainties in the data, there are two uncertainties to consider.
So called \textit{aleatoric} uncertainties are uncertainties inherent to the data or data collection procedure, e.g. noise. These cannot be removed or reduced by adding more data points. 
Second, \textit{epistemic} uncertainties are uncertainties that are introduced with our ML model for the estimator $\hat f(x) \approx f(x)$. An example of this type of uncertainty is the uncertainty related to the trained values of the parameters (e.g. the weights of a neural network) of the ML algorithm. One way to estimate such uncertainties in neural networks is by applying {\it dropout} of nodes that stays turned on during inference~\cite{gal2015dropout}. As dropout randomly disables a fraction of the network parameters, the output of the network will again change at each inference step. The spread of the output can then be interpreted as a measure of model uncertainty.

\subsection{Collection of Machine Learning Models}
\label{sec:PhenoAI:modelcoll}

The core of this initiative is formed by a collection of trained ML models. These models, whether they are neural networks or other types of models, should be stored together with meta data which allows them to be read by an end-user package. 
Such an end-user package would allow external users to quickly use ML models and access the predictions made by these models.

The format of the trained ML model can differ based on which library was used to create the trained model. For scikit-learn~\cite{scikit-learn} for example the authors suggest~\cite{sklearn-persistence} to serialise the entire model object with the \texttt{joblib} package in Python. For more neural network based packages a more natural choice would be to store the model in an .onnx-file. Especially .onnx-files guarantee a cross-library support of trained networks, giving a strong guarantee that information encoded in trained networks will still be usable in the coming future.

As to where to store these trained models, we think we can best look at the current best practice in the machine learning research community. There they use primarily GitHub for this. Trained models are often stored together with their training and production code and instructions in a README file on how to use the included codes. As most papers use data from publicly available sources (e.g. the MNIST data set~\cite{lecun-mnisthandwrittendigit-2010} or credit card data set~\cite{creditcard-dataset}), data is most of the time not included, but instead referred to.

We suggest to adopt this practice for high energy physics, with the only exception being for the data. Most of the already published papers in our field generate their own data. This makes it impossible to just cite the data location; the data needs to be stored somewhere as well. As the standard platform to publish data in particle physics phenomenology is Zenodo we suggest using that as a default location to store the data sets used for creating machine learning algorithms. This includes both training and testing data.


For this collection of models to work the inclusion of meta data is crucial. It should contain for example what the inputs for the model should be and what its output(s) represent(s). It will furthermore contain information that will make the use of the model more robust against mistakes. By including information on the boundaries of the training region, the package's code can warn the user if a model is used outside of its intended range. The specifics of the meta data and the exact storage format are currently in the process of being developed.

The advantage of this methodology is that the trained model can be bundled together with the meta data, so that the model can easily be communicated by the creating user, and easily be used by the external user. In this way efficient communication of for example high-dimensional results becomes possible. To encourage this, a searchable library of publicly available models will be created, providing a single go-to location for neural networks in high energy physics.


\subsection{Collection of Training Data}
\label{sec:PhenoAI:datacoll}

Alongside a trained model with meta data, each model instance should also come with a publicly accessible data set. For this we have set up webpage (\url{https://www.phenoMLdata.org}) and a Zenodo group\footnote{\url{https://zenodo.org/communities/phenoml_database/}}, in which the high dimensional training data of each of the trained models will be published. We intend to also create a link with a tool like SPOT~\cite{Diblen:2018mib}, which makes online and codeless visualisation of high dimensional data possible. A link to the data and the ML model could be added to the arXiv or {\sc inSPIRE} entries of a HEP publication. 
This publishing of data is in our view essential to fully understand the trained models and to speed up the improvement of existing models.

\subsection{Collection of  Code to build Machine Learning models}
\label{sec:PhenoAI:codecoll}

To stimulate the adoption of machine learning in high energy physics even further each trained model will also have its training (and all other relevant) code published on platforms like GitHub. This code can then serve as example or as best-practice show case for both basic (e.g. how to train a network) to more complicated machine learning cases (e.g. how to extract epistemic and aleatoric uncertainties from a neural network).

\section{Examples and best-practice}
\label{sec:PhenoAI:trained_models}

\subsection{Examples of past phenomenological  studies  }
Over the last years there are several examples where the use of ML techniques have shown 
to be very useful for phenomenology studies. Typically these methods have been applied 
to replace the event generation plus detector simulation chain using neural networks or Gaussian processes. This is usually the bottleneck of recasting and doing global fits of new physics models using LHC data.   

The first use case was in the study of coverage properties on the constrained minimal supersymmetric standard model (cMSSM) parameter space inferred from a Bayesian posterior 
and the profile likelihood based on an ATLAS experiment sensitivity study \cite{Bridges:2010de}.
The use of a shallow neural network allowed a fast prediction of the cMSSM mass spectrum gaining a factor of $\sim 10^4$ with respect to run a supersymmetry (SUSY) spectrum calculator as softsusy in sampling the cMSSM parameter space. 

In a similar manner Buckley et al. \cite{Buckley:2011kc} employed both a Bayesian deep neural network and a supported vectorial machine to interpolate between a grid of points in the cMSSM parameter space and therefore get fast predictions for the signal predicted by the model in the context of an ATLAS analysis.

Other phenomenological applications of ML in the SUSY context have been the development of SCYNeT \cite{Bechtle:2017vyu} and SUSY-AI \cite{Caron:2016hib} packages. The first one uses neural network regression for a fast evaluation of the profile likelihood ratio using the 11-dimensional phenomenological minimal supersymmetric Standard Model (pMSSM) as an input. The authors have applied it to a global fit of the model including LHC data speeding up enormously the inference. Instead SUSY-AI does classification using a random forest algorithm which is trained with input data based in an analysis done by the ATLAS collaboration of the 19-dimensional pMSSM. Besides predicting whether a point of the model is excluded or not, it provides the epistemic uncertainty in the classification.

A recent application of deep neural networks have been developed to predict production 
cross sections of SUSY particles at the LHC. This is the case of DeepXS \cite{Otten:2018kum} which employs deep neural networks for a fast prediction of electro-weakino production cross-sections at the next-to-leading order in the pMSSM context. In this case the gain in speed is of order $10^7$ compared with prospino \cite{Beenakker:1996ed}
.

A further recent development has been the proposal of using deep neural networks for learning the full experimental information contained in the likelihood functions being used for the statistical inference in searches and measurements at the LHC \cite{Coccaro:2019lgs}. The likelihoods learnt this way, so-called DNNLikelihoods, would allow a complete and framework-independent distribution of the physics analysis results, also enabling a precise combination with other experimental likelihoods, whenever the correlations among parameters are know. 

Finally Kvellestad et al. \cite{Kvellestad:2018crh} investigated the performance of 
deep neural networks to learn the signal mixture estimation of a ditau signal coming from a pair of degenerate Higgs bosons of opposite CP charge in the context of a Two-Higgs-Doublet model. They found a $\sim20$ \% improvement in the estimate of the uncertainty of signal mixture estimates, compared to estimates based on fitting, say, standard discriminating kinematic variables. 

In addition to deep neural networks, Gaussian processes have been applied as fast predictors 
of LHC analyses efficiencies in the context of the reconstruction of a natural SUSY scenario \cite{Bertone:2016mdy} and dark matter simplify models \cite{Bertone:2017adx} using LHC simulated data. 

So far there is a very limited number of works which have followed somehow the 
prescription in line of what is suggested in Sec.~\ref{sec:PhenoAI:project} to publish results. 
This is the case of the DeepXS (\url{https://github.com/SydneyOtten/DeepXS}) and DNNLikelihood
(\url{https://github.com/riccardotorre/DNNLikelihood}) projects where weights of the trained models and python scripts with neural networks implementations have been published on GitHub so the user can train the data himself. Furthermore training data are also available in Zenodo \footnote{Links to the training data can be found at \url{https://www.phenomldata.org}}. 
In the case of SUSY-AI pickle files containing the trained weights are available (for details see \url{https://www.susy-ai.org}), whereas data were made public by the ATLAS collaboration.

To the best of our knowledge the estimation of aleatoric and epistemic uncertainties of neural networks has not been applied in the HEP community.


\subsection{Learning $W_R$ Boson Production and Decay Rates}

\begin{figure}[t!]
\begin{center}
\subfigure[]{\includegraphics[width=0.49\textwidth]{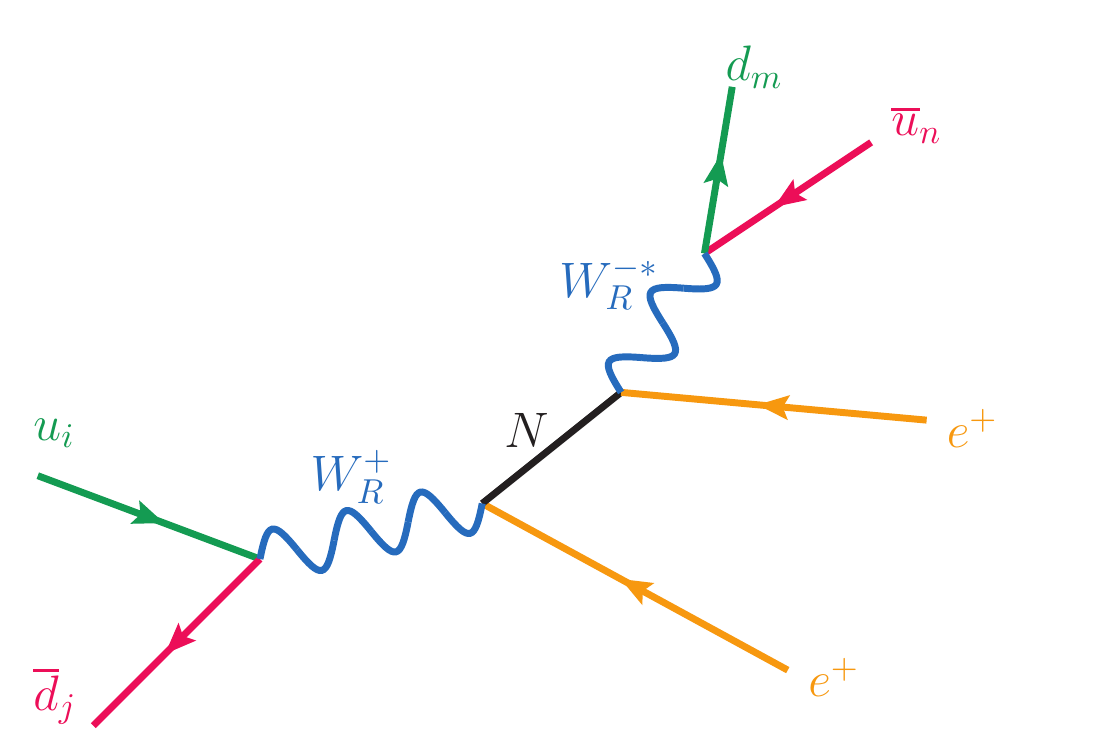}\label{fig:PhenoAI:lrsm_xsec_diagram}}
\subfigure[]{\includegraphics[width=0.49\textwidth]{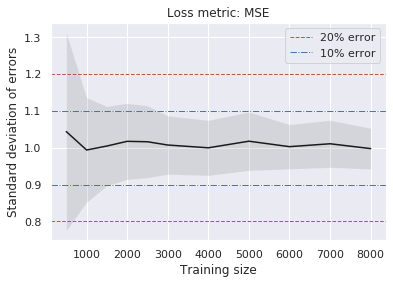}\label{fig:PhenoAI:lrsm_loss}}
\end{center}
\caption{(a) Diagram within the Left-Right Symmetric Model (LRSM) containing a heavy neutrino and two new gauge bosons. (b) Spread of the prediction error made by the trained machine learning algorithm on the LRSM cross sections.}
\label{fig:PhenoAI:lrsm_xsec}
\end{figure}

In light of its successful explanation of LHC data,
the SM remains the best description of nature at high energies and short distances.
Despite this success, there exists several experimental and theoretical
motivations for the existence of new physics.
One such example is the discovery~\cite{Ahmad:2002jz,Ashie:2005ik} of nonzero neutrino masses:
in order to accommodate neutrino masses in a gauge-invariant and renormalizable manner,
the SM must be extended by new particles and new couplings~\cite{Ma:1998dn}.
Such models that achieve this, known collectively as Seesaw models, 
can be tested at a variety of laboratory-based experiments, 
including the LHC and its potential successors~\cite{Cai:2017jrq,Cai:2017mow,Benedikt:2018csr,Abada:2019ono}.

In practice, only benchmark and limiting cases of full, realistic, UV-complete Seesaw models
are tested at the LHC due to the often cumbersome number of free parameters in the theory.
For example: in the Left-Right Symmetric Model (LRSM)~\cite{Pati:1974yy,Mohapatra:1974hk,Mohapatra:1974gc,Senjanovic:1975rk,Senjanovic:1978ev},
which predicts the existence of a new $W_R$ gauge boson and heavy Majorana neutrinos
as shown in Fig.~\ref{fig:PhenoAI:lrsm_xsec_diagram},
one typically assumes that $W_R$ bosons can only decay to one charged lepton flavor and one heavy neutrino.
Such a scenario is unlikely to be realized in nature as it is weakly motivated and 
not even observed in SM $W$ boson decays.
Were one able to efficiently interpolate and extrapolate from constraints on parameter space benchmarks,
then LHC searches could be more fully utilized
and subsequently make more general statements on new physics models.
We explore this possibility here using ML techniques.
As a first step, we attempt to quantify how efficiently a typical, off-the-shelf
deep learning neural network can learn the production and decay cross sections for a $W_R$ boson
in the LRSM as a function of several (five) relevant mass and coupling parameters.
Knowing then the acceptance and selection efficiencies as a function of inputs,
which can be derived from Monte Carlo simulation, then one can in principle 
derive constraints on a fuller LRSM parameter space beyond that which was directly constrained.

\subsubsection{Model and Model Inputs}
For our purposes, it suffices to stipulate that the LRSM~\cite{Pati:1974yy,Mohapatra:1974hk,Mohapatra:1974gc,Senjanovic:1975rk,Senjanovic:1978ev}
is a gauge-extended scenario that postulates that 
the universe, in the UV limit, respects the gauge and parity symmetries
$\mathcal{G}_{\rm LRSM}={\rm SU}(3)_c\otimes{\rm SU}(2)_L\otimes{\rm SU}(2)_R\otimes{\rm U}(1)_{B-L}\otimes\mathcal{P}_X$ .
Here, ${\rm SU}(2)_R$ is a copy of the SM ${\rm SU}(2)_L$ gauge group
and describes maximally parity violating, right-handed interactions.
In addition, $\mathcal{P}_X$ is a discrete parity that ensures $L\leftrightarrow R$
field exchange symmetry.
To protect LRSM, one RH neutrino $(N_R)$ for each fermion generation is required.
The residual U$(1)$ that protects against anomalies conserves
the difference between baryon and lepton numbers $(B-L)$.
After Left-Right symmetry breaking at a scale $v_R\gg v_{\rm EW}\approx 246$ GeV,
the RH and $B-L$ gauge sector breaks down to the SM hypercharge gauge group,
${\rm SU}(2)_R\otimes{\rm U}(1)_{B-L}\otimes\mathcal{P}\to
{\rm U}(1)_{Y}$. This in turn reduces to QED when EWSB occurs.

In the mass basis, the relevant interaction Lagrangian that describes 
massive $W_R^\pm$ bosons from the ${\rm SU}(2)_R$ gauge group 
coupling to RH quarks and leptons is given by
\begin{eqnarray}
    \Delta\mathcal{L} =
    \frac{g_R}{\sqrt{2}}W_{R \mu}^{-}
    \sum_{i,j}\left[\overline{d}_L^i V^{\rm R}_{ij} \gamma^\mu P_R u_R^j\right] 
    +
    \frac{g_R}{\sqrt{2}}W_{R \mu}^{-}
    \sum_{i,j}\left[\overline{e}_L^i Y_{ij}^R \gamma^\mu P_R N_R^j\right] 
    + {\rm H.c.}
\end{eqnarray}
The mixing matrices $V^R$ and $Y^R$ describe the mixing between 
RH quarks and RH leptons with their respective mass eigenstates, 
in analogy to the CKM and PMNS matrices.
We ignore quark mixing and approximate both the CKM matrix 
and $V^R$ with $3\times3$ identity matrices.
The gauge couplings $g_L,~g_R$ controls the strength of LH,~RH currents and 
set $g_R=g_L\approx0.65$ in accordance with LR symmetry.

We consider at $\sqrt{s}=14$ TeV the canonical LRSM signature
featuring the production of same-sign leptons and two light quarks
via an $s$-channel $W_R$ and $N_1$, shown in Fig.~\ref{fig:PhenoAI:lrsm_xsec_diagram} and given by~\cite{Keung:1983uu}:
\begin{equation}
    u_i \overline{d_j} \to W_R^\pm \to \ell_1^\pm N_1
    \to \ell^\pm_1 \ell^\pm_2 d'_i \overline{u'_j}.
    \label{eq:lrsm_ML_DYX_ssllX_XLO}
\end{equation}
We decouple heavy neutrinos $N_2$ and $N_3$
and vary discretely the three mixing and two mass parameters
\begin{eqnarray}
Y_{e N_1},~Y_{\mu N_1},~Y_{\tau N_1}&\in&\{10^{-4},10^{-3},10^{-2},10^{-1},1\},
\\
~m_{N_1}&\in&\{15,30,45,60,75,100,150,300, \nonumber
\\
    & & \qquad 450,600,750,1000,1500,3000,4500\}~{\rm GeV},
\\
M_{W_R}&\in&\{1, 2, 3, 4, 5, 6, 8, 10, 12.5, 15, 17.5, 20\}~{\rm TeV}.
\end{eqnarray}
Altogether, we compute a total of $N=22.5$k cross sections,
which constitutes our data set.
We simulate the full $2\to4$ process at leading order
and do not make the narrow width approximation.
This means that a number of interesting kinematic limits are covered,
including:
non-resonant 
$W_R$ production when $M_{W_R}\gtrsim \sqrt{s}$~\cite{Ruiz:2017nip,Nemevsek:2018bbt},
the production of long-lived 
$N_1$~\cite{Nemevsek:2018bbt,Cottin:2018kmq},
and boosted regimes such as when $m_{N_1}\ll M_{W_R}\ll\sqrt{s}$~\cite{Mitra:2016kov,Mattelaer:2016ynf}.

\subsubsection{Computational Setup}
To simulate the process in Eq.~\ref{eq:lrsm_ML_DYX_ssllX_XLO},
we use \texttt{MadGraph5\_aMC@NLO}~\cite{Alwall:2014hca}
in conjunction with the \texttt{EffLRSM} UFO libraries~\cite{Mitra:2016kov,Mattelaer:2016ynf}.
The NNPDF 3.1 NLO + LUXqed parton densities~\cite{Bertone:2017bme}
are used and evolved using \texttt{LHAPDF6}~\cite{Buckley:2014ana}.
Total widths for $W_R$ and $N_{1}$ are computed on the fly
for each parameter space point~\cite{Artoisenet:2012st,Alwall:2014bza}.
The collinear factorization $\mu_f$ scale is dynamically to be 
half the sum of the transverse energy $E_T^k=\sqrt{p_T^{k2}+m_k^2}$
of all final-state particles $k$.
No generator-level phase space cuts are applied.

\subsubsection{Cross Section Learning}
Using the generated data we trained a 3-layer neural network with elu~\cite{DBLP:conf/iclr/2016} activation functions. As preprocessing we applied a base-10 logarithm to the input coupling strengths and to the cross sections that we aimed to predict. The data was z-score-normalized. The network was trained for 1000 epochs or until no mean squared error improvement was shown in 100 epochs (whichever occurred first).

Training was performed over random selections of the data in sizes of [500, 1000, 1500, 2000, 2500, 3000, 4000, 5000, 6000, 7000, 8000] events. 80\% of the data was used for training, the remaining 20\% for testing. For each of the training sizes the experiment was repeated 10 times (each with a new randomised selection of the total data). The test data was used to determine the relative error in the prediction of the cross section. This allowed us to determine the error made by the algorithm as function of the training size and, more importantly, when it becomes acceptable compared to the error made by the workflow through which the data was generated in the first place. The results of this procedure can be found in Figure~\ref{fig:PhenoAI:lrsm_loss}.

\begin{figure}
\begin{center}
\subfigure[]{\includegraphics[width=0.49\textwidth]{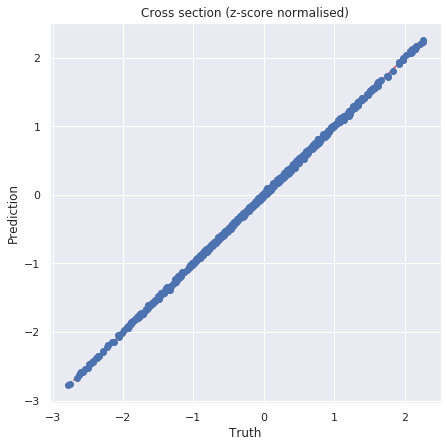}\label{fig:PhenoAI:predictionplot}}
\subfigure[]{\includegraphics[width=0.49\textwidth]{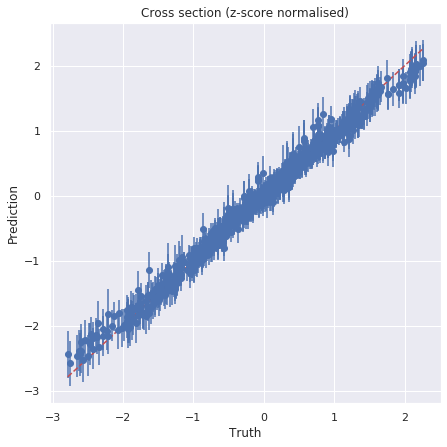}\label{fig:PhenoAI:predictionplot_errors}}
\end{center}
\caption{Truth-prediction plots for the trained algorithm without Monte Carlo Dropout (a) and with Monte Carlo Dropout (rate: 0.2) (b) (tentative). The cross sections are z-score normalised to help the training of the algorithm.}
\label{fig:PhenoAI:predictionplots}
\end{figure}

A more direct indication of the performance of the algorithm can be made in the form of a truth-prediction plot, in which the prediction of the algorithm is plotted against the value that the algorithm ought to predict. For a perfect algorithm the predictions would form a perfect diagonal line in such a plot. As seen in Figure~\ref{fig:PhenoAI:predictionplot}, after training we
find good agreement between predicted and truth rates. 

When applying an algorithm such as the one we trained, the user is not only interested in the prediction of the algorithm, but also in the prediction uncertainty due to the algorithm. To take all related sources of uncertainty into account, one also has to determine the uncertainty due to the model configuration and training (i.e. the epistemic uncertainty), as explained in Section~\ref{sec:PhenoAI:uncertainties}. Using the Monte Carlo Dropout technique discussed in Section~\ref{sec:PhenoAI:uncertainties}, it is possible to determine the event-by-event uncertainty on the predictions made by the algorithm. Although work on estimating the uncertainty (including the aleatoric uncertainty) is ongoing, a tentative result of this can be seen in Figure~\ref{fig:PhenoAI:predictionplot_errors} where the dropout rate was set to 0.2. To compensate the network capacity for this dropout rate, the width of the network layers is increased from [50, 50] to [60, 60].

From comparing Figure~\ref{fig:PhenoAI:predictionplot} and Figure~\ref{fig:PhenoAI:predictionplot_errors} we see that the trained algorithm has a reduced accuracy when the epistemic uncertainty is determined alongside a prediction. This is expected as the inclusion of dropout can degrade the performance of the algorithm. However, by altering the configuration of the neural network it might be possible to counteract this loss in performance. To find this configuration, 
in future iterations of the network
we plan on performing a hyperparameter scan for our model. Additionally, as our cross section data has some intrinsic uncertainty, as least from the Monte Carlo process through which it was calculated, we also plan on including aleatoric uncertainty determination.

\subsection{Learning the production cross sections of the Inert Doublet Model.}\label{sec:PhenoAI:idm}

One of the simplest extensions of the Standard Model (SM) is the addition of a second Higgs doublet in the so-called Two-Higgs-Double Models (2HDM).
If a $\mathcal{Z}_2$ parity is further imposed, the model can easily avoid the bounds from flavor changing neutral currents and provide a dark matter candidate. This realisation of the 2HDM is known as the Inert Doublet Model (IDM)~\cite{Deshpande:1977rw}, where the SM doublet  ($\mathbf{H_1}$) is $\mathcal{Z}_2$-even and the new doublet ($\mathbf{H_2}$) is odd.
The scalar potential in this case is given by:
\begin{equation}
\begin{split}
    V=\mu_1^2|\mathbf{H_1}|^2+\mu_2^2|\mathbf{H_2}|^2+\lambda_1|\mathbf{H_1}|^4+\lambda_2|\mathbf{H_2}|^4+\lambda_3|\mathbf{H_1}|^2|\mathbf{H_2}|^2
    \\
    +\lambda_4|\mathbf{H_1}^{\dagger}\mathbf{H_2}|^2+\frac{\lambda_5}{2}[(\mathbf{H_1}^{\dagger}\mathbf{H_2})^2+\mathrm{h.c.}] \,.
\end{split}    
\end{equation}
The above parameters are chosen so only the SM Higgs, $h$, acquires a vacuum expectation value (vev), thus maintaining the $\mathcal{Z}_2$ symmetry unbroken.
After imposing the correct values for the Higgs mass and vev, the model contain five free parameters which we choose to be the masses of the new scalars ($H^0$, $A^0$ and $H^\pm$):
\begin{align}
M_{H^0}^2=\mu_2^2+\frac{1}{2}(\lambda_3+\lambda_4+\lambda_5)v^2\,, \label{eq:PhenoAI:mH0} \\
M_{A^0}^2=\mu_2^2+\frac{1}{2}(\lambda_3+\lambda_4-\lambda_5)v^2\,, \label{eq:PhenoAI:mA} \\
M_{H^\pm}^2=\mu_2^2+\frac{1}{2}\lambda_3 v^2\,, \label{eq:PhenoAI:mHpm} 
\end{align}
and the two couplings:
\begin{equation}
  \lambda_2 ~\mbox{ and }~ 
  \lambda_L \equiv\lambda_3+\lambda_4+\lambda_5\,.
  \label{eq:PhenoAI:lambdas}
\end{equation}
In the above expressions, $v=246$\UGeV\ is the Higgs vev.

Searches for the inert scalars at the LHC are particularly challenging due to their small (electroweak) production cross sections. Since the new states are $\mathcal{Z}_2$-odd, they are pair produced at the LHC and the following 8 processes can be relevant for LHC searches:
\begin{equation}
    \label{eq:PhenoAI:IDMxsections}
   p p \rightarrow H^{0} H^{0}, \ A^{0} A^0, \ H^{0} A^0, \ H^0 H^+, \ A^0 H^+, \ H^+ H^-, \ H^0 H^- \ \mathrm{and} \ A^0  H^-.
\end{equation}
See \cite{Belanger:2015kga} and references therein for a discussion of LHC signatures and limits.  
The goal of this project is to train a Deep Neural Network to accurately predict the above (leading order) cross sections given as input the five free parameters of the model. In other words, to create a function that maps the free parameters of the IDM to their corresponding production cross sections: 
\begin{equation}
    g_{\phi}: x_{\rm{IDM}} \equiv \left(M_{H^0},M_{A^0},M_{H^\pm},\lambda_L,\lambda_2\right) \to \mathbf{\sigma}_{\rm{IDM}}
\end{equation}
where $\sigma_{\rm{IDM}}$ represents a vector containing the 8 cross section values. 

As a first step, 50000 samples were generated following the method of jittered  sampling \cite{Bellhouse:jittered},  from a parameter space chosen as:
\begin{equation}
  50<M_{H^0},\,M_{A^0},\,M_{H^\pm} <3000\UGeV; \quad  -2\pi<\lambda_{2},\:\lambda_{L}<2\pi. 
\end{equation}
All the cross sections were computed at leading order using \textsc{MadGraph}~2.6.4~\cite{Alwall:2014hca} and the IDM UFO implementation from the FeynRules data  base~\cite{Goudelis:2013uca,Belanger:2015kga}.
Since the expected integrated luminosity at the High-Luminosity LHC is about $3\Upb^{-1}$, we imposed a lower limit on the cross sections of our  data set of $\sigma_{\rm{min}}=10^{-7}\Upb$ by discarding the cross sections below this limit. Afterwards, the remaining data was divided as training and test data in a 70:30 split.

An efficient training of the neural network requires some re-scaling of the input variables. For this, we followed the prescription and recommendations in~\cite{Otten:2018kum} and pre-processed the model parameters via a z-score transformation: 
\begin{equation}
    x'_{\rm{IDM}} = \frac{x_{\rm{IDM}} - \mu (x_{\rm{IDM}})}{\sigma (x_{\rm{IDM}})}\,,
\end{equation}
where $\mu (x_{\rm{IDM}})$ and $\sigma (x_{\rm{IDM}})$ are the mean and standard deviation of $x_{\rm{IDM}}$, respectively. In addition the corresponding cross sections were logarithmically rescaled as:
\begin{equation}
    \sigma'_{\rm{IDM}} = \log \left[\frac{\sigma_{\rm{IDM}}}{\min (\sigma_{\rm{IDM}})} \right].  
\end{equation}

The hyperparameters of the training algorithm were set as follows. As initializer of the neural network weights we chose He normal. In each hidden layer, we set LeakyReLU as an activation function. In order to obtain an approximation of the Bayesian uncertainties~\cite{gal2015dropout} as Monte Carlo dropout  a ``permanent'' dropout layer was implemented after each hidden layer, where ``permanent'' means that the dropout is present not only during training, but also for inferences. To take into account the pre-processing of the target values we used a custom loss function that minimises the mean absolute percentage error (MAPE) of the original cross sections:
\begin{equation}
    L(\sigma'_{\rm{true}}, \sigma'_{\rm{pred}}) = \frac{1}{N}\sum_{i=1}^N |1-\exp(\sigma'_{\rm{pred}} - \sigma'_{{\rm{true}}})|\,,
\end{equation}
where N is the batch-size which we choose to be 32. Furthermore, we applied the Adam optimizer with an initial learning rate of $\alpha_{\rm{init}}=10^{-3}$ and the EarlyStopping callback with a patience of 50. After 500 epochs have ended or EarlyStopping has terminated the iteration, the learning rate was divided by 2 and the training continues until 10 of those iterations were completed. To choose the best configuration, we ran a scan over the rest of the hyperparameters: the number of hidden layers, the number of artificial neurons, $\lambda$ of the L2 regularization term and the dropout fraction, and trained a neural network with each combination for the $p p \rightarrow H^{0} H^{0} $ process.  Finally, we trained one neural network for each of the remaining production processes using the configuration that better minimised the MAPE for the first process. This configuration is formed by 6 hidden layers with 192 artificial neurons, $\lambda=10^{-5}$ and a dropout fraction of 1\,\%. The training  data set, the code and the trained neural networks are presented in the GitHub repository  \url{https://github.com/SydneyOtten/IDM_XS} fulfilling the criteria from sections \ref{sec:PhenoAI:modelcoll}, \ref{sec:PhenoAI:datacoll} and \ref{sec:PhenoAI:codecoll}.

In order to test the performance of the neural network, 100 sample predictions were drawn for each point in the test data and their mean $\mu(\sigma_{pred})$ and standard deviation std$(\sigma_{pred})$ was computed. From this, we computed the relative error (RE), 
\begin{equation}
\mathrm{RE}=\left|\frac{(\sigma_{pred}-\sigma_{true})}{\sigma_{true}}\right| ,
\end{equation}
which quantifies the distance between $\mu(\sigma_{pred})$ and the true cross section $\sigma_{true}$, and the coefficient of variance (CV), 
\begin{equation}
\mathrm{CV}=\frac{\mathrm{std}(\sigma_{pred})}{\mu(\sigma_{pred})},   
\end{equation}
which describes the estimation of the Bayesian uncertainty.

\begin{figure}[t]
\centering     
\subfigure[$p p \rightarrow A^{0}H^{+}$]{\label{fig:PhenoAI:IDMerrorsa}\includegraphics[width=78mm]{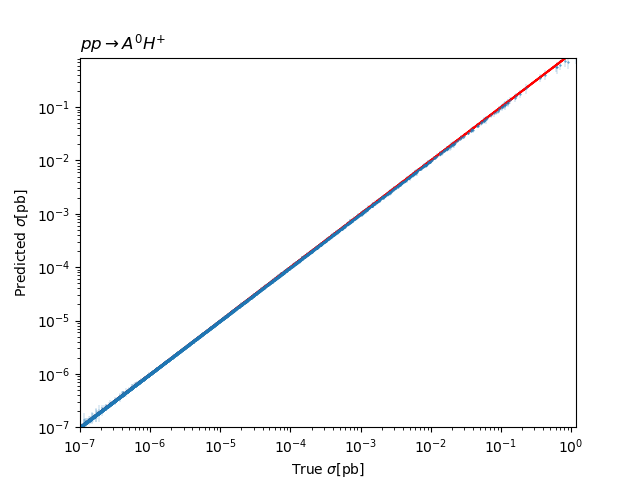}}
\subfigure[$p p \rightarrow A^{0}A^{0}$]{\label{fig:PhenoAI:IDMerrorsb}\includegraphics[width=78mm]{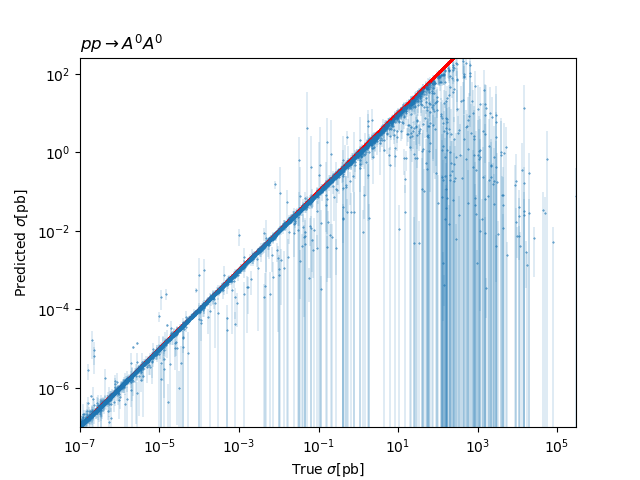}}
\caption{Mean predicted vs.\ true cross section for the processes with (a) the best and (b) the worst performance. The error bars correspond to 1~std from the mean prediction in each point.}\label{fig:PhenoAI:IDMerrors}
\end{figure}

\begin{figure}[t]
	\centering
	\subfigure[$p p \rightarrow A^{0}H^{+}$]{\label{fig:PhenoAI:Massvsxsectiona}\includegraphics[width=80mm]{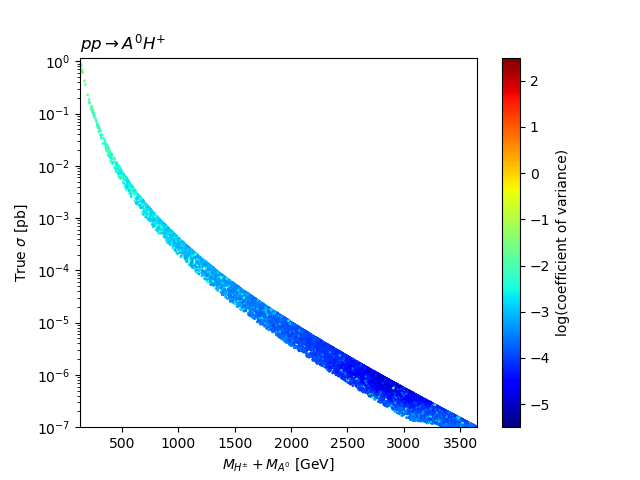}}
\subfigure[$p p \rightarrow A^{0}A^{0}$]{\label{fig:PhenoAI:Massvsxsectionb}\hspace*{-4mm}\includegraphics[width=80mm]{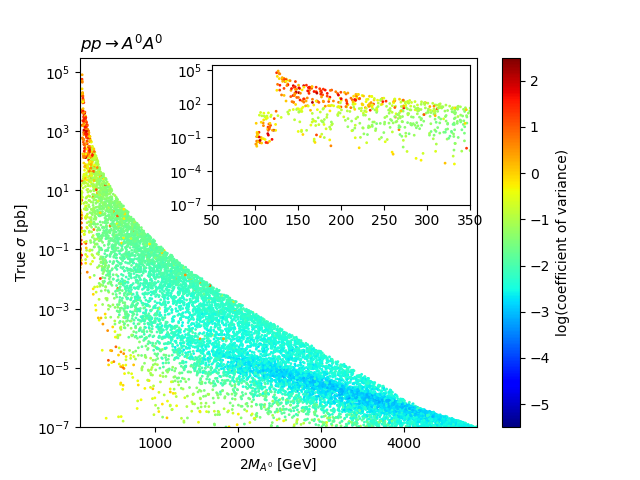}}
	\caption{True cross section vs.\ effective mass for the processes with (a) the best and (b) the worst performance. In color, the logarithm of the coefficient of variance is shown.  }\label{fig:PhenoAI:Massvsxsection}
	\end{figure}

The results are summarised in Table~\ref{fig:PhenoAI:summaryerrors}. For the processes of associated production of two different inert scalars, we obtained quite good results. The best one overall is for $A^{0}H^{+}$ production, 
with $\mu(\mathrm{RE})\approx 0.005$, $\mu(\mathrm{CV})\approx 0.03$, and the 1~std interval around the mean predicted cross section containing the true value for 99.97\% of the test points.
For the pair-production processes, however, the outcomes are not ideal: we observe large REs and CVs, specially the regions with large cross sections. The worst case is $A^{0}A^{0}$ production 
with $\mu(\mathrm{RE})\approx 0.1$ and  $\mu(\mathrm{CV})\approx 0.19$, and 1~std around the mean prediction containing the true value for only 95.09\% of the test points. 
The mean predicted vs.\ the true cross section for these two cases is shown in Fig.~\ref{fig:PhenoAI:IDMerrors}. We see here that 1.~$\sigma(pp\to A^0A^0)$ reaches much higher values than $\sigma(pp\to A^0H^+)$ and 2.~the largest uncertainties arise for the highest cross sections. 
To understand this further, we plot in Fig.~\ref{fig:PhenoAI:Massvsxsection} the true cross section vs.\ the mass of the final state, with CV shown as color code.   
We observe that as the cross sections get larger so does the CV, a fact that is more notorious for the pair-production processes, which reach much higher values, specially when $2M_{A^0}\approx M_{h}$ where $h$-mediated production becomes kinematically allowed. Moreover, as expected, the cross sections peak towards low masses; in this region the density of points is rather low, which is also a cause of larger uncertainties. This suggests that the target values (i.e.\ the values of the cross sections) of the training sample should be more evenly distributed. Nevertheless, there is the positive conclusion that, 
in general, we observe a direct proportionality between the relative error and the standard deviation of the predictions. This is very important in order to ensure a correct interpretation of the uncertainty on the prediction.  

\begin{center}
\begin{table}[]
    \centering
    \begin{tabular}{|c|c|c|c|c|c|c|c|c|}
    \hline
          & $H^{0} H^{0}$&$A^{0} A^0$&$H^+ H^-$&$H^{0} A^0$&$H^0 H^+$&$A^0 H^+$&$H^0 H^-$&$A^0H^-$\\ \hline
        
        $\mu(\mathrm{RE})$ & 0.0303 & 0.1049& 0.2259 & 0.0058 & 0.0076 & 0.0048   & 0.0057 & 0.0072 \\ \hline
        $\mu(\mathrm{CV})$ & 0.0850 & 0.1880& 0.1508 & 0.0272 & 0.0402 & 0.0276  & 0.0276 & 0.0287 \\ \hline
        within 1 std & 0.9833  & 0.9509 & 0.9812& 0.9981 & 0.9995 & 0.9997  & 0.9886  & 0.9817 \\ \hline
       
    \end{tabular}
    \caption{Summary of the accuracy of the predictions of the trained neural network from the test data: 
    mean relative error, $\mu(\mathrm{RE})$, 
    mean coefficient of variance, $\mu(\mathrm{CV})$, and
    fraction of test points whose true values lie within 1~std from the mean correspondent prediction, denoted as ``within 1~std'',  for the eight production processes.}
    \label{fig:PhenoAI:summaryerrors}
\end{table}    
\end{center}

In summary, we obtained first results from trained neural networks that predict the LHC production cross sections for inert scalars in the IDM, with an estimation of the Bayesian uncertainty. The next step of this project will focus on reducing this uncertainty and, by making sure that the true value is inside the 1~std uncertainty interval of the prediction, reducing the relative error of the predictions over the full parameter space for all the processes under consideration. One way to achieve this regards implementing Dropout-based Active Learning~\cite{Tsymbalov_2018}.

\subsection{Global fits of Gambit Zenodo data}

The GAMBIT collaboration has released the data resulting from their 
global fits of a variety of beyond Standard Model models \cite{Athron:2017kgt, Athron:2017qdc, Athron:2017yua, Athron:2018ipf, Athron:2018hpc, Athron:2018vxy, Hoof:2018ieb, Chrzaszcz:2019inj} in Zenodo. These data can be trained using deep learning methods to predict, for instance, dark matter and LHC observables or likelihoods related with those observables for a posteriori interpolation. 

One example is the seven-dimensional Minimal Supersymmetric Standard Model (MSSM7) \cite{Athron:2017yua} for which  we have created a machine learning model in the form of two stacked deep neural networks to perform a regression on the MSSM7 combined likelihood. A total of 22.6 million samples were used for the training and evaluation of the models. Data exploration reveals that $\approx 595000$ of those samples have a likelihood of 0 whereas all the other samples range from $\approx-450$ to $-255$. The great void between -255 and 0 enhances the difficulty for a single neural network to perform well in every likelihood region. This particular inhomogeneity of the data raises the standard deviation from $\approx 7$ when excluding the zero likelihoods to $\approx 42$ when including them. Thus, it was no surprise to find an extraordinarily well performing deep network when excluding the zeros from the training procedure. However, when training a neural network on the full  data set, we achieved an accuracy of 100\,\% for identifying zero likelihoods. We have implemented a stacking mechanism that combines the knowledge of two deep neural networks by merging them into a deep hybrid network.

\subsection*{Deep Hybrid Network Architecture and Training}

The two neural networks comprising the hybrid architecture were constructed and trained with \verb+Tensorflow+ and \verb+Keras+. Their most important difference is due to the data they were being trained on:
\begin{description}
\item [Net A] is trained on the full data set including the samples with a likelihood of 0.
\item [Net B] is trained on the full data set excluding the samples with a likelihood of 0.
\end{description}
The stacking mechanism is therefore simple: firstly, the input is processed by net A. If it predicts a 0, the final prediction is 0. If it does not predict 0, query net B and give its prediction as the final result.
Nets A and B share most of their hyperparameters. Thus, if not explicitly mentioned otherwise, the subsequent network features are true for the procedures of both nets:
\begin{description}
\item [Data Preprocessing:] The input, as well as the output, are z-score normalised, i.e. for sample $x_i$, it is transformed into $x_i'$ and
\begin{equation}
x_i'=\frac{x_i-\bar{\mu} (x)}{\sigma (x)},
\end{equation}
where $\bar{\mu}$ and $\sigma$ are the mean and standard deviation. However, for the parameters $Q$ and $\rm{sgn}(\mu)$, $\bar{\mu}=0$ and $\sigma=1$. After normalising the data, it is split into three parts: first of all, 5\% of the $2.26\cdot 10^7$ samples are stored as test samples the neural networks will not see at all during their training. The other 95\% are then again split into training and validation samples with a ratio of 9:1 respectively. The training samples are the ones that the loss function gets to compare to the model predictions while the validation samples are only used as a monitor for the performance. However, as will be explained in the training paragraph, the Network will be slightly biased towards the validation set.
\item [Network Topology:] The input is processed by 8 layers of fully connected neurons with 64 (net A) or 100 (net B) neurons with the \verb+selu+ activation function. The initial weights are drawn from a normal distribution with $\mu=0$, $\sigma_{\rm{net A}}=0.125$ and $\sigma_{\rm{net B}}=0.1$.
\item [Training:] The loss function measuring the deviation of the model predictions and true values is minimised by the \verb+ADAM+\,\cite{2014arXiv1412.6980K} optimiser with default values beside the learning rate and a batch-size of 180000 and has been customised with respect to the data preprocessing. Our loss function is a modification of the mean absolute error (MAE):
\begin{equation}
\rm{Loss}\,(\hat{y_i}(x_i),y_i(x_i))=\frac{1}{N}\sum_{i=1}^N\sigma(y)\cdot\left|\hat{y_i}-y_i\right|,
\end{equation}
where $\hat{y_i}$ is the predicted and $y_i$ is the true label for input $x_i$. Additionally, a learning rate scheduling with $\alpha_i=0.01$, $\alpha_f=10^{-5}$ and a factor of 1.7 dividing the learning rate for each iteration. One iteration is equivalent to 2500 epochs of training or until the \verb+EarlyStopping+ routine has ended the iteration with a patience of 200 epochs. For each iteration, the model parameters giving the best validation loss are loaded into the architecture and the optimisation continues which causes the bias mentioned earlier.
\end{description}

\subsection*{Performance measurement}

\begin{figure}[t]
	\centering
    \includegraphics[width=0.8\textwidth]{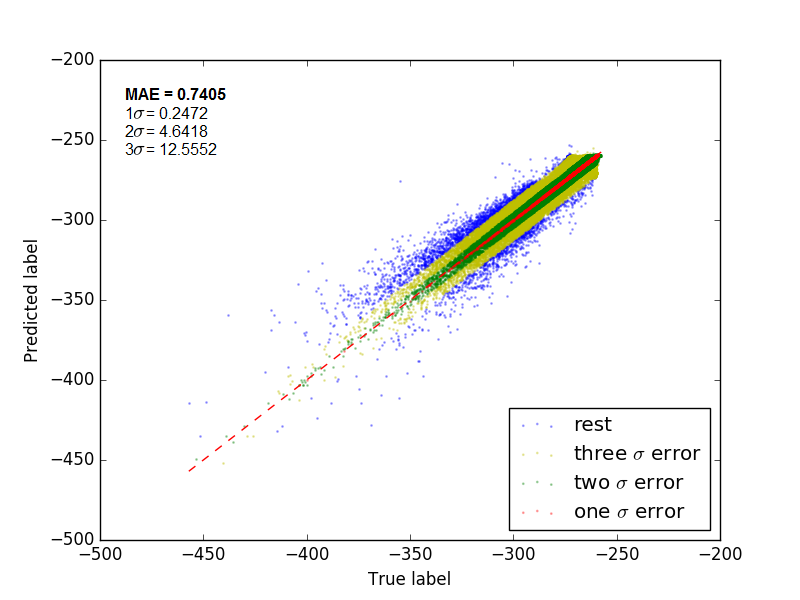}
    \caption{A validation plot of net B, showing the true label $y_i$ on the x-axis and the predicted label $\hat{y_i}$ on the y-axis with a mean absolute error (MAE) of 0.7405. The test points are additionally showing the error bounds: $1\sigma$ in red, $2\sigma$ in green, $3\sigma$ in yellow and the rest in blue.}
    \label{fig:PhenoAI:gambit}
\end{figure}

To measure the performance of the model, we evaluate both neural networks by comparing their predictions to 5\% of the corresponding data sets (the test sets).  The test set of net A consists of 1.13 million points, including 50650 samples whose likelihood is 0. Although the training procedure for net A is a typical regression, its function in the stack is to classify whether the parameter space point corresponds to a likelihood of 0 or not. Therefore, the only interesting measure for net A is the binary classification accuracy which was 100\,\%. For net B, an evaluation plot is shown in Figure \ref{fig:PhenoAI:gambit}. 

\section{Conclusions}
\label{sec:PhenoAI:conclusion}
 Machine Learning (ML) 
 can find various applications in HEP phenomenology. 
 In this contribution, the use of ML based regression and classification of information for high-dimensional HEP models was discussed 
 with emphasis on
 best practices for a sustainable reuse of phenomenological information. 
 To give an example, phenomenological 
 studies of 
 the likelihood of a particular HEP model should publish the multidimensional records of all likelihood evaluations.
 Such data can be used to train ML models which allow to evaluate the likelihood of this HEP model in its full dimensionality. 
The ML code and the trained ML model should then be published with appropriate meta data specifying input and output variables, boundaries of the training regions, etc., to allow an easy reuse.
We propose to build a HEP-wide framework storing such information.
 
As proof of concept and feasibility, we showed various examples of applications such as 
ML-based regression of cross sections in left-right symmetric models (LRSM) and in the two-higgs doublet model (2HDM) and the regression of likelihoods for the minimal supersymmetric standard model (MSSM7) as provided by the GAMBIT collaboration.




%% file: darkml/darkmachines.main.tex
\graphicspath{{darkml/}}


\definecolor{darkred}{rgb}{0.8,0.2,0.3}
\definecolor{darkblue}{rgb}{0.15, 0.2, .85}

\def\GeV{~\textrm{GeV}}
\def\be{\begin{equation}}
\def\ee{\end{equation}}
\newcommand{\beq}{\begin{eqnarray}}
\newcommand{\eeq}{\end{eqnarray}}
\def\ba{\begin{align}}
\def\ea{\end{align}}
\newcommand{\tn}{\textnormal}
\def\nn{\nonumber}
\newcommand{\comment}[1]{\textcolor{darkred}{[#1]}}
\newcommand{\MB}[1]{{\color{blue}[{\bf MB:} #1]}}
\newcommand{\LH}[1]{{\color{green}[{\bf LH:} #1]}}
\newcommand{\RR}[1]{{\color{red}[{\bf RR:} #1]}}
\newcommand{\JS}[1]{{\color{orange}[{\bf JS:} #1]}}
\newcommand{\RV}[1]{{\color{brown}[{\bf RV:} #1]}}


\def\arraystretch{1.25}
\setlength{\tabcolsep}{12pt}

\makeatletter
\def\l@subsubsection#1#2{}
\def\l@subsubsubsection#1#2{}
\makeatother

\setcounter{secnumdepth}{4}

\chapter{Model-Independent Signal Detection: A Challenge using Benchmark Monte Carlo Data and Machine Learning}
{\it M.~van Beekveld, S.~Caron, A.~De Simone, A.~Farbin,  L.~Hendriks, A.~Jueid,  A.~Leinweber,  J.~Mamuzic,  E.~Mer\'enyi,  A.~Morandini,  C.~Nellist,  S.~Otten,  M.~Pierini,  R.~Ruiz de Austri, S.~Sekmen,  J.~Schouwenberg, R.~Vilalta, M.~White}

 \label{sec:darkmachines}

\begin{abstract}
We discuss model-independent signal detection algorithms, with a particular focus on approaches that are based on unsupervised machine learning. We also offer a set of simulated LHC events, corresponding to $10\, {\rm fb}^{-1}$ of data. These events can be used as a benchmark dataset, for example for the comparison of signal detection algorithms. We explain the main features, the data format and describe the use of this data for an upcoming data challenge. The data is available at the webpage
 \url{https://www.phenoMLdata.org}.
\end{abstract}

\section{Introduction and Goals}
\label{sec:intro}

\paragraph*{Problem}
The Standard Model (SM) has been tremendously successful in describing particle physics phenomena. Nevertheless, many questions still remain unanswered, e.g.~the origin of neutrino masses, the nature of dark matter, or the dynamics of electroweak symmetry breaking. Therefore, it is commonly accepted that physics beyond the SM (BSM) is needed in order to provide answers to the questions not addressed in the SM. A key ingredient for the journey towards a new physics discovery is handling the huge amount of complex experimental data collected at LHC. LHC data was initially analyzed for various signals that were predicted by high energy models that extend the SM. Typical examples are supersymmetry (SUSY) or models with extra dimensions. Since these searches did not show any significant deviations from the SM, the LHC search strategy was expanded by using so-called "simplified models" and "effective models". For simplified models, a certain production and a decay of a new hypothetical particle  is assumed, and the model is tested using LHC data by optimizing data selection criteria on the energy, momenta and types of particle predicted by the model. For effective models, a new effective interaction is added to the SM Lagrangian and the new interaction is typically constrained with the measurement of SM processes.
A sign of new particles typically shows up as an overproduction of events (compared to the SM) in a specific data-selection where the number of events expected from SM processes is compared to the number of measured events in statistical tests. A hint of new physics requires that the "SM-only" hypothesis is highly disfavoured. Often the test is quantified with the help of a p-value defined as the probability that a given result (or a more significant result) occurs under the SM hypothesis. A typical requirement for the discovery of an \textit{expected} signal (such as the Higgs particle) is $p<3 \times 10^{-7}$ corresponding to 5 standard deviations (5$\sigma$).\\
To date, no signal of BSM physics has been found at the LHC. However, the new physics could look different than generally assumed. This project deals with the question of how to search for a signal in collider data without adopting a specific signal hypothesis. 

\paragraph*{Attempts}

A few attempts have been made to systematically search for new physics without signal assumption by scanning specific observables, such as the sum of the transverse momenta, or the invariant mass. Scans have been done with the help of model-independent (i.e.~unsupervised) algorithms to locate anomalies. Such {\textit general} searches without an explicit BSM signal assumption have been 
been performed by the D\O\ Collaboration~\cite{Abbott:2000fb,Abbott:2000gx,Abbott:2001ke, Abazov:2011ma} at the Tevatron using an unsupervised multivariate signal detection algorithm termed SLEUTH,
by the H1 Collaboration~\cite{Aktas:2004pz,Aaron:2008aa} at HERA using a 1-dimensional  signal detection algorithm, 
and by the CDF Collaboration~\cite{Aaltonen:2007dg, Aaltonen:2008vt} at the Tevatron (using again 1-dimensional algorithms). A version of these 1-dimensional signal detection algorithms used in general searches is known as BUMPHUNTER in the HEP community~\cite{Choudalakis:2011qn}.
At the LHC, versions of such searches have been performed 
by the ATLAS Collaboration at $\sqrt{s}=13$~\UTeV{}~\cite{Aaboud:2018ufy}, and preliminary versions have been performed  by the ATLAS and CMS Collaboration at $\sqrt{s}=7$ and $8$~\UTeV{}.  Here, the ATLAS experiment proposed that
the observation of one or more significant deviations in some phase-space region(s) can serve as a trigger to perform 
dedicated and model-dependent analyses where 
these `data-derived' 
phase-space region(s) can be used as 
signal region(s). 
Such an analysis can then determine the level of significance by testing the SM hypothesis in these signal regions in a second dataset. Since the signal region is known also control selection can be defined to determine the background expectations in the signal region(s).\\
The field of machine learning (ML), sitting at the intersection of computational
statistics, optimization, and artificial intelligence, has witnessed unprecedented progress over the past decade. Research in ML has recently led to the development
of new and enhanced anomaly detection methods that could be used and extended for applications employing LHC or astroparticle data.
Examples of such outlier detection algorithms recently proposed for HEP include density-based methods \cite{DeSimone:2018efk}, model-independent searches with multi-layer perceptrons \cite{DAgnolo:2018cun} , autoencoders \cite{Farina:2018fyg, Blance:2019ibf,Hajer:2018kqm}, variational autoencoders \cite{Cerri:2018anq,Otten:2019hhl} or ML extended bump-hunting algorithms~\cite{Metodiev:2017vrx,Collins:2018epr}.

\paragraph*{Methodology} 
This contribution aims to initiate a comparison of signal detection algorithms. To this end, we supply a benchmark dataset containing simulated high-energy collision data. Furthermore, we provide a (non-exhaustive) list of methods that may be employed to extract a possible signal from this dataset in a model-independent and/or unsupervised way.

\section{Data Description}
\label{sec:data}

\subsection{Data generation procedures}

We generate LHC events corresponding to a center-of-mass energy of 13~TeV. Events for the background and signal processes are generated at leading order (LO) with up to two additional partons in the matrix element using the event generator MG5$\_$aMC@NLO~v6.3.2~(Madgraph) and versions above~\cite{Alwall:2014hca} with the NNPDF PDF set\cite{Ball:2017nwa} using $5$ flavors in the definition of the proton. Madgraph is interfaced to Pythia 8.2~\cite{Sjostrand:2014zea},  that handles the showering of the matrix element level generated events. The matching with the parton shower, needed in the case when one or more additional jets are generated in Madgraph, is done using the MLM merging prescription \cite{Mangano:2002ea}. Then, a quick detector simulation is performed with Delphes~3~\cite{deFavereau:2013fsa,Cacciari:2011ma}, using a modified version of the ATLAS detector card. Pileup is not included in this dataset. A repository of the data scripts that are used to generate the events is on GitHub
\footnote{\url{https://github.com/melli1992/unsupervised_darkmachines}}.\\
The final state objects, as described in Table~\ref{tab:objects}, are stored in a one-line-per-event text file (see below Section~\ref{subsec:format} for details). An event consists of a variable number of objects. An event is stored when at least one of the following requirements are fulfilled:
\begin{itemize}
    \item At least one (b)-jet with transverse momentum $p_T > 60$ GeV and pseudorapidity $|\eta| < 2.8$, or
    \item at least one electron with $p_T > 25$ GeV and $|\eta| < 2.47$, except for $1.37 < |\eta| < 1.52$, or
    \item at least one muon with $p_T > 25$ GeV and $|\eta| < 2.7$, or
    \item at least one photon with $p_T > 25$ GeV and $|\eta| < 2.37$.
\end{itemize}
Of course, these are unrealistic trigger requirements, but we aim to create a flexible data set that allows for different types of studies that might need different selection criteria. The $\eta$-restriction on the electrons models a veto in the crack regions as often applied in ATLAS analyses. Such a veto can also be applied to photons by the user. Note that for the processes with the largest cross sections ($W^{\pm}/\gamma/Z + {\rm jets}$ and QCD jet production) we have applied cuts on $H_T >100$~GeV and $600$~GeV respectively to make the data generation manageable. The observable $H_T$ is defined as the scalar sum of the transverse momenta of all jets (with $p_{T,j_i} > 20$~GeV and $|\eta_{j_i}| <2.8$):
\begin{eqnarray}
\label{eq:defHT}
    H_T = \sum_i |p_{T,j_i}|.
\end{eqnarray}
Therefore, if one includes any of these processes in their analysis, one must make sure that the same cuts are also applied to the the other processes, which impacts the cross sections that are indicated in Table~\ref{table:procs} (and therefore the event weights). \\
The requirements on the final states objects that are stored are
\begin{itemize}
    \item ($b$-)jet: $p_T > 20$ GeV and $|\eta| < 2.8$,
    \item electron/muon: $p_T > 15$ GeV and $|\eta| < 2.7$,
    \item photon: $p_T > 20$ GeV and $|\eta| < 2.37$.
\end{itemize}
This means that, for example, a jet with $p_T = 10$~GeV is not included in the dataset. The detector simulation as performed by Delphes removes any electrons with $|\eta| > 2.5$, as the reconstruction efficiency is set to $0$ beyond that point.  \\
The scale choice is set dynamically by Madgraph during the event generation. The resulting cross sections are not reweighted with any of the available higher-order and/or resummed cross sections. All relevant SM (background) processes that have been generated are summarized in Table~\ref{table:procs}. For each process, the total number of generated events ($N_{\rm tot}$) is at least the number that is needed for 10 fb$^{-1}$ of data ($N_{10\, {\rm fb}^{-1}}$). \\

\begin{table}[t]
\centering
\begin{tabular}{|c|c|}
\hline
\textbf{Symbol ID} & \textbf{Object} \\ \hline
\texttt{j} & jet \\ \hline
\texttt{b} & $b$-jet \\ \hline
\texttt{e-} & electron ($e^-$)\\ \hline
\texttt{e+} & positron ($e^+$)\\ \hline
\texttt{m-} & muon ($\mu^-$)\\ \hline
\texttt{m+} & antimuon ($\mu^+$)\\ \hline
\texttt{g} & photon ($\gamma$)\\ \hline
\end{tabular}
\caption{Definition of symbols used for final-state objects. Only $b$-quark jets are tagged, no $\tau$- or $c$-jets have been defined.}
\label{tab:objects}
\end{table}

\begin{table*}
\centering
\begin{tabular}{ |l|l|l|l| }
\hline
\multicolumn{4}{ |c| }{\textbf{SM processes}} \\
\hline
Physics process & Process ID & $\sigma$ (pb) & $N_{\rm tot}$ ($N_{10\,{\rm fb}^{-1}}$) \\ 
\hline
$ p p \rightarrow j j$ \, & njets & 19718$_{H_T > 600 \, \textrm{GeV}}$ & 415331302 (197179140) \\
$ p p \rightarrow W^{\pm}(+2j)$ & w\_jets & 10537$_{H_T > 100 \, \textrm{GeV}}$ & 135692164 (105366237) \\  
$ p p \rightarrow \gamma(+2j)$  & gam\_jets & 7927$_{H_T > 100 \, \textrm{GeV}}$ & 123709226 (79268824) \\
$ p p \rightarrow Z(+2j)$ & z\_jets & 3753$_{H_T > 100 \, \textrm{GeV}}$ & 60076409 (37529592) \\
$ p p \rightarrow t\bar{t} (+2j)$ & ttbar & 541 & 13590811 (5412187) \\
$ p p \rightarrow W^{\pm}t (+2j)$ & wtop  & 318 & 5252172 (3176886) \\
$ p p \rightarrow W^{\pm}\bar{t}(+2j)$ & wtopbar & 318 & 4723206 (3173834) \\
$ p p \rightarrow W^+W^{-}(+2j)$ & ww & 244 & 17740278 (2441354)  \\
$ p p \rightarrow t+$jets$(+2j)$ & single\_top  & 130 & 7223883 (1297142)\\
$ p p \rightarrow \bar{t}+$jets$(+2j)$ & single\_topbar & 112 & 7179922 (1116396) \\
$ p p \rightarrow \gamma\gamma(+2j)$ & 2gam & 47.1 & 17464818 (470656) \\
$ p p \rightarrow W^{\pm}\gamma(+2j)$ & Wgam & 45.1 & 18633683 (450672) \\
$ p p \rightarrow ZW^{\pm}(+2j)$ & zw & 31.6 & 13847321 (315781) \\
$ p p \rightarrow Z\gamma(+2j)$ & Zgam & 29.9 & 15909980 (299439) \\
$ p p \rightarrow ZZ(+2j)$ & zz & 9.91 & 7118820 (99092)  \\
$ p p \rightarrow h(+2j)$ & single\_higgs & 1.94 & 2596158 (19383) \\
$ p p \rightarrow t\bar{t}\gamma(+1j)$ & ttbarGam & 1.55 & 95217 (15471) \\
$ p p \rightarrow t\bar{t}Z$ & ttbarZ & 0.59 & 300000 (5874)  \\
$ p p \rightarrow t\bar{t}h (+1j)$ & ttbarHiggs & 0.46 & 200476 (4568) \\
$ p p \rightarrow \gamma t (+2j)$ & atop & 0.39 & 2776166 (3947) \\
$ p p \rightarrow t\bar{t}W^{\pm}$ & ttbarW & 0.35 & 279365 (3495)\\
$ p p \rightarrow \gamma \bar{t} (+2j)$ & atopbar & 0.27 & 4770857 (2707)  \\
$ p p \rightarrow Zt (+2j)$ & ztop & 0.26 & 3213475 (2554) \\
$ p p \rightarrow Z\bar{t} (+2j)$ & ztopbar & 0.15 & 2741276 (1524) \\
$ p p \rightarrow t\bar{t}t\bar{t}$ & 4top & 0.0097 & 399999 (96) \\
$ p p \rightarrow t\bar{t}W^{+}W^{-}$ & ttbarWW & 0.0085 & 150000 (85)  \\
\hline
\end{tabular}
\caption{Generated background processes (first column) with the corresponding identification (second column), the LO cross section $\sigma$ in pb (third column) and the total number of generated events $N_{\rm tot}$ (fourth column). In the last column, we also indicate the  number of events corresponding to $10$ fb$^{-1}$ of data ($N_{10\,{\rm fb}^{-1}}$).}
\label{table:procs}
\end{table*}

\noindent For the BSM scenarios (signal events) we have chosen two SUSY channels: gluino ($\tilde{g}$) pair and lightest stop ($\tilde{t}_1$) pair production. The production channels and decays are 
\begin{eqnarray}
    p p \rightarrow \tilde{g} \tilde{g}, \quad \tilde{g} \rightarrow t \bar t \tilde{\chi}^0_1, \nonumber \\
    p p \rightarrow \tilde{t}_1 \tilde{t}_1,  \quad \tilde{t}_1 \rightarrow t \tilde{\chi}^0_1. \nonumber 
\end{eqnarray}
For the gluino events, we used a simplified model in which the lightest SUSY particle is a 1~GeV neutralino. The considered masses of the gluino are indicated in Table~\ref{table:bsm_procs}. All other SUSY particle are set to 4.5~TeV. For the stop production scenarios, we assume a more realistic SUSY scenario with a varying lightest neutralino ($\tilde{\chi}^0_1$) mass. The masses of $\tilde{t}_1$ and $\tilde{\chi}^0_1$ are provided in Table~\ref{table:bsm_procs}. Again for this scenario, all other SUSY masses are set to 4.5~TeV. \\
We include a second BSM model corresponding to a leptophobic topcolor $Z'$ model~\cite{Harris:2011ez}, where an on-shell $Z'$ boson is produced that subsequently decays into a pair of top quarks:
\begin{eqnarray}
    p p \rightarrow Z'\rightarrow t\bar t.
\end{eqnarray}
The masses of the $Z'$ are provided in Table~\ref{table:bsm_procs}. In Table~\ref{table:bsm_procs}, one may find the process ID, cross sections $\sigma$, and total number of generated events $N_{\rm tot}$ of the BSM processes mentioned above.  \\
Generally, the processes with lower cross sections are harder to extract out of the background events, as such processes result in a lower number of signal events. A notable exception that is present in the BSM dataset is the scenario where the lightest stop mass is 220~GeV (process ID stop\_01). Although the cross section of this production scenario is relatively high, the signal events are nearly indistinguishable from the background events due to their topology, making it extremely difficult to separate the signal events from the background events. 

\begin{table*}
\centering
\begin{tabular}{ |l|l|l|l| }
\hline
\multicolumn{4}{ |c| }{\textbf{BSM processes}} \\
\hline
Physics process & Process ID & $\sigma$ (pb) & $N_{\rm tot}$ ($N_{10\,{\rm fb}^{-1}}$)  \\ 
\hline
$ p p \rightarrow \tilde{g}\tilde{g} \, (1 \, {\rm TeV})$ & gluino\_01 & 0.20 & 50000 (2013) \\
$ p p \rightarrow \tilde{g}\tilde{g}\, (1.2 \, {\rm TeV})$ & gluino\_02 & 0.05 & 50000 (508) \\
$ p p \rightarrow \tilde{g}\tilde{g}\, (1.4 \, {\rm TeV})$ & gluino\_03 & 0.014  & 50000 (144) \\
$ p p \rightarrow \tilde{g}\tilde{g} \, (1.6 \, {\rm TeV})$ & gluino\_04 & 0.004 & 50000  (44) \\
$ p p \rightarrow \tilde{g}\tilde{g} \, (1.8 \, {\rm TeV})$ & gluino\_05 & 0.001 &50000 (14)  \\
$ p p \rightarrow \tilde{g}\tilde{g}\, (2 \, {\rm TeV})$ & gluino\_06 & 4.8$\cdot 10^{-4}$ & 50000 (5) \\
$ p p \rightarrow \tilde{g}\tilde{g}\, (2.2 \, {\rm TeV})$ &  gluino\_07 & 1.7$\cdot 10^{-4}$ & 50000 (2) \\
\hline
$ p p \rightarrow \tilde{t}_1\tilde{t}_1 \, (220 \, {\rm GeV})$, $m_{\tilde{\chi}^0_1} = 20\, {\rm GeV}$ & stop\_01 & 26.7 & 500000 (267494) \\
$ p p \rightarrow \tilde{t}_1\tilde{t}_1 \, (300 \, {\rm GeV})$, $m_{\tilde{\chi}^0_1} = 100\, {\rm GeV}$ & stop\_02 & 5.7 & 500000  (56977)\\
$ p p \rightarrow \tilde{t}_1\tilde{t}_1 \, (400 \, {\rm GeV})$, $m_{\tilde{\chi}^0_1} = 100\, {\rm GeV}$ & stop\_03 & 1.25 & 250000 (12483) \\
$ p p \rightarrow \tilde{t}_1\tilde{t}_1 \, (800 \, {\rm GeV})$, $m_{\tilde{\chi}^0_1} = 100\, {\rm GeV}$ & stop\_04 & 0.02 & 250000 (201) \\
\hline
$ p p \rightarrow Z' \, (2 \, {\rm TeV})$ & Zp\_01  & 0.38 & 50000 (3865)\\
$ p p \rightarrow Z' \, (2.5 \, {\rm TeV})$ & Zp\_02 & 0.12 & 50000 (1221) \\
$ p p \rightarrow Z' \, (3 \, {\rm TeV})$ & Zp\_03 & 0.044 & 50000  (443) \\
$ p p \rightarrow Z' \, (3.5 \, {\rm TeV})$ & Zp\_04 & 0.018 & 50000 (179)\\
$ p p \rightarrow Z' \, (4 \, {\rm TeV})$ & Zp\_05 & 0.008 & 50000 (81) \\
\hline
\end{tabular}
\caption{Generated signal processes (first column) with the corresponding identification (second column), the LO cross section $\sigma$ in pb (third column) and the total number of generated events $N_{\rm tot}$ (fourth column). In the last column, we also indicate the number of events corresponding to $10$ fb$^{-1}$ of data ($N_{10\,{\rm fb}^{-1}}$). }
\label{table:bsm_procs}
\end{table*}

\subsection{Description of the data format}
\label{subsec:format}

\begin{figure}
\centering
\includegraphics[width=.45\linewidth]{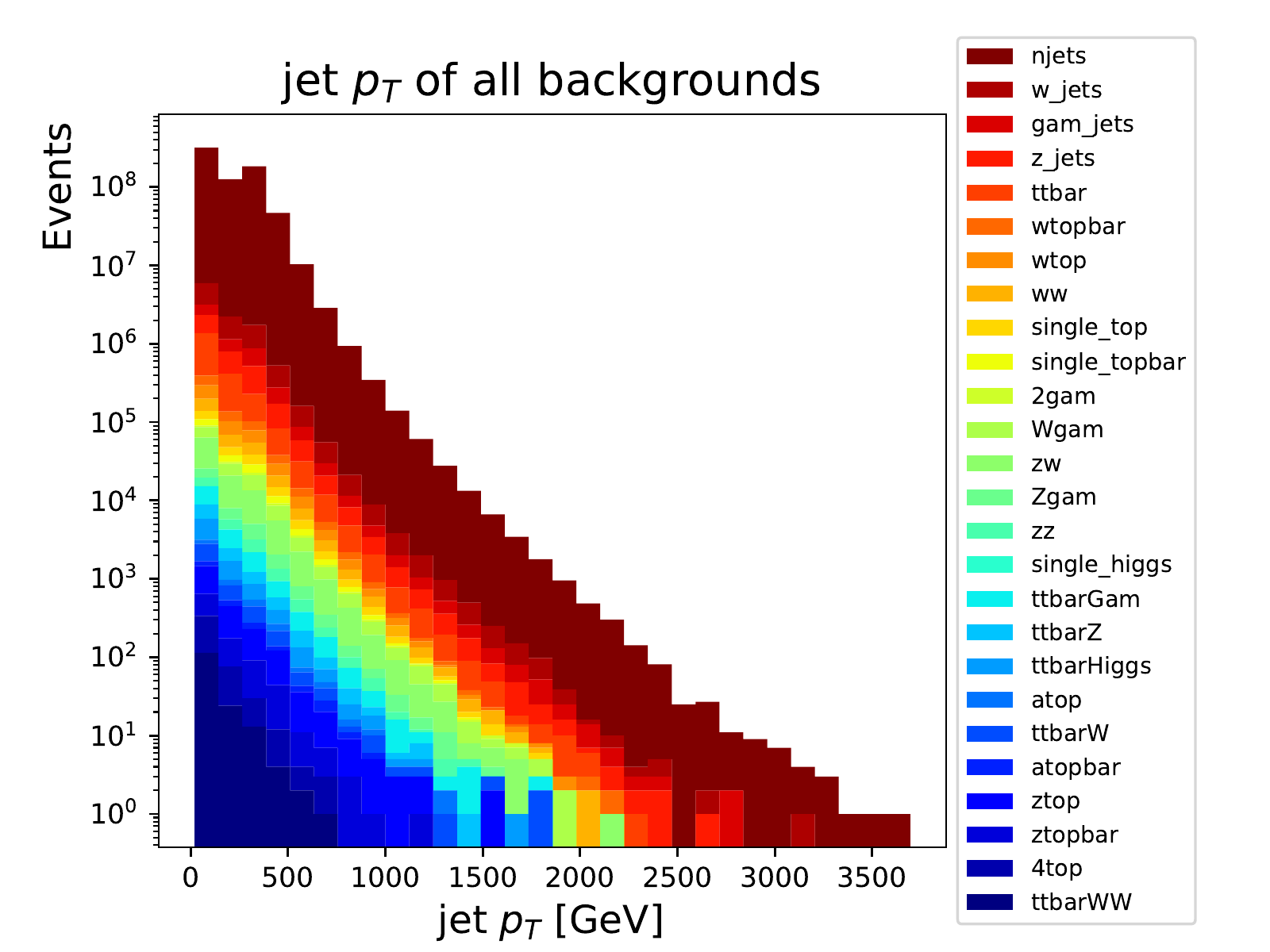}
\includegraphics[width=.45\linewidth]{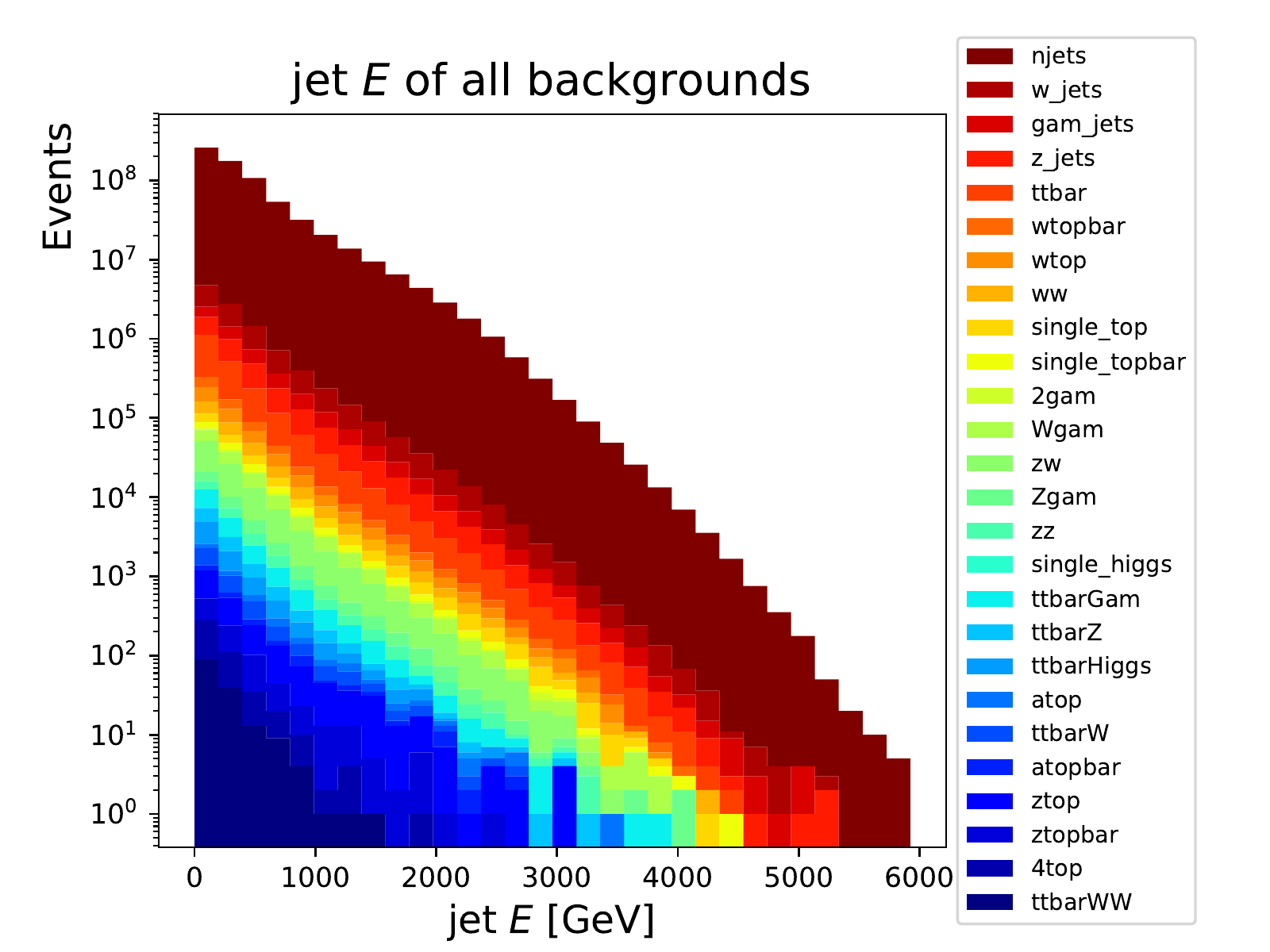}
    \caption{Transverse momentum $p_T$ (left) and energy $E$ (right) in GeV of the jets for all backgrounds.}
    \label{fig:jetpt}
\end{figure}

\begin{figure}
    \centering
    \includegraphics[width=.45\linewidth]{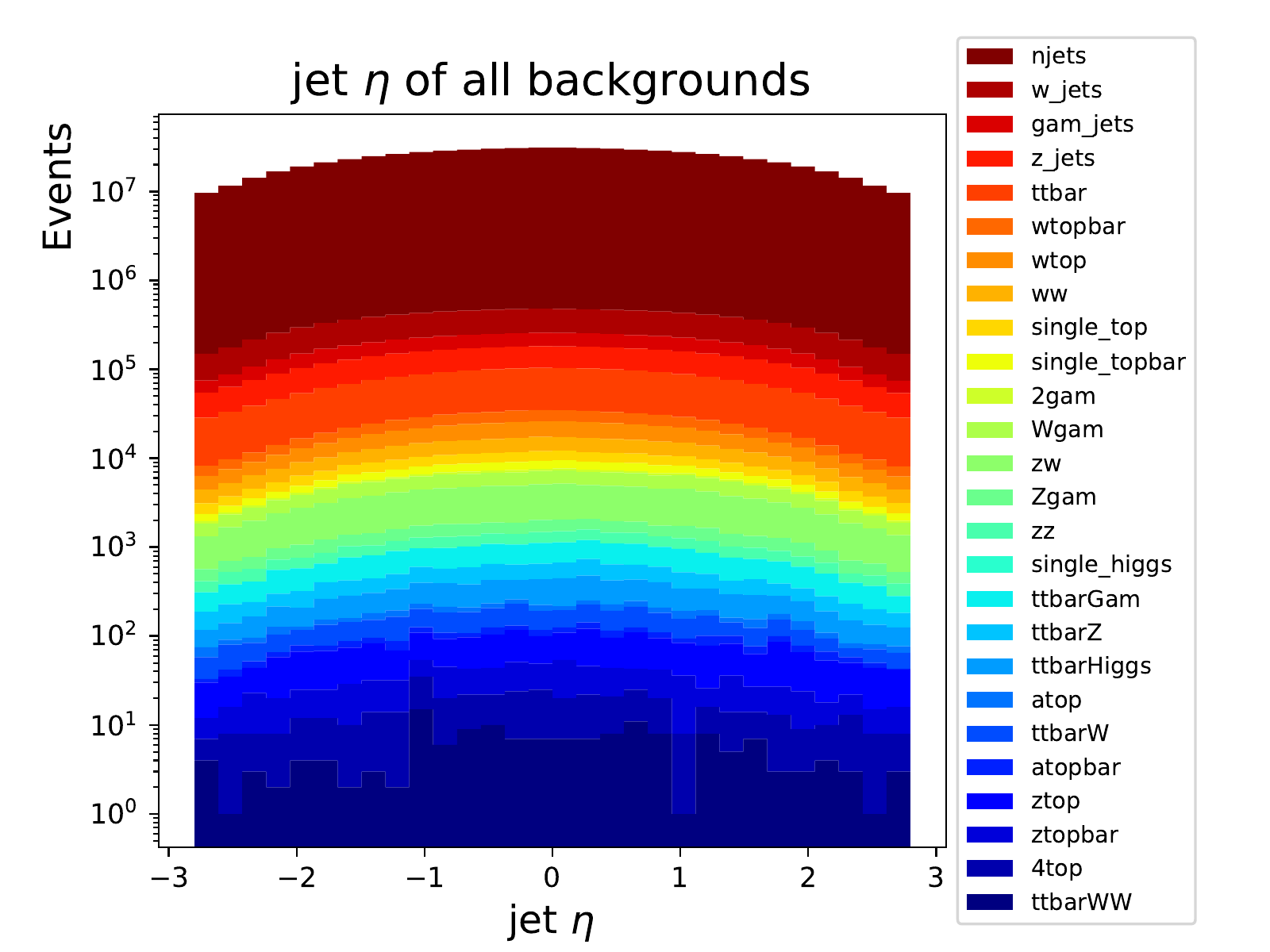}
\includegraphics[width=.45\linewidth]{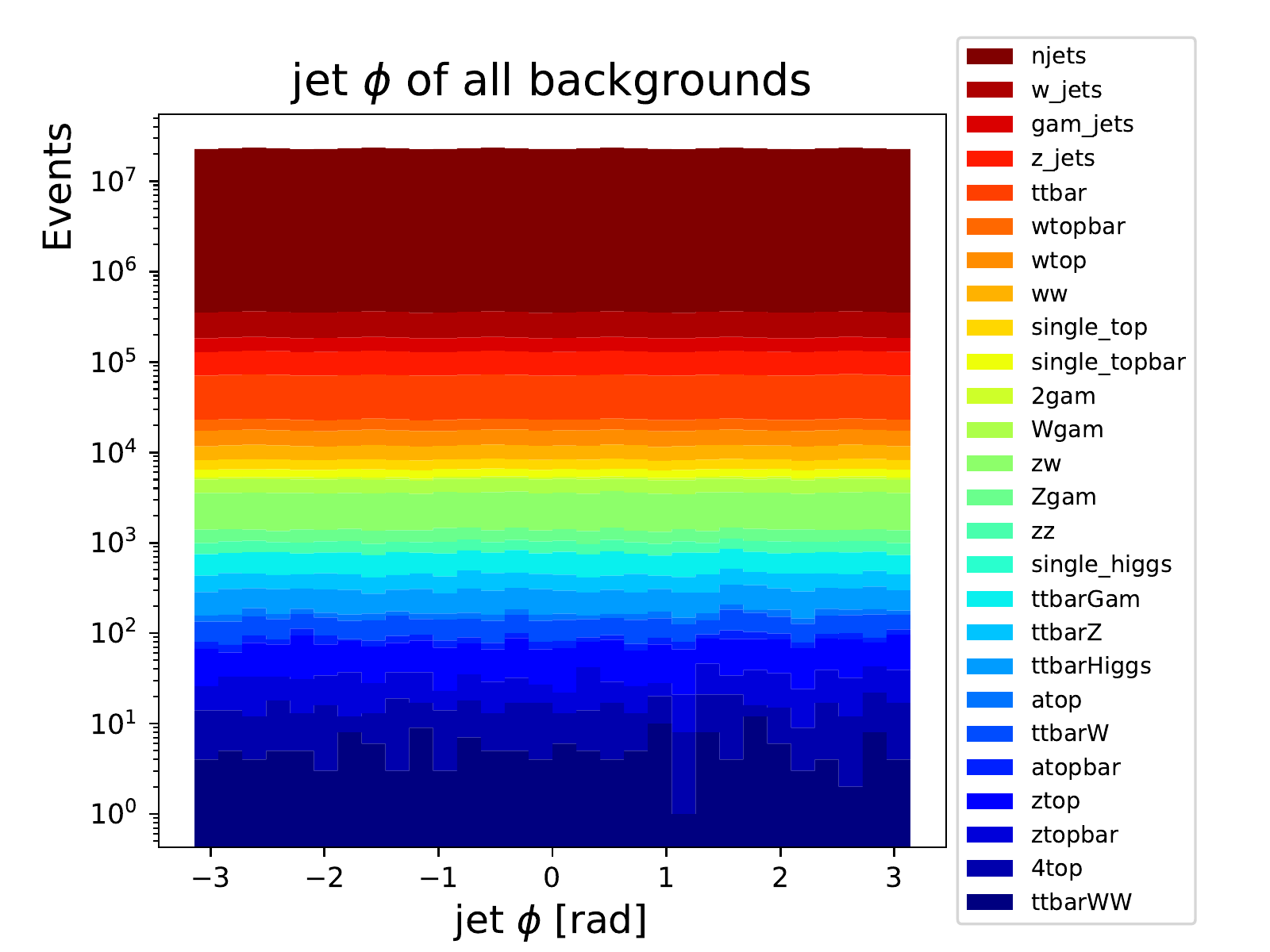}
    \caption{Pseudorapidity $\eta$ (left) and azimuthal angle $\phi$ (right) of the jets for all backgrounds.}
    \label{fig:jeteta}
\end{figure}

\begin{figure}
\centering
\includegraphics[width=.45\linewidth]{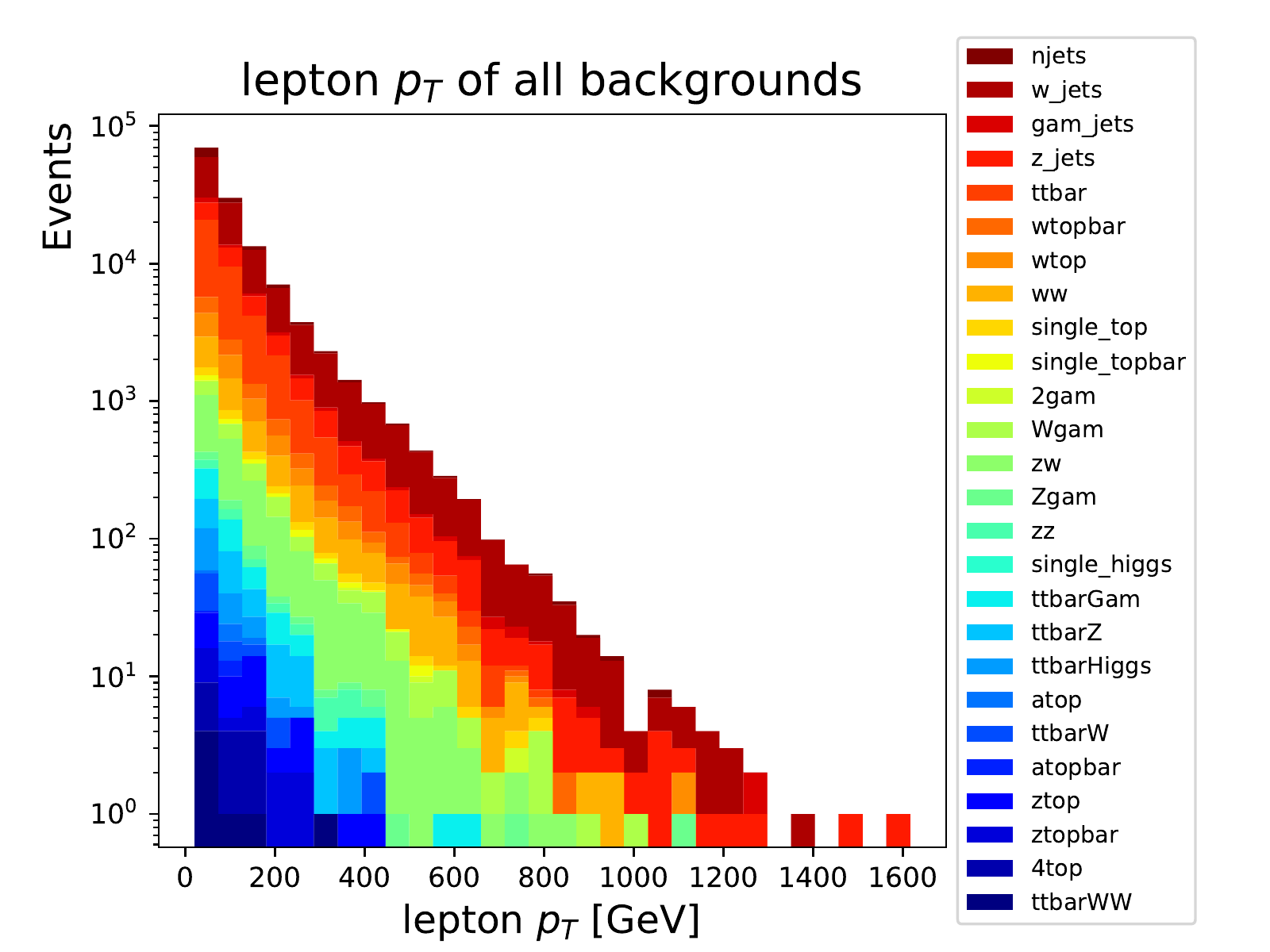}
\includegraphics[width=.45\linewidth]{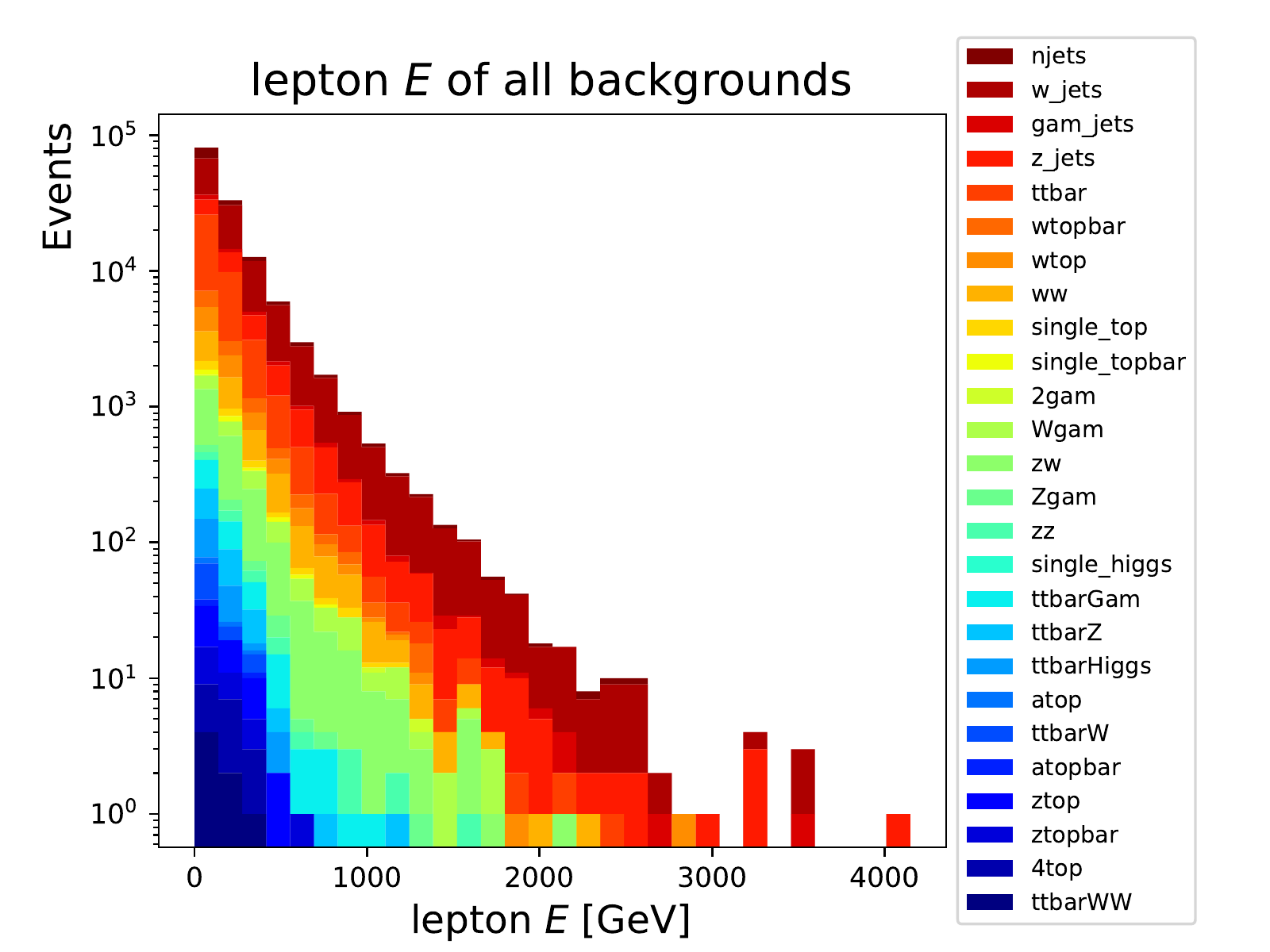}

    \caption{Transverse momentum $p_T$ (left) and energy $E$ (right) in GeV  of the leptons ($e^+$, $e^-$, $\mu^+$, $\mu^-$) for all backgrounds. }
    \label{fig:leptonpt}
\end{figure}

\begin{figure}
\centering
\includegraphics[width=.45\linewidth]{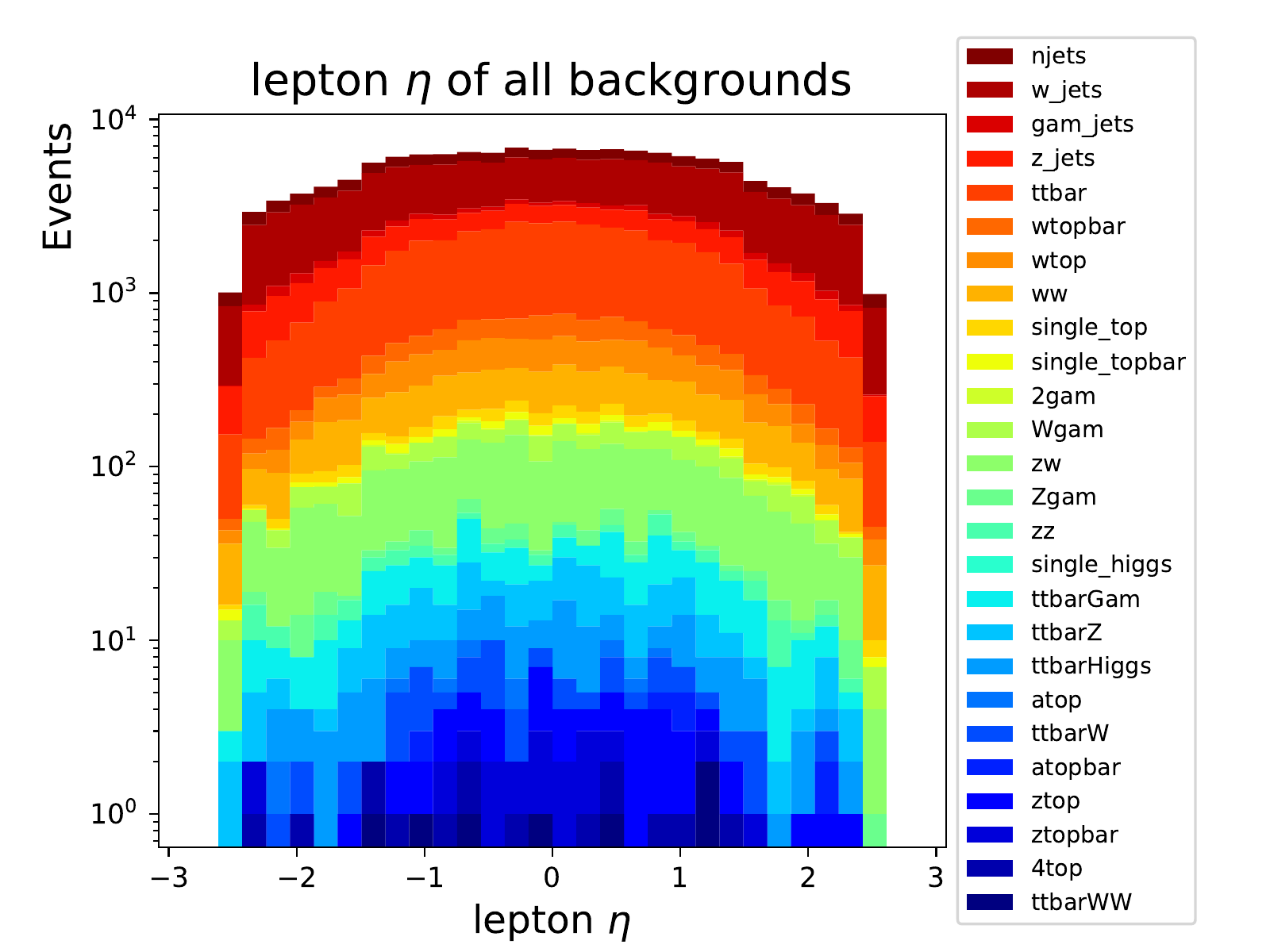}
\includegraphics[width=.45\linewidth]{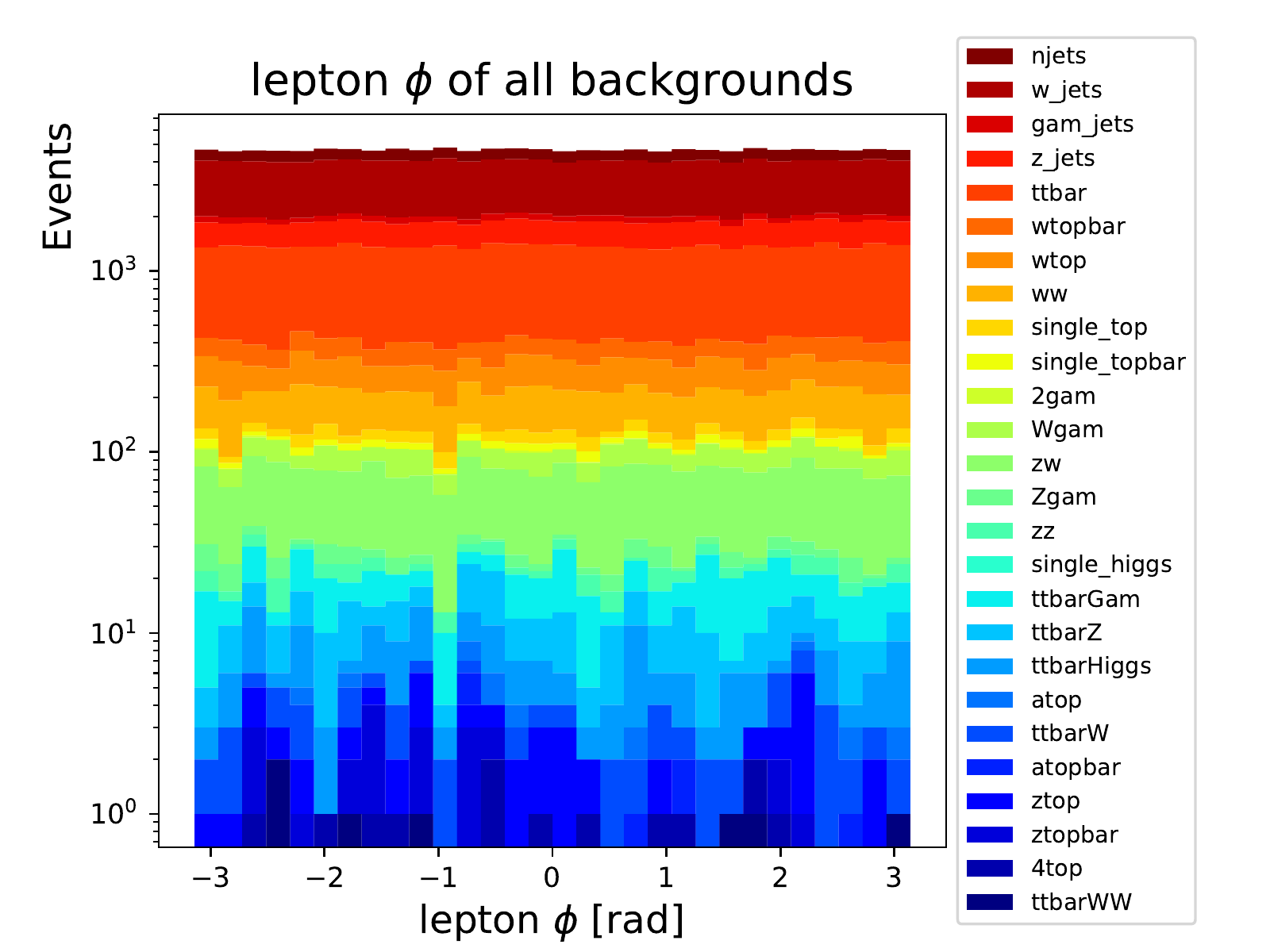}
    \caption{Pseudorapidity $\eta$ (left) and azimuthal angle $\phi$ (right) of the leptons ($e^+$, $e^-$, $\mu^+$, $\mu^-$) for all backgrounds.}
    \label{fig:leptoneta}
\end{figure}

\begin{figure}
    \centering
    \includegraphics[width=.45\linewidth]{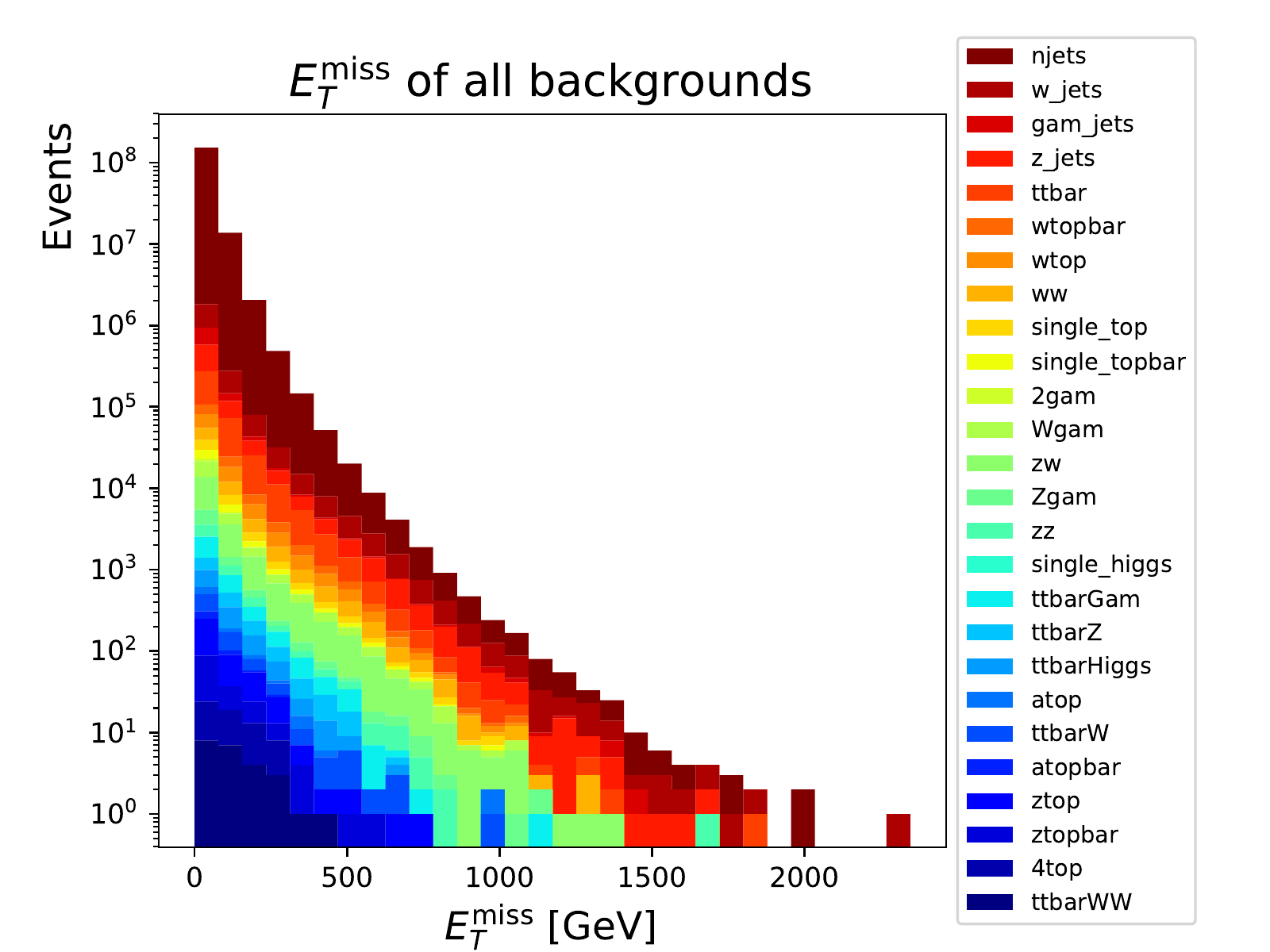}
    \includegraphics[width=.45\linewidth]{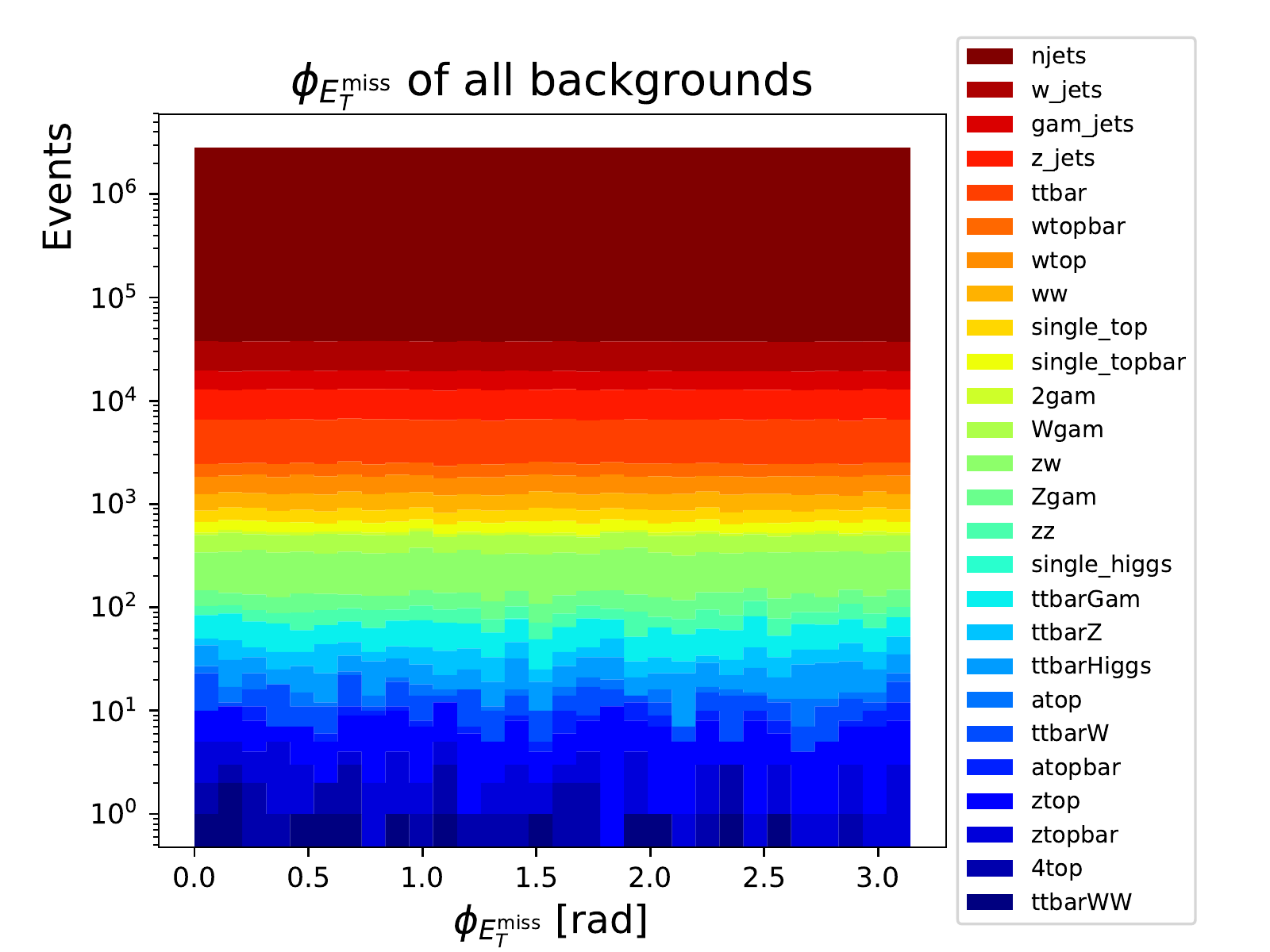}
    \caption{Missing transverse energy $E_T^{\rm miss}$ in GeV and azimuthal angle $\phi_{E_T^{\rm miss}}$ for all backgrounds. }
    \label{fig:metpt}
\end{figure}

\begin{figure}
\centering
\includegraphics[width=.45\linewidth]{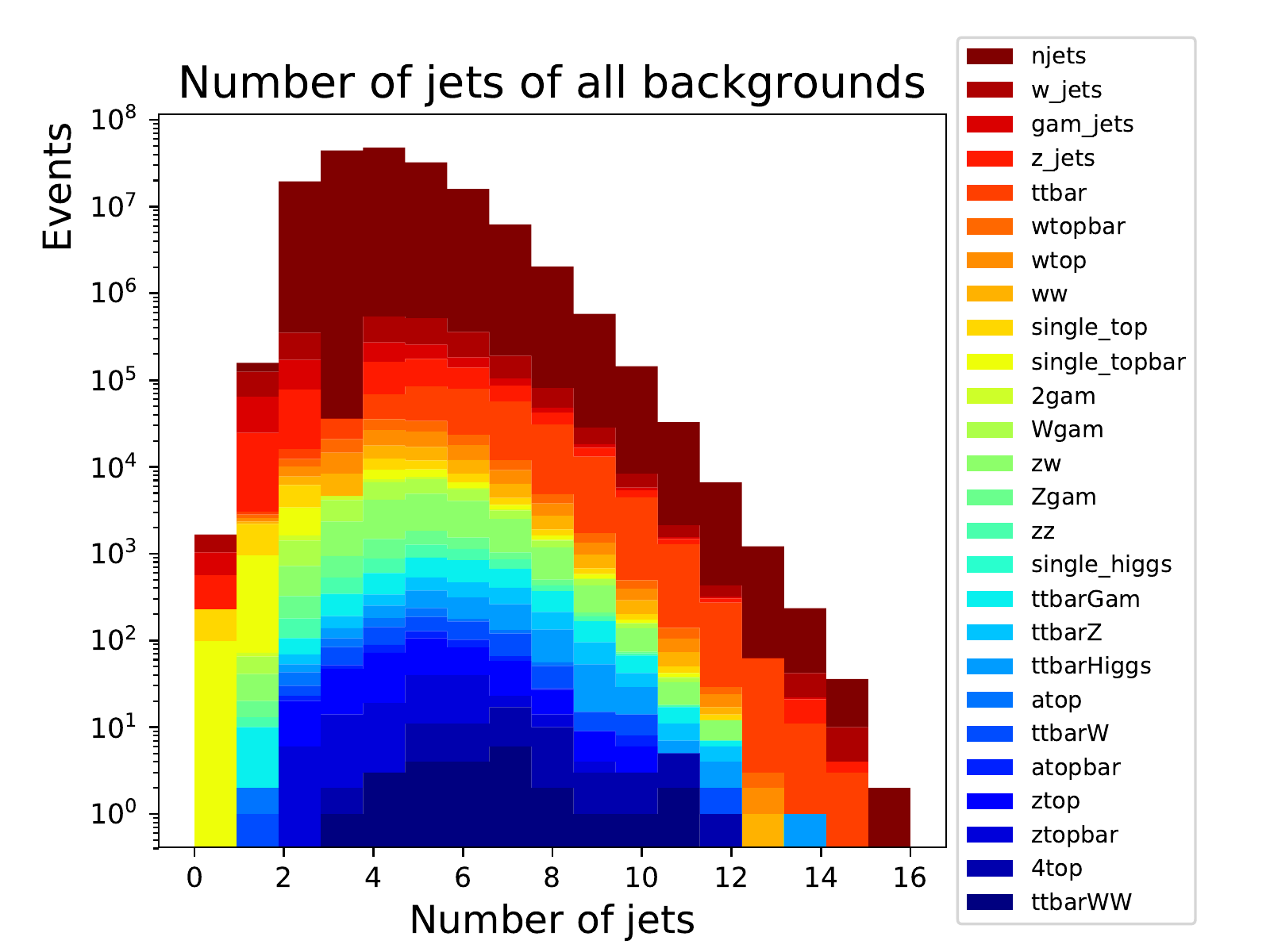}
\includegraphics[width=.45\linewidth]{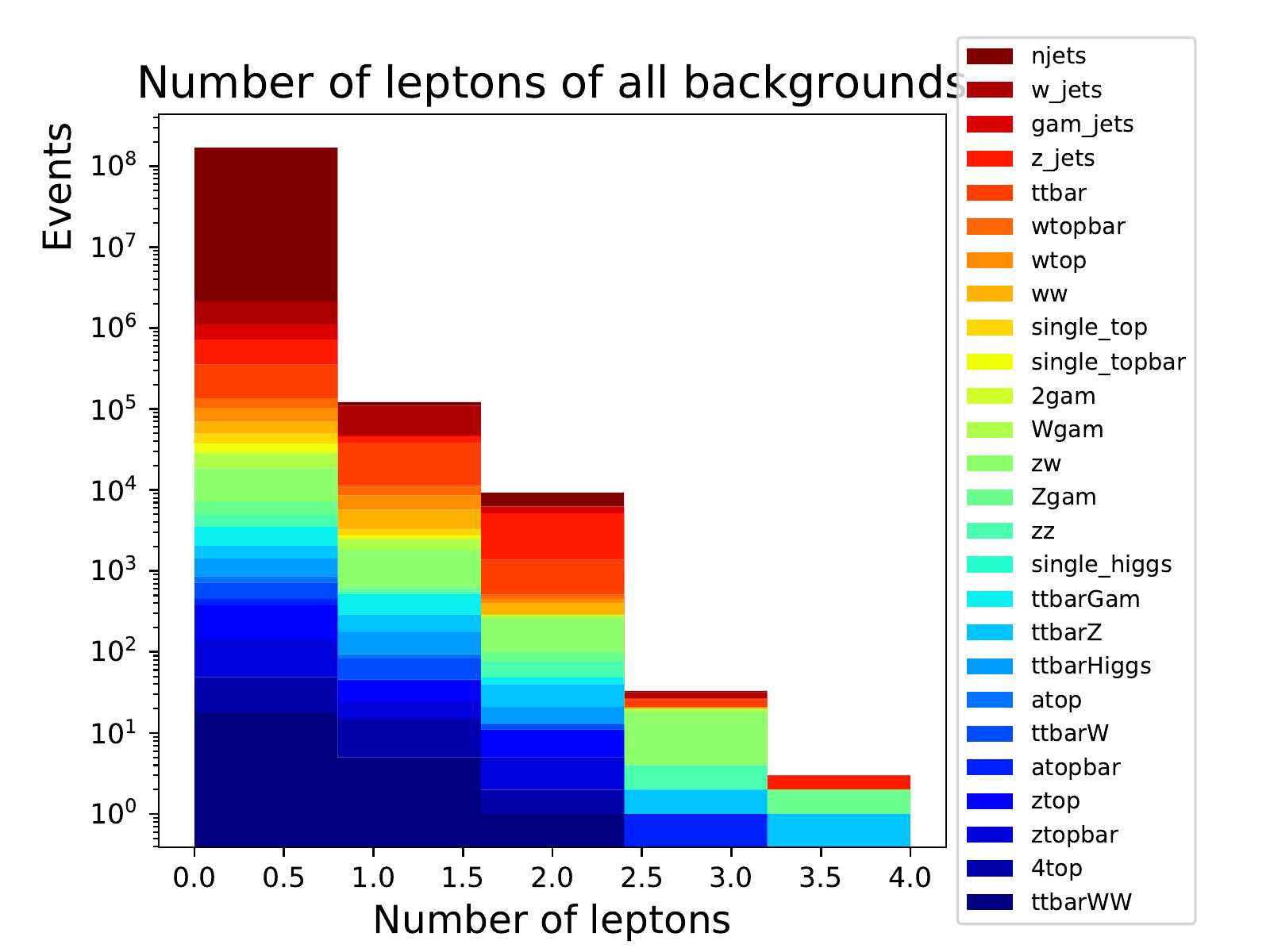}
    \caption{Number of jets (left) and leptons (right).}
    \label{fig:njets}
\end{figure}

\begin{figure}
\centering
\includegraphics[width=.6\linewidth]{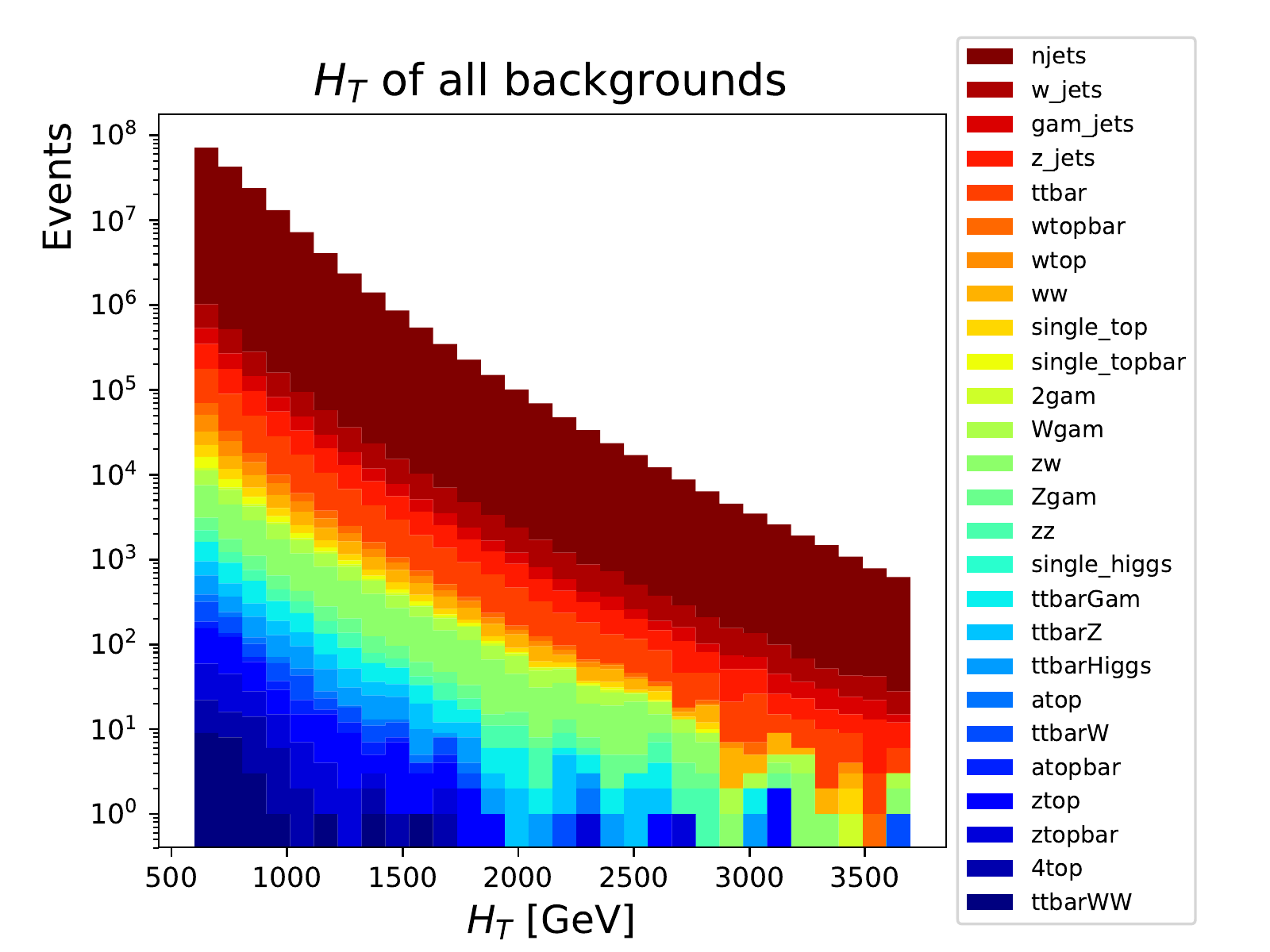}
    \caption{The scalar sum of the jet transverse momenta $H_T$ in GeV (see Eq.~\eqref{eq:defHT}) for the all backgrounds with $H_T > 600$~GeV imposed. }
    \label{fig:ht}
\end{figure}

The data are provided in a one-line-per-event text format (CSV file), where each line has variable length and contains 3 event-specifiers, followed by the kinematic features for each object in the event. The format of CSV files are:
\begin{center}
\texttt{event ID; process ID; event weight; MET; METphi; obj1, E1, pt1, eta1, phi1; obj2, E2, pt2,
eta2, phi2; 
} \ldots
\end{center}
The \texttt{event ID} is an event specifier. It is an integer to identify the generation of that particular event, included for debugging purposes only. The \texttt{process ID} is a string referring to the process that generated the event, as mentioned in Tables~\ref{table:procs} and \ref{table:bsm_procs}. The event weight $w$ is defined as
\begin{eqnarray}
    w = \frac{\sigma}{N_{\rm lines}}\times \left(10\,{\rm fb}^{-1}\right),
\end{eqnarray}
with $\sigma$ the cross section for a particular process, and $N_{\rm lines}$ the number of events in a single CSV file. With the release of this contribution we provide files for $N_{\rm lines} = N_{10\,{\rm fb}^{-1}}$ (with $N_{10\, {\rm fb}^{-1}}$ in Table~\ref{table:procs}), such that all weights are $1$. Additionally, when $N_{10\,{\rm fb}^{-1}} < 20000$, we provide a second CSV file with $N_{\rm lines} = 20000$. These conclude the event specifiers of each line in the CSV file. \\
Concerning the kinematic features, the \texttt{MET} and \texttt{METphi} entries are the magnitude $E_T^{\rm miss}$ and the azimuthal angle $\phi_{E_T^{\rm miss}}$ of the missing transverse energy vector of the event. The $E_T^{\rm miss}$ is based on the truth $E_T^{\rm miss}$, meaning the transverse energy of those objects that genuinely escape detection. The object identifiers (\texttt{obj1}, \texttt{obj2},\ldots) are strings identifying each object in the event, using the identifiers of Table~\ref{tab:objects}.
Each object identifier is followed by 4 comma-separated values fully specifying the 4-vector of the object: \texttt{E1}, \texttt{pt1}, \texttt{eta1}, \texttt{phi1}. The quantities \texttt{E1} and \texttt{pt1} respectively refer to the full energy $E$ and transverse momentum $p_T$ of \texttt{obj1} in units of MeV.
The quantities \texttt{eta1} and \texttt{phi1} refer to the pseudo-rapidity $\eta$ and azimuthal angle $\phi$ of \texttt{obj1}. \\
As an example, an event corresponding to the final state of the  $t\bar{t}+2j$ process with two $b$-jets (with $E = 33.2$~GeV and $E=55.8$~GeV) and one jet (with $E=100.4$~GeV) reads:
\begin{center}
94;ttbar;1;112288;1.74766;b,331927,147558,-1.44969,-1.76399;j,100406,85589,-0.568259,-1.17144;b,55808.8,54391.4,-0.198215,1.726
\end{center}
In Figures~\ref{fig:jetpt}-\ref{fig:leptoneta} we show the (stacked) distributions of the kinematic variables $E$, $p_T$, $\eta$, and $\phi$ of the jets and leptons in all of the generated background processes. In Figure~\ref{fig:njets} we show the number of jets $N_{\rm jet}$ and leptons ($N_{{\rm lepton}}$) for the generated backgrounds. The $E^{\rm miss}_T$ and $\phi_{E^{\rm miss}_T}$ distributions are shown in Figure~\ref{fig:metpt}, and the $H_T$ distribution is shown in Figure~\ref{fig:ht}. Note that only for Figure~\ref{fig:ht}, we have filtered out the events with $H_T < 600$~GeV. For the other Figures, we show the events for all values of $H_T$ for most backgrounds, except for the ones with tags njets ($H_T > 600$~GeV), w\_jets, gam\_jets and z\_jets ($H_T > 100$~GeV). We stress again that for any analysis, the same kinematic cuts on \emph{all} the background and signal events should be made. 

\subsection{Data storage}
The generated MC data is stored in the form of ROOT files (including all stable hadrons) and in CSV files including only the information as described above.
The CSV files corresponding to 10 fb$^{-1}$ of data per process are available in \url{ https://www.PhenoMLdata.org} for further validation. We encourage the community to explore the data, and report any issue to the authors of the proceedings. In the near future we plan to extend the dataset and to make the full set of ROOT files available, which currently take about 150 TB of disk space.  

\subsection{Benchmarking}
The dataset presented in this paper is the result of an effort started back in 2018 in one of the working groups of the DarkMachines initiative \footnote{\url{https://www.darkmachines.org/}}.
We plan to use this dataset as a benchmark dataset to open a challenge, addressed to both particle physics and computer science communities. 
The challenge will aim at stimulating these communities to design and employ new methods/algorithms for detecting and characterizing signals  in datasets featuring degenaracy, high dimensionality, and low signal-to-noise ratios, such as those faced when searching for new physics at the LHC.\\
\noindent Anomaly detection datasets used in ML are e.g.~credit card fraud detection data \cite{credit_card}. Other challenges similar in spirit have been previously ran, e.g.~in 2006 teams of theorists compared LHC data analysis approaches with mock datasets \footnote{see e.g. http://public-archive.web.cern.ch/en/Spotlight/SpotlightOlympics-en.html} and a QCD-oriented LHC Olympics 2020~\cite{shih_2019} \footnote{The LHC Olympics 2020 is a  low statistics dataset and a challenge to study anomalies in jets, i.e.~to build an ``anomaly jet detection algorithm'' with inputs being the kinematical features of stable hadrons. }
On the other hand, we provide  data set with very high statistics and 
including event-level features as well as full 4-vectors features, 
with several potential use cases. \\
\noindent The dataset in this paper (corresponding to an integrated luminosity of 10 fb$^{-1}$ of data) can be used by the interested readers for training and validation, using the BSM signal samples provided. For the challenge, we will provide a statistically independent dataset, where signal events are included. These signal events are generated e.g.~by one or more (undisclosed) types of BSM processes. The goal of the challenge will be to identify and characterize such signals. The submitted solutions will be judged and ranked according to specified metrics, based on the classification performance of the proposed algorithm with respect to the dataset with true labels assigned. More details will be provided when officially opening the challenge, 
and in a follow-up paper.

\section{Approaches to the problem}
\label{sec:approaches}

The task at hand is to distinguish background from signal events. Since signal events are very similar to background events in terms of their topology, it is usually impossible to identify them by looking at individual events. Therefore, one needs to take into account effects that only appear when examining distributions in a collection of events. 
Since the signal and background events can be viewed as samples drawn from an (unknown in the case of signal) multi-dimensional probability distribution, and we only have a finite amount of data, we are restricted to statistical investigations,
e.g.~in the form of a hypothesis test against the null hypothesis that
the given dataset does not contain any signal. In this section, we aim to give some examples for a signal detection algorithm. \\
In order to maximize the power of the test, it may be very helpful
to transform the low-level (raw) features of the events into high-level ones. This crucial step of feature selection/engineering can be performed by using unsupervised learning techniques~\cite{Albertsson:2018maf}, e.g.~by creating low-dimensional (latent) model of the data. \\
\noindent There are at least four different approaches to design the signal detection algorithm and train it on data.

\begin{enumerate}[label=(\alph*)]
\item 
    Training the algorithm on real data, possibly being a mixture of signal and background.
    This is necessary when a reliable or accurate model for the background is not available.
    It will then be tested on another independent sample of real data.
    \item 
    Training the algorithm on computer-generated backgrounds.
    It will then be tested on real data.
     \item 
    Training the algorithm by two-sample comparison of background data and real data
    (e.g.~\cite{DeSimone:2018efk, DAgnolo:2018cun})
    \item
 Training the algorithm on a specific signal and background. 
    This is what is typically done at LHC. Another possibility would be to train the algorithm on a large number of possible signals with a large variety.
\end{enumerate}
\noindent In all cases, the outcome can be reduced to constructing one or more   variables which maximize the power to discriminate signal from background (e.g.~ the probability of being an outlier, see Fig~\ref{fig:ande1}). \\
\noindent In the cases where the training involves a real dataset, it is possible to replace it by a mock dataset where signals of various kinds are injected. This is done to validate the algorithm and assess its performance to spot outliers. \\
\noindent In the approaches involving the background-only data, one should keep in mind that the simulated data are not a perfect description of the LHC data, and mismodeling may show up as fake new physics signals. \\
\noindent Several traditional ML techniques and various deep learning techniques enable the design of an algorithm that promise to serve our purpose of identifying new physics from LHC data: Kernel Density Estimation~\cite{scott2015multivariate}, Gaussian Mixture Models~\cite{mclachlan1988mixture},  Flow models~\cite{kingma2016improved}, Variational Autoencoders~\cite{Kingma:2013hel} and GANs~\cite{NIPS2014_5423}. In the subsequent subsections we have listed four overlapping approaches to the problem of finding new physics: anomaly detection, clustering, dimensional reduction and density estimation, all of whom could potentially be supported by the above-mentioned ML techniques.
 
\subsection{Anomaly detection}
Anomaly detection generally describes the process of identifying unexpected events in a dataset. With the aid of ML tools, this can be achieved in a supervised, semi-supervised or unsupervised manner. Since, in a model-independent scenario, we do not have labels for new physics signals,  we are in principle only interested in the unsupervised approach. Nevertheless, one could label the SM expectation of some observable. A special feature of this label is that it does not correspond to an individual event, but to a collection of events. This allows us to employ all the possible approaches mentioned above, while still being unsupervised with respect to the signal. A successful anomaly detection algorithm would then be able to tag the signal events as outliers. A potential problem of the approaches is that very rare SM events may also be part of the collection of outliers. \\

\begin{figure}
    \centering
    \includegraphics[width=0.8\textwidth]{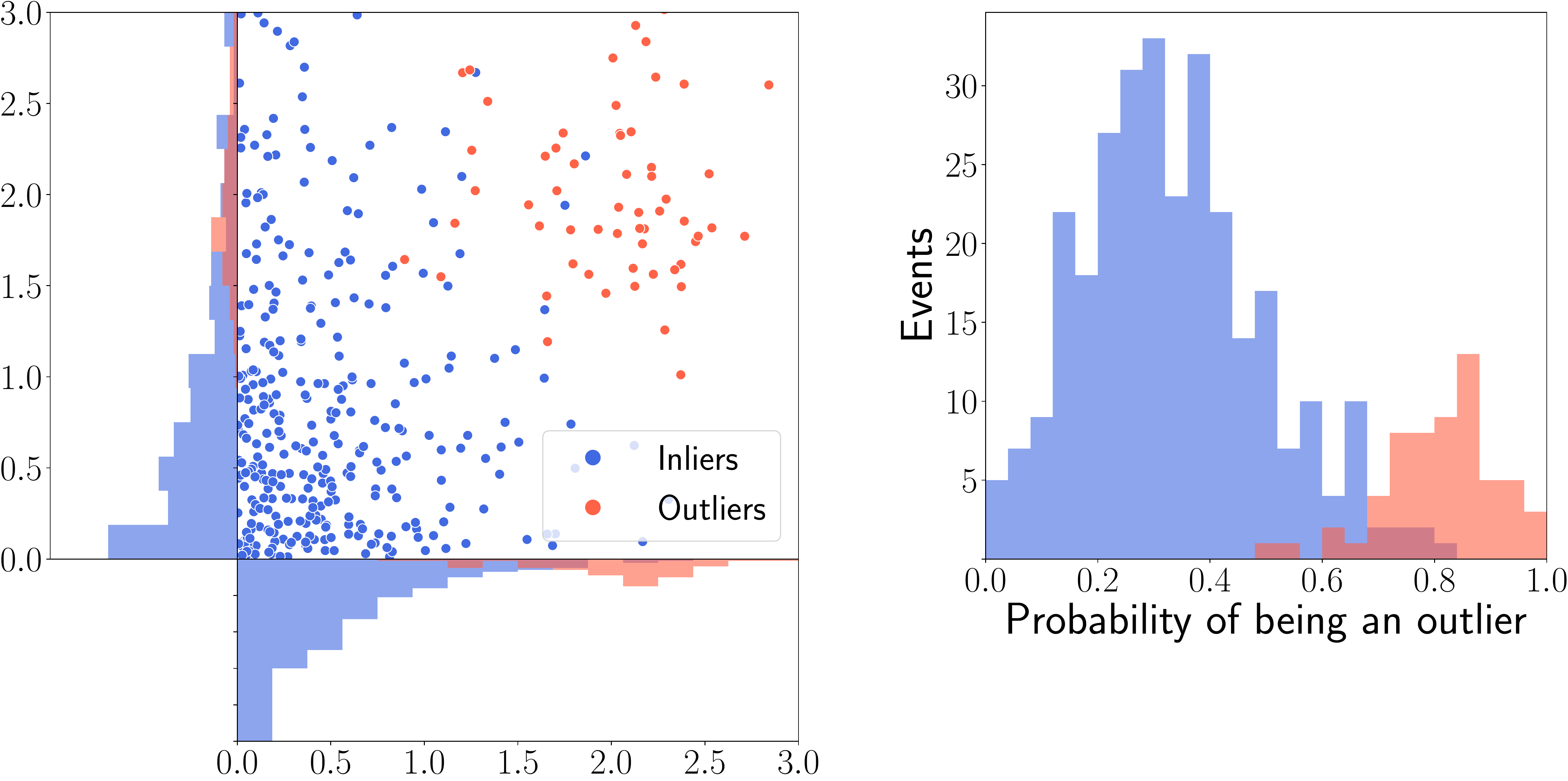}
    \caption{\emph{Left:} A narrow Gaussian anomaly centered around $(2,2)$ (in red) is added to an exponentially-distributed background (in blue). \emph{Right:} The probability of belonging to the signal events (outliers) is assigned to each point of the dataset and we can perform a counting. In this case, higher probabilities are correctly assigned to the outliers.}
    \label{fig:ande1}
\end{figure}

\noindent We present an instructive toy example in Figure~\ref{fig:ande1}. We simulated data from 
a background expectation distributed  exponentially and we combined it with a narrow Gaussian signal anomaly. In order to give an anomaly score to the points we trained the Local Outlier Factor (LOF)~\cite{breunig2000lof} on a background-only simulation, and subsequently used it on the dataset containing both inliers and outliers (this would correspond to approach (b) mentioned above).
Despite its simplicity, this example shows two interesting characteristics. First, it is clear that feature selection is important, since the variable on the $x$-axis is discriminating, while the variable on the $y$-axis  is less discriminating.
This is because the exponential distribution of the background has a different variance in the two directions. Second, the example has the characteristics that it is difficult to separate an anomaly from the background with a simple selection on one of the two plotted variables. 
The purpose of anomaly detection in this context is not to find \emph{all} anomalous points, but to be able to reliably state when a point (or a set of points) is anomalous and worth studying.
The LOF gives a score to all points in order to assess how much they differ from the background. On the right-hand side of Figure~\ref{fig:ande1}, we see that most outliers have a high probability of being part of the signal, and not belong to the background. 

\noindent Once all points are assigned an anomaly score, one may compare the distribution of such scores to a validation set containing only SM events.  Therefore, we use the framework of a two-sample test, aimed at detecting statistically significant differences in the score distributions of inliers and outliers.

\subsection{Clustering}
We expect data amenable for analysis to lack in class labels (e.g. it is not known if the data is a signal event); it will then be necessary to extract information in an unsupervised fashion. A solution is to invoke clustering techniques~\cite{Han11,Hastie09}, where the goal is to group the data into clusters, each cluster bearing certain unique properties. Specifically, the goal is to partition the data such that the average distance between objects in the same cluster (the average intra-distance) is significantly less than the distance between objects in different clusters (the average inter-distance). Several approaches have been developed to cluster data based on diverse criteria, such as the cluster representation (e.g.~flat, hierarchical), the criterion function to identify sensible clusters (e.g.~sum-of-squared errors, minimum variance), and the proximity measure that quantifies the degree of similarity  between data objects (e.g.~Euclidean distance, Manhattan norm, inner product). Our goal is to experiment with a variety of clustering approaches to gain a better understanding of the type of patterns emerging from clustering 
structures. \\
\noindent In order to analyze clusters to identify novel groupings that may point to new physics, one approach is to use what is known as \textit{cluster validation}~\cite{Theodoridis08}, where the idea is to assess the value of the output of a clustering algorithm by computing statistics over the clustering structure. Clusters with high degree of \textit{cohesiveness}, where events within the group are sampled from regions of high probability density, are particularly relevant for analysis. In addition, one could carry out a form of \textit{external cluster validation}~\cite{Dom02}, where the idea is to compare the output clusters to existing, known classes of particles. While finding clusters resembling existing classes may serve to confirm existing theories, clusters bearing no resemblance to known classes can potentially drive the search for new physics models.

\subsection{Dimensionality Reduction}
 Data stemming from LHC arrive in copious amounts, are highly dimensional, and lack class labels; clustering can be useful to find patterns hidden in the data, a task whose importance has been highlighted in the previous section.  Unfortunately, highly dimensional data create a plethora of complications during the data analysis process. Two possible solutions exist: we can either pre-process the data through dimensionality reduction techniques~\cite{Kohonen1982}, or we can make use of specialized approaches ~\cite{doi:10.1002/widm.1062}. \\
Dimensionality reduction can be done through feature selection, by determining which features are most relevant, i.e.~those that possess a high power to discriminate signal from background. This may come with some information loss, but it is commonly the case at the LHC that only a subset of information is needed to distinguish among different types of data. Another approach is to invoke principal component analysis:  the data is transformed while eliminating cross correlations among the new features; the resulting subset can be further analyzed to filter out irrelevant features. \\
Another promising direction is to use ML to attain a reduced representation of the data by performing non-linear transformations~\cite{Hinton06,Bengio13}. This approach can have a strong impact in the search for new physics since it implements data transformations that can unveil hidden patterns corresponding to new particle signals.

\subsection{Density estimation}
Events produced at the LHC (either real or simulated) can be thought as samples drawn from an unknown  probability density function (PDF) that characterizes the complex physical processes leading to the generation of the events themselves. The PDF of a new physics signal might be different from the PDF of the SM. However, also the estimated PDFs of the SM, and the one from real experimental data may be different.
Spotting and analyzing the differences in these two densities can provide a great deal of information about the underlying process (i.e.~the true physical model) that generates the signal events. \\
\noindent However, estimating the PDF reliably starting from the raw data is far from trivial, especially if the number of features is high. This constitutes an active field of research in data science and, depending on the specific task, different approaches may be suitable~\cite{scott2015multivariate, doi:10.1002/wics.1461}. One such approach is  kernel density estimation, which estimates the PDF by a sum of kernel functions (e.g. multivariate Gaussians) centered around each data point~\cite{Silverman86}. \\
\noindent Assuming density estimation can be performed accurately, there are several ways to use it for model independent unsupervised analysis. For instance, one can compare the PDFs of real and simulated data from the LHC to detect differences. They point towards interesting signal regions, which can be used in order to guide further scrutiny. 
Furthermore, one could also perform clustering and anomaly detection in a way independent from the approaches mentioned before, see e.g. ~\cite{doi:10.1002/widm.30, Nachman:2020lpy}. \\
\noindent One difficulty in applying density estimation on the dataset described in this work is the fact that the events change in dimensionality because the number of objects is not the same in every event.  Additionally, there are both continuous data (for example energy and angles) and categorical data (object symbol). To circumvent these issues, one might try to map events to a different parameter space, a potential methodology is described in  Ref.~\cite{Otten:2019hhl}. \\

\section{Conclusions}
In this paper we have described a dataset aimed at constituting a benchmark for future model-independent studies of new physics detection at the LHC. We described the details of the data generation and the data format, which allow the user to easily handle the data with any programming language. We encourage the community to acquire familiarity with this dataset, which 
will also form the basis for a signal detection challenge to be announced soon. The challenge will be addressed to both computer scientists and particle physicists, fostering fruitful collaborations between them. Furthermore, we outlined some approaches, inspired by machine learning, to the problem of signal identification in background-dominated situations, like the ones commonly faced in high-energy physics.\\
With a benchmark dataset such as the one described in this paper it is possible to test and compare different techniques and algorithms for signal detection. We believe the effort of designing and comparing new algorithms tailored to the needs of high-energy physics will prove very useful for the future of the field.

\let\darkred\undefined
\let\darkblue\undefined
\let\GeV\undefined
\let\be\undefined
\let\ee\undefined
\let\beq\undefined
\let\eeq\undefined
\let\ba\undefined
\let\ea\undefined
\let\tn\undefiened
\let\nn\undefined
\let\comment\undefined
\let\MB\undefined
\let\LH\undefined
\let\RR\undefined
\let\JS\undefined
\let\RV\undefined